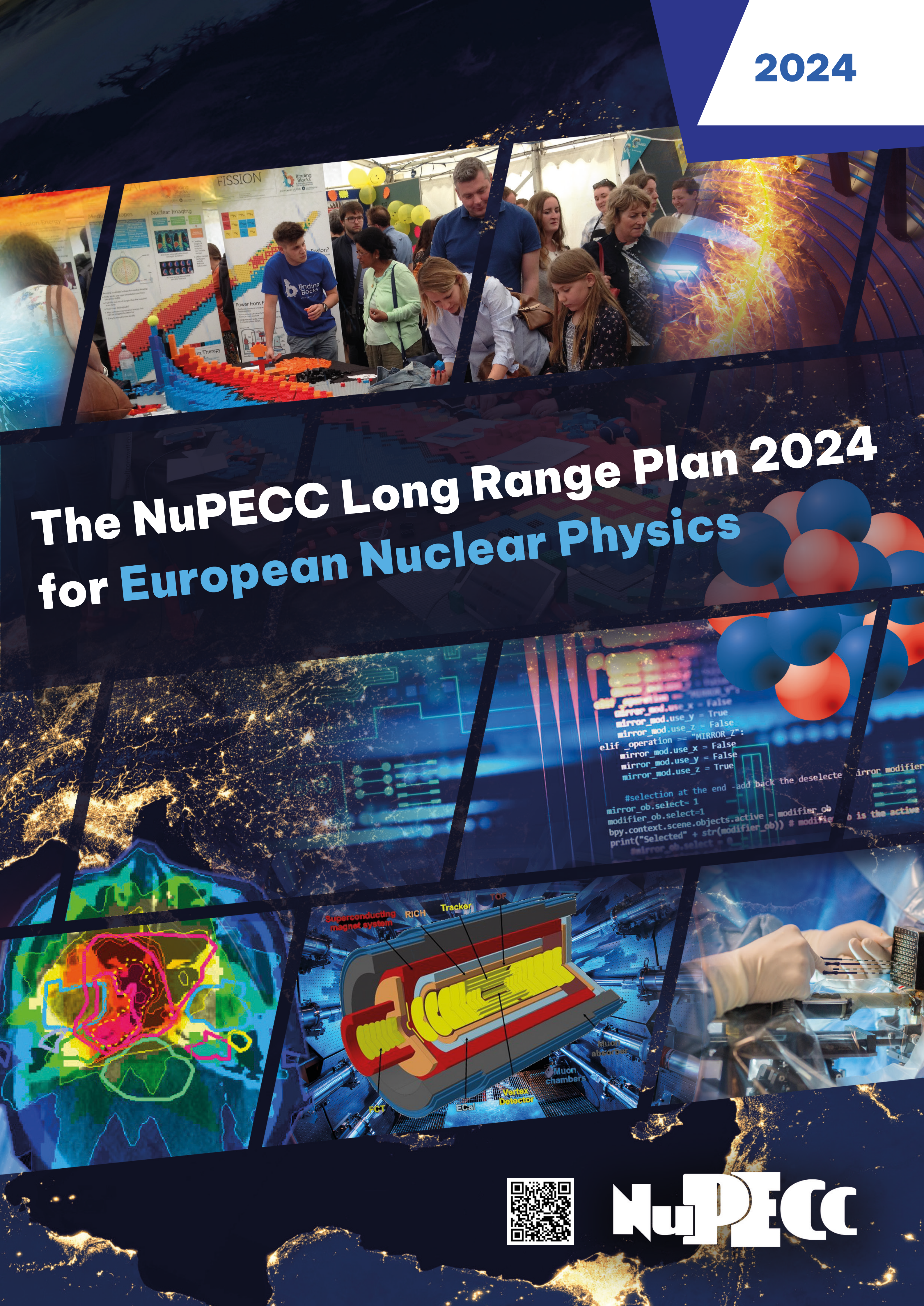

# Foreword

The Nuclear Physics European Collaboration Committee (NuPECC) http://nupecc.org/ hosted by the European Science Foundation represents today a large nuclear physics community from 23 countries, 3 ESFRI (European Strategy Forum for Research Infrastructures) nuclear physics infrastructures and ECT* (European Centre for Theoretical Studies in Nuclear Physics and Related Areas), as well as from 4 associated members and 10 observers.

As stated in the NuPECC Terms of Reference one of the major objectives of the Committee is: “on a regular basis, the Committee shall organise a consultation of the community leading to the definition and publication of a Long Range Plan (LRP) of European nuclear physics”. To this aim, NuPECC has in the past produced five LRPs: in November 1991, December 1997, April 2004, December 2010 and November 2017.

The LRP, being the unique document covering the whole nuclear physics landscape in Europe, identifies opportunities and priorities for nuclear science in Europe and provides national funding agencies, ESFRI and the European Commission with a framework for coordinated advances in nuclear science. It serves also as a reference document for the strategic plans for nuclear physics in the European countries.

NuPECC decided to launch the process of creating a new Long Range Plan for European nuclear physics in May 2022 aiming to publish the document in 2024 http://nupecc.org/?display=lrp2024/main. The whole process of elaboration of the LRP2024 was supervised by a Steering Committee composed of recognized experts in different sub-fields of nuclear science and of representatives of major nuclear physics facilities, including the chairs of the Astroparticle Physics European Consortium (APPEC) and the European Committee for Future Accelerators (ECFA). The Committee has also invited two observers from the Nuclear Science Advisory Committee (NSAC), USA and the Asian Nuclear Physics Association (ANPhA).

The bottom-up approach which has always played an essential role in the LRPs, was strengthened by the Steering Committee launching an open call in 2022 for contributions to the LRP 2024. The received 159 contributions, submitted by more than 400 individual scientists, collaborations, research infrastructures, and institutions in Europe composed a solid basis for the further analysis and elaboration of the LRP by 11 Thematic Working Groups (TWG). The TWG covered a large set of topics relevant to the development of nuclear physics. Compared to previous versions of the LRP, TWGs on detectors, tools, and topics of current interest were added.

Two intense and productive working meetings, hosted by GSI/FAIR in Darmstadt, Germany in October 2023 and in February 2024, were dedicated to the drafting of the LRP recommendations and finalisation of the LRP chapters, respectively. A draft of the full LRP2024 was presented and discussed with the nuclear physics community at a dedicated three-day Town Meeting organised by IFIN-HH/ELI-NP in Bucharest, Romania in April 2024. The Town Meeting, as well as numerous remarks received before and after it, allowed for improving and completing the initial LRP draft. Finally, the LRP 2024 document was approved by NuPECC at its meeting in June 2024 in Lund, Sweden.

The current study, "NuPECC Long Range Plan 2024 for European Nuclear Physics," is the outcome of this cooperative work.

The report’s Executive Summary, which contains NuPECC’s recommendations for the advancement of nuclear physics research in Europe, is followed by in-depth chapters on the scientific and organisational aspects of fundamental and applied nuclear physics research.

Europe has a leading position in nuclear physics research with a wealth of forefront infrastructures and a high-level workforce. It is through the collaborative effort of the European community that it can maintain such a position and advance it further. We strongly hope that this plan will persuade the European financial agencies to look for ways to achieve the goals specified in the suggestions, especially those that exceed a single nation’s capacity.

We would like to express our deep gratitude to the 29 members of the Steering Committee, 35 members of NuPECC, and more than 250 Thematic Working Group conveners and members for their invaluable contributions and strong implication in the whole two-year-long process of the elaboration of LRP2024.

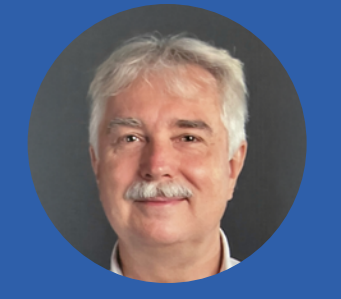

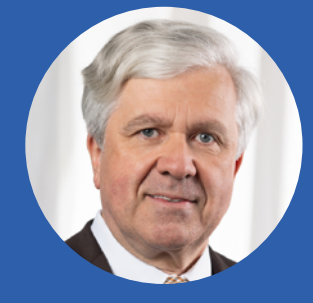

***Marek Lewitowicz***
*Chair of NuPECC*

***Eberhard Widmann***
*Deputy Chair of NuPECC*

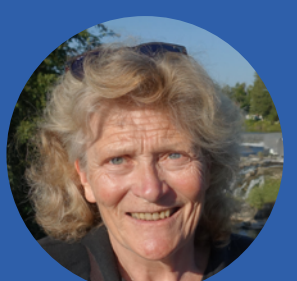

***Gabriele-Elisabeth Körner***
*NuPECC Scientific Secretary*

# NuPECC LRP2024
# Executive Summary

# Introduction

## What is modern nuclear physics?

Nuclear physics is the study of the atomic nucleus, its constituents, structure, reactions and the properties of strongly interacting matter in its various forms. It is a key basic scientific field that investigates the properties of matter at the subatomic level. This domain of research affects not only our fundamental understanding of nature but also has many peaceful applications in all areas of modern life. Research in nuclear physics originally started in Europe in the late 19th century. Now, in the 21st century, Europe is still at the forefront of nuclear physics research and applications. This leading European role is due to a rich and diverse landscape of research institutions and infrastructures in all European countries.

The present Long Range Plan for European nuclear physics summarises progress in the field over the last decade. It provides an outlook on expected developments in the next decade, and presents recommendations for scientific institutions, policymakers, and research funding organisations.

Modern nuclear physics is made up of a wide range of subfields, such as hadron physics, strongly interacting matter under extreme conditions, nuclear structure and reactions, nuclear astrophysics, fundamental interactions and symmetries, and nuclear science applications.

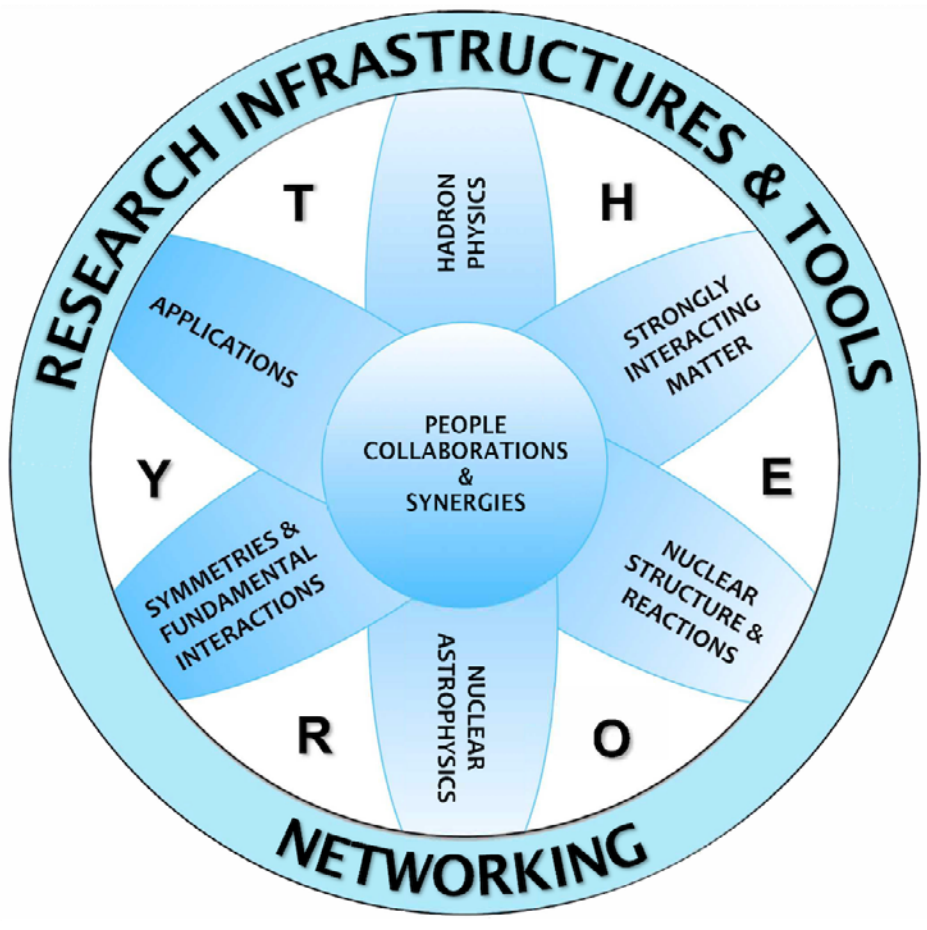

*Fig. 1.1: Nuclear physics research is divided into six subfields well-grounded in fundamental theory. To make progress in these subfields, advanced infrastructures, experimental and theoretical tools such as artificial intelligence and quantum computing and a highly educated workforce are required.*

The goal of nuclear physics and, more broadly, nuclear science is to comprehend the fundamental forces of nature that underlie phenomena involving the atomic nucleus and its constituents. To be more specific, the research programme is guided by the search for answers to the following scientific questions and challenges:

- How does the majority of the visible mass of the universe emerge from the almost massless quarks?
- What are the properties of the quark-gluon plasma, what is the qualitatively novel state of nuclear matter at extreme conditions of temperature and density?
- How do nuclei and nuclear matter emerge from the underlying fundamental interactions?
- What shapes can nuclei take, how do nuclear shells evolve, and what role do nuclear correlations play?
- What are the limits of the existence of nuclei, and what phenomena arise from open quantum systems?
- What are the mechanisms behind nuclear reactions and nuclear fission?
- How can we better understand the synthesis of heavy elements and the chemical evolution of the visible universe?
- What is the role of the strong interaction in stellar objects?
- What can nuclear physics teach us about the limits of the Standard Model of Particle Physics?
- How might nuclear physics strengthen its role in society's sustainable development?

## Nuclear Physics and Society

In 2015, the United Nations adopted the 2030 Agenda for Sustainable Development to secure peace and prosperity for people and the planet. The agenda comprises 17 Sustainable Development Goals (SDGs) which are intended as a call for action to all governments across the globe, but research communities can also make a significant contribution.

Nuclear physics and its applications in Europe play a major role in the domains of #7 energy, #3 health, and #9 space. The field of nuclear physics, with its educational, political and economic influence, can support numerous overarching objectives, including #4 high-quality education, #5 gender equality, #10 reduced disparities, #12 responsible consumption and production, and #13 climate action. Nuclear physics techniques such as isotopic markers to study plants and the water cycle have strong effects on #2 zero hunger, #6 clean water, #14 life below water, and #15 life on land. The responsible treatment of nuclear waste from medical and energy applications addresses #11 sustainable cities and #12 responsible consumption. The nuclear physics-based monitoring of non-proliferation aims to address #16 peace. Finally, the strong collaborative nature of nuclear physics in particular in Europe supports #17 partnership.

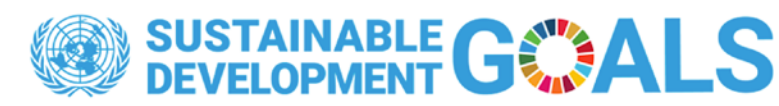

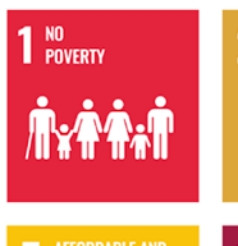

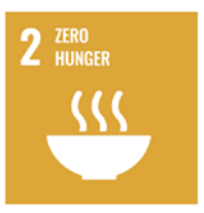

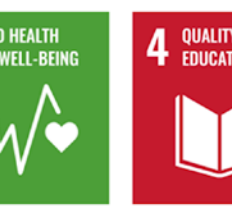

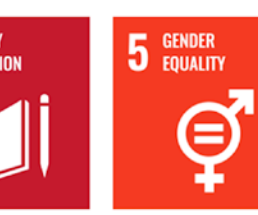

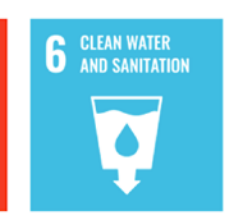

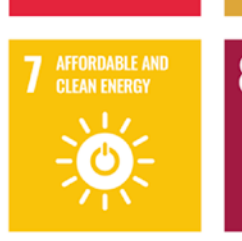

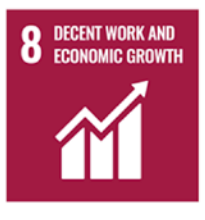

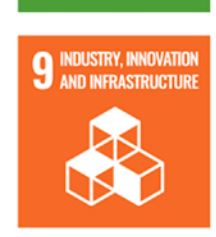

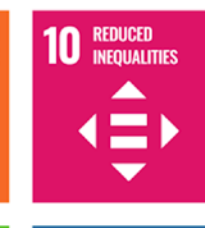

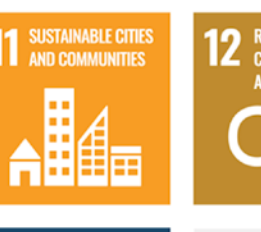

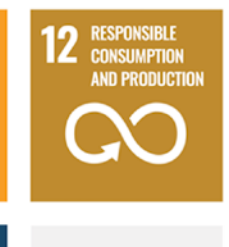

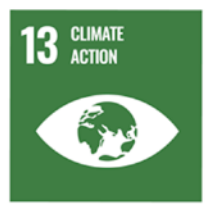

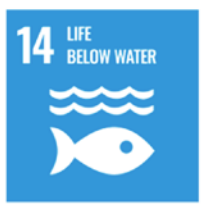

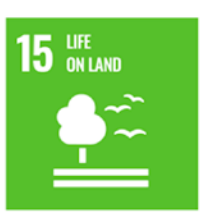

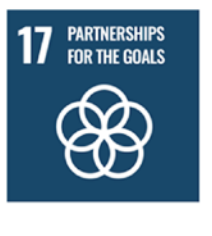

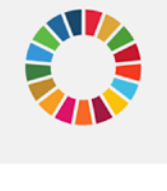

*Fig. 1.2: The nuclear science research community contributes to all of the 17 Sustainable Development Goals (SDGs) of the United Nations.*

Thus, nuclear science and technology have benefited human progress, culture and our understanding of our delicate environment in general, as well as health, economic growth, and security in nations all over the world. The overall impact of nuclear science on the environment and society is the subject of the NuPECC 2022 report "Nuclear Physics in Everyday Life".

# European landscape of nuclear physics

## People

According to a recent NuPECC survey, more than 5300 scientists work in the field of nuclear physics in Europe. Approximately 1200 of these work on nuclear physics theory and 4100 on experimental nuclear physics. There are approximately 1800 PhD students, 1000 post-docs, and over 2500 scientists with permanent positions.

It is of utmost importance to inspire and invest in the next generation of nuclear scientists, drawing on the talent and potential of the entire society. This is essential, not only within nuclear physics, but to ensure capacity-building across the wide range of disciplines and impact areas relying on the nuclear sciences. The investment of European funding bodies and institutions in nuclear physics is essential to inspire the public in nuclear science and its impacts, to educate and train the next generation of nuclear scientists and support equitable career progression with an inclusive approach to diversity across academic, industrial and vocational career paths. In order to further develop the pool of knowledge for future generations in nuclear science, the nuclear physics community explores these areas of knowledge, ensures their understanding and development, and also communicates them to – and develops them jointly with – the next generations, through outreach, education and training.

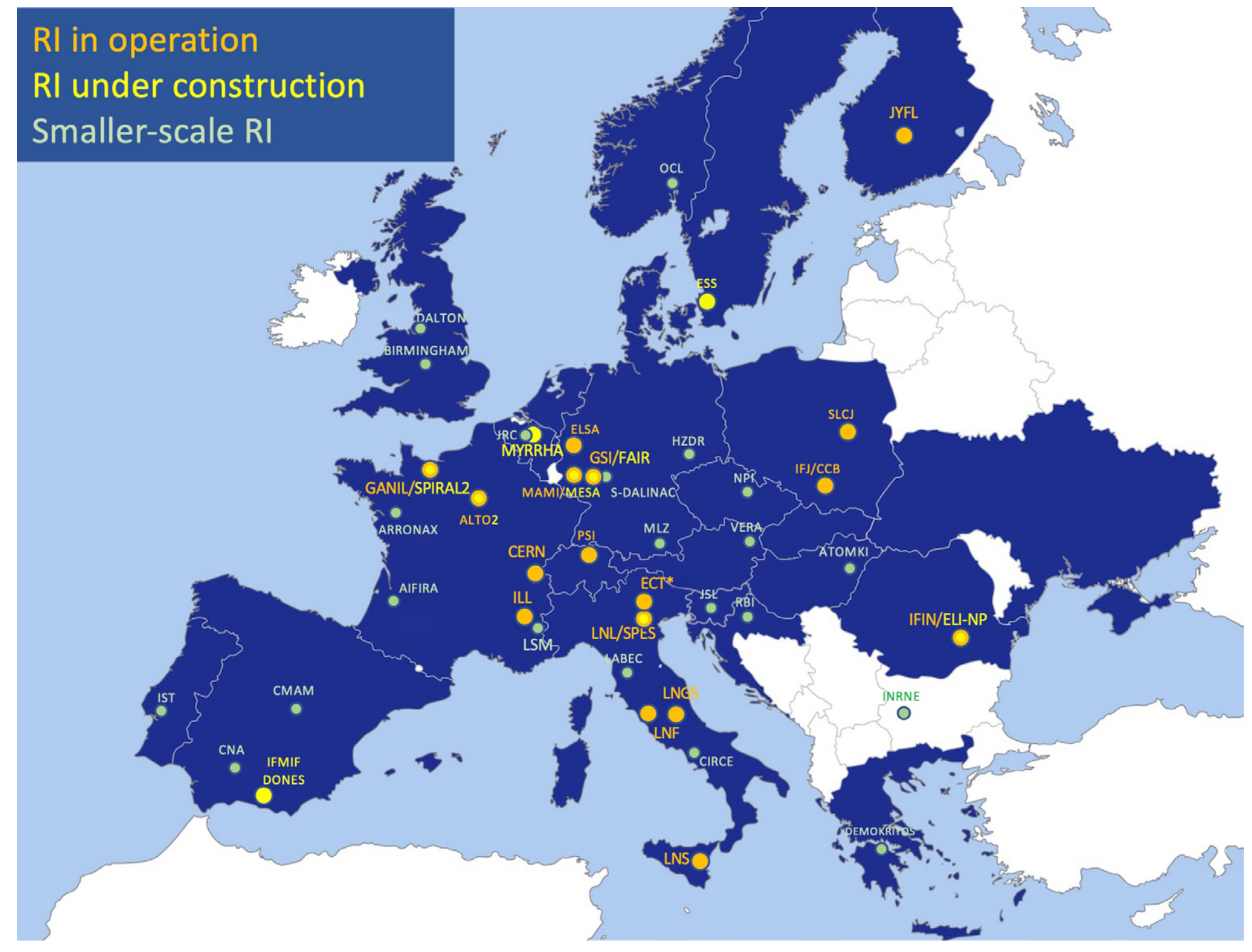

*Fig. 1.4: European landscape of nuclear physics infrastructures.*

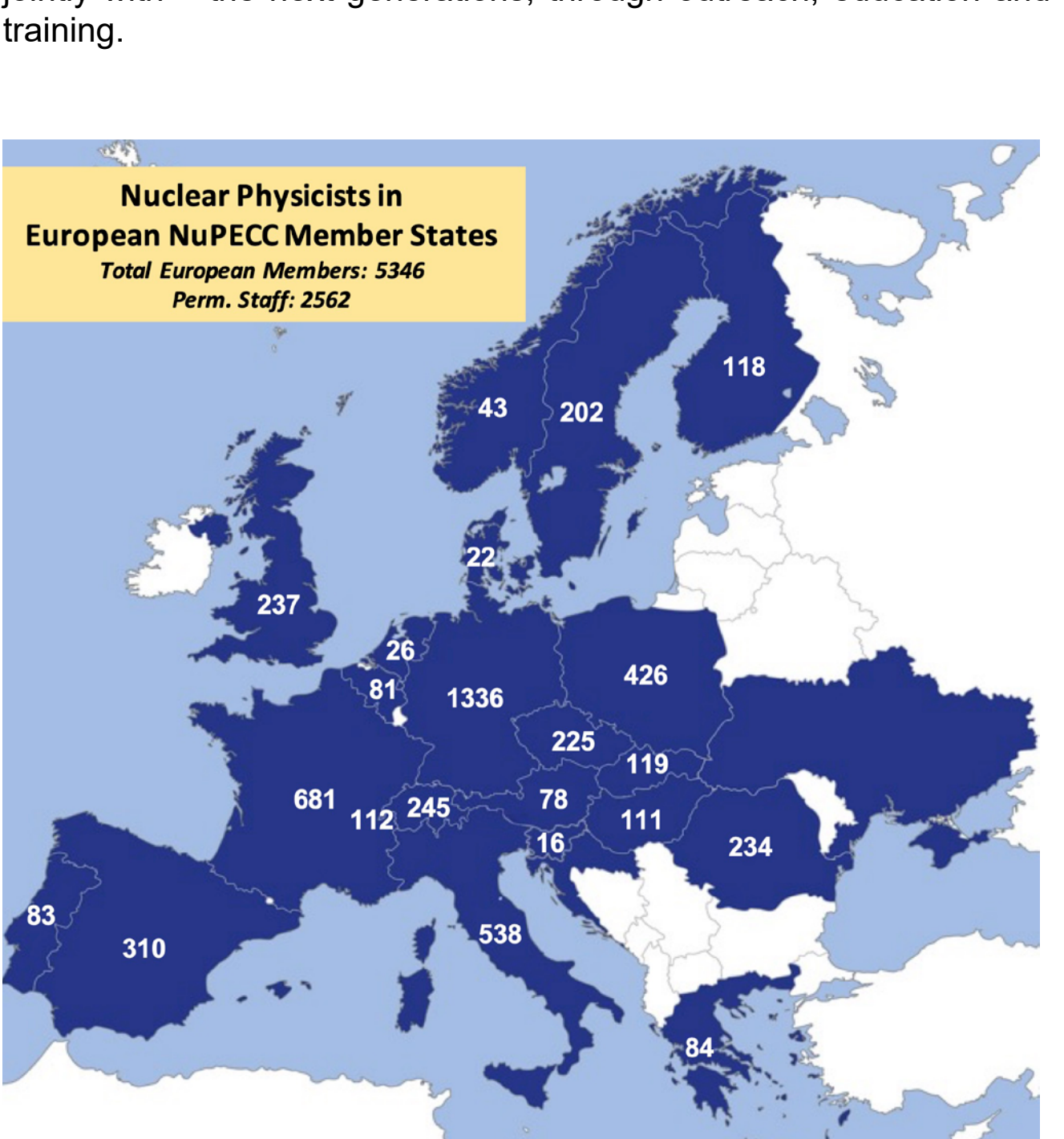

*Fig. 1.3: Nuclear physicists in the European NuPECC Member countries and the Associated Member CERN (source: NuPECC survey 2021 and 2023).*

## Infrastructures

Scientific questions and challenges are answered by the European nuclear science community by carrying out experimental and theoretical investigations within a rich ecosystem of research infrastructures and institutes of varying sizes, ranging from ESFRI-scale facilities to small university departments. Each area of this ecosystem has an important role to play in the delivery of nuclear science in Europe. The landscape of European Research Infrastructures is shown in the Fig. 1.3. The huge range of nuclear science phenomena and applications presented in this Long Range Plan requires, in turn, a very large range of facilities. These facilities offer both stable and radioactive ion beams, electromagnetic probes, neutron sources, elementary particles, computing infrastructure or other instrumentation essential to delivering nuclear science. Some of them offer low-energy beams with dedicated characteristics for analysis and applications. Others produce unique short-lived radioactive nuclei, explore the high-energy frontier of nuclear physics or provide high-intensity beams to explore rare nuclear processes. All these types of facility find their natural place in the European research landscape and are necessary. Often, access to these Research Infrastructures is facilitated by EU-funded programmes, such as STRONG2020, EURO-LABS, ChETEC-INFRA, OFFERR, RADIATE, RADNEXT and SANDA. Such programmes guarantee that all potential users gain access to state-of-the-art facilities, irrespective of their status.
All the sub-fields of nuclear physics exploit several large-scale and more than 15 smaller-scale infrastructures in Europe. Among them, large infrastructures from the European Strategy Forum for Research Infrastructures (ESFRI) Roadmap play a particularly important role in federating and shaping nuclear physics research and the scientific community.

# Recommendations for Nuclear Physics Infrastructures

The NuPECC Long Range Plan 2024 resulted in the following main recommendations for infrastructures of importance for nuclear physics:

● The first phase of the international **FAIR** facility is expected to be operational by 2028, facilitating experiments with SIS100 using the High-Energy Branch of the Super-FRS, the CBM cave and the current GSI facilities. Completing the full facility including the **APPA, CBM, NUSTAR** and **PANDA** programmes will provide European science with world-class opportunities for decades and is highly recommended.

● At **GANIL/SPIRAL2** the Super-Separator Spectrometer S³ is in an advanced stage of completion and the low-energy **DESIR** facility and heavy-ion injector **NEWGAIN** will be operational from 2027/28. The refurbishing of the cyclotrons will ensure their operation for the next decades. Timely completion and full exploitation of these GANIL/SPIRAL2 projects is recommended. The plan for the progression of the infrastructure towards a high-intensity reaccelerated RIB facility of up to 100 MeV/u should be actively pursued.

● Nuclear physics opportunities at **CERN** constitute a world-leading research programme. The construction of **ALICE 3** as part of the **HL-LHC** plans is strongly recommended. Continued support for exploitation and new developments is recommended to maximise the scientific output of **ISOLDE, n_TOF, SPS fixed-target programme** and **AD/ELENA**. As the roadmap for the post-LHC future of CERN is developed, a strategy should be prepared to secure future opportunities for continuing world-leading nuclear-physics programmes that are unique to CERN.

● At **ELI-NP** studies will focus on addressing key topics, such as laser-driven ion and electron acceleration. Implementing the gamma beam system to achieve the full completion of the facility and allow breakthrough results in the field of nuclear photonics is of high importance and is strongly recommended.

● Timely completion of the **SPES** facility and continuing coordinated efforts in developing the **ALTO, IGISOL, ISOLDE, SPES and SPIRAL ISOL facilities** in Europe will be key to maintaining their world-leading position in many areas of radioactive isotope science and are strongly recommended. Extending these efforts towards future facilities, such as **ISOL@MYRRHA, TATTOOS@PSI, and RIB@IFIN-HH**, together with the development of common instrumentation, will secure Europe's leading position for radioisotope production, separation and acceleration techniques, and create new avenues for the future. They should therefore be actively pursued.

● The exploitation of large-scale **stable beam** facilities such as **FAIR/GSI, GANIL/SPIRAL2, IFIN, JYFL-ACCLAB, LNL, LNS, NLC (SLCJ and IFJ-PAN)**, as well as smaller ones such as tandems, underground facilities and AMS systems, should be maximised. It is recommended that synergies between all these facilities, irrespective of size, be reinforced. Developments of novel and more intense beams and capabilities are also recommended to open new opportunities for basic science and applications.

● It is strongly recommended to complete the **AGATA** gamma-ray tracking array to its full configuration as a key instrument for studying atomic nuclei in both stable and radioactive ion beam facilities.

● Exploitation and optimisation of the European **lepton beam facilities**, including **ELSA, MAMI, and S-DALINAC**, are necessary in order to realise their full physics potential. The completion of the **MESA** facility and the **High-Intensity Muon Beams** project at **PSI** is recommended.

● Neutron facilities play a significant role in fundamental nuclear research and applications, producing unique and valuable experimental results. The new **NFS** facility, located at SPIRAL2, provides a highly intense neutron flux of fast neutrons, attracting a broad scientific community. It is crucial and strongly recommended to maintain the operation of exceptional neutron facilities like **ILL** and **n_TOF** at CERN. The **ESS** facility and the future infrastructure **IFMIF-DONES** will provide advanced tools for interdisciplinary research and their unique capabilities to serve advances in nuclear physics should be explored.

● Theory groups and centres should be strongly supported throughout Europe to ensure the fundamental contribution of theory to nuclear physics. An important role is played by the European Centre for Theoretical Studies (**ECT***, Trento, Italy), which is a unique centre dedicated to theoretical nuclear physics and related areas. Stronger pan-European support is needed to ensure that ECT* activities continue to play a strategic role in the development of nuclear physics in Europe.

● Collaboration with **non-European infrastructures** should be fostered in all areas of nuclear research to seize unique scientific opportunities and synergies complementing scientific programmes based in Europe. In particular, European participation in the construction of the **ePIC** experiment at the future international flagship facility **EIC** is recommended.

# International and Interdisciplinary Context

Nuclear Physics is embedded in the landscape of the European Research Area (ERA). Close collaboration with neighbouring fields of science, in particular with the particle and astroparticle physics communities in the framework of Joint ECFA-NuPECC-APPEC Activities (JENAA) has resulted in several new initiatives like JENA Seminars, specialised working groups and Expressions of Interest on various research topics of joint interest.
Research in nuclear physics is a worldwide effort and the European nuclear physics community is heavily involved in scientific programmes at overseas laboratories in North America (US and Canada), Asia (China, India, Japan and Korea) and South Africa. The exchange of information and strengthening of collaboration between international partners is facilitated through the associated membership of NuPECC in the research infrastructures CERN, RIKEN Nishina Centre of Japan, iThemba Labs in South Africa and in Israel, and by their status as permanent observers in the Committee of sister organisations ALAFNA in South and Latin America, ANPhA in Asia, CINP in Canada and NSAC in the US. In the international context, IUPAP plays an important role in the integration of the nuclear physics community across continents with its Commission 12 on nuclear physics and Working Group 9 on nuclear physics research infrastructures.

**Further fast development of nuclear physics in Europe will require investment in people, infrastructures, experimental techniques and theoretical approaches. Each sub-field of nuclear physics has its own specific priorities and requirements leading to the recommendations specified in the following parts of this Executive Summary.**

# Recommendations for Fundamental Nuclear Physics

## Hadron Physics

The goal of hadron physics is to understand the rich and complex features of the strong interaction. How does the major part of the visible mass of the universe emerge from the almost massless quarks? Can massless gluons form massive, exotic matter? What is the role of strong interactions in stellar objects, and in precision tests of the Standard Model? Answering these questions requires a diverse set of experimental and theoretical approaches. European hadron physicists play a leading role by conducting experiments at facilities within Europe, with great success, but also at the global level. These facilities, their planned upgrades and the approved flagships PANDA at FAIR, Germany and ePIC at EIC, USA, open new avenues

### Existing facilities

We recommend continuing support of the successful hadron physics programmes in Europe and the participation of European groups at global facilities. Particularly important hadron physics facilities are

- **AMBER** at CERN
- **ELSA** in Bonn, **HADES** at GSI, **MAMI and MESA** in Mainz, Germany
- **Jefferson Laboratory** in Newport News, USA

Furthermore, we recommend the support of ongoing hadron physics activities at the multi-purpose facilities Belle II, BESIII and those at the LHC.

### Future flagships

We recommend the expedited completion of the antiproton experiment PANDA and the support of European groups to contribute to the electron-ion experiment ePIC. By virtue of their different beam species and energy regimes, PANDA and ePIC will explore complementary aspects of physics. Over the next ten-years these two next-generation experiments must be made ready to launch.

- **PANDA**: The physics programme, including the prospect of unravelling exotic matter, remains unique and compelling. PANDA will strengthen the European position on the global scene and act as a unifying force for the community. We therefore recommend support for its construction and for the development of instruments, software and analysis tools.

- **ePIC**: Here, European researchers will be able to explore unknown features of quarks and gluons inside nucleons and nuclei. We recommend supporting the participation of European groups in ePIC and reinforcing scientific and technological activities which synergise with European projects.

### Theory / Computing

We recommend the support of theory groups at universities and research centres such as ECT* to prepare the community to benefit from European investments in supercomputing and quantum computing infrastructure. Theorists play an essential role in interpreting experimental results but also in providing input and predictions for new experiments. To match experimental progress, sophisticated approaches need to be developed. In lattice QCD, the rapid evolution of computational techniques and hardware calls for new algorithms and software. Similarly, quantum computing requires appropriate algorithms and tests on quantum hardware. Support for theoretical groups in terms of positions and career prospects is, thus, essential for progress in hadron physics.

## Strongly Interacting Matter at Extreme Conditions

Ultra-relativistic heavy ion collisions aim at producing and studying the quark-gluon plasma (QGP), which is the qualitatively novel state of nuclear matter at extreme conditions of temperature and density. Different collision energies achieve the QGP at different temperatures and densities.

The experimental focus is to discover in microscopic detail the material properties of the QGP at the highest temperature reached at the LHC, and to find the expected onset of the first-order phase transition at finite baryon density at FAIR.

Given the long timescales necessary for the R&D and construction of these experiments, a sustained research effort is required to advance the development of the next-generation experiments in parallel with the ongoing exploitation of existing facilities and detectors.

The priorities in this multi-pronged endeavour can be summarised as follows:

### Future flagship facilities and experiments

- **ALICE 3** at **CERN** is a completely new dedicated high-energy nuclear physics experiment based on innovative detector concepts essential for continuing the scientifically leading role of Europe in high-energy nuclear physics after 2035. The programme relies on innovative R&D that will benefit neighbouring fields of nuclear and particle physics. Strong support for R&D should be provided to maintain the opportunity of installing ALICE3 for Run 5 at the LHC.

- To investigate nuclear matter at high baryonic density, the timely completion of **SIS-100** at **FAIR** and the completion of the **CBM** experiment are of utmost importance. Efforts should continue to support R&D activities related to advanced **CBM** silicon vertexing and tracking devices.

- To exploit physics opportunities at the **CERN LHC** after 2035 (Run 5 and 6), the **LHCb Upgrade2** and the fixed-target setup will have a strong impact on the heavy ion programme. **ATLAS** and **CMS** will play an important role in the characterisation of high-momentum transfer processes up to the end of the LHC programme in Run 6.

- The **NA60+** detector at the **SPS** will address the remaining open questions in the electromagnetic and charm sector at the SPS with unprecedented event rates. R&D and construction for this detector deserve strong support.

### Support for existing facilities and experiments

- To maximise scientific output from the significant investment in current detector upgrades at the **LHC**, the continuation of the heavy-ion programme with Runs 3 and 4 (up to 2029) should receive full support. Timely support for the further **ALICE** upgrades in long shutdown 3 will provide a unique opportunity to enhance the physics reach in Run 4.

- With its Upgrade I detector and with the new particle-identification subdetectors to be installed during long shutdown 3, **LHCb** is equipped to pursue a unique fixed target programme at the LHC and to perform competitive measurements for Pb-Pb systems in collider mode. The exploitation of these opportunities should receive full support.

- The full exploitation of the existing detectors and facilities, in

particular **HADES** and **R3B** at **SIS-18/SIS-100**, should receive full support.

● The full exploitation of **NA61** at **SPS** should receive full support.

### Theory developments

● Theoretical work in the field of heavy-ion collisions should be guaranteed continuous support, both in its phenomenological aspects (theoretical support needed to interpret the results and to provide feedback to the experimental programme) and from first principles (quantum chromodynamics).

● Collaboration should be particularly encouraged and nurtured in theoretical centres such as ECT* to strengthen the relation between heavy-ion physics and neighbouring fields, including astrophysics and particle physics. Such collaboration would stimulate novel ways of computing and analysing data, as well as improving the interplay between theory and experiment.

## Nuclear Structure and Reaction Dynamics

The main challenges in Nuclear Structure and Reaction Dynamics in the next decade will be to answer the following questions: How do nuclei and nuclear matter emerge from the underlying fundamental interactions? What is the limit of nuclear existence and which phenomena arise from open quantum systems? How do nuclear shells evolve across the nuclear landscape; what kind of shapes can nuclei take, and what is the role of nuclear correlations? What are the mechanisms behind nuclear reactions and nuclear fission? How can we probe the equation of state with nuclear structure observables, such as resonances? How can nuclear structure and reaction dynamics contribute to astrophysics, hadron physics and fundamental symmetries?

### Support for existing facilities and experiments

● To ensure complementarity in experimental programmes, it is essential to actively support large- and small-scale facilities guaranteeing access to the whole community, allowing detector testing and exploratory experiments in preparation for more complex future experiments, and playing a key role in the training of new generations of physicists.

● Coordinated effort amongst the **ISOL facilities** in Europe has been key to securing a world-leading position in many areas of radioactive beam science. Reinforcing this collaboration on radioisotope production, separation and acceleration techniques, together with the exploitation of common instrumentation and a stream of new ideas, will secure the leading position of Europe in the future.

● To push the frontiers of spectroscopy and lifetime measurements at the limits of energy and production, exceptional resolution and high efficiency for gamma-ray spectroscopy is essential. Therefore, the full completion of the European flagship gamma spectrometer **AGATA-4π** (with ancillaries) is essential. AGATA is and will remain the major workhorse for nuclear structure gamma-spectroscopy and nuclear astrophysics precision physics, at both radioactive and stable ion-beam facilities.

### Future flagship facilities and experiments

● Unique insights into Nuclear Structure and Reaction Dynamics can only be obtained via the urgent completion of the **FAIR** facility (including the **NUSTAR** Low-Energy-Branch), **SPIRAL2, SPES, ELI-NP, ISOL@MYRRHA, and ISOLDE upgrades**, as unique laboratories for studying reactions of very exotic nuclei, and for the exploration of the nuclear chart towards the driplines.

● Europe's world leadership in the use of heavy ion storage rings - as key precision instruments for the study of nuclear masses and radii, nuclear resonances, isomers, reactions and fission - should be maintained by the construction of **future rings at FAIR and HIE-ISOLDE.**

### Theory developments

● It is mandatory to establish efficient interfaces between theories based on different degrees of freedom, to assess and reduce theoretical uncertainties, to improve the efficiency of many-body methods for a good description of spectroscopic observables, to improve time-dependent methods and reaction calculations, and to advance methods like Bayesian inference in combination with new computational techniques (e.g. Artificial Intelligence, Quantum Computing).

● Nuclear theory is crucial for interpreting experimental results and guiding future research. Excellence programmes to train, attract and keep talent within the field should be pursued. Theory centres should be strongly supported throughout Europe, in particular the **ECT*** and emerging virtual access facilities, which provide theory results for experimentalists (e.g. the Theo4Exp VA facility in the **EURO-LABS** project).

## Nuclear Astrophysics

Nuclear astrophysics is the study of nuclear processes in astrophysical objects such as stars, covering the wide range of physical scenarios found in space. Traditionally, nuclear processes have been the underlying scientific link between different observations of the same object. For example, our knowledge of the sun has only in the last few years seen tremendous progress in our understanding of the processes from its core to its atmosphere.

Gravitational wave telescopes have opened a new window to astrophysics. Now transient phenomena such as neutron star or black hole mergers, kilonovae, supernova explosions and gamma-ray bursts can be studied using multiple channels: from gravitational waves to electromagnetic radiation, neutrinos, or other particle emissions. These multi-messenger studies need a nuclear physics foundation. Here is a tremendous opportunity to study the synthesis of chemical elements, the evolution of their abundance over the life span of the universe, and, crucially, the properties of dense nuclear matter in the laboratory and in astrophysical scenarios. Experimentally, the understanding of neutron-capture nucleosynthesis requires frontier radioactive-beam facilities. Theoretical descriptions are the key for nuclear properties far from stability, for fission and for neutron capture. Nuclear astrophysics benefits significantly from progress in neighbouring fields. The next generation of gravitational wave telescopes, both space and ground-based (LISA and the Einstein Telescope) will strongly boost nuclear astrophysics research. The same is true for space-borne observatories like COSI, eXTP and e-ASTROGAM, and for new ground- and space-based telescopes.

● Nuclear astrophysics research is distributed over many institutions and thus benefits greatly from support for linking research activities and networks throughout Europe. We recommend strengthening nuclear astrophysics networks in Europe (e.g. **ChETEC-INFRA**) and making them sustainable. Such networks are needed to connect to international topical networks, e.g. in the U.S. and Asia.

● In order to facilitate nuclear astrophysics studies, nuclear data, and also nuclear reaction data, need to be evaluated with uncertainties and made accessible.

### Support for existing facilities and experiments

● Small-scale facilities are key infrastructures for nuclear astrophysics research. The activities connecting their work such as direct measurements to the lowest possible energies, as well as indirect measurements, should be further supported.

● European underground laboratories (**LNGS Bellotti Ion Beam Facility** and **Felsenkeller**, including its planned upgrade at DZA) play an essential role in nuclear astrophysics and should be supported.

● Dedicated facilities at large laboratories should be fully exploited, such as the **CRYRING** and **ESR** storage rings at FAIR, which open important new physics cases, and **n_TOF** at CERN.

### Future flagship facilities and experiments

● Radioactive beam facilities in Europe (in particular the **Super-FRS** at FAIR, including the Low-Energy-Branch, the upgrade of **ISOLDE**, and **SPIRAL2**) are needed to study exotic nuclei involved in explosive stellar events. We strongly recommend the completion of the ongoing construction and upgrades at these large-scale European facilities.

● A large (> 10 MV) **AMS** system is currently missing in Europe. It would provide the high abundance sensitivity required, e.g. for the most demanding astrophysical applications such as 60Fe.

### Theory developments

● To achieve a better understanding of the element synthesis and chemical evolution in the Cosmos, microscopic models for nuclear structure, decay and reactions as well as the equation of state of dense matter are needed,

● Access to large and fast supercomputing facilities in Europe is essential to performing microscopic nuclear physics calculations as well as multidimensional astrophysics simulations with improved numerical resolution, advanced neutrino transport methods, and due inclusion of rotation and magnetic fields.

● **ECT*** in Trento is an essential place for training and networking in nuclear astrophysics, in particular for theory.

## Symmetries and Fundamental Interactions

Fundamental interactions and symmetries can be studied by powerful low-energy probes. As such, precision measurements are complementary to collider searches for new physics. Pioneering techniques are under development to produce, manipulate, cool and trap a diverse range of particles, including radioactive nuclei, neutrons, antiprotons, pions, muons, exotic atoms, and highly charged ions. Also, the development of low background, ultra-high resolution spectroscopy techniques is essential. Low energy, precision experiments require dedicated efforts and often depend on access to extended beam times at our research facilities.

### Support for existing facilities and experiments

● The multidisciplinary research infrastructures **ILL, FRM-II** and **PSI** provide unique opportunities for fundamental physics at their cold and ultracold neutron beamlines. Their efforts for infrastructure upgrades should be supported. The long-term operation of ILL should be ensured beyond 2033.

● Continued support should be granted for the development and commissioning of facilities for the production, storage and trapping of heavy highly charged ions, such as **ESR, CRYRING** and **HITRAP** at **GSI/FAIR**, and high-energy **EBITs** at other laboratories.

● The **AD/ELENA** physics programme at CERN should be strongly supported over the long term, including running experiments, planned projects, and potential new proposals.

● In general, customised instrumentation and beam time availability should be guaranteed for fundamental tests at RIB facilities like **ISOLDE, GANIL-SPIRAL2,** and **JYFL-ACCLAB/IGISOL.**

● **Direct neutrino mass measurements** will soon be limited by systematic uncertainties. To overcome this limitation, **cross-disciplinary efforts** aimed at integrating available and novel technologies should be supported. Multiple and **complementary experimental searches** for neutrino-less double beta decay have to be encouraged as they can reach into the inverted hierarchy in the next decade.

### Future flagship facilities and experiments

● **Specialisation** of upcoming Radioactive Ion Beam facilities such as **ISOL@MYRRHA** and **DESIR at GANIL-SPIRAL2** should be regarded as an opportunity not to be missed.

● The **HIMB upgrade** at **PSI** will allow for improved measurements and new experiments in both fundamental and applied physics with muons. The project should be vigorously pursued and executed to allow for these exciting new possibilities with the first beam expected in 2028.

● At **ESS**, a fundamental neutron physics beamline should be installed as soon as possible.

● The future **CR** and **HESR** at **FAIR** would provide unique opportunities by extending the storage ring programmes with highly charged ions to high energies and, therefore, their completion should be vigorously pursued.

### Theory developments

● Initiatives to develop an inclusive theoretical framework fostering sustainable connections between nuclear theory, quantum chemistry, atomic and molecular physics and particle physics must be encouraged and vigorously supported. Effective Field Theories, combined with many-body microscopic approaches which rely on the degrees of freedom of the single nucleons, are an example of such theoretical frameworks that can offer a bridge between the very different scales involved in these problems, and also provide robust predictions of nuclear quantities.

They are essential for the emergence and application of new powerful probes, including radioactive molecules, thorium clock, or quantum-logic spectroscopy in highly charged ions. In parallel, efforts should be directed towards enhancing the precision of traditional probes, like the determination of $V_{ud}$ in nuclear beta decay, which is currently limited by theoretical uncertainties.

● To enhance the discovery potential of various experiments, a precise theoretical description of different nuclei is essential. For example, improved sensitivities of QED tests in highly charged ions and exotic systems require improved knowledge of nuclear polarisation. Another key challenge concerns rates of neutrinoless double beta decays, which directly relate to nuclear matrix elements.

Theoretical predictions obtained with nuclear structure models should be validated by employing the same models to reproduce experimental electroweak-interaction benchmarks.

● The strong interdisciplinary programme of the **ECT*** in Trento, which should be strongly supported, provides a unique meeting ground for members of the various fundamental-symmetry subcommunities.

# Recommendations for Applications and Societal Benefits

To ensure peace and prosperity for people and the planet, in 2015 the United Nations adopted the 2030 Agenda for Sustainable Development, which led to 17 Sustainable Development Goals (SDGs). These are intended as a call to action for all governments across the globe, but also a call for all research communities to contribute. Nuclear science must critically assess where it can contribute to them and engage fully in such opportunities. The nuclear science community contributes to all SDGs but more specifically, it directly addresses some of these goals (#3 Good health and well-being, #7 Affordable and clean energy, #9 Industry, innovation and infrastructure, #13 Climate action) or indirectly (#4 quality education, #5 gender equality, #10 reduced inequalities) through innovative and collaborative approaches. We highlight here how nuclear science in Europe can have the highest impact.

● For nuclear applications to have societal relevance for the foreseeable future, it is essential to continue work on **improving nuclear data**, both the measurement and the evaluation of nuclear data. These are needed to support research in the fields of energy, health, space, and materials science.

● The interdisciplinary nature of nuclear science applications needs a workforce with a diverse background. This requires **capacity building** in the fields of radiochemistry and radiobiology, as well as maintaining nuclear application competencies in Europe. For this purpose, it is essential to maintain and develop the landscape of smaller-scale facilities along with the large-scale facilities, crucial for nuclear applications.

● The wide array of nuclear application research fields requires **adapted beam access models** reflecting the dynamic developments in their fields. Adequate funding for the current smaller and larger facilities, as well as auxiliary platforms, is required for health, space, heritage science and materials research applications.

● New generations of nuclear energy sources and the management of nuclear waste through partition and transmutation depend on sustained technological developments in the present facilities, as well as on the **completion of MYRRHA and IFMIF-DONES.**

● The potential of novel medical radionuclides can only be achieved by upscaling the production capacity in Europe to clinically relevant activities. This could be achieved by the **enhanced use of the MEDICIS separator** at CERN, the **expansion of the PRISMAP project,** and the **completion of the medium and high-energy accelerators and radionuclide mass separators ISOLPHARM** at **SPES, ISOL@MYRRHA, IMPACT-TATTOOS** at **PSI,** and **SMILES** at **Subatech.**

● Europe's capacity in space dosimetry, radiobiology and radiation hardness testing requires the sustained effort of the present irradiation facilities as well as the completion of the first galactic cosmic ray simulator **in Europe at GSI/FAIR.**

● There is an increased need for **isotope-sensitive techniques** in environmental, heritage and material science for which a sufficient capacity should be ensured. This can be achieved through the sustained operation of research reactors for neutron activation analysis, improved sample preparation, novel technologies (e.g. ICP-MS-CRIS, MIXE) and the improvement of the present **AMS** facilities. The installation of a high-energy **AMS** in Europe (>10 MV) is recommended.

# Recommendations for Nuclear Physics Tools

Advancement in the understanding of fundamental physics is intimately related to progress in the development of tools for experimental and theoretical investigations. These tools are used for detector R&D, detector operation, data acquisition and analysis, theoretical interpretation of experimental results and genuine theoretical developments.

The tremendous progress in the field of nuclear physics has led to the pressing need for appropriate numerical tools aimed at addressing the most relevant experimental, theoretical and technological challenges, such as those encompassed by the Joint ECFA-NuPECC-APPEC (JENA) initiatives. To this end, the advent of algorithms based on Machine Learning (ML) and Artificial Intelligence (AI) techniques, and the fast progress in the field of Quantum Computing (QC) has opened an entire new world of possibilities.

## Detectors and experimental techniques

Efforts in the field of detector R&D in nuclear physics should be reinforced as follows:

● Development of a roadmap for detector R&D dedicated to the specific needs of low-energy nuclear physics and applications in radiation monitoring and heritage science must be supported.

● Strengthening of the collaborative effort in developing cutting-edge detector technology for identified applications in accelerator experiments, with respective activities in high-energy particle physics and other adjacent research fields.

● Building new expertise, training the next generation of detector R&D experts and strong collaboration with industry are vital for the continuous development and sustainability of the manifold tools in the European nuclear science community. This is also one of the major goals of projects within the Horizon Europe programme of the European Union. It must be pursued and supported with the highest priority as an integrated building block of all scientific and technological activities.

The following strategic actions are needed to ensure the continuation and progress of experimental nuclear physics in Europe:

● Cutting-edge developments in the overall read-out chain, from front-end electronics to the data acquisition systems, to the use of heterogeneous computing resources and advanced algorithms, also exploiting artificial intelligence techniques;

● Enhanced precision and efficiency in high-resolution laser spectroscopy and mass spectrometry to study the structure of rare isotopes and test fundamental symmetries. Improve existing ion-trap setups, ion manipulation and decay detection techniques for these experiments.

● Establish infrastructures to ensure the provision of stable and radioactive targets, such as a dedicated mass separator for providing radioactive samples and targets – forecast to be built at PSI.

● Europe needs to secure a strategic supply of stable enriched isotopes for fundamental research and applications, such as by the installation of a European Electro-Magnetic Ion Separation facility, providing material of the highest enrichment in rare stable isotopes;

● In some areas of experimental nuclear physics, the following specific actions are recommended:
- Successful development of polarised frozen spin targets and high-performance jet targets for use inside accelerator rings;
- The production of lead tungstate crystals should be continued;
- The scope of radioactive molecule production and research should be broadened in Europe to further this growing field of research.

The ever-growing demands of future nuclear physics experiments will require continuous and focused R&D activity with the following aims:

● To enhance the performance, radiation hardness and environmentally friendly features of the current detector technologies based on gaseous and solid sensing materials;

● To develop novel efficient neutron detectors to replace those based on $^{3}$He;

● To develop new materials for scintillators, p-type segmented Germanium detectors and new particle identification techniques.

## Numerical tools, techniques and resources

● **Develop, maintain and disseminate optimised algorithms and codes.** Promote freely- available software packages used by international collaborations. Encourage structured software collaborations. Develop interfaces between system software layers designed by IT experts and core software produced by nuclear physicists. Provide adequate funding for software development.

● **Provide long-term career perspectives for software developers in the field**. Acknowledge the growing need for software developers and IT experts. Strengthen this community by training early-career researchers. Enable institutional recognition through awards and career advancement.

● **Educate and train in software development.** Encourage collaborative software development through the use of code versioning, container technologies, CI/CD, AI/ML, workflow management, etc.

● **Invest in software frameworks**. Create multi-architetural software. Support parallelism and performance of algorithms on heterogeneous architectures (CPUs, GPUs, FPGAs, etc.).

● **Support the current effort to provide a solid basis for systematic comparisons of experimental results to theoretical predictions.** Test different approaches, introduce dedicated tools (e.g. RIVET) for the comparison and physical interpretation of heavy-ion collisions data, the determination of hadron structure, etc.

● **Call for more long-term storage solutions for gauge ensembles for lattice QCD.** Relevant and timely with the advent of exascale facilities.

● **Facilitate and strengthen access by nuclear physics researchers to large HPC centres.** Address computing needs for both theory and experiment. Allocate funding for enhanced GPU clusters (in particular for AI and ML) within established HPC centres across Europe.

● **Support virtual access infrastructures.** Disseminate theoretical results for the community at large. Capitalise on and further develop initiatives from the EURO-LABS or STRONG-2020 projects.

## Machine learning (ML) and artificial intelligence (AI) in nuclear physics

● **Transform ML prototypes into applications for production.** Advance from current proof-of-concept ML projects towards practical applications usable in nuclear physics projects.

● **Train scientific foundation models**. Invest in training and fine-tuning of models tailored for scientific purposes, such as GPT models specialised for nuclear science.

● **Develop research into explainable AI.** Enhance transparency and interpretability in scientific AI applications in nuclear physics and adjacent fields.

## Quantum computing (QC) in nuclear physics

● **Formulate a strategy for the quantum-classical interface**. Develop common open-source libraries for integration of classic and quantum software frameworks.

● **Facilitate access to quantum platforms.** Ensure access to state-of-the-art quantum platforms by bridging the gap between academic institutions and private companies, and forming agreements with national HPC centres.

● **Establish a European network on quantum activities related to nuclear physics.** Foster cooperation and knowledge exchange among researchers from different institutions and countries, allowing e.g. student exchanges and joint fellowships.

# Recommendations for Open Science and Data

Open science and Findable, Accessible, Interoperable, Reusable (FAIR) data offer an important opportunity for the nuclear physics community to uphold the highest research standards and enhance its societal impact, by treating the scientific production process as a strategic asset. The nuclear physics community should vigorously endorse and adopt open science practices, be actively involved in shaping the necessary policies, and lead the way in their implementation. The results of the ESCAPE and OSCARS EU projects should be fully deployed by and for the nuclear physics community.

The richness and diversity of the community present major challenges in applying common FAIR data principles across a variety of data (theoretical, experimental, software, and scientific publication) and research ecosystems (small to large experiments, short- to long-life collaborations, multi-site travelling detector facilities). To reach this goal of establishing a scalable standard widely adopted by the community, a substantial investment in resources for its implementation and training of the next generation of data specialists is essential.

Progress in fundamental nuclear physics research relies on the availability of quality-assured nuclear data. Additionally, advancements in nuclear technologies and their various applications are greatly influenced by the prompt integration of cutting-edge scientific knowledge into these databases. To achieve both objectives, it is essential to recognise that the expertise, research facilities, and best practices required for nuclear data development extend beyond the capabilities of any single field or application. A coordinated and collaborative effort at both the national and international levels, accompanied by significant investment, is therefore imperative.

● The creation and adoption of open science policies and guidelines addressing the pillars of open science such as open data, open source software, and open hardware, as well as promoting best practices within individual institutes and research infrastructures should be strongly encouraged.

● It is strongly encouraged that open access publishing, strategies and infrastructure for data and software publication, as well as training of researchers in open science practices, be pursued.

● Strong support is encouraged for the application of the FAIR principles and common scientific computing frameworks: encourage training and investment in human resources for data management (data officers, data curators,…) at the various levels (institutions, laboratories, collaborations) to effectively advance the open data practices.

● The creation of coordination bodies to pursue standardisation of the Data Life Cycle to ensure data FAIRness should be supported. The development of guidelines and tools for researchers and collaborations should be undertaken for an effective implementation of these practices.

The primordial role of software in the reproducibility of scientific results should be formalised through systematic software publications and software and computing sessions in workshops and conferences. Software collaborations should be encouraged to improve structuration and oversight, and enable institutional recognition through awards and career advancement.

## Infrastructures for effective open science

Strong investment in federated infrastructures relying on technology standards being promoted in nuclear physics and other disciplines for data cataloguing and data management, data access, data preservation, user analysis and reproducibility should be pursued. The various stakeholders (researchers, collaborators, research infrastructures, institutes) should actively contribute to joint activities with the scientific community and follow the technical developments in the field. The implication of the nuclear physics community in existing European cross-domain initiatives should be strengthened and future activities within Joint ECFA-NuPECC-APPEC (JENA) activities should be initiated to implement scalable infrastructures, favour economies of scale and adapt to the nuclear physics community-specific and diverse needs.

● Combine the forces of the European nuclear physics research and applications communities to establish a comprehensive European nuclear data programme with well-defined priorities defined by stakeholders, as well as sustainable funding to fulfil the needs in nuclear structure and dynamics, astrophysics and applications.

● Support dedicated efforts to train the next generation of nuclear data experts in data evaluation and the use of AI/ML methods and modern data-driven technologies.

● Strengthen the cooperation between nuclear data curators, data scientists, database programmers and AI/ML experts and international organisations (IAEA, OECD/NEA).

# Recommendations for Nuclear Science – People and Society

It is recommended that European funding bodies and institutions see nuclear science as a critical societal investment to inspire the public in nuclear science and its impacts; educate and train the next generation of nuclear scientists; and support equitable career progression with an inclusive approach to diversity across academic, industrial, and vocational career paths.

### Outreach

We recommend that funding agencies, national and international bodies and the community of European nuclear physicists emphasise this critical societal investment to the public, bringing nuclear science and its impacts to its attention. This can be facilitated through:

● establishing and equipping a European network for outreach, resourced by national and transnational funding schemes through research-linked and earmarked funding for outreach; and

● strengthening the provision and support of digital outreach projects and using them as a link to inspire face-to-face and extracurricular activities.

### Education

We recommend that national educational accreditation bodies, funding agencies, universities and educational institutions, in collaboration with the community of European nuclear physicists, work to embed nuclear science across all levels of education, highlighting its interdisciplinary nature and impact. This will require:

● the development and resourcing of a European network of science educators across nuclear sciences, to showcase the possibilities of the field, based on the latest teaching methodologies and guided by research; and

● resourcing research-informed and curriculum-linked training and teaching resources for science teachers, funded by dedicated national and transnational funding.

### Training

We recommend that the community of European nuclear physicists, in collaboration with funding bodies and other stakeholders, resource and support the training of new generations of nuclear scientists to provide the broad skills base required across experimental and theoretical nuclear physics research, needed by all disciplines and industries in our society relying on the expertise, techniques and skills from the nuclear sciences. This includes the provision of training for technical and engineering staff as well as interdisciplinary researchers.

It is further recommended to create new opportunities for increased international mobility and short-term exchange of early-career and technical personnel across institutions to enhance knowledge exchange and skills transfer.

### Diversity

The Nuclear Physics community supports respectful, inclusive and safe work and training environments in academic, industrial, and vocational nuclear science careers.

● We recommend that the network of research organisations, funding agencies, scientific collaborators and conference committees sign up to and promote a diversity charter, such as the one prepared by NuPECC together with APPEC and ECFA.

● The nuclear physics community and its stakeholders should further identify a body in Europe to take charge of collating and providing an overview of the monitoring of diversity across Nuclear Science in Europe. This information should then underpin recommendations and policies adapted to the nuclear physics community.

### Careers

We recommend that equitable and inclusive career development be further prioritised by stakeholders across the European nuclear physics community, giving recognition and visibility to the critical contributions of early career researchers who are the future of nuclear physics and its impact on society. This includes efforts to:

● support tenure track programmes giving highly qualified early career researchers (ECR) the opportunity to lead their own group and establish scientific independence (e.g. permanent staff position openings for ECR, ERC Starting Grants);

● provide opportunities for participation in national and transnational training workshops for early career researchers; opportunities to contribute to prestigious committees and to attend at highly visible conferences, including through further early career awards; to establish career centres for early career researchers at universities, institutes, and research infrastructures, offering support and training in career development.

# Hadron Physics

**Coordinators:**
**Constantia Alexandrou** (University of Cyprus, Cyprus)
**Karin Schönning** (Uppsala University, Sweden)

**NuPECC Liaisons:**
**Diego Bettoni** (INFN Ferrara, Italy)
**David Ireland** (University of Glasgow, UK)

**WG Members:**
- **Luis Alvarez-Ruso** (IFIC Valencia, Spain)
- **Pietro Antonioli** (INFN Bologna, Italy)
- **Gilberto Colangelo** (University of Bern, Switzerland)
- **Annalisa D'Angelo** (University of Roma Tor Vergata, Italy)
- **Luigi Del Debbio** (University of Edinburgh, UK)
- **Achim Denig** (JGU and Helmholtz Institute Mainz, Germany)
- **Gernot Eichmann** (University of Graz, Austria)
- **Jeremy R. Green** (DESY, Zeuthen, Germany)
- **Bernhard Ketzer** (University of Bonn, Germany)
- **Jan Matousek** (Charles University Prague, Czech Republic)
- **Silvia Niccolai** (IJCLab Orsay, France)
- **Elena Perez del Rio** (Jagiellonian University Cracow, Poland)
- **Catarina Quintans** (LIP Lisbon, Portugal)
- **Marc Vanderhaeghen** (JGU and Helmholtz Institute Mainz, Germany)

# Introduction

Hadrons, i.e. systems of quarks and gluons held together by the strong interaction, constitute the bulk of the mass of the visible universe. Only a tiny fraction of its mass is due to the Higgs mechanism; the overwhelming contribution comes from the strong interaction, which dynamically creates mass from almost massless constituents. But how exactly do the properties of hadrons emerge from the quarks and gluons? This question remains to this day one of the unsolved problems of physics. The answer is multi-faceted and complex, and can only be found by a coherent and quantitative description of the strong interaction and its emergent phenomena. Hadron physics aims to explore and understand all aspects of the strong interaction - a tremendous quest in itself. In addition, the importance of hadron physics for other fields cannot be overestimated: understanding the strong interaction is a prerequisite for precision experiments in nuclear, particle and atomic physics, and astrophysics.

The theoretical foundation of our understanding of strong interactions is Quantum Chromo-Dynamics (QCD), an SU(3) gauge theory that, together with the SU(2)xU(1) of electroweak interactions, forms the Standard Model (SM) of particle physics. At very high energies, asymptotic freedom results in a decrease in the coupling among quarks and gluons, allowing for a perturbative treatment of QCD. The predictions based on this approach have been rigorously and successfully tested. However, at low energies, the strong coupling increases to a point where perturbative QCD breaks down. In this regime, the strong interaction and its emerging phenomena become remarkably rich and complex. Achieving a full understanding of the non-perturbative regime requires a variety of approaches, on the theoretical side as well as on the experimental. A common Modus Operandi to gain deeper insights into an object is to study its behaviour when brought slightly out of its natural state. A hadron may be excited into a higher energy state, take part in and be investigated in scattering processes, or interact with another hadron. These three approaches constitute the principal hadron physics topics, namely: hadron spectroscopy, hadron structure and hadron interactions. In addition, recent conceptual and technical developments have established hadron physics as a precision tool, with far-reaching applications.

In the realm of hadron spectroscopy, primary enquiries revolve around understanding the observed hadron spectra in terms of the quark and gluon degrees of freedom, and to search for forms of matter beyond the simplest mesons (quark-antiquark pairs) and baryons (triple-quarks). Examples of such exotic forms of matter are multiquarks, glueballs and hybrids. Hadron structure aims to construct a comprehensive picture of the spatial distribution and motion of quarks and gluons inside a hadron. Through investigations of hadron interactions, we gain insights into the strong interaction dynamics and relevant degrees of freedom at different energy scales, as well as how the residual interaction forms more complex systems, such as atomic nuclei. Furthermore, hadron interactions play an important role in macroscopic stellar objects, such as neutron stars. Precision hadron physics provides a tool for testing the SM and for searching for physics beyond it. The richness and complexity of the research field of hadron physics present both challenges and opportunities.

The challenges arise due to the non-perturbative nature of the strong interaction, which makes quantitative predictions from first principles difficult. In the dawn of hadron physics, it was even considered questionable whether quantum field theory (QFT) was suitable to describe hadron physics. However, it became evident that it was the perturbative approach that had to be re-evaluated. Once this became clear, the theory community delved into exploring and devising new methodologies, such as dispersion relations, the S-matrix approach, and current algebra. Nowadays, these approaches have been seamlessly integrated into QFT and remain essential tools for a plethora of applications, extending far beyond the realm of strong interactions. On the experimental side, the challenge lies in capturing the richness of the objects and phenomena emerging from the strong interaction, that manifests in the tremendously diverse world of hadrons: sizes around a femtometre, lifetimes spanning from $10^{-24}$ s to $10^{30}$ y, fluctuations of one state into another, decays that can be both very complex and very rare. One single experiment cannot address all our questions. Instead, experimental hadron physics requires a combination of large-scale, multi-purpose facilities and smaller experiments with very specific aims and varying degrees of precision.

Addressing the challenges of a non-perturbative theory has opened immense opportunities: Firstly, it has led to a conceptual revolution in our way of thinking about QFTs and facilitated the development of new theoretical tools. The revolution revolves around the insight that one single theory can encompass both asymptotic freedom at high energies, and confinement of its constituents into bound states at low energies. Secondly, key theoretical methodologies have been developed, including effective field theories (EFTs), dispersion relations, functional methods and the lattice theory approach to gauge QFTs. The impact of these advancements has been profound: EFTs have gained a central role in our conceptual view of QFTs and their applications are so widespread that it is difficult to find research areas where they do not play a role. This is particularly pronounced in hadron physics, where EFTs are ubiquitous, and in high-energy physics where they are essential in searches of physics beyond the Standard Model (BSM). Dispersion relations and functional methods, in particular those associated with the Bethe-Salpeter and Dyson-Schwinger equations, have become fundamental to QFTs and their importance in hadron physics is increasing. The lattice theory approach has been widely adopted and is applied across diverse fields, ranging from quantum gravitational interactions to solid-state systems. Lattice QCD both relies on and fosters the development of more powerful hardware systems, including quantum computers, alongside increasingly efficient algorithms. Substantial investments and concerted efforts by the lattice QCD community in hadron physics have yielded significant progress, propelling the field into the precision era.

The demand for a variety of experimental approaches also brings opportunities: it is possible to perform cutting-edge hadron physics both at dedicated hadron experiments and at facilities that were originally designed for other fields of physics. The present and future experimental landscape exploit different probes, such as beams of photons, leptons, hadrons and heavy ions, and a variety of fixed targets or colliding beams. Hadron spectroscopy can be explored with lepton beams, hadron beams and to some extent also heavy-ion beams. The choice of beam and target depends on the spectrum of interest: Electron-positron collisions are optimal for vector meson production, a photon beam on a proton target facilitates light baryon spectroscopy and searches for hybrid mesons, while pion and kaon beams are suitable for light and strange hadron spectroscopy, respectively. B-meson factories provide access to heavy charm hadrons and a future antiproton beam offers opportunities to search for glueballs and to conduct precision studies of high-spin hadrons. Hadron structure is primarily investigated through electromagnetic processes with lepton beams, either in elastic lepton-hadron scattering and deep inelastic scattering, or electron-positron annihilation. Depending on the momentum transfer of the process, the structure can be described either in terms of spatial distributions or with functions related to the quarks and gluons. Polarised beams and targets, as well as self-analysing decays of the final state particles, allow access to spin degrees of freedom. Hadron interactions are predominantly studied using hadron beams, although certain aspects can also be investigated with beams of heavy ions or radioactive nuclei. In recent years, secondary interactions of hadrons produced with lepton, photon or meson beams have also proved to be a viable tool.

Hadron physics at the precision frontier provides a means to search for BSM physics. This requires the highest possible beam intensities. Fixed-target experiments at electron accelerators benefit from extremely clean conditions and have made significant strides in constraining the parameter space for dark matter. The next generation of these accelerators will enable unprecedented precision in determining parity-violating asymmetries in polarised electron-proton or electron-electron scattering. Electron-positron accelerators have been proven invaluable in providing the necessary input to hadronic corrections of the anomalous magnetic moment of the muon. Hadron colliders, where various hadrons are produced in abundance, serve as a hunting ground for rare hadron decays.

In the following sections, we outline the highlights of hadron physics in the areas of spectroscopy, structure, interactions and precision measurements.

# Hadron Spectroscopy

The multi-faceted features of the strong interaction manifest themselves in the excitation spectra and decay modes of strongly bound systems. In modern experiments, properties of hadrons such as their mass, lifetime or width, spin, and other quantum numbers, can be measured. Most well-established hadrons conform to valence quark-antiquark *($q\bar{q}$)* and three-quark *(qqq)* configurations. Therefore, it came as a surprise when several experimental collaborations discovered hadrons that defied this conventional framework. These unconventional, or exotic, hadrons can take the form of either large-size molecule-like states bound by residual interactions, akin to the deuteron, or compact multi-quark states, or even exhibit explicit gluonic degrees of freedom, such as hybrid mesons or glueballs (Fig. 2.1). A coherent understanding of the excitation spectra of hadrons necessitates the combined effort of experiment and theory: ever-more precise data and advanced amplitude analyses, combined with progress, in theory using e.g. lattice QCD simulations (see Box 2.4), functional methods (see Box 10.1), or effective field theories (Box 2.5).

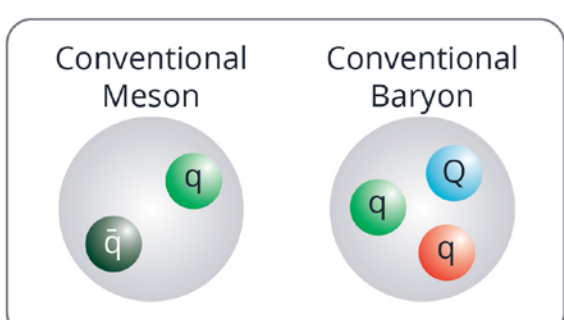

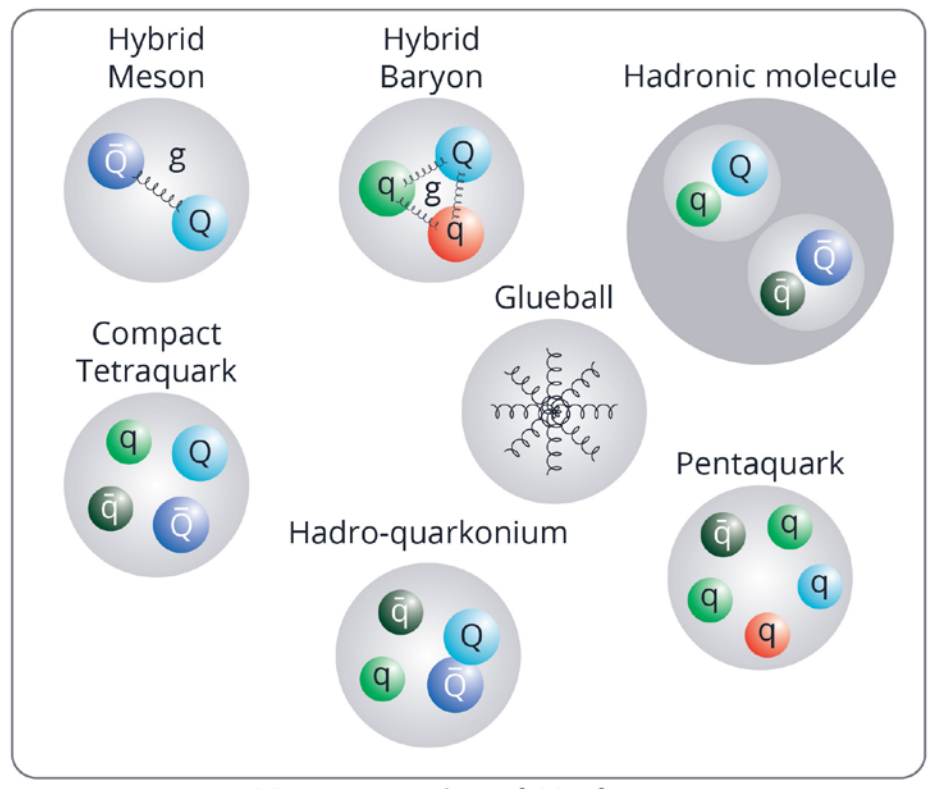

*Fig. 2.1 Valence content of conventional and unconventional (exotic) hadrons*

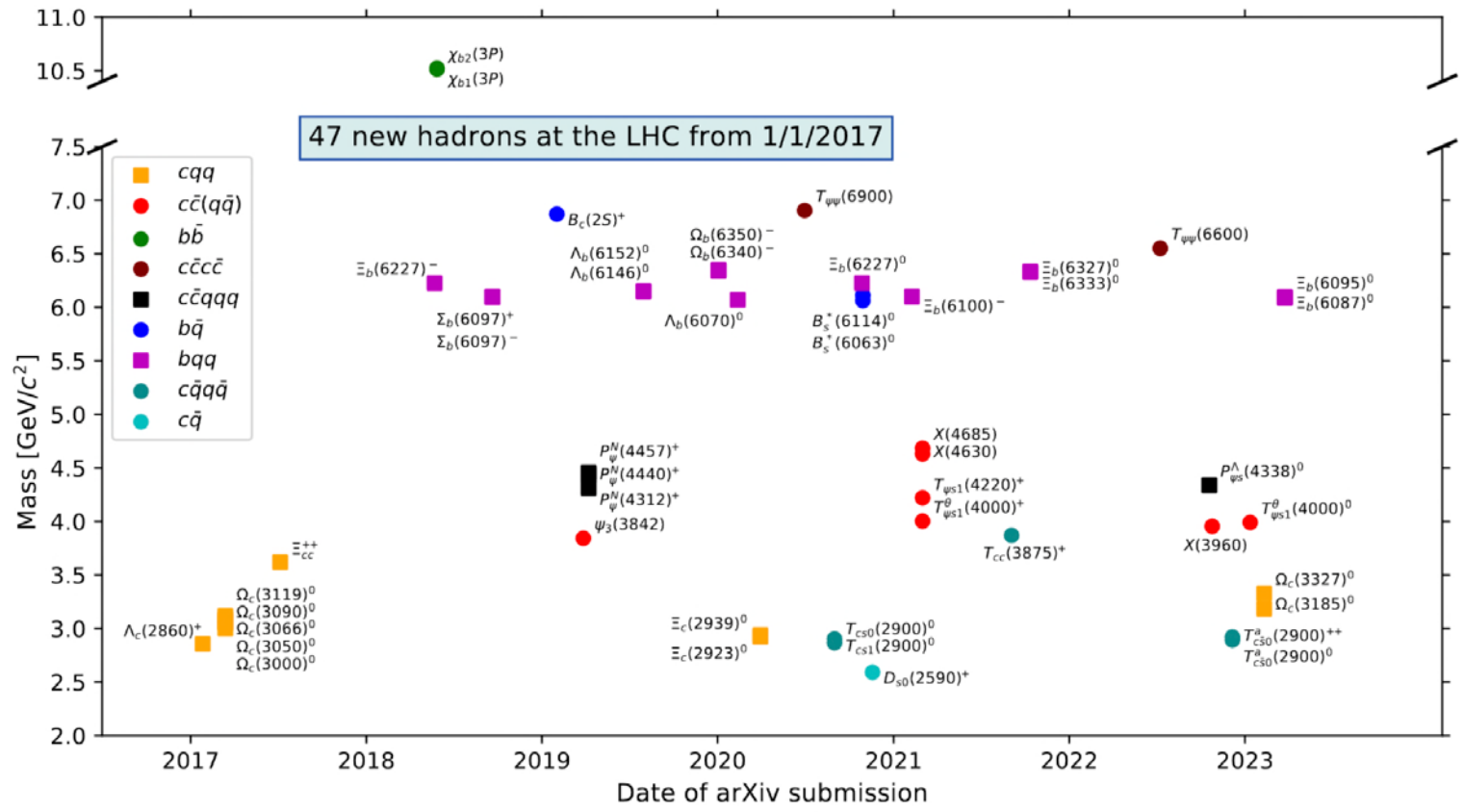

*Fig. 2.2: Hadrons discovered at the LHC since the NuPECC LRP 2017. Picture credit Patrick Koppenburg, LHCb, CC by 4.0.*

## Recent achievements and highlights

Substantial advancements have been made in hadron spectroscopy since the NuPECC LRP of 2017. COMPASS (CERN) and BESIII (China), as well as experiments at ELSA (Bonn), MAMI (Mainz) and Jefferson Lab (US) have published compelling new findings on hadrons containing light and strange quarks. In the heavy-quark sector, numerous discoveries have been made in recent years, with LHCb (CERN) playing a prominent role (Fig. 2.2). In addition, BESIII and Belle / Belle II (SuperKEKB, Japan) contributed with important results. Many of the newly discovered states are manifestly exotic.

### Light mesons

Recently, the wealth of data from COMPASS helped to resolve a decade-long debate on the existence of states with spin-exotic quantum numbers $J^{PC} = 1^{-+}$ with masses below 2 GeV. A coupled-channel analysis of the $\eta\pi$ and $\eta'\pi$ final states by the Joint Physics Analysis Center (JPAC) showed that the signals at 1.4 GeV and 1.6 GeV can be attributed to single resonance pole, the $\pi_1$(1600) (Fig. 2.3), as confirmed by including $p\bar{p}$ annihilation and $\pi\pi$ scattering data in the analysis. The $\pi_1$ (1600) is the prime candidate for a hybrid meson, where gluons explicitly contribute to the quantum numbers. The $\eta_1$ (1855), a possible isoscalar partner to the $\pi_1$ (1600), was recently reported by BESIII in *J/ψ* radiative decays. In the $1^{++}$ sector, COMPASS found that the $a_1$ (1420), a distinct resonance-like signal inconsistent with any quark-model, can be explained by a triangle singularity (Fig. 2.4). The fact that resonance-like signals can arise from such rescattering effects in multi-body final states may also have implications for several new states discovered in the heavy-meson sector.

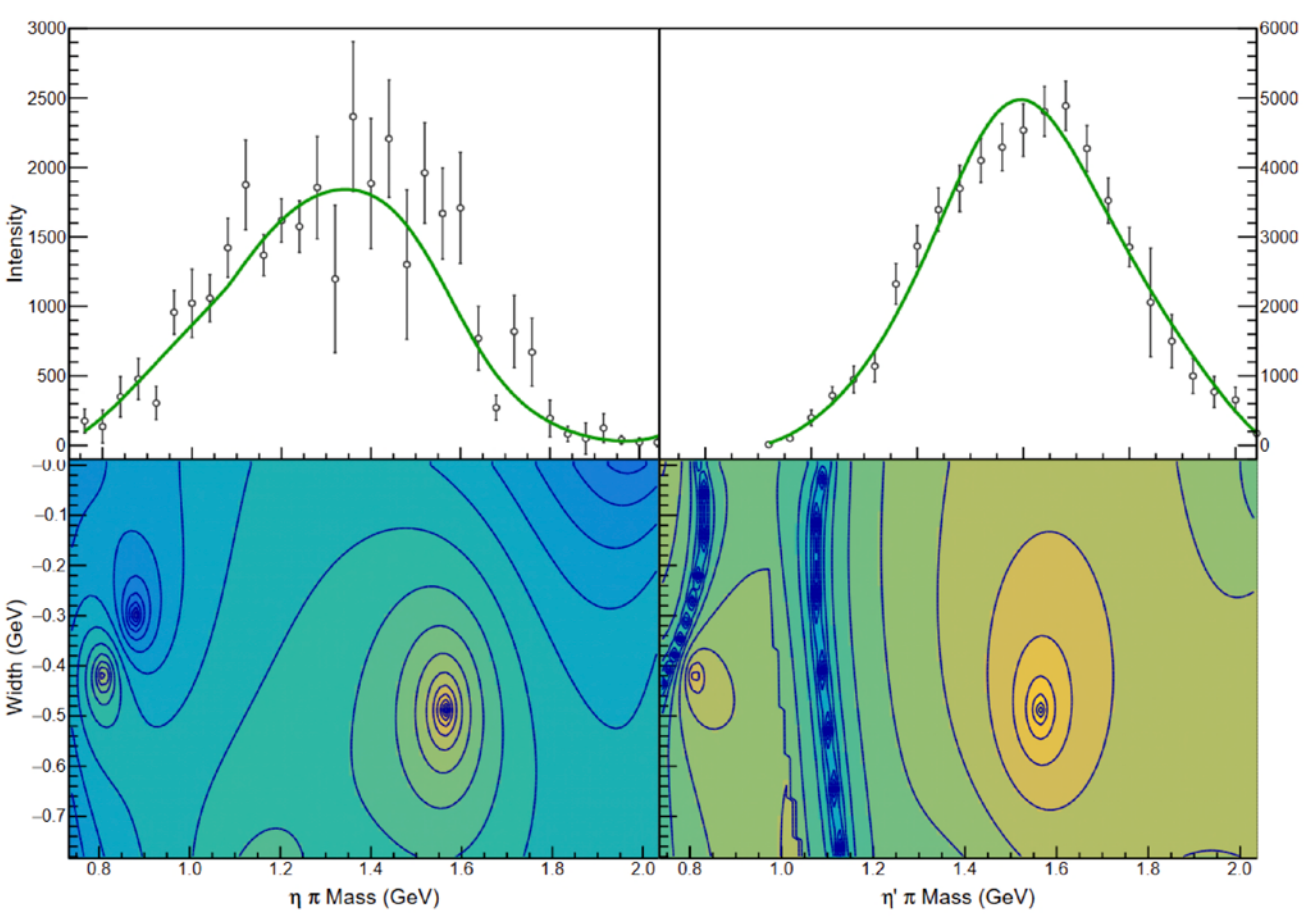

*Fig. 2.3: Pole structure of eta pi and eta' pi final states. Picture credit Alessandro Pilloni/JPAC, Phys. Rev. Lett. 122, 042002 (2019)*

First-principle calculations of light-hadron spectroscopy have substantially advanced in recent years. Lattice QCD is now able to address hadron resonances decaying into many different two-hadron final states, including the lightest hybrid resonance with quantum numbers $J^{PC} = 1^{-+}$. Lattice QCD can also be applied to study $\pi\pi$ and $K\pi$ scattering, providing insights into the elusive $\sigma$, $\kappa$ and $a_0/f_0$(980) scalar mesons. The framework for analysing three-body decays has been established with initial lattice QCD calculations in the meson sector.

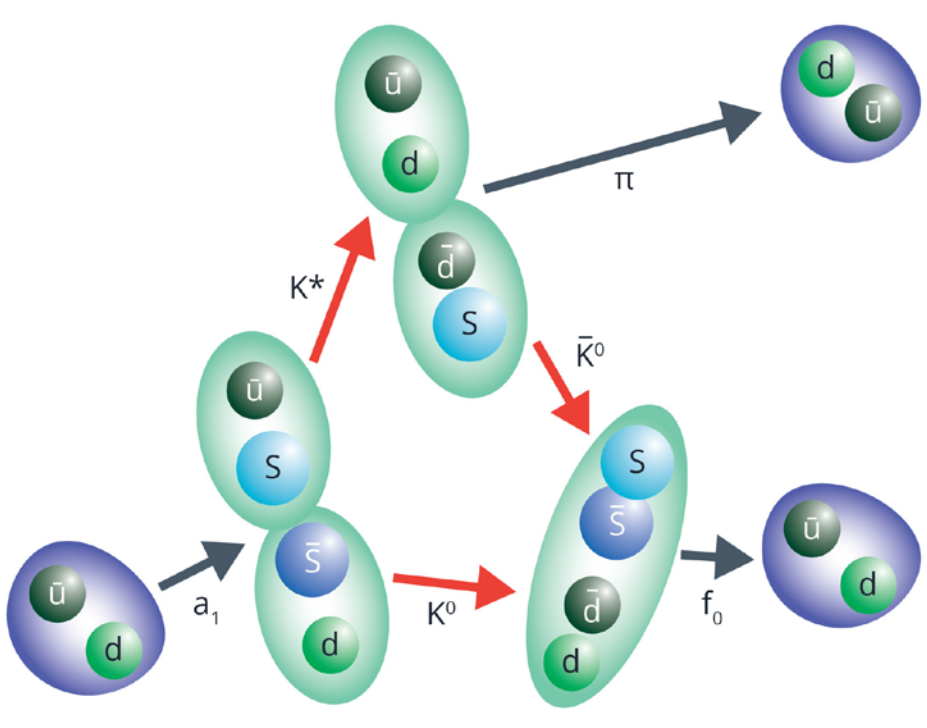

*Fig. 2.4: Triangle singularity diagram for the $a_1$ (1420).*

An open question pertains to the existence of glueballs. According to theory predictions, the lightest glueballs should be scalar, tensor and pseudoscalar states. Since these can mix with conventional mesons, their identification is difficult. However, a signature for glueballs is their expected preference for gluon-rich processes, such as radiative *J/ψ* decays. Such decays are being studied at BESIII, where partial-wave analyses on new, high-precision data reveal a strong production of the $f_0$ (1710) and $f_0$ (2100), as well as the $f_2$ (2340)

## Heavy exotic mesons

While chiral symmetry and relativistic effects complicate the properties of light hadrons, heavy quarks offer a cleaner environment for studying exotic states. This manifests in the surprising discoveries of the so-called *XYZ* states that do not fit into the conventional $q\bar{q}$ picture. The first to reveal itself among these states was the *X*(3872), now designated $\chi_{c1}$(3872), by Belle in 2003. The $\chi_{c1}$(3872) mass is extremely close to the $D(\bar{D}^*)$ threshold, indicating a strong molecular component with a small binding energy. Subsequent discoveries include the *Y*(4230), nowadays referred to as the *Ψ*(4230), observed by Belle, BaBar and BESIII. This state has been interpreted either as hadro-quarkonium, comprising a *J/ψ* accompanied by a light-quark cloud, as a compact tetraquark made of diquark-antidiquark pairs interacting via coloured forces, as a hadronic molecule with the heavy-light mesons *D* and $D_1$ (2460) as clusters, or as a mixture of various components. Each scenario leaves different imprints on the properties of these particles, like their line-shapes or decay patterns, which remain to be investigated. The Z states, e.g. the charged charmonia $Z^{\pm}$ (3900) and $Z^{\pm}$ (4430), cannot be pure quark-antiquark states and are thus inherently exotic. The family of hadrons has continued to expand: the first discovery of a doubly open-charm tetraquark $T_{cc}(cc\bar{u}\bar{d})$ in 2021 by LHCb, was followed by observations of states with double hidden charm ($T_{c\bar{c}c\bar{c}}$) and open strangeness ($T_{cs}$). As shown by the CMS experiment, exotic hadronic states can be produced and detected in ultra relativistic ion collisions, offering novel diagnostic tools for elucidating their production mechanism.
A conceptually clean system, extensively investigated within lattice QCD, consists of two heavy anti-bottom quarks and two light quarks. For the quantum numbers $J^P = 1^+$, independent lattice calculations revealed that such a four-quark system forms a QCD-stable tetraquark with relatively large binding energy, around 100 MeV below the $B\bar{B}^*$ threshold. Replacing one of the light quarks with a strange quark yields a QCD-stable tetra-quark with a binding energy of around 30 MeV.

Theory highlights include predictions of heavy exotic meson decay patterns and line-shapes using EFTs, which can be tested experimentally. Functional methods have demonstrated how four-quark systems dynamically generate meson-meson and diquark-antidiquark components, where the former is typically dominant. Lattice QCD and Born-Oppenheimer calculations indicate that doubly open-bottom states like $bb\bar{u}\bar{d}$ can form stable tetraquarks.

## Light and strange baryons

Quark models anticipate the existence of 200 light and strange baryons with masses below 3 GeV, yet only a third of these have been experimentally observed. Efforts to identify these "missing" states and the pursuit of exotic hybrid baryons have been the main goals of experiments at JLab, MAMI and ELSA. In recent years, high-quality photoproduction data has facilitated the determination of the properties of many light baryons, including the eight newly discovered ones reported in the NuPECC LRP 2017.
Furthermore, recent data from in particular MAMI and ELSA, has contributed to establishing the $N(1895)\frac{1}{2}^-$ and to resolving the longstanding puzzle with the large difference in the η-decay branching ratio of $N(1535)\frac{1}{2}^-$ and $N(1650)\frac{1}{2}^-$.
Notably, CLAS at JLab has identified a second $N'(1720)\frac{3}{2}^+$ resonance, which has almost the same mass but with very different electro-couplings and hadron decay branching fractions, as compared to the already established $N(1720)\frac{3}{2}^+$. Important information on the internal structure of, for example, the Roper resonance, the $N(1520)\frac{3}{2}^-$, and the $N(1535)\frac{1}{2}^-$ has been extracted from the dependence of their electro-couplings on the virtual photon four-momentum transfer squared. Significant theoretical advances alongside quark-model studies include functional methods illuminating the role of relativity and diquarks in the baryon spectrum, as well as lattice QCD description of decays to three-hadron final states. The latter are expected to be applied in the baryon sector too in the near future.

The hyperon spectrum plays a pivotal role in bridging the gap between the relativistic light-quark and non-relativistic heavy-quark sectors. However, it is to this day poorly understood. χPT analyses of new sub-threshold $\bar{K}N$ scattering data from AMADEUS and KLOE-2 suggest a solution characterised by two poles; one narrow close to the $\bar{K}N$ threshold, denoted the *Λ*(1405), and one broad at lower mass, referred to as the *Λ*(1380). Recent lattice calculations resulted in a finite-volume spectrum of the *Λ*(1405) at an unphysical pion mass, also indicating the necessity of two poles. Extrapolation from the physical pion mass using χPT confirms the lattice result. New photoproduction data of $K^+\Lambda$(1405) and $K^+\Sigma^0$ from ELSA show cusp structures close to the $K^*\Sigma$ threshold that may indicate similarities to the LHCb pentaquarks close to the $D^*\Sigma_c$ threshold.

## Heavy baryons and pentaquarks

The heavy-baryon sector has progressed significantly since the latest LRP. Almost 30 previously undiscovered heavy baryons with charm and bottom quarks have been identified, including the first double-charm baryon $\Xi_{cc}^{++}$ at LHCb. Many of the states predicted by quark models remain to be found, promising prospects for further discoveries at LHCb and Belle II. Particularly noteworthy is the observation of several $P_c^+$ pentaquark candidates by LHCb, showcasing quark content $c\bar{c}uud$ and $c\bar{c}uds$ and decaying into *J/ψp* or *J/ψΛ*. Since 2017, the number of such states has increased from two to five. Despite extensive theoretical efforts to understand the nature of these states, there is currently no consensus and the interpretations range from rescattering effects or virtual states to hadronic molecules, baryo-charmonium or compact pentaquarks. Remarkably, GlueX has so far not observed any clear signal of pentaquarks in photoproduction.

# Next steps and future

Hadron spectroscopy is subject to major experimental endeavours worldwide, poised to significantly change the nuclear physics landscape in the years ahead. Anticipated breakthroughs on the spectrum of ordinary and exotic hadrons are expected from CLAS12 and GlueX at JLab, BESIII, AMBER at CERN-SPS, ELSA, MAMI, the LHC experiments at CERN, Belle II, and in the long-term future, PANDA at FAIR as well as ePIC at EIC. These facilities utilise complementary probes and production mechanisms, crucial for understanding the symmetries governing the meson and baryon multiplets: photoproduction with polarised beams and target (MAMI, ELSA, GlueX), electroproduction (CLAS12); meson beams (AMBER, HADES, J-PARC); $e^+e^-$ annihilation (BESIII, Belle II); *ep* (ePIC @ EIC), *pp* and heavy-ion collisions (LHC); and $p\bar{p}$ annihilations (PANDA), all of which are needed for studying different properties.
In the light meson sector, establishing the spin-exotic $\pi_1(1600)$ as a hybrid meson calls for measurements of its decay modes and branching fractions. In a similar way, confirming the $\eta_1(1855)$ and identifying its expected $\eta_1'$ partner state is imperative and therefore new results from COMPASS, AMBER, BESIII, GlueX and CLAS12 are eagerly awaited. Since other hybrid mesons are expected to have non-exotic quantum numbers, it is important to also understand the ordinary $q\bar{q}$ light-meson spectrum and identify supernumerary states. The spin and parity of two glueball candidates above 2 GeV, seen in radiative $J/\psi$ decays by BESIII, will be determined thanks to new, high-precision $J/\psi$ data. Ultimately, $p\bar{p}$ annihilation at PANDA will provide access to all possible quantum numbers for glueballs, hence facilitating the comprehensive mapping of the full glueball spectrum including high-spin states.
Progress in elucidating the surprisingly narrow exotic states in the charm sector is expected to continue in the coming few years. Belle II has a unique possibility to study *XYZ* production in both *B* decays and $e^+e^-$ annihilations, providing insights into the nature of the inherently exotic *Z* states, which must consist of at least four quarks. Belle II and BESIII will further study *Y* states, some of which are hybrid candidates. The endeavour by Belle II to measure the width of the $\chi_{c1}(3872)$ to sub-MeV precision is noteworthy, although discerning the nature of non-vector states requires precise, model-independent line-shape measurements that are only possible in antiproton beam scans as planned with PANDA.

In the light baryon sector, numerous open questions persist regarding the existence of missing states, parity doublets, hybrid baryons or diquark correlations. Photoproduction measurements of a comprehensive set of polarisation observables at MAMI and ELSA will serve to further constrain partial-wave analyses and lead to a unique identification of excited nucleons. The investigation of $\gamma n$ scattering alongside $\gamma p$, is of paramount importance to unravel the isospin dependence. Electroproduction with CLAS12 and the pion beam at HADES at GSI will offer complementary insights into the structure of nucleon resonances.

In the heavy baryon sector, LHCb is expected to remain at the forefront, with a substantially increased data sample during Run-3 fostering a high discovery potential for doubly charmed heavy baryons in addition to the $\Xi_{cc}^{++}$. The doubly open heavy-flavour tetraquark $T_{bc}$ and baryons containing bottom quarks such as the $\Xi_{bc}$ may also be within reach. Single and possibly double-charm baryons will also be studied by the Belle II experiment in the next decade with extremely high luminosity. Finally, the hidden-charm pentaquark candidates reported by LHCb will be investigated in photo- and electroproduction at JLab. The ePIC experiment at the future EIC will be able to access charm, bottom open, hidden-flavour mesons and pentaquarks via quasi-real photoproduction. Complementary insight into the nature of heavy-flavour exotics will be provided by the LHC experiments, in particular ALICE-3, by studying their production in ultra-relativistic heavy-ion collisions and by femtoscopic techniques.

While substantial progress has been made and is expected to continue in the light and heavy quark sectors, corresponding advances have been absent for multi-strange hadrons. The coming decade holds promise to change this picture by dedicated efforts in experiments, such as AMBER and JLab. AMBER will use a high-energy charged kaon beam to meticulously map the strange meson spectrum, one of the goals being the search for strange members of exotic multiplets. At JLab, there are plans to extend the physics reach of Hall D using a $K_L^0$ beam and proton and deuteron targets. Reactions with $K_L^0$ beams are poised to shed light on the multiplets of $\Lambda$, $\Sigma$, $\Xi$ and $\Omega$ hyperons, but also light scalar mesons like the $\kappa$, which are believed to have strong four-quark components. A pressing question in the hyperon sector pertains to the existence of exotic strange configurations, as exemplified by the two-pole structure of the $\Lambda(1405)$. A forthcoming experiment at ELSA aims at investigating hyperons and specifically addressing the question of two-pole structures in the spectrum. Furthermore, future luminosity upgrades at CLAS12 are expected to facilitate hyperon spectroscopy, with the potential to discover several excited double-strange $\Xi^*$ and the photoproduction of the $\Omega$ baryon.

The study of excited and exotic hyperons remains a principal goal of PANDA, where $p\bar{p}$ annihilations hold the distinct advantage that all excited multi-strange hyperons and antihyperons can be produced in simple, parameterisable two-body reactions.

Progress in understanding QCD in the strong coupling regime hinges on precision. The new generation of experiments provides huge data samples thanks to increased luminosities and enhanced acceptance and efficiency of the detectors. Advanced statistical methods and sophisticated analysis techniques are imperative in exploiting the resulting data sets and achieving increased precision. Machine learning techniques, already commonly used in data reconstruction, are gaining importance in complex analysis tasks like model selection and amplitude fitting. Theoretical advances in lattice QCD and functional methods are timely and complementary to experimental findings. The close collaboration between experimentalists and theorists in analysing and interpreting the data has proven extremely fruitful and should be extended in the coming years. A comprehensive understanding of ordinary and exotic hadrons from experiment and theory is paramount in unravelling one of the great puzzles of the SM, namely how the strong interaction forms matter.

# Hadron Structure

Unravelling the structure of hadrons in terms of quarks and gluons (partons) is imperative to understanding the strong interaction. Lepton-nucleon scattering with the exchange of a virtual photon serves as the primary tool for this endeavour. The four-momentum squared $Q^2$ of the photon is inversely proportional to the spatial resolution at which the nucleon is probed. Hadron structure provides a benchmark for non-perturbative theory calculations or quantities such as electric charge radii, electric and magnetic polarisabilities, and mesons-photons couplings. At high $Q^2$, the process can be factorised into a perturbative and a non-perturbative part, simplifying the theoretical description. Correlations among the spatial location, momentum and spin of the quarks and gluons yield detailed information on their contribution to the nucleon mass and spin. The information is encoded in structure functions, which can be accessed by various scattering processes:

- Form Factors (FFs) give access to the transverse charge and magnetic distributions.

- Parton Distribution Functions (PDFs) describe the momentum and spin distributions of quarks and gluons.

- Generalised Parton Distributions (GPDs) encode the correlation between the transverse position and longitudinal momentum of the partons.

- Transverse Momentum Dependent PDFs (TMDs) link the longitudinal and transverse momenta of the partons.

Obtaining a comprehensive three-dimensional picture of hadron structure requires the knowledge of all these distributions.

## Recent achievements and highlights

### Electromagnetic FFs

Elastic electron-scattering ($eN \rightarrow eN$) experiments probe space-like electromagnetic FFs. Lattice QCD calculations of FFs have primarily been performed up to $Q^2 \sim 1$ GeV$^2$, and recently using physical values of the QCD parameters. Unstable hadrons pose challenges for elastic scattering studies, but can be accessed in the timelike region ($q^2 = -Q^2 > 0$), either through $e^+e^-$ annihilations or Dalitz decays. Time-like FFs are complex functions with a relative phase, which can be measured in weak, parity-violating hyperon decays. Using this approach, BESIII has determined the time-like FFs of $\Lambda$ and $\Sigma^+$. From these, space-like quantities, such as the charge radius, can be calculated using dispersion relations.

## Gravitational FFs

In contrast to the proton charge densities, little is known about the mass density of the proton. As dominated by the energy carried by gluons, the mass density is difficult to access in electron scattering since gluons do not carry electromagnetic charge. GlueX and Hall C at JLab have investigated the mass density using a small colour dipole, through threshold photoproduction of $J/\Psi$. These studies found the mass radius to be smaller than the electric charge radius.

## Polarisabilities

Nucleon polarisabilities characterise the internal response of particles to an external field and are accessed via Compton scattering. Recent progress includes the measurements on the proton with unprecedented precision by A2 at MAMI. Virtual Compton scattering provides access to generalised polarisabilities, which depend on $Q^2$ linked to the spatial scale of the induced deformation. While theory expects a monotonic $Q^2$ dependence, recent results by A1 at MAMI and VCS-I at JLab reveal a structure at the $3\sigma$ level. Meson polarisabilities are vital for studying chiral symmetry and its breaking. These can be accessed in Primakoff reactions, e.g. with a pion beam interacting with the electric field of nuclei. The latest result by COMPASS for $\pi^-$ is consistent with theory predictions, unlike earlier measurements from other experiments. Pion polarisabilities have also been measured in real and virtual photon-photon fusion into a pion pair with BESIII, from which generalised pion polarisabilities can be extracted using a dispersive formalism.

## PDFs

PDFs quantify the probability of finding a parton in a hadron with a given momentum fraction $x$. They can be accessed in deep inelastic scattering (DIS) and provide information about the mass and spin distributions. High-precision PDFs are essential to interpret precision tests of SM and BSM at the upgraded LHC; the uncertainties from the PDFs would otherwise be dominant when the expected increased luminosity is achieved. Fig. 2.7 shows the kinematic space explored by present and future facilities to measure the three collinear PDFs.
The decomposition of the proton spin is progressing though polarised DIS double-spin asymmetry measurements at COMPASS and JLab. These provide information about the helicity structure of quarks. Furthermore, recent RHIC measurements of the parity-violating spin asymmetry in the production of longitudinally polarised $W$ bosons demonstrated, for the first time, a clear asymmetry between the helicity distribution of $\bar{u}$ and $\bar{d}$. RHIC also provided evidence of a positive gluon spin contribution within an $x$ range complementary to that previously studied by COMPASS. There have been tremendous strides in lattice QCD computations of the lower Mellin moments of PDFs, yielding a first complete decomposition of the proton spin (see Box 2.4). New methods based on factorisation theorems applied to hadron matrix elements at large momenta allow the direct computation of PDFs, GPDs, and TMDs. Results on PDFs including flavour decomposition have emerged, and are already being used to constrain phenomenological analyses.
Gluon saturation, predicted by models like colour glass condensate, is a fundamental aspect and it needs to be established whether the gluon density will saturate at low $x$ or continue to rise. A recent STAR result on forward $\pi^0$ $\pi^0$ correlations corroborates this, in line with an older HERA measurement.

The precision of the PDFs as extracted from LHC data now matches NNLO theory calculations, further improving our knowledge. With emerging $N^3LO$ calculations for PDFs, corresponding precision is also required for hard processes.

## GPDs

GPDs offer a universal description of nucleon structure as a function of the longitudinal momentum and the transverse position of the partons. They can be accessed through high momentum transfer scattering processes of electrons off nucleons in exclusive reactions. The most straightforward process is the Deeply Virtual Compton Scattering (DVCS), in which the proton remains intact and emits a high-energy photon. However, GPDs are also accessible in hard meson electroproduction (DVMP) and other exclusive final states.

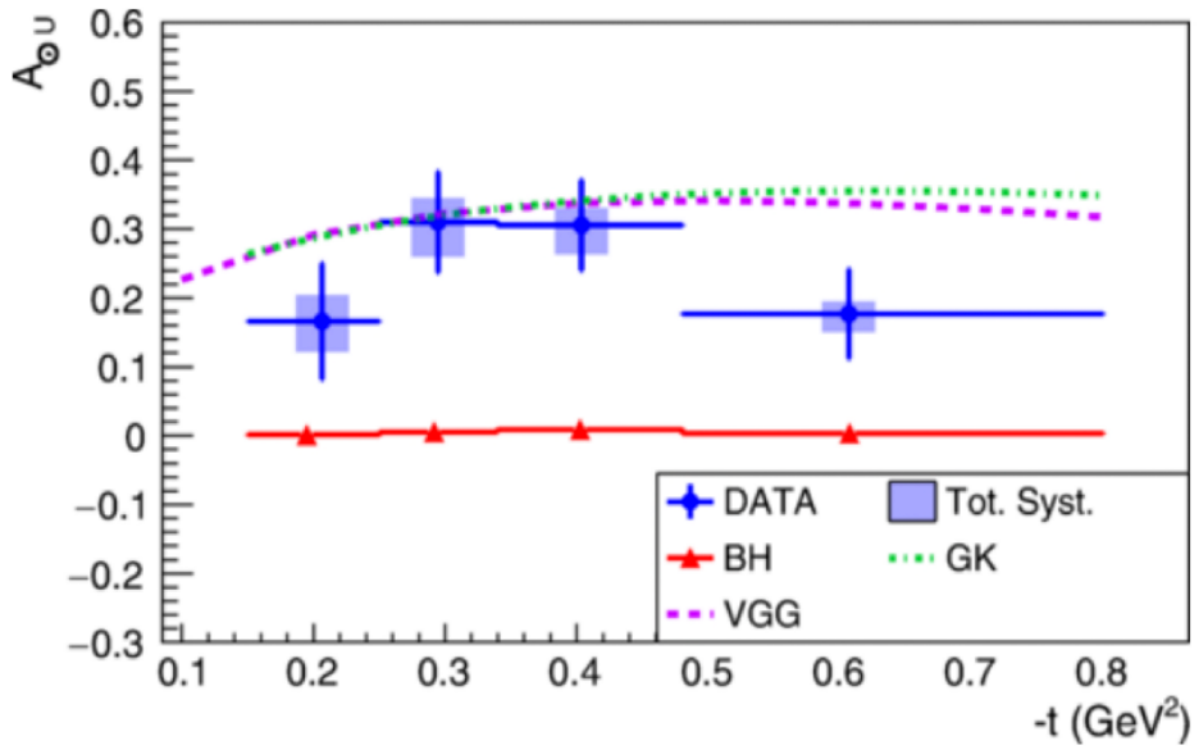

***Fig. 2.5: First measurement of the beam spin asymmetry for Timelike Compton Scattering (CLAS12) [PRL. 127 (2021) 26, 262501]: the fact that models reproducing this asymmetry are also reproducing DVCS data provides evidence of the universality of GPDs.***

The first generation of DVCS measurements, pioneered by HERMES and JLab, enabled femtometre-scale proton tomography. This revealed that valence quarks concentrate in the centre of the proton, while sea quarks spread towards its edges. A new era has begun with the JLab 12-GeV upgrade: Hall A and CLAS12 have recently conducted experiments dedicated to proton GPDs. The DVCS programme for neutrons offers access to the contribution from the angular momenta of the quarks to the nucleon spin. Recent results on timelike Compton scattering provided an experimental test of the universality of GPDs (Fig. 2.5) and paved the way for the study of the mechanical properties of the proton. Experiments on unpolarised and polarised proton and deuterium targets are ongoing or planned at JLab. Combining proton and neutron measurements, the quark-flavour dependence of the GPDs can be extracted. Furthermore, a recent DVMP measurement from CLAS12 with a nucleon resonance opens the perspective to study transition GPDs. High-luminosity DVCS and DVMP experiments at JLab (Hall C) are underway. The sea-quark region is explored by COMPASS, whose data are still being analysed to access T-odd GPDs. In lattice QCD, the new developments in the calculation of PDFs have been extended and have led to the exploration of new methods for GPDs. Various European groups are also involved in the extraction of GPDs, via either local or global fits.

## TMDs

The TMD formalism provides access to spin-orbit correlations and to fundamental quantities like the tensor charge of the nucleon. The transverse structure of the nucleon is explored through semi-inclusive DIS (SIDIS) and Drell-Yan (DY, $q\bar{q} \rightarrow l^+ l^-$) experiments, as well as at $pp$ and $e^+e^-$ colliders. Successful extractions of quark unpolarised TMDs from global fits have propelled this research field into the precision era, with the quark unpolarised TMDs known at the $N^3LL$ accuracy.

### Box 2.1: Proton Radius

**The proton radius can be extracted with high-precision electron scattering experiments, but also inferred from atomic hydrogen Lamb shift measurements where the proton radius enters as a hadronic correction. With muonic hydrogen, the precision can be improved by at least a factor of ten due to the muon being 200 times heavier than the electron. This was exploited for the first time more than a decade ago at PSI in Switzerland. At the time, it came as a great surprise when the muonic hydrogen measurements yielded a significantly smaller proton radius, as compared to values extracted with electron scattering and electron hydrogen. The 5.6 σ deviation led to the so-called proton radius puzzle and sparked an intensive debate about its underlying causes. In particular, the comparison with ordinary hydrogen spectroscopy raised suspicions of lepton universality violation, suggesting BSM physics. Driven by this conundrum, a series of new hydrogen spectroscopy measurements have been carried out. Most of these confirm the smaller proton radius value as measured in muonic hydrogen spectroscopy. Furthermore, the PRad electron scattering experiment at JLab reported a proton radius value consistent with that from muonic hydrogen. A re-analysis of old electron scattering data, using a dispersive framework respecting unitarity and analyticity, also yielded a radius in agreement with the value obtained with muonic hydrogen. The current experimental status is summarised in Fig. 2.8.**

**The focus of current research is to thoroughly investigate all the systematic effects involved in precision experiments. A series of new high-precision lepton scattering measurements are either ongoing or under preparation, including MUSE at PSI, AMBER, PRES at MAMI, ULQ² at Sendai/Japan, and PRad-II at JLab, as illustrated in Fig. 2.6. Novel techniques are being employed, such as muon beams (MUSE, AMBER), high-pressure active time-projection chambers (AMBER, PRES) or calorimeters (PRad and PRad-II) instead of magnetic spectrometers, and very low beam energies to reach ultra-low momentum transfers. The MAGIX experiment at MESA with its low beam energies, gas-jet target and high-resolution spectrometer will provide ultra-clean experimental conditions and, thus, be uniquely suited for proton radius measurements.**
**Furthermore, new developments in muonic atom spectroscopy are planned in the next decade to improve the accuracy of Lamb shift measurements in muonic hydrogen by a factor of five. The goal is to extract the 1S hyperfine splitting in muonic hydrogen to an unprecedented accuracy of 1 ppm.**
**Extending the electron scattering measurements to light nuclei will allow for comparisons of the extracted radii with the corresponding values in muonic atoms. Even more importantly, they will be essential for data-driven evaluations of hadronic uncertainties in muonic systems. In this regard, programmes for polarisability measurements on nucleons will significantly reduce hadronic uncertainties. In a similar spirit, experimental programmes with positron beams are planned both at DESY and at JLab to measure the two-photon contribution.**
**The pursuit of unravelling the low-energy hadron structure, in close collaboration with EFT theory specialists, has the potential to extract fundamental constants and observables with unprecedented precision from planned future measurements in atomic spectroscopy. Ultimately, this provides a new handle to probe the quantum level of nature with the utmost accuracy.**

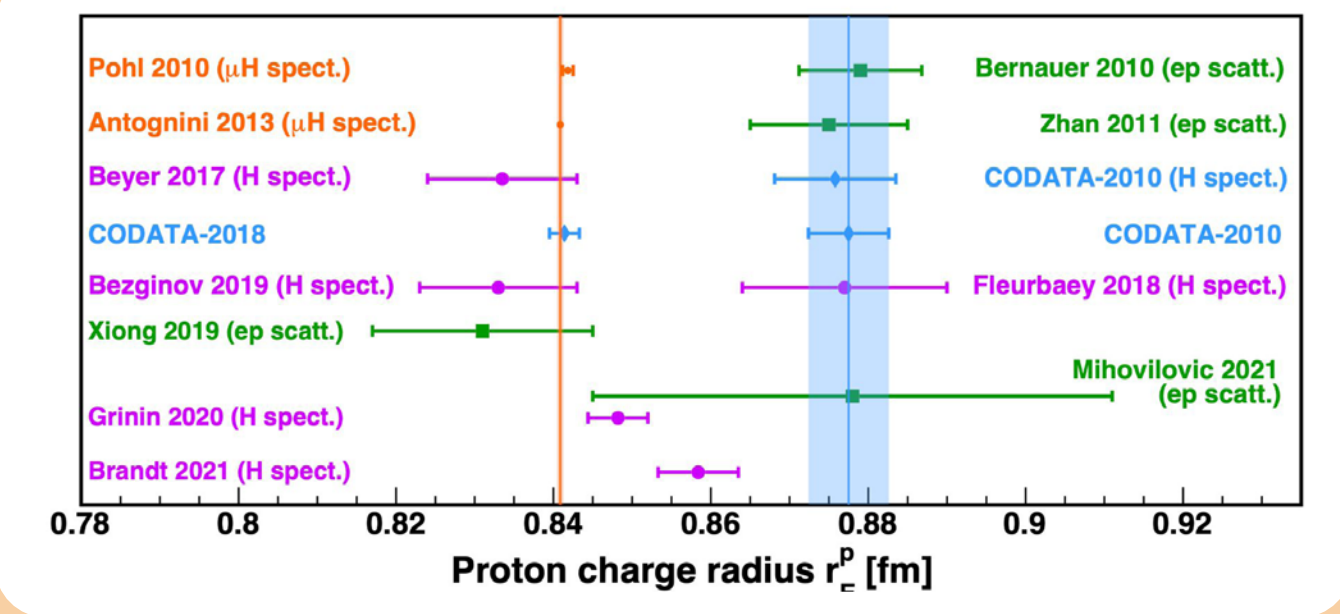

*Fig. 2.8: The proton charge radius determined from ep elastic scattering, hydrogen spectroscopic experiments, as well as world-data compilation from CODATA since 2010. The muonic spectroscopic measurements are shown in orange filled circles, ordinary hydrogen spectroscopic results are shown in purple filled circles, electron scattering measurements are shown in green squares, and the CODATA compilations are shown as blue diamonds ([from:https://doi.org/10.48550/arXiv.2302.13818](https://doi.org/10.48550/arXiv.2302.13818) ). Picture credit J. Zhou (Duke University).*

Ongoing efforts aim to extend the progress to polarised TMDs, which contain information on otherwise elusive parton correlations, like spin-orbit effects and the nucleon tensor charge.
COMPASS, CLAS12 and to some extent, HERMES are the main players in TMD experiments. Analyses are ongoing for different observables (multiplicities, unpolarised cross section, and spin observables), final states (pions, kaons, and protons) and targets (hydrogen, deuteron, and heavier nuclei). In particular, measurements have started with final-state kaons, accessing the strange-quark contribution. DY results from COMPASS confirm the well-known QCD prediction, which is a crucial test of the TMD approach. COMPASS has also collected pioneering data on hadron production in DIS off hydrogen and transversely polarised deuterons. Studies of TMDs using lattice QCD are yielding promising results.

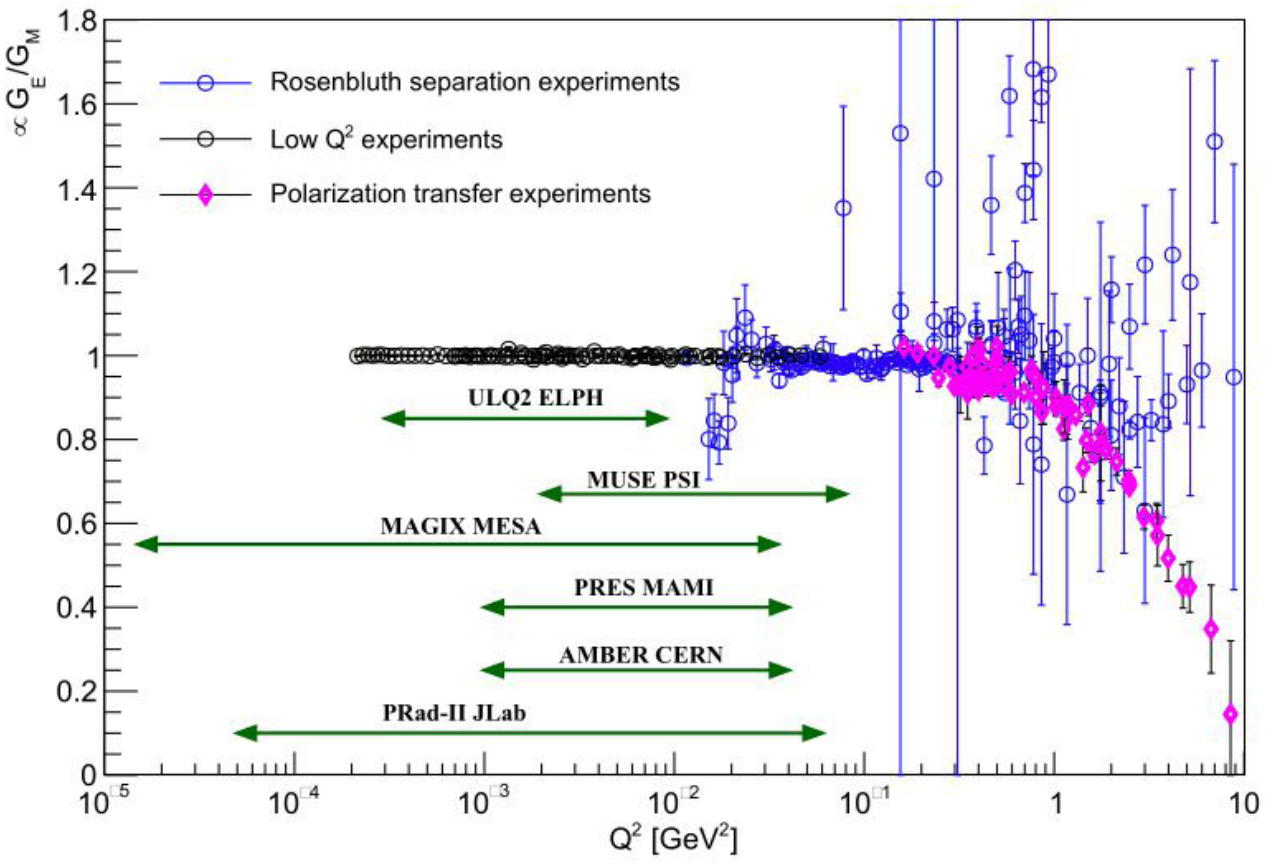

*Fig 2.6: World data on the ratio $\mu G_E/G_M$, where experiments that measured only GE used a parametrisation of existing data for $G_M$. At high $Q^2$, the difference between the Rosenbluth-separation experiments and the polarisation-transfer ones is evident. The $Q^2$ coverage of upcoming proton-radius experiments (see Box 2.1) is also shown. (Figure courtesy of M. Atoui)*

## Next steps and future

The theoretical and experimental exploration of the three-dimensional (3D) structure of the nucleon will continue during the coming decade, on all fronts discussed. A prominent example is two-photon exchange (TPE), a commonly suggested explanation for the long-standing discrepancy between the FFs as measured with the Rosenbluth and the polarisation-transfer methods (see Fig. 2.6).

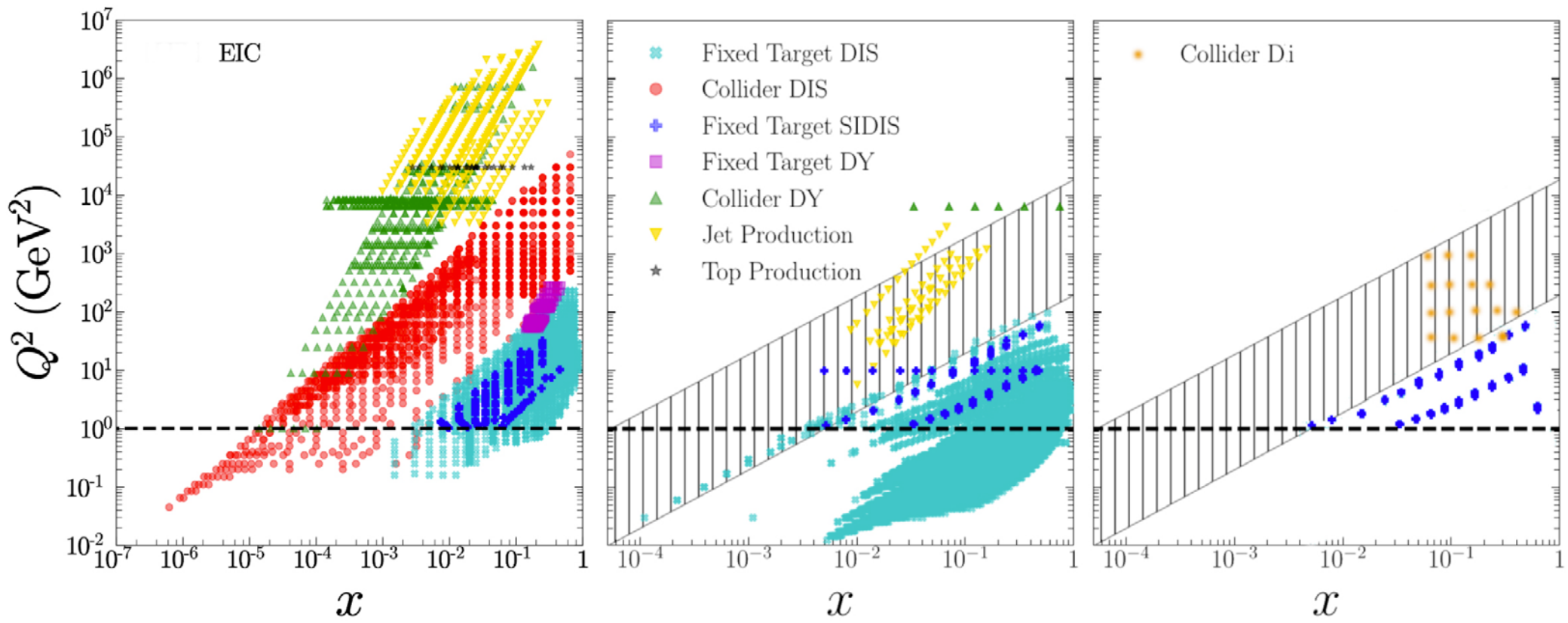

***Fig. 2.7: The kinematic coverage in the ($x$,Q²) plane of the hadronic cross-section data for the processes commonly included in global QCD analyses of collinear unpolarised, helicity, and transversity PDFs. The extended kinematic ranges to be attained by the EIC are also displayed [ Based on arXiv: 2006.08636 [hep-ph] using the arXiv distribution license; published in PPNP 121 (2021) 103908].***

The TPE hypothesis will be tested by the TPEX experiment proposed at DESY, where elastic cross sections with electron and positron beams will be measured and compared. The experiment aims to achieve better accuracy than the previous OLYMPUS experiment. In experimental investigations of FFs, the space-like region is being extended to the domain up to $Q^2$ = 15 GeV² using proton and neutron data to unravel the quark-flavour decomposition of FFs. Future experiments at CERN, JLab, MAMI, MESA, PSI and ULQ2 will probe the very low-$Q^2$ region for precision measurements of the proton radius (see Box 2.1 and Fig. 2.6). In the time-like region, strange and charm baryons will be studied with large data samples from BESIII and Belle II, providing information on the role of heavy flavour. The recently upgraded HADES experiment will give access to the low-$Q^2$ transition FFs of light baryons and hyperons through Dalitz decays. In the longer term, PANDA will probe the high-$Q^2$ region through $\bar{p}p\to e^+e^-$ and $\bar{p}p\to\mu^+\mu^-$ processes.
Recent $J/\Psi$ production experiments at JLab, accessing gravitational FFs, paved the way for a deeper understanding of the role of gluons in the emergence of hadron mass. These studies will be followed by ePIC at EIC. A Compton scattering experiment on light nuclei is underway at HIGS (TUNL, USA), to further increase the precision of proton and neutron polarisabilities. The Virtual Compton scattering experiment VCS-II at JLab will investigate an unexpected observed behaviour of generalised proton polarisability. GlueX has collected data on the Primakoff production of pion pairs and the ongoing analysis aims at revealing the neutral pion polarisability for the first time. For kaon polarisabilities, a first measurement was proposed by AMBER via the Primakoff reaction $K^-Z\to K^-\gamma Z$. An upgrade of the M2 beamline at CERN would enable this and other unique precision measurements. AMBER phase 2, with its kaon programme, would provide unique insights into low-energy QCD.
The ePIC experiment, planned at the approved EIC facility that will be built during the time span of this LRP, opens new horizons for hadron structure. The unique electron-nucleus scattering programme of ePIC will allow comprehensive studies of hadron structure, including a detailed study of nuclear PDFs. ePIC will significantly improve the database of both fragmentation functions and PDFs over a wide range of low-to-medium $x$. From Fig. 2.7, where the kinematic coverage in the ($x$,$Q^2$) plane is displayed, it is evident that the EIC will play a prominent role in the future for both transversity and helicity.

Key contributions to PDFs, TMDs, and GPDs will be provided by already collected data from COMPASS and upcoming campaigns at JLab, the final years of operations at RHIC, and at AMBER, where the pion and kaon PDFs will be measured through DY and charmonium production. The latter will provide additional insights into the emergence of hadron mass as well as aiding global PDF fits.
At the TeV scale, the LHC experiments offer access to the PDFs at the smallest values of $x$ with measurements of forward production of photons (ALICE FoCal upgrade), heavy flavour mesons (LHCb) and jets (all experiments) to explore saturation effects. Dedicated measurements of gluon fragmentations that are important for particle production are planned for runs 3 and 4 of LHC. Using a new generation of polarised gas targets, the LHCspin project will explore an uncharted kinematic region, enabling access to heavy quarks to characterise nucleon spin structure.

At JLab, several electron scattering experiments on various targets will facilitate a comprehensive investigation in the valence-quarks region with unprecedented precision. The luminosity will be three orders of magnitude higher than the previous generation of experiments in Europe. The TDIS experiment at JLab will illuminate the elusive mesonic content of the nucleon, and serve as an important precursor for meson structure studies via spectator tagging at EIC.
Ongoing efforts to increase the luminosity of CLAS12 by two orders of magnitude pave the way for GPD measurements using the Double-DVCS reaction, with a virtual photon in the final state. This will unravel the dependence of the GPDs on an additional kinematic variable, not accessible by DVCS. The HERMES beam-charge asymmetry measurements of DVCS have demonstrated the potential of studying GPDs with a positron beam. An accelerator R&D effort is therefore ongoing at JLab to produce a highly-polarised positron beam. The future ePIC experiment at EIC will be ideal to pursue the measurement of GPDs via DVCS and DVMP in the sea-quark/gluon regime.
While several quark TMDs have been extracted from experimental data in recent years, gluon TMDs are mostly unknown. The EIC programme, spanning deep into the gluon sea, is complementary to the ongoing experiments at RHIC, JLab and CERN. It builds upon the exploration of the gluon-dominated kinematics previously conducted at HERA, with the benefit of a much higher luminosity.
The investigation of hadron structure will continue worldwide, exploiting a set of complementary facilities in Europe and elsewhere. European scientists play particularly strong roles at JLab and EIC (US) and contribute significantly to BESIII (China) and Belle II (Japan). The prospects for new, ground-breaking insights into the puzzling structure of visible matter are therefore excellent.

## Hadron Interactions

Quarks play a crucial role in the evolution of the universe and its structures, making hadron interactions a link between macroscopic and microscopic scales. This manifests in massive stellar objects, such as neutron stars. Their extreme density leads to a rapid increase of the chemical potential at the centre, which should allow for the conversion of nucleons into hyperons. However, this relieves the Fermi pressure, which softens the equation of state (EoS). As a result, the maximum mass reduces to about 1.4 solar masses, which contradicts the observation of neutron stars with masses around two solar masses. This conundrum, illustrated in Fig. 2.9, is referred to as the *hyperon puzzle.*
A major component in the neutron star EoS is the strong interaction at the femtometre scale. It has been argued that a repulsive core in the two-body nucleon-hyperon (*NY*) or hyperon-hyperon (*YY*) interaction, in combination with three-body forces (e.g. *NNY* and *NYY*) would stiffen the EoS and suppress the presence of hyperons inside neutron stars.

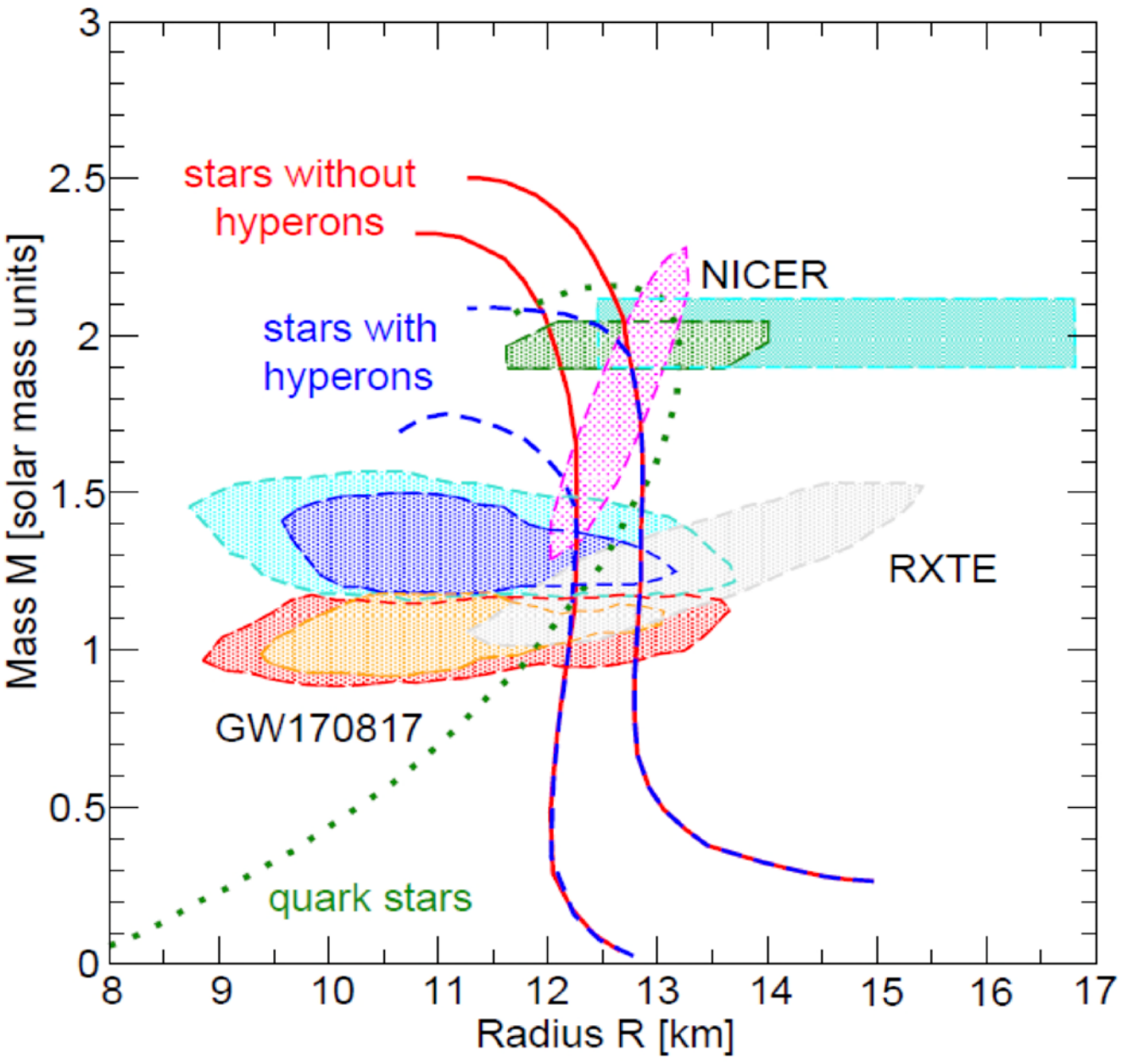

*Fig. 2.9: The blobs represent the mass and radius constraints from the gravitational wave detectors LIGO and Virgo, as well as NASA's Neutron star Interior ExploreR (NICER), while the dashed and solid lines represent the predictions obtained with various EoS models, with and without hyperons. Picture credit I. Vidana (Universita di Catania).*

In experiments, the short life-time of the hyperons make them unfeasible as beams or targets and they are therefore difficult to study in conventional scattering experiments. Instead, interactions involving unstable hadrons can be explored in final-state or secondary interactions, and with the femtoscopy technique. These approaches, in combination with modern, high-intensity facilities, enable a significant advance in our understanding of the *YN* interaction. Furthermore, spectroscopic measurements of hypernuclei provide a precision instrument for hyperon interactions in few-body and many-body systems.

Other examples of hadron interactions as an instrument to understand the interplay between microscopic and macroscopic scales are dark matter searches, the origin of cosmic rays and the properties of neutrinos. Since the next-generation experiments will predominantly use nuclear targets, it is essential to have a reliable understanding of hadron interactions. Significant deviations from the *impulse approximation* — wherein a probe interacting with a nucleus only sees one nucleon, behaving as if it were unbound — have long been observed in the quenching of the axial-vector coupling of nuclei. The same is true for the EMC effect, *i.e.* the observation that the cross section for deep inelastic scattering of an atomic nucleus is different from that of the same number of free nucleons. It is a challenge to understand these multi-hadron effects theoretically and it will require efforts using both nuclear chiral effective field theory and lattice QCD.

In lattice QCD calculations, the primary method to study hadron interactions is to use finite-volume quantisation conditions, which are well understood for any two-particle system. It has also become standard to study resonances and shallow bound states using the rigorous approach of identifying poles in scattering amplitudes. A second method used primarily by the HAL QCD collaboration, is to obtain a hadron-hadron potential, which eventually yields scattering observables. $\chi$EFT provides a description of *NN* interactions incorporating chiral symmetry with input from $\pi N$ scattering. It has been developed up to N$^4$LO, exploiting the rich database in the nuclear sector, and leads to a good description of low-energy *NN* scattering.

# Recent highlights

## Hyperon-nucleon interaction:

Recently, there has been a remarkable advance in the study of the *YN* interaction: At CLAS at JLab and E40 at J-PARC, studies of secondary and/or final state interaction have provided extensive data on low-energy $\Lambda N$ and $\Sigma N$ scattering. Together, these data provide precise constraints to $\chi$EFT calculations. Results from $\chi$EFT for the $\Sigma N$ and $\Lambda N$ cross sections now show some agreement with experimental results at E40 at J-PARC and other experiments, see Fig. 2.15 in Box 2.5. In parallel, two-particle momentum correlation functions of e.g. $\Lambda p$, $\Lambda\Lambda$, $\Sigma p$, $\Xi p$ and $\Omega p$ have been obtained with the femtoscopy technique using *pp* and heavy ion collision data from STAR and ALICE. Part of these data are shown in Fig. 2.10 and they provide important benchmarks for lattice QCD. Calculations of *YN* potentials at near-physical quark masses using the HAL approach have produced a qualitative agreement with femtoscopy correlation functions as shown in Fig. 2.10. In addition, state-of-the-art lattice QCD calculations of baryon-baryon interactions using the finite-volume approach and with increased scrutiny of systematic errors show that the deuteron is unbound at heavy pion mass.

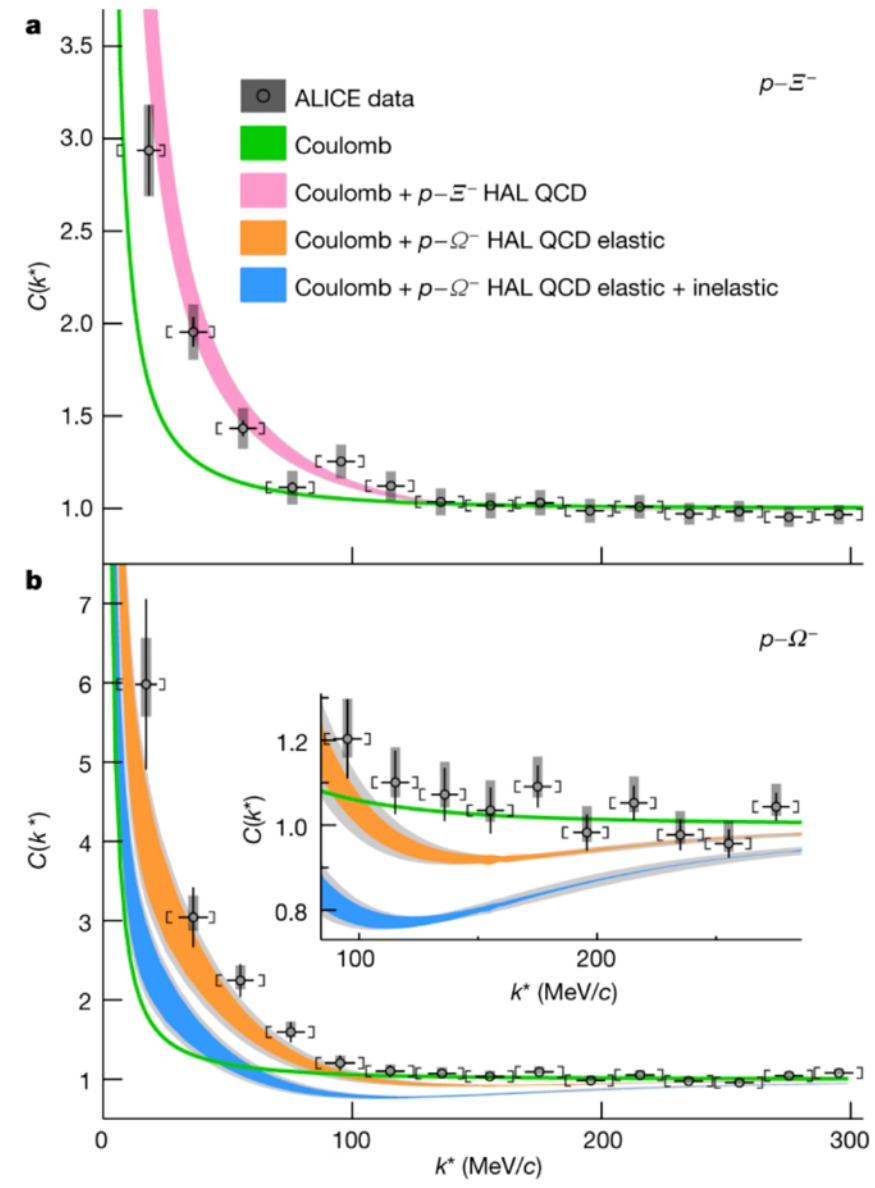

*Fig. 2.10: Femtoscopy study of $p\Xi^-$ and $p\Omega^-$ correlations from ALICE and comparison with results based on baryon-baryon potentials calculated using lattice QCD from HAL QCD. Figure from Nature 588, 232-238 (2020).*

## Hypernuclei

Following recent advances in the understanding of the YN interaction, the next step is to describe a more complex system. For instance, the aforementioned N$^2$LO *YN* potentials derived from EFT calculations have been utilised to predict properties of light hypernuclei. However, the limited database for hypernuclei, to date including only ~40 single-strange and six double-strange hypernuclei, has hindered large-scale, systematic validation of theory predictions. This said, the hypertriton serves as a litmus test for the theoretical framework of many-body systems with strangeness. Accurate knowledge of its properties constitutes the initial step towards a reliable description of heavier hypernuclei. The binding energy and lifetime of the hypertriton have been measured by different experiments, such as HypHI and STAR, and even more recently with high precision by ALICE. The latter measurement reveals a lifetime close to that of a free $\Lambda$, indicating a loosely bound $\Lambda$ . New precision measurements, including branching ratios of various decay channels, pave the way for a systematic description of strange nuclei. Further insights are obtained from the $\Lambda$ separation energies in mirror hypernuclei such, as $^{4}_{\Lambda}H$ and $^{4}_{\Lambda}He$, probing isospin symmetry.

## Dibaryons

Dibaryons provide another means in studying to the baryon-baryon interaction, in particular using dibaryons with a compact hexaquark structure rather than deuteron-like states of colour-neutral constituents. Hexaquarks are also of interest in the context of neutron stars, since they are bosons and therefore not affected by the Pauli principle when formed in dense nuclear matter. The first hexaquark candidate found in experiment is the $d^{*}$(2380), discovered by WASA at COSY. Further searches with photon beams with A2 at MAMI are ongoing. Recent studies of spin observables in photon-beam experiments provide support for the hexaquark interpretation, but detailed information on electric and magnetic multipole moments, form factors, etc. will be crucial to establish its inner structure. Lattice QCD studies include the deuteron, the $d^{*}$(2380), as well as the $H$ dibaryon, $N\Omega$, $\Omega\Omega$ and several bound states of charm and bottom baryons.

## Meson interactions

Meson-baryon interactions provide an important component to understanding the strong interaction in general, but are also crucial in order to correctly interpret, e.g. hadron spectroscopy. The strange $\Lambda$(1405) resonance is a prominent example: it has been found to be dynamically generated by the kaon-nucleon interaction. Kaon-nucleon interactions are experimentally accessible in kaon beam experiments, through kaon-nucleon femtoscopy, as demonstrated by ALICE, and in kaonic atoms (see chapter 6 on symmetries).

## Hadronic three-body interactions

Three-body forces are a crucial component in the EoS of neutron stars and have long been investigated with EFTs and Fadeev equations. Also, in lattice QCD, there has been a considerable effort to develop methods for three-particle scattering. After development of three-particle quantisation conditions, the first lattice QCD calculations were carried out, initially with three $\pi^{+}$ but recently also combinations of pions and kaons. Functional methods, based on approximate solutions of Dyson-Schwinger equations, Bethe-Salpeter equations and Fadeev equations, have led to substantial progress over recent years.

# Future prospects

The need for precise and accurate measurements of the hypertriton properties is recognised by the scientific community worldwide, and several ongoing and planned experiments are devoted to this endeavour. Prominent examples are the WASA-FRS experiment at GSI/FAIR and the P73/P77 collaborations at J-PARC, which will perform precision lifetime measurements. Also, ALICE data from LHC runs 3 and 4 can be utilised for this purpose and also to measure three-body YNN interactions with the femtoscopy technique. The hypertriton binding energy can be measured at MAMI, Jefferson Lab, ALICE and J-PARC E07, where the extension of J-PARC will enable γ-ray spectroscopy.

Heavy-ion collisions and radioactive beams give access to previously unexplored territory on the hypernuclear chart, far from the β-stability line. This is in contrast to emulsion experiments or γ-ray spectroscopy with Germanium detectors, which require stable targets. With these new techniques, FAIR has the potential to take a leading role through the NUSTAR experiments Super-FRS and R3B. Furthermore, R3B will probe the halo structure using radioactive beams and HADES at GSI/FAIR will study the production of light hypernuclei in detail.

Double-strange hypernuclei, in particular their excitation spectra, probe three-body $\Lambda\Lambda N$ interaction, as well as $\Lambda\Lambda$-$\Sigma\Sigma$-$\Xi N$ mixing. The current data bank is scarce, but ongoing efforts with the emulsion experiment E07 at J-PARC, the heavy-ion experiments ALICE, STAR and the future CBM at FAIR, have the potential to drastically change this picture. In the future, ALICE 3 will be able to search for charm-nuclei, such as the charm-deuteron. However, the techniques applied at these experiments only give access to ground-state hypernuclei. PANDA is the only planned experiment where the excitation spectrum of light double-strange hypernuclei can be measured in the future.

Charged hyperons can be captured in an atomic orbit of a nucleus and form a hyperatom. When the hyperatom de-excites, it emits X-rays that can be measured in Germanium detectors. Hyperatoms are studied at J-PARC, where the hyperons are produced by a kaon beam impinging on an extended target where also the re-scattering takes place. For target materials with atomic number Z below 50, the emitted X-rays escape the target and can be detected. In the future PANDA experiment, the stored antiproton beam will also result in high hyperon production rates for very thin targets. These hyperons will be captured in a likewise thin secondary target from which X-rays can also escape when the target consists of heavy elements. This makes PANDA unique for studies of the hyperon-nucleus potential in neutron-dense environments.

Properties of the $d^{*}$(2380) dibaryon, such as electric quadrupole moments, magnetic octupole moments and transition form factors, will be studied within the experimental programmes at MAMI, ELSA and JLAB. The results will be important to establishing the nature of the $d^{*}$(2380) and the role of dibaryons in the EoS of neutron stars. Searches for other dibaryons are also ongoing. Cosmic ray antinuclei can be a signature of exotic reactions, such as dark matter annihilations. Interpreting data from astroparticle experiments, such as AMS and GAPS, requires a solid understanding of all processes involved in the production of antinuclei. This can be achieved with data from ongoing and future ventures with ALICE and LHCb, AMBER, NA61/Shine, Belle II and FAIR.

The recent advances and the continued progress expected in femtoscopy and baryon-baryon scattering measurements from ALICE, STAR, JLab, J-PARC and the future CBM have inspired the theory community. Lattice QCD calculations will yield scattering amplitudes for $NN$, $YN$ and $YY$, providing considerable inputs for models and EFTs. To be comparable with experimental results, the systematic uncertainties of these calculations need to be studied carefully. These efforts include using ensembles generated with physical pion masses and investigation of finite lattice spacing and volume effects. Furthermore, mitigating the signal-to-noise problem will require large data samples and/or new algorithms.

There are a number of lattice QCD calculations of meson-meson scattering amplitudes and resonance poles, while baryon-baryon and baryon-meson computations are still typically done using gauge ensembles simulated with unphysically heavy pions. Within the next five years, we expect more studies on meson-baryon scattering including systems with strangeness, the $\Delta$-resonance and $N\pi$ scattering lengths that can shed light on the nucleon σ-term and phenomenological investigations. An ambitious goal is to make robust predictions of the spectra and matrix elements of two-nucleon systems but also for $YN$ and $YY$ systems. These efforts include the full error budget with controlled systematic uncertainties that will require light pion masses, fine lattice spacings, and large volumes. Since many resonances, *e.g.* the Roper, can couple to three or more particles, and since three-baryon interactions are important for (hyper)nuclear physics, there are considerable efforts to develop methods for three-particle scattering. Such calculations and further extensions to three-body systems can have a significant impact on our understanding of the structure of the lightest nuclei and provide inputs to nuclear theories.

A formalism to study $YN$ interactions in chiral EFT has been successfully developed at $N^{2}$LO. In order to reduce the remaining systematic errors, higher-order potentials are going to be developed in the coming years, leading to accurate predictions for cross sections, the phase diagramme and hypernuclear properties. Future developments of the techniques used for functional methods have the potential to access hadronic scattering amplitudes over a wide range of kinematics in the coming years. The interplay of the aforementioned theoretical approaches will shed new light on our understanding of hadron interactions and our interpretation of forthcoming experimental results. Together, these advances will provide an important step forward to solve the hyperon puzzle of neutron stars.

## Hadronic effects in precision physics and rare processes

Though the SM has been extremely successful in describing fundamental particles and their interactions at high energies, it falls short of providing solutions to a number of elusive puzzles: Why is there so much matter in the universe, but almost no antimatter? There should be dark matter - but what does it consist of? What is the nature of neutrinos? The search for BSM physics using precision probes has taken centre stage in subatomic physics and will remain a crucial hunting ground for years to come. In this endeavour, hadron physics plays a central and interconnecting role in the measurement and interpretation of precision observables. Prominent examples of precision probes are the anomalous magnetic moment of the muon, referred to as the *muon g-2*, rare hadron decays, the overall quest to understand the flavour structure of quarks and neutrinos, as well as searches for electric dipole moments (EDMs). In all aforementioned examples, the precision of the SM/BSM predictions and the subsequent comparison to measurements is limited by our quantitative understanding of the strong interaction. By investigating strong interaction, we will not only expand the frontier in precision physics, but also obtain a new diagnostic toolkit to understand the strong interaction.

The mixing of flavours is one of the great mysteries of the SM. The hierarchical structure of the Cabibbo-Kobayashi-Maskawa (CKM) matrix, which accounts for quark mixing, is very different from the one of the Pontecorvo-Maki-Nakagawa-Sakata (PMNS) for neutrinos, with relatively large off-diagonal elements. In addition, charge-lepton mixing is absent in the SM. Hadron decays and hadron oscillations probe the flavour structure, where BSM effects would manifest as differences between precise measurements and SM predictions.

Precision probes in abundant and rare hadron decays can reveal the violation of fundamental symmetries, such as charge conjugation and parity (CP). CP violation is one of the necessary criteria for *Baryogenesis* - a long-standing yet empirically unverified explanation for the matter-antimatter asymmetry of the universe. Rare decays can also unveil unknown physics in the form of charge-lepton flavour violation or emission of light particles such as axions and dark photons. Oscillations in neutral baryon-antibaryon systems would indicate baryon number violation, another necessary criterion for Baryogenesis. Neutral meson oscillations are an established probe of CP violation, while neutrino oscillations offer access to the PMNS matrix.

BSM physics could also manifest itself in low-energy nuclear experiments, in the form of anomalous nuclear EDMs, and is a frequently suggested explanation for the *muon g-2* anomaly (see Box 2.2). If effects of BSM processes are found, they would imply the existence of hitherto unknown particles. Hence, searches for these effects provide a complementary hunting ground for ongoing efforts at the high-energy frontier. However, nuclear and hadron physics facilities also offer direct searches for low-mass BSM particles. These searches are well-motivated and are so far not accessible by other means. Maximising the sensitivity of the measurements requires a solid understanding of hadronic effects. To this day, no BSM particle has been discovered experimentally. Hence, if such particles exist, they are unlikely to interact with SM particles. For this reason, BSM particles constitute appealing candidates for dark matter.

### Box 2.2: Hadronic contributions to the muon g-2

**The muon *g-2* has attracted a lot of attention, due to the long-standing large discrepancy between theory prediction and measurements, including the recent precise result from Fermilab. The theoretical uncertainties are dominated by hadronic effects such as the hadronic vacuum polarisation (HVP), and can be calculated using two different approaches: one *data-driven*, where experimentally measured hadronic cross sections and meson form factors are used as input to a dispersive analysis, and one based on *lattice QCD* calculations. The recent Fermilab measurement yields a discrepancy concerning the data-driven prediction at the level of ~5.0 standard deviations (σ). The lattice approach, however, leads to a different conclusion: In 2021, the BMW collaboration published the first calculation of the leading hadronic contribution, HVP, with uncertainty at the sub-percent level. The central value was found to be about 2 σ larger than the data-driven calculation, and only 1.7 σ away from the most recent experimental world average. While the full BMW result needs to be confirmed by other, independent lattice calculations, some components have already been confirmed by other collaborations, including European-based groups within ETM and CLS-Mainz. These components concern the contribution to HVP known as the short and intermediate time windows and can also be evaluated using the data-driven approach. Currently, all lattice QCD data are at tension with those from the data-driven approach within the intermediate time window (Fig 2.12). Moreover, a new measurement of the cross section $e^+e^- \rightarrow hadrons$ by the CMD-3 collaboration is found to be significantly higher than in all previous experiments. In the data-driven approach, this would lead to an HVP value between the BMW result and the experimental measurement. The discrepancy between the CMD-3 measurement and the previous ones is large, mainly in the vicinity of the ρ resonance, and cannot be attributed to a statistical fluctuation. The only possible explanation is that either the CMD-3 measurement, or that of all others, suffers from a yet unknown systematic bias. Hence, the muon *g-2* can only be used as evidence for BSM physics if two open issues are resolved: i) the discrepancy between data-driven approaches and lattice QCD results; and ii) the larger value measured by CMD-3 as compared to all other experiments. There is a lot to be learned from all these endeavours: in addition to the outcome of the comparison itself, we will sharpen our tools and thereby reduce the uncertainties in the calculations of the relevant hadronic matrix elements.**

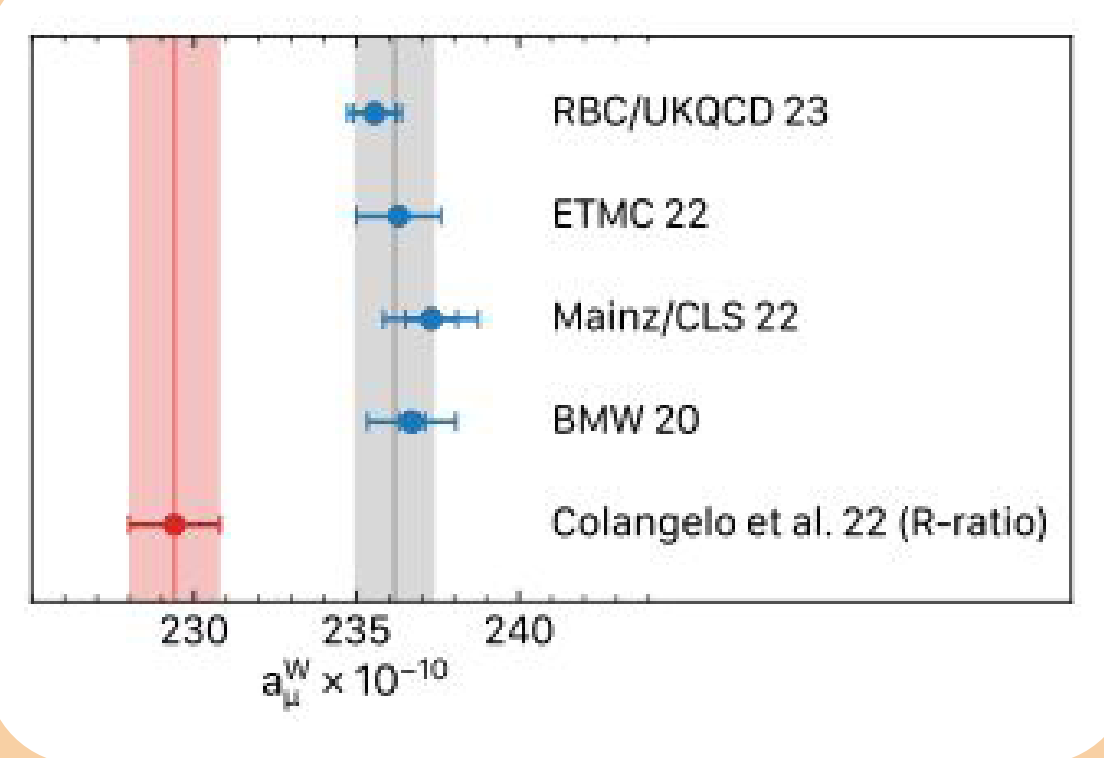

*Fig. 2.12: Contribution to the muon anomalous magnetic moment in the intermediate energy window showing 3.8σ deviation between data-driven (red band) and average lattice QCD results (grey band).*

## Box 2.3: The $\Lambda$ hyperon decay parameter

**The weak, two-body decay of a hyperon is self-analysing: the interference between the parity-violating and the parity-conserving decay amplitudes results in a measurable tendency of the daughter baryon to be emitted in the direction of the spin of the mother hyperon. This effect is quantified in terms of the Yang-Lee parameters $\alpha$, $\beta$ and $\gamma$ and can be exploited in hyperon spectroscopy, polarisation studies, heavy-ion physics and tests of charge conjugation and parity (CP) conservation. The decay parameter $\alpha$ of the $\Lambda$ hyperon was measured in several experiments during the 1960s and 1970s using the proton polarimeter technique and the resulting world average of $\alpha = 0.642 \pm 0.013$ was unchallenged for several decades. However, the insight that hyperon-antihyperon pairs can be spin polarised and entangled inspired the development of a different and more precise technique. When applied on new, large data samples from the BESIII experiment, the precision was greatly improved, opening a new avenue for precision CP tests in baryon decays. The first BESIII measurement was also surprising by being significantly different from the previous world average. This finding motivated a re-analysis of $\Lambda$ production in CLAS data that resulted in a value more consistent with the new result from BESIII. More recently, BESIII data from numerous other, independent reactions and significantly larger data samples confirmed the larger value, and the new world average from PDG is calculated solely based on BESIII and CLAS data.**

**The 2022 world average being 18% larger than the one from 2018, has implications for a large number of polarisation measurements, partial wave analyses and studies involving subsequent decays into $\Lambda$ hyperons.**

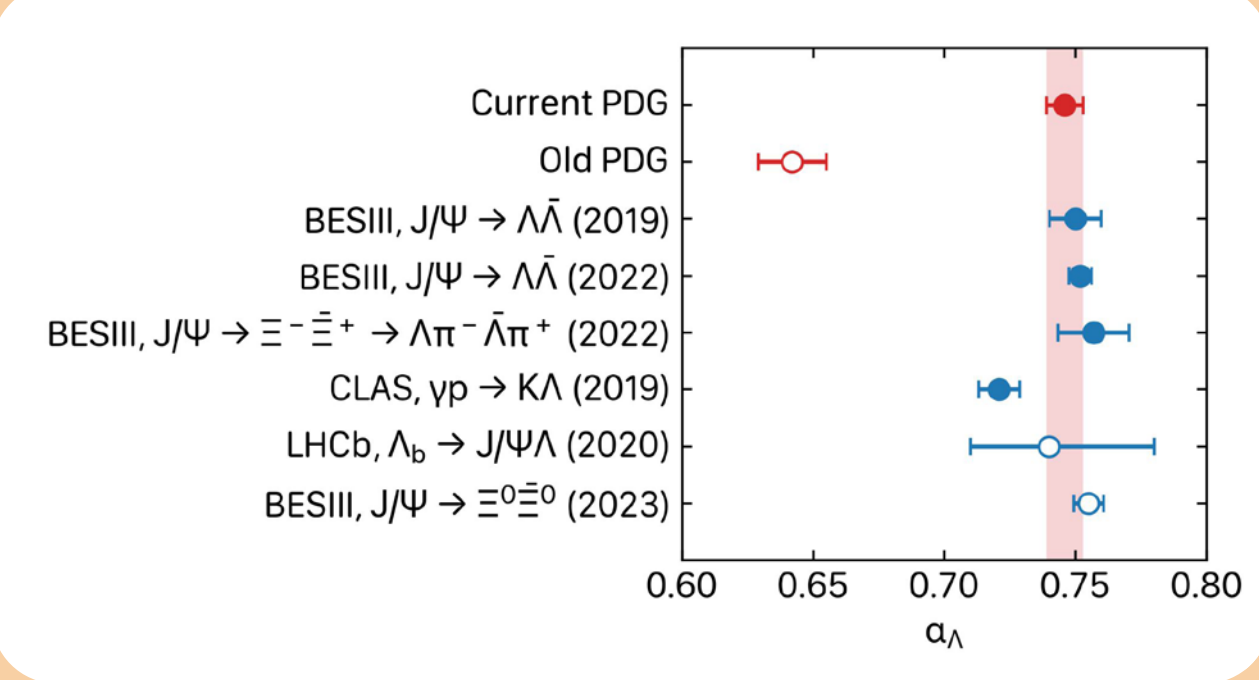

*Fig. 2.11: The PDG world average of the $\Lambda$ decay parameter for all experiments before 2018 (red), more recent measurements from BESIII, CLAS and LCHb (blue) and the 2023 world average from PDG (Prog. Theor. Exp. Phys. 2022, 083C01 (2022 + 2023 update)).*

# Recent achievements and highlights

### Hadron decays

The SM is under scrutiny by the ongoing dedicated kaon decay programme at NA62 (CERN) and KOTO (J-PARC), complemented by measurements at BESIII, Belle II and LHCb. The latter three offer access to rare decays of heavy flavour mesons. In the baryon sector, BESIII and LHCb are studying rare decays of charged $\Sigma$ hyperons, searching for lepton-number violation and flavour-changing neutral currents.

CP violation is well-established in the strange and bottom meson sector and has recently also been observed in *D*-meson decays by the LHCb. However, all observed effects are consistent with the SM. Spin-carrying baryons provide an additional angle to CP violation, which is particularly interesting for hyperons where spin properties are straightforward to access in their weak, self-analysing decays. The recently collected large samples of spin-polarised and entangled hyperon-antihyperon pairs from BESIII, has resulted in a series of high-precision CP tests (see Box 2.3). In particular, sequential decays of multi-strange hyperons enable the separation of strong, CP-conserving contributions on the one hand, and weak/BSM, possibly CP-violating contributions on the other. This separation increases the sensitivity by several orders of magnitude. However, all results so far are consistent with CP symmetry.

In cases where strong and weak contributions cannot be separated, the extraction of relevant quantities from experimental data requires the input of hadronic matrix elements. Lattice QCD is the main source of information, complemented by heavy quark effective theory, soft-collinear effective theory or chiral perturbation theory in the case of light quark systems. As QCD uncertainties are reduced, a more precise treatment of QED corrections becomes important.

### Neutrino interactions and oscillations

The existence of neutrino oscillations has established that lepton flavour symmetry is broken, providing the so far only direct indication of BSM physics. Precise determination of oscillation parameters in reactor and accelerator experiments, such as T2K and NOvA, has been undertaken in recent years. This experimental programme critically relies on our understanding of neutrino interactions with nucleons and nuclei, which is improving thanks to recent data from T2K and MINERvA. Extraction of the nucleon axial form factor and parton distribution functions have been pursued, as well as weak meson production. Electron- and meson-nucleon-scattering data are also relevant to developing realistic simulations of neutrino interactions at the detectors.

### Dark Matter searches

In a recent measurement by the ATOMKI collaboration, an excess was observed in the angular distribution of $e^+e^-$ pairs emitted in the transition of the $^8$Be, $^4$He and $^{12}$C nuclei. This observation led to the hypothesis of a Feebly Interacting Particle (FIP), a dark matter candidate that decays into an $e^+e^-$ pair. This particular FIP candidate is referred to as the X17, due to the invariant mass of the $e^+e^-$ pair. Other searches for FIPs have been conducted by the NA64 and NA62 experiments at CERN, so far without any beeing discovered. The general absence of discoveries of dark matter candidates in the low to medium energy ranges has constrained the parameter space of several proposed models. Precise tests of fundamental symmetries are also important in this regard. For example, nEDMs are sensitive to mirror matter, a dark matter candidate restoring parity, while oscillating EDMs can be induced by axions and axion-like particles.

# Future prospects

## Radiative corrections in hadronic processes

Computation of radiative corrections are not only important for the muon *g-2*, but also for other hadronic quantities where high precision matters. Currently, there are no established reliable methods to calculate these corrections other than chiral perturbation theory that applies only at low energies. Recently, there have been promising developments in the formulation of a dispersive formalism for radiative corrections for $e^+e^- \to hadrons$. The first results are promising and motivate further efforts.

The precision of contemporary lattice QCD calculations is reaching a level where QED corrections are needed. However, it is a challenge to include QED in lattice QCD. Recently, several approaches have been developed and are applied to calculate isospin-breaking corrections.

## Hadron decays

In the light meson sector, the searches for rare decays will be further pursued by kaon decay experiments like NA62 and KOTO, but also by the future JLab Eta Factory (JEF), addressing $\eta/\eta'$ decays. The latter will allow us to determine transition form factors of relevance for hadronic light-by-light scattering in muon *g-2* studies. The $\pi^+$ decay programme planned by the PIONEER experiment at PSI aims to test lepton universality and determine the matrix element $V_{ud}$ with high precision. Equally important is the detailed study of heavy-quark observables. The excellent statistical precision expected at LHCb and Belle II can probe lepton flavour universality violation in $b \to s$ and $b \to c$. These experiments will also search for flavour-changing neutral currents, perform CKM-unitarity tests and search for sources of CP violation.

The studies of precision probes in hyperon decays will continue at BESIII, Belle II and LHCb. BESIII has its focus on strange and to some extent single-charm baryons, while Belle II and LHCb can access a broad range of heavy baryons with bottom and charm quarks. These studies will enable searches for rare decays involving for example, flavour changing neutral currents, as well as CP tests. The improved sensitivity to CP violation achieved in decays of polarised and entangled hyperon-antihyperon pairs can be exploited in the future by using longitudinally-polarised $e^+e^-$ beams, or with the expected huge data samples from PANDA.

Experimental progress calls for corresponding theoretical advances. More precise lattice QCD computations with controlled systematic errors are eagerly awaited. Systematically improvable approaches are valuable but the convergence of various perturbative expansions needs to be better understood. In searches for BSM physics, Wilson coefficients accompanying higher dimensional operators in an effective theory can then be extracted model-independently.

## Neutrino interactions and oscillations

The future HyperKamiokande and DUNE experiments aim to establish CP violation in the lepton sector. This calls for realistic modelling and precise measurements of neutrino scattering cross sections over a broad range of energy scales. The CERN Forward Physics Facility will provide the neutrino beam necessary to probe not only the Deep Inelastic Scattering regime but also the quark-hadron transition region. Complementary measurements are possible at electron scattering facilities e.g. MAMI and JLab, and at hadron scattering experiments, such as ProtoDUNE at CERN. The NA61/SHINE experiment at CERN is expected to play a key role in reducing the uncertainties in the neutrino flux in hadron scattering experiments. All these efforts will bring this field into the precision era. Furthermore, accurate determination of nucleon axial form factors with controlled systematics will provide valuable input for the ongoing neutrino physics programmes. As the formalism for resonances develops, the $N \to N^*$ and $N \to \Delta$ axial transition form factors will complement such input.

## Dark Matter searches with PADME

A complete programme of measurements exploiting low-energy $e^+e^-$ annihilation will be conducted by the PADME experiment at the Laboratori Nazionali di Frascati (LNF) in Italy. The linear accelerator at the LNF can deliver positron beams in the energy range 250-550 MeV, enabling the production of FIPs, axion-like particles and potential light scalar particles, such as a dark Higgs, $h'$. The latter would be responsible for the mechanism that generates the mass of the dark photon, A'. The PADME experiment will be unique since no other existing or planned facilities can deliver positron beams in this energy region. The outcome of the analysis of the first energy scan performed in 2022 will shape upcoming investigations.

## Precision Nuclear Physics with MESA

In 2025, the new Mainz Energy-Recovering Superconducting Accelerator (MESA) and its three experiments MAGIX, P2 and DarkMESA will launch at the Johannes Gutenberg University in Mainz. This will provide the basis for a rich physics programme in precision nuclear, hadron and particle physics for decades to come.

The MAGIX spectrometer will make possible a new generation of electron scattering experiments, ranging from nuclear cross section measurements of astrophysical relevance and measurements of nucleon electromagnetic form factors, to searches for BSM particles. The P2 experiment will perform the most precise measurement of the weak charge of the proton by parity-violating electron scattering with a spin-polarised beam. By confronting a P2 measurement (see Fig. 2.13) with the SM prediction, mass scales for BSM particles up to 49 TeV can be probed. Furthermore, a measurement of the parity-violating asymmetry using nuclear targets can probe the neutron skin thickness of nuclei and therefore allow extraction of the EoS nuclear matter, crucial for the understanding of neutron stars (see section on hadron interactions). Finally, the beam dump of P2 can be used as a target for the production of dark matter particles. For their direct detection, a dedicated calorimeter experiment will be placed behind the beam dump, thus providing the world's best sensitivity for light dark matter particles.

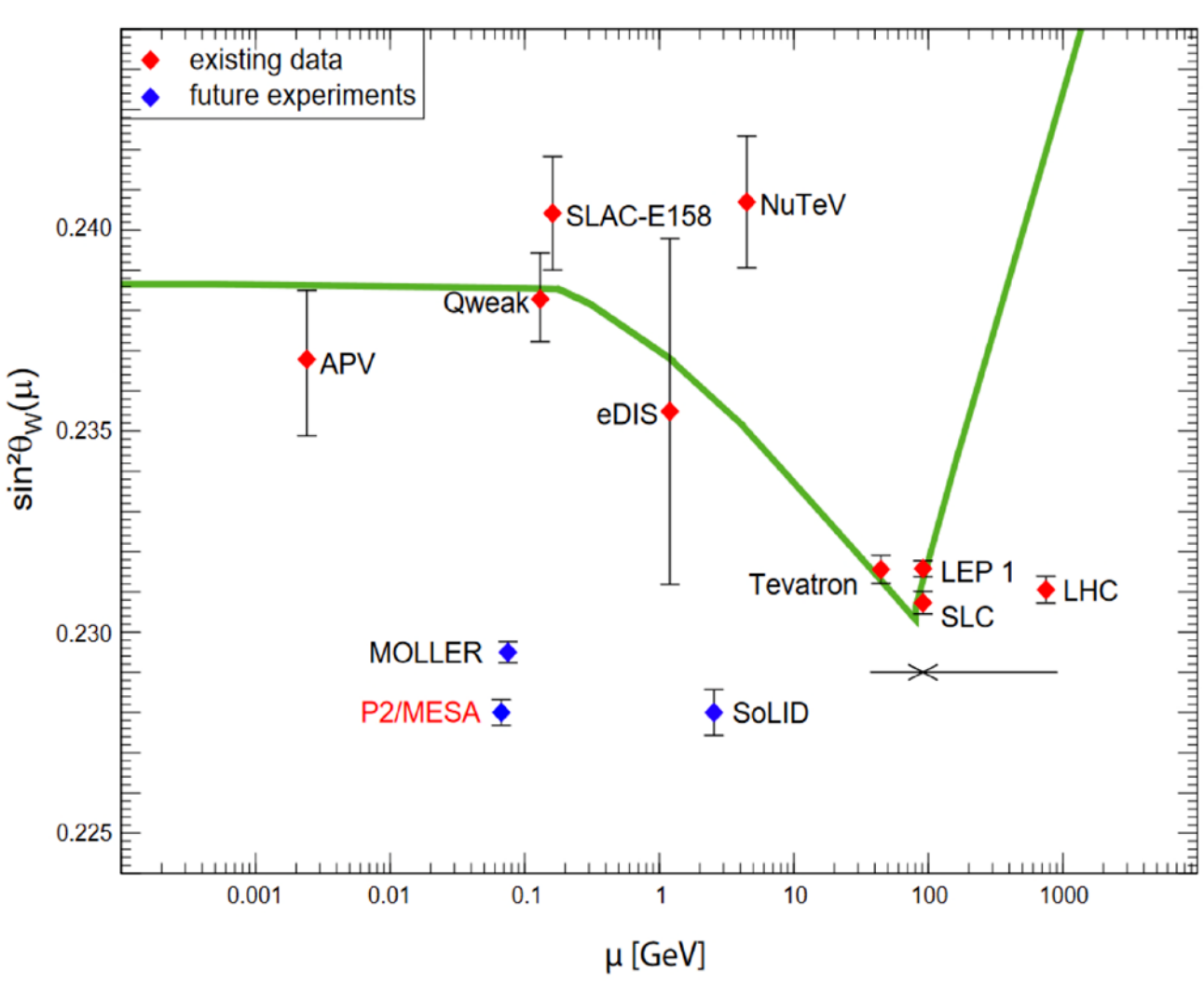

*Fig. 2.13: Scale dependence of the electroweak mixing angle estimated within the SM (green curve) together with existing experimental determinations (red points). Shown is the expected accuracy of the future P2 experiment at MESA together with future results from the MOLLER and SoLID experiments at JLab (blue data points). A comparison of the P2 experiment with the SM prediction allows one to probe the mass scales of BSM particles of up to 49 TeV.*

## Box 2.4: Lattice QCD

**The lattice QCD community in Europe consists of several large collaborations and smaller groups, all contributing to an international research endeavour. Major European collaborations include groups from outside Europe and comprise the ALPHA, BMW, CLS, ETMC, HotQCD, QCDSF and UKQCD collaborations, each using its own simulation and analysis codes. These collaborations generate and analyse gauge ensembles that include physical values of the QCD mass parameters, known as physical pion ensembles. The analysis of these ensembles has led to significant improvements in the precision and predictive power of lattice QCD calculations across all domains of hadron physics: i) in spectroscopy, many QCD-stable hadrons, below strong-interaction thresholds, are determined with sub-percent precision and include a complete uncertainty quantification; ii) in hadron interactions, the finite volume approach has successfully identified poles in scattering amplitudes and has been extended to beyond meson-meson scattering to study systems with baryons or with three mesons; iii) remarkable progress has been achieved in hadron structure, where besides Mellin moments, a new formalism has been developed for directly computing GPDs and TMDs; iv) in precision physics and searches for BSM physics, lattice QCD is providing valuable in put as exemplified by the computation of the hadronic contributions to the muon *g-2* with unprecedented accuracy highlighted in Box 2.2.**
**The impact of lattice QCD in our understanding of nucleon structure is underscored by recent calculations of the spin carried by quarks and gluons in the nucleon. These calculations revealed that half the spin of the nucleon is carried by sea quarks and gluons. The results, presented in Fig. 2.14, help resolve a 30-year puzzle of nuclear physics. With European investment in high-performance computing, enabling the construction of exascale machines, we anticipate further acceleration of this progress in the coming years.**

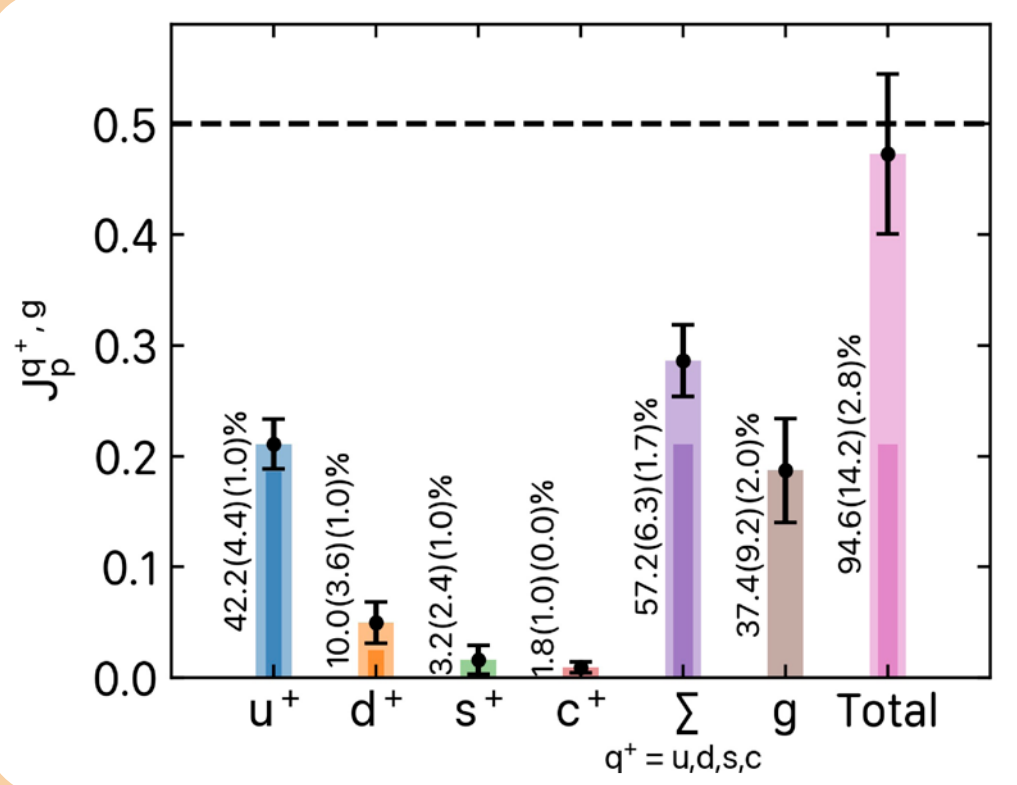

*Fig. 2.14: The decomposition of the proton spin. Whenever two overlapping bars appear, the inner bar denotes the purely connected contribution while the outer one is the total contribution. Individual quark and gluon contributions are given in the $\overline{MS}$ scheme at 2 GeV. The dashed line denotes the spin sum. Figure adapted from C. Alexandrou et al. (ETMC) Phys. Rev. D 101 (2020) 9, 094513.*

## Box 2.5: Effective Field Theories

**The concept and techniques of effective field theory (EFT) provide a powerful framework for the study of hadronic properties and interactions. An EFT only considers the degrees of freedom relevant for describing a particular physical phenomenon at a given scale length or energy scale, while any substructure and degrees of freedom at shorter distances / higher energies are disregarded.**

**The EFT Lagrangian, respecting all pertinent symmetries, is characterised by an infinite number of initially unknown low-energy constants (LEC) to be determined directly from QCD or from experimental data. Underlying power-counting rules establish a hierarchy among these terms, enabling predictions with a certain precision by including contributions up to a specific order. A systematic improvement of the accuracy in determining an observable quantity is achieved by going to higher orders, albeit at the cost of introducing an increasing number of LECs. Assessing the uncertainty associated with the truncation of the expansion is an important and often challenging task.**

**Chiral EFT (χEFT) describes interactions among baryons. It has long been established for nuclear interactions, benefitting from the extensive nucleon-nucleon data (see the nuclear structure chapter). In recent years, advancements in χEFT calculations of hyperon-nucleon interactions have been made, as shown in Fig. 2.15, partly driven by the prospect of a significantly improved data base in the strangeness sector.**

**Chiral Perturbation Theory (χPT) is a low-energy realisation of QCD that successfully describes the interactions of hadrons composed of light (*u,d,s*) quarks. χPT can also be combined with other methods, valid even beyond its domain of applicability, such as functional methods or dispersion relations, to provide a rigorous constraint at low energy. The synergy between χPT and lattice QCD is highly productive: χPT allows the determination of LEC from lattice QCD that might be difficult to obtain from experiment, while χPT often provides the correct analytic form for the continuum and chiral extrapolations of lattice QCD calculations. Furthermore, χPT helps constraining excited state contamination of nucleon matrix elements determined by lattice QCD.**

**In the infinite quark mass limit, interactions become independent of flavour and spin. Heavy quark EFT relies on this symmetry of QCD to study the properties of charmed and bottomed hadrons. Examples of EFTs applied to doubly heavy exotic hadrons are non-relativistic QCD and Born-Oppenheimer EFT, both working with the heavy quarks as degrees of freedom. In contrast, the combination of this heavy quark spin-flavour and chiral symmetries gives rise to Heavy Hadron χPT. The latter employs light and heavy hadrons as degrees of freedom and is valuable for the study of hadronic molecules.**

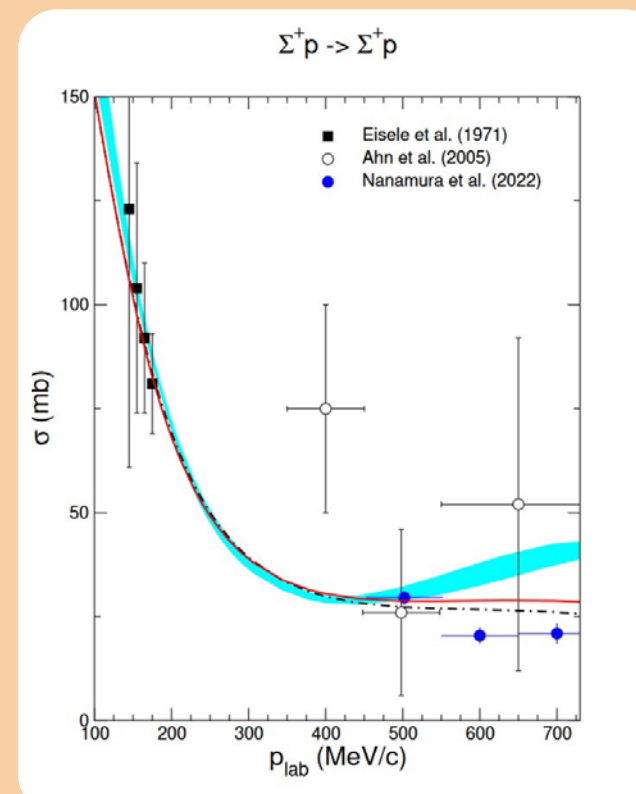

*Fig. 2.15: Chiral EFT calculations of the cross section of elastic $p\Sigma^+ \to p\Sigma^+$ scattering as a function of the momentum in the lab system. The new data come from the E40 experiment at J-PARC and the curves represent calculations with a semi-local momentum-space (SMS) chiral YN potential at NLO (dashed-dotted) and $N^2LO$ (solid), described in Haidenbauer et al. in Eur. Phys. J. A 59 (2023) 63. The 2022 data points come from the E40 experiment at J-PARC. Also included are results from NLO calculations by the same group in 2019 (cyan band). Picture credit J. Haidenbauer.*

# Perspectives

## European computational infrastructures

Access to large pan-European supercomputing resources was provided by the Partnership of Advanced Computing in Europe (PRACE) from 2012 to 2022 and after that by the EuroHPC Joint Undertaking (JU), which currently includes three pre-exascale and two exascale systems, see chapter 6 (Infrastructure). The computational resources provided by pan-European machines, supplemented by national computing facilities, have led to tremendous progress in lattice QCD calculations. Provided the algorithms and codes needed for these computer architectures are developed, progress is expected to accelerate during the next decade. In parallel to classical computing, there has been a revolution in quantum computing. Lattice gauge theories are being used to develop algorithms and error-mitigation approaches. Quantum computers do not only have the potential to solve problems that classical computers cannot handle, but they also enable the development of new energy-saving technologies. Six quantum computers were funded by EuroHPC JU in 2023 to drive science and innovation, see chapter 8. Hadron physicists are therefore in a good position to prepare the community to benefit from these new technologies.

**European Centre for Theoretical Studies (ECT*)**, Trento, Italy
**ECT*** is the only European centre dedicated to theoretical nuclear physics in the broadest sense. It constitutes a platform for a wealth of workshops and training schools and is complementary in scope and activities to research facilities based at universities, research centers and experimental laboratories.

## European experimental facilities

**CERN**, Switzerland: The new, dedicated hadron physics experiment **AMBER** will perform unique measurements in structure and spectroscopy, using muon and hadron beams with momenta up to 300 and 100 GeV/c, respectively, working on fixed targets. The programme will go substantially beyond that of the predecessor COMPASS. The **LHCb** experiment is specialised in b-quark physics from proton-proton collisions at energies up to 13 TeV. Despite being originally designed for flavour physics, LHCb has also contributed significantly to the field of heavy meson spectroscopy and the search for exotic hadrons. Hadron spectroscopy can also be carried out with the proton and heavy-ion beams of the **ALICE** experiment. In recent years, ALICE has proved successful in hyperon femtoscopy.

**ELSA** in Bonn, Germany: Polarised and unpolarised electron beams, operating at energies up to 3.2 GeV, serve two hadron physics experiments: **CBELSA/TAPS** and **BGO-OD**. This enables a successful spectroscopy programme including light mesons, light baryons and single-strange baryons.

**MAMI** and **MESA** in Mainz, Germany: The high intensity of the MAMI electron and photon beams, with the additional option of beam polarisation, provides excellent conditions for precision structure studies of nucleons and nuclei, as well as hadron spectroscopy. The new MESA facility and its experiments **MAGIX, P2** and **DarkMESA** will become operational in 2025, providing the basis for a rich physics programme in the fields of low-energy nuclear, hadron, and particle physics.

**HADES** at GSI, Germany: Originally designed for heavy-ion physics, HADES has increased its scope by an upgrade, with several new detectors. The hadron physics programme is a part of the FAIR Phase 0 initiative and was launched in 2022. The projected pion beam campaign and a planned joint initiative with the CBM experiment at FAIR will further increase its scope.
The aforementioned facilities are described in more detail in Chapter 6 (Infrastructures). Together, they offer a broad spectrum of research opportunities for European hadron physicists and attract collaborators from all parts of the world.

The future European large-scale project **PANDA at FAIR**, Darmstadt, Germany, will use a uniquely intense stored beam of antiprotons in the range 1.5 - 15 GeV/c, which corresponds to the production of strange and charm hadrons. The $\bar{p}p$ / $\bar{p}A$ annihilations, in combination with a large-acceptance, versatile detector, will enable a broad physics programme including hadron spectroscopy, new aspects of hadron structure, interactions between anti-hadrons and nuclei, precision probes in hadron decays as well as hyperatom and hypernuclear studies. Though the PANDA experiment is delayed, individual detector systems are already in operation within the FAIR Phase 0 initiative. Prominent examples are the joint PANDA@HADES hyperon programme using PANDA tracking planes, and precision physics at MAMI with the PANDA backward end-cap calorimeter.

## Global perspectives

In addition to hadron and multi-purpose facilities within Europe, the European hadron community relies on long-term participation in global experiments to thrive. European research groups play a crucial role in the design, construction, operation and science harvesting at these facilities, the most important of which are described below.

**Belle II @ SuperKEKB**, Tsukuba, Japan is a new-generation B-factory and $e^+e^-$ collider that has the potential to expand its hadron physics scope further. In addition to hadron spectroscopy, the multi-purpose detector and the ongoing data campaign offer investigations of hadron structure, hadron decays and hadron interactions.

**BESIII @BEPC-II**, Beijing, China is a multi-purpose $e^+e^-$ collider experiment optimised in the $\tau$-charm region. The physics programme of BESIII includes spectroscopy of charmonium and light mesons, hadron structure and searches for physics beyond the SM.

**GlueX@ JLab**, US, operates with a beam of real, polarised photons up to 12 GeV. It is designed for hadron spectroscopy with emphasis on searches for hybrid mesons; its future activities include the proposed intense $K_L$ beamline that will enable strange hadron spectroscopy.

**CLAS12@JLab**, US, operates with polarised or unpolarised electron beams up to 11 GeV. The broad physics programme of CLAS12 includes the structure and interactions of nucleons, nuclei, and mesons.

The future **ePIC@EIC**, Brookhaven National Laboratory, US, will be the first detector to be built at the future Electron-Ion Collider (EIC). In EIC, highly polarised electrons will collide with beams of high-energy heavy ions, polarised light ions and protons at CMS energies between 29 and 140 GeV. By inclusive and semi-inclusive DIS, as well as exclusive processes, ePIC will image gluons and quarks inside protons and nuclei. The construction of the facility as well as of the detector is planned to take place during the period of this LRP and start to collect data for physics in the early 2030s. A substantial participation from Europe (30% of ePIC Collaboration institutions are currently from Europe) is expected in the design, construction and commissioning of the ePIC detector.

# Hadron Physics Recommendations

The goal of hadron physics is to understand the rich and complex features of the strong interaction. How does the major part of the visible mass of the universe emerge from the almost massless quarks?

Can massless gluons form massive, exotic matter? What is the role of strong interactions in stellar objects, and in precision tests of the Standard Model? Answering these questions requires a diverse set of experimental and theoretical approaches. European hadron physicists play a leading role by developing theoretical approaces and by conducting experiments at facilities within Europe with great success, but also at the global level. These facilities, their planned upgrades, and the approved flagships PANDA at FAIR, Germany and ePIC at EIC, USA, open new avenues for ground-breaking discoveries.

## Existing facilities

We recommend the continuing support of the successful hadron physics programmes in Europe and the participation of European groups at global facilities. Particularly important hadron physics facilities are:

- **AMBER** at CERN, Switzerland

- **ELSA in Bonn, HADES at GSI, and MAMI and MESA** in Mainz, Germany

- **Jefferson Laboratory** in Newport News, USA

Furthermore, we recommend the support of ongoing hadron physics activities at the multi-purpose facilities Belle II, BESIII and the LHC.

## Future flagships

We recommend the expedited completion of the antiproton experiment PANDA, and the support of European groups in contributing to the electron-ion experiment ePIC. By virtue of their different beam species and energy regimes, PANDA and ePIC will explore complementary physics aspects. In a ten-year perspective, these two next-generation experiments must be made ready to launch.

- **PANDA**: The physics programme, including the prospect of unravelling exotic matter, remains unique and compelling. PANDA will strengthen the European position on the global scene and act as a unifying force for the community. We, therefore, recommend support for its construction and for the development of instrumentation, software and analysis tools.

- **ePIC**: European researchers will be able to explore unknown features of quarks and gluons inside nucleons and nuclei. We recommend supporting European groups to play significant roles in ePIC, reinforcing scientific and technological activities, which synergise with European projects.

## Theory / Computing

We recommend the support of theory groups at universities and research centres to prepare the community for the benefits from European investments in supercomputing and quantum computing infrastructure.

Theorists play an essential role in interpreting experimental results but also in providing input and predictions for new experiments. To match experimental progress, sophisticated approaches need to be developed. In lattice QCD, the rapid evolution of computational techniques and hardware calls for new algorithms and software. Similarly, quantum computing requires appropriate algorithms and tests on quantum hardware. Support for theoretical groups in terms of positions and career prospects is, thus, essential for progress in hadron physics.

# Properties of Strongly Interacting Matter at Extreme Conditions of Temperature and Baryon Number Density

**Coordinators:**
**Laura Fabbietti** (TU München, Germany)
**Urs Achim Wiedemann** (CERN Geneva, Switzerland)

**NuPECC Liaisons:**
**Gert Aarts** (Swansea, UK and ECT* Trento, Italy)
**Raimond Snellings** (Utrecht University, The Netherlands)

**WG Members:**
- **Roberta Arnaldi** (INFN Torino, Italy)
- **Andrea Dainese** (INFN Padova, Italy)
- **Stefan Flörchinger** (Jena University, Germany)
- **Tetyana Galatyuk** (GSI and TU Darmstadt, Germany)
- **Tuomas Lappi**, (JYFL-ACCLAB, Jyväskylä, Finland)
- **Yen-Jie Lee** (MIT, Cambridge, USA)
- **Giulia Manca**, (Cagliari University, Italy)
- **Alexander Milov** (Weizmann Institute, Israel)
- **Luciano Musa** (CERN Geneva, Switzerland)
- **Piotr Salabura** (Jagiellonian University, Krakow, Poland)
- **Carlos Salgado** (University of Santiago de Compostela, Spain)

# Introduction

The electromagnetic, weak and gravitational forces between two objects become weaker with increasing distance. In contrast, the strong force gets stronger with increasing distance ("confinement") and it decreases at short distances ("asymptotic freedom"). These characteristic properties of the strong nuclear force enabled us to identify Quantum Chromodynamics (QCD) as the underlying theory in the early 1970s and they have been central to progress in nuclear physics, hadron physics and particle physics ever since. QCD governs the interactions between quarks and gluons (known as partons). Partons carry colour charges, but the strong force confines these partons into colourless bound states (known as hadrons). QCD confinement implies that a coloured parton cannot be isolated from a colourless hadron. Moreover, QCD implies a vacuum condensate ("chiral condensate") that is responsible for generating most of the mass of standard matter in our Universe.

These fundamental properties of the strong nuclear force have profound consequences for the properties of strongly interacting matter at extreme conditions of temperature and density. In contrast to other known plasmas the physical degrees of freedom that govern the macroscopic properties of strongly interacting matter change with increasing temperature or density from hadrons to partons. The system transits from hadronic matter to the quark gluon-plasma (QGP) and the chiral condensate melts.

Phase transitions are a well-known and abundant phenomenon in solid state physics. However, within our modern understanding of the Standard Model of particle physics, we know of only two phase transitions involving fundamental quantum fields. The one is the QCD phase transition that deconfines quarks and gluons and that restores chiral symmetry by melting the chiral condensate. The other is the electroweak phase transition in which the Higgs vacuum expectation value melts. According to standard cosmology, both phase transitions played central roles in the hot early Universe: the electroweak phase transition took place at a temperature of ~100 GeV where mass generation occurred via the Higgs mechanism, and the strong phase transition took place at a temperature of ~155 MeV where mass generation occurred via chiral symmetry breaking. In today's Universe, the transition to a deconfined partonic system may occur in the interior of ultra-massive neutron stars. The scientific challenge of understanding the behaviour of fundamental quantum fields at extreme temperature and density has thus profound implications for our understanding of the world we live in.

The QCD transition temperature of ~155 MeV corresponds to trillions of degrees kelvin occurring at ~30 microseconds after the Big Bang. The critical matter density for this transition is a few times normal nuclear matter density. This makes the QCD phase transition the only transition of fundamental quantum fields that is within reach of laboratory experiments. It has long motivated controlled experimentation with deconfined QCD matter in (ultra-)relativistic heavy-ion collisions at hadron colliders and at dedicated fixed-target experiments. The fundamental questions addressed by this research at the high-temperature and at the high-baryon-number-density frontier are interrelated but characteristically different.

**Nuclear Collisions at the highest collider energies** produce a QGP that is almost symmetrical in matter and antimatter (baryo-chemical potential $\mu_B$~0), has the highest initial temperature and lives for the longest time. These conditions are ideal for testing QGP thermal properties and QGP hydrodynamic behaviour. They also make possible the production of highest abundancy hard penetrating and out-of-equilibrium probes that test the equilibration and attenuation mechanisms of the QGP and electroweak processes carrying direct information about the earliest stages of the collision. At small $\mu_B$, the QCD phase transition is known from ab-initio calculations to be a smooth cross-over, and research focuses on characterising the fundamental matter properties of the high-temperature phase.

**Nuclear Collisions at fixed target energies** produce QCD matter with a significant excess of matter over antimatter i.e. with sizeable baryo-chemical potential $\mu_B$. Experimentally, the value of $\mu_B$ can be varied by changing the collision energy. While theory predicts a first-order phase transition at sufficiently large $\mu_B$, the location of the so-called QCD critical end point at which this first-order phase transition changes to a crossover is not known. The QCD critical end point is a much-searched-for hallmark of the QCD phase diagram. In addition, nuclear collisions at lower fixed-target energies can contribute to our understanding of the QCD equation of state in a regime of relatively low temperature and high baryo-chemical potential relevant for the modelling and understanding of neutron stars and neutron-star mergers.

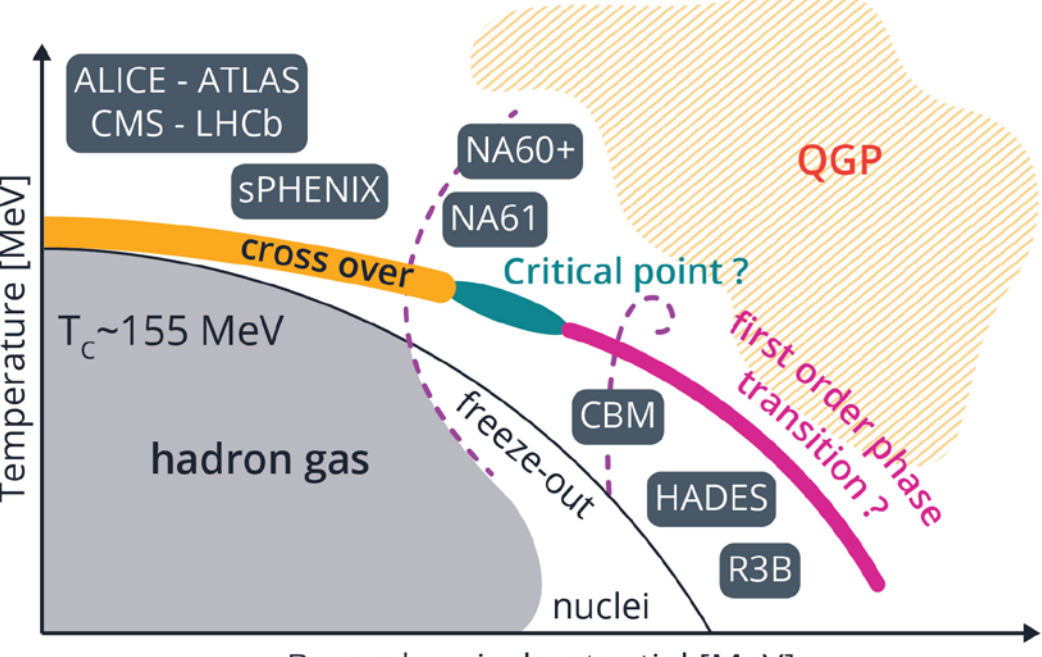

***Fig. 3.1: Schematic QCD phase diagram indicating the regions in the T-$\mu_B$ plane that are accessed experimentally by collider experiments and by experiments at lower fixed-target energies.***

In the following sections, we summarise the substantial progress made in collider and fixed-target experiments since the last NuPECC Long Range Plan and we highlight the fundamental questions that are now coming into experimental and theoretical reach. We emphasise that both recent and future progress is enabled by the interdisciplinary interplay between many technological, computational, experimental and theoretical developments pursued by a highly diverse, world-wide community.

## Box. 3.1: The space-time evolution of relativistic heavy ion collisions

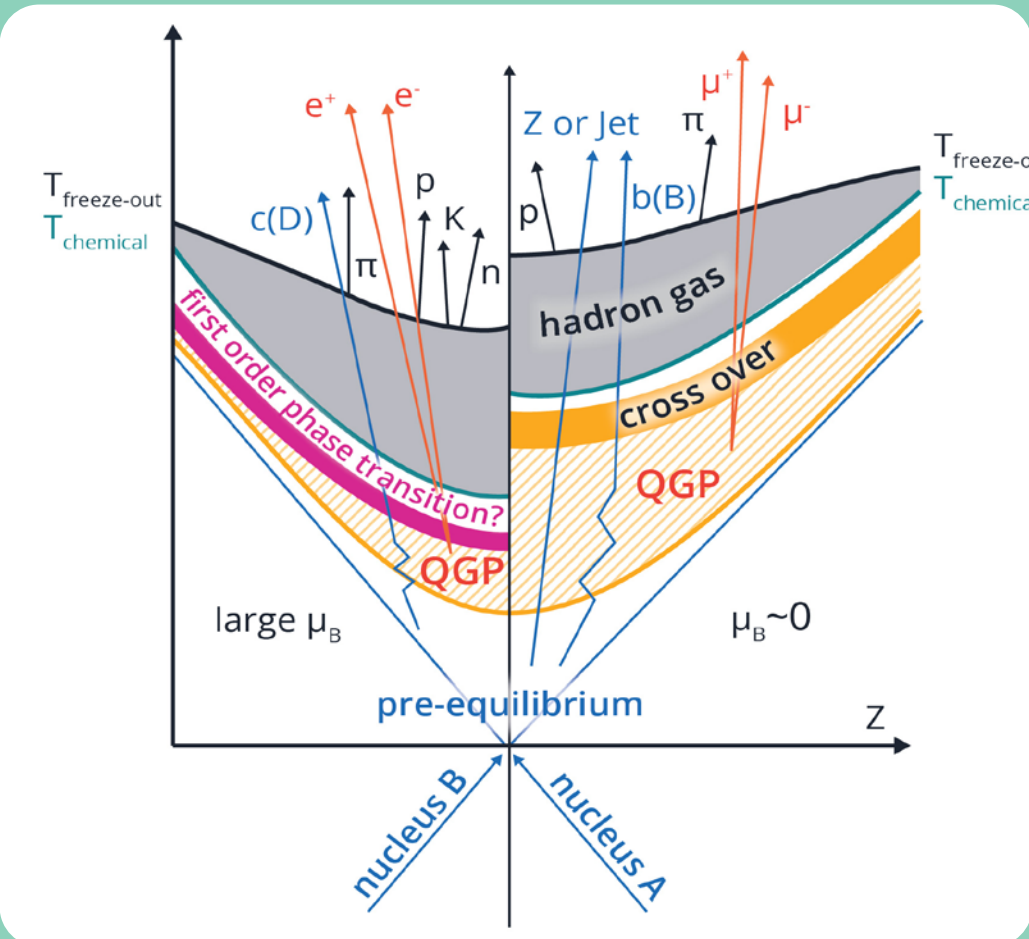

***Fig. 3.2: The space-time evolution of heavy ion collisions at fixed target energies (left side of the plot) and at LHC collider energies (right side of plot). Nuclei accelerated along the beam-direction z up to relativistic energies propagate along the diagonal in the space(z)-time(t) plane and collide as Lorentz-contracted pancakes. In the earliest non-equilibrium stage, the over-dense system evolves rapidly towards a Quark Gluon Plasma (QGP). The QGP then expands longitudinally with velocities up to the speed of light ("Bjorken boost-invariance"), and the extreme initial transverse density gradients drive an explosive transverse expansion. Due to this expansion, the QGP cools before transiting into a hadronic phase. Hadro-chemical abundances are frozen early in the hadronic phase (chemical freeze-out $T_{chem}$). The expanding hadron gas maintains kinetic equilibrium up to a lower freeze-out temperature $T_{fo}$, at which hadrons decouple and free-stream to the detector. By experimentally varying the centre-of-mass energy of the collision, QGPs with characteristically different baryo-chemical potential and different initial temperature can be studied. These QGPs live and cool over different time scales and result in different hadro-chemical distributions in the final state. The QCD first order phase transition at finite $\mu_B$ can only be tested at sufficiently low fixed target energy, while conditions for testing the properties of the QGP are most favourable at collider energies where the QGP is initially the hottest and longest lived and where hard penetrating probes are available for its detailed characterisation.***

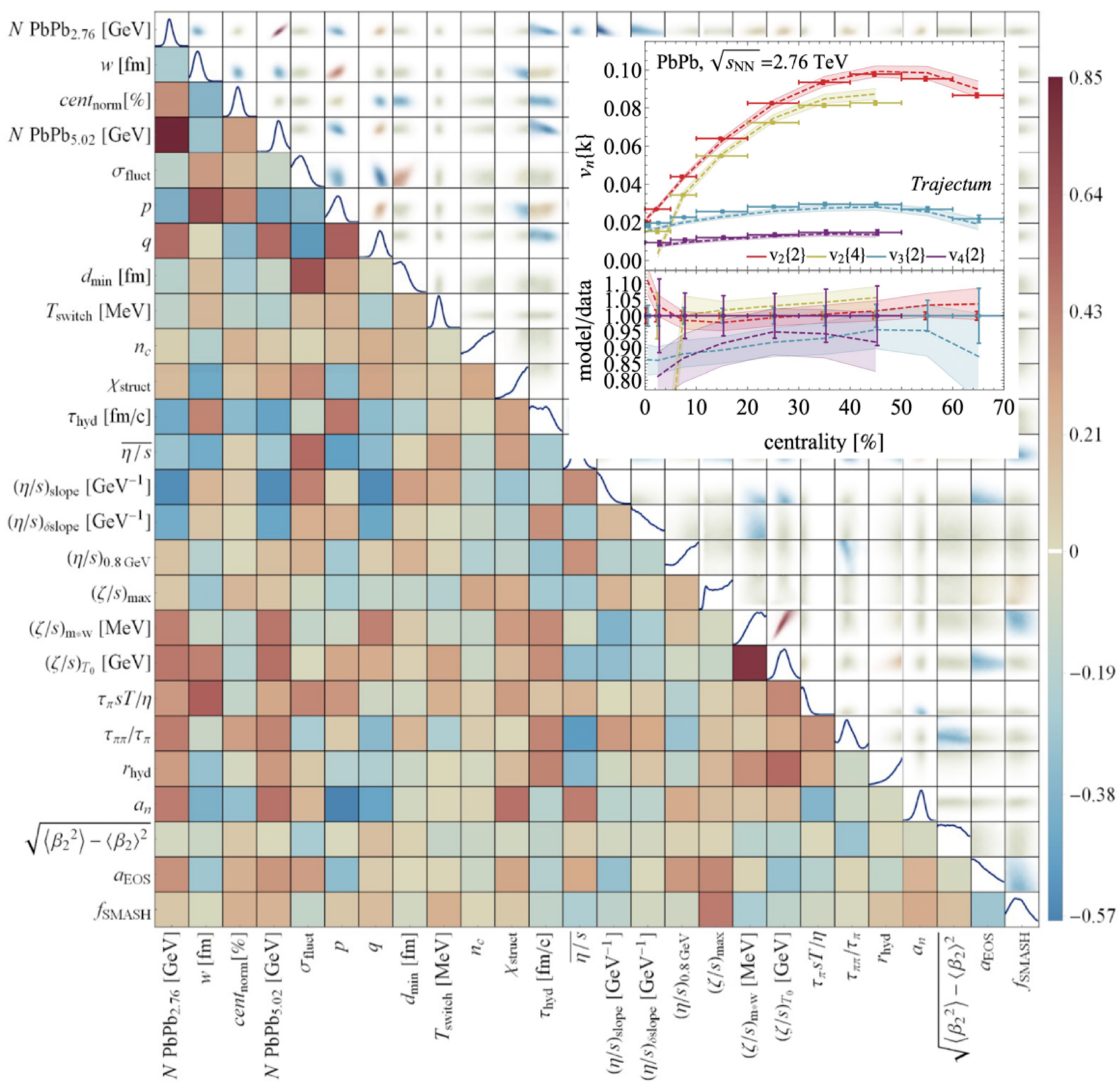

*Fig. 3.3: Results of a Bayesian inference analysis using 670 individual data points to constrain the space-time evolution and material properties of the QGP produced in Pb-Pb collisions at $\sqrt{s_{NN}}$ = 2.76 TeV [Giacalone, Nijs, v.d.Schee, Phys.Rev.Lett. 131 (2023) 02]. The correlation matrix displays all 26 model parameters. The inset shows, as an example, the comparison of the model with data on the centrality dependence of the flow anisotropies $v_n$.*

Box 3.1 summarises the qualitative understanding of how an (ultra-) relativistic heavy-ion collision evolves in space time from the earliest time at which the colliding nuclei overlap up to the time at which hadrons decouple from the hadronic phase (“freeze-out”). There are essentially three handles for testing this picture experimentally:

1. The initial conditions of the collision can be controlled experimentally by varying the centre-of-mass energy, the atomic number of the colliding nuclei and the impact parameter of the collision. In this way, the baryo-chemical potential, the volume and lifetime within which QGP is produced and the transverse shape and transverse pressure gradients with which the QGP expansion is initiated are varied.

2. The measurement of bulk particle production (hadronic final state at low transverse momentum) provides information on the thermodynamic and hydrodynamic properties of the produced matter. This includes information about the kinetic freeze-out temperature $T_{fo}$, the chemical freeze-out temperature $T_{chem}$ and the hydrodynamic response to pressure gradients.

3. The measurement of hard and electromagnetic probes. These are out-of-equilibrium processes that are produced early in the collision. Hard hadronic processes participate incompletely in the equilibration processes and provide information on how individual quarks and gluons interact in the QGP. Electromagnetic radiation from the QGP is of interest since what is emitted in the early high temperature phase remains unaffected by the further evolution.

As depicted in Box 3.1, the space-time evolution of relativistic heavy ion collisions is a multi-stage process. To infer detailed properties of the QGP phase and the hadronisation of this partonic system, the use of experimental and theoretical constraints on all stages of the collision is required. One important development in the field is therefore to formulate complete dynamic models that encode the best theoretical understanding of all stages of the collision and that can be over-constrained with the wealth of data collected by experiments.

Recent years have seen significant progress in implementing this strategy. For illustration, Fig. 3.3 shows the example of a Bayesian inference analysis in which more than 500 data points from light-flavoured bulk hadron production in Pb-Pb collision at the LHC (ALICE and CMS experiments) are used to constrain 20 parameters within the definition of the initial conditions, in the dynamics of the pre-equilibrium state, in the properties of the equation of state (EoS), in the transport properties and relaxation times of the fluid-dynamic QGP evolution and in the transition of the system to final state hadrons. In particular, these measurements reveal a non-trivial temperature-dependence of the QGP viscosity-over-entropy ratio which can be compared directly to calculations of that quantity in finite temperature quantum field theory. The phenomenologically extracted value is close to the value $\hbar/(4\pi\, k_B)$ predicted in the strong coupling limit for a large class of non-abelian plasmas. The same analysis yields first data-driven constraints on more subtle properties of QGP hydrodynamics, such as constraints on the relaxation times $\tau_\pi$ with which viscous shear excitations decay in the QGP and on the temperature dependence of transport coefficients.
These examples illustrate how the increased precision and wealth of data collected at the LHC yield step-by-step more detailed information about the fundamental properties of the QGP.

The picture resulting from these and many other studies is that of a collision system that equilibrates extremely quickly on time-scales lower than 1 fm/c. Equilibration creates a QGP with close to minimal dissipative properties ("perfect liquid") at an initial temperature well above the QCD transition temperature. The initial extreme transverse pressure gradients drive a collective transverse expansion that reaches transverse velocities of 2/3 of the speed of light. The system evolves close to equilibrium for more than 10 fm/c and expands into a volume of about 5000 $fm^3$ per unit of rapidity. These values are very large compared to the typical QCD time and length scales of sub-fm and they justify thinking of central heavy ion collisions as macroscopic equilibrated systems that undergo chemical freeze-out at a temperature of ~ 155 MeV and kinetic freeze-out at a temperature of ~ 100 MeV respectively.

## Experimental highlights from the high-temperature frontier

Heavy-ion collisions are an integral part of the physics programme of the LHC. The ALICE detector was conceived to be dedicated to the study of heavy-ion collisions. The ATLAS and CMS experiments participated in the programme from the very beginning and the LHCb experiment participated from the end of the first Run (2012). Since the previous LRP, Run 2 of the LHC has been completed (2015-2018) with centre-of-mass energies per nucleon-nucleon pair ($\sqrt{s_{NN}}$) larger by almost a factor two compared to Run 1 (Pb+Pb at 5.02 TeV, p+Pb at 5.02 and 8.16TeV). In addition, a small Xe+Xe sample at 5.44 TeV was also delivered and LHCb studied fixed-target collisions at centre-of-mass energies of ~100 GeV using the SMOG noble-gas target system. The Run 2 integrated luminosities were higher by about one order of magnitude compared to Run 1. The larger data samples and collision energies made possible the study of QGP properties with unprecedented precision and with additional and rarer observables and probes, in particular multi-particle correlations, heavy quarks and jets. The surprising finding that modified-hadronisation and collective effects are also present in small collision systems like proton-proton and proton-nucleus, was followed up with very detailed experimental studies during Run 2.

The spatial asymmetry created by the geometrical overlap of two colliding nuclei is converted to an anisotropic momentum distribution. This anisotropy is quantified via the Fourier components $v^2$, $v^3$, $v^4$, $v^5$… of the hadron azimuthal distribution. The dominant second-order harmonic in non-central collisions is called elliptic flow coefficient and indicated with $v^2$. For light-flavoured hadrons, these flow harmonics are known with high precision and they provide one of the most sensitive constraints on QGP transport properties (see inset in Fig. 3.3). The high-resolution vertex detectors of the LHC experiments enabled important measurements of the flow coefficients for charmed hadrons (see Fig. 3.4) during Run 2. The flow of charmed hadrons is smaller than that of light-flavoured hadrons, indicating a mass hierarchy that provides important constraints on a heavy flavour transport property, namely the charm-quark spatial diffusion coefficient $D_S$ in the QGP.

Quarkonia, with different hadron sizes and binding strengths, are probes of the Debye screening lengths in the QGP. Measurements of bottomonium production in Run 2 show a hierarchy of increasing suppression with decreasing binding energy for the 1S, 2S and 3S states. Charmonium states are also sensitive to the parton recombination involving two heavy quarks. At low hadron momentum, suppression of *J/Ψ* in Pb-Pb compared to pp collisions is much lower at LHC energies than at RHIC energies. The Run 2 measurements down to zero momentum at central rapidity are even consistent with no suppression of the yield. This observation and the elliptic flow of *J/Ψ* (see right panel of Fig. 3.4) points to a novel dynamic of hadronisation that involves the recombination of charm and anti-charm quarks into pairs.

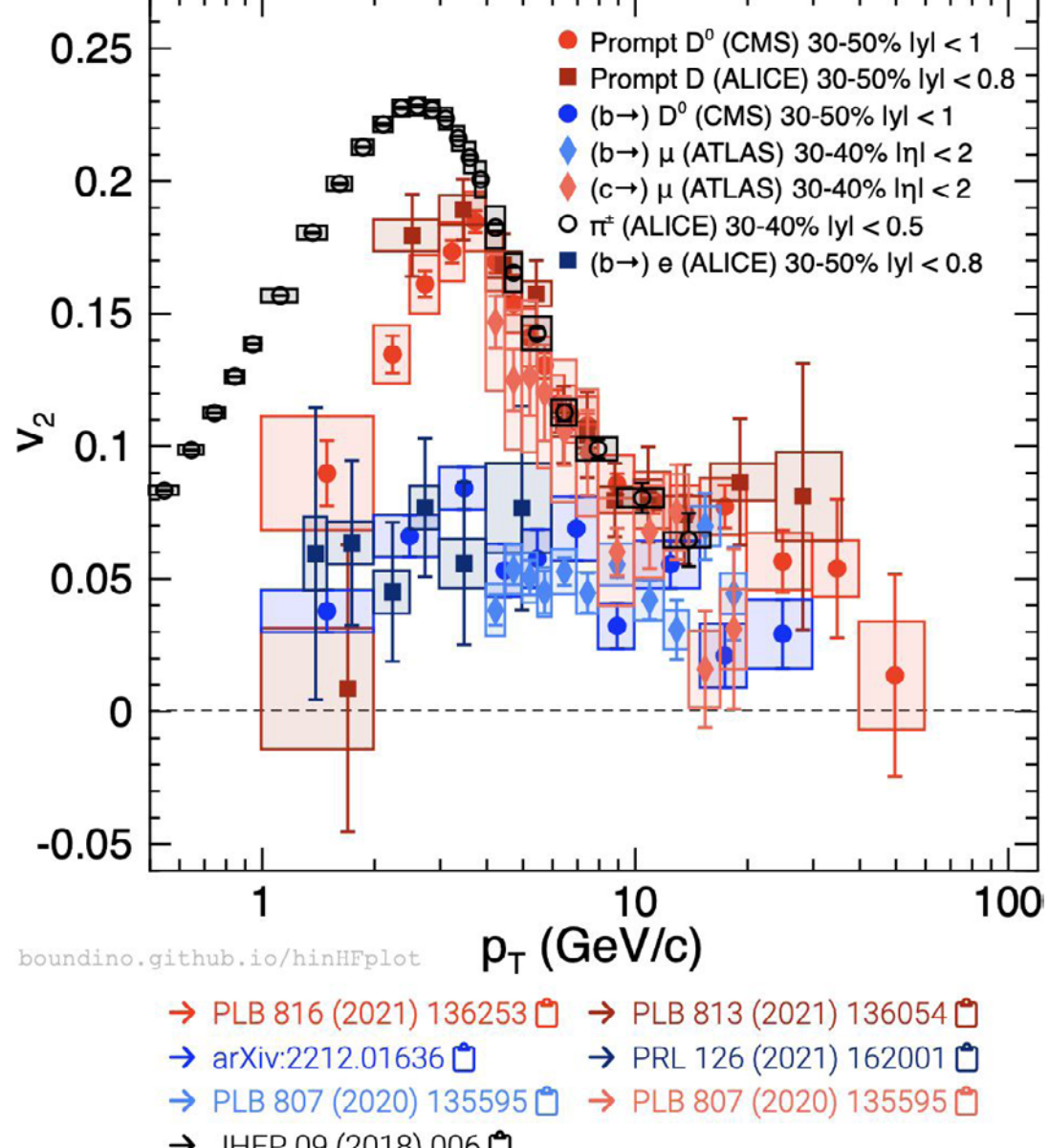

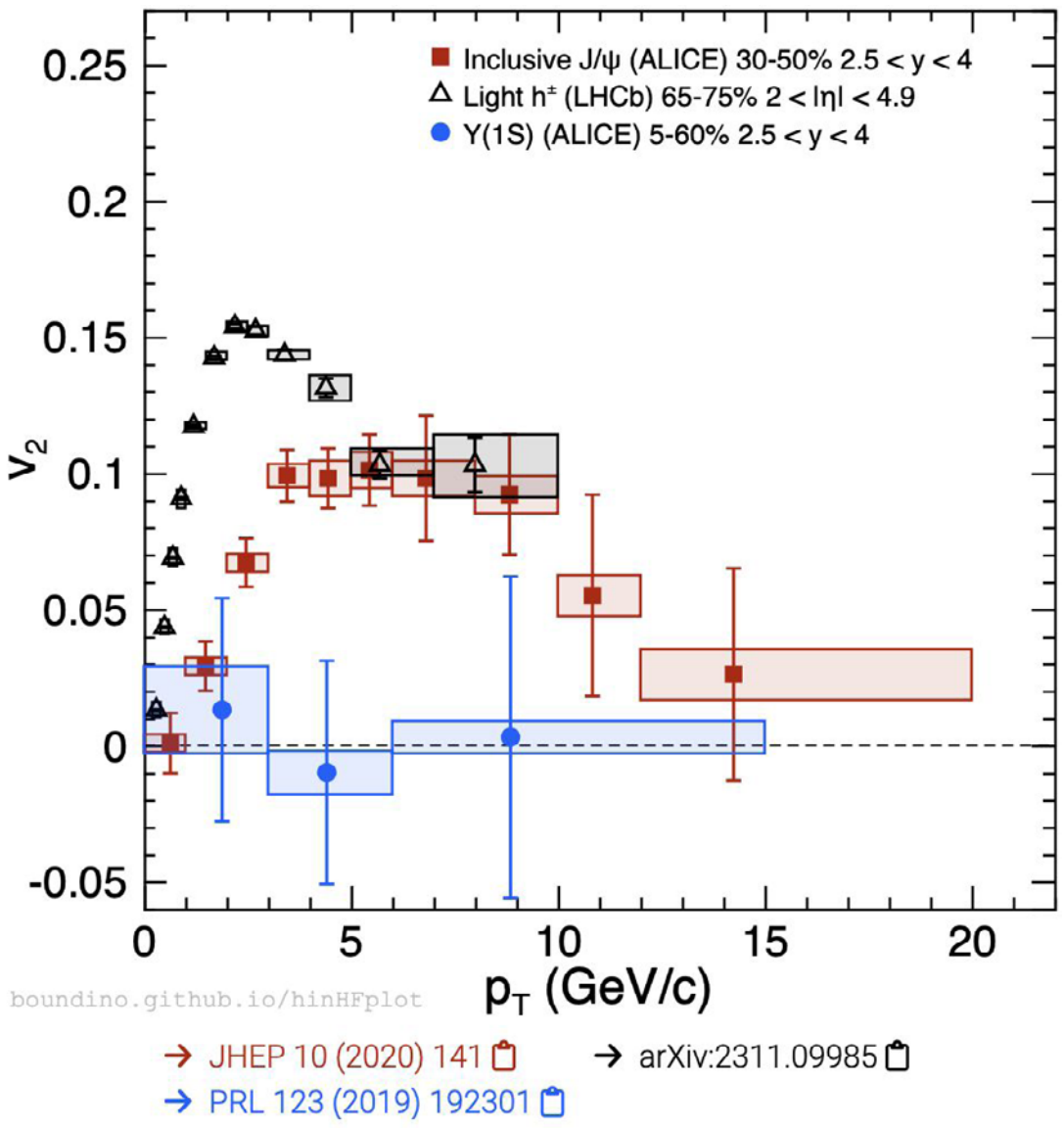

***Fig.3.4: Elliptic flow $v^2$ is a hallmark of the collectivity in nucleus-nucleus collisions. The plot compares, as a function of $p_T$, the elliptic flow measured in Pb-Pb collisions at $\sqrt{s_{NN}}$= 5.02 TeV for light-flavour hadrons, charm and beauty hadrons at midrapidity (left) and for light-flavour hadrons, charmonium and bottomonium at forward rapidity (right).***

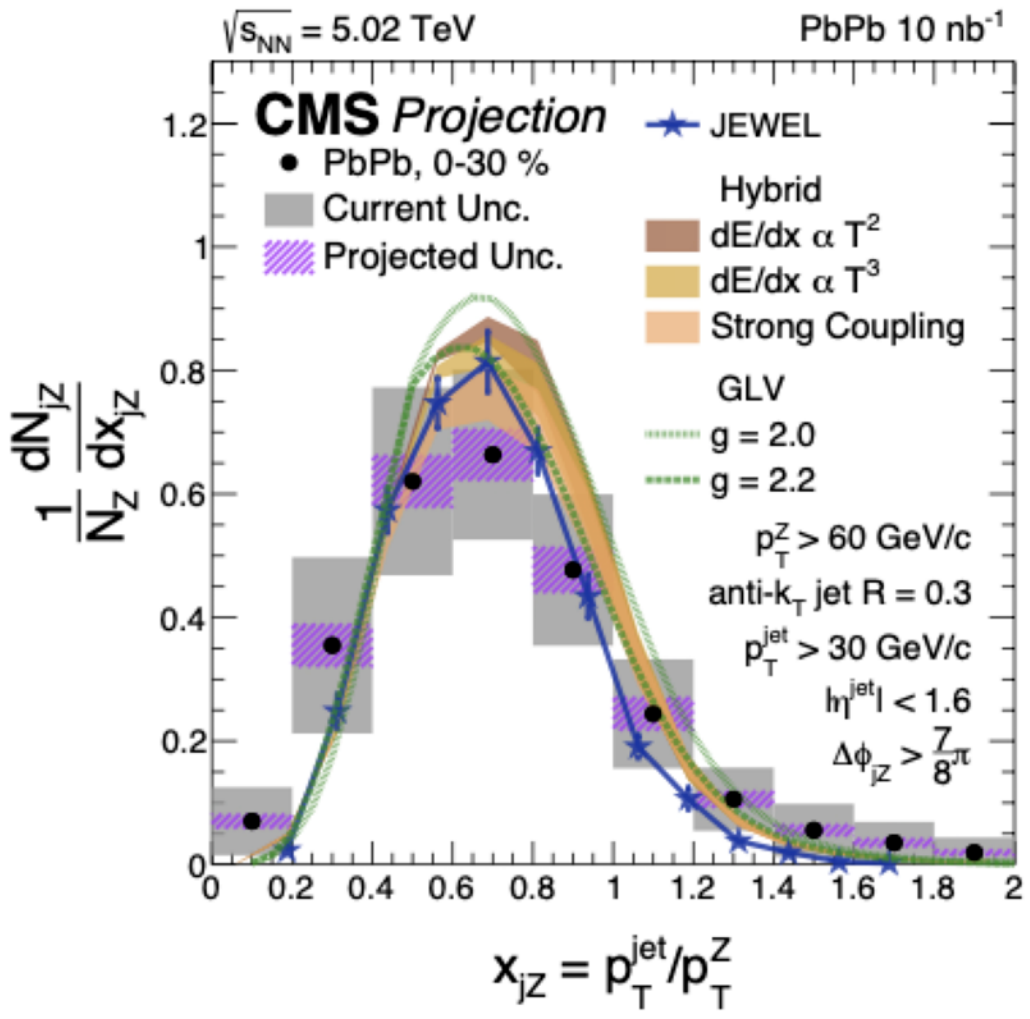

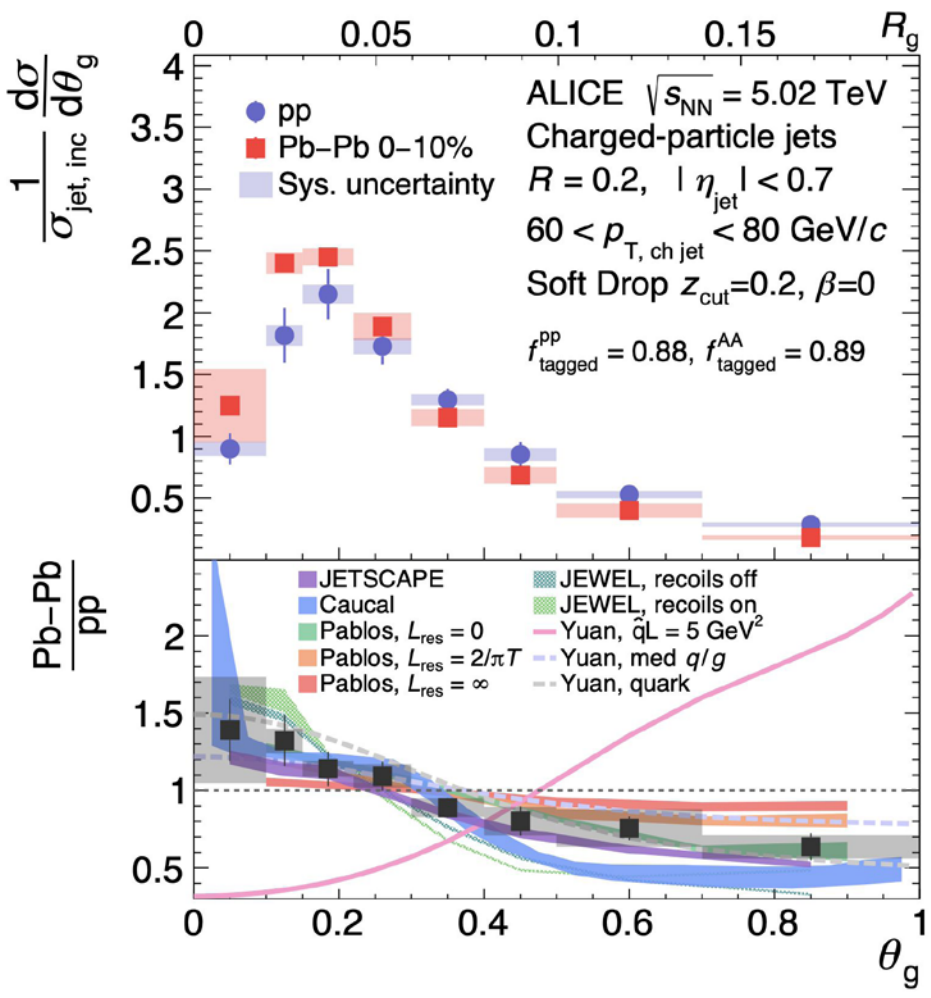

*Fig. 3.5: Progress in jet quenching measurements. Measuring the transverse momentum of jets recoiling against Z-bosons (left) provides a novel tool for characterising jet quenching [CMS, PAS FTR-18-025]. The jet angularity θg distribution (right) shows that the splittings within the jet substructure are more collimated for jets measured in Pb-Pb collisions than for those measured in pp collisions [ALICE, Phys. Rev. Lett. 128 (2022) 102001].*

High momentum-transfer interactions between partons in the nuclei produce energetic quarks and gluons that can be used as probes of the QGP. At the LHC, "jet quenching" was observed in the nuclear modification of single inclusive jet and hadron spectra, in the enhanced dijet $p_T$ asymmetry and boson-jet $p_T$ imbalance and in a multitude of jet substructure measurements. In particular, Z-bosons have now been used as recoil triggers to better control experimentally the conditions with which jets are initiated in the collision (see left panel in Fig. 3.5). Modifications of inclusive jet fragmentation function, radial shape and substructure indicate that narrow jets with fewer constituents are less quenched (see right panel in Fig. 3.5). The QGP was found to enhance low $p_T$ particles inside the jet cone and to broaden photon-tagged jet shapes. Parton flavour dependence of jet quenching has been studied with jets and hadrons. High $p_T$ gluons lose more energy than quarks due to colour factors. This effect has been studied with photon-tagged jets and heavy quark jets. By comparing the suppression of beauty and charm mesons, a significant mass dependence has been conclusively observed at low $p_T$ with the LHC Run 2 data. In general, these data show the telltale signs of medium-modified QCD parton showers via which highly energetic out-of-equilibrium probes evolve in the QGP towards equilibrium.

**Heavy-ion-like behaviour in small collision systems ?** Heavy-ion collisions demonstrate hadrochemical and kinetic equilibration and collectivity. One of the most surprising discoveries at the LHC is that these signals persist from ion-ion down to the smallest proton-ion and proton-proton collision systems (see Fig. 3.6). This system size dependence raises the fundamental question of the size of the smallest droplet of matter in which the QGP can be formed. Another conceptual challenge arising from these measurements is that the final state interactions that reveal themselves in signs of collectivity should also leave imprints on the medium-induced energy loss of jets. So far, however, these latter have not been observed. The LHC discovery of collectivity in small collision systems has thus opened a challenging window for both future experimentation and for a theoretical understanding of the onset of collectivity that unifies the medium-modification of soft bulk matter and hard processes.

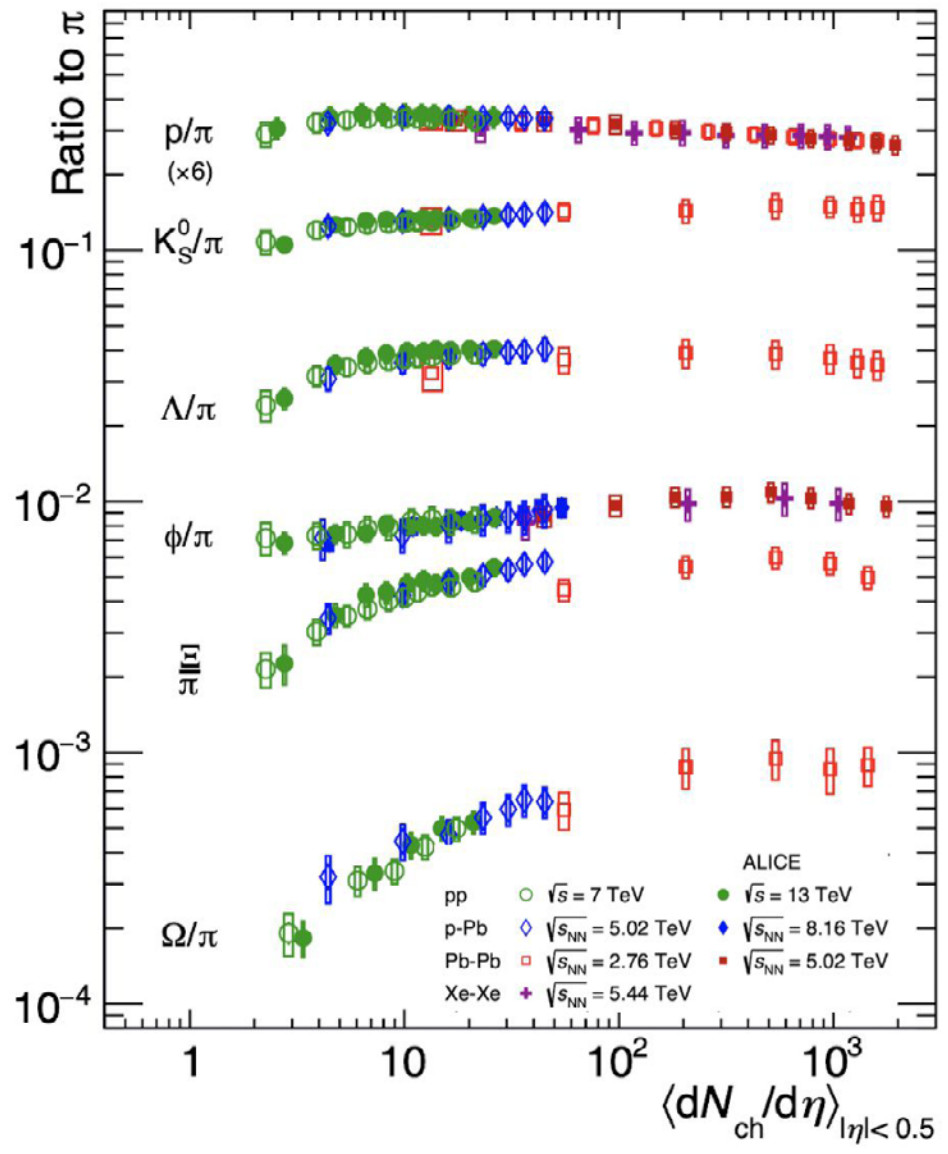

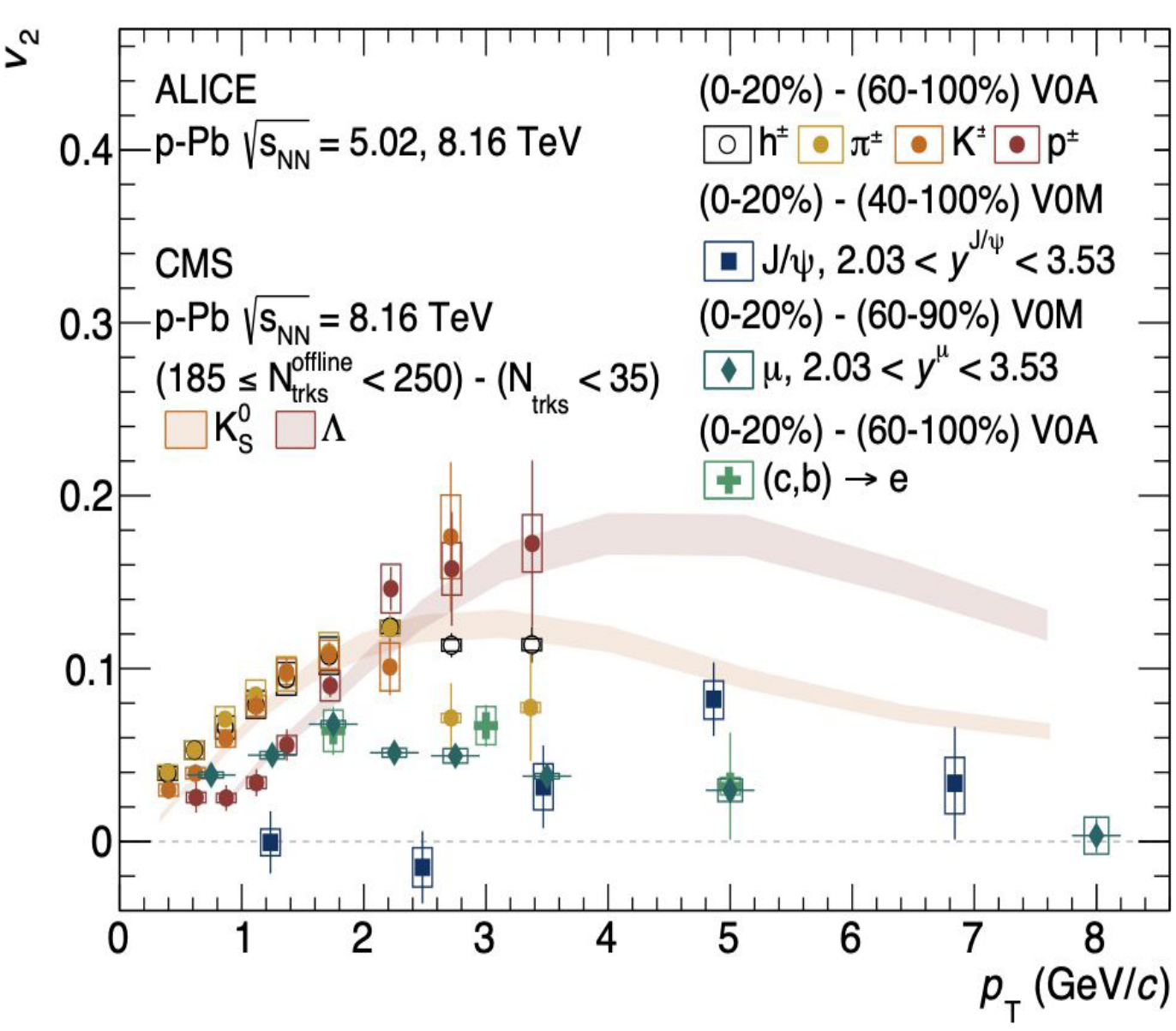

*Fig. 3.6: Relative hadronic abundances (left panel) and elliptic flow (right panel) are signals of chemical and kinematic equilibration that are observed in the smallest hadronic collision systems and whose strength changes smoothly with system size [ALICE Coll., arXiv:2211.04384; CMS Coll, PRL121(2018)082301].*

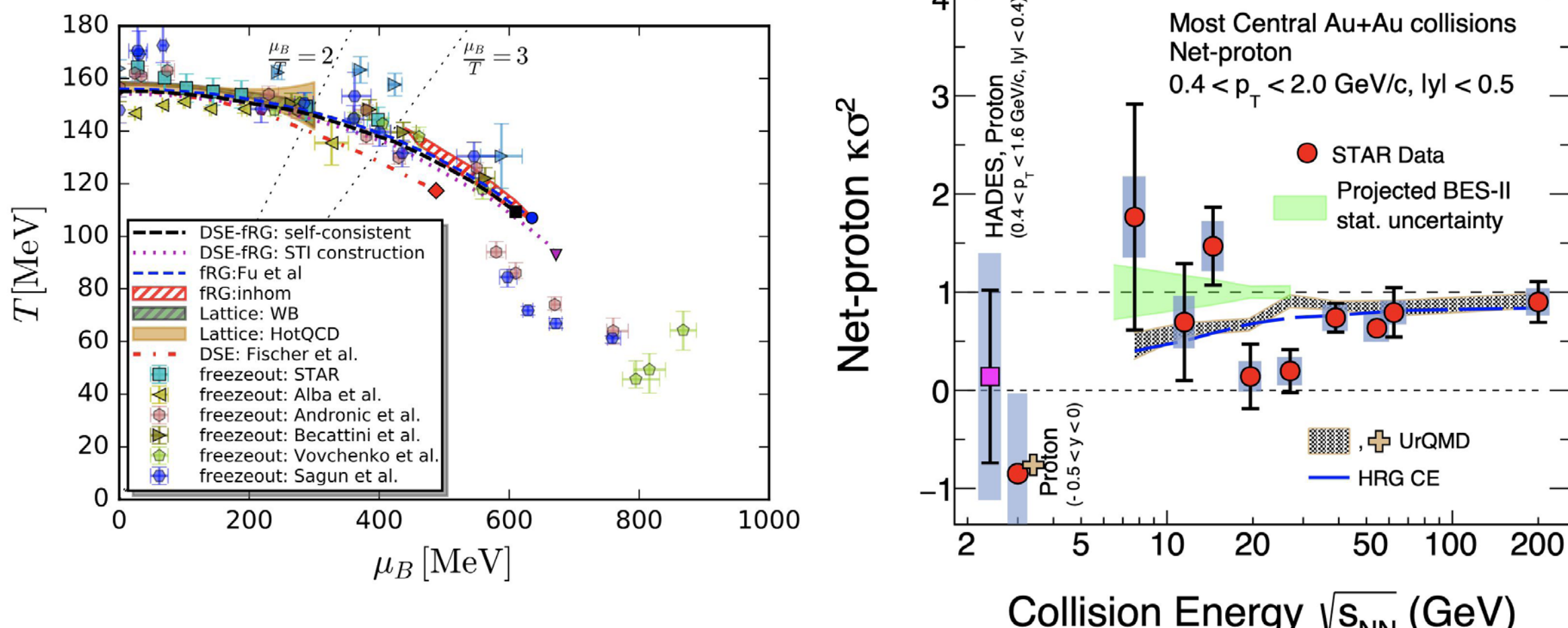

**Fig 3.7: (Left panel) Overview of calculation of the temperature and baryo-chemical potential of different phases of QCD matter. The red area indicates a prediction for the critical point region [Gao, Pawlowski, Phys.Lett.B 820 (2021) 136584]. (Right panel) Measured net-proton fluctuations as a function of the collision energy [STAR Phys.Rev.Lett. 128 (2022) 20, 202303].**

## Experimental highlights from the high-net-baryon density frontier

A detailed understanding of the microscopic properties of QCD matter in the region from moderate to high $\mu_B$ and lower temperatures is complementary to the heavy-ion programmes at RHIC and LHC. From studying QCD on the lattice it is known that the temperature-driven smooth transition at vanishing net-baryon densities is indeed a crossover. At large values of net-baryon densities ($\mu_B$> 3T) lattice QCD is currently not applicable but several model calculations do predict a line of first order phase transition (see Box 3.1). In this context, the existence of a second order chiral critical end point is predicted.

Recently, significant progress has been made in establishing phase structures of QCD matter at high baryon-number-densities (see Fig. 3.7, left) employing various QCD-related methods, like Dyson-Schwinger equations, Functional Renormalisation Group, holographic gauge-gravity correspondence etc. The location of the critical points is predicted in the region of (T, $\mu_B$) = (~100-120 MeV; 450 - 650 MeV) which is in reach of fixed target experiments and shown in Fig. 3.7 (left panel).

Event-by-event fluctuations, such as higher-order cumulants of multiplicity distributions of conserved quantities (net-baryon, net-charge, and net-strangeness), are sensitive to the correlation length inside the medium, and thus to the critical point. For instance, the right panel of Fig. 3.7 shows a recent compilation of the net-proton cumulants of the fourth-order measured in different experiments, as a function of the centre-of-mass energy. Non-monotonous behaviour in this quantity could suggest critical fluctuations in the QCD matter close to the phase boundary. Measurements of this kind, including measurements of higher order cumulants, will be further investigated by HADES, CBM, NA61/SHINE and by MPD at NICA.

One experimental handle on studying chiral symmetry restoration is to establish the degeneracy of spectral functions of chiral partners, such as $\rho$(770) and a1(1260). Relevant experimental observables are invariant mass distributions of thermal dileptons in the low mass region (M<1 GeV/c²), related to $\rho$-meson in-medium distribution, and in the mass region (1.1<M<1.5 GeV/c²) related to the chiral rho-a1 mixing.

The NA60 experiment has measured, in In+In collisions at $\sqrt{s_{NN}}$ =17.3 GeV, a $\rho$-meson spectral function strongly modified by the medium, in agreement with a microscopic many body theory, predicting a significant broadening. Recent results by the HADES collaboration have shown that the $\rho$(770) spectral function is strongly broadened in a baryon-dominated regime (see Fig. 3.8, left panel). A salient feature of the calculations used to explain these results is a strong coupling of $\rho$-meson to baryons demonstrated by an independent measurement of the Dalitz decay shown in Fig. 3.8 (right panel).

The slope of the dilepton invariant mass spectrum is closely related to the temperature of the thermal medium. In contrast to the slope of hadron and dilepton transverse-mass spectra, this observable is not blue-shifted and thus easier to interpret. HADES has measured in Au+Au collisions the dilepton excess above known hadronic sources and has determined an average temperature of <T> = 72 ± 4 MeV. This should be compared to the higher temperature of <T> = 205 ± 12 MeV extracted by NA60 for the fireball produced in In+In collisions at $\sqrt{s_{NN}}$= 17.3 GeV. Precision measurements of this spectrum at various centre-of-mass energies are of interest since they might be sensitive to the crossing of a first-order phase boundary.

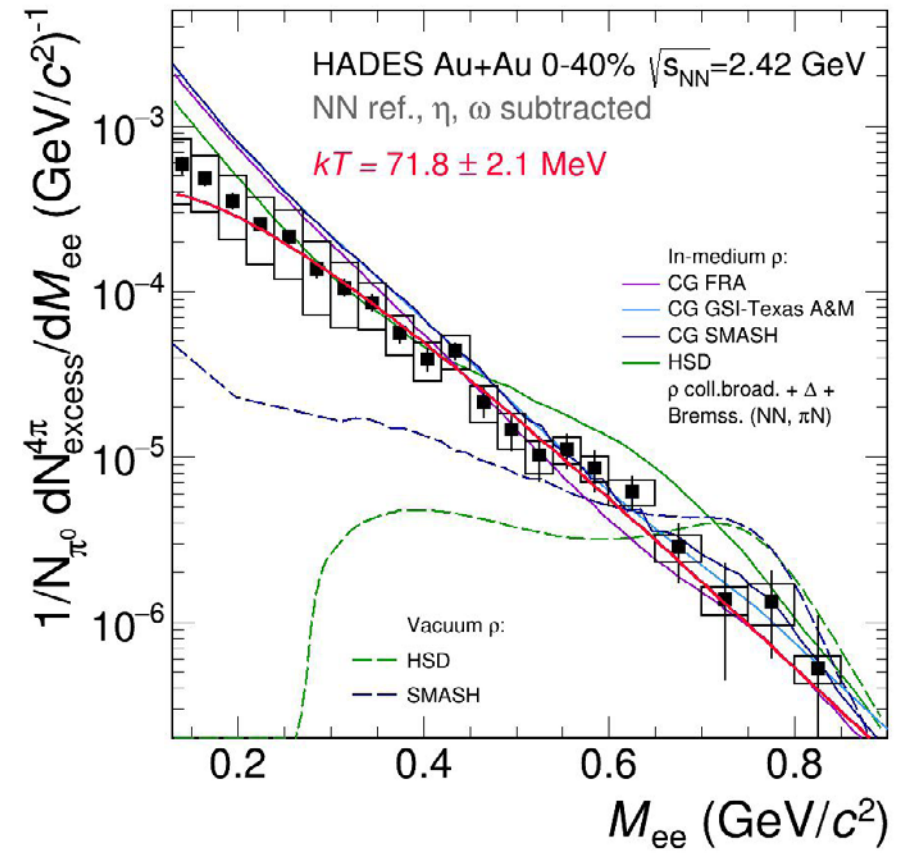

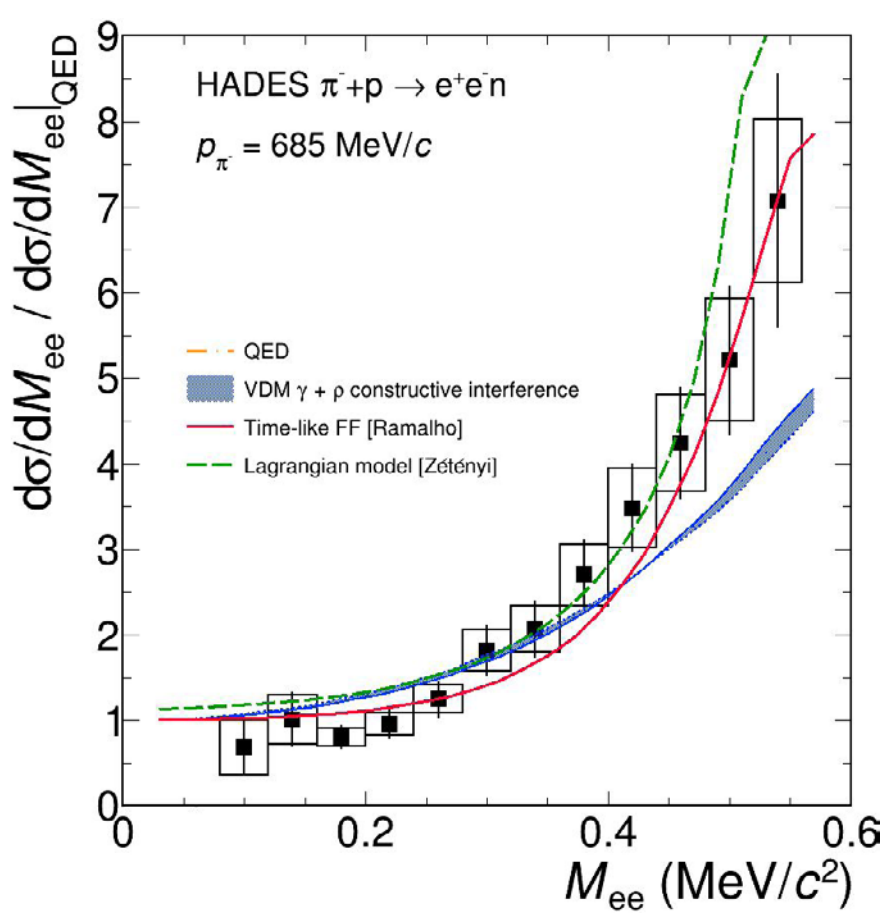

**Fig 3.8: (Left panel) Dilepton excess yield in the low invariant mass region identified as the modified $\rho$- spectral function. Different theoretical calculations are compared to the data measured by HADES in AuAu collisions at = 2.42 GeV [HADES, Nature Phys. 15, 1040 (2019)]. (Right) Invariant mass of pairs from $N^*$->$N\gamma^*$->$Ne^+e^-$ normalised by the QED prediction for point-like transitions [HADES Coll., arXiv:2205.15914 ] .**

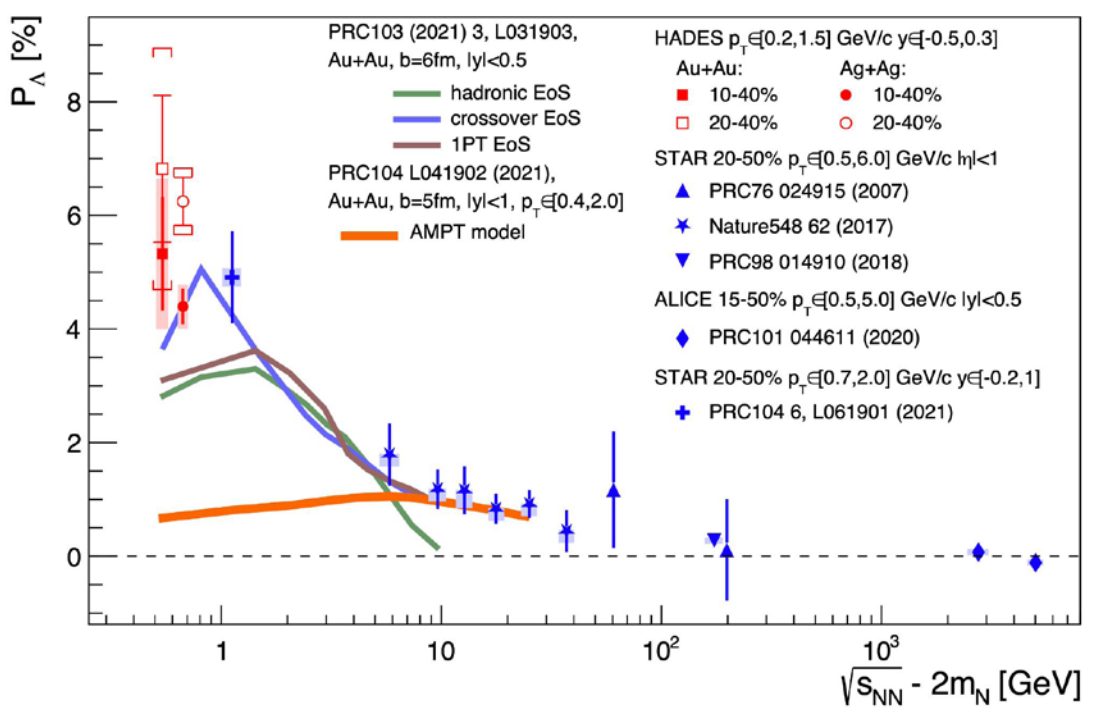

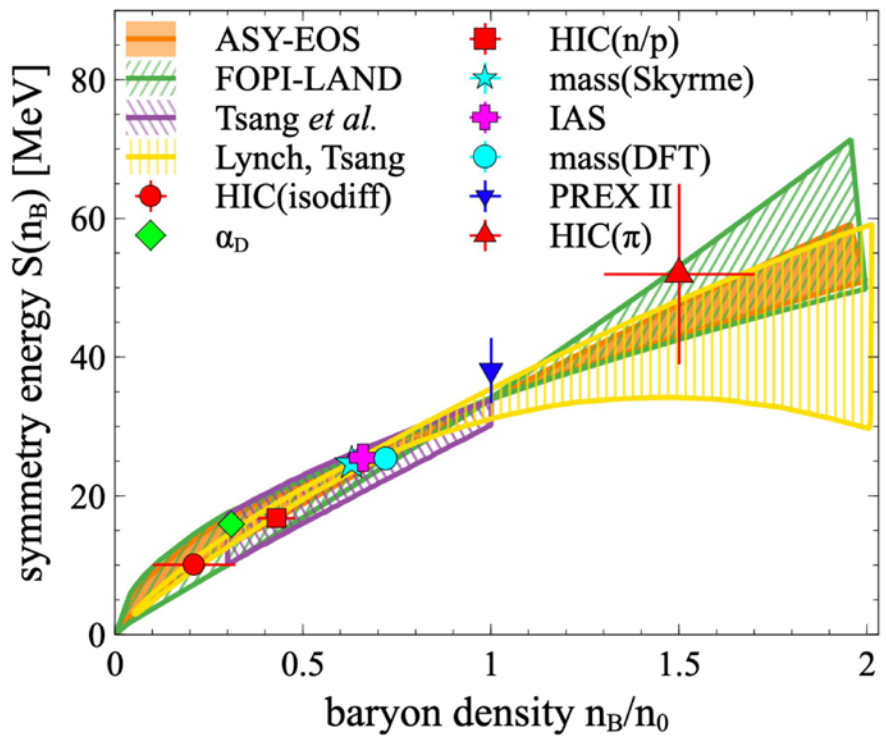

*Fig. 3.9: (Left panel) Overview of polarisation measurement of the L hyperon as a function of excess energy compared to several theoretical calculations based on different equation of states [HADES, Phys.Lett.B (2022) 137506]. (Right panel) Symmetry energy as a function of the baryonic density extracted from heavy ion collisions experiments at low and intermediate energies [A.Sorensen, et al. Prog. Part. Nucl. Phys. 134 (2024) 104080].*

Recently, measurements of production yields of hyperons with multi-strange content and their polarisation have been employed to test various theoretical predictions based on hydrodynamic and (microscopic) transport-model calculations. Figure 3.9 (left panel) shows the excitation function of polarisation compared to 3-d fluid calculations based on different equations of state for a hadronic, crossover and 1-order phase transition and calculations assuming direct connection between the polarisation vector and the thermal vorticity in thermal equilibrium. As one can see, the largest signal is measured at low energies (HADES) but the data quality is still not sufficient to discriminate between models and this calls for future high precision data.

Constraints to the equation-of-state (EoS) of nuclear matter in the region of sub- to supra-saturation densities can also be obtained employing heavy-ion collision from Fermi energies (GANIL, INFN-LNS) to (ultra-)relativistic energies (GSI, CERN), specifically at the Lawrence Berkeley National Laboratory (LBNL), at Brookhaven National Laboratory (BNL) at the Michigan State University (MSU) and at GSI/FAIR. The focus of the experimental and theoretical efforts of the last years has been determining the (baryon-) density dependence of the symmetry energy, which sets apart the EoS of neutron-rich nuclear matter from the symmetric one (see the right panel of Fig. 3.9). The relevant observables in this branch of heavy-ion collision physics at supersaturation densities are the flow of charged pions, protons and neutrons, and the sensitivity to the EoS is pinned down by the detailed comparison to several flavours of transport models (e.g. pBUU, RVUU, SMF, or IQMD, dcQMD, ImQMD, AMD). These models have been calibrated and tested against each other in extensive benchmark campaigns in past years (TMEP: Transport Model Evaluation Project). The collision energy ranges from 200 AMeV to several AGeV with the intent to maintain a nucleonic degree of freedom during the evolution of the system and to test its isospin-dependent behaviour by selecting different isotopes as targets and projectiles.
Other observables which provide access to the symmetry energy as, for example, the measurement of the neutron skin of neutron-rich nuclei or giant dipole resonance effects, are discussed in chapter 3, “nuclear spectroscopy”.

# Exploiting future opportunities

With the CERN LHC, the CERN SPS and FAIR, Europe hosts leading research facilities for the study of matter under conditions of extreme temperature and density. Europe collaborates within a world-wide context that includes, notably, the Relativistic Heavy Ion Collider (RHIC) in Brookhaven USA, as well as research facilities planned in Japan and China which drive science and innovation, see chapter 8. Hadron physicists are therefore in a good position to prepare the community to benefit from these new technologies.

## The experimental high temperature frontier: LHC Run 3 and 4 (2022-32)

### Exploiting detector upgrades

The next decade represents a precision era for high-energy-density QCD with small and large colliding systems. With the LHC Runs 3-4 (2022-2032), an increase of the delivered luminosity by a factor of about 10 compared to Run 2 is planned for all experiments in p+Pb and Pb+Pb collisions. The LHC discovery of QGP-like effects in small systems will be explored in great detail with high-multiplicity pp collisions at top LHC energy (13.6 TeV, currently), p+Pb collisions, and with a short O+O run. p+O collisions will also be used to constrain models of primary cosmic-ray interactions.
For this ambitious programme, the ALICE and LHCb detectors were successfully upgraded before Run 3, while the ATLAS and CMS detectors will be upgraded before Run 4. All four experiments will have improved tracking precision and data taking rates (see Fig. 3.10).

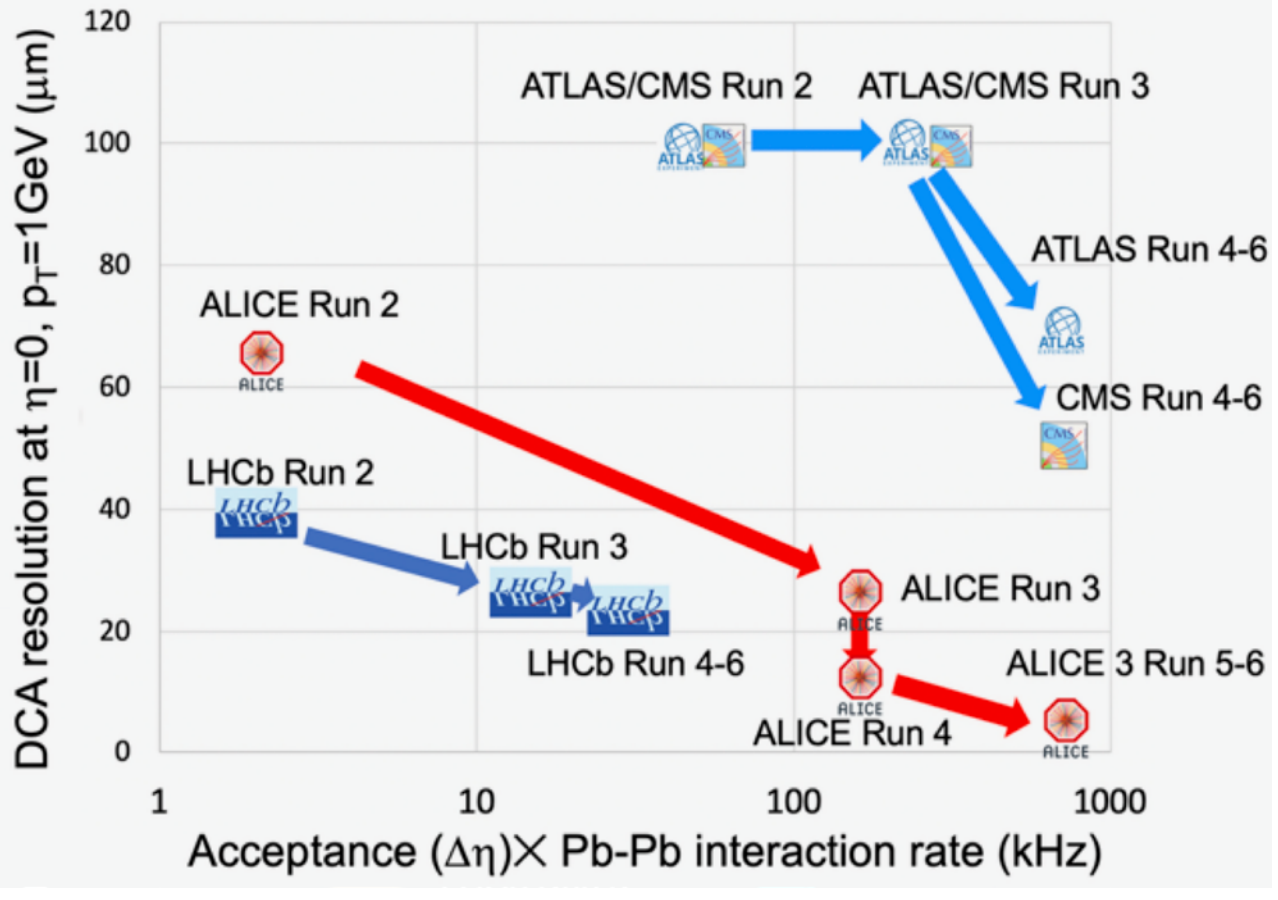

*Fig. 3.10: Improved performance of future LHC detectors. In this particular view, the resolution on the track distance of closest approach to the primary vertex (which is central, for example, for heavy flavour measurements) is shown together with the effective acceptance, defined as pseudorapidity acceptance multiplied by interaction rate. Note that LHCb will gain full reconstruction performance up to most central Pb-Pb collisions in Run 5.*

**ALICE**

The new inner tracking system improves spatial resolution tracking by a factor close to three and precise tracking is now also possible for forward-rapidity muons. Thanks to the major ALICE readout upgrade, the increase in inspected luminosity will reach a factor of ~100 for a broad range of measurements. Between Run 3 and Run 4, ALICE plans upgrades including a new ultra-light and smaller-radius pixel detector (ITS3) and a forward calorimeter (FoCal). The ITS3 is based on the technology of large-area curved monolithic active pixel detectors, which will set new state-of-the-art high-precision vertex detection down to 15 $\mu$m for particles with 1 GeV/c produced at central rapidity.

These upgrades will enhance performance for heavy-flavour mesons and baryons, as well as thermal di-electrons radiated by the QGP. FoCal includes a high-readout-granularity electromagnetic calorimeter coupled to a hadronic calorimeter and it will provide access to gluon densities in nuclei with $x$ down to $10^{-6}$ using forward direct photon measurements in p+Pb collisions.

**ATLAS**

The ATLAS apparatus is undergoing a series of general upgrades which enhance its capability to perform heavy ion physics. The new small wheel is a part of the muon trigger system that enhances the muon coverage and reduces fake rates. During the long shutdown 3 the new all-silicon inner tracking system with 10 times more readout channels will extend ATLAS tracking coverage to ±4 units of pseudorapidity and will improve the precision of the vertexing. Improvements are foreseen in the calorimeter electronics and data acquisition system. In particular, the high-granularity timing detector will be used as a part of the trigger system.

Besides general improvements in the ATLAS detector, upgrades are ongoing for heavy-ion specific studies. The zero-degree calorimeter that plays a crucial role in ultra-peripheral collision studies has been refurbished for Run 3 and equipped with a new reaction-plane detector. During the long shutdown 3 it will be replaced with a new-generation detector devised by a collaboration between ATLAS and CMS teams.

**CMS**

The CMS detector's capabilities will be significantly enhanced by the MIP Timing Detector (MTD) from Run 4 onwards. This detector will offer high timing precision and cover a pseudorapidity range up to six units. It includes barrel and endcap timing layers, utilising LYSO:Ce crystal scintillators and low gain avalanche diodes (LGADs) respectively. The MTD will enable charged hadron identification over a wide acceptance for the first time, providing unique opportunities in understanding collision dynamics and collective behaviour, and will enhance heavy flavour particle reconstructions by enabling precise charm hadron collectivity measurements.

**LHCb**

With its recent upgrade, LHCb has improved resolution on the impact parameter by 40% (20 $\mu$m) and can reconstruct tracks in Pb-Pb events in centralities up to 30% for the first time. LHCb has also completed the fixed target SMOG2 setup. This allows the exploration of centre-of-mass energies up to 115 GeV, which explore the region between the SPS and the RHIC regimes. The LHCb collaboration plans another upgrade between Run 3 and Run 4 to include a RICH Cherenkov detector to improve hadron identification, a Calorimeter and possibly additional scintillator counters to improve the acceptance for low momentum tracks and thus allow reconstructing converted photons.

**sPHENIX**

Before Brookhaven's Relativistic Heavy Ion Collider RHIC is transitioned to the Electron-Ion-Collider EIC, the sPHENIX experiment at RHIC will characterise, with greatest accuracy, the rarest and hardest probes accessible at RHIC energies. The sPHENIX programme also anticipates collecting a large data set of proton-Au collisions. The combined analysis of data from sPHENIX and LHC will allow for the comparison of rare QGP signatures measured over an unprecedentedly large range of a factor 30 in centre-of-mass energy. This provides a very valuable complementary access to understanding the dependence of jet quenching and heavy-quark properties on the temperature and density of the QGP.

## Physics aims

Future LHC runs aim at learning how collective phenomena and macroscopic properties, involving many degrees of freedom, emerge from strong-interaction physics in the non-perturbative regime at the microscopic (quark, gluon) level. This includes the characterisation of the long-wavelength properties of the QGP with unprecedented precision, as well as the study of microscopic parton dynamics underlying QGP properties and the development of a unified picture of QCD particle production from small (pp) to larger (pA and AA) systems. Future LHC runs will also aim at further probing nuclear parton densities in a broad $(x,Q^2)$ range, and studying the strong interaction between hadrons and the formation of molecules and light nuclear states.

Macroscopic properties of the QGP. The long-wavelength behaviour of hot and dense QCD matter can be described in terms of fluid- and thermodynamic concepts. It is experimentally investigated mainly using measurements of low-momentum (< 5 GeV/c) hadron production and flow patterns, as well as of electromagnetic radiation. Among the macroscopic properties of the QGP that will benefit strongly from the large luminosity increase, the temperature will be, for the first time at the LHC, determined with an accuracy of about 10% by measuring thermal radiation. LHC will perform heavy flavour flow measurements in the future with an accuracy currently available only for light-flavour flow. This will permit, for instance, reducing current uncertainties in extracting the heavy-quark diffusion coefficient $D_s$ by approximately a factor of 2 (see Fig. 3.11).

Microscopic structure and inner workings of the QGP. Hard processes resolve the QGP constituents and test jet-medium interactions at the microscopic level. Multi-differential jet measurements are one of the main avenues for these investigations. This includes in particular the Z-jet recoil measurements with 4 times reduced uncertainties and novel jet-substructure studies. Measurements of the production of charmonium and bottomonium states with different binding energies give access to a well-defined set of length scales in the QGP. LHC determines their modification in a colour-deconfined medium via the characterisation of the mechanisms of melting and regeneration.

Nuclear parton densities and search for saturation. High-luminosity p-Pb and Pb-Pb runs, will provide largely improved precision and kinematic coverage for measurements of the PDFs in nuclei (see right-hand panel of Fig. 3.11), from the high-$Q^2$ and x~ $10^{-3}$-$10^{-1}$ region, with Z, W, dijets, and possibly top quarks, down to the small-x region below $10^{-4}$ with forward photons, Drell-Yan and heavy quarks, where non-linear QCD evolution and phase-space saturation could set in.

Particle production and QCD dynamics from small to larger systems. High-precision studies of rare probes are essential to addressing outstanding open questions on the existence of a QGP in small collision systems. Example studies include the comparison of heavy-quark and quarkonium flow in small and large systems and the searches for thermal radiation and partonic energy loss, an outstanding puzzle given that the observed flow requires significant final-state interactions.

Strong interaction between hadrons and formation of light nuclear states. LHC will allow precise measurements of light and hyper nuclei production, as well as of hadron-hadron interactions. Elucidating the production mechanisms of nuclei has an impact on the searches of

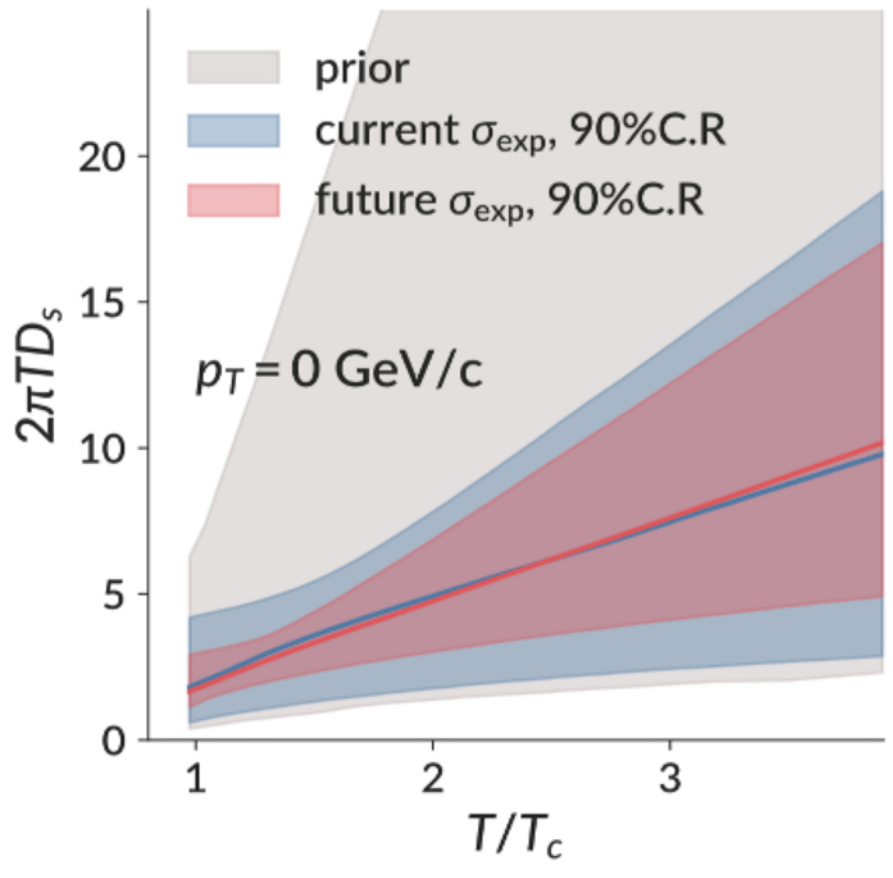

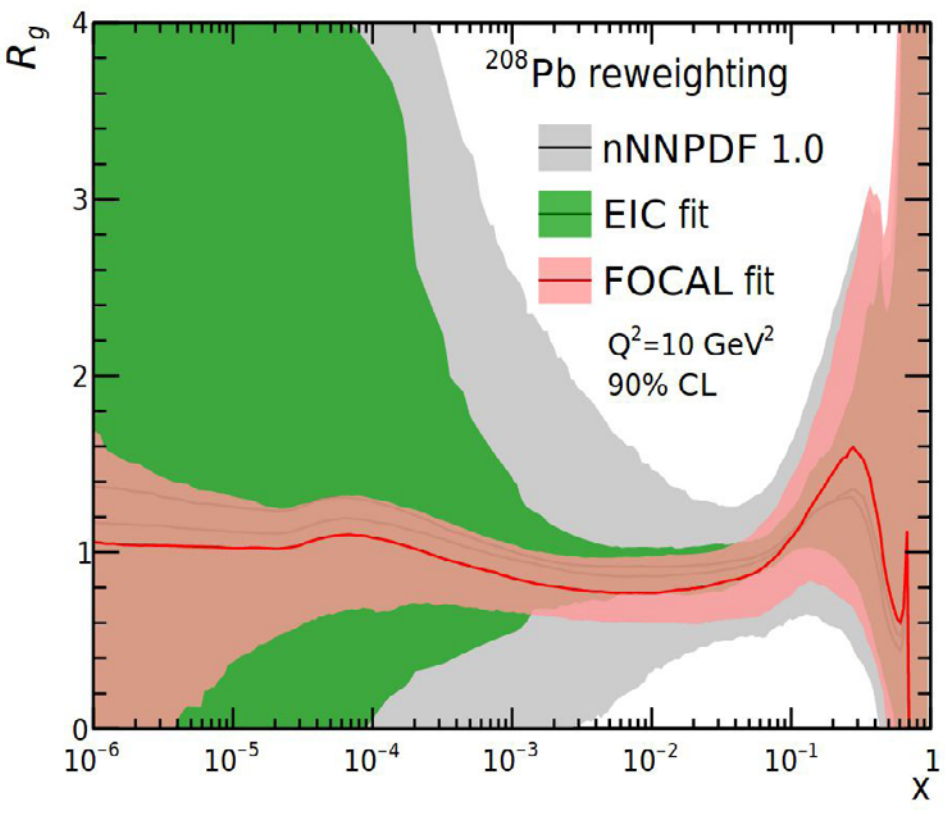

*Fig. 3.11: (Left panel) Expected improvement of the heavy-quark diffusion coefficient $D_s$ due to the LHC data from Run 3 and Run 4. Right panel) ALICE upgrade projection for the measurement of nuclear modification of the gluon distribution function as a function of $x$ [CERN Yellow Rep.Monogr. 7 (2019) 1159-1410].*

Dark Matter annihilation into nucleus-antinucleus pairs in Space, for which nuclei produced in cosmic-ray interactions is the main background. The studies of proton-hyperon correlations in momentum space will improve significantly during Run 3 and will be extended to multi-strange hyperons, charm baryons, and to hadron triplets. The results are expected to have implications for the equation of state of matter in dense neutron stars.

# The experimental high temperature frontier: LHC Runs 5 and 6 (2035 - 41)

## Preparing for the next physics harvest

Despite the strong progress expected from the upgraded experiments and the large increase in integrated luminosity, several crucial questions on the QGP and its properties will remain unanswered after LHC Runs 3-4. These questions include the time evolution of the QGP temperature, the expected restoration of chiral symmetry, the degree of thermalisation of charm and beauty quarks and the underlying microscopic mechanisms, and the hadronisation of heavy quarks from the QGP. The exploitation of the LHC as a heavy-ion collider throughout its entire High-Luminosity phase, up until Run 6, is a unique opportunity to address these questions.

## Detectors and physics aims for Runs 5 and 6 at the LHC

For the LHC Runs 5 and 6 (2035-2041) the ALICE collaboration proposes a new, next-generation detector called ALICE 3. A proposed second major LHCb upgrade will grant access to central Pb-Pb collisions and to fixed-target collisions on a polarised target. The ATLAS and CMS detectors will undergo their major upgrades before Run 4 and should exploit the full integrated luminosity for heavy ions that the LHC will deliver throughout Run 6.

The ALICE 3 detector consists of a vertexing and tracking system over a large pseudorapidity range $(-4 < \eta < +4)$, complemented by multiple sub-detector systems for particle identification including silicon time-of-flight layers, a ring-imaging Cherenkov detector with high-resolution readout, a muon identification system and an electromagnetic calorimeter. Unprecedented pointing resolution of 3-4 $\mu m$ at $p_T$=1 GeV/c at midrapidity in both the transverse and longitudinal directions can be achieved by placing the first layers at 0.5 cm from the beam axis on a retractable structure to leave sufficient aperture for the beams at injection energy. ALICE 3 also aims at a superb time resolution of 20-30 ps and low-gain avalanche d (LGAD devices have already demonstrated this capability). CMOS sensors with an additional gain layer or silicon photomultipliers coupled to a thin resistive layer could also achieve the same performance and are currently being investigated.

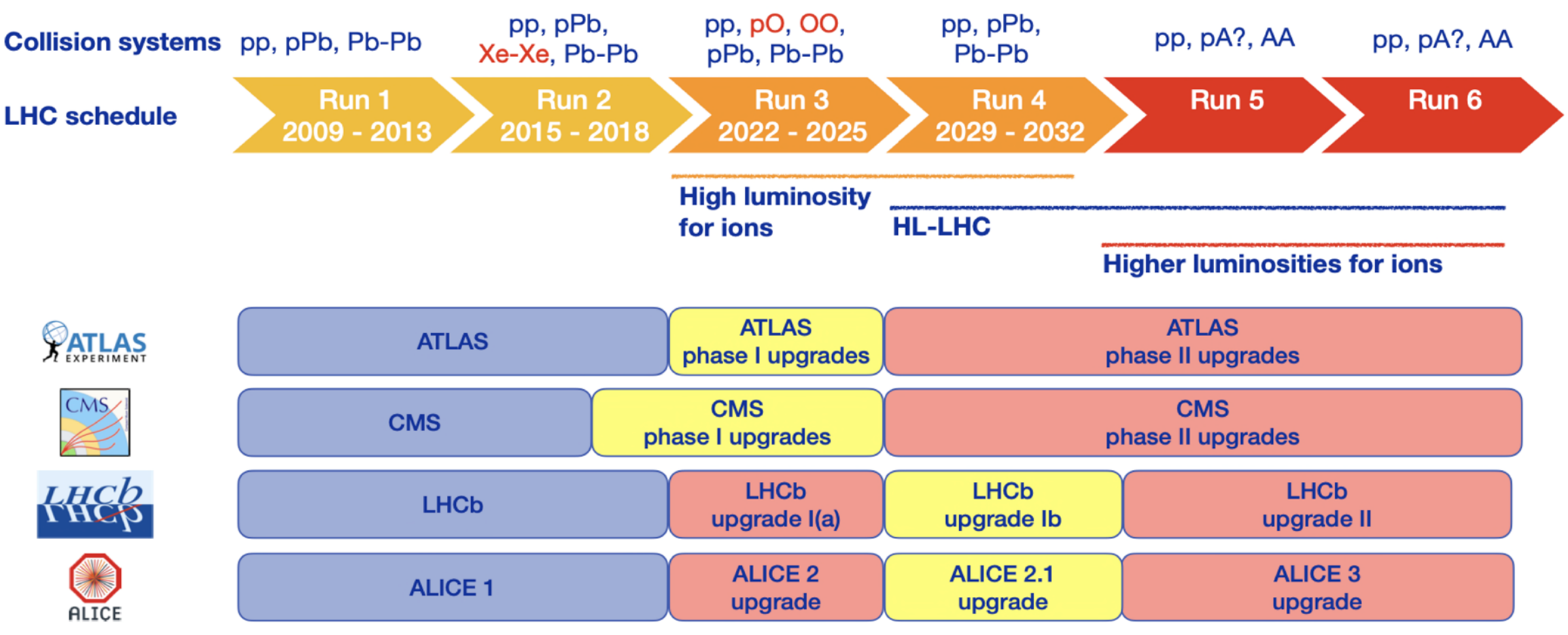

*Fig. 3.12: Timeline of current LHC experiments and upgrades. [J. Klein, QM2022, Acta Phys. Polon. Supp. 16 (2023) 1,25]*

This next-generation apparatus will focus on novel studies of the QGP, including the low-$p_T$ production of charm and beauty hadrons, multi-charm baryons and charm-charm correlations, and the precise multi-differential measurements of di-electron emission to probe the mechanism of chiral-symmetry restoration and the time-evolution of the QGP temperature. The programme aims to collect an integrated luminosity of about 35 $nb^{-1}$ with Pb-Pb collisions and 18 $fb^{-1}$ with pp collisions at top LHC energy. The technology developed for ALICE 3 can also be applied at other facilities for nuclear and high-energy physics (SPS/NA60+, EIC/ePIC, FCC-ee).

Both LHCb and SMOG are planning a further upgrade for Run 5 to cope with a substantial increase in luminosity and pile up. An additional scintillating fibre detector on the side of the magnet will allow for the reconstruction of low-energy hadrons and converted photons. A novel UT detector based on pixel technology will increase the VELO-UT matching efficiency up to almost 90%. An additional silicon tracker in the innermost region (Mighty Tracker) will permit reconstructing the most central Pb-Pb collisions.
The SMOG2 programme will include a polarised target with the aim of studying the inner structure of the proton in a way complementary to the studies at the EIC.

## The experimental high net-baryon density frontier

During the period of this Long Range Plan, RHIC will be transitioned to the Electron-Ion-Collider, and experiments at GSI will be transitioned to the FAIR facility currently under construction. The CERN SPS will continue to provide heavy ion beams to the experiments, possibly exploiting high beam intensities (up to ~$10^7$ ions/spill) and several nuclear species (see Fig. 3.13). Experiments at the high net-baryon density frontier will thus be prepared and performed at FAIR and at the CERN SPS.
The future exploration of the QCD phase structure aims at systematic measurement of excitation functions and system size dependencies, including measurements of event-by-event fluctuations, thermal radiation of photons and dileptons, and the production of charmed and (multi-)strange hadrons and the production of hypernuclei. To scan the QCD phase structure for the crossover region, first order deconfinement and chiral phase transition, it is important to perform these measurements for beam energies ranging from full SPS down to SIS100 and to SIS18. Collision systems including nucleus-nucleus and proton nucleus should be studied, with proton-proton data providing a mandatory reference.

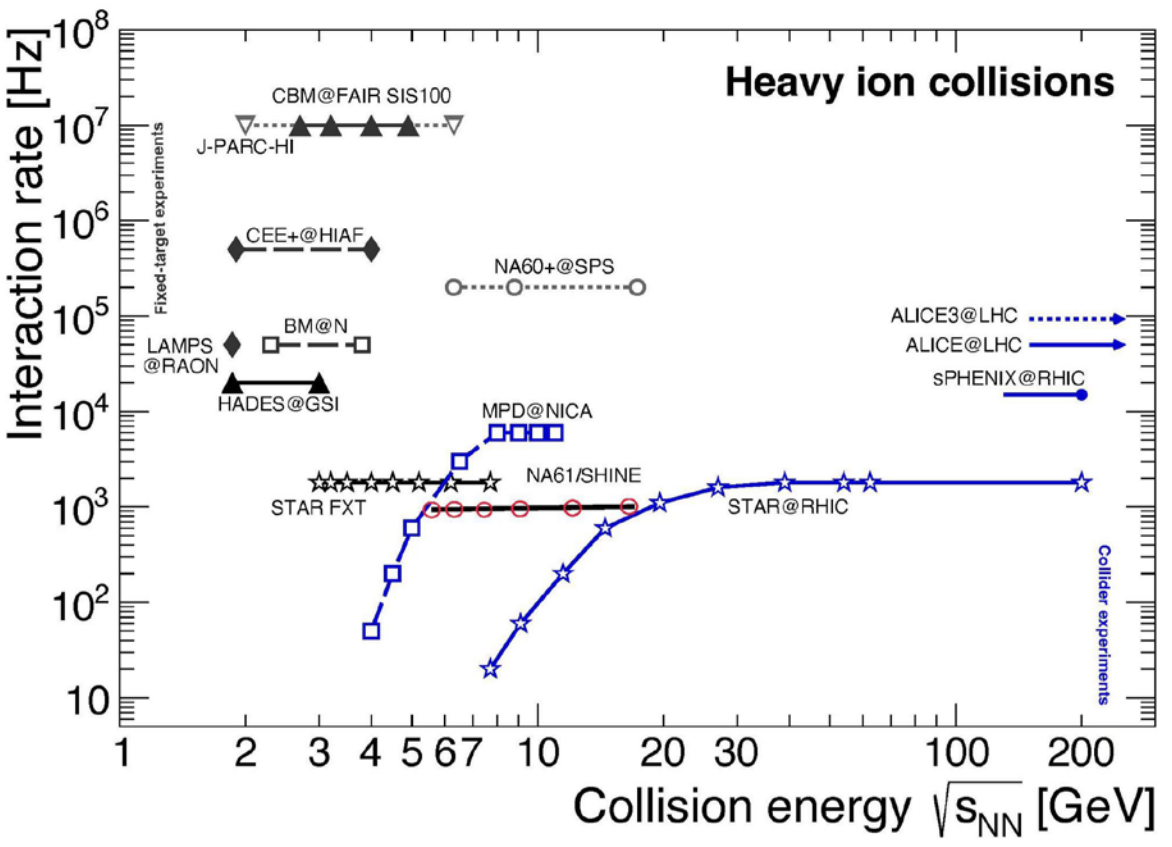

*Fig. 3.13: (Left) Available or expected interaction rate reached by existing or foreseen experiments, as a function of $\sqrt{s_{NN}}$ [EPJA 53 3 (2017) 60, NPA 982 (2019)].*

## Exploiting the novel capabilities of CBM and NA60+ and the role of HADES at FAIR and NA61/SHINE at SPS

Compared to previous studies, the CBM and NA60+ experiments will have unique capabilities in dealing with a high interaction rate for nuclear collisions (up to 10 MHz for CBM and ~150 kHz for NA60+, see Fig. 3.13). This will allow the study, with high precision, of rare probes so far accessible only at top SPS ($\sqrt{s_{NN}}$ = 17.3 GeV) and SIS18 ($\sqrt{s_{NN}}$ = 2.42 GeV) energies. For instance, both CBM and NA60+ plan to measure the rare thermal dilepton signal with uncertainties as low as 3 - 5%. Thus, together with high precision ALICE 3 measurements, the entire QCD phase diagram will be mapped out with dilepton observables (Fig. 3.14).

Searches for the signal of degeneracy of chiral partners $\rho$(770) and $a_1$(1260) in the dilepton spectra will focus on the 1.1<M<1.5 GeV/c² invariant mass range. An increase of the dilepton yield in this region, due to chiral mixing, would signal chiral symmetry restoration. The search for such direct signals of the chiral mixing will be pursued in fixed target experiments at high baryon number density (NA60+, CBM, HADES) and in collider experiments at the high temperature frontier (ALICE, ALICE 3, LHCb-II). Another interesting application of dilepton radiation will be provided by the measurement of very soft dileptons with masses and momenta below 100 MeV to extract the electrical conductivity of QCD matter in both the baryon dominated environments (HADES and CBM) and the pion dominated ones (ALICE/ALICE 3 at the LHC). Measuring the production of open heavy flavours in nucleus-nucleus collisions at low collision energies, it is possible to access the charm diffusion coefficient, extract information on charm thermalisation, on the hadronisation mechanism and eventually measure the total charm cross-section.

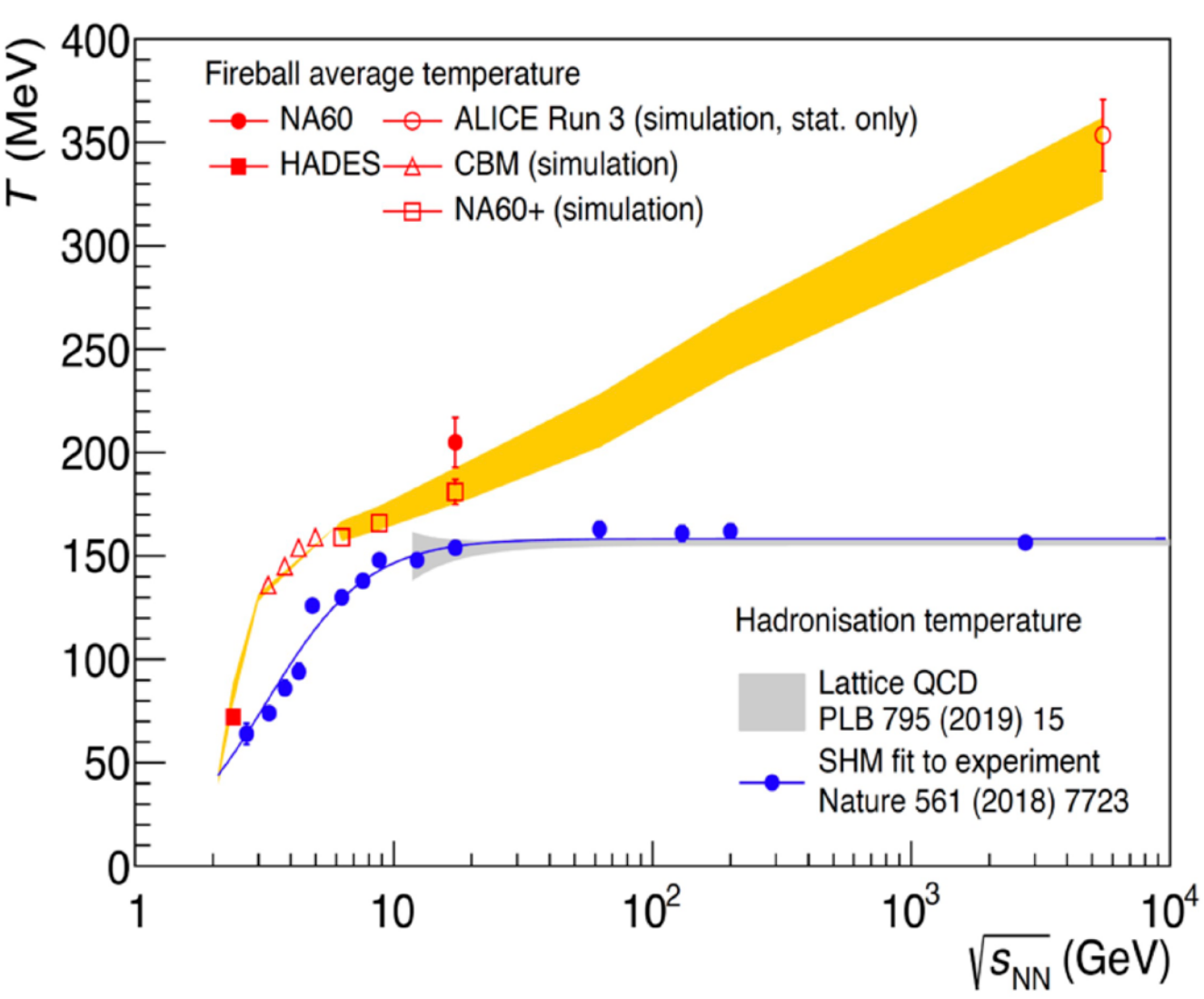

***Fig. 3.14: Excitation function of the invariant-mass slope parameter (red symbols) and of the chemical freezeout temperature (blue symbols). Temperatures extracted from dilepton spectra measured by HADES [HADES, Nature Phys. 15, 1040 (2019)] and NA60 [AIP Conf. Proc. 1322, 1 (2010)] (solid read symbols), anticipated performance by CBM, NA60+ and ALICE (after Run4) (open red symbols). The yellow band represents the temperature as obtained from fireball model calculations with in-medium electromagnetic spectral functions [R. Rapp and H. van Hees, Phys. Lett. B753, 586 (2016)] (for > 6 GeV) and from a coarse-grained transport approach [T. Galatyuk, P. M. Hohler, R. Rapp, F. Seck, and J. Stroth, Eur. Phys. J. A 52, 131 (2016)] (for < 6 GeV).***

NA60+ measurements of charm production will be sensitive to the interactions of charm hadrons in the hadronic phase and to the degree of thermalisation of the heavy quarks in such a medium. NA60+ plans to perform multi-differential open charm production at low SPS $\sqrt{s_{NN}}$. Quarkonium suppression, already addressed in great detail at top SPS, RHIC, and LHC energies, will also have a pivotal role in the NA60+ physics programme. Before NA60+ starts operation, NA61/SHINE aims to perform a first direct measurement of open charm production in Pb-Pb collisions at top SPS energies and a first exploratory measurement at lower SPS energy.

Studies of the open- and hidden charm production and in-medium propagation in proton-induced reactions with an overlapping energy for CBM and NA60+ will provide complementary results and enable us to understand charm production and interaction at low energies. This might shed more light on the question of intrinsic charm content of nucleon and origin of nucleon mass.

EoS studies of neutron rich nuclear matter and the symmetry energy at saturation density will be provided via measurements of dipole resonance strength distributions, neutron skin thickness, cluster formation in a broad range of proton/neutron asymmetry. At higher densities, measurement of elliptic flow of protons and neutrons in heavy ion collisions will be studied at GSI-SIS18 by ASY-EOS II (at R3B) and HADES experiments with their greatly improved set-ups. This experimental effort will be supported by a network of transport model calculation experts aiming to investigate the sensitivity of the experimental observables to the nuclear EoS at larger densities and to improve codes by simultaneous description of new high-precision multi-differential experimental data. Isospin dependence of short-range proton-neutron (SRC) correlations, important for the understanding of neutron star cooling, will be investigated in short-lived neutron-rich nuclei and different mass ranges. Complementary to R3B, HADES will also measure SRC with particular focus on studying the possible impact of correlated pairs in sub-threshold strangeness production. Whether such a process would enhance the production of mesons near the free threshold is still a matter of theoretical debate.

## Timeline of experiments at the high-density frontier

The FAIR facility plans to have completed its start configuration in 2028. This will include the Super Fragment Separator (S-FRS), SIS100 and CBM detector. FAIR Phase-0 was launched in 2018 to bridge the physics programme from SIS18 to SIS100 energy ranges. Experiments in baryon dominated matter frontier are - and will be - carried out by R3B, HADES and CBM experiments. R3B is designed for kinematically complete measurements of various reactions with radioactive beams with energies of 0.5 to 1.6 GeV/nucleon provided by S-FRS. Versatile design of R3B and unprecedented intensity of S-FRS will make possible studies of the equation of state of baryonic matter (see Fig.3.9, right panel) with large neutron content.

HADES plans a beam energy scan of 200-800 MeV/nucleon with Au+Au collisions to study hadron and dilepton production, particle correlations, flow and fluctuations (higher order cumulants) in the vicinity of the nuclear liquid - gas phase transition. A significant number of reference measurements for the CBM heavy-ion programme at SIS100 will be conducted in p+Ag collisions at kinetic beam energy of 4.5 GeV. The HADES detector will be moved to the FAIR fixed-target experimental hall once the CBM detector is installed there. The HADES physics programme at the CBM cave is complementing the beam energy scan programme of the CBM experiment by providing acceptance at mid- and backward rapidity at beam energies below 4 AGeV, thus providing an important sensitivity to cold-matter medium effects.

The start of operation of the SIS100 synchrotron and CBM experiment is envisaged for late 2028. Within the first three years of operation CBM aims at measuring, with unprecedented precision, the excitation function of hadron and dilepton production in A+A collisions for beam energies in the range of $\sqrt{s_{NN}}$ = 2.7 – 4.9 GeV.

### Experiments at the CERN SPS

The CERN-SPS offers unique opportunities for studying high-density matter at beam energies between the top SIS100 energy of 29 * Z/A GeV/u and 400 * Z/A GeV/u. This gives access to a region of the QCD phase diagram at finite but lower µB.

NA60+ is a proposed fixed target experiment at the CERN SPS that will bring unique high-rate capabilities to the study of electro-magnetic processes and charm production in nucleus-nucleus and proton-nucleus collisions at SPS energies. Its setup is based on a silicon vertex telescope coupled to a muon spectrometer that will study heavy ion collisions in the centre-of-mass-energy range between 6.3 ($E_{lab}$= 20 A GeV) and 17.3 GeV ($E_{lab}$= 158 A GeV). NA60+ has submitted a Letter of Intent to the CERN SPS review committee (SPSC), and it pursues a strong R&D programme with the aim of taking data after CERN Long Shutdown 3, in 2029.

NA61/SHINE is a multi-purpose experiment that investigates hadron production in nuclear collisions. So far, NA61/SHINE has systematically searched for signals of the onset of deconfinement, using a large variety of ion species (Be+Be, Ar+Sc, Xe+La, Pb+Pb) at different energies, ranging, in centre-of -mass energy, between 5 and 17 GeV. The experiment has undergone a significant upgrade of both the vertex detector, in order to reach 1 kHz interaction rate, and of the electronic hardware to significantly increase the acquisition capabilities.

## The theory frontier

Ultra-relativistic heavy-ion collisions pose qualitatively novel challenges for our understanding of QCD (see also chapter 8.2 on computing). As the collision dynamics include many stages in which characteristically different physics occurs, the full exploration of the data collected in recent and future experimental campaigns requires, in particular, progress on the following topics:

#### Pre-equilibrium dynamics of fundamental quantum fields

The rapid space-time evolution of the system towards an equilibrium state calls for improved understanding of non-equilibrium dynamics. Recent progress in understanding mechanisms of rapid equilibration and hydrodynamisation in weakly and strongly coupled quantum field theories needs to be developed further.

#### QCD thermodynamics and hydrodynamics

Fundamental properties of the equilibrium state are calculable from first principles in quantum field theory. At vanishing baryon chemical potential, computer simulations permit solving QCD at finite temperature on a discretised space-time lattice. However, such ab-initio calculations based on first principles in QCD are extraordinarily hard and resource-intensive. So far, they exist only for an important but small set of quantities and they do not extend e.g. to QCD transport coefficients. Also, because of the sign problem, lattice QCD techniques do not extend to QGPs at finite baryon chemical potential. Powerful approximations allow us to address a broader set of physics questions that are not yet accessible in lattice QCD. This includes perturbative methods in finite temperature field theory at vanishing and finite µB. At lower temperature, effective field theories that are based on hadronic degrees of freedom and that respect the symmetries of QCD are needed.
Further progress in this direction relies on unwavering support for both resource-intensive future lattice QCD simulations and complementary field theory-based approaches that exploit powerful approximations.

#### QCD of hard and electromagnetic processes in the QGP

QCD perturbation (pQCD) theory allows us to calculate the initial production rate of “hard” high-momentum transfer processes, such as jets, from first principles. pQCD-inspired calculations have also been key to our current phenomenological understanding of jet-QGP interactions. These lead to medium-induced modifications of initial production rates that are characterised experimentally by a broad suite of “jet quenching” observables. In the coming decade, experiments at the LHC will significantly increase the accuracy of detailed jet quenching measurements. Further advances in the theoretical understanding of this phenomenon and in its phenomenological simulation in medium-modified parton showers should be strongly supported to fully exploit the physics potential of hard probe measurements at the LHC.

### Modelling of ultra-relativistic heavy ion collisions

is an essential step in interfacing experimental data with firm QCD predictions. Modelling of the global space-time and temperature evolution, and over-constraining this modelling with as many data as possible, is essential for understanding the cauldron within which detailed properties of the QGP are subjected to experimentation. The challenges of modern modelling include, amongst others, a theory-guided modelling of the initial pre-equilibrium dynamics based on insights from QCD effective kinetic theory and kinetic transport models, a fluid dynamic evolution that encodes information about QCD thermodynamics, the modelling of hadronisation and of hadronic elastic and inelastic interactions based on detailed knowledge of hadronic physics. In addition, hard penetrating and electro-weak processes need to be modelled throughout these multi-stage simulations. Increased experimental data sets enable more detailed data-theory comparisons that warrant more detailed theoretical input formulated in more complex simulation codes. This complexity now requires support for larger theory collaborations, the use of Bayesian analysis techniques and a tighter interplay between theory and experiment. Applications of quantum computing to central computational problems in the theory of heavy ion collisions should be supported, as they could be potential game changers in the long run.

The theory programme should be supported via excellence programmes to train, attract and keep talent within the field. Theory centres should be strongly supported throughout Europe, in particular the European Centre for Theoretical Studies (ECT*, Trento, Italy), which is the unique European centre dedicated to theoretical nuclear physics in the broadest sense and complementary in scope and activities to existing research facilities based at universities or experimental laboratories.

## Forthcoming Detector Challenges and New Instrumentation

Detector R&D is a key tool for the successful planning of the next generation of nuclear physics experiments exploring the QCD phase space. In particular, the following technologies (see chapter 8.1 ‘Detectors and experimental techniques’ for details) are recognised as essential.

● Precise vertexing and tracking accuracy in high particle density by means of MAPS detectors

● Ultra-fast (20-30 ps time resolution) silicon detectors for time-of-flight-based particle identification based on either LGAD or CMOS sensors with an additional gain layer or SiPM with resistive layers.

● RICH detectors for high-momentum particle identification optical-quality aereogels or fluorocarbon gases as an active medium.

● High-granularity calorimeters which sample showers via silicon pixel detectors (MAPS).

## Recommendations: Strongly Interacting Matter at Extreme Conditions

Ultra-relativistic heavy ion collisions aim at producing and studying the quark gluon plasma, which is the qualitatively novel state of nuclear matter at extreme conditions of temperature and density. Different collision energies achieve the QGP at different temperatures and densities.

The experimental focus is to discover in microscopic detail the material properties of the QGP at the highest temperature reached at the LHC, and to find the expected onset of the first-order phase transition at finite baryon density at FAIR. Given the long timescales necessary for the R&D and construction of these experiments, a sustained research effort requires advancing the development of the next generation experiments in parallel with the ongoing exploitation of existing facilities and detectors.

The priorities in this multi-pronged endeavour can be summarised as follows:

### Future flagship facilities and experiments

● ALICE 3 at CERN is a completely new dedicated high-energy nuclear physics experiment based on innovative detector concepts that will be essential for continuing a scientifically leading role for Europe in high energy nuclear physics after 2035. The programme relies on innovative R&D that will benefit neighbouring fields of nuclear and particle physics. Strong support for R&D should be ensured to maintain the opportunity of installing ALICE 3 for Run 5 at the LHC.

● To investigate nuclear matter at high baryonic density, the timely completion of SIS100 at FAIR and the completion of the CBM experiment are of utmost importance. Efforts should continue to support R&D activities related to advanced CBM silicon vertexing and tracking devices.

● To exploit physics opportunities at the CERN LHC after 2035 (Runs 5 and 6), the LHCb Upg2 and the fixed-target setup will have a strong impact on the heavy ion programme. ATLAS and CMS will play an important role in the characterisation of high-momentum transfer processes up to the end of the LHC programme in Run 6. Efforts should be made and support given to these initiatives; R&D and construction for the LHCBb Upg2 detector deserves full support.

● The NA60+ detector at the SPS will address the remaining open questions in the electromagnetic and charm sectors at the SPS with unprecedented event rates. R&D and construction for this detector deserves strong support.

### Support of existing facilities and experiments

● To maximise scientific output from the significant investment in current detector upgrades at the LHC, the continuation of the heavy-ion programme with Runs 3 and 4 (up to 2029) should receive full support. Timely support for the further ALICE upgrades in long shutdown 3 will provide a unique opportunity for enhancing the physics reach in Run 4.

● With its Upgrade I detector and with the new particle-identification subdetectors to be installed during LS3, LHCb is equipped to pursue a unique fixed target programme at the LHC and to perform competitive measurements for Pb-Pb systems in collider mode. The exploitation of these opportunities should receive full support.

● Exploitation of the existing detectors and facilities, in particular HADES and R3B at SIS18, should receive full support.

● NA61 at SPS should be fully exploited.

**Theory developments**

● Theoretical work in the field of heavy-ion collisions should be guaranteed continuous support, both in its phenomenological aspects (theoretical support needed to interpret the results and to provide feedback to the experimental programme) and from first principles (quantum chromodynamics).

● Collaboration strengthening the relationship between heavy-ion physics and neighbouring fields including astrophysics and particle physics; linking to novel ways of computing and data analysis; or improving the interplay between theory and experiment, should be particularly encouraged and nurtured.

# Nuclear Structure and Reaction Dynamics

**Conveners:**
**Silvia Leoni** (Università degli Studi di Milano and INFN, Italy)
**Tomás R. Rodríguez** (Universidad Complutense de Madrid, Spain)

**Liaisons:**
**Adam Maj** (IFJ-PAN, Kraków, Poland)
**Jelena Vesić** (Jožef Stefan Institute, Ljubljana, Slovenia)
**Ihor Kadenko** (Taras Shevchenko National University of Kyiv, Ukraine)

**WG Members:**

- **Dieter Ackermann** (GANIL Caen, France)
- **Alejandro Algora** (IFIC Valencia, Spain)
- **Sonia Bacca** (JGU Mainz, Germany)
- **Saul Becerio** (Novo, La Coruña, Spain)
- **Michael Block** (JGU Mainz, Germany)
- **Michal Ciemala** (IFJ-PAN Kraków, Poland)
- **Gianluca Colò** (University of Milano, Italy)
- **Ruben De Groote** (KU Leuven, Belgium)
- **Alessia Di Pietro** (LNS Catania, Italy)
- **Timo Dickel** (GSI Darmstadt, Germany)
- **Evgeny Epelbaum** (Ruhr Universität Bochum, Germany)
- **Bogdan Fornal** (IFJ-PAN Kraków, Poland)
- **Hans Fynbo** (University of Aarhus, Denmark)
- **Liam Gaffney** (University of Liverpool, UK)
- **Andreas Heinz** (Chalmers University, Göteborg, Sweden)
- **Morten Hjorth-Jensen** (Univ. Oslo & MSU, Norway)
- **Guillaume Hupin** (IJCLab Orsay, France)
- **Beatriz Jurado** (LP2I, Bordeaux, France)
- **Michal Kowal** (University of Warsaw, Poland)
- **Javier Menéndez** (University of Barcelona, Barcelona, Spain)
- **Daniele Mengoni** (University of Padova, Italy)
- **Tamara Niksic** (University of Zagreb, Zagreb, Croatia)
- **Janne Pakarinen** (JYFL-ACCLAB Jyväskylä, Finland)
- **Sorin Pascu** (IFIN-HH Bucharest, Romania)
- **Silvia Piantelli** (INFN Sezione di Firenze, Italy)
- **Zsolt Podolyak** (University of Surrey, UK)
- **Olivier Sorlin** (GANIL Caen, France)
- **Paul D. Stevenson** (University of Surrey, UK)
- **Marine Vandebrouck** (Irfu, CEA Saclay, France)
- **Carl Wheldon** (University of Birmingham, UK)
- **Jonathan Wilson** (IJCLab Orsay, France)
- **Kathrin Wimmer** (GSI Darmstadt, Germany)
- **Andreas Zilges** (Universität zu Köln, Germany)

Nuclear physics aims at understanding the structure and dynamics of atomic nuclei, as well as properties of nuclear matter including the beta-equilibrated matter inside neutron stars, through the synergy of theoretical and experimental research. Key open questions include the emergence of the nuclear chart and the boundaries of nuclear existence, the evolution of nuclear structure across the nuclear landscape, the complexity of nuclear shapes and spectra, decay rates of unstable nuclei, physical processes that govern all nuclear reactions and fission, and the study of exotic structures related to weakly-bound and open quantum systems, among others. While interesting on their own, these and other nuclear physics topics also figure importantly in atomic, hadron, particle, and astrophysics.
The advent of cutting-edge facilities, reaching out into hitherto unexplored regions of the Segré chart towards the limit of stability, and progress in nuclear theory provide a very substantial discovery potential for complex phenomena arising from fundamental constituents.

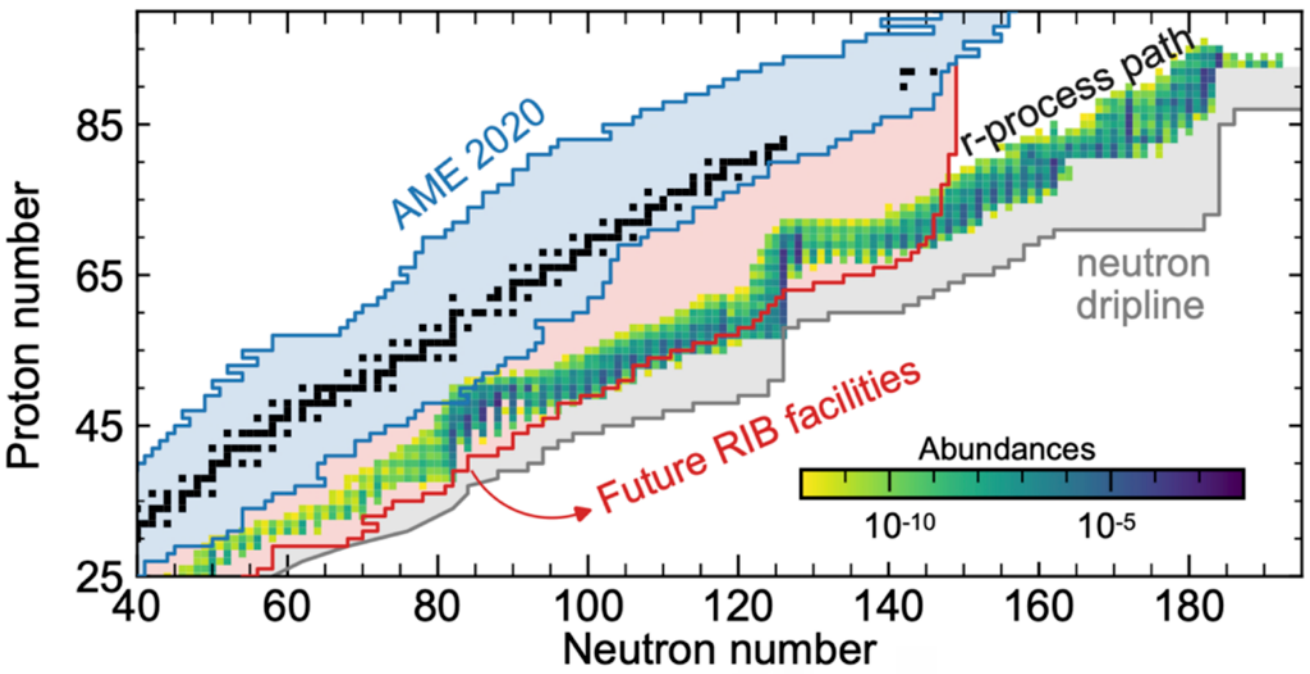

*Fig. 4.1: Nuclear chart showing stable nuclei (black squares), the limit of known masses from the 2020 Atomic Mass Evaluation (from Chinese Physics C 45 030003 (2021)) (blue region) and the reach of future radioactive ion-beam facilities (red region). Gray line represents the neutron drip-line (taken as heaviest nucleus in each isotopic chain with $S_{2n} > 0$) obtained by combining the predictions of 11 global models via Bayesian model averaging (from Physical Review C 101, 044307 (2020). Colour-code represents the (logarithm of the) abundances predicted at neutron freeze-out from a nuclear network calculation, simulating the occurrence of the r-process nucleosynthesis in a binary neutron star merger (Adapted from Physical Review C 102, 045804 (2020))*

# Nuclear interactions and nuclear models

The theoretical description of the atomic nucleus presents several intertwined problems, i.e. nuclear interactions are not univocally defined and are intricate, and the equations derived from quantum mechanics are hard (or impossible) to solve exactly. Therefore, approximations and/or models are unavoidable. Additionally, very demanding computing resources and complex programming tools are required (see Chapter Nuclear Physics Tools). State-of-the-art models that describe the nuclear structure aim at: a) using effective interactions or Energy Density Functionals that are increasingly accurate, or more rooted in QCD, or (possibly) both; b) providing approaches to the many-body problem that are systematically improvable and might extrapolate to exact solutions; c) computing an extensive list of observables in the widest possible range of nuclei; and/or, d) quantifying theoretical uncertainties. For example, the current ab initio nuclear structure methods are based on the rationale of effective theories. The starting nuclear interaction relates to QCD through effective field theory (EFT) methods and is then evolved (with unitary transformations) into a more suitable form to be eventually used in many-body calculations. A prerequisite for ab initio methods, besides using EFT-based nuclear interactions, is that these steps must be systematically improvable, and the uncertainties must be quantified [cf. a), b) and d) above].

## Nuclear interactions from effective field theory

On the fundamental level, nuclear properties ultimately emerge from the highly complex many-body dynamics of interacting quarks and gluons. As a more feasible alternative to the first-principles description of nuclear systems in terms of QCD, the problem must be simplified by using the appropriate effective degrees of freedom and effective interactions (see Box 2.5 in Chapter Hadron Physics). EFTs are directly connected to QCD and permit the exploitation of separations between energy (or distance) scales. The so-called Chiral EFT ($\chi$EFT) is the most widely used framework in nuclear physics nowadays: it is a low-resolution version of QCD based on its symmetries and the approximate chiral symmetry, and utilises protons, neutrons, pions and, sometimes, the lowest lying nucleon excitations (i.e., the $\Delta$ resonance) as active degrees of freedom. The interactions between nucleons can be derived from the chiral Lagrangian using perturbation theory. This expansion – in powers of the pion mass over a breakdown scale associated with the non-resolved physics – entails a power counting scheme that produces hierarchical 2N, 3N, etc. nucleon interactions as well as electroweak currents that depend on coupling constants. This framework is internally consistent, but discussions are underway regarding its ultimate capability to converge and/or alternate power counting schemes. The low-energy constants (LECs), although they should eventually be derived from the underlying QCD, are so far found by fitting to 2N, 3N, $\pi$N and, sometimes, many-nucleon experimental data. This leads to a variety of effective Hamiltonians that are not yet accurate enough to provide fully consistent predictions; to improve on this, there is a blooming of advanced Bayesian methods aimed at quantifying various types of uncertainties.
The $\chi$EFT expansion of the 2N force has already been pushed to fifth order (N4LO), allowing for a high-precision description of 2N scattering data below the pion production threshold. In contrast, 3N forces are still a challenge, facing conceptual problems (e.g., consistent regularisation) and computational issues (fitting more than 15 LECs will require the development of efficient emulators for 3N scattering), yet to be solved. Other open questions in $\chi$EFT are the comparison between expansions with and without explicit $\Delta$ isobars, combinatorial enhancements of many-body interactions for A-nucleon systems, resummation of relativistic corrections, large-Nc constraints merging with dispersion theory, or conceptual issues such as the renormalisation and power counting of the chiral expansion.
Nuclear phenomena characterised by lower momentum scales below ~100 MeV/c are insensitive to pion dynamics and can be studied using a simplified EFT framework with nucleons as the only degrees of freedom. This formulation, $\pi$-less EFT, is particularly efficient for studying universal aspects of few-body systems close to the unitary limit, halo nuclear systems or nuclear equation of state at low density.

## The nuclear many-body problem

Over the years, nuclear ab initio methods have undergone significant advancements: previously limited to a corner of the nuclear chart, these methods can now address heavier nuclei, although still presenting difficulties when going far from magic numbers. To achieve this, techniques like the similarity renormalisation group (SRG) have been crucial in setting up EFT Hamiltonians to be ready to use in many-body calculations. Additionally, a wealth of exact or systematically improvable many-body techniques, which differ in their formulation, numerical resolution and convergence behaviour, are being developed to enlarge the range of applicability and reduce the uncertainties of ab initio methods.
These methods are either based on the variational principle (e.g. quantum Monte Carlo or no-core shell model), or rely on expansion techniques (e.g. self-consistent Green's function, in-medium SRG (IMSRG), coupled-cluster theory, many-body perturbation theory), or take a combination of both (e.g. valence space IMSRG). Computational issues, such as scaling with the number of nucleons in the system or handling 3N and higher many-body interactions, play a crucial role in the range of applicability of the different methods. Furthermore, each of these tools possesses unique capabilities. Only a few of them can compute, on their own, the spectrum of a nuclear system, albeit often limited to a few low-lying states and, likewise, the extension to study deformed systems or heavy nuclei is limited to a few ab initio methods. Nuclear properties such as high-energy excitations, clustering, or reaction dynamics (to name a few) also remain a challenge. Basic nuclear properties like charge radii clearly display the uncertainties related to different Hamiltonians and many-body methods that we have already alluded to. Reducing these large uncertainties (compared to more phenomenological interactions) also in absolute binding energies and other observables is a goal for the near future.

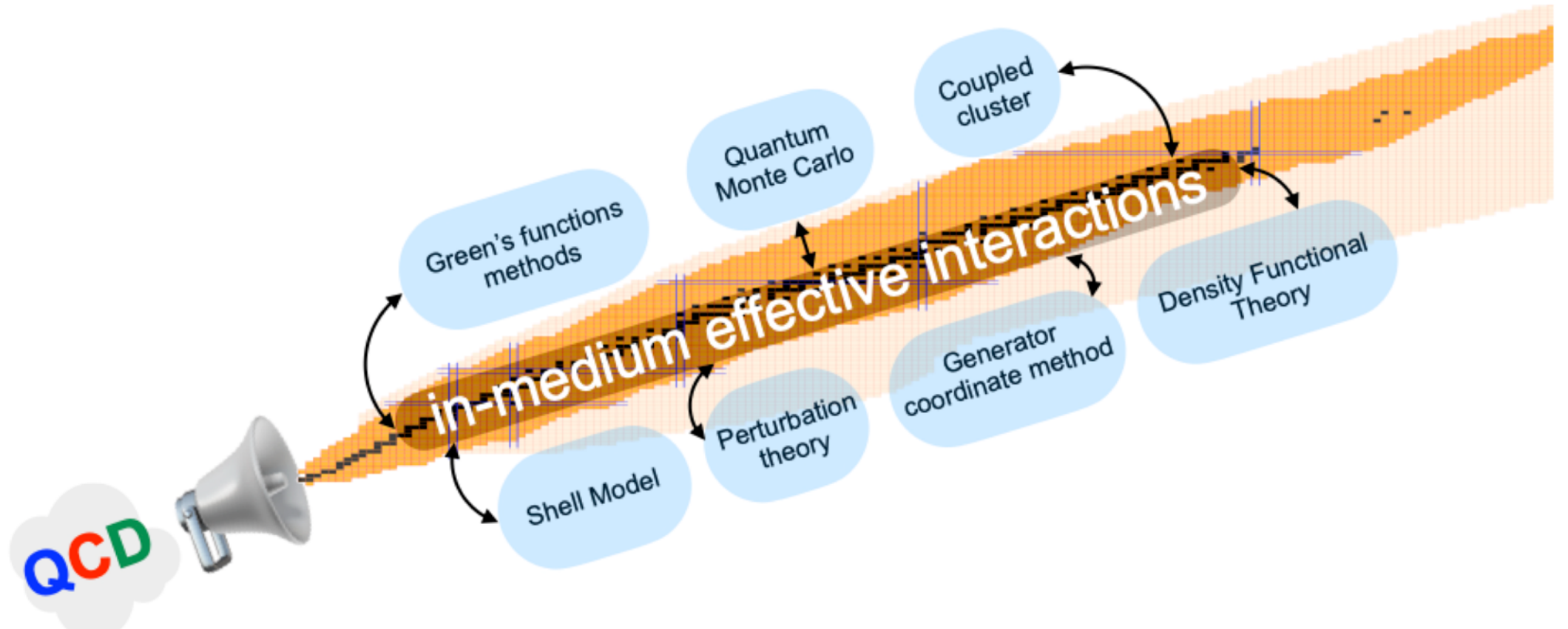

*Fig. 4.2: Nuclear effective theories can be derived from, or informed by, quantum chromodynamics (QCD) using techniques that produce effective nuclear interactions in the many-body system (e.g. the similarity renormalisation group) and using controlled methods to solve the quantum many-body problem. Nuclear effective theories should aim, in the near future, at providing systematically improvable methods and reliable error estimates. At present, still much empirical/phenomenological input is used.*

Given the present limitations of ab initio methods, spectroscopy studies across medium-mass and heavy nuclei are usually performed with the nuclear shell model (NSM) and/or energy density functional (EDF) techniques that well describe the evolution of magic numbers, the coexistence of different deformed shapes, low-energy gamma-decay spectroscopy, beta and double beta decays of nuclei.

For the NSM, most of these phenomena require large configuration spaces; defining effective interactions encompassing more than one major oscillator shell as well as their numerical diagonalisation is still challenging.

EDFs serve as the primary or only option for extensive microscopic calculations that cover the entire chart of nuclides and/or include states with increasing excitation energy. Their predictive power diminishes as we venture further from the regions where input data is obtained for parameter fitting, such as approaching the drip lines. Techniques like Bayesian inference and machine learning can play a pivotal role in understanding the sensitivity of predictions to the training data used for EDFs. Developing more general functionals or even meta-models may become necessary to accommodate all observed data if tensions appear, that is if current models are not flexible enough (see Chapter Nuclear Physics Tools).

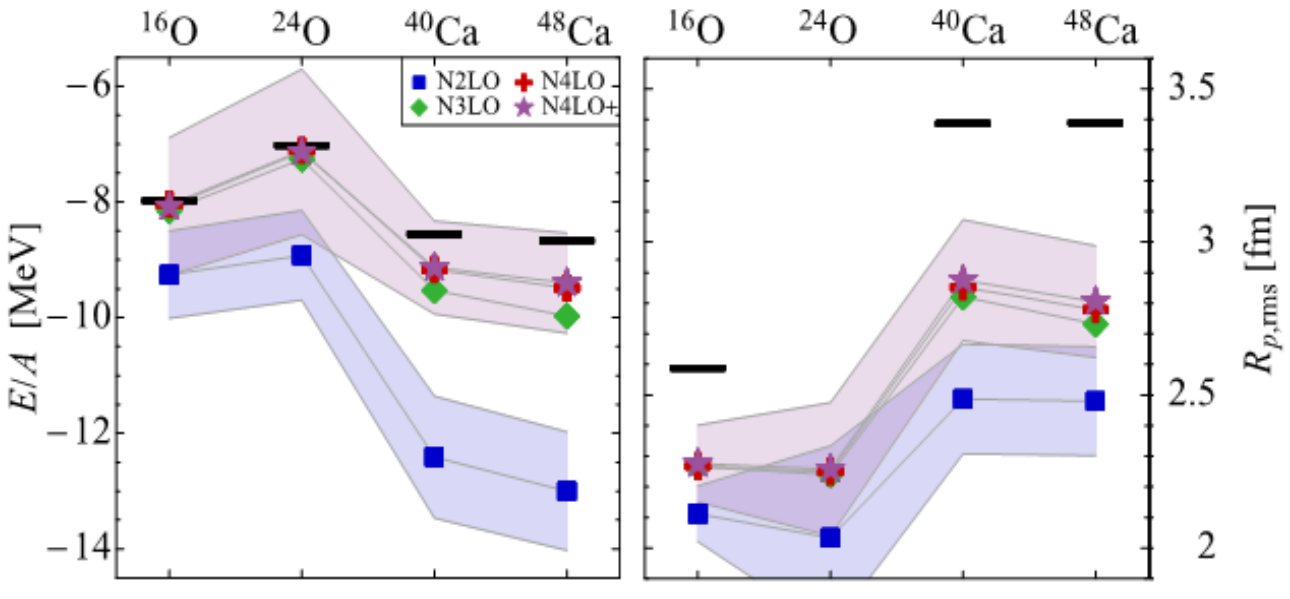

*Fig. 4.3: Ground-state energies and proton radii for doubly magic oxygen and calcium isotopes obtained with state-of-the-art ab initio methods based on chiral effective field theory interactions. The error bands show the chiral truncation uncertainties at the 95% confidence level obtained with the pointwise Bayesian model. Black bars indicate the known experimental data. (Adapted from Physical Review C 106, 064002 (2022)).*

Within the nuclear EDF framework, it is essential for the quantitative description of the low-lying spectra in atomic nuclei to include correlations originating from the restoration of broken symmetries and fluctuations of collective coordinates, as well as further correlations like those associated with particle-vibration couplings. These correlations can be considered by using, for example, the projected generator coordinate method (PGCM, also known as multi-reference EDF) and many-body methods on top of EDFs like the quasiparticle-vibration coupling model or QPVC. Modern computer codes incorporate several collective coordinates together with the restoration of the particle number, parity, and angular momentum. However, the computational complexity of the PGCM and related methods increases rapidly with the number of collective coordinates, thus precluding systematic large-scale calculations. In addition, the description of odd nuclei within the EDF framework is still behind that of even nuclei. Hence, computer codes should be extended and optimised and new methods should be devised to reduce the computational burden of EDF and NSM calculations.

Bridging ab initio and phenomenological domains is becoming feasible. Hence, methods like valence-space IMSRG or the PGCM perturbation theory combine many-body techniques traditionally used with phenomenological approaches with interactions and operators derived with ab initio techniques. Furthermore, it is crucial to envision grounding nuclear EDF in ab initio approaches, akin to what is done for density functional theory (DFT) in Coulomb systems. Understanding these models more coherently, perhaps under the effective theory umbrella, should be envisioned.

During the past decade, new developments in the field of time-dependent (TD) methods have been released, e.g. unrestricted 3D calculations with full effective interactions and complete treatment of superfluidity at the mean-field level. Although applications of the TD methods (e.g. collective vibrations, fusion, deep inelastic collisions and fission) have provided important physical information, some shortcomings related to the treatment of quantum aspects are also present. Since atomic nuclei are finite many-body quantum systems, quantum fluctuations in collective space are large and simple mean-field theory underestimates them. Furthermore, quantum tunnelling in collective space is completely absent in the TD-DFT framework. It is necessary to improve the microscopic TD description beyond the independent particle picture to resolve these issues. Hence, because of the computational complexity of TD methods, most applications rely on stationary state methods.

# Nuclear Shells and Shapes

## Shell structure and evolution

One of the cornerstones that underpin our understanding of nuclei is based on the shell structure picture, where nucleons lie on single-particle orbitals provided by a mean field generated by the interacting nucleons themselves. In the last decades, significant developments in theory have allowed us to make more rigorous predictions on the energy evolution of single-particle orbitals, as a function of neutron or proton number. In parallel, extended experimental investigations have triggered the discovery of significant changes in the nuclear shell structure, when moving away from the stability valley. Ground state properties, like masses, radii, and electric and magnetic moments, give access to shell gaps and deformations. High precision mass in the ground and excited metastable states (isomers) can be obtained with Penning traps and multi-reflection time-of-flight (MR-TOF) techniques. Electron scattering on radioactive nuclei could provide access to nuclear radii and form factors in exotic systems. Precise values of spin and parities of excited levels, as well as magnetic moments, can be measured with beta-decay of laser-polarised nuclei which can reach part-per-million accuracy using the novel b-NMR technique.

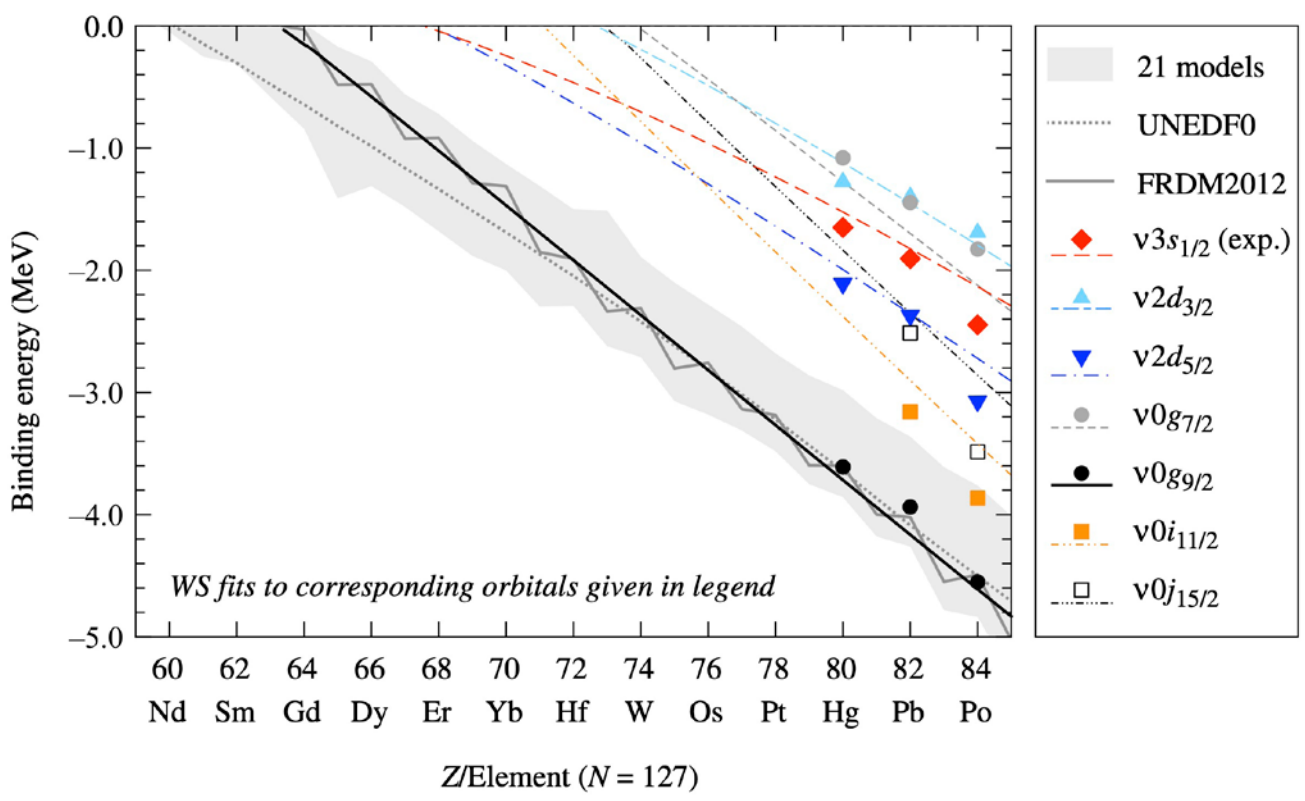

*Fig. 4.4: Evolution of the neutron single-particle orbitals (in the form of binding energies) for N=127 nuclei. (Taken from Physical Review Letters 124, 062502 (2020))*

The spectroscopic strength of nucleon removal reactions has become a key tool for probing the robustness of sub-shell closures. In addition, high-precision particle and gamma spectroscopy, often taking advantage of the extra sensitivity given by isomeric states, has started to pin down in great detail the properties of nuclear wave functions in exotic regions, thanks to the availability of sophisticated devices like AGATA, HISPEC/DESPEC DEGAS, HISPEC/DESPEC AIDA, the future GRIT, etc., and exploiting a variety of reactions based on stable ions, intense neutron beams and radioactive high-intensity beams at existing and new/upgraded facilities (e.g., HIE-ISOLDE, GANIL, ILL, LNL-SPES, ISOL@MYRRHA and GSI/FAIR).

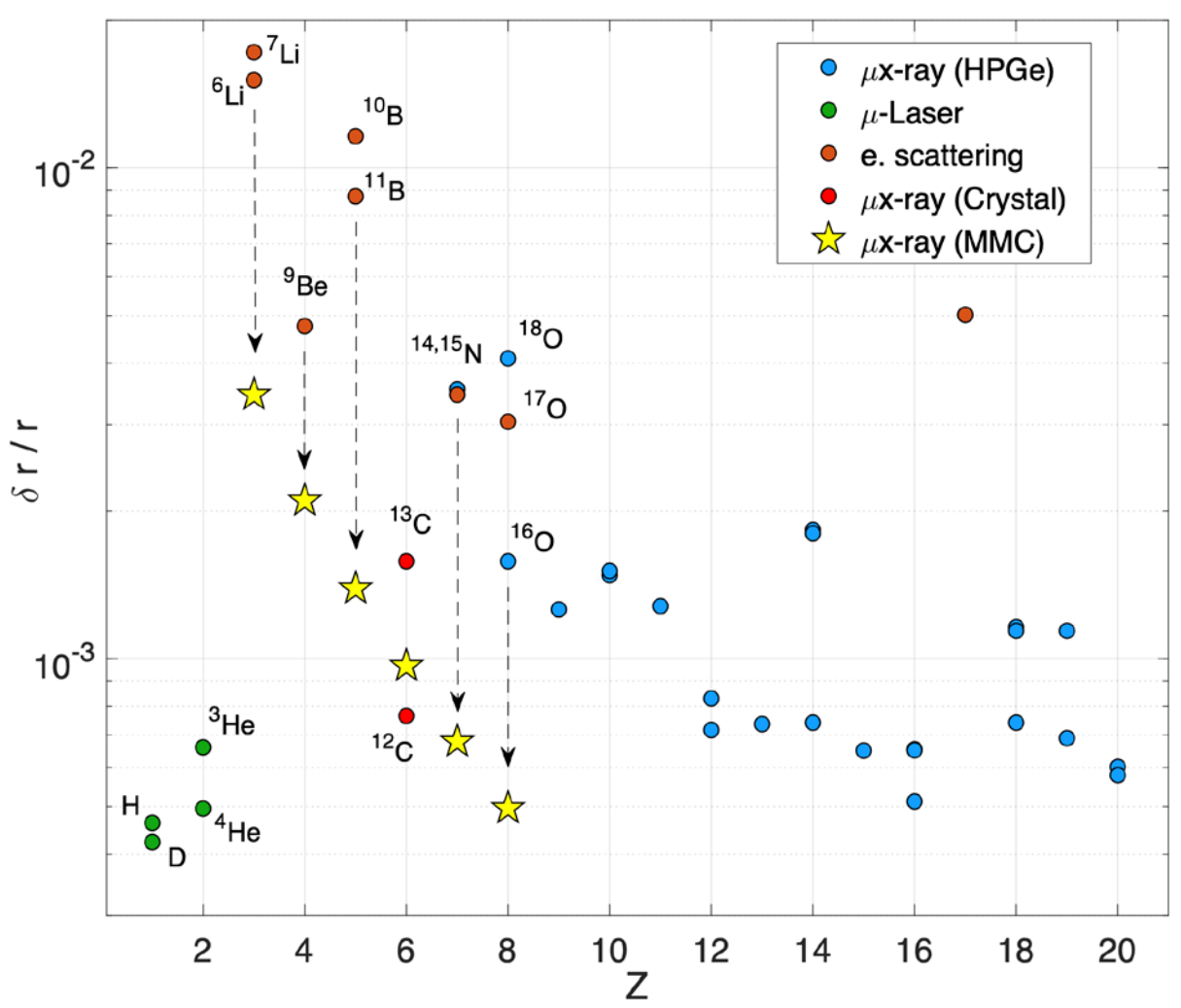

*Fig. 4.5: Relative uncertainty in the most precise RMS charge radii of the light stable nuclei. The stars denote the achievable precision when $E_{2P-1S}$ transitions in muonic atoms are measured with 10 ppm accuracy (see also Box 6.2).*

In the region of light-, medium-mass nuclei, the observed disappearance in exotic nuclei of classical shell closures at N=8, 20 and 28, and the appearance of new magic numbers at N=16 in O and N=32, 34 in neutron-rich Ca isotopes, have been attributed to changes in the level spacing and order caused by the complex interference between the various components entering two- and three-nucleon interactions (e.g. central, spin-orbit and tensor). Although the spin-orbit component is recognised to play a crucial role in the creation of most of the magic numbers (via the energy splitting between spin-orbit partner states), significant deviations from the expected global trend of a proton or neutron spin-orbit splitting, as a function of the mass number, have been observed and attributed to the density dependence of the spin-orbit force (e.g. between $^{36}$S and $^{34}$Si as well as in $^{132}$Sn) and the effect of the tensor force (between $^{36}$S and $^{40}$Ca), so far poorly constrained and often omitted in models. Other cases must be studied to quantify the effect of the tensor force, exploiting

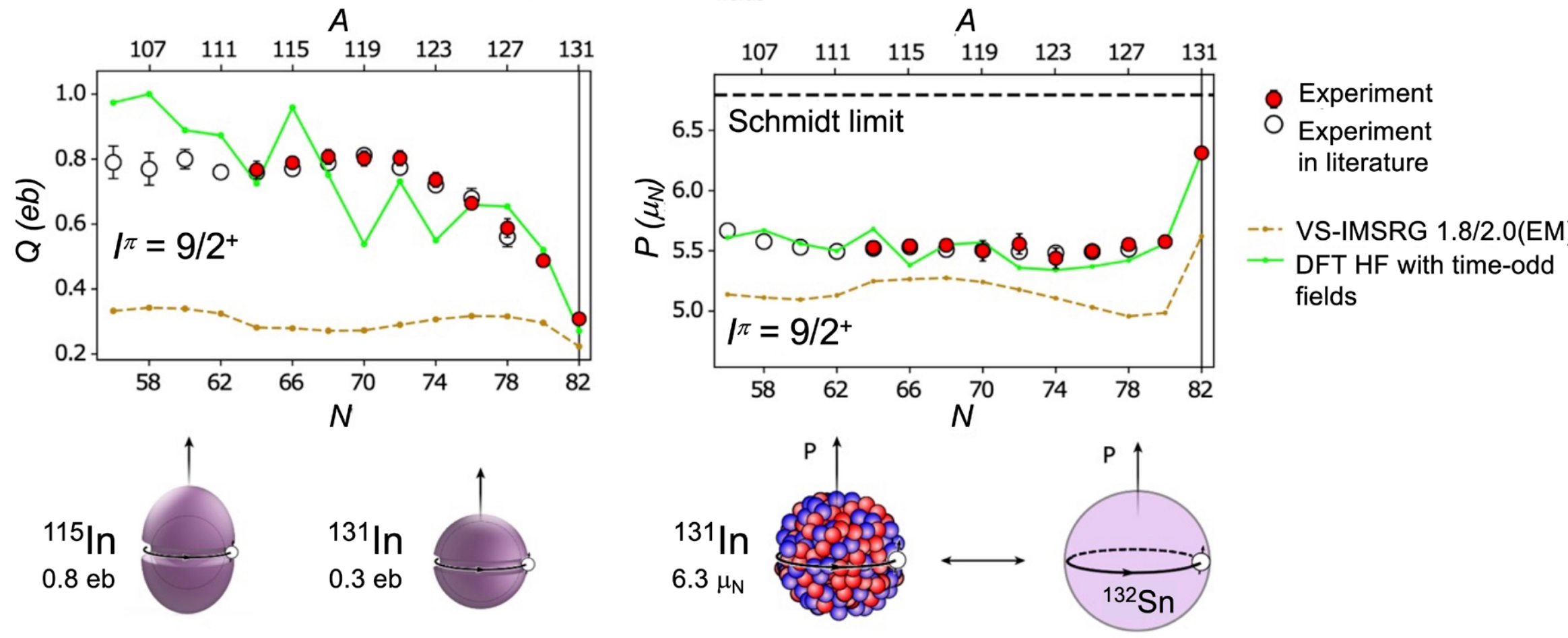

*Fig. 4.6: Top: evolution of nuclear electromagnetic properties for the 9/2+ ground states of In isotopes (A=105-131), from laser spectroscopy studies (electric quadrupole moment (left), magnetic dipole moments (right)). Bottom: Cartoon representations of the evolution of collective properties of these isotopes. Left: the quadrupole polarisation gradually reduces to a single-proton-hole value at N = 82; Right: the magnetic dipole moments abruptly approach the value for a single proton hole in a $^{132}$Sn core at N = 82, as the dominant effect changes from charge to spin distribution. (Adapted from Nature 607, 260(2022)).*

## Box 4.1: Beta-decay of the doubly-magic nucleus $^{100}$Sn

**In our present understanding of nature, the weak interaction is one of the four known fundamental interactions. This interaction governs the beta-decay process. For that reason, beta decay can be used as a tool for fundamental studies of the weak interaction in the nuclear medium, but not only. The beta-decay process also constitutes a relevant tool for studies related to nuclear structure, nuclear astrophysics, and many practical applications. In this realm the beta decay of $^{100}$Sn is a very special one. $^{100}$Sn is the last particle bound N = Z double magic nucleus, and its beta decay shows the largest Gamow-Teller strength detected so far in the nuclide chart. The special character of this decay is related to the shell structure around at N=Z=50, and the large available Q window. Studying this decay remains very challenging, because even though remarkable developments have improved dramatically the primary beam intensities in the last decade, the low production cross section of this exotic nucleus remains a limiting factor. Future studies should improve the knowledge of the populated states in the beta decay and clarify if there is fragmentation of the Gamow-Teller strength employing setups of higher efficiency. State-of-the-art theoretical studies show that this decay also holds the key to a better understanding of the quenching of the $g_A$ constant in the nuclear medium that can impact, e.g., astrophysical calculations and neutrinoless double beta decay. This study remains a flag experiment of existing and upcoming radioactive beam facilities worldwide.**

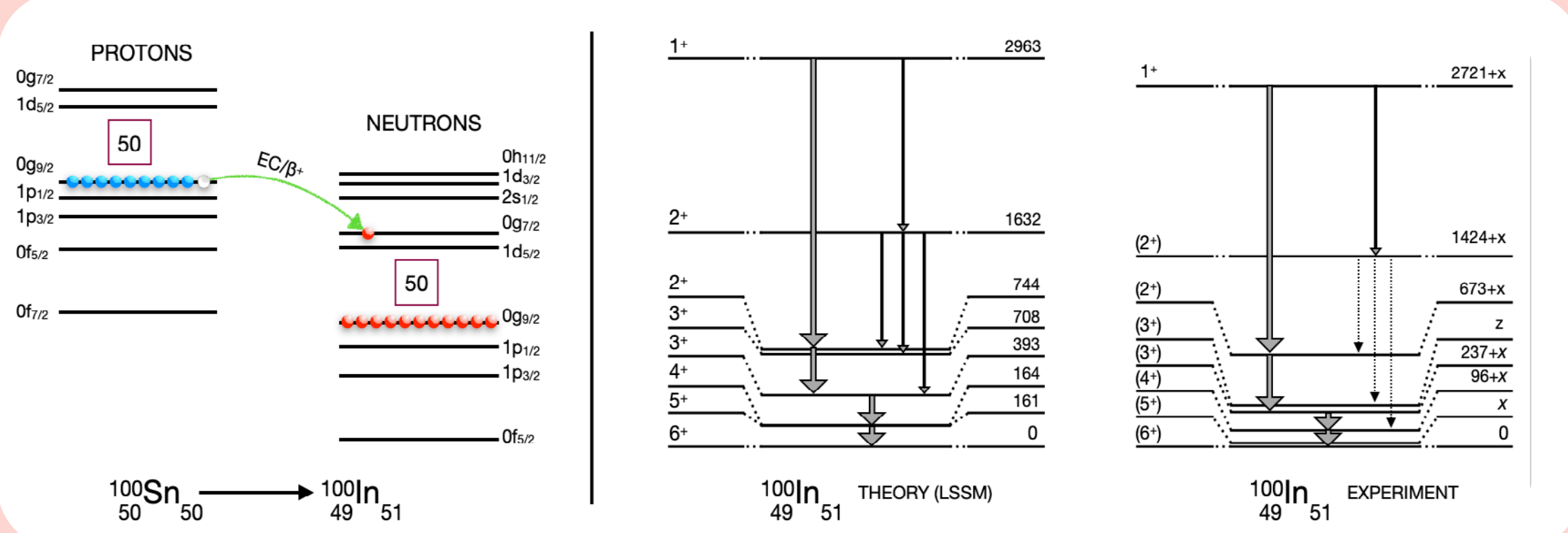

***Left panel:*** *Shell structure relevant in the beta decay of $^{100}$Sn.* ***Right panel:*** *Populated levels in $^{100}$In from the beta decay of $^{100}$Sn. Due to the limited statistics, the level scheme is based on comparisons with theory (Large Scale Shell Model calculations). Fixing the $1^+$ state energy is an important step to obtain a more precise experimental value of the Gamow-Teller strength of this unique decay.*

transfer reactions at HIE-ISOLDE with the superconducting solenoid, or at SPES-LNL and GANIL/LISE using segmented charged-particle detectors (e.g., GRIT) in combination with very-efficient gamma-ray setups (e.g. AGATA or EXOGAM).

On the proton-rich side of the valley of stability, shell evolution is more ambiguous as the regions of extreme isospin lie beyond the drip line. Nevertheless, studies of $^{12}$O, the mirror of $^{12}$Be at N=8, show that the Z=8 shell closure disappears similarly to the neutron equivalent. Attacking the evolution of the shell at Z=20 is a task for upcoming facilities. Two and/or four neutron removal from $^{36}$Ca together with high-resolution invariant mass spectroscopy of the decay products of $^{34-32}$Ca at the future FAIR facility could provide answers to the possible symmetry in shell evolution.

In heavier nuclei, the quest for new magic numbers is still open. Earlier predictions for sub-shell closures at the harmonic oscillator numbers 40 and 70 have not been confirmed - very large ground state deformation (in $^{80}$Zr) or triaxial deformation (in $^{110}$Zr) are now expected. Theoretical predictions also diverge around $^{78}$Ni, some suggesting a reduction of the N=50 gap and others a firm shell closure, unaffected by the neutron-to-proton asymmetry. Experimental investigations are currently at the beginning in this region which is also the starting point of the astrophysical r-process. The latter can be seriously affected by shell changes, in terms of neutron binding and beta decay properties. On the proton-rich side, $^{100}$Sn is the last nucleon-bound nucleus with equal proton and neutron magic numbers, so far experimentally investigated by its beta decay only. Data on excited states, collective excitation and single-particle occupancies are on the horizon by exploiting direct reactions, while neighbouring nuclear systems, e.g. $^{102}$Sn and $^{98}$Cd, are already being investigated. Nuclear structure information at the proton drip line, especially in the mass region along the drip line up to the A=100 range, is also much needed to address the multitude of open questions on the stellar nucleosynthesis path in the rp-process towards its actual endpoint. Among others, capture reactions on isomeric states are very relevant, since they may cause a significant change in the reaction flow. Such nuclear structure aspects can also be investigated via in-beam gamma-ray spectroscopy, using devices such as JUROGAM3 or AGATA coupled to MARA at JYFL.

The heaviest doubly magic nuclei, $^{132}$Sn and $^{208}$Pb and their immediate neighbours are experimentally rather well studied - the first is easily produced as a fission fragment, and the second is stable. Surprisingly, in both cases, experimental knowledge finishes a few protons below, with their southeast quadrants in the Segrè chart hardly explored, due to the challenges in production, purification, and identification of these rare isotopes. New reaction approaches, like multi-nucleon transfer reactions with stable and secondary beams or in-flight fragmentation of secondary beams and two-step reactions, are currently being considered to reach these areas. Only at the HISPEC/DESPEC experiments at the future Low Energy Branch of the FAIR facility, will in-beam reaction studies with such heavy fast beams become possible, providing a world-unique asset to the European nuclear physics community. Both mass regions are linked to the r-process nucleosynthesis of the 2nd and 3rd r-process yield peaks, providing additional motivation for their study.

Beyond Z=82 and N=126, the next shell closures are in the "island of stability" of superheavy nuclei, predicted to occur at Z=114, 120 or 126 and N=172 and 184. Experimental findings, mainly achieved by decay spectroscopy after separation (DSAS) for lighter superheavy nuclei (SHN), established the presence of deformed shell gaps at Z=100 and N=152, giving indications for a second region of this kind at Z=108 and N=162. Such results, together with advanced theory efforts, suggest extended isotopic and isotonic chains of strongly bound nuclides. New facilities presently developed promise an extended investigation towards high Z and N numbers and eventually the synthesis of new elements and isotopes (see section *Towards the limit of nuclear existence*).

## Box 4.2: Shape Coexistence in Pb isotopes

**A complete picture of atomic nuclei can only emerge from complementary experimental techniques and systematic studies which provide stringent benchmarks for nuclear theory.**

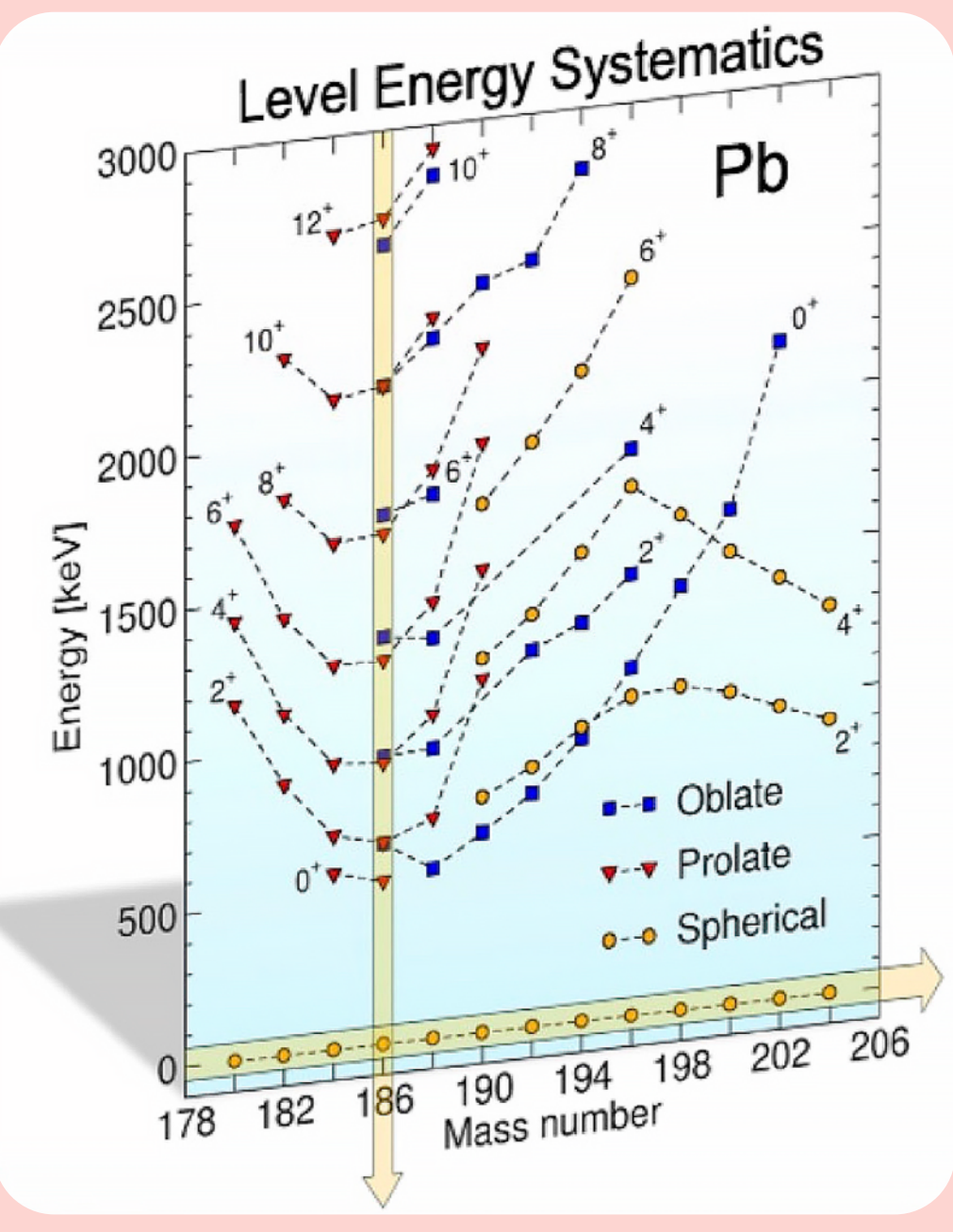

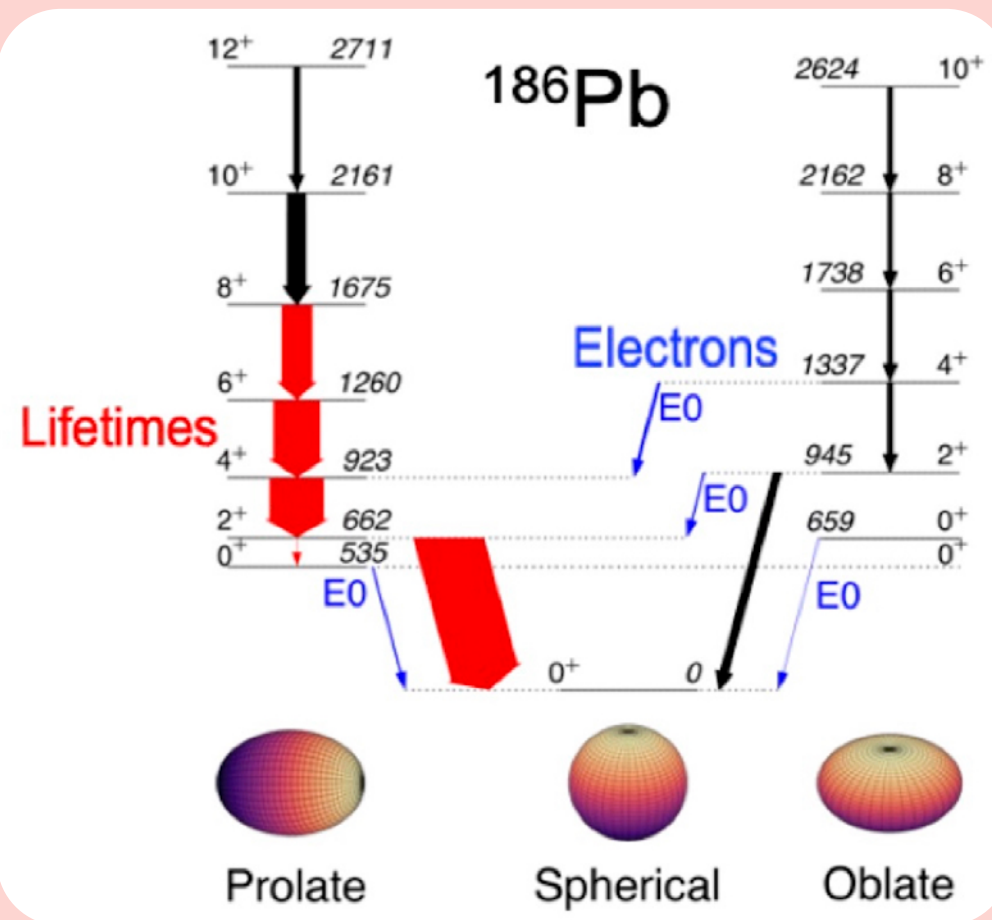

### Pb isotopes are archetypal examples:

**Isotope shift measurements** from the neutron-deficient region up to the N=126 neutron shell gap have probed mean squared charge radii of ground states and isomeric states, confirming spherical ground states in Pb.

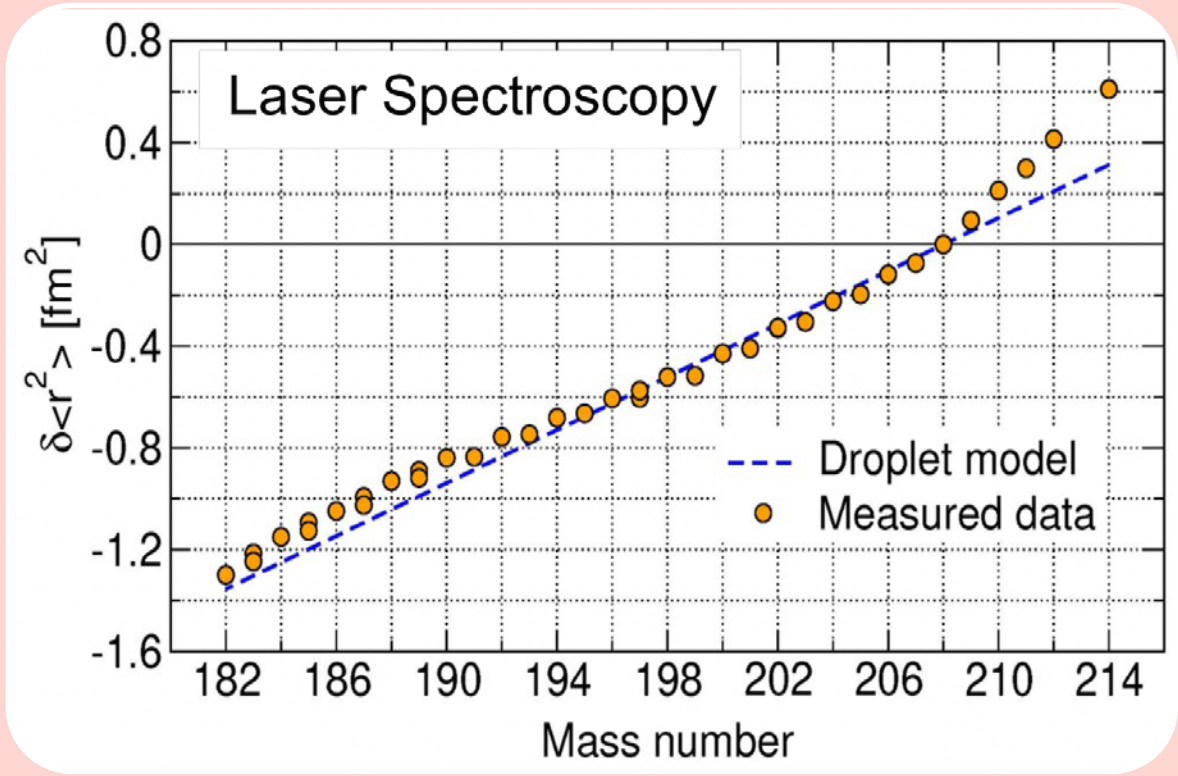

**Alpha-decay fine-structure measurements** have revealed a triple shape-coexistence scenario in the neutron mid- shell nucleus 186Pb, with the three lowest states, with spin zero, residing in local minima of the nuclear potential characterized by different shapes (spherical, prolate and oblate).

**In-beam γ-ray spectroscopy** have probed the existence of rotational bands built on the deformed shapes. Together with in-beam conversion-electron spectroscopy and combined with lifetime measurements, the shapes of the deformed band-head states have been firmly assigned.

Future complementary measurements are still needed to obtain a more detailed picture:
**Coulomb excitation** experiments to probe quadrupole deformation and degree of shape mixing, **transfer reaction** to examine underlying intrinsic configurations, more sensitive **electron spectroscopy** to assess charge radii of the excited, short-lived, 0+ states. Innovative detection techniques and developments of radioactive beams are mandatory to extend systematic studies towards the most exotic regions of the nuclear chart.

## Nuclear shape evolution and shape coexistence

Many nuclear structure aspects (e.g. rotations and vibrations) can be naturally understood if we endow the atomic nuclei with *intrinsic shapes*. Shapes of different varieties emerge because of the nuclear interactions and ultimately depend on the details of the wave function describing a nuclear state. Therefore, they may evolve not only as a function of the number of nucleons along isotopic/isotonic chains but also within states belonging to the same nucleus. Inferring the shape of a nucleus constitutes one of the most active fields of research because of its potential in benchmarking nuclear structure models and for the impact of shapes on decay properties of astrophysical-relevant nuclei. The shape of a nucleus can be probed from direct measurements of the electric quadrupole moment, from electromagnetic transitions and related patterns, from measurements of nuclear radii as well as decay studies and beta strength in daughter nuclei.

Self-conjugated N=Z systems are of special interest due to the enhanced proton-neutron pairing caused by the occupation of the same orbitals by protons and neutrons. In the region of Kr isotopes, with (N, Z) ≈ 34, 36, shape transitions and coexistence of different shapes (oblate and prolate) are expected, and their experimental study will heavily rely on Coulomb excitations mainly at intermediate-energy, with lifetime measurements from fusion-evaporation reactions also providing valuable information.

On the neutron-rich side, the Zr (Z ≈ 40) region, around A ≈ 100, has been found to exhibit a sudden onset of deformation, with features typical of a quantum phase transition, and the appearance of multiple shapes, including triaxial ones (namely with unequal axes). Available data for isotopes ranging from the closed shell nucleus $^{90}$Zr to $^{110}$Zr show a major change in spectral characteristics beyond neutron number N=56. This calls for extended high-precision spectroscopy over the entire region below Zr, which will become available by employing intense radioactive beams from ISOL facilities and state-of-the-art detection systems. The A$\sim$190 transitional elements, around the N = 126 closed shell, are also of special interest due to the predicted prolate-to-oblate shape transition, with triaxiality playing a larger role at higher Z. In the Os chain, the critical point between prolate and oblate shapes is $^{196}$Os (an almost perfect γ-unstable/triaxial rotor), while for tungsten the observed turning point is $^{190}$W. The shape transitional region has not yet been reached for Hf isotopes with Z=72 and future studies, using multi-nucleon transfer and fragmentation techniques, will be crucial in reaching more neutron-rich, key nuclei.

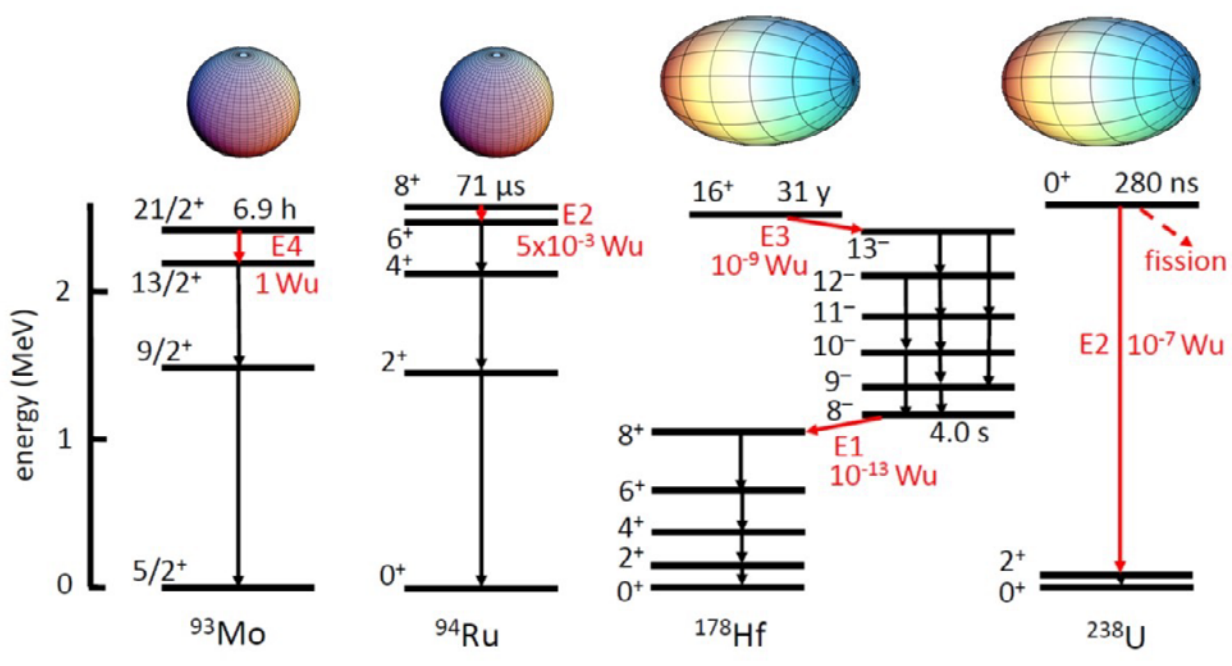

*Fig. 4.7: Isomeric states are long-lived nuclear excitations, with lifetimes ranging from sub-nanoseconds to years. The decay of isomers is inhibited by at least one of three constraints: (i) a large change in the magnitude of the angular momentum (spin isomers); (ii) a large change in the direction of the angular momentum (K isomers); (iii) a significant change in the shape of the nucleus (shape isomers). The level of inhibition gives information on the wave functions of the involved states. Experimentally, isomeric states provide additional sensitivity, as their decay can be studied after isotope and element selection, far from the place where they were synthesised. Information on the most neutron-rich nuclei produced at FAIR will be obtained from HISPEC/DESPEC isomer decay experiments at the focal plane of spectrometers and at the storage ring.*

Triaxiality has become a central issue in several other mass regions, its appearance also being related to the action of the monopole tensor part of the nuclear force. The strong impact of triaxiality on the re-interpretation of collective phonon excitations in Sm/Er nuclei calls for high-precision Coulomb excitation measurements with tracking arrays like AGATA, both with stable beams and in HISPEC/DESPEC experiments at FAIR. Triaxiality is also crucial for the calculations of nuclear matrix elements in rare decays, such as 0νββ decay in key nuclei like $^{76}$Ge. Recent measurements at colliders (RHIC and LHC) have also provided an alternative access route to study the intrinsic deformation of all nuclei, both even and odd.

Nuclei can also possess pear shapes, arising from the coupling between pairs of nucleons occupying close-lying orbitals with $\Delta j=\Delta l=3$. This leads to octupole deformation and spontaneous breaking of rotational and reflection symmetry in the intrinsic nuclear frame. Multistep Coulomb excitation can help distinguish stable octupole deformation from octupole vibration, as recently done for Ra and Rn isotopes. Such a distinction is also relevant when searching for permanent electric dipole moments (EDM), which could indicate substantial CP violation, calling for a revision of the standard model which predicts values that are too small to account for the matter-antimatter asymmetry in the Universe (see more in Chapter Symmetries and Fundamental Interactions).
The atomic isotope with the best limit on its EDM is currently $^{199}$Hg, but the presence of octupole deformation in nuclei could enhance EDMs if CP violation occurs within the nucleus. Direct measurements of odd-mass nuclei relevant for atomic EDM measurements will be pursued, including in transuranium elements, which are candidates for the largest asymmetric deformations of all nuclei. Higher-order multipole deformations may also be explored in the future.

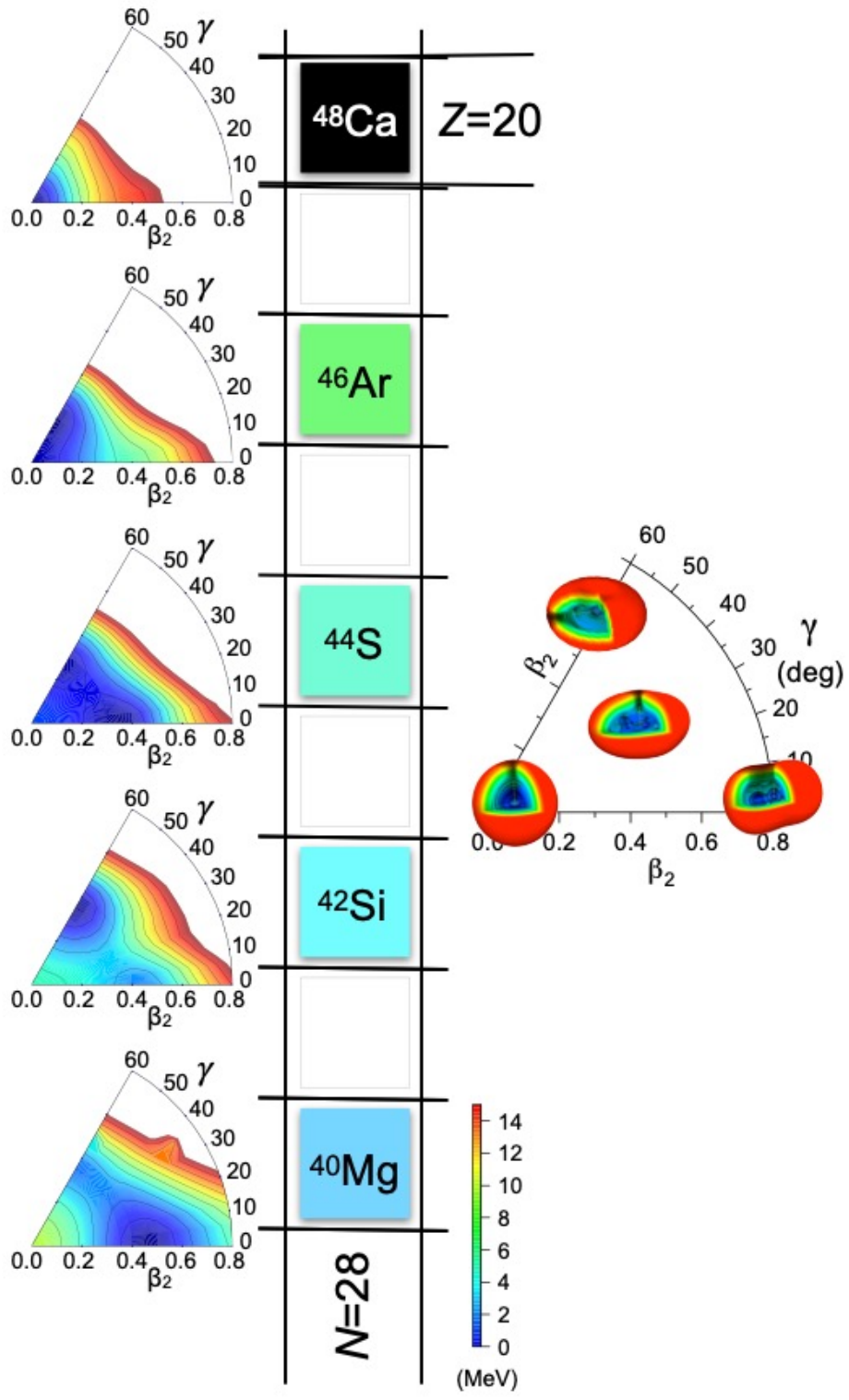

*Fig. 4.8: Breaking of magic numbers and emergence of deformation. In rather light systems, moving along N=28, a shape change from oblate to prolate is predicted to occur from $^{46}$Ar to the hard-to-reach $^{40}$Mg, passing through a scenario of multiple shapes in $^{42}$Si, which will serve as a strong benchmark for microscopic theories.*

Detailed spectroscopic studies, employing a variety of experimental observables and techniques (from laser spectroscopy to in-beam gamma and electron spectroscopy), have now confirmed the appearance of different intrinsic shapes, at the same spin and similar energies, over the entire nuclear chart. Understanding the phenomenon of shape coexistence and shape isomerism has been a long-standing challenge for nuclear structure theory, and microscopic approaches are now becoming computationally available in light and medium-mass nuclei. Besides the well-established regions of shape coexistence near the major shell and subshell gaps of Pb, Sn and Zr, with A~190, 110 and 100, other regions have become relevant, for which intensified experimental and theoretical efforts are needed.

Among light systems, Mg and Si isotopes at A~20-30 are predicted to show multiple shapes, with a strongly deformed excited $0^+$ state and one of the largest monopole strengths to the ground-state $0^+$ already observed in $^{24}$Mg. In $^{28}$Si, the ground state has been determined to be oblate, the first excited $0^+$ state was suggested to be prolate, and a superdeformed band with yet another shape may be present. In the Ni isotopes, triple-shape coexistence has also been proposed in $^{64,66,68}$Ni. Hints of shape coexistence in $^{66}$Fe were recently obtained, while the evolution beyond N=40 remains unknown. Additional candidates for triple-shape coexistence have been suggested in Ca, Zr, and Cd. Near N=50, low-lying deformed states have been reported in $^{79}$Zn and $^{81,82}$Ge, and tentatively in $^{82}$Zn, thus providing the first hints of shape coexistence near the very exotic double shell closure of $^{78}$Ni, where deformed $0^+$ states are expected to appear at low excitation energy. All above mass regions will be prime targets for future research with high-intensity stable and radioactive ion beams, exploiting enhanced sensitivities of the currently developed detection systems.

The appearance of superdeformed rotational bands at high spins is also a special case of shape coexistence. No clear connection with low-spin shape coexistence has been made so far in medium-heavy mass nuclei, because of the sparseness of experimental information on highly deformed structures at low spins. The completion of the AGATA tracking array, specially designed to enhance sensitivity to high-multiplicity gamma cascades, together with fusion reactions with intense ISOL beams, will push towards a unified microscopic description of shape-coexistence phenomena from zero to high spins. Searches for Hyper-deformed structures in expected exotic Cd, Ba and Yb isotopes will also become possible.

# Nuclear Correlations

## Few-nucleon resonances and clustering

Few-body systems serve as a unique laboratory within nuclear physics, establishing connections with other fields such as cold atoms and atomic and molecular physics. An intriguing recent observation is attributed to the existence of a narrow resonance that could be associated with a tetraneutron in the vicinity of the reaction threshold, which opens the path for the study of multi-neutron interacting systems as pairs, quartets or even sextets. Almost all theoretical calculations that correctly describe well-studied few-body systems predict that this does not correspond to a 4n system. Consequently, there is a pressing need to provide a consistent description of data.

The narrow $\alpha$-cluster resonances present in the $^{12}$C and $^{16}$O nuclei are essential to produce the fine-tuned $^{16}$O/$^{12}$C ratio that allows the development of life on Earth. Based on experimental observations, Ikeda speculated on the appearance of cluster states close to the corresponding emission thresholds in nuclei. This phenomenon is believed to be a common feature of open quantum systems and its existence has also been suspected recently for one-nucleon states and two-neutron or two-proton clusters. The archetype for two-neutron clustering is the $^{11}$Li halo nucleus, the ground state of which is about 300 keV below the two-neutron emission threshold. Several near-threshold states have recently been discovered and their tiny gamma-decay branches ($\sim 10^{-5}$ – $10^{-3}$), probed by highly efficient arrays like AGATA, will help to understand the mechanism leading to the occurrence of such narrow resonances, in particular the impact of the continuum on the shell structure. It would also be extremely interesting to see if this conjecture applies to a system of four neutrons above a given core nucleus. The systematic appearance of narrow two-neutron resonances close to the two-neutron emission threshold could also play an important role in nuclear astrophysics, as it could speed up neutron-capture cross sections in the rapid-capture nucleosynthesis. The future R$^3$B beam line at GSI-FAIR coupled with the high-granularity NeuLAND neutron array will provide ample opportunities to study such phenomena in light neutron-rich nuclei.

Finally, tritium and $^3$He clustering are expected to take over the $^4$He clustering in neutron-rich or neutron-deficient nuclei respectively. The search for these fermionic cluster systems and their ability to form quasi-molecular states when further nucleons are added is of major interest and will challenge nuclear physics for the next decade. Such studies can be carried out using transfer reactions with RIB in which light particles are detected by GRIT and/or ACTAR-TPC detectors.

## Short-range correlations and quenching of single-particle states

Atomic nuclei are subject to many correlations of pairing, quadrupole and tensor origin. Nuclear correlations are responsible for remarkable features such as superfluidity and deformation, as well as the development of soft or giant collective modes. These correlations, which are of short (SRC), medium or long-range (LRC), also lead to a reduced occupancy of single-particle states, as compared to their expected occupancy by (2j+1) nucleons. A reduction factor of about 40% has been determined, for instance in low-energy transfer reactions. An additional quenching factor is expected in nuclei with large neutron-to-proton imbalance, in which the nucleons in the minority are expected to be more involved in SRC, for example for the removal of a neutron from a neutron-deficient nucleus. Thus far, transfer reactions and quasi-free scattering (QFS) reactions have shown a possible modest additional quenching at both sides of the stability valley. However, the actual precision of measurement does not permit a probe of this additional effect of SRC on the quenching of spectroscopic factors. More precise investigations can be carried out at the GANIL/LISE and CERN/HIE-ISOLDE facility for transfer reactions, and at the FAIR/R3B facility using QFS reactions.

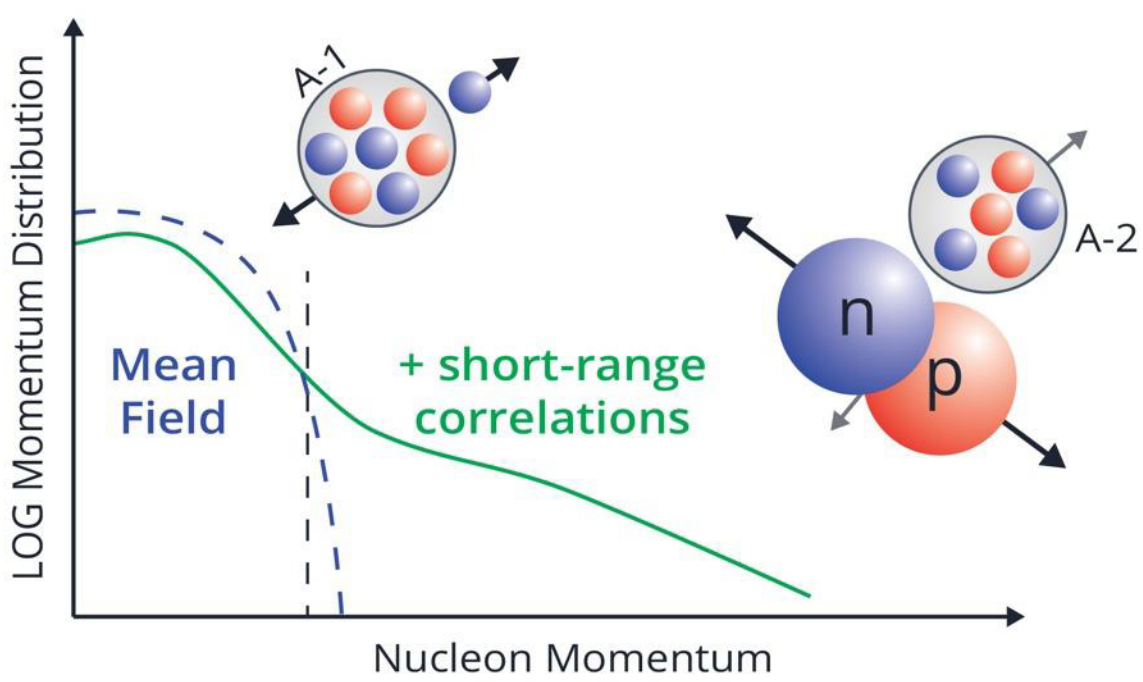

*Fig. 4.10: The strong interaction between proton and neutrons leads to the formation of close-proximity pairs in nuclei, characterised by high relative momentum and low centre-of-mass momentum, significantly deviating from the momentum distribution in a mean-field picture. The existence of SRC has been proven in (e,e') experiments, but nucleon-nucleon collisions at relativistic energies offer the chance to study the isospin dependence of this phenomenon with higher cross sections. This will allow for a better understanding of the influence of SCRs on the evolution of nuclear structure, but also on properties of nuclear matter, the nucleon-nucleon interaction and possibly the connection between nuclear physics and QCD. The FAIR/R$^3$B and HADES facilities can perform complementary complete-kinematics experiments at sufficiently high beam energies. Typical experiments will use (p,3p) or (p,pd) reactions and a liquid-hydrogen target.*

*Fig. 4.9: The central part shows the existence of $\alpha$-cluster states (red colour) in the $^{12}$C and $^{16}$O nuclei, very close to the 3$\alpha$ (above) and $\alpha$ (below) emission thresholds respectively. In the left panel, a narrow resonant state is shown in $^{15}$F about 50 keV above the two-proton emission threshold, while narrow states are also found close to the two-neutron separation energy in the bound $^{11}$Li and in the unbound $^{26}$O nuclei. Dash lines indicate the particle emission thresholds.*

## Pairing and Isospin symmetry

While proton-proton and neutron-neutron pairing (both with isospin T=1) are by far the dominant contributions to the nuclear superfluidity throughout the nuclear chart, the region near the N=Z line may be the scene of emerging neutron-proton (np) pairing, particularly in the T=0 channel. Pair-transfer reactions such as (p,$^3$He) and ($^3$He,p) have been used to study np pairing in the stable sd-shell nuclei and were recently extended to the f-shell nuclei, up to the N=28 shell closure. The results point towards a superfluid phase in the isovector np T=1 channel, whereas the isoscalar np T=0 pairing remains elusive: either the isoscalar strength is mainly concentrated in the aligned configuration with maximum J value (so far rarely studied), or, more likely, the large spin-orbit splitting of high-L orbits leads to a fragmentation of the T=0 strength to higher energies (hard to be reached experimentally). Two-nucleon transfer experiments on heavier N=Z nuclei in the p-shell and approaching the g-shell are the next logical steps for studying further the isovector vs. isoscalar pairing competition. Theoretical works also point to the prevalence of alpha clustering over T=0 pairing, a feature that should be searched for in the ground state components of N=Z nuclei as well.

In atomic nuclei, isospin symmetry originates from the charge independence of the strong interaction that considers protons and neutrons as representations of the same particle, the nucleon. Even after removing the Coulomb contribution (the main mechanism responsible for isospin symmetry-breaking effects), mirror nuclei with the same mass A but N and Z interchanged, present small mirror energy differences (MED) in energy levels (~10-100 keV). MED can be used to probe the nuclear wave function in unique ways, from the effect of the proximity of the continuum on weakly bound orbitals to nuclear shapes, while the MED evolution along rotational bands yields insight into spatial correlations and spin alignments. The persistence of mirror symmetry between the unbound $^{12}$O and its bound partner $^{12}$Be reveals that the loss of N=8, Z=8 magicity appears at both edges of the Segrè chart, while the large MED (~500 keV) between the $0^+_2$ states in $^{36}$S and $^{36}$Ca points to coexistence phenomena. A change of shape was also proposed, for the first time, between the mirror $^{70}$Kr and $^{70}$Se. Future investigations will also encompass mirror reactions, which are of crucial importance to nuclear astrophysics.

# Towards the limit of nuclear existence

## Driplines

The proton and neutron drip lines, where nuclear binding is vanishing marking the limits of the nuclear landscape, can be met by adding and removing neutrons from the nuclear system. Phenomena which are found in the vicinity of the limit of stability for extreme N/Z ratios, such as exotic clustering, neutron skin, halos, change of magic numbers etc., can at the same time be present at excitation energies near the particle threshold.

In terms of progress in mapping the borders of the chart of nuclei, while the location of the proton drip line has been experimentally confirmed up to neptunium (Z=93), the neutron drip line has been reached only up to Ne (Z=10), and for heavier elements its identification relies on theoretical predictions. Due to the slowly varying binding energy on the neutron-rich side, the neutron drip line is expected to extend far out into unknown territory and to extreme neutron-to-proton asymmetries, with novel exotic features to be expected.

At the proton dripline, proton radioactivity occurs when the nucleus is energetically unstable to proton emission. Due to the Coulomb and centrifugal barriers, the proton emission is not instantaneous, and its lifetime provides information on the structure of the nucleus on the single-particle orbital from which the proton is emitted. In the case of proton emission from excited states, internal gamma radiation can

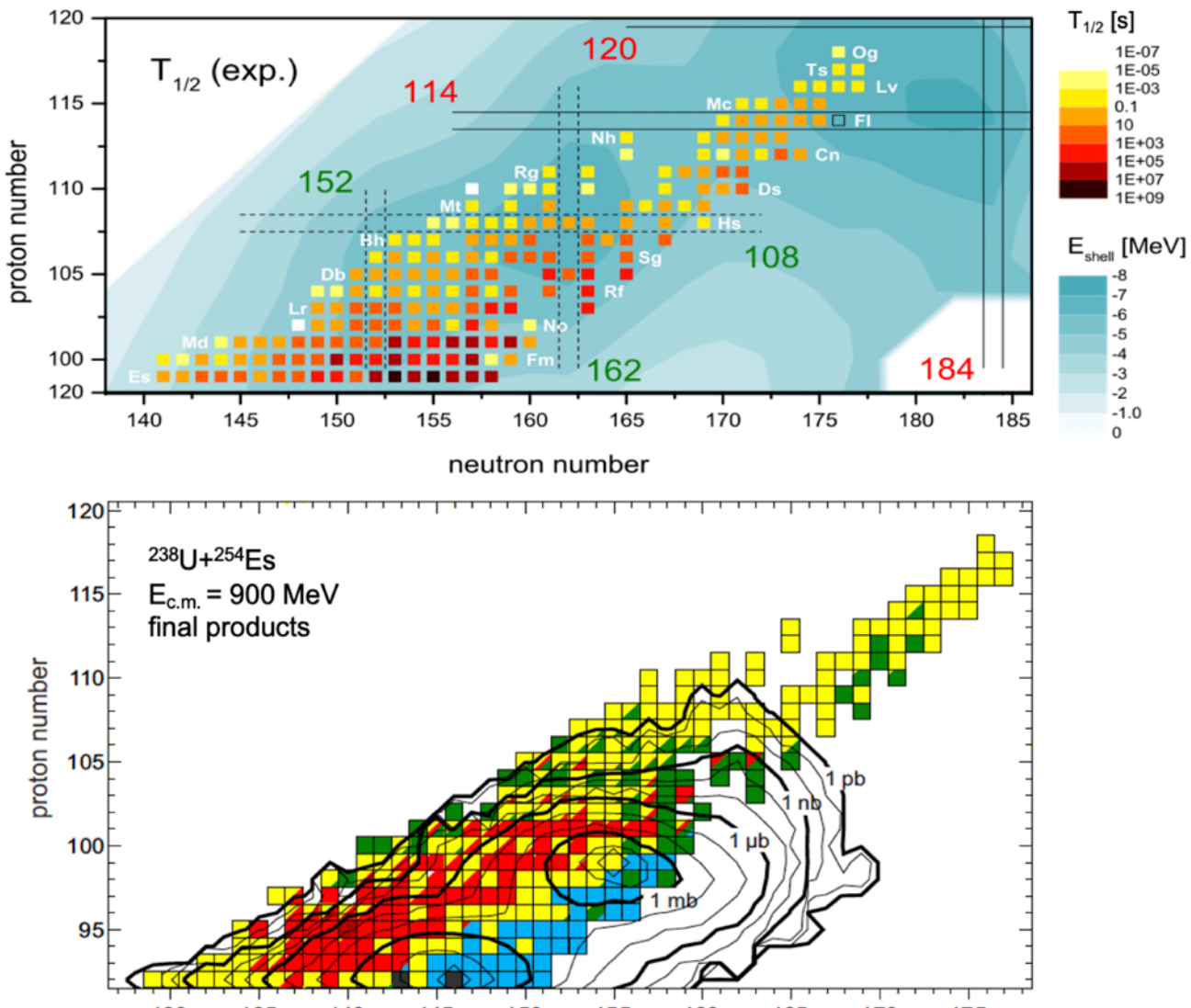

*Fig. 4.11: Top panel: part of the Segrè chart of nuclides displaying experimental half-lives for the up-to-date known nuclides, ranging from einsteinium (Z=99) to oganesson (Z=118) isotopes. The deformed subshell closures and the predicted spherical shell closure are indicated (calculations from Prog. Part. Nucl. Phys. 292, 58 (2007)).*

*Bottom panel: Production cross sections of final products (after evaporation) in the multinucleon transfer $^{238}$U+$^{254}$Es reaction at $E_{c.m.}$=900 MeV, based on Langevin calculations. The contour lines are drawn over an order of magnitude of the cross section down to 1 pb. (Adapted from Physical Review C 99, 014613 (2019)).*

compete, so proton radioactivity can also be used to study nuclei beyond the drip line. Fundamental aspects of quantum tunnelling as well as the coupling of (quasi) bound quantum states with the continuum are involved in the description of proton emitter systems, of which about 20 cases have been discovered thus far. The heaviest known proton emitter is $^{185}$Bi, and the strong interest of the community is in higher mass proton emitters as well as more neutron-deficient isotopes. A special case is also two-proton radioactivity, so far observed only in a few cases but predicted to be widespread in light and mid-heavy nuclei beyond the proton dripline. On the neutron-rich side, neutron radioactivity (i.e. emission of neutrons from isomeric states, never observed so far) could also be searched for. Fragmentation and fusion-evaporation reactions are best suited for the synthesis of nuclei on the proton dripline, and sophisticated separation systems are required, combined with advanced set-ups (digital data processing, TPC, charged and gamma-ray detection systems), such as those currently available at JYFL (MARA, RITU, JUROGAM3) and those that will become available at FAIR and SPIRAL2. The use of AGATA at RITU, in particular, will result in an extremely powerful setup for gamma spectroscopy of proton dripline and N=Z nuclei.

## Superheavy nuclei

The region of the heaviest atomic nuclei, living on the high-Z edge of the Segrè chart only thanks to quantum mechanics, extends into the realm of the unknown. Our understanding of superheavy nuclei (SHN) is really limited due to the challenges for their experimental investigation, mainly because of extremely low production rates and often short half-lives. In the next ten years, typical one-atom-per-week synthesis rates for the heaviest species are expected to be boosted by about one order of magnitude with new accelerators and upgraded facilities, compared to current state-of-the-art machines. This is the case for the SPIRAL2 LINAC, recently online and planned to be fully operational by 2030 with the new injector NEWGAIN, and for the HELIAC accelerator at GSI/FAIR, with similar design goals. These next-generation accelerators, existing and new recoil separators, advanced detection instrumentation and target systems will enable the synthesis of new elements beyond Og (Z=118). This will be supported by advanced theory predictions based on our recent understanding of lighter SHN. At higher Z and N values, Macroscopic-Microscopic and EDF theory approaches significantly diverge, resulting in considerable uncertainty

## Box 4.3: Advanced tools and opportunities for SHE research

**With S³ at GANIL/SPIRAL2 coming online soon and the HELIAC project at GSI/FAIR, the “island of stability” comes closer in reach, including advanced detection technologies for nuclear and atomic spectroscopy, chemistry and synthesis.**

**Decay Spectroscopy After Separation (DSAS) – SIRIUS/SEASON Two installations for particle and photon detection, to be installed in the focal plane of the separator installation S³ of GANIL/SPIRAL2, have the potential to extend our knowledge on the nuclear structure of the heaviest nuclear species to the edge of the region of spherical SHN.**

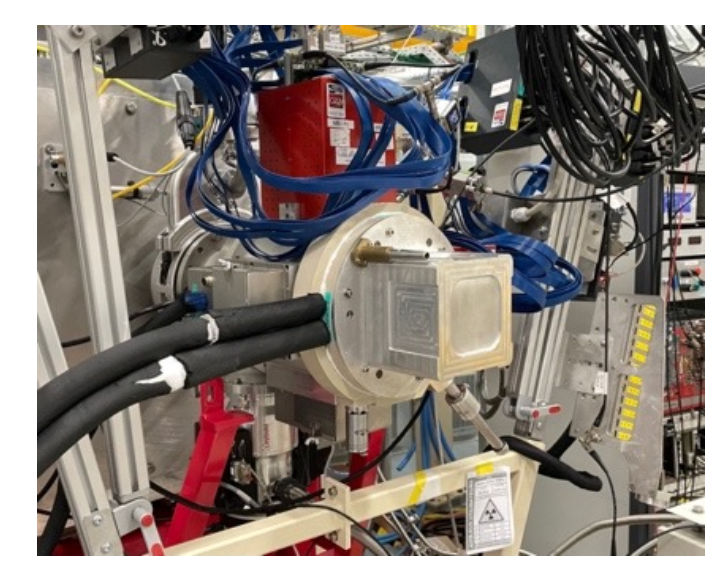

*SIRIUS mounted at S³*

**Laser spectroscopy, mass measurements and DSAS – SHIPTRAP/S³-LEB Ground state properties and atomic structure of the heaviest species are at present accessible in the Fm/No/Lr region at GSI/FAIR at SHIPTRAP. One of the workhorses for the coming decade in this field of research will be the low energy branch of S³, coming online in the next couple of years, followed by the low-E project DESIR.)**

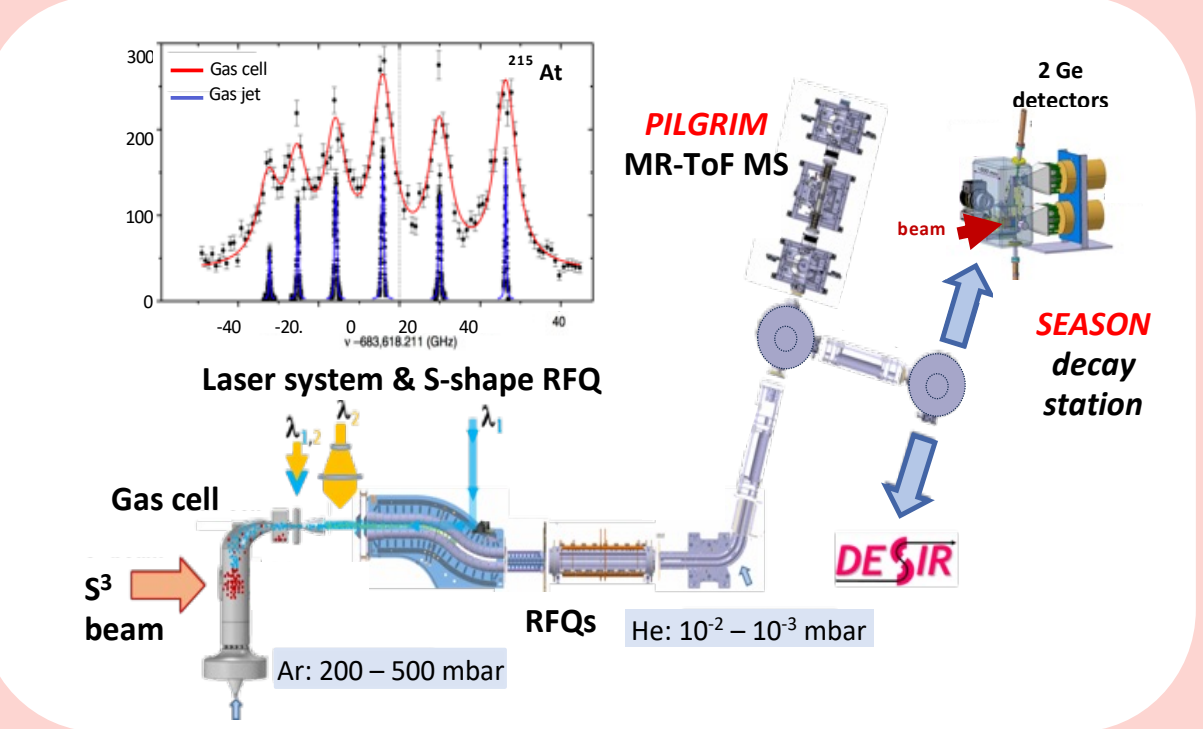

*S³-LEB with gas-stopping cell laser spectroscopy instrumentation, Multi-Reflection ToF Mass Spectrometer (MRToF MS PILGRIM) and the SEASON decay station*

*SHE gas-phase chemistry with COMPACT at TASCA (courtesy A. Yakushev)*

**Chemistry - TASCA Relativistic effects coming important for the heaviest Z and superheavy elements cause chemical properties to deviate from those expected based on simple extrapolations. Chemical studies for the heaviest elements at TASCA of GSI/FAIR thus probe the foundations of the architecture of the periodic table.**

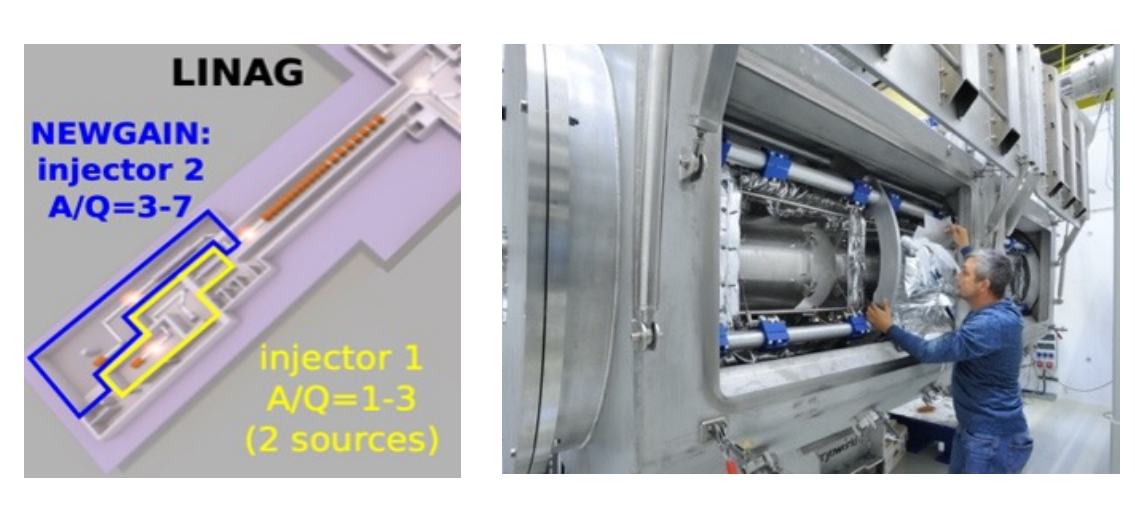

*The NEWGAIN project at GANIL-SPIRAL2 (left), HELIAC advanced demonstrator (CM1) (right)*

**SHE/SHN synthesis – NEWGAIN and HEILIAC The two heavy-ion accelerator projects promise an increase in beam intensities for all projectiles needed for SHE/SHN research by at least an order of magnitude. The SPIRAL2 LINAC, being operational for ions with A/q=3, will be upgraded with the new injector NEWGAIN capable of accelerating ions with A/q=7. The HELIAC project with similar objectives is following a staged approach with the first cryo-module (CM1) built and tested.**

in fundamental properties such as the maximum number of protons and neutrons that can assemble into a bound system and persist for a sufficient duration to be detected (at present ~$10^{-5}$ s); ground state masses; nucleon and Q-alpha separation energies; shell energies and deformations and fission barriers. Evaluation of decay mode competition involving alpha decay, spontaneous fission and electron capture is therefore essential, since they show particular sensitivity to the decaying state’s quantum character and its complexity.

The investigation of SHN’s nuclear structure features (e.g. the development of deformation towards spherical nuclei on the ”island of enhanced stability”, the single-particle structure and formation of spin-, shape- and *K*-isomers, so far mostly restricted to the fermium-nobelium region and the behaviour of odd and odd-odd nucleon numbers in these nuclei and their influence on stability) will thus be extended to heavier and more neutron-rich SHN as well as to the next subshell closures at Z=108 and N=162. Detection capabilities will be enhanced for electron capture decays and longer-lived nuclei. More detailed studies of fission will come into focus to improve our understanding of fission on a microscopic level to improve predictions for nuclei on the island of enhanced stability. Progress in nuclear structure studies in this mass region could be improved with AGATA in a 3π configuration coupled to a gas-filled separator such as RITU at JYFL.

Extensive research also focuses on the intricacies of reaction mechanisms leading to SHN with remarkably low production cross sections. Several aspects are considered, from the study of evaporation residue cross-sections relevant to the creation and decay of these nuclei to the analysis of competing channels, the investigation of hindrances that affect the fusion process and the examination of fission fragment mass distributions, Total Kinetic Energy (TKE) and spin properties.

In recent years, laser spectroscopy and mass spectrometry have been extended to heavy and superheavy nuclei. Laser spectroscopy has meanwhile reached nobelium, allowing us to gain insights into atomic properties and to access nuclear properties such as the spin, the g.s./isomer dipole and quadrupole moment derived from hyperfine spectroscopy, and changes of mean-square charge radii via isotope-shift measurements.

The accurate determination of ionisation potentials via Rydberg states has provided additional information on the electronic structure of heavy elements, achieved thus far for all actinides except Fm, Md, and Lr. This gap will be closed in the coming years. Novel concepts such as laser-resonance chromatography are being developed to extend the reach of laser spectroscopy to Rf. Methodological developments comprise in-gas-jet laser spectroscopy (for increased spectral resolution, crucial for quadrupole moment determination from hyperfine spectroscopy) and the determination of octupole moments with ion-trap-based approaches. Laser ionisation techniques combined with mass spectrometry will allow state-selected nuclear spectroscopy experiments, for example the preparation of long-lived isomers. At present, precise mass measurements are possible for nuclides in the Db isotope region. They will be extended toward heavier nuclei and along longer isotopic chains in the future. Mass spectrometry also gives access to (long-lived) isomeric states as a complementary mode to nuclear spectroscopy. High-resolution mass spectrometry offers options for the detection and unambiguous identification of long-lived SHN, independent of their decay mode.
Chemical properties of superheavy elements (SHE) may also deviate from those expected by simple extrapolations from their lighter homologs due to the influence of relativistic effects, increasing with increasing Z. Chemical studies of SHE thus probe the foundations of the architecture of the periodic table. Experiments are performed with single-atom quantities of the superheavy elements and their compounds in the gas phase as well as in the aqueous phase. These, together with relativistic quantum chemistry calculations, help to further improve our understanding of how relativity governs the physicochemical properties of heavy elements in general. In the coming years, the examination of the chemistry of elements beyond Z=114, approaching the heaviest so far synthesised element with Z=118 is envisaged, employing novel setups. Latest predictions expect element 118, Og, which is the heaviest member of the noble gas group, to be neither noble nor a gas (at standard pressure and temperature).

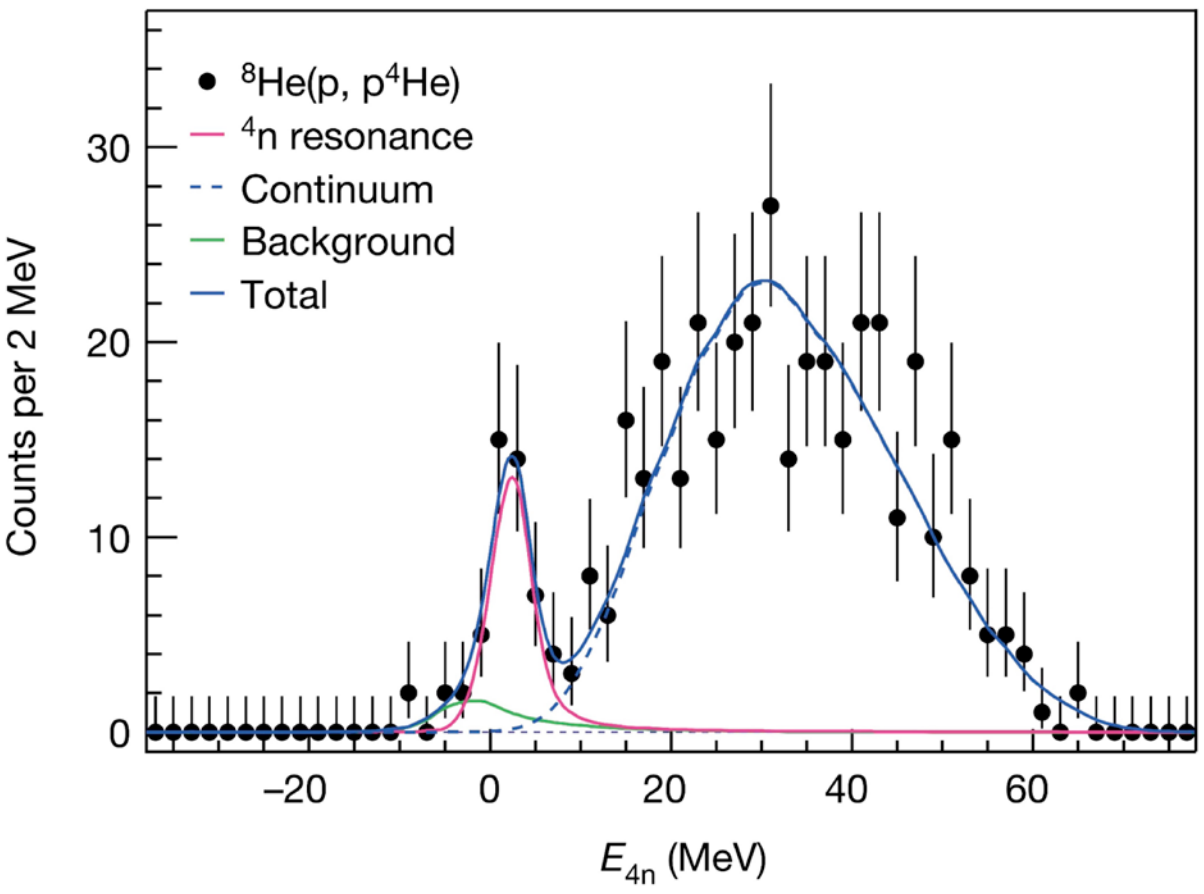

*Fig. 4.12: Missing-mass spectrum of the four-neutron system extracted from the $^{8}$He(p, p$^{4}$He) reaction. The different curves represent a Breit–Wigner resonance (pink), a non-resonant continuum (dashed blue), the background from two-step processes (green) and the total sum (solid blue). (Taken from Nature 606, 678(2022))*

# Nuclear reactions and open quantum systems

Reactions between nuclei are the principal method from which information about nuclear properties is derived. A wide range of nuclear reactions exist, from direct processes like elastic and inelastic scattering, charge-exchange/transfer reactions, and breakup/knockout reactions, to compound nucleus processes, fusion, fission, deep inelastic collisions, and/or multifragmentation reactions. A good understanding of nuclear reaction mechanisms is needed to extract details of the underlying nuclear structure. Reactions themselves are interesting in their role in the creation of new elements and isotopes in stars and the laboratory, and as an example of an interacting many-body quantum system in which the interplay between single-particle and collective dynamics is in evidence, and where the advancement of quantum theories of open quantum systems can be developed.

Few-body dynamics also present an opportunity to investigate the nucleon-nucleon and many-nucleon interactions with experimental data. Hence, the difference in proton-proton (p-p) and neutron-neutron (n-n) scattering lengths contributes to understanding the charge-symmetry breaking of nuclear forces. The Coulomb-free p-p scattering length can be extracted from cross-sections measured with the Trojan Horse method using the quasi-free p+d→p+p+n reaction. The observed difference in proton-proton and neutron-neutron scattering lengths suggests a lower charge symmetry breaking of nuclear forces than predicted so far. This methodology can be used to better constrain the current existing models of charge symmetry breaking and Coulomb corrections, filling out our basic understanding of low-energy NN scattering. This approach can also be used to access the n-n scattering length by measuring the quasi-free n+d→n+n+p reaction. Similarly, the intricate dynamics of breakup reactions of light nuclei with photonuclear reactions (A>3 in the final channel), involving final state interactions, can serve as an experimental tool to probe the accuracy of fundamental parameters governing nuclear interactions (such as the scattering length) connected with the possible existence of a tetraneutron in the continuum.

One central question concerns how the memory of pairwise interactions between nucleons is preserved in reaction observables. Current experimental precision allows for the detection of multiple particles -including neutrons- in the final state. The challenge is to recover the initial configuration of the nucleus using either a stationary reaction model or dynamical propagation, considering the presence of final state interaction in the energy regime. While state-of-the-art techniques are being developed, achieving a model-independent interpretation of data remains an outstanding challenge. For very light systems, ab initio methods (Faddeev-Yakobuvsky, hyperspherical harmonic, stochastic variational methods, configuration interaction methods) can compute both the structure of the bound nuclear system and its continuum within the same framework. This has led to a successful hybrid approach where ab initio inputs, such as scattering length and asymptotic normalisation coefficients, are used as inputs for reaction models adapted to the experimental conditions. However, covering the energy range of the experiment with a many-body technique remains a significant challenge, resulting in deficiencies in the absorption cross-section obtained from ab initio methods.

For heavier systems, the present and future interests in three-body, two-step reactions are multifold, for example the applications of indirect or surrogate methods for nuclear astrophysics as a spectroscopic tool to study clusterisation and/or the structure of neutron-rich systems, and/or the study of breakup dynamics of halo/weakly-bound nuclei. Theoretical methods -such as the Ichimura-Austern-Vincent method- have been successful in predicting inclusive cross-sections for non-elastic-breakup (NEB) of two-body projectiles. However, more efforts should be devoted to expanding the theory to include exclusive cross-sections and four-body reactions to investigate two-neutron-halo breakup dynamics. The dynamic core excitation of halo nuclei has been observed in elastic and inelastic scattering of one-neutron-halo $^{11}$Be. Calculations of dynamic core excitations are feasible in the case of two-body systems using continuum discretised coupled channels with core excitation (XCDCC) calculations. Exploratory studies have indicated the importance of core excitations in two-neutron halo nuclei and it would be desirable to extend reaction theories and experiments to investigate these processes in three-body systems.

Optical potentials play a crucial role in interpreting nuclear data. They enable the projection of the intricate nuclear dynamics into a single-open channel while preserving the unitarity of the full scattering S-matrix through the inclusion of a complex absorptive part in the effective potential. In essence, the transformation of the target-projectile system to its final states and its internal structure is effectively incorporated to calculate the scattering wave function of interest.

## Box 4.4: Dynamic core excitations in nuclear-halo systems

**For collisions involving halo nuclei at low energies ($E_{c.m.} \simeq Vc$ ) the collision time, $\tau_{coll} \simeq (5\text{-}10) \times 10^{-22}$ s, is large enough to let the halo nucleus charge distribution readapt during the collision. In this slow breakup scenario, it is required a description of the scattering which takes into account the distortion of halo nucleus continuum wave functions. This is done in the continuum-discretized coupled-channels (CDCC) calculation accounting for the coupling of the halo degrees of freedom, and in the extended CDCC calculation (XCDCC), for the coupling to both the halo and the core degrees of freedom, hence, the dynamical polarization of the core. This effect is essential to reproduce the experimental data as shown in the figure. Dynamic core excitation calculations for two-nucleon halo systems will be further explored in the future.**

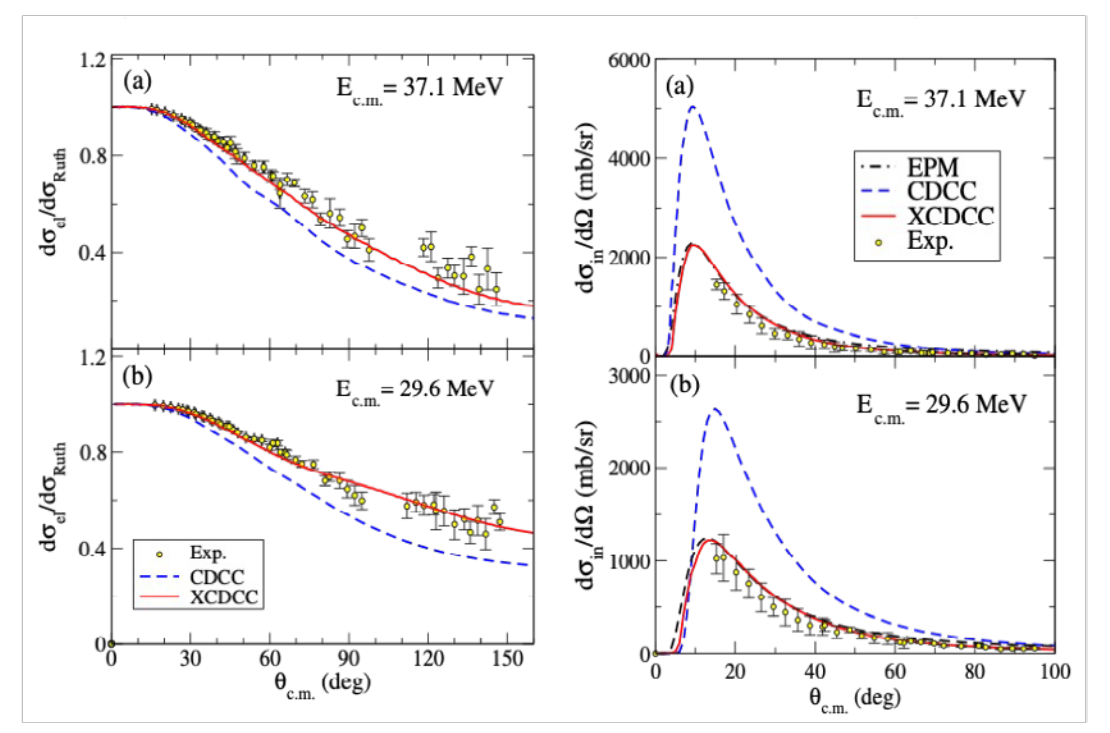

*(Left) Measured elastic scattering angular distribution of $^{11}Be+^{197}Au$, and (right) measured inelastic differential cross section of $^{11}Be+^{197}Au$, populating the 1/2- bound excited state in $^{11}Be$. The curves represent different models (equivalent photon method (EPM), continuum-discretized coupledchannels (CDCC), and extended CDCC (XCDCC)). Adapted form Physical Review Letters 118, 152502 (2017)).*

Statistical analyses of parameters used in phenomenological functionals fitted to data have revealed that a significant portion of errors, reaching tens of per cent, can be attributed to the determination of these parameters. Furthermore, efforts have been made to compute optical potentials from ab initio calculations. However, these calculations often exhibit suboptimal performance compared to phenomenology and further developments must be performed. Future efforts will need to be devoted to developing the formalism and the computational methods to use microscopic optical potentials generally in inelastic scattering, transfer, break-up, and/or pick-up reactions. Specific efforts will also be devoted to studying the impact of the nonlocality of the optical potential in these reactions.

Fusion dynamics at energies much lower than the Coulomb barrier is a particularly challenging topic with implications in astrophysics and the approach towards the superheavy island of stability. The cross-sections involved are extremely difficult to measure as are their predictions by theory. The modelling, from capture via fission competition to the eventual survival as evaporation residue, relies on solid experimental data to disentangle partly contradictory model approaches. High-intensity accelerator facilities like SPIRAL2 at GANIL whose first phase is already operational, or the HELIAC project at GSI, will push, together with state-of-the-art and novel detection instrumentation, the cross-section limits down by more than an order of magnitude. The astrophysical S-factor, introduced to eliminate the exponential behaviour from the cross section due to the tunnelling effect, shows a maximum for reactions with $Q_{value}$<0 and then decreases at lower energies due to the closing of the fusion channel. It is expected that the S-factor does not decline with decreasing centre-of-mass energy for reactions with $Q_{value}$>0. The experimental problem is even more challenging if, in addition to the exponential behaviour of the cross-section, there are resonances -as in carbon $^{12}C+^{12}C$ burning- further complicating the S-factor extrapolation. In these cases, Bayesian methodologies should be applied to extract the cross-section. Future experiments, like the ones foreseen at LUNA and INFN-LNL, will address these important issues. On the other hand, for many reactions indirect methods like the Trojan Horse Method are the only suitable way to reach the astrophysical relevant energy region, and theory should be improved to reliably extract two-body cross-sections from three-body measurements.

The intermediate energy domain (20-100 MeV/u) presents open questions concerning the reaction mechanism, such as the persistence of clustering at high excitation energy and low density. Particle-fragment correlations allow reconstruction of the population of excited levels close to the particle separation energy, thus permitting the investigation of the survival of structure effects at high excitation energies. Open questions also concern the interplay between the interaction time and the N/Z equilibration time when N/Z asymmetric partners are involved or low-density configurations are reached, with implications on the symmetry energy. These kinds of studies can also bring information on the in-medium properties of nuclear interaction, which is still not well known, on the mean free path of nucleons in nuclei and the effective mass splitting. Also, the limiting temperature, related to the amount of excitation energy which can be stored in an excited nucleus before its disintegration, and the corresponding caloric curve, is a subject which deserves investigation as a function of the mass and of the N/Z of the nuclei. Multifragmentation events are the main framework in which the low-density phase transition of nuclear matter can be studied, leading to the extraction of the nuclear caloric curve. Multifragmentation has been associated with a mechanical instability, but chemical instabilities due to proton fluctuations for asymmetric matter may take place. The use of radioactive beams in this energy region (e.g., FRAISE@LNS) can be a tool to explore all the cited topics in more extreme conditions.

Multinucleon transfer reactions (MNT) have been identified as the most promising approach to producing heavy neutron-rich isotopes in the terra incognita. MNT reactions with secondary beams with energies above the Coulomb barrier and instrumentation to detect unambiguously the reaction products must be explored to extend their reach. These experiments will be crucial to steer and optimise developments at upcoming facilities like the Super-FRS at FAIR. For heavy species close to the N=126 shell closure and superheavy nuclei (SHN), MNT is suggested as an alternative approach to populate neutron-rich nuclei which cannot be reached by fusion-evaporation reactions. The region south of $^{208}Pb$ is already the focus of worldwide efforts while first attempts to use $^{238}U$ projectiles on actinide targets have also been undertaken. Models like the Langevin calculations predict production cross sections which would allow for detailed spectroscopy, e.g. for nuclei in the fermium (Z=100) region with N extending beyond the N=162 subshell closure.

On the theory side, an in-depth understanding beyond the widely used GRAZING code has still to be achieved. State-of-the art models differ by an order of magnitude when computing cross-sections at the µbarn level. Recent signs of progress include the combination of microscopic theories, such as time-dependent Hartree-Fock with particle number projection as input for GRAZING, but a more robust theory is still an open question.

In-flight fragmentation is a versatile tool for producing radioactive beams at high kinetic energies. While those energies present challenges, e.g., for Doppler correction, they offer unique advantages, as reactions of radioactive beams in inverse kinematics allow for complete kinematics experiments. A key advantage is that reaction products are mostly focused on forward direction and almost all reaction products can be detected by covering the forward angles in the laboratory system. Moreover, at high kinetic energies coupled with appropriate large-acceptance dipole magnets, full identification of the nuclear charge and mass numbers even for heavy reaction products becomes possible. Unbound systems close to the continuum can also be studied, using missing mass or relative energy measurements. Finally, in-flight fragmentation reactions using very thick targets are under investigation to reach an order of magnitude higher production yields for neutron-deficient isotopes, by exploiting two-step reactions. Recent in-flight fragmentation experiments include investigation of the possible existence of the tetraneutron, study of the unbound $^{28}O$, or investigation of nuclear shell structure around the islands of inversion.

The use of (p, pN) reactions has been developed into a highly versatile tool, which will remain a key experimental approach for the foreseeable future. Here, high beam energies are needed to reduce final-state interactions and thereby provide clean experimental probes. This is essential because the intensities of radioactive beams at the limits of existence are very limited and data with as little background as possible are thus needed. In Europe, at FAIR the R3B and EXPERT setups are optimised for such experiments. Alternatively, reaction studies in storage rings, like the ESR at GSI/FAIR, can be performed under extremely clean conditions with very thin gas-jet targets and high luminosity, because the beam passes the target about a million times for one filling of the ring.

## Nuclear fission

Nuclear fission is a complex phenomenon resulting from a collective motion of large amplitude during which a heavy nucleus deforms until it divides into two excited fragments that, after particle evaporation and/or decay, produce two product nuclei in more stable states. It provides a rich laboratory for studying nuclear structure, dynamical and statistical properties. Moreover, it is relevant for understating how the heaviest elements were produced in the universe and for the recycling of nuclear matter in astrophysical nucleosynthesis processes. Fission is also widely used as a production mechanism for exotic nuclei and hence of relevance for the development of radioactive ion beam facilities. Ultimately, fission has important societal applications as both a powerful, low-carbon source of energy and a generator of radioisotopes for medical purposes (see more in Applications and Societal Benefit).

Fissioning systems can be produced with different initial conditions depending on whether the nucleus is in the ground state (spontaneous fission) or an excited state produced by a reaction (neutron or charged-particle induced fission, photofission, beta-delayed fission, heavy-ion reactions, ...). The fission process is still far from being understood and several key open questions have not yet been fully answered, such as: How is it decided whether a nucleus fissions or not? What are the effects that rule the sharing of nucleons between the fragments? How are excitation energy and angular momentum generated intrinsically and how are they propagated through the process?
Intense experimental and theoretical programmes are being carried out to answer the above questions. Precise measurements of neutron-induced fission cross sections and fission probabilities are relevant to investigation of the first step of the fission process where it is determined whether the nucleus fissions or not. Fission fragment mass and charge yields permit the study of how nucleons are shared between fragments during the progression to scission. The kinetic energies of the fragments, as well as the multiplicities and distributions of neutrons and gamma rays, give valuable information on the fragment deformation at scission, the sharing of the excitation energy and the generation and sharing of the angular momentum. The measurement of these quantities for different nuclei with different initial conditions can be achieved in experiments in direct kinematics using various projectiles (neutrons, photons and charged particles) and target nuclei, as well as in inverse kinematics, where the heavy nucleus is the accelerated beam which interacts with a light target nucleus at rest. This technique makes possible access to fissioning systems far from the valley of stability. The challenge for the next few years is to reduce the uncertainty of neutron-induced cross sections to less than a few per cent by combining independent measurements at different facilities such as JRC-Geel, nTOF/CERN and NFS/GANIL. On the other hand, high-precision fission probabilities will be measured in experiments in inverse kinematics at GANIL, GSI/FAIR with the high-quality beams provided by the ESR and CRYRING storage rings as well as at HIE-ISOLDE with the Isolde Solenoidal Spectrometer. The measurement of fission-fragment yields, kinetic energies as well as the multiplicities and distributions of neutrons and gamma rays for different nuclei with different initial conditions can be achieved in many complementary experiments in direct kinematics using various projectiles and target nuclei. Neutron-induced measurements will be carried out at ILL, JRC-Geel, NFS and ALTO. Charged-particle-induced fission will be investigated at the Oslo cyclotron. High precision mass and charge identification of fission fragments can be obtained with a Penning trap and an MR-TOF for spontaneous fission at GSI, charged-particle-induced fission at IGISOL, neutron-induced at IGISOL and SARAF, and photon-induced fission at ELI-NP. Additionally, the combination of inverse kinematics and large acceptance recoil spectrometers makes it possible to identify isotopically both the light and heavy fission fragments over wide ranges of actinides (including very short-lived nuclei), with complementary programmes on the way at GSI, GANIL, and LNL.

Statistical and semi-empirical models of fission are still very useful tools for comparison with experimental data and for applications in nuclear technology. However, the description of fission dynamics is a challenging theoretical problem. Hence, mic-mac approximations and/or nuclear energy density functional theory (DFT) are used to compute multi-dimensional PES along multipole deformations as well as collective inertias. Quantum tunnelling in spontaneous fission is usually described using a quasi-classical approach for the action integral, while the dynamic evolution of fission-induced systems is dealt with by classical multi-dimensional Langevin equation of motion, semi-classical time-dependent DFT and/or time-dependent generator-coordinate-method techniques. However, a complete microscopic description of fission is not available yet and many improvements may be addressed shortly, e.g. the development of energy density functio-nals and computer codes tailored for fission; consistent treatment of quantum and statistical fluctuations, adiabatic and non-adiabatic processes, quantum tunnelling and inclusion of dissipation in microscopic theories; restoration of quantum numbers (particle-number and angular momentum); description of fission trajectories in the full time-dependent DFT manifold. Finally, another important challenge is the application of techniques from reaction theory to the fission problem.

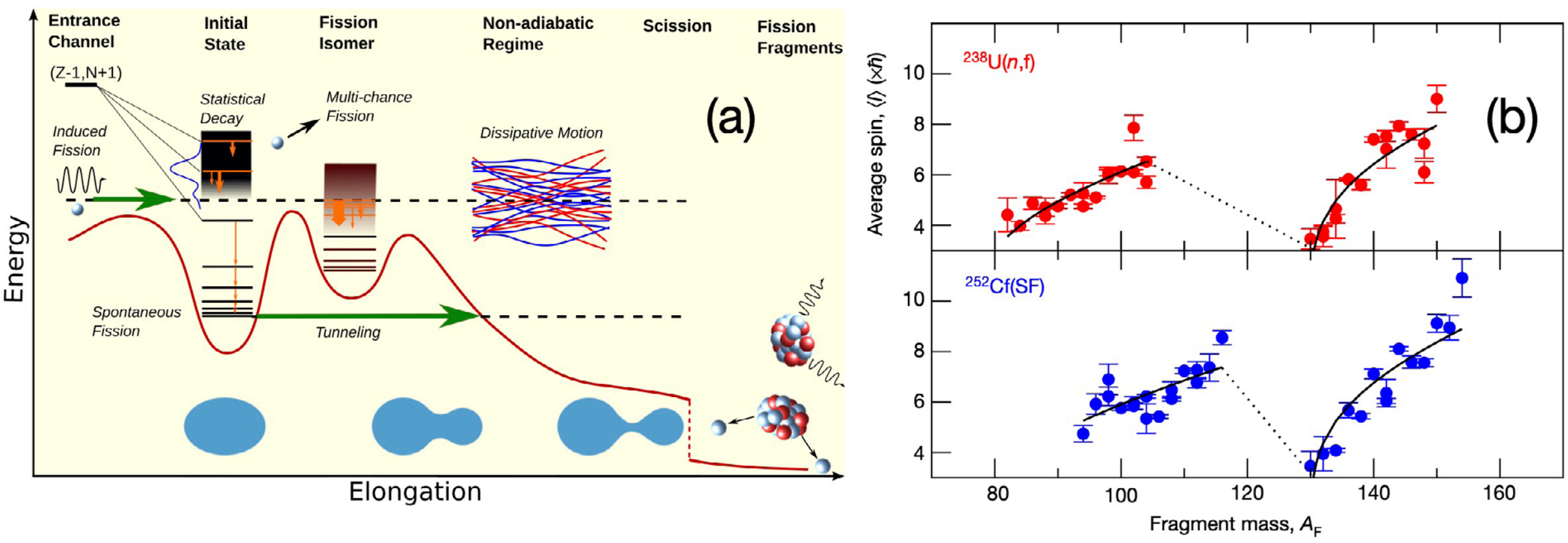

*Fig. 4.13: Left: The fission process is governed by the potential energy of the fissioning nucleus as a function of its (multidimensional) elongation. Such a potential energy surface (PES) shows a ground state minimum and other more deformed shape-isomeric minima/saddle points, followed by a continuous descent towards scission. Then, the two nascent fragments repulse each other due to the Coulomb interaction, reach their equilibrium shapes as they move apart, and, finally, de-excite by evaporating neutrons and/or radiating photons, and eventually beta-decay. Fission can involve excitation energies above or below the fission barrier (e.g. in spontaneous fission). In the latter case quantum tunnelling is needed to cross such a barrier (Journal of Physics G: Nuclear and Particle Physics 47 (2020) 113002) Right: Dependence of average spin on fragment mass in fast-neutron-induced fission of $^{238}U$ and the spontaneous fission of $^{252}Cf$. (Taken from Nature 590, 566 (2021)).*

# Nuclear resonances

Resonances are collective excitations where many of the nucleons that compose the nucleus are involved in the excitation. These modes help to understand the nuclear structure and properties of nuclear matter and its Equation of State (EoS). Their measurement is still a challenge, especially in exotic nuclei, and many questions are still open: how do their properties evolve with the isospin, the deformation, or the excitation energy? Could the EoS be constrained with giant resonances and how do these constraints complement multi-messenger studies with various probes?

The isoscalar giant monopole resonance (ISGMR) corresponds to a compression-dilatation mode of the nucleus and is often referred to as "breathing mode". The ISGMR centroid energy relates to the nuclear EoS through the incompressibility of the nucleus, which can be linked to the incompressibility of nuclear matter $K_\infty$. Recent calculations have shown that beyond-mean-field effects (like quasiparticle-vibration coupling (qPVC) and/or subtracted second random phase approximation (SSRPA)) are needed to provide a consistent description of the ISGMR resonance in Sn and Pb isotopes and soft breathing modes in neutron-rich nuclei. Additionally, new theoretical approaches have emerged recently: the ab-initio QRPA and the PGCM approach (see Nuclear Interactions and Nuclear Models). These calculations of ISGMR are compared in light and mid-mass closed and open-shell nuclei, which makes possible investigation of the role of superfluidity from an ab-initio standpoint.

On the experimental side, European laboratories have pioneered ISGMR measurements in exotic nuclei using active targets. This experimental programme is still ongoing thanks to the new-generation active target ACTAR at GANIL-SPIRAL2. More precise measurements of the monopole strength in exotic nuclei are expected in the coming years using active targets and storage rings. They will provide new constraints on $K_\infty$ and should shed light on the possible existence of a soft monopole mode.

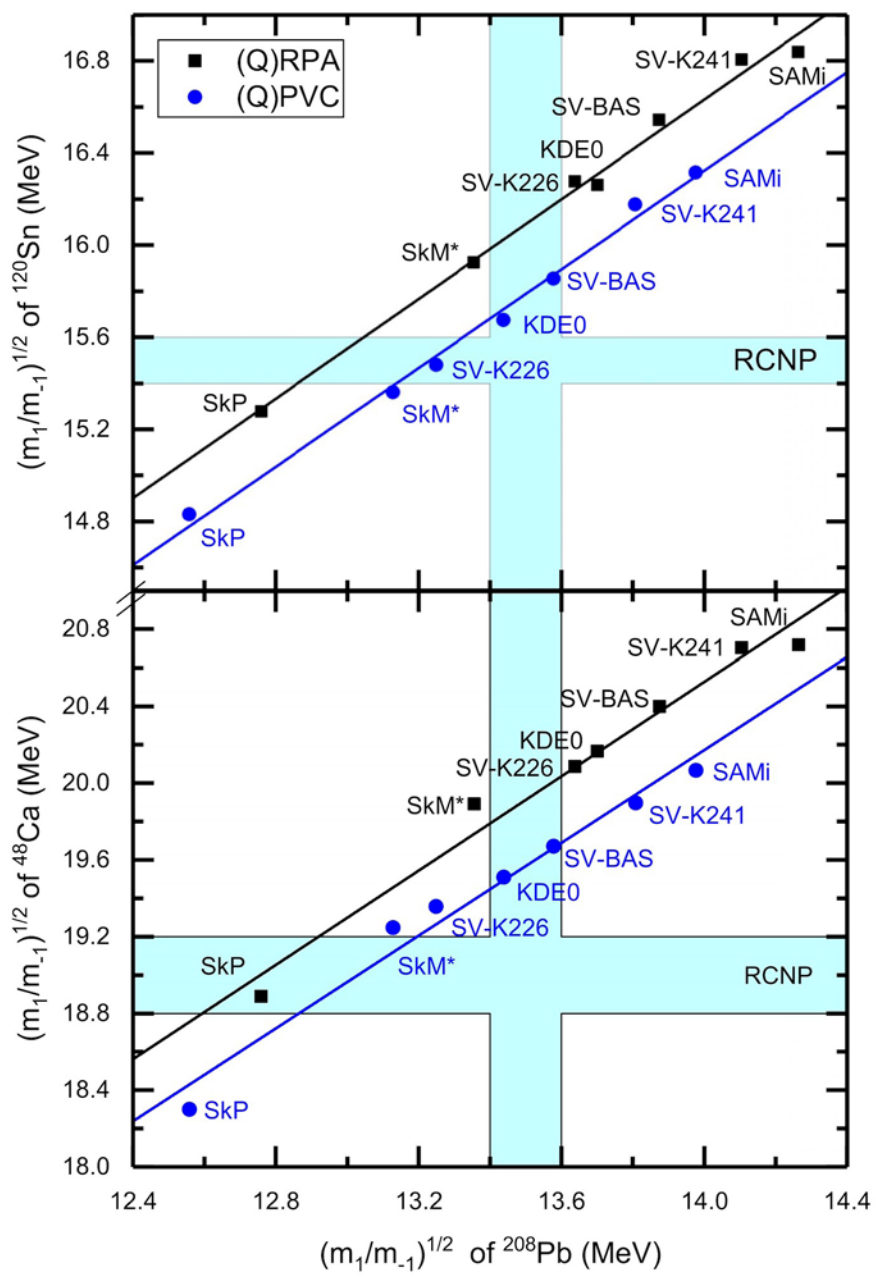

*Fig. 4.14: The ISGMR energies in $^{208}$Pb vs. the ones in $^{120}$Sn (upper panel), and $^{48}$Ca (lower panel). These are calculated by (Q)RPA (black square), and by (Q)RPA+(Q)PVC (blue circle) using 7 different Skyrme parameters. The regression lines are obtained by a least-square linear fit of the (Q)RPA results and (Q)RPA+(Q)PVC results, respectively. The experimental data and their uncertainties are displayed by means of cyan-coloured bands. (Taken from Physical Review Letters 131, 082501 (2023)).*

The giant dipole resonance (GDR) corresponds to the oscillation of protons against neutrons. While for light to medium-mass nuclei ab initio calculations are available, for heavy nuclei other methods are more suitable. If the nucleus is axially deformed, the GDR strength function splits into two components corresponding to oscillations of neutrons versus protons along and perpendicular to the symmetry axis of the nucleus due to deformation. The dependence of the GDR strength function on the nuclear deformation is being studied in hot nuclei produced in fusion reactions and their residue after evaporation. These kinds of experiments will continue, e.g. at LNL with the PARIS scintillator detection array coupled with the AGATA spectrometer. GDR experiments will also allow the study of the abrupt changes from oblate to prolate deformations in the mass region $100<A<130$ (Jacobi shape transition) or changes from elongated triaxial to very elongated left-right asymmetric shapes at high angular momentum around the fission limit (Poincare shape change). The gamma decay of the isoscalar giant quadrupole resonance (ISGQR), corresponding to a vibration of the nuclear surface with protons and neutrons oscillating in-phase, is expected to provide a unique probe of the resonance wave function. So far explored in $^{208}$Pb only, being hindered by the competing neutron and GDR emission, the gamma decay of the ISGQR should be investigated in the future in other systems using AGATA-PARIS combined setups. QRPA calculations including pairing correlations and axial deformations are the standard approaches in the study of giant resonances, but further development of the theoretical tools should include triaxial, octupole deformations, and finite temperature effects.

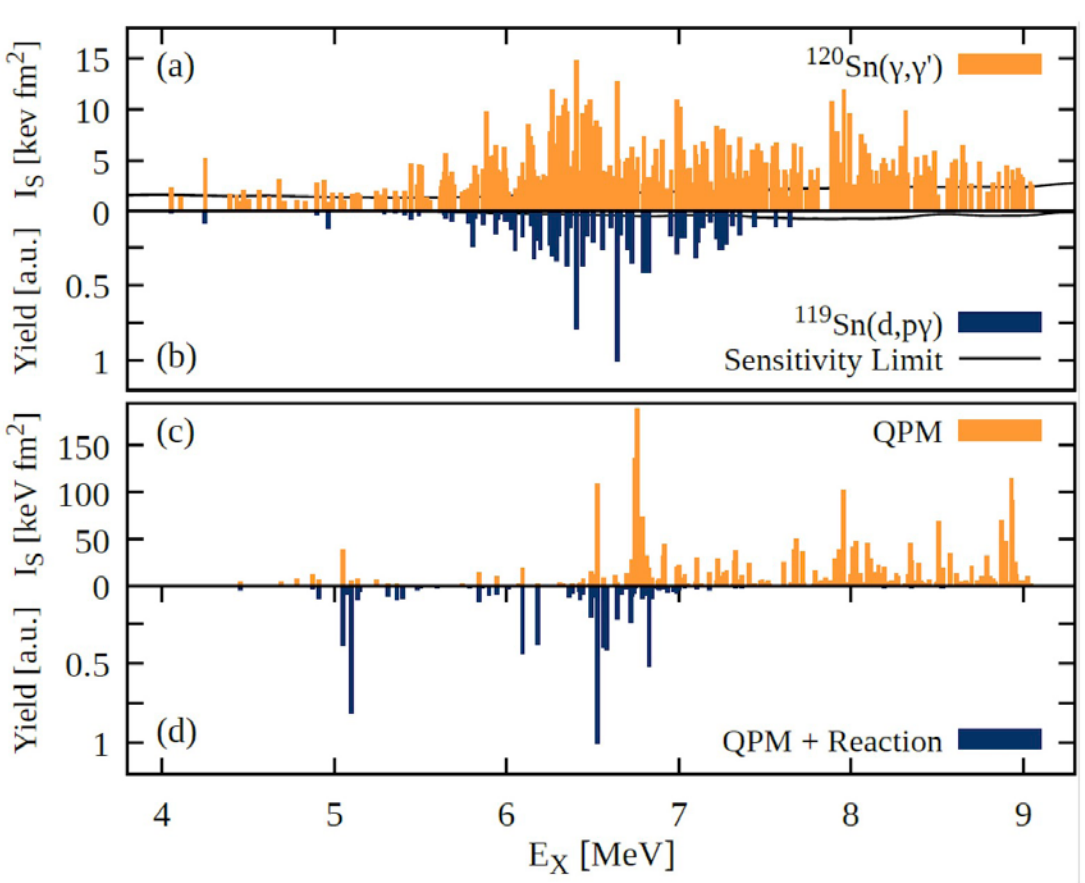

*Fig. 4.15: Experimental investigations of the Pygmy Dipole Resonance (PDR) in $^{120}$Sn via (a) photon scattering $(\gamma,\gamma')$ experiments, not sensitive to the collective/single-particle nature of the excitations, and (b) neutron-transfer experiments, showing an enhancement of single-particle excitations in the low energy part of the PDR. Comparisons with calculations by the quasiparticle-phonon model (QPM) (c) and QPM combined with reaction theory for the (d,pγ) reaction (d) show qualitative agreement and give first insights into details of the microscopic structure of the PDR. (Adapted from Physical Review Letters 127, 242501 (2021)).*

The low-energy counterpart of the GDR is frequently called the Pygmy Dipole Resonance (PDR) because it carries only a few per cent of the total electric dipole strength in a nucleus. However, this increase of the cross section at relatively low energies can have an important impact on the survival of nuclei in a stellar photon bath and may help to constrain certain parameters in the nuclear EoS. In a simplified geometrical picture, the PDR is often described as a collective out-of-phase oscillation of a neutron skin against an isospin-saturated proton-neutron core, and its strength is predicted to increase in neutron-rich exotic nuclei. The first studies of the PDR in the past focused on the electromagnetic response. Recently, the results of hadron scattering experiments showed that only a part of the dipole strength below the GDR is structurally different from the GDR. Neutron scattering offers the possibility to further address the PDR structure problem since it is sensitive to the nucleus. The use of $(n, n'\gamma)$ reactions, e.g. at the neutrons for science (NFS) facility at GANIL-SPIRAL2, will reveal the nature of the PDR at the nuclear surface and highlight the role of the protons in the PDR excitation. Moreover, the validity of the skin-os-cillation picture can be tested by comparing experiments that examine the single-particle character of the excitation mode, like (d, p$\gamma$) neutron-transfer reactions, with other experiments that are not sensitive to the structure of the excitations, such as photon scattering ($\gamma$, $\gamma$') (also at ELI-NP, in the future). Another important aspect is whether the PDR survives at finite temperatures in hot nuclei. This question will be studied in stable nuclei and in nuclei produced in fusion-evaporation reactions, as recently done in Ni isotopes at CCB Krakow and IFIN-HH Magurele.

In general, studies of giant resonances and/or PDR that decay emitting gamma-rays will greatly profit from detectors such as PARIS and AGATA at facilities like GSI/FAIR, GANIL, and LNL-SPES, and from kinematically complete measurements at R3B.

Finally, the knowledge of electric dipole states close to the neutron threshold is of paramount importance in refining models of photon strength functions (PSF), a quantity that describes the average probability of emitting or absorbing $\gamma$ radiation with a given $\gamma$-ray energy. The Brink-Axel hypothesis is widely used in nuclear astrophysics to model electromagnetic decay processes and assumes that the electromagnetic decay rate does not depend on the absolute excitation energy, nor on the properties of the states involved but only on transition energy. This means that the PSF is independent of the excitation energies. However, recent experiments with monochromatic photon beams have shown the breaking of the Brink-Axel hypothesis. An enhancement of E1 strength at low energies is not only predicted for neutron-rich nuclei but a similar effect is expected on the proton-rich side. Unambiguous experimental observations are still missing, but the advent of new radioactive beams may permit detailed studies in the future.

# Nuclear equation of state (EoS)

The nuclear equation of state (EoS) describes the relationship between the energy per nucleon *(E/A)* (or pressure *(P)*), temperature *(T)* and density for nuclear matter. It is customary to introduce the total density $\rho=\rho_p+\rho_n$, and the isospin asymmetry $\delta=(\rho_n-\rho_p)/(\rho_n+\rho_p)$ and writes the EoS as $E(\rho,\delta)=E(\rho,\delta=0)+S(\rho)\,\delta^2$, $S$ being the symmetry energy. In turn, the symmetric matter EoS and $S$ can be expanded around the saturation density $\rho_0$ as $E(\rho,\delta=0)=E(\rho_0)+1/2K_\infty\,(\rho-\rho)^2/(3\rho_0)^2+O[(\rho-\rho_0)^3]$, and $S(\rho)=J+L(\rho-\rho_0)/3\rho+1/2K_{sym}\,(\rho-\rho_0)^2/(3\rho_0)^2+O[(\rho-\rho_0)^3]$. The values of the different parameters entering the EoS can be constrained experimentally, from theory and/or from astrophysical observations.

The experimental probes to investigate the EoS and the symmetry energy can be classified according to the density region they are testing. The binding energies of nuclei near the dripline contain information about the asymmetric part of the EoS but do not allow for direct extraction of the density behaviour of the EoS. Neutron skins, which represent the differences between neutron and proton radii, have emerged as significant indicators of the density dependence of the EoS around and slightly below the saturation density. Extracting neutron densities through parity-violating electron scattering has received considerable attention. This method is virtually model-independent, as the weak charge of the neutron (proton) is close to 1 (0). Experiments conducted on $^{208}$Pb and $^{48}$Ca at JLab have already provided interesting results. At MESA the most precise extraction of the neutron radius in $^{208}$Pb will be achieved with the P2 apparatus. Moreover, the elastic scattering between protons and radioactive beams in a storage ring at FAIR might be useful.

The energy of the Giant Monopole Resonance (GMR) is strongly correlated with the nuclear incompressibility ($K_\infty$) (see previous section). Moreover, further research is necessary to understand the compressibility of deformed nuclei and the trend of compressibility in neutron-rich nuclei, since this is crucial for elucidating the role of the EoS in astrophysical scenarios such as core-collapse supernovae and neutron star mergers. Regarding the symmetry energy, dipole modes, in conjunction with the neutron skin, play a vital role in extracting the density behaviour. Alongside the well-known GDR and PDR (see previous section), the total dipole polarisability has emerged as a valuable tool for investigating this density dependence. The dipole polarisability can be measured through small-angle proton scattering with minimal (if any) model dependence. Other probes, such as gamma rays or heavy ions, can be used to investigate the nuclear dipole response. This is essential for testing the isoscalar versus isovector and surface versus volume character of the dipole states and their relevance in extracting the symmetry energy. Extending these techniques to the case of inverse kinematics is crucial to going toward exotic nuclei. Other resonances, such as isovector quadrupole or isobaric analogue, have also demonstrated their ability to provide complementary information about the nuclear EoS.

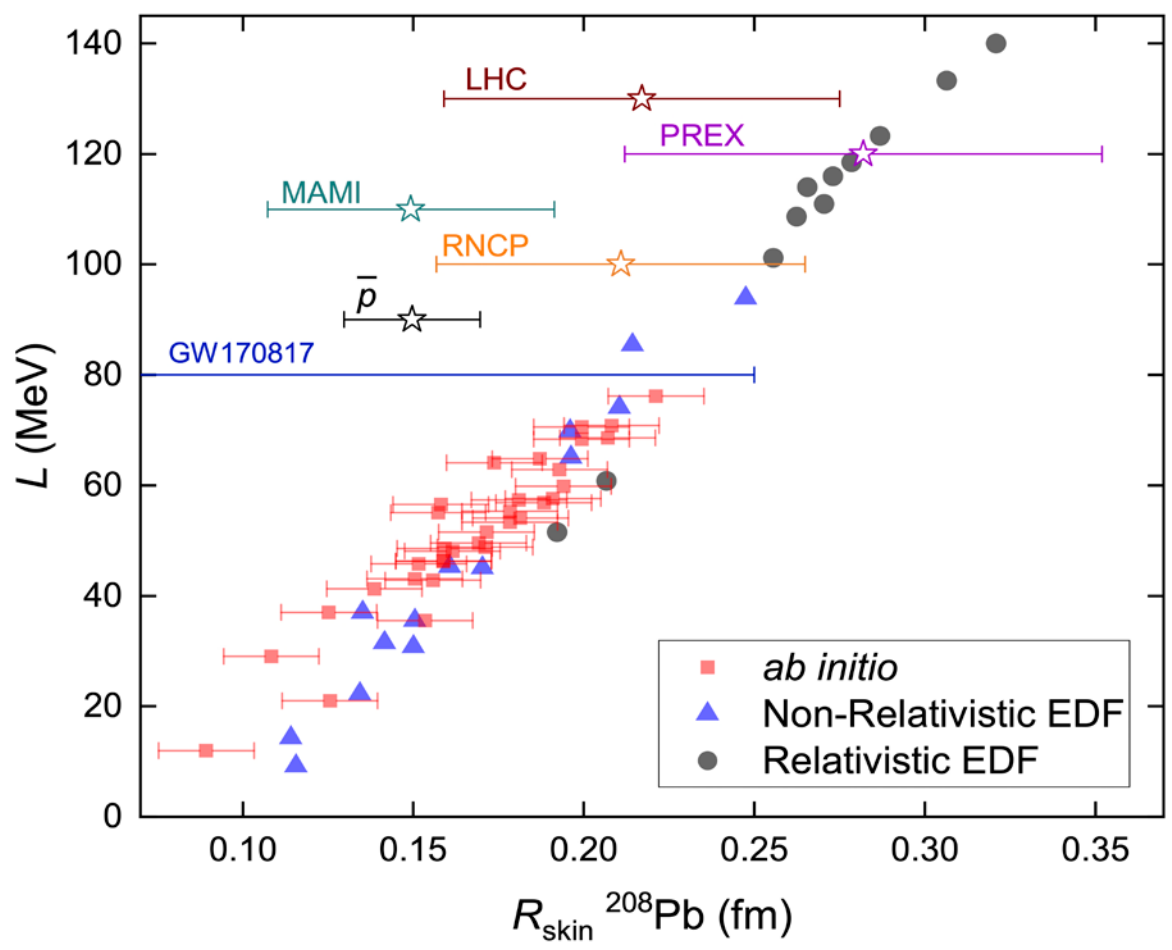

***Fig. 4.16: Correlation of the neutron skin $R_{skin}$ (for $^{208}$Pb) vs the slope of the symmetry energy L. Relativistic and non-relativistic mean-field calculations are indicated with open symbols, while ab initio results using the 34 non-implausible samples are indicated with filled circles. The experimental extractions of $R_{skin}$ shown in the figure are from PREX, MAMI, RCNP, LHC, and GW17081736. Only the x-value matters for the exp. points, and the y position has been adjusted based only on readability. All these results involve modelling inputs, as the neutron skin thickness cannot be measured directly. (Adapted from Nature Physics 18, 1198(2022)).***

The sub-saturation region can be accessed with heavy ion collisions (e.g., INDRA-FAZIA and INDRA-VAMOS at GANIL, and CHIMERA at LNS). The main probes are associated with the isotopic composition of the ejectiles, which is strongly influenced by the isospin equilibration time in charge asymmetric reactions versus the characteristic reaction time. Therefore, to test the symmetry energy below saturation density it will be useful to use radioactive beams at intermediate energies to maximise the N/Z difference in the entrance channel.

Beyond the saturation density, the main probes are the elliptic flow of neutrons and Light Charged Particles (LCP) (e.g. ASYEOS and FOPI at GSI). Better probes are expected to be the K+/K0 ratio (e.g. HADES at GSI), especially in the range 1.5$\rho_0$-3$\rho_0$, because kaons are produced at the early stage of the collisions where the density due to the compression is very high, and they do not interact significantly with the system. Also, short-range correlations (e.g. R3B at FAIR) are useful in this density range. Beyond 3$\rho_0$ the elliptic flow is no longer a good probe, but the collective flows of protons and other LCPs keep their validity (see the programmes of HADES, R3B, CBM at GSI/FAIR, and RIKEN and FRIB outside Europe). Supra-saturation region, besides being the least known, is the most important one for astrophysics. It is the region where synergies between heavy ion reactions and the observation of gravitational waves associated with the merging of neutron stars are evident. This multimessenger approach allows for strong tightening of the constraints on the EoS in a very wide range of ρ (see also chapters Strongly Interact Matter and Nuclear Astrophysics).

The EoS cannot be directly accessed from the experiments, so theoretical models are the essential tool for extracting information. They can provide either mild or strong biases depending on the specific case.

Now, uncertainties in the ab initio nuclear Hamiltonians hinder straightforward calculations of the EoS. Neutron matter calculations, particularly at low density, currently exhibit greater reliability than symmetric matter or high-density calculations. Nevertheless, significant progress in our understanding of chiral forces is anticipated in the coming years.

DFT stands as the sole microscopic model capable of connecting the EoS with ground-state and excited-state properties of finite nuclei, as well as certain observed properties of compact objects like neutron stars. Over the past few decades, various tensions among different DFT approaches, such as nonrelativistic or covariant formulations, have been resolved. "Agnostic" meta-models that encode only the EoS parameters have already been developed and can be further extended in the future.

Despite attempts based on time-dependent DFT, there is a gap between the microscopic models just mentioned above and the phenomenological transport models used for intermediate-energy and relativistic heavy-ion reactions. These models allow the simulation of the experimental observables with different input parametrisations for the EoS. The Transport Model Evaluation Project, aiming at comparing the predictions of different codes run in the same conditions, will be a valuable tool to properly study the EoS, since the results obtained by the different models present many discrepancies. Particularly, the description of pion emission in the fireball should be improved, as well as the inclusion of three-body forces in effective potentials.
Finally, advanced statistical techniques such as Bayesian inference or machine learning are being used to deduce the EoS from experimental observations obtained with different probes. These methods can be applied separately to the results of various heavy-ion experiments, to astrophysical observations (neutron star masses, radii, and tidal polarisabilities), or other constraints, including masses, radii, neutron skins, and giant resonance energies. Efforts to widen the applications of these methods should be envisaged.

# Synergies with astrophysics, fundamental symmetries, and hadron physics

## Impact in nuclear astrophysics

Nuclear physics input (S-factors, beta-decay half-lives and beta-delayed neutron emission probabilities of neutron-rich nuclei, masses, neutron-capture probabilities, fission, neutrino-nucleus interactions, …) are key quantities to understanding the synthesis of elements in the Universe and stellar dynamics (see Chapter Nuclear Astrophysics). Intense experimental efforts will be made to measure such properties in future facilities, but not all data will be accessible via experiments. Hence, theory must provide this missing nuclear physics input. Models must be extended to provide reliable predictions of reaction cross-sections, nuclear masses, beta-decay half-lives including forbidden transitions, and gamma and multi-nucleon emission above the neutron separation threshold.

Concerning beta-decay experiments, measuring beta strength functions, half-lives and beta delayed (multi-)neutron emission probabilities in neutron-rich nuclei will be of paramount importance to probe nuclear structure models (e.g. with the total absorption technique, TAGS). Measuring gamma-neutron competition above the neutron separation energy can provide constraints on neutron capture cross-sections of very exotic nuclei and can also be relevant in the study of collective phenomena populated in the decay like the Pygmy resonance, with astrophysical impact.

## Nuclear physics and physics beyond the standard model

Nuclear structure and reaction dynamics play a crucial role in several quests related to the search for physics beyond the standard model of particle physics (see Chapter Symmetries and Fundamental Interactions).

In long-baseline neutrino experiments (e.g. DUNE, Hyper-K, ORCA, PINGU), broad neutrino beams hit a nuclear detector producing charged particles. Neutrino interactions with nuclei need to be accurately modelled to extract oscillation parameters such as neutrino masses and the neutrino mixing charge-parity (CP)-violating phase that is related to the matter-antimatter asymmetry. Several theoretical frameworks are being explored, ranging from ab-initio methods to density functional theory. Their validity needs to be first tested on precise electron scattering data on nuclei. At low energies, neutrino-nucleus scattering is also relevant for the searches of supernova neutrinos. Additionally, coherent elastic neutrino scattering can be used to learn about physics beyond the standard model and to extract the neutron-skin thickness of a target nucleus as an alternative to parity-violating electron scattering experiments (CREX/PREX and future MREX in the USA and Europe, respectively). The matter/antimatter asymmetry will also be studied with the beta decay of polarised trapped ions within JYFL-MORA. Finally, the total absorption technique used in beta-decay experiments has proved its usefulness in the description of the reactor antineutrino spectrum and the understanding of the reactor anomaly. In this context, the so-called spectrum distortion remains a question that should be addressed.
Dark matter is sought in direct detection experiments (e.g. XENONnT at LNGS) where nuclear targets are used. The scattering cross section depends on nuclear structure functions that must be calculated from nuclear theory and will be needed to extract dark matter properties in case of detection.

Another sensitive tool to check the completeness of the standard model and set tight constraints on its extensions is superallowed beta decays, where the unitarity of the Cabibbo-Kobayashi-Maskawa (CKM) quark mixing matrix is tested. The main limiting factor comes from the nuclear structure uncertainty on superallowed transitions; modern approaches, such as ab-initio methods, will shed light on this issue. Together with the free neutron decay, nuclear beta decays allow for constraining non-standard scalar and tensor charged-current interactions. Polarised nuclei can be used to determine the b-asymmetry parameter in mirror nuclei with great precision, which will also contribute to constraining the unitarity of the CKM matrix. Future facilities DESIR/SPIRAL2 and ISOL@MYRRHA at SCK CEN in Belgium will employ polarised nuclei and be able to address all these topics when completed. These constraints are complementary to those from accelerators such as the LHC.

Finally, neutrinoless double beta decay ($0\nu\beta\beta$) is a proposed decay of a nucleus that only emits two electrons. The detection of this baryon minus lepton number violating process will prove that neutrinos are their own (Majorana) antiparticle. The decay half-life of $0\nu\beta\beta$ depends on nuclear matrix elements that need to be reliably known both to plan the reach of current and future experiments (LEGEND, nEXO, DARWIN, NEXT, …) and to fully exploit a $0\nu\beta\beta$ measurement. Ab-initio calculations have recently shown their potential to provide consistent matrix-element values for the lightest double beta nuclei - including theoretical error bars - and they will be extended to heavier systems. Also in this context, a better understanding of the quenching of the gA constant in the nuclear medium is necessary. In addition, systematic calculations with both ab-initio and phenomenological methods are also needed to correctly understand the degrees of freedom of the problem and to extract information from surrogate processes such as double charge-exchange (DCE) reactions, muon capture on nuclei and/or second-order electromagnetic decays. These processes will be experimentally accessible, e.g. DCE reactions within the NUMEN project at LNS.

## Box 4.5: Neutrinoless double beta decay research

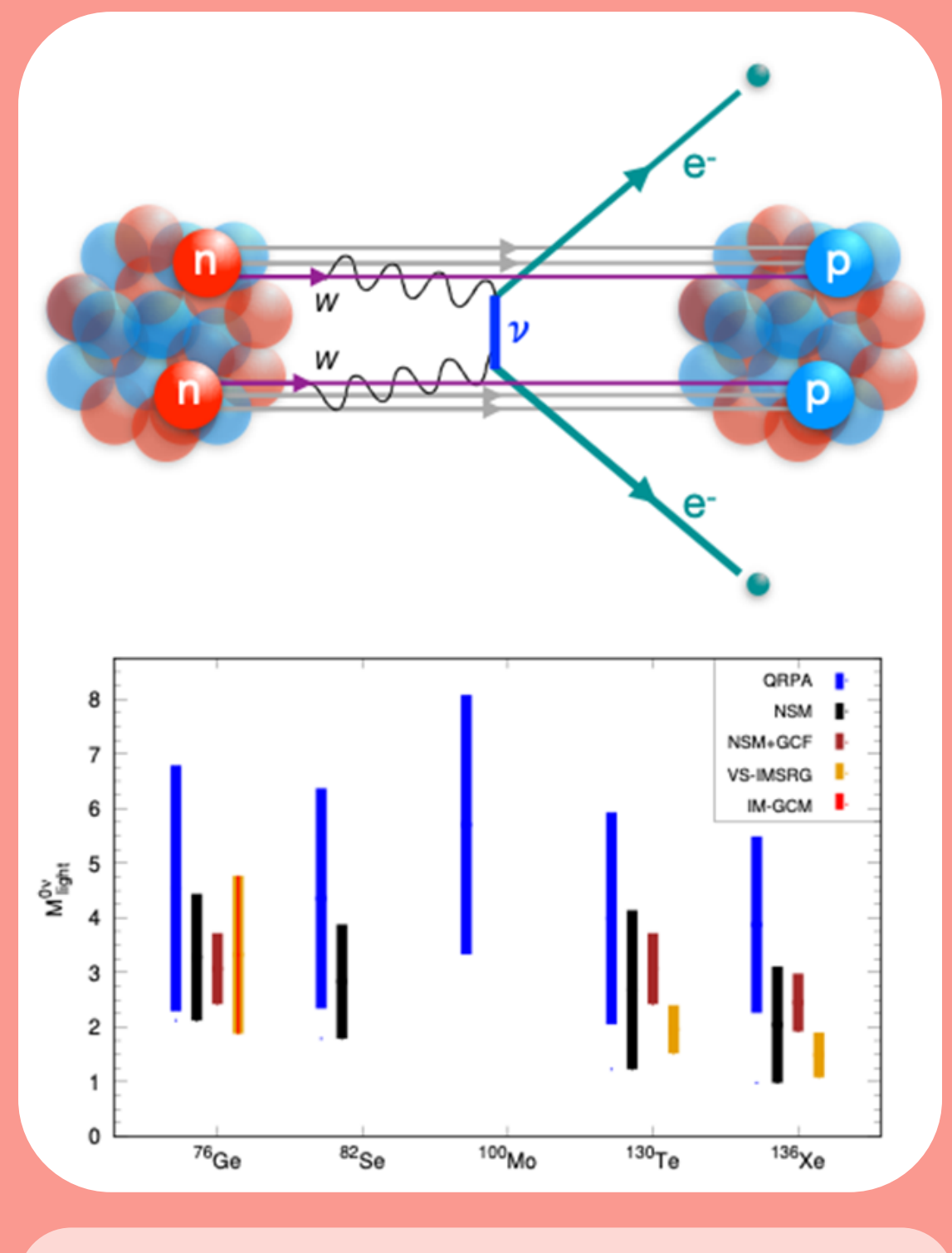

***Upper panel: Schematic view of the $0\nu\beta\beta$ decay through the exchange of a Majorana neutrino. Lower panel: NMEs for $0\nu\beta\beta$ decay candidates calculated with different many-body methods.***

**In the hypothetical neutrinoless double beta decay ($0\nu\beta\beta$) two neutrons in the nucleus transform into two protons, and only two electrons are emitted. This process is forbidden in the Standard Model of particle physics, as two matter particles are created without any antimatter. $0\nu\beta\beta$ is expected to be mediated by the exchange of a neutrino (see upper panel) that must be its own antiparticle (i.e., a Majorana particle). Hence, the detection and subsequent study of $0\nu\beta\beta$ would deeply impact our understanding of neutrinos, physics beyond the Standard Model, and why the observed universe is formed by matter. Searching for this extremely rare decay is the goal of several ongoing and future experimental programs. However, a main uncertainty for predicting $0\nu\beta\beta$ half-lives comes from the nuclear matrix elements (NMEs) connecting the initial and final nuclear states. NMEs are calculated with many-body techniques routinely used to describe nuclear structure (e.g., nuclear shell model, energy density functional, QRPA) but the predictions differ up to a factor of 3-4 for the isotopes of interest (see lower panel). Ab-initio methods (Coupled cluster, In-medium SRG, PGCM) are starting to study $0\nu\beta\beta$ with a more consistent assessment of the uncertainties and will become essential to provide reliable NMEs in the next few years. Another promising way is to extract NMEs with surrogate processes like double-charge exchange reactions (e.g., NUMEN project) or second-order electromagnetic transitions, using the theoretical correlations between the NMEs of $0\nu\beta\beta$ and these other processes.**

# Recommendations for nuclear structure and reaction dynamics

● **Unique insights into nuclear structure and shell evolution as well as input and constraints for nuclear astrophysics** can only be obtained via the urgent completion of the *FAIR facility, including the Low-Energy-Branch, SPIRAL2, SPES, ELI-NP*, as unique laboratories for studying reactions of **very exotic nuclei**, and for exploration of the **nuclear chart towards the driplines**. To ensure **complementarity in experimental programmes**, it is essential to give strong support to all other large facilities (and their upgrades), as well as to small-scale and university-based facilities which guarantee access to the whole community, allow detector testing and exploratory experiments in preparation for the most complex future experiments and play a key role in the training of new generations of physicists (see more in chapter Research Infrastructures).

● **To push the frontiers of spectroscopy and lifetime measurements** even for low-intensity secondary beams and small cross sections, superb in-beam resolution and high efficiency in gamma-ray spectroscopy are essential. The *full completion of the European flagship gamma spectrometer AGATA-4π (with ancillaries)* which is, and will continue to be, the major workhorse for nuclear structure and nuclear astrophysics precision physics at both radioactive and stable ion-beam facilities (see more in chapter Nuclear Physics Tools Detectors and Experimental Techniques) is therefore essential.

● Heavy-ion storage rings are key precision instruments for **future studies on nuclear masses and radii, reactions, nuclear resonances, isomers and fission**. The world leadership of Europe in the use of heavy ion storage rings should be maintained by supporting experiments in the existing low energy storage rings of *GSI/FAIR (ESR and CRYRING)* and, in a larger time frame, the construction of future storage rings at *GSI/FAIR and HIE-ISOLDE* (see more in chapter Research Infrastructures and chapter Nuclear Physics Tools Detectors and Experimental Techniques).

● It is essential to vigorously pursue the **development of a unified theoretical description of all nuclei and nuclear matter** based on systematic theories of strong interactions at low energies, advanced few- and many-body methods, as well as a **consistent description of nuclear reactions**. Theoretical calculations are crucial for interpreting experimental results and guiding future research. *Theory centres should be strongly supported throughout Europe and a larger number of new positions for young researchers in theory should be opened.* We recommend increasing the support given to the European Centre for Theoretical Studies *(ECT*)*, where new ideas are born through the broad workshop programme, and to support emerging *virtual access facilities* providing theory results for experimentalists (as Theo4Exp VA facility in the Eurolabs project, see more in chapter Research Infrastructures and chapter Nuclear Physics Tools Detectors and Experimental Techniques). *Synergetic programmes connecting nuclear structure and reaction dynamics with other related fields, like nuclear astrophysics and fundamental symmetries, should also be strongly supported.*

● Enriched rare stable isotopes (ESI), most prominently $^{48}$Ca, are essential ingredients **in the investigation of nuclei at the limits of stability**. Therefore, with the shortage/absence of some important ESI ($^{48}$Ca crisis), it is essential to develop a *European strategy for a secure supply of enriched isotopes* for European research facilities, including a hitherto non-existent European facility for electromagnetic isotope separation (see chapter Nuclear Physics Tools Detectors and Experimental Techniques).

# Nuclear Astrophysics

**Conveners:**
**Anu Kankainen** (JYFL-ACCLAB Jyväskylä, Finland)
**Jordi José** (UPC, Barcelona, Spain)

**NuPECC Liaisons**
**Daniel Bemmerer** (HZDR and TU Dresden, Germany)
**Sandrine Courtin** (IPHC, Strasbourg, France)

**WG Members:**
- **Umberto Battino** (University of Hull, Hull, UK)
- **Andreas Bauswein** (GSI Darmstadt, Germany)
- **Sonja Bernitt** (GSI Darmstadt, Germany)
- **Carlo Bruno** (University of Edinburgh, Edinburgh, UK)
- **Cristina Chiappini** (Leibniz Institute for Astrophysics, Potsdam, Germany)
- **Rosanna Depalo** (UNIMI Milan, Italy)
- **Cesar Domingo Pardo** (University of Valencia, Valencia, Spain)
- **Jenny Feige** (Museum für Naturkunde, Berlin, Germany)
- **Stephane Goriely** (Université Libre de Bruxelles, Brussels, Belgium)
- **Francesca Gulminelli** (ENSI & LPC, Caen, France)
- **Marcel Heine** (IPHC, Strasbourg, France)
- **Gabor Kiss** (ATOMKI, Debrecen, Hungary)
- **Ann-Cecilie Larsen** (University of Oslo, Oslo, Norway)
- **Yuri A. Litvinov** (GSI, Darmstadt, Germany)
- **Maria Lugaro** (IAU, Konkoly, Hungary)
- **Jerôme Margueron** (IP2I Lyon, France)
- **Uwe Oberlack** (University Mainz, Mainz, Germany)
- **Francois de Oliveira** (GANIL, Caen, France)
- **Rosario Gianluca Pizzone** (University of Catania, Catania, Italy)
- **Konrad Schmidt** (HZDR Dresden, Germany)
- **Nicolas de Séréville** (IJCLab Orsay, France)

# Introduction

Nuclear astrophysics is a multidisciplinary research field combining both experimental and theoretical nuclear data with astrophysical modelling of stellar events to compare and understand observations made via astronomy or cosmochemistry. The key research questions are:

- Where and how are chemical elements formed in the Universe?

- What are the nuclear processes that drive the evolution of stars and what is their impact on the evolution of galaxies and the Universe?

- What is the nature of matter in the extreme conditions of compact astrophysical objects such as mergers or pulsars?

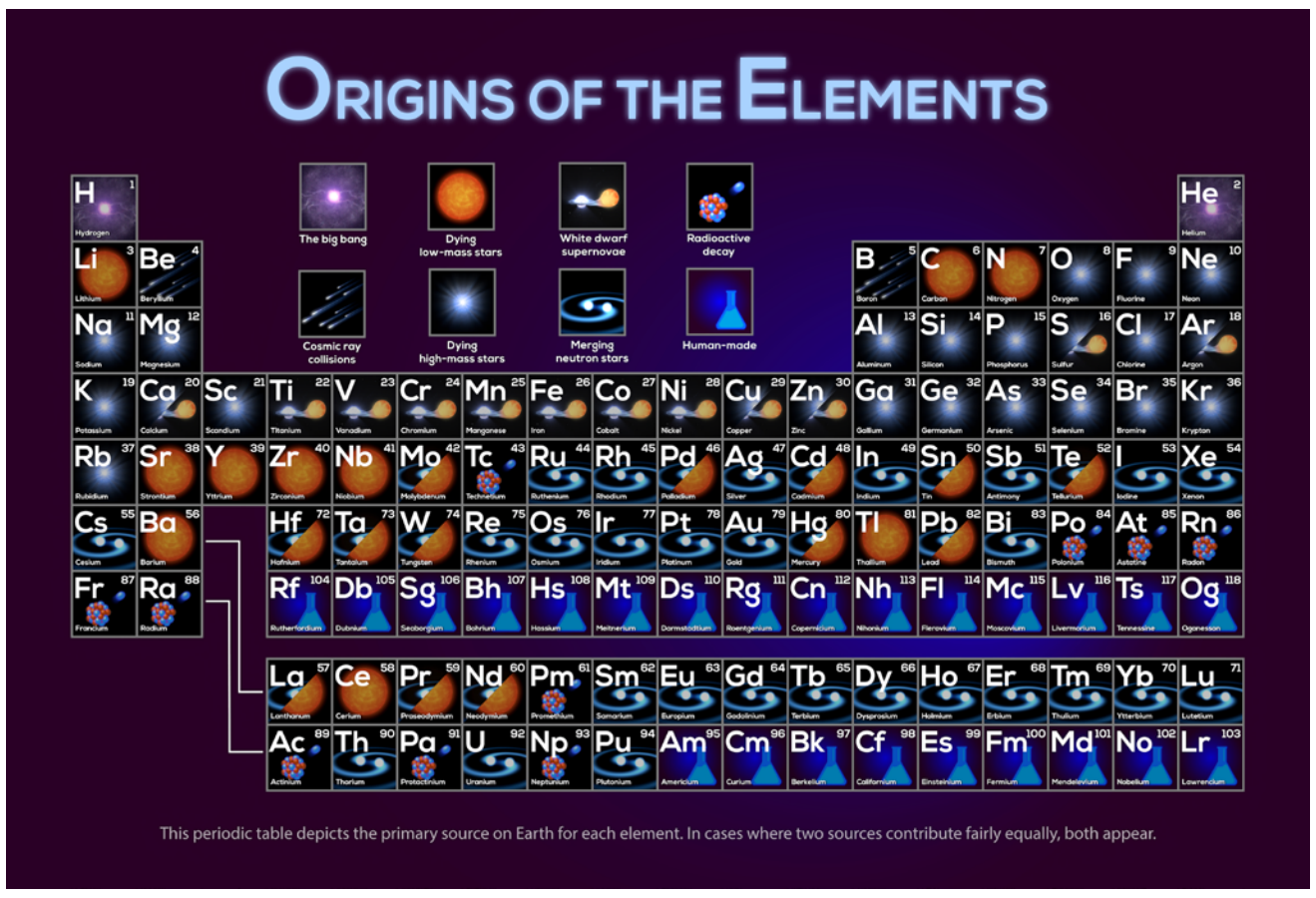

*Fig. 5.1: Origin of chemical elements. Credit: NASA's Goddard Space Flight Center.*

## Cosmic origins

Almost all chemical elements have astrophysical origins as shown in Fig. 5.1. Hydrogen and helium were already produced in the Big Bang nucleosynthesis. The first stars were formed around 100 million years after the Big Bang when the gas clouds contracted and enabled high enough temperatures to initiate fusion (stellar burning). Stellar evolution and the different burning stages a star goes through depend on its initial mass (see Fig. 5.2). All stars spend most of their lifetime in the so-called main-sequence fusing hydrogen to helium. Low-mass stars like our Sun stay in the main sequence for billions of years, after which they evolve to red giant stars via helium burning. In the end, low-mass stars will eject their outer envelopes (planetary nebulae) and leave a hot white dwarf (WD) as a remnant. Massive stars evolve much faster, on the timescales of millions of years. They go through a handful of nuclear-burning stages in addition to hydrogen and helium burning. Core-collapse supernovae resulting from massive stars spill out freshly produced nuclear material into the interstellar space and leave either a neutron star (NS) or a black hole (BH) as their remnant.

The evolution of intermediate-mass stars depends on the ignition of their core. Stellar remnants can further lead to nucleosynthesis in explosive conditions, either via accreting material from, or merging with, a companion star.

In astrophysics, all elements heavier than helium are called metals. Their fraction in stellar matter, metallicity, increases over time in the Universe. This change of elemental and isotopic abundances over time is called chemical evolution and is sometimes referred to as the life cycle of matter. Material synthesised during stellar evolution is distributed to the interstellar medium via mass lost in the form of planetary nebulae or in explosive events, such as supernovae. This enriches the gas clouds in which new stars will form, which will have a higher metallicity than the previous stars. Massive stars evolve much faster than low-mass stars and therefore play an essential role in producing the heavier elements that polluted the early Universe. Despite their slow stellar evolution, low- and intermediate-mass stars are more abundant than massive stars and therefore also have an important contribution to the chemical enrichment at later times.

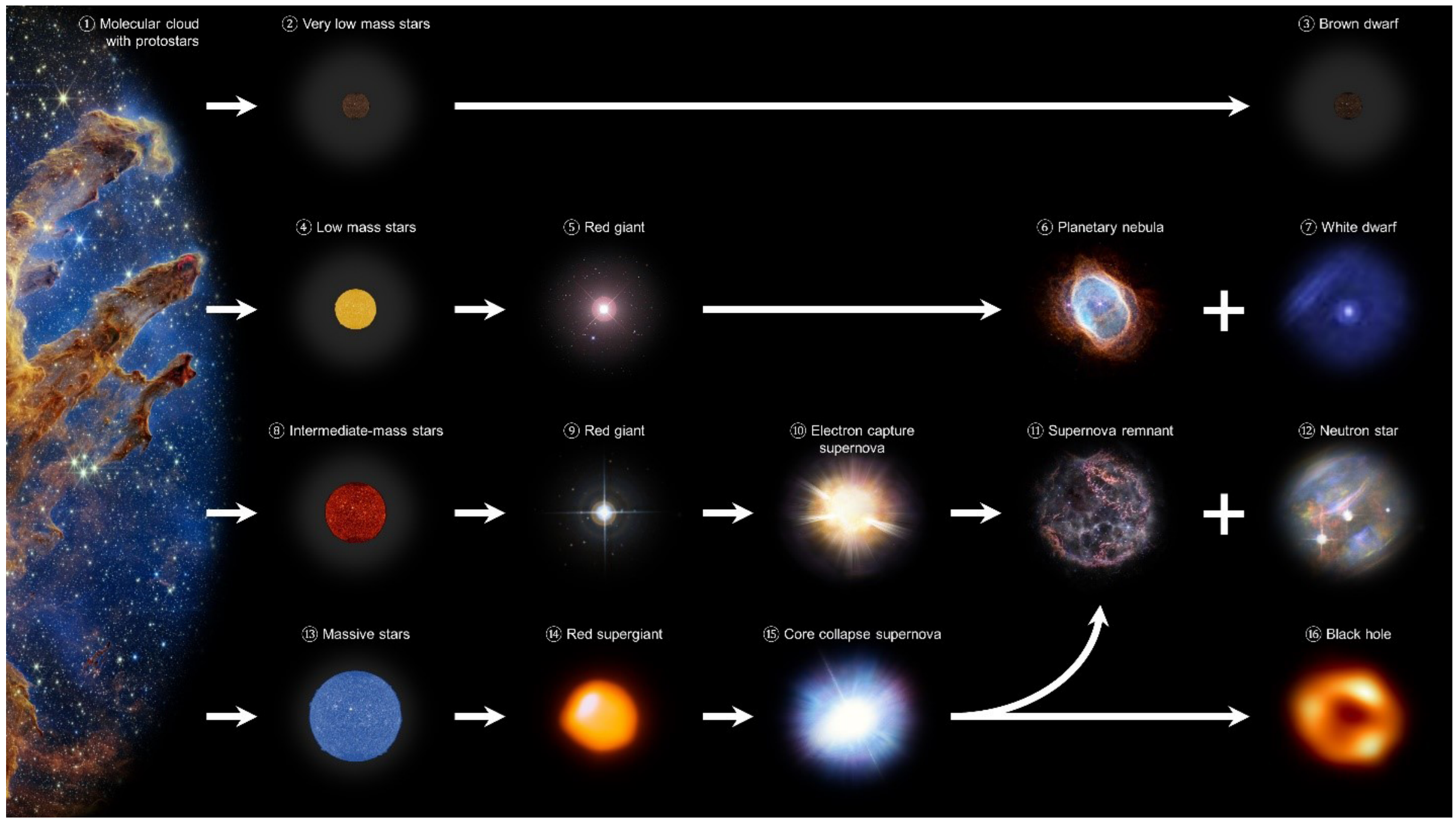

***Fig. 5.2: Stellar evolution for low- and intermediate-mass and massive stars. Recent research suggests that intermediate-mass stars can also evolve toward thermal explosion and a white dwarf. Credits:** ① NASA, ESA, CSA, STScI, J. DePasquale, A. Koekemoer, A. Pagan: Pillars of Creation (NIRCam Image). ②, ③, ④, ⑧, ⑬ NASA, GSFC, SDO: Small Prominences, **adapted in colour.** ⑤ BAO, F. Espenak**: Arcturus.** ⑥ NASA, ESA, CSA, STScI: Southern Ring Nebula (NIRCam Image). ⑦ NASA, ESA, H. Bond, M. Barstow: The Dog Star, Sirius A, and its white dwarf companion. ⑨ DSS2, D. Ford: R/B-band color composite of Iota Sculptoris. ⑩ CTIO, SOAR, NOIRLab, AURA, A. M. Geller: Artist's interpretation of the calcium-rich supernova 2019ehk. ⑪ NASA, ESA, CSA, STScI, D. Milisavljevic, I. De Looze, T. Temim: Cassiopeia A (NIRCam Image). ⑫ ESA, Hubble, NASA, M. Zamani: Moving heart of the Crab Nebula. ⑭ ALMA, ESO, NAOJ, NRAO, E. O'Gorman, P. Kervella: Betelgeuse captured by ALMA. ⑮ NOIRLab, NSF, AURA, M. Garlick, M. Zamani: Artist's impression of a Luminous Fast Blue Optical Transient. ⑯ ETHC: First image of our black hole*

The Sun, with an age of 4.6 billion years, has a rich spectrum of elements. While most astronomical observations yield abundances for elements, the measurements of individual isotopic abundances require high spectral resolution. Otherwise, they are inferred based on their known abundance ratios in the solar system and from cosmic rays, which sample abundances from the Galaxy. Abundance data based on spectroscopic observations of the Sun, mass spectroscopy of solar-wind particles and cosmic rays, as well as primordial meteorites agree well with each other overall, with a few notable differences. The solar-system abundance pattern (see Fig. 5.3) has distinct features. Elements up to the iron peak can be synthesised via stellar fusion reactions, and many lighter elements are produced in exploding white dwarfs in binary systems (see Fig. 5.1). Elements above A~60 are mainly produced via slow (s) and rapid (r) neutron-capture processes; however, there are also other processes that contribute. The different astrophysical processes on the chart of nuclides and the required nuclear data to better understand these processes are summarised in Box 5.1.

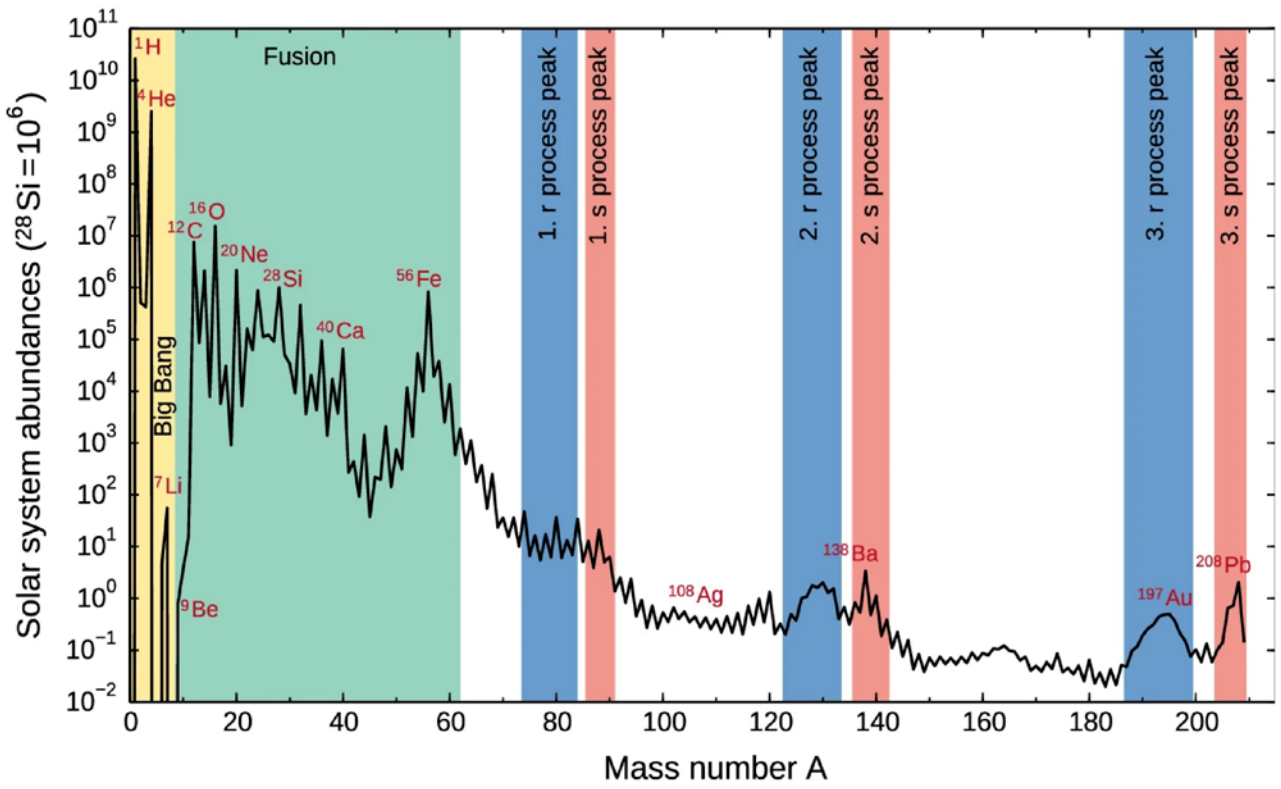

*Fig. 5.3: Solar system abundances based on Lodders (2021) as a function of mass number A (A=Z+N), with Z being the number of protons and N the number of neutrons), normalised to Si at 10$^6$. Figure adapted from Arcones & Thielemann, Astron. Astrophys. Rev. 31, 1 (2023).*

## Box 5.1: Nuclear astrophysics: A field with high demand for nuclear physics data

**Nuclear physics plays a key role in astrophysics applications, in particular for nucleosynthesis processes as well as the composition and structural properties of neutron stars. Astrophysical processes along with the nuclear data required are shown in the enclosed figure 5.14. These include the Big Bang nucleosynthesis, production by Galactic cosmic rays, the intergalactic propagation of ultra-high energy cosmic rays (UHECR), hydrostatic and explosive burning stages of stellar evolution (from H- to Si-burning), the rapid proton-capture in x-ray bursts or exploding massive stars as well as the different nucleosynthesis processes responsible for the production of elements heavier than iron, such as the slow (or s-process), intermediate (i-process) and rapid (r-process; potentially preceded by an α-capture process) neutron-capture processes, as well as the p-process responsible for the neutron-deficient stable nuclei. Finally, it also includes the composition and structural properties of the crust of neutron stars for which the nuclear equation of state (EoS) is a key ingredient (as it is for the explosion of massive stars as supernovae).**

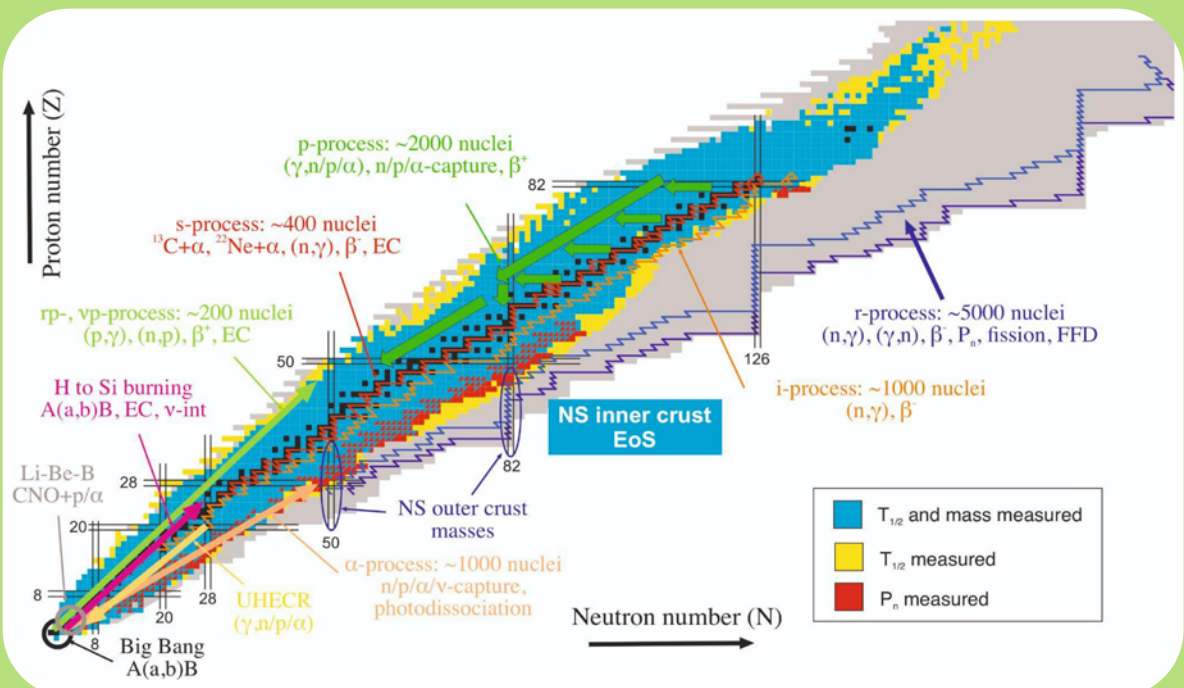

*Fig. 5.14: Chart of nuclides showing astrophysical processes and relevant nuclear data. The black squares correspond to stable or very long-lived nuclei, the blue and red squares to the nuclei for which, respectively, atomic masses and/or beta-delayed neutron emission probabilities ($P_n$) have been measured. Typically, for nuclei with measured masses, the half-lives ($T_{1/2}$) are also known. Yellow squares indicate nuclei with measured $T_{1/2}$ but unknown masses. The grey region illustrates the reach of bound nuclei as calculated using a mass model.*

## Emerging opportunities with new observations require nuclear data

Progress in nuclear astrophysics is accelerated by observations and increased computing power and methods for astrophysical simulations. The first multimessenger observations from a binary neutron star merger GW170817 and its kilonova yielded unique data for the r-process and the origin of the heaviest elements, and new multimessenger observations are anticipated. The neutron-star and black-hole gravitational-wave observations by the LIGO (Laser Interferometer Gravitational-Wave)/Virgo collaborations, as well as the Neutron Star Interior Composition Explorer (NICER) measurements from pulsars, constrain the nuclear equation of state (EoS) and challenge stellar models. The James Webb Space Telescope (JWST), launched in December 2021, has already provided tantalising observations of redshifted galaxies in the very early Universe, showing that stars and galaxies were forming and evolving earlier than expected. Undoubtedly, JWST will rapidly expand our understanding of the very early Universe as well as provide new observations from astrophysical objects with unprecedented precision. Large spectroscopic surveys in the Milky Way are also enlarging the observed parameter space by unveiling rare stellar populations.

Astrophysical models require a broad range of nuclear physics data (see Box 5.1). Nuclear physics provides essential inputs to the modelling of a large number of astrophysical processes and phenomena, associated with different astrophysical sites. Ideally, the reaction rates for these different processes should be based on experimental data; however, theoretical models are fundamental in providing the various predictions needed, for example, for the hundreds of experimentally unknown nuclei involved in the r-process. In the following, we will describe the different astrophysical sites and their rich nucleosynthesis processes, focusing on challenges and opportunities for nuclear physics in astrophysics.

## Big Bang nucleosynthesis

Primordial nucleosynthesis, also known as Big Bang Nucleosynthesis (BBN), occurred in the first 20 minutes after the Big Bang. BBN is one of the three main pieces of evidence for Hot Big Bang cosmology, probing very early cosmological times. In particular, BBN predicts the primordial abundances of light nuclides, such as deuterium, $^3$He, $^4$He, and $^7$Li, as a function of a single free parameter η (the baryon-to-photon ratio), provided that the neutron lifetime, the number of neutrino families, and approximately 12 key nuclear reactions that were active during BBN are known and fixed. The agreement between primordial abundances predicted by BBN and the same abundances obtained from astronomical observations is excellent for $^{2}$H and $^{4}$He but not for $^{7}$Li. Furthermore, the value of the η parameter that can be inferred from a combination of astronomical observations and BBN predictions is in excellent agreement with a completely independent estimate via Cosmic Microwave Background (CMB) measurements, probing an epoch 380,000 years after the Big Bang. Overall, BBN theory is a resounding success for science. However, the precision for the BBN prediction still lags behind the CMB precision and can affect the conclusions.

Among primordial isotopic abundances, deuterium provides the most stringent constraint to the baryon-to-photon ratio η. Astronomical observations have reached per cent-level precision (D/H = (2.527 ± 0.030) × $10^{-5}$ from Cooke et al. 2018), and a comparable precision is now required from nuclear input parameters. During BBN, deuterium is produced by p(n,γ)d and is mainly destroyed by the d(p,γ)$^{3}$He, d(d,p)t and d(d,$^{3}$He)n reactions. The d(p,γ)$^{3}$He reaction cross section is used to carry most of the uncertainty on BBN predictions of the deuterium abundance. Recently, the LUNA collaboration determined the d(p,γ)$^{3}$He cross section with 3% precision in the INFN Gran Sasso National Laboratory (LNGS), see Fig. 5.4. A similar precision is now also required on the d(d,p)t and d(d,$^{3}$He)n cross sections. These high-precision measurements are crucial to fully exploring the interplay between CMB and BBN.

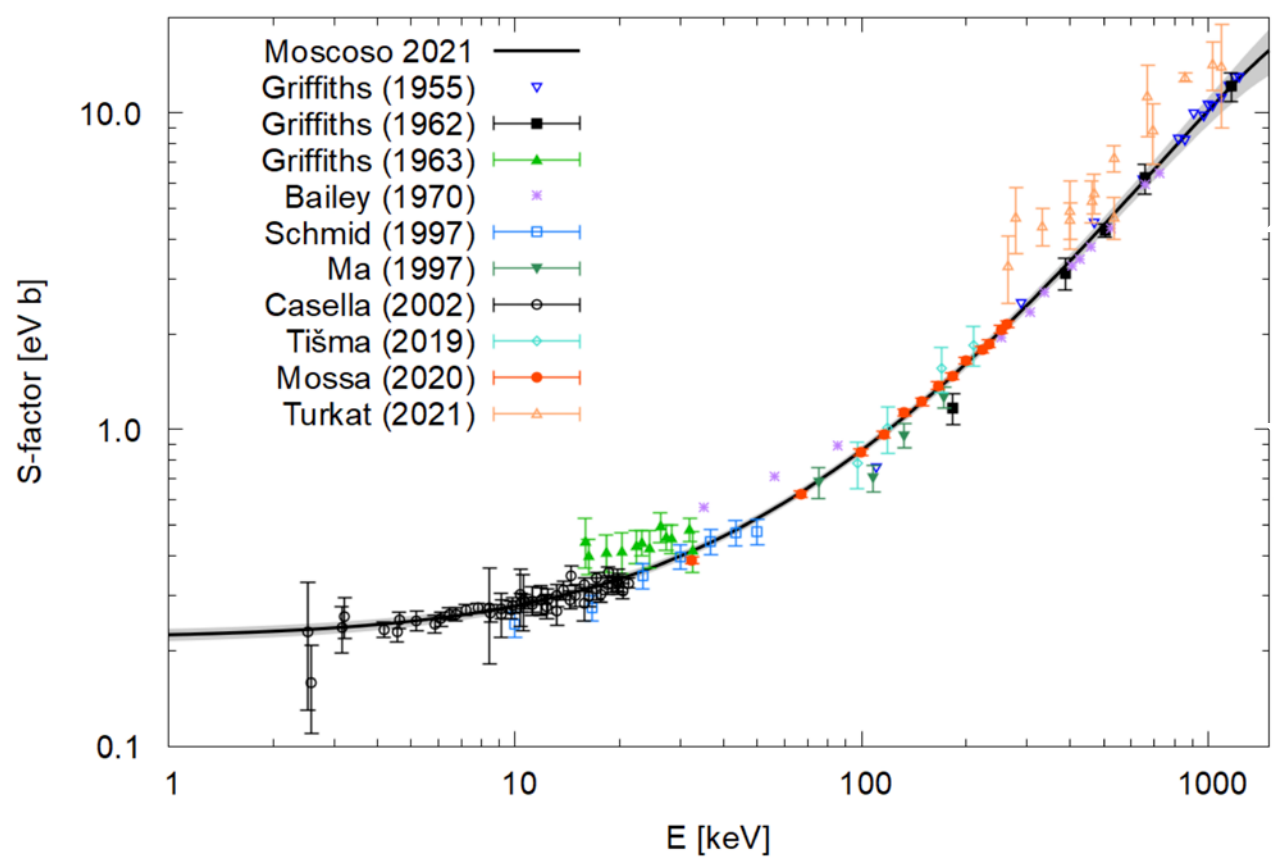

*Fig. 5.4: The astrophysical S factor, of the d(p,γ)$^{3}$He reaction. Results from various experiments are plotted together with a fit (black solid line).*

## Precise nuclear physics measurements point to other solutions for the lithium problem

While BBN models have been incredibly successful, one unresolved issue – the lithium problem – persists. The $^{7}$Li abundances predicted by BBN are three times higher than abundances measured in pristine metal-poor halo stars. Over the last decade, the nuclear astrophysics community has worked to constrain the cross sections of all reactions producing and disrupting $^{7}$Li, using both direct and indirect techniques. Recent examples are the $^{7}$Be(n,p) and $^{7}$Be(n,α) reactions that were studied both directly at CERN n_TOF and indirectly with the Trojan Horse Method (THM), ruling out some possible nuclear solutions to the lithium problem. Looking at the future, potential solutions may be found in other fields such as, for example, stellar observations, or stellar models, or even physics beyond the Standard Model of particle physics. Accurate nuclear data will be needed to prove or disprove their predictions, and also in preparation for the ESO's Extremely Large Telescope era when precise lithium measurements in other galaxies could add more pieces to the puzzle.

## Toward high-precision measurements for BBN

Looking to the future, ultra-high precision measurements (<<5% overall uncertainty) of reaction cross-sections playing a central role in BBN will be critical. Such levels of precision are technically very challenging. This demanding objective will require both a quantitative improvement of existing methods and the development of new techniques to carry out nuclear reaction studies. For example, nuclear burning (*e.g.* d(d,p)t and d(d,$^{3}$He)n discussed above) could be studied not only in underground labs but also via laser-induced plasmas. Techniques, such as the Coulomb Explosion of cryogenically cooled gas mixtures, allow the generation of plasma with a Maxwell-Boltzmann distribution at astrophysical temperatures. Preliminary studies have already been carried out worldwide (e.g. OMEGA and NIF in the USA). The petawatt laser system at ELI-NP opens new opportunities for such studies in Europe. In addition, developments at PALS-Praha (Czech Rep.), FLAME at LNF, and I-LUCE at LNS are promising. In addition to shedding light on BBN, these facilities may also play a role in other astrophysical contexts, *e.g.*, stellar nucleosynthesis.

Low-energy ion storage rings affecting recirculating stable or radioactive ion beams on a pure, windowless, ultra-thin internal target also offer unique new opportunities to study nuclear fusion in BBN. Europe is a world leader in storage-ring science with the GSI/FAIR laboratory (Germany) and its unique, newly commissioned ultra-low energy CRYRING. Storage rings have the potential to be revolutionary beyond BBN reactions, for explosive nucleosynthesis and beyond (see next sections). In this context, a low-energy storage ring at an ISOL facility, as proposed for CERN/ISOLDE, would offer the highest beam intensities for stored radioactive beams.

# Nucleosynthesis in low- and intermediate-mass stars

## Hydrogen burning - pp chains and CNO cycles

All stars begin their life cycle by burning hydrogen into helium. This so-called main-sequence phase takes the majority of the stellar lifetime. Compared to massive stars, low- and intermediate-mass stars (M <~ 8 $M_{Sun}$) spend a much longer time in the main sequence and contribute to the chemical evolution of the interstellar medium over longer timescales, in the order of billions of years. The Sun is an example of such a low-mass star. It creates 99% of its energy via the so-called pp chains, which effectively convert four protons into one $^{4}$He nucleus; around 1% of the released energy is produced via CNO cycles. The CNO cycles play a larger role in stars with higher metallicity than the Sun and become dominant in stars with M > 1.5 solar masses, which reach higher temperatures. While the physics related to the pp chains is rather well constrained, the $^{3}$He(α,γ) and $^{7}$Be(p,γ) reactions in the pp-chains and the CNO cycles require further experiments to better constrain the solar abundances.

The recent detection of solar CNO neutrinos by the Borexino experiment in LNGS (see Fig. 5.5) marked a big step forward in the investi-gation of the properties of the solar core. In conjunction with $^{7}$Be and $^{8}$B neutrinos, CNO neutrinos provide new constraints to the Standard Solar Model and give insights into the metallicity of the solar core, assuming that the cross sections of all reactions of the CNO cycle are known to high accuracy. The main source of uncertainty is the cross section of the $^{14}$N(p,γ)$^{15}$O reaction, which is not directly accessible at solar energies with current facilities. New measurements helping to improve extrapolations are still required, both with direct and indirect approaches. This reaction is part of the LUNA-MV scientific programme at the Bellotti Ion Beam Facility of LNGS.

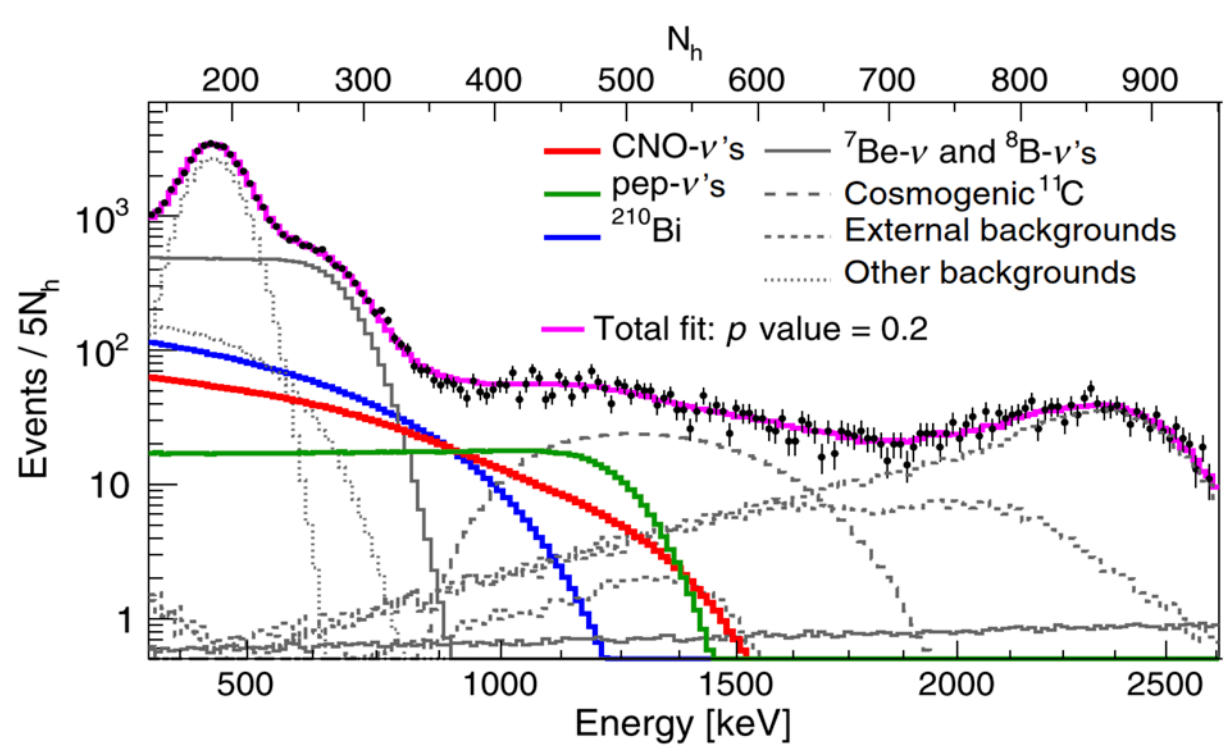

*Fig. 5.5: Spectral fit of the Borexino Phase-III solar neutrino flux data. Figure adopted from Appel et al. Phys. Rev. Lett. 129, 252701 (2022).*

## Helium burning

Hydrogen-burning produces $^{4}$He nuclei in the stellar core. After hydrogen in the core has been exhausted, the helium core contracts and temperature increases. Helium burning starts with the so-called triple-alpha process, which is a quasi-simultaneous fusion of three $^{4}$He nuclei into $^{12}$C. The triple-alpha reaction is dominated by a 0+ resonance known as the Hoyle state, which is essential to account for the stellar carbon abundance. Detailed studies on the excited levels of $^{12}$C have yielded important information on the role of other potential resonances. While the resonances in $^{12}$C are rather well constrained, their radiative widths and therefore the decay to the ground state, have remained a focus of experimental and theoretical efforts. Taking into account the importance of the Hoyle state, such refined revisions and measurements and precise ab-initio calculations combined with machine learning techniques are indispensable. Helium burning continues from $^{12}$C via further alpha captures to $^{16}$O and $^{20}$Ne. Of these, $^{12}$C(α,γ)$^{16}$O is of crucial importance in nuclear astrophysics (see Box 5.2) as it has a strong impact on the abundances of heavier elements in many astrophysical sites. $^{12}$C(α,γ) is presently under study at the ERNA recoil separator. In the long term, it could also be investigated in new underground facilities, such as the Bellotti Ion Beam Facility at LNGS, and Felsenkeller.

### Box 5.2: Helium burning and the critical reaction in nuclear astrophysics: $^{12}$C(α,γ)$^{16}$O

**$^{12}$C(α,γ)$^{16}$O is one of the most important reactions for nuclear astrophysics as the carbon and oxygen abundances are used in almost all nucleosynthesis models for different astrophysical environments. It could also help in the interpretation of astrophysical observations. Despite its importance, the reaction has remained poorly constrained as it is complicated and the cross section decreases steeply toward lower energies, being around $10^{-17}$ b at helium-burning energies (around 0.3 MeV). This has been beyond the reach of experiments. Different compilations for the reaction rate differ from each other up to around 50%. Efforts should be made to reach lower energies and improve the precision of this crucial reaction.**

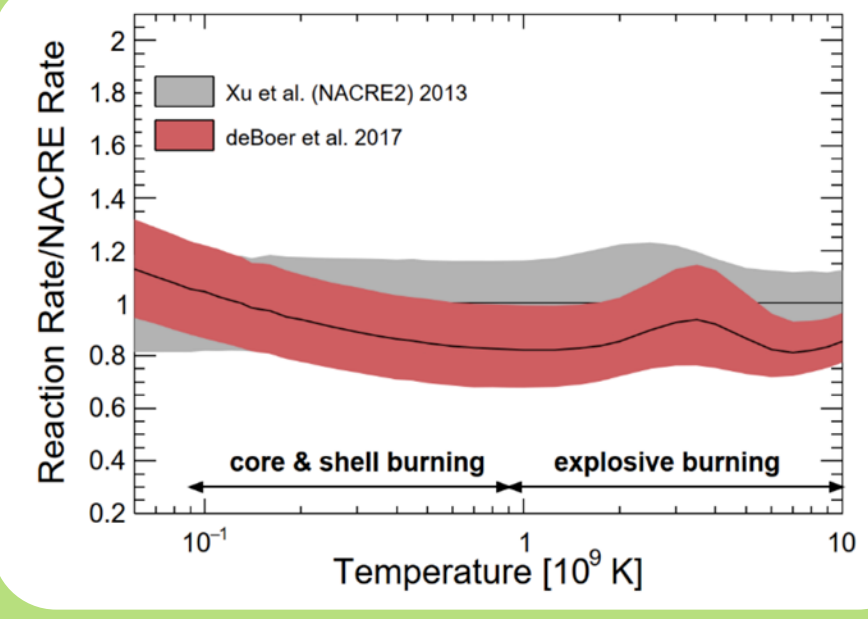

*Fig. 5.15: Comparison of the $^{12}$C(α,γ)$^{16}$O reaction rate and its uncertainty calculated in de Boer et al. (2017) (red band, dash-dotted central line) and from the NACRE2 compilation (grey band, solid central line). Temperature regions relevant for core and shell helium burning and for explosive helium burning are indicated.*

## Asymptotic Giant Branch phase

Low- and intermediate-mass stars finish their evolution as compact white dwarfs and predominantly eject nucleosynthesis products via stellar winds during the asymptotic giant branch (AGB) phase. In this phase, burning in hydrogen and helium shells alternate. Depending on the initial stellar mass, during the AGB phase, light elements such as carbon, nitrogen, fluorine and sodium can be made by charged-particle reactions activated at the bottom of the AGB envelope or in the He intershell region. Additionally, AGB stars are a fundamental generator of heavy elements beyond iron via the slow neutron-capture process (s-process).

## Hot bottom burning produces light elements in AGB stars

Stars with an initial mass larger than ~4 $M_{Sun}$ may be characterised during the AGB phase by a high enough temperature at the bottom of the convective envelope to burn hydrogen over a timescale comparable with the mixing timescale. This causes the so-called hot bottom burning (HBB), which efficiently produces lithium and converts $^{12}$C into $^{14}$N through a partial CN cycle. Alpha-capture reactions on $^{6,7}$Li and $^{10,11}$B could lead to the production of CNO elements in first-generation stars, starting from primordial material. These reactions might help to understand lithium depletion in stars and could be tackled at LUNA-400 kV, and possibly also at other small-scale facilities.

Over the past years, big efforts have been made to constrain the cross sections of CNO, NeNa and MgAl cycle reactions at astrophysical energies, using both direct and indirect methods. Nowadays, the largest uncertainty is carried by (p,α) reactions on $^{17}$O, $^{18}$O, $^{19}$F, $^{23}$Na and $^{27}$Al. Among them, $^{23}$Na(p,α) and $^{27}$Al(p,α) will be tackled in the next few years at LUNA-400 kV. Moreover, the $^{23}$Na(p,α) will be studied indirectly using the THM.

## Neutron sources in AGB stars

AGB stars are the major site of the slow neutron-capture process (s-process), which produces around half of the heavy-element abundances. The s-process takes place at relatively low neutron densities (~$10^{6}$ -$10^{12}$ $cm^{-3}$) and temperatures (~0.1-0.4 GK). Such conditions are found in the hydrogen and helium shells around the inert carbon-oxygen core in thermally-pulsing AGB stars. Hydrogen-shell burning in AGB stars lasts for about 10,000 years and accumulates its burning ashes (helium) in the helium shell underneath. This eventually triggers explosive helium-shell burning ("He flash"), which lasts only around 50 years. The helium burning causes the hydrogen shell to expand and cool, and the hydrogen shell burning is seized. It is reignited after the helium-shell burning is finished and the star contracts.

Recent high-resolution observations of barium stars have confirmed that $^{13}$C nuclei are the major neutron source in AGB stars. The $^{13}$C(α,n)$^{16}$O reaction takes place in a so-called "carbon pocket", located in the top layers of the helium shell. The protons diffused into the helium shell are captured by $^{12}$C abundant in the shell, followed by the decay of the produced $^{13}$N nuclei into $^{13}$C. The exact mechanism responsible for the diffusion of protons across the convective boundary is still a matter of debate and might require multidimensional hydrodynamic simulations. Different convective boundary mixing processes in the stellar models lead to remarkably different sizes of the $^{13}$C pocket, directly impacting the total amount of s-process elements produced. The reaction rate of $^{13}$C(α,n)$^{16}$O has recently been measured with high precision in the Gamow window in the Bellotti IBF underground facility and at JUNA as well as by THM using the $^{13}$C($^{6}$Li,n$^{16}$O)d reaction.

A complementary minor neutron source in AGB stars is the $^{22}$Ne(α,n)$^{25}$Mg reaction activated during the He-flash. At the elevated temperatures of the thermal pulse, $^{22}$Ne is produced starting from $^{14}$N through a sequence of alpha capture reactions: $^{14}$N(α,γ)$^{18}$F(β+)$^{18}$O(α,γ)$^{22}$Ne. Significant uncertainties are still affecting the $^{22}$Ne(α,n)$^{25}$Mg reaction. Indeed, this is reflected in the contradictory results of the latest redetermination of both the $^{22}$Ne(α,n)$^{25}$Mg and the $^{22}$Ne(α,γ)$^{26}$Mg reaction rates, competing in the temperature range of interest for carbon burning in massive stars (~1 GK), as shown in Fig. 5.6. A good scientific opportunity to settle these discrepancies could be represented by future direct measurements of these reactions in underground laboratories, taking advantage of the extremely reduced cosmic-ray-induced background. In addition, relevant reactions could be studied in inverse kinematics.

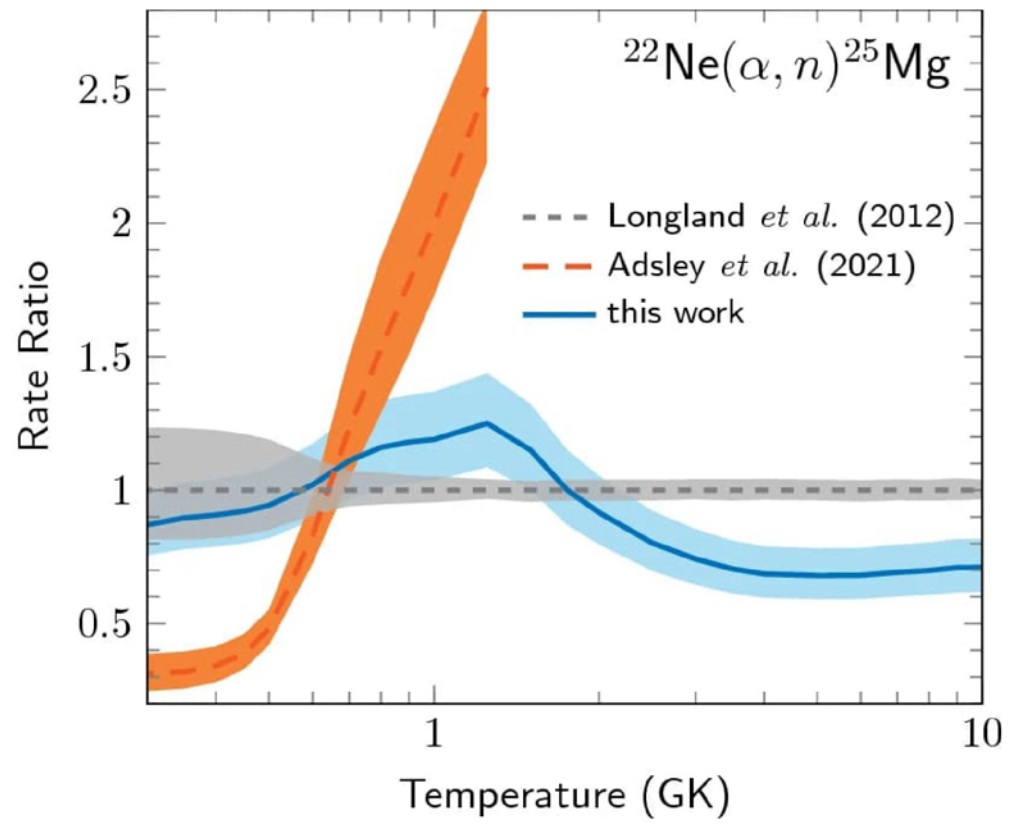

*Fig. 5.6: Thermonuclear reaction rate for the s-process neutron source reaction $^{22}$Ne(α,n)$^{25}$Mg by Wiescher et al. 2023 ("this work", blue solid line). Figure adopted from Wiescher et al., Eur. Phys. J. A 59, 11 (2023).*

## Slow neutron-capture process (s-process) produces heavier elements

The s-process proceeds close to stable nuclei as the neutron captures are slow and usually cannot compete with beta decays except at branching points, such as $^{85}$Kr, $^{95}$Zr, and $^{134}$Cs. The abundances of subsequent isotopes in the reaction chain are highly sensitive to the ratio of neutron-capture and β-decay rates. If these rates are known, the neutron density can be inferred from the abundance ratio. Furthermore, if the neutron density is already known, neutron-capture cross-section measurements can be used to determine the effective stellar half-life and thus, the temperature of the star. High-precision laboratory measurements of the isotopic anomalies present in meteoritic silicon carbide grains from AGB stars as well as spectroscopic observations of barium stars are very important to constrain the s-process conditions.

The main challenge in neutron-capture measurements of s-process branching points is related to the radioactivity of the target. The recent measurement of $^{79}$Se(n,γ) at CERN n_TOF used a sample of only 1018 atoms with several ~10 MBq activity. Other recent measurements include $^{94}$Nb(n,γ), $^{147}$Pm(n,γ) or $^{204}$Tl(n,γ), all of them achieved with only tiny sample amounts and large intrinsic sample radioactivity. Such measurements on highly radioactive samples have eventually become feasible thanks to important efforts on sample production, commonly involving the high reactor flux of ILL-Grenoble and the radiochemistry laboratory of PSI-Villigen, as well as upgrades in the CERN n_TOF facility and a significant innovation in the related measuring devices and techniques (see Fig. 5.7). However, experimental limitations in most of these measurements still hinder the cross-section measurement in the full stellar energy range of interest, i.e. up to 100 keV. Short-lived species, such as $^{85}$Kr, $^{95}$Zr or $^{134}$Cs, still remain impossible to access with the most advanced techniques.

Further developments in the next few years will help to overcome many of the above-mentioned limitations and provide access to most of the remaining branching-point nuclei where no data exists yet. Upgrades of the n_TOF facility, such as the new NEAR station for activation measurements, are expected to provide invaluable information in the higher neutron energy range (10 - 100 keV). Joint efforts with RIB facilities, such as ISOLDE, and dedicated laboratories for isotope separation and sample preparation, are also expected to significantly advance the field. In the future, experiments using inverse kinematics might open new avenues to enable the first direct neutron-capture cross-section measurements on many radioactive isotopes.

In addition to the branching points, nuclei produced only via the s-process (s-only nuclei) are of special interest. Many of these s-only or branching-point nuclei have remained inaccessible with existing techniques, or the uncertainties in their cross sections are still too large for a reliable astrophysical interpretation. The measurements of s-only isotopes like $^{82}$Kr, $^{86}$Sr, $^{100}$Ru or $^{198}$Hg would help to further refine and benchmark more consistently the performance of state-of-the-art AGB stellar models. Further measurements on radioactive isotopes, e.g., $^{85}$Kr, $^{95}$Zr, $^{185}$W, $^{186}$Re, could provide new and more accurate constraints on the physical conditions of the s-process, as well as for nuclear cosmochronology.

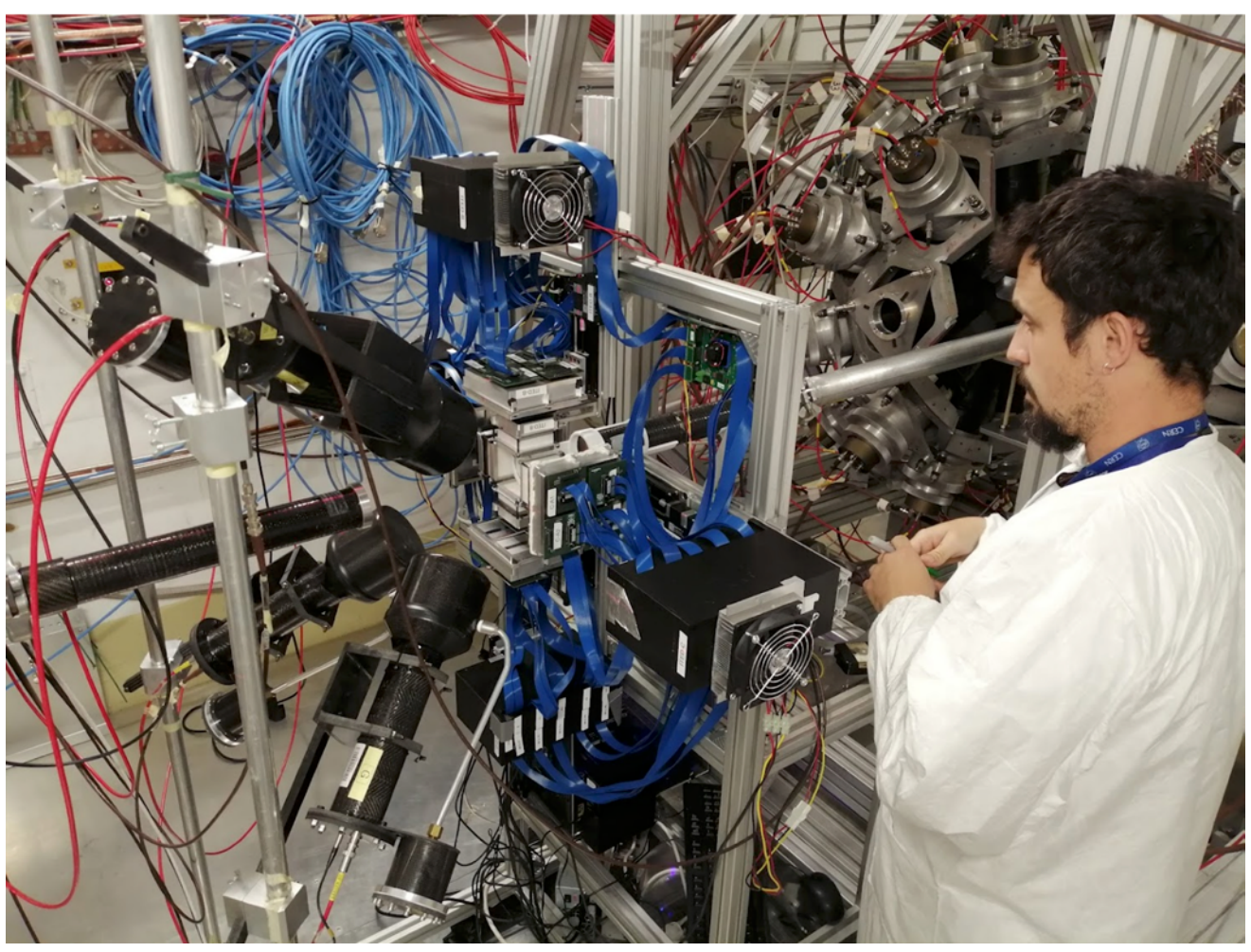

*Fig. 5.7. Experimental set-up at CERN n_TOF. Array of four C6D6 detectors (left) with gamma-ray imaging capability (middle), and total-absorption calorimeter with 40 BaF2 crystals (right).*

## Intermediate neutron-capture process (i-process)

The intermediate neutron-capture process (i-process) takes place at intermediate neutron densities and traverses between the s- and r-process paths on the nuclear chart (see Box 5.1). It was proposed in 1977 by Cowan and Rose but did not receive much attention until 1999 when Asplund et al. observed extreme enhancements in strontium, yttrium and zirconium elements in the post-AGB star called Sakurai's object. The observed element distribution was very difficult to explain in terms of the traditional s-process. Currently, more and more observations point towards the existence of the i-process, such as Carbon-enhanced metal-poor (CEMP) stars in the Galactic halo showing signatures of both s- and r-elements. As the s- and r-processes take place in different astrophysical sites and on different time scales, it is hard to explain why such old stars have both barium (s-process) and europium (r-process) elements.

The modelling of the i-process is in its infancy and nuclear inputs are required to comprehend it, especially concerning the unstable nuclei adjacent to the valley of beta stability. Nuclei involved in the i-process are more accessible experimentally than those in the r-process. Most of the masses and beta-decay half-lives can be measured with existing and new radioactive beam facilities. Major missing nuclear data are the neutron-capture rates for neutron-rich nuclei a few neutrons away from stability. Here, the nuclear physics community has a great opportunity to provide indispensable data through direct neutron-capture measurements in future storage rings combined with a neutron source. In the meantime, indirect methods like the beta-Oslo method, inverse-kinematics Oslo method, the surrogate method and the shape method can help to constrain the neutron-capture rates. Also, the impact of isomers (astromers, see Box 5.3) remains completely uninvestigated. The need for detailed spectroscopy of neutron-rich nuclei is therefore potentially of vital importance.

# Massive stars and their explosions

Massive stars ($M > 8 – 10\ M_{Sun}$) have much shorter lifetimes, of the order of millions of years. Stars with initial masses < 20 – 30 $M_{Sun}$ explode as type II, or core-collapse, supernovae (CCSN) and likely leave neutron stars as remnants. Stars above this mass range that have lost their H (He) envelopes are thought to be the progenitors of type Ib and Ic supernovae. Whether the latter explosions leave a neutron star or a black hole as remnants is debated, our knowledge of post-main sequence mass loss and matter fall-back during the blast being incomplete. Pair instability supernovae have been predicted for population III stars within 130 – 250 $M_{Sun}$. They are driven by the production of electrons and positrons during the energetic collisions between nuclei and gamma rays, which result in a temporary reduction of the internal radiation pressure that supports the core of a supermassive star against collapse.

These stars can enable higher core temperatures, i.e. more advanced fusion stages, in addition to the hydrogen (H) and helium (He) fusion. While most of their stellar lifetimes are still spent on core H and He burning, these differ from that of the low- and intermediate-mass stars in a number of aspects: i. H-burning is fully dominated by the CNO cycles in massive stars; ii. He burning produces more oxygen than in low-mass stars. The subsequent burning stages (carbon to silicon) occur on increasingly shorter timescales. Several techniques are used to identify supermetal-poor stars more efficiently in our Galaxy (e.g. Gaia Data Release 3) and provide invaluable constraints to massive star nucleosynthesis.

## Pre-explosive evolution

### Weak s-process in massive stars

Weak s-process takes place in massive stars mainly during core He-burning and is responsible for the synthesis of light elements with A < 90 (C-shell burning also contributes, to some extent, to the s-process). The main neutron source for this weak s-process is the reaction $^{22}$Ne(α,n)$^{25}$Mg. This rate is affected by significant uncertainties, in particular in the temperature range of interest for C-burning in massive stars, T ~ 1 GK. Direct measurements of this reaction could also help to improve our understanding of s-process element production in massive stars.

### Carbon, neon, oxygen and silicon burning in massive stars

After H- and He-burning, burning stages are dominated by fusion reactions of intermediate-mass nuclei, like $^{12}$C and/or $^{16}$O. The reaction probability is extremely low (picobarn levels) due to the high Coulomb barrier through which the nuclei have to tunnel. Precise experimental determination of the fusion cross-sections is thus highly challenging. Nuclear structure effects (e.g. cluster resonances in $^{12}$C+$^{12}$C) may make the situation even more complex.

To accomplish high-accuracy measurements of C fusion at extremely low signal-to-background ratios, direct coincident gamma-particle detection techniques as well as indirect experiments have been performed, notably at small-scale facilities, leading to first high-precision results in the astrophysical region. A broad overlap of both approaches with future measurements (e.g. LUNA collaboration, STELLA) will contribute to minimising the systematic uncertainty. The $^{12}$C+$^{16}$O reaction has more open exit channels and is even more challenging. First-principle calculations give precise information on cluster states in the compound system. Experiments capable of measuring these states in heavy ion fusion reactions for C-burning need to be designed to determine the role of such resonances, which could be massive. The role of extreme branching of the final states of the $^{12}$C+$^{12}$C reaction at resonance energy and the existence of stronger resonances than previously thought was investigated in stellar dynamics simulations. Recoil mass separators (e.g., ERNA at CIRCE) are envisaged to investigate $^{12}$C(α,γ), $^{12}$C+$^{12}$C, and $^{12}$C+$^{16}$O reactions.

Neon burning follows C-fusion when the core temperature reaches 1.2 GK. This stage is dominated by $^{20}$Ne(γ,α)$^{16}$O and $^{20}$Ne(α,γ)$^{24}$Mg. O-burning takes place when the core temperature reaches 1.5 – 2.6 GK and is dominated by $^{16}$O+$^{16}$O, a very challenging reaction. However, its cross section does not show as many resonances as $^{12}$C+$^{12}$C. Experiments have not yet reached the centre of the Gamow window for this reaction. Theoretical estimates are needed to extrapolate the rate to lower energies and better experimental data are needed. Si-burning is the final fusion stage in stars. It takes place when the core temperature reaches 2.7 – 3.5 GK and consists of a complex network of light particle captures and photodisintegration reactions. Toward the end of Si-burning, nuclear statistical equilibrium conditions are reached, under which abundances are largely determined by nuclear binding energies, but the neutron excess, density and temperature also affect the final abundances in the stellar core.

### Explosive evolution: Core-collapse supernovae

During Si-burning, a massive star continually increases its core mass, which is sustained by electron-degenerate pressure. But when the core exceeds the Chandrasekhar limit (~1.4 $M_{Sun}$), it has no other nuclear energy source available to maintain the pressure and the star experiences a gravitational collapse. During its early stages, the collapse is accelerated by two effects: i. electron captures onto nuclei remove the number of electrons that were contributing to the pressure; ii. at the achieved very high temperatures, iron peak nuclei are photodisintegrated into lighter and less stable species, removing the energy content that could have provided pressure. In less than a second, a core with a size of several thousand kilometres collapses into a proto-neutron star.

When the density reaches ~$10^{12}$ g cm$^{-3}$, the neutrino diffusion time becomes larger than the collapse time and neutrinos become trapped in a neutrinosphere. When the inner core reaches nuclear densities, it bounces and powers a shock wave that propagates outwards. This prompt shock loses energy rapidly by dissociating iron peak nuclei into free nucleons. When the shock reaches the neutrinosphere, additional electron captures on free protons also remove energy from the shock, giving rise to a strong burst of electron neutrinos. Ultimately, the shock stalls at ~100 – 200 km from the centre of the star, in the outer core. It is believed that this stalled shock is revitalised by neutrinos and antineutrinos that emerge from the hot and dense proto-neutron star. After the shock is reborn, the strong electron neutrino and antineutrino fluxes drive a continuous flow of protons and neutrons from the region near the proto-neutron star surface, known as neutrino-driven wind.

The exact mechanisms by which neutrinos give rise to a successful explosion are still debated. In many cases, computer models have been successful in reproducing the light curve and the production of elements during a CCSN once the shock has been revitalised.

### Radionuclides $^{26}$Al, $^{44}$Ti and $^{60}$Fe carrying information from massive stars and supernovae

$^{26}$Al ($T_{1/2}$ ~ 0.72 Myr) is an important radioisotope mainly produced in massive stars. The far-reaching significance of the cosmic abundance of $^{26}$Al includes the interpretation of its detection in the Galaxy via gamma-ray spectroscopy and its presence in meteoritic grains. In massive stars, $^{26}$Al is mainly synthesised through $^{25}$Mg(p,γ)$^{26}$Al. Destruction channels involve p- and n-capture reactions. The main uncertainties that affect $^{26}$Al production in massive stars involve nuclear reaction rates, in particular the $^{25}$Mg(p,γ)$^{26}$Al rate and its branching ratio to the ground state of $^{26}$Al, and the $^{26}$Al(p,γ)$^{27}$Si rate, and those related to stellar models: the size of the convective core, the mass-loss rates, and the mixing processes in the radiative zones of the stars. Neutrinos from the collapsing stellar core also affect $^{26}$Al production directly via the $^{26}$Mg($\nu_e$,e$^-$) reaction, and indirectly by providing additional protons for the $^{25}$Mg(p,γ) reaction. The contribution of neutrino-induced reactions to $^{26}$Al synthesis is very sensitive to the uncertain neutrino-energy spectrum. Direct measurements on $^{26}$Al(n,α) and $^{26}$Al(n,p) have been made at CERN n_TOF and EC-JRC, while additional indirect measurements (THM) are also planned.

COMPTEL mapped the Galactic $^{26}$Al (1.809 MeV) emission, revealing hotspots potentially associated with our Galaxy's spiral arms, the Cygnus region, the Vela SNR, and OB associations. INTEGRAL/SPI published the first resolved spectroscopy of this emission, verifying the Galaxy-wide distribution through measurement of the line's Doppler shift due to Galactic rotation. The integrated 1.809 MeV flux is believed to map CCSN and massive star wind activity in our Galaxy over the last $10^6$ yr, including >$10^4$ combined SN and Wolf-Rayet (WR) stars.

Whereas $^{26}$Al is thought to be ejected to the ISM both during the hydrostatic phase of massive stars and its final supernova, $^{60}$Fe ($T_{1/2}$ ~ 2.6 Myr) is assumed to result from the supernova only. Mapping both $^{26}$Al and $^{60}$Fe is a key science objective of the future NASA's Compton Spectrometer and Imager (COSI) mission. Comparison of the two emissions will allow us to determine the evolutionary stages of different regions of the Galaxy.

$^{60}$Fe has also been discovered in terrestrial and lunar samples demonstrating the clear exposure of Earth to recent (<10 Myr) cosmic explosions, presumably a close-by supernova event. The identification of extremely rare radioactive isotopes from recent nucleosynthesis events is possible only with the aid of dedicated accelerator mass spectrometry (AMS) facilities. Currently, the only facility worldwide that can perform such challenging measurements is the 14-million-volt tandem accelerator at the Australian National University (ANU). The construction of a European high-energy AMS facility would be essential.

$^{44}$Ti is another radionuclide that provides physical constraints for the explosion mechanism and information on the supernova interior physics. Similarly to $^{26}$Al, $^{44}$Ti can be detected from the characteristic gamma rays following its decay. Three-dimensional velocity maps of the $^{44}$Ti ejecta have been measured from the supernova remnant Cassiopeia A in the NuSTAR observing campaign. To date, Cas A has been the only galactic supernova remnant (SNR) where decay of $^{44}$Ti has been observed, both in its line at 1.157 MeV and in the lines at 68/78 keV of its beta-decay daughter $^{44}$Sc. In addition, INTEGRAL detected the 68/78 keV lines from SN1987A. COSI will search for $^{44}$Ti-rich galactic SNRs at the best sensitivity yet, which may result in the discovery of ~10 SNRs in $^{44}$Ti. Sensitivity studies on $^{44}$Ti and $^{56}$Ni production during CCSN indicate that many reaction rates need to be improved. These include $^{13}$N(α,p)$^{16}$O, $^{17}$F(α,p)$^{20}$Ne, $^{52}$Fe(α,p)$^{55}$Co, $^{56}$Ni(α,p)$^{59}$Cu, $^{57}$Ni(n,p)$^{57}$Co, $^{56}$Co(p,n)$^{56}$Ni, $^{39}$K(p,γ)$^{40}$Ca, $^{47}$V(p,γ)$^{48}$Cr, $^{52}$Mn(p,γ)$^{53}$Fe, $^{57}$Co(p,γ)$^{58}$Ni, and $^{39}$K(p,α)$^{36}$Ar.

Interestingly, the observed $^{44}$Ti has systematically implied a larger initial $^{44}$Ti mass than predicted by the standard CCSN models. Recently, three-dimensional CCSN models have provided more reliable nucleosynthesis predictions than spherically symmetrical (1D) models, particularly in Ti/Fe yields (see Fig. 5.8). However, some challenges remain (e.g. models of 12 – 14 $M_{Sun}$), which do not result in successful explosions. Observational data are increasing and help in turn to develop and constrain the models.

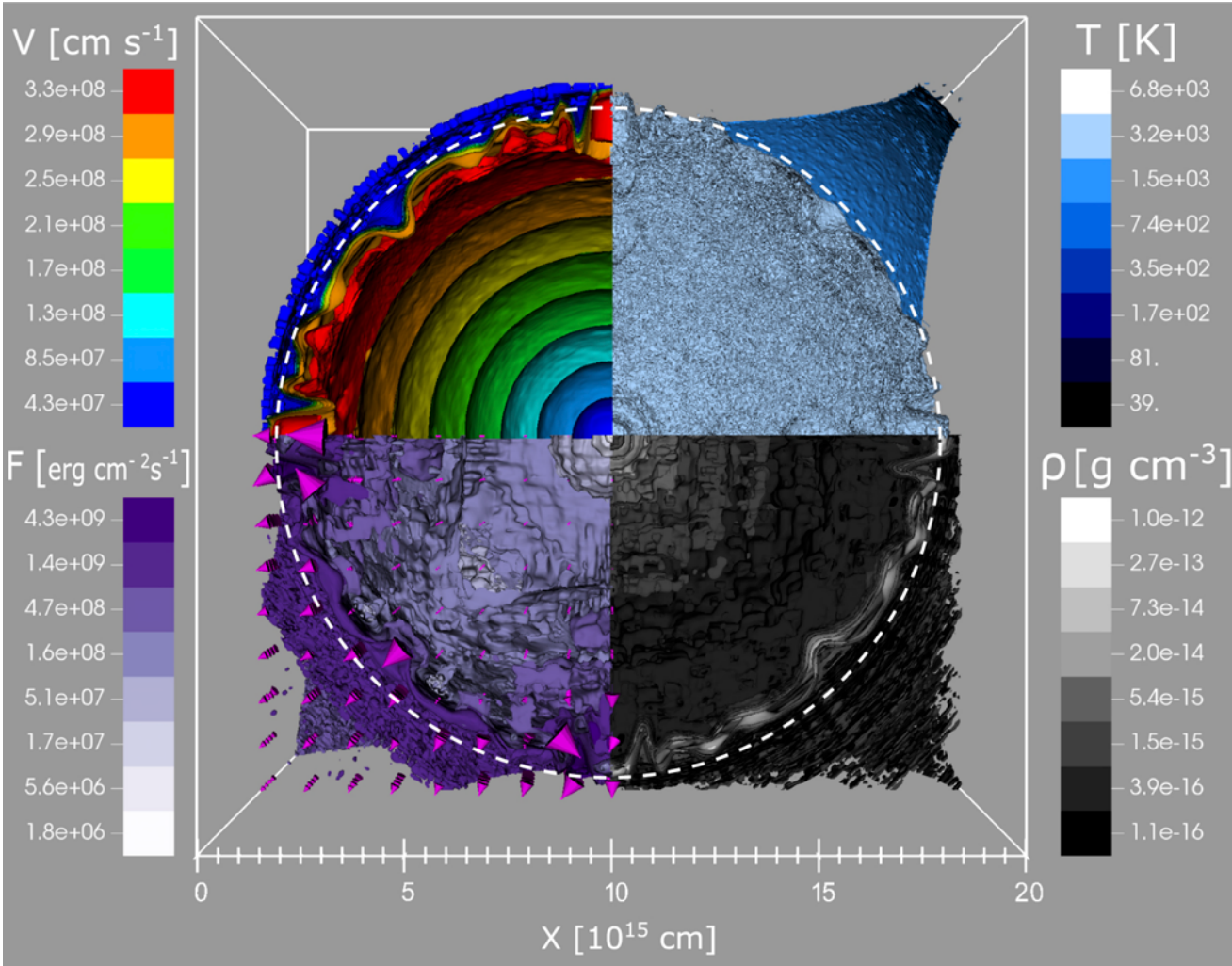

*Fig. 5.8: Snapshot of the 3D simulation of a pulsational pair-instability supernova. The four subplots show radial velocity V, gas temperature T, radiation flux F, and gas density ρ. Figure adopted from Chen et al. ApJ 955 (2023) 39.*

## Core collapse and electron captures

Electron captures (EC) play an essential role in CCSN dynamics. In particular, the evolution of the lepton fraction during the collapse determines important aspects of the evolution. Sensitivity studies have shown that the most important nuclei are extremely neutron-rich around the neutron gaps N=40 and N=50. Reliable estimates of the rates require beyond mean-field techniques. Present models comprise temperature-dependent Quasiparticle Random Phase Approximation (RPA), hybrid models combining RPA rates with thermal occupations calculated within the Shell Model Monte Carlo, and most recently second RPA augmented with particle-vibration couplings. However, the total EC rates for nuclei can vary by an order of magnitude and it is essential to benchmark the calculations against experimental data.

## p-process

The p-process is called upon to explain the production of stable neutron-deficient nuclides heavier than iron as observed in the solar system and up till now in no other galactic location. Various scenarios have been proposed to account for the bulk p-nuclide content of the solar system, as well as for deviations concerning the bulk p-isotope composition of some elements discovered in primitive meteorites. The p-isotopes are mainly produced by (γ,n), (γ,p), and (γ,α) reactions, at stellar temperatures ~ 2 – 3 GK, on s- and r-nuclei. The p-nuclides are mostly produced in the final explosion of a massive star as a CCSN in pre-explosive O-burning episodes, or the explosion of a binary system in a type-Ia supernova.
The p-process includes a network of about twenty thousand reactions, with only a small fraction of experimentally known reactions. Nuclear uncertainties stem from the determination of the photodisintegration rates. Proton or radiative capture and photoemission measurements exist at low energies only for a restricted number of stable nuclei, so that theoretical estimates remain poorly constrained. Reaction predictions have been found to deviate rather significantly from measurements of the radioactive targets, demonstrating the need to further measure and test proton or alpha captures on neutron-deficient nuclei. Also needed is an improved determination of the α-nucleus optical potential (OP) far below the Coulomb barrier for A>140 nuclei as well as the nucleon-nucleus OP, photon strength function and nuclear level densities of neutron-deficient nuclei. In this context, storage rings at GSI/FAIR are important. The stored radioactive beams decelerated into the Gamow window can collide with pure hydrogen or helium targets. An ideal facility for such experiments would be a storage ring capable of accumulating post-accelerated ISOL beams. Elastic scattering of radioactive beams on silicon foils containing large $^4$He quantities is planned at ISOLDE. Photodisintegration reactions on stable targets are envisioned at ELI-NP.

## ν-process

The huge energy released as neutrinos during the collapse of the central layers of a massive star is expected to induce the synthesis of nuclei in the outer layers of the star. Neutrinos can interact with matter via inelastic neutral-current neutrino scattering, involving μ and τ neutrinos. Electron neutrinos may interact with nuclei through charged-current interactions. These interactions can populate excited nuclear levels that later decay by the emission of light particles, which may induce reactions with other nuclei, contributing to nucleosynthesis.

Neutrino-induced nucleosynthesis may occur before or after the arrival of the shock wave. Simulations suggest that the ν-process may contribute to the synthesis of some rare species, such as $^{11}$B, $^{19}$F, and $^{138}$La. Results are sensitive to uncertainties in neutrino interaction cross sections, average neutrino energies of each flavour, total neutrino luminosity, and to details of the explosion models.

## νp-process

Most simulations of the core collapse of a massive star show that the induced neutrino-driven wind is proton-rich. The wind launched at T > 10 GK consists of free neutrons and protons in nuclear statistical equilibrium. Nucleosynthesis in this νp-process occurs in four different stages: (i) expansion and cooling through the temperature range T ~ 10 – 5 GK causes all neutrons to combine with protons, leaving a composition consisting of α-particles and an excess of protons; (ii) additional cooling, down to T ~ 5 – 3 GK, allows the α-particles to form heavier nuclei, increasing the abundance of $^{56}$Ni, $^{60}$Zn, and $^{64}$Ge; (iii) when T ~ 3 – 1.5 GK, the charged-current interaction between electron antineutrinos and the very abundant protons produces free neutrons, which allow fast (n,p) reactions on $^{56}$Ni, $^{60}$Zn, and $^{64}$Ge and subsequently (p,γ) reactions that extend the nuclear flow toward heavier nuclides on the neutron-deficient region. (iv) Finally, when T ~1.5 GK, (p,γ) reactions freeze out and (n,p) reactions and β+-decays transform the heavy nuclei into stable, neutron-deficient nuclides. This νp-process is halted by cooling below T ~1 GK.

While some studies suggest that the νp-process may account for the synthesis of light p-nuclides up to $^{108}$Cd, the issue is still debated. The νp-process is sensitive to details of the explosion, the mass, and the rotation rate of the proto-neutron star. Key uncertainties originate from neutrino luminosities and average energies. Most (n,p) rates of interest are based on the Hauser–Feshbach model. Changing these rates strongly affects the synthesis of A > 100 p-nuclei. A recent study has identified the key uncertainties including the 3α reaction, $^{56}$Ni(n,p)$^{56}$Co and several (n,p) and (p,γ) reactions.

### r-process

The rapid neutron capture process can also take place in neutrino-driven winds of core-collapse supernovae, but it is unclear if supernova ejecta are sufficiently neutron-rich to produce the heaviest r-process nuclei. The related nucleosynthesis is thus often called the weak r-process. However, supernovae with strong magnetic fields have the potential to create heavier r-process elements.

# Neutron-star mergers

The first detection of gravitational waves from a binary neutron star (NS) merger GW170817 by the LIGO and Virgo observatories marked a significant milestone in astrophysics. Although only the inspiral epoch was discernible in the GW data, it allowed inferring constraints on the cold equation of state above nuclear density from measurements of the NS tidal deformability, a characteristic parameter that depends on the properties of matter. Together with the electromagnetic counterpart, GW170817 provided the first definite identification of an astrophysical site where the r-process takes place. This extraordinary event, together with the very recent x-ray measurements by NICER, has notably enriched our knowledge of nuclear physics in extreme environments, with considerable progress made since the last NuPECC Long Range Plan.

## Neutron stars and Equation of State

The most straightforward connection between the astrophysical observations of compact stars and the underlying microscopic properties concerns the functional relation between the pressure and the baryonic density, the so-called nuclear Equation of State (EoS). The hydrostatic equilibrium imposes a one-to-one correspondence between the EoS and all the static properties of neutron stars, such as the radius or the tidal deformability (see Box 5.4). In turn, the EoS can be theoretically calculated, imposing beta-equilibrium to the energy functional of baryonic matter that is calibrated on nuclear ab-initio calculation and experimental nuclear observables.

The field has advanced spectacularly in the past five years. The Bayesian inference has enabled simultaneous consideration of constraints coming from nuclear theory, nuclear experiments and astrophysical observations. With this method, quantitative and well-controlled uncertainties can be assessed in the nuclear physics parameters, especially the ones connected to the density dependence of the symmetry energy. For the analysis of astrophysical data, nuclear predictions of pure neutron matter provide essential constraints. Recently, remarkable progress has been made in ab-initio nuclear modelling and the development of two- and three-body chiral interactions at next-to-next leading order based on a reliable power counting scheme. This approach provides a quantitative understanding of systematic uncertainties coming from the nuclear theory side up to its breakdown density.

On the experimental side, the very recent PREX-II and CREX results on the neutron skin thickness of $^{208}$Pb and $^{48}$Ca have triggered an intensive discussion in the nuclear physics community because no model has been able to reproduce them simultaneously. In the next few years, these measurements sensitive to the symmetry energy should be readdressed to produce precise enough data for the estimation of

### Box 5.4: Neutron stars and Equation of State

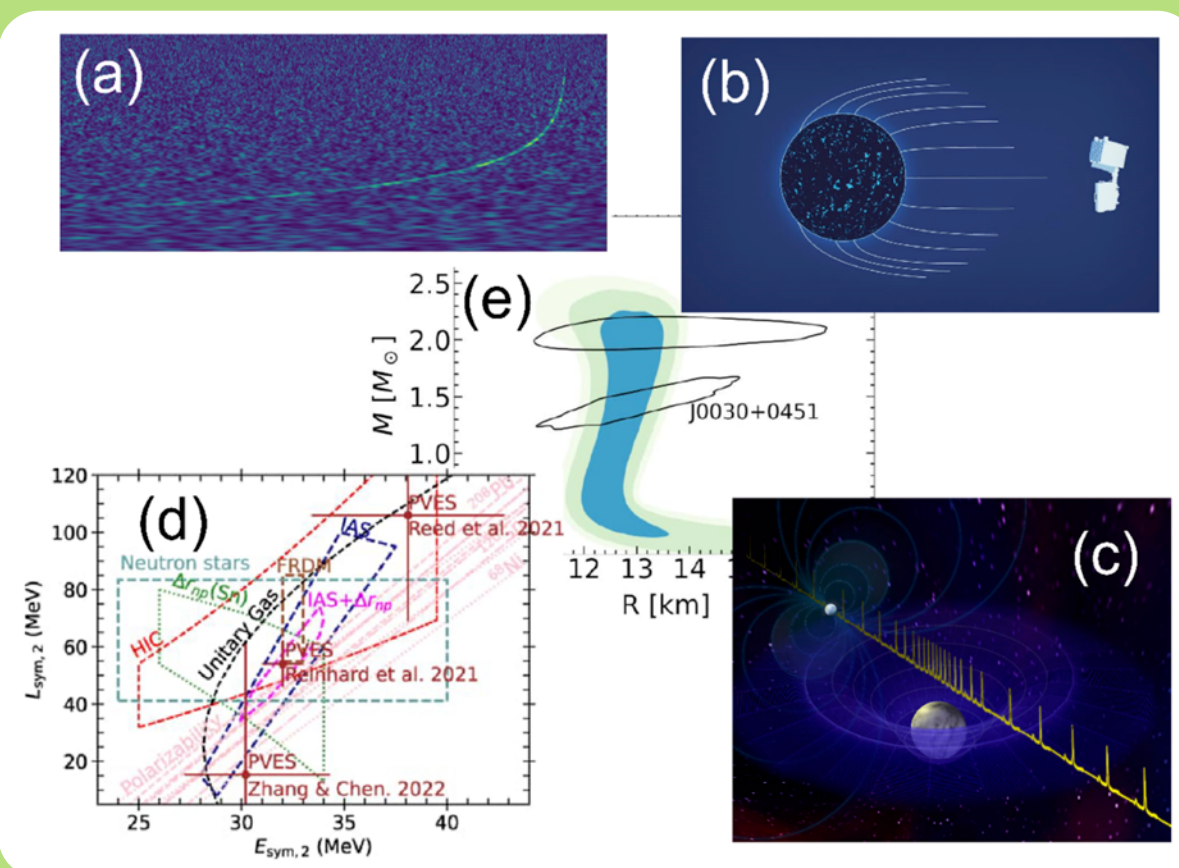

*Fig. 5.17: Astrophysical observations and nuclear physics constrain the EoS of nuclear matter and the mass-radius relation of neutron stars. (a) Gravitational wave signal from GW170817. Credit: LSC/Alex Nitz. (b) NICER observes X-ray light from the surfaces of neutron stars. Credit: NASA's Goddard Space Flight Center Conceptual Image Lab. (c) Artist's impression of a pulse from a massive neutron star. Credit: BSaxton, NRAO/AUI/NSF (d) Correlation between the symmetry energy and its slope L constrained by experiments, theory and observations, reprinted from Carlson et al. PRC 107, 035805 (2023) with the permission of APS. Copyright (2023) by the APS. (e) Probability density plots of NS mass M as a function of radius R, adapted from Dinh Thi et al. Universe (2021).*

**General relativity imposes a one-to-one correspondence between the Equation of State of nuclear matter and the static properties of neutron stars. Nuclear theory and experiments have been used to directly constrain the neutron-star properties, together with the information coming from astrophysical observations. Gravitational waves and x-rays exclude extremely large and short radii respectively. Radio timing reveals the existence of very massive neutron stars, above twice the mass of the Sun. Ab-initio nuclear theory and a wealth of different nuclear experiments constrain the density dependence of the symmetry energy, for example collective modes, mass and skin measurements, dipole polarisability (to cite just a few), put stringent constraints on the lower density part of the star, while heavy-ion collisions (HIC) also reach out to densities of several times saturation density. In the past few years, the combined consideration of these different constraints within Bayesian techniques has permitted controlled predictions on the mass-radius relation of a neutron star, hence the maximal compactness that baryonic matter can sustain. Improved future observations (Virgo-nEXT, ET, CE, Athena, XRISM) and experiments (FAIR, FRIB, RIBF), together with the expected progress of ab-initio theory, will further tighten the present constraints thus unveiling the longstanding question of the internal structure of neutron stars and the possible existence of new phases of ultra-dense matter.**

the nucleonic EoS and to assess the composition of the neutron-star inner core. In addition, other observables with sensitivity to the nuclear EoS, such as isospin transport in heavy ion collisions (HICs), isovector and charge-exchange resonances and the electric dipole polarisability from Coulomb excitation, for densities up to saturation density, and particle production and flow in intermediate-energy HICs up to several times saturation density, should be explored. Nuclear spectroscopic data for more neutron-rich isotopes and neutron-rich heavy-ion collision data at relativistic energies at FAIR will be essential to constrain the extrapolation of nuclear models.

## Merger dynamics

The measurement of the tidal polarisability of GW170817 has provided unique information on the internal structure of neutron stars. Since this observation, at least three more events (GW191219, GW200105 and GW200115) involving at least one NS have been discovered. Due to the projected increase in the sensitivity of the existing detector network of the LIGO, Virgo, and KAGRA facilities, dozens of such observations are expected in the coming years, including several cases with an electromagnetic counterpart. The analysis of these datasets will have a significant impact on nuclear physics. Further upgrades, LIGO Voyager and Virgo nEXT are planned to fill the gap before the onset of next-generation detectors projected for mid-2030s. Third-generation interferometers, such as the Einstein Telescope, in which the European nuclear physics community is largely involved, can also help to detect higher-frequency, post-merger signals. These are expected to provide much richer information on the structure of ultra-dense matter than the tidal deformability, as the post-merger remnant probes higher densities and finite temperatures where possibly additional degrees of freedom may occur (hyperons, deconfined quark matter, etc.).

From the nuclear physics viewpoint, modelling the post-merger gravitational-wave signal is probably the most ambitious but also the most important challenge of the next decade. Hydrodynamic, fully general relativistic 3D simulations with accurate treatment of the neutrino radiation are nowadays accessible but further developments are needed, particularly concerning the numerical resolution and neutrino transport. These numerical codes require high-performance computing techniques and run on supercomputing facilities. In recent years they have reached a high degree of maturity and confidence, both from the numerical side (e.g. inclusion of magnetic fields, more sophisticated temperature-dependent equations of state, multidimensional neutrino transport) and concerning the prediction of observable signals (e.g. reliable gravitational waveforms for binary neutron-star mergers). From the nuclear physics perspective, the equations of state that serve as inputs in those codes must be temperature-dependent and cover the full range of isospin asymmetries, from symmetric matter to neutron matter. To calibrate such models, a plethora of nuclear data will be needed including excited-state information from spectroscopic studies and in-medium modification of light clusters from heavy ion collisions.

## Rapid neutron capture process (r-process) produces the heaviest elements in nature

The rapid neutron capture process (r-process) proceeds mainly via neutron captures and beta decays along extremely neutron-rich nuclei (see Box 5.1). It involves more than 5000 nuclei and rapidly changing astrophysical conditions, thus setting a challenge for nuclear astrophysics studies. The imprint of the r-process history in the universe is revealed through the abundance patterns of various elements and their isotopes in stars. Whereas the heaviest naturally occurring elements, such as uranium and thorium, are solely produced by the r-process, for many of the lighter elements both the s- and the r-process contribute. The r-process abundance pattern is usually determined as s-process residuals, i.e. subtracting the rather well-constrained s-process contribution from the total abundances. The r-process abundance patterns can then be compared with the calculations, which depend on both astrophysics and nuclear physics parameters.

The astrophysical site of the r-process has been debated for a long time. The multimessenger observations from GW170817 confirmed that it takes place at least in NS mergers. The fits to the kilonova light curves were in agreement with the matter heated as one would expect from the r-process. The observation of direct absorption lines is difficult due to Doppler broadening and incomplete atomic data (opacities); however, indications of absorption lines belonging to the abundantly produced r-process element Sr have been identified. The association of certain spectral features with a few other elements has been proposed. The multi-messenger signals, which combine information on the initial state and energetics of the collision from GWs with the features in the Kilonova lightcurves, are a highly anticipated input and can significantly advance the field.

In addition to NS mergers, other r-process sites have been proposed (magnetorotational supernovae, collapsars, hypernovae, etc.) but their contribution to the total r-process abundances is an open question. Observations from the Milky Way and its dwarf galaxy satellites point toward prompt and delayed r-process sources. The timescales of these two sources are compatible with CCSN (prompt) and NS mergers (delayed). In addition, the detection of interstellar $^{60}Fe$ and $^{244}Pu$ in terrestrial samples can place constraints on r-process frequency and production yields over the last few 100 Myr. For such long-lived nuclides, accelerator mass spectrometry facilities provide high-precision data on their abundances in the studied samples.
Many r-process nuclei will remain experimentally inaccessible and nuclear theory plays an essential role. Realistic modelling of the r-process path is a formidable theoretical task which combines state-of-the-art hydrodynamical simulations with complete nuclear reaction networks. Nuclear data are needed for more than 5000 different nuclei (see Box 5.1), including masses, neutron-capture and beta-decay rates. Fission plays a role in many r-process scenarios, circulating material back to lower-mass regions, serving additional neutrons, and impacting the kilonova light curves. Consistent and microscopic large-scale calculations benchmarked on the largest possible pool of experimental data are needed, with controlled treatment of theoretical uncertainties and their propagation. The development of high-performance computing and controlled machine-learning approaches is a key aspect of future developments.

Beta-decay half-lives ($T_{1/2}$) define the initial elemental abundances and the pace of the process, while beta-delayed neutron emission probabilities ($P_n$) alter the decay path towards stability and supply additional neutrons for late captures. Remarkable progress has been made in recent years, with 26 half-lives, 83 $P_{1n}$, and 55 $P_{2n}$ values experimentally determined for the first time at the RIBF facility of the RIKEN Nishina Centre through the international BRIKEN collaboration (see Fig. 5.9). The know-how gained from this experiment, which included many European physicists and equipment, will pave the way for exploring more exotic neutron-rich nuclei in future experiments at European facilities, e.g. at JYFL, ISOLDE, GANIL, SPES, and FAIR.

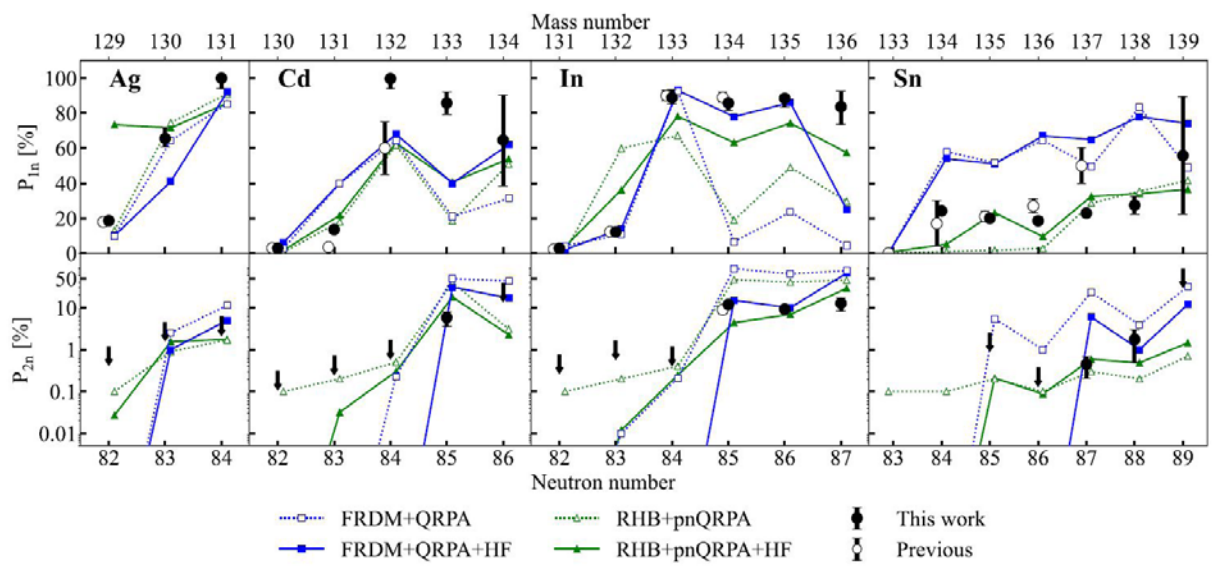

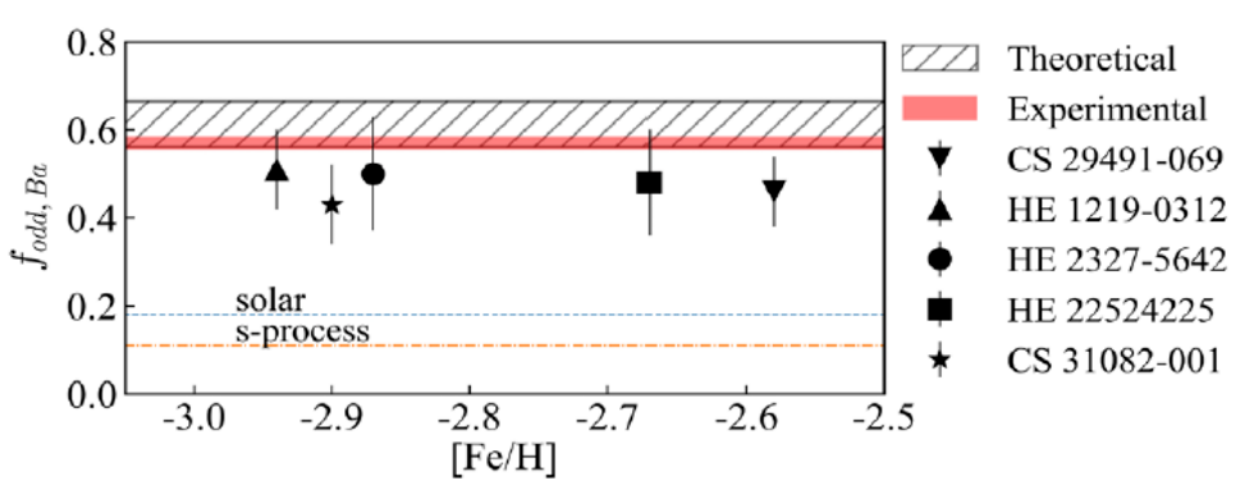

*Fig. 5.9: a) Measured one- and two-neutron emission probabilities in the region of the 2nd r-process peak. b) Odd-mass Ba isotopic fraction for five r-process enhanced stars (r-II) calculated using experimental and theoretical neutron-emission probabilities. Figures reprinted with permission from Phong et al., Phys. Rev. Lett. 129, 172701 (2022). Copyright (2022) by the American Physical Society.*

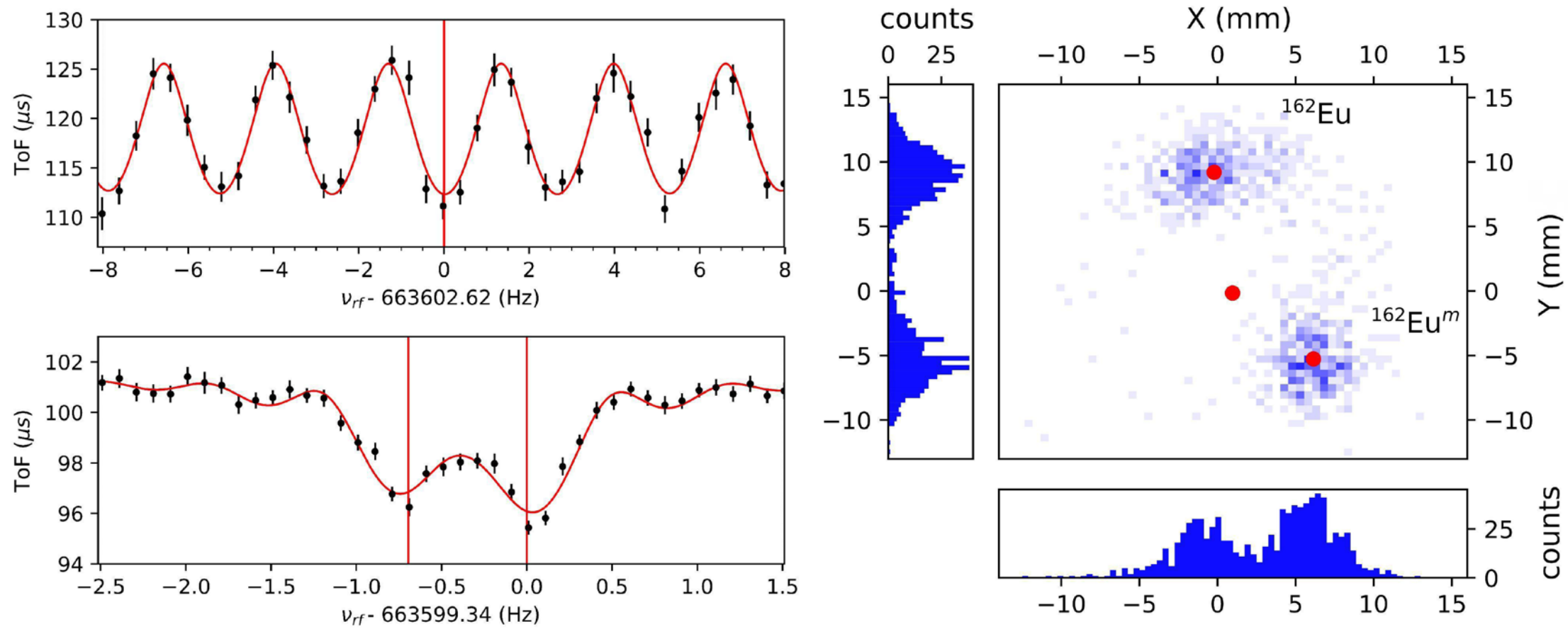

*Fig. 5.10: The phase-imaging ion cyclotron resonance (PI-ICR) technique has enabled measurements of low-lying isomeric states. The 156(3) keV isomer in 162Eu was not resolvable using the conventional TOF-ICR technique with a 400-ms excitation time (top left). With the PI-ICR technique, the ground and isomeric states are easily resolved with a 600-ms accumulation time. Resolving the states with the TOF-ICR technique required a 1600 ms excitation time (bottom left). Figures reprinted with permission from Vilén et al., Phys. Rev. C 101, 034312 (2020). Copyright (2020) by the American Physical Society.*

The main r-process path is largely determined by the masses in addition to the astrophysical conditions such as temperature and neutron density. The neutron-capture reaction rates depend sensitively on the Q-values, i.e. masses. Multi-Reflection Time-of-Flight mass spectrometers (MR-ToF-MS) have shown their versatile applicability for short-lived radioactive nuclides. The phase-imaging ion cyclotron resonance (PI-ICR) technique has become the standard method in Penning-trap mass spectrometry, allowing superior resolving power compared to previous methods and enabling measurements of low-lying isomeric states with excitation energies below 100 keV (see Fig. 5.10). Recent studies on the role of isomers in astrophysics have shown their important role in many astrophysical scenarios. These astrophysically important isomers, known as astromers (see Box 5.3) can potentially play a role in the kilonova light curves, either delaying or speeding the energy output (decay) rate compared to the ground-state beta-decays. For shorter-lived nuclei (down to a few hundred microseconds), isochronous mass spectrometry (IMS) can be used for mass and lifetime measurements in a storage ring, such as the ESR at GSI. The sensitivity and precision of the IMS have recently been boosted by combining the revolution frequency with a semi-destructive measurement of the velocities of every ion stored in the ring. With the new method, it was possible to determine the masses with a relative precision of about $2\times10^{-6}$ for nuclides with production rates smaller than two particles per week.

The beta-decay and mass measurements have rather well pinned down the region of fission fragments; however, with the improved yields at new radioactive beam facilities and more sensitive and faster measurement methods, the known region can be pushed further. The third r-process abundance peak at around A=195 still includes large uncertainties due to lack of experimental data in particular masses. Multinucleon-transfer reactions provide an alternative way to produce heavy exotic nuclei. Several projects are ongoing.

Neutron-capture cross-sections relevant to the r-process cannot be directly measured owing to the very exotic nature of the nuclei involved. Thus, alternative approaches such as the Oslo and beta-Oslo methods can be applied to constrain and validate the level densities and gamma-strength functions needed for the reaction-rate calculations. These indirect techniques will provide a significant amount of valuable information to guide, constrain and validate theoretical models in the coming years. Finally, surrogate reactions in inverse kinematics, such as (d,p) exploiting multi-circulation of stored exotic ions in storage rings has recently been demonstrated for heavy nuclei at the ESR of GSI. An alternative solution could be a low-energy ring at a post-accelerator of an ISOL facility, where the beam can be delivered and accumulated immediately at the required energy. A free-neutron target coupled to such a storage ring could enable direct neutron-induced reaction studies.

### Box 5.3: Astromers - astrophysically important nuclear isomers

**Astrophysical reaction network codes have traditionally only considered reactions with nuclear ground states but many nuclei have long-lived excited states known as isomers. Currently, around 2000 isomers are known. Many isomers play an important role in nuclear astrophysics and are called astromers. The most famous example is the isomer in $^{26}$Al, strongly affecting its effective half-life in stellar conditions. In the r-process, the population of isomers can impact the reaction flow and the kilonova light curve. Sensitivity studies have shown that several isomeric states are strongly populated in the r-process and are potential astromers.**

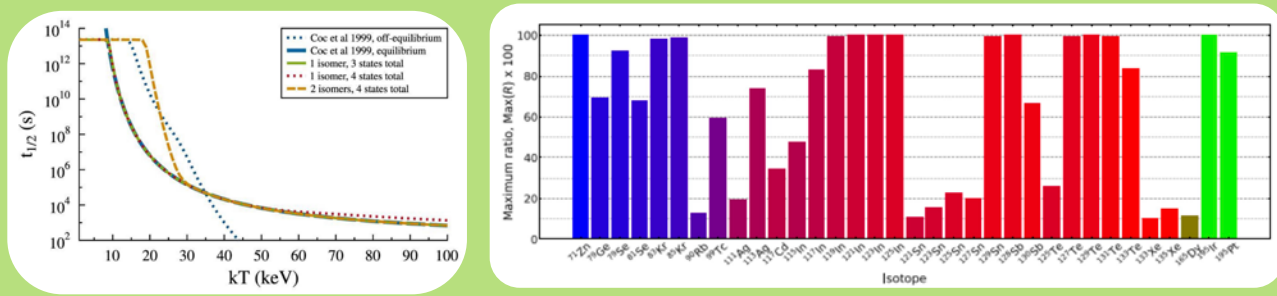

*Fig. 5.16: Left: Effective half-life of $^{26}$Al with an astromer depends on temperature. Figure adopted from Laird et al., J. Phys. G: Nucl. Part. Phys., 50, 033002 (2023). Right: Strongly populated isomers in the r-process can act as astromers. The colour indicates approximately the r-process abundance peak the nucleus resides in: first (blue), second (red), or third (green) peak. Figure adopted from Misch et al., Astrophys. J., 913, L2 (2021).*

## Exploding White Dwarfs

### Type Ia Supernovae

Type Ia supernovae (SNIa) occur 2 – 3 times per century in our Galaxy. They are identified by the lack of H in their spectra and the presence of a prominent absorption feature near 6150 Å due to Si II. This limits the amount of H that can be present in the expanding atmosphere of the star (MH ≤ 0.03 – 0.1 $M_{sun}$), while the detection of Si reveals the presence of intermediate-mass elements resulting from incomplete nuclear processing. An example of the remnant of a SNIa is shown in Fig. 5.11.

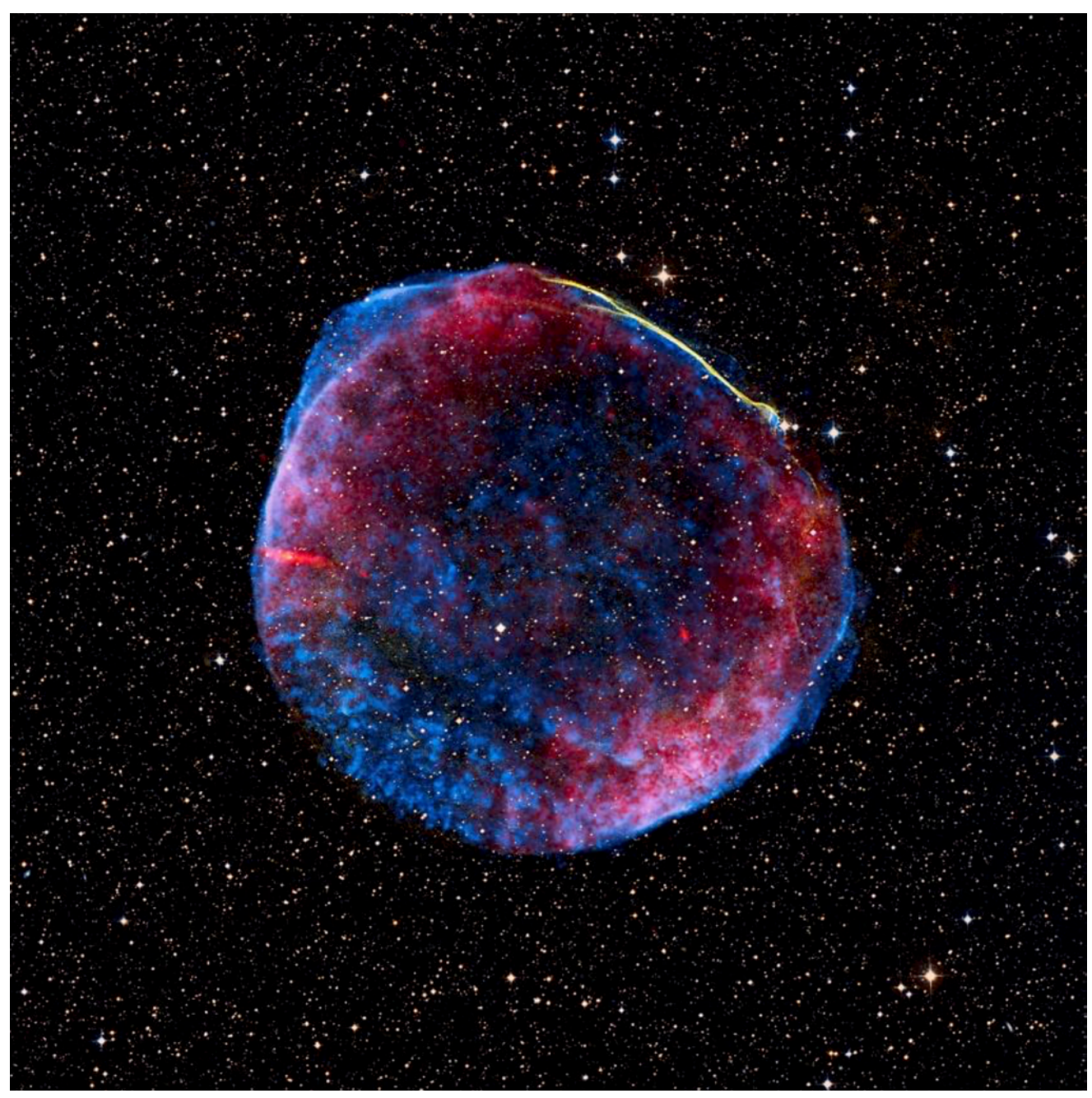

***Fig. 5.11: Composite image of the SN 1006 remnant, likely a type Ia supernova, located at 7100 light-years from Earth. The object is 65 light-years across. Data has been obtained from NASA's Chandra X-ray Observatory (X-rays), the University of Michigan's 0.9 m Curtis Schmidt telescope at the NSF's Cerro Tololo Inter-American Observatory (optical), the Digitised Sky Survey (optical), NRAO's Very Large Array and Green Bank Telescope (radio).***

The energy released in a typical SNIa can be inferred from the overall kinetic energy of the expanding ejecta, K ~$10^{51}$ erg (with velocities ~5000 - 10000 km $s^{-1}$) and the energy integrated over the light curve, $E_{rad}$ ~$10^{49}$ erg. Photometrically, SNIa shows a fast increase in luminosity in ~20 days, reaching a peak value of ~$10^{10}$ $L_{sun}$, followed by a steep decline in ~30 days, and later, by a smoother decline over ~70 days. Peak luminosity depends on the amount of $^{56}$Ni synthesised (0.1 – 1 $M_{sun}$). The late light curve is powered by the decay chain $^{56}$Ni → $^{56}$Co → $^{56}$Fe, with two different slopes, attributed to the different half-lives of $^{56}$Ni ($T_{1/2}$ = 6.1 days) and $^{56}$Co ($T_{1/2}$ = 77.3 days). The first unambiguous detection of the 847 keV and 1238 keV $^{56}$Co lines was reported from SN 2014J, the closest SNIa observed since the dawn of γ-ray astronomy. More observations are required to assess the predicted emission of the 158 keV and 812 keV $^{56}$Ni lines, for which only upper limits exist. NASA's COSI will be launched in 2027, with enough sensitivity to detect $^{56}$Co decay lines from SNIa up to 20 Mpc, as well as the 511 keV emission, providing constraints on progenitor models, explosion geometries, and positron emission. There are hints for $^{44}$Sc lines from historic SNIa which are non-conclusive. COSI's full-sky search for the $^{44}$Ti (the parent nucleus of $^{44}$Sc) line at 1.157 MeV will also reveal whether this isotope can provide information not only on CCSN but also on SNIa.

About 70% of all observed SNIa display similar spectral features, peak luminosities, light curve shapes and characteristic timescales. This means SNIa as cosmological probes to infer properties of the large-scale structure of the universe (e.g. the acceleration of the expansion rate) can be used. These similarities favour a dominant progenitor system and explosion mechanism, such as a 1.4 $M_{sun}$ CO-rich white dwarf (WD) that is disrupted by the explosion. However, the increasing number of peculiar SNIa has raised interest in different explosion mechanisms. Two main scenarios have been proposed: a single-degenerate, in which a low-mass star transfers H- or He-rich matter onto a WD, and a double-degenerate, involving the merger of two WDs. The major drawback faced by the single-degenerate scenario is the difficult evolutionary pathway to increase the WD mass and reach the Chandrasekhar limit. The double-degenerate scenario faces the scarcity of candidates detected. However, recent analyses of high-mass WD data in the GAIA Data Release 2 have led to a new double WD merger rate that increases support to the double-degenerate channel. Nucleosynthesis in SNIa depends critically on the peak temperature achieved and the density at which the explosion initiates. The abundance pattern of the ejecta is the result of different burning regimes: normal and α-rich freeze-out from nuclear statistical equilibrium in the inner regions of the star and incomplete Si-, O- and C/Ne-burning in the outer layers. Efforts have focused on understanding the nature of the ignition front that propagates and incinerates the star. Typical densities at which C is expected to ignite near the centre of a WD are around $10^9$ g $cm^{-3}$ and favour the generation of a subsonic (deflagration) burning front, catalysed by the thermal conduction of the degenerate electron plasma. However, these models result in a severe overproduction of some n-rich species, such as $^{54}$Cr and $^{50}$Ti, the presence of big clumps of $^{56}$Ni in the photosphere around peak luminosity and the absence of chemical stratification in the ejecta. Explosion models based on delayed (supersonic) detonation fronts have been more successful in matching the observational features of SNIa. In these models, an early deflagration front that propagates and pre-expands the star subsequently switches into a detonation front. However, the physical mechanism that drives this deflagration-detonation transition is not fully understood.

The sensitivity of the predicted SNIa nucleosynthesis to variations of both thermonuclear reaction and weak interaction rates has been thoroughly investigated. Main uncertainties affecting SNIa yields include $^{12}$C(α,γ), $^{12}$C+$^{12}$C, $^{20}$Ne(α,p), $^{20}$Ne(α,γ), and $^{30}$Si(p,γ), together with several weak interaction rates, most notably $^{28}$Si(β+)$^{28}$Al, $^{32}$S(β+)$^{32}$P, and $^{36}$Ar(β+)$^{36}$Cl.

## Classical and Recurrent Novae

Classical novae (CN) are also stellar explosions occurring in binary systems, consisting of a WD (usually, CO- or ONe-rich) and a low-mass main sequence (or a more evolved) companion. They exhibit a sudden increase in optical brightness in 1 - 2 days, reaching peak luminosities ~$10^4$-$10^5$ $L_{sun}$. About 50 CN are expected to explode every year in our Galaxy, constituting the second most frequent class of stellar thermonuclear explosions after X-ray bursts. Unfortunately, detection of Galactic CN is often hampered by interstellar extinction, and only a fraction is observed. Figure 5.12 shows Nova GK Persei as an example. Neither the WD nor the binary system are destroyed by the explosion. CN are therefore expected to recur, typically after $10^4$-$10^5$ yr. The subclass of *recurrent novae* (i.e., novae that have been observed in outbursts more than once) undergo an explosion after 1 - 100 yr, likely implying very massive WDs and high mass-accretion rates. CN are also characterised by a modest ejected mass ($10^{-7}$ - $10^{-4}$ $M_{sun}$), at moderate velocities (~1000 km $s^{-1}$).

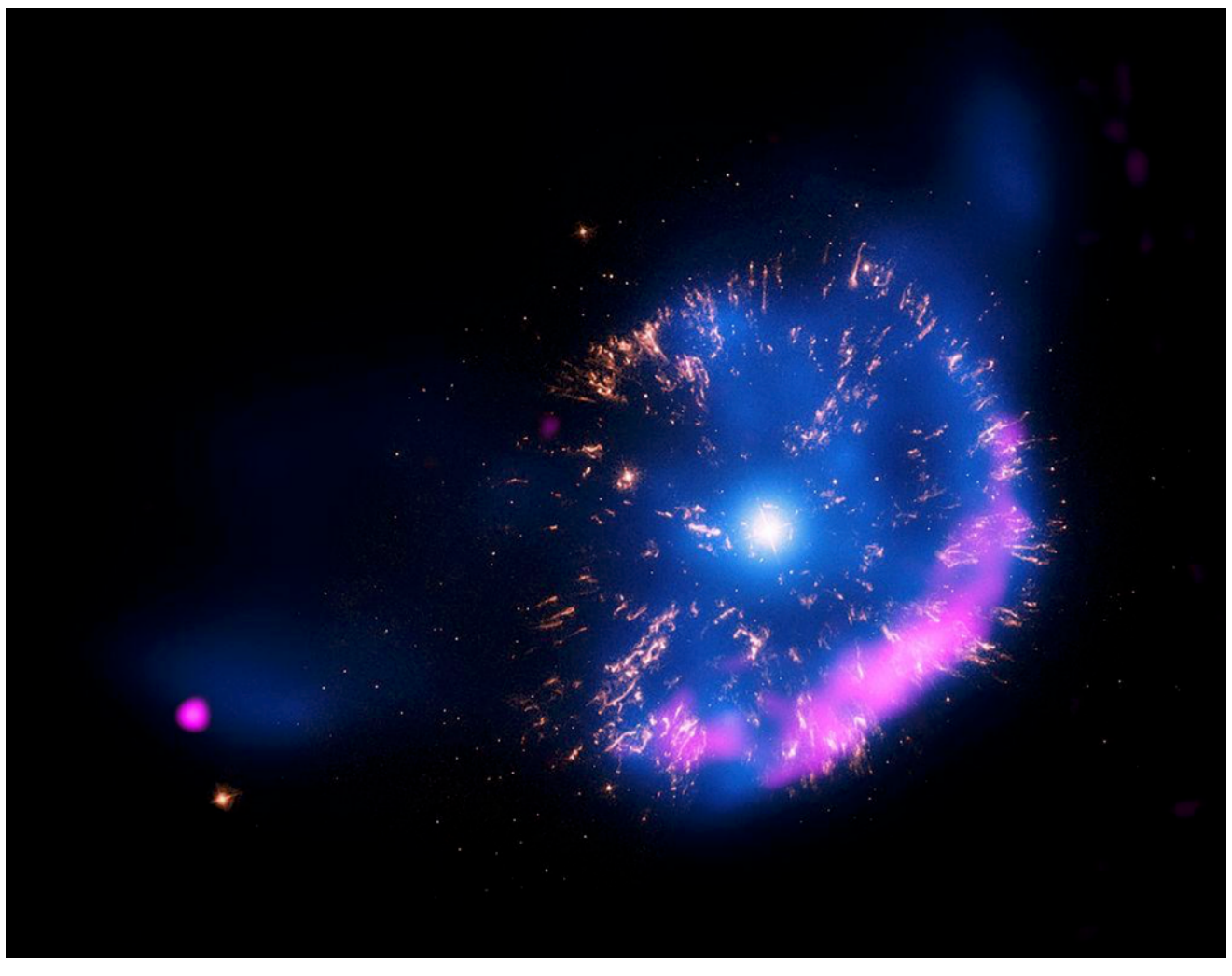

***Fig. 5.12: Composite image of the shell ejected by Nova GK Persei (1901), that combines X-ray observations from Chandra (blue colour), optical data from NASA's Hubble Space Telescope (yellow), and radio data from the National Science Foundation's Very Large Array (pink). The shell is located 1500 light-years away, as seen between 2000 and 2013. The shell contains about $10^{-4}$ solar masses and is about 1 light-year in diameter.***

The strength of a CN depends on four parameters: the mass and initial luminosity of the WD that hosts the explosion and the metallicity and mass-transfer rate from its stellar companion. Only envelopes with CNO-enhanced abundances (with metallicities Z ~ 0.2 - 0.5) can account for the main observational properties of *fast novae*. The origin of such enhancements is a matter of debate. Several mechanisms have been proposed but none has proved fully successful. Promising results have been obtained when relaxing the constraints imposed by spherically symmetric models. Indeed, 2D/3D simulations of mixing at the core-envelope interface revealed that Kelvin-Helmholtz instabilities can lead to self-enrichment of the accreted envelope with core material at levels that agree with observations.

The main nuclear path in CN runs close to the valley of stability and is driven by proton-capture reactions and $\beta^+$-decays. The large amounts of $^{13}$N, $^{14,15}$O, and $^{17}$F synthesised through the hot-CNO cycle translate into large amounts of their daughter nuclei $^{13}$C, $^{15}$N, and $^{17}$O in the ejecta, constituting the main contribution of novae to the Galactic abundances. Several species may provide potentially detectable γ-rays, including $^{18}$F, that powers the prompt γ-ray emission at and below 511 keV, and the longer-lived $^{7}$Be and $^{22}$Na that decay when the envelope is already optically thin to γ-rays, powering line emission at 478 keV and 1275 keV respectively. Future instruments like NASA's COSI may be able to confirm such predictions (for instance, line sensitivity at 1275 keV will be improved by an order of magnitude over existing instruments). An open issue is how much $^{7}$Li can be produced by novae and how much CN contribute to its Galactic content. There is a solid basis for the detection of blueshifted $^{7}$Be II absorption lines in multiple CN (e.g. V339 Del, V5668 Sgr, V2944 Oph), and recent models with shear mixing provide a promising framework to account for these observations.

There is, in general, good agreement between the abundance patterns inferred from observations and simulations, with a predicted nucleosynthetic endpoint around Ca. However, only elemental abundances are inferred spectroscopically. Better perspectives are offered by analyses of presolar meteoritic grains. Infrared and ultraviolet observations have revealed dust-forming episodes in the shells ejected during CN and several grains of a putative nova origin have been identified, in particular SiC and graphite grains characterised by low $^{12}$C/$^{13}$C and $^{14}$N/$^{15}$N ratios, high $^{30}$Si/$^{28}$Si, and close-to-solar $^{29}$Si/$^{28}$Si, and high $^{26}$Al/$^{27}$Al and $^{22}$Ne/$^{20}$Ne ratios (with $^{22}$Ne attributed to *in situ* $^{22}$Na decay). Several studies aimed at identifying the most uncertain reactions for nova nucleosynthesis have been performed. Most of these reactions have already been successfully re-evaluated in recent years. Main uncertainties affect $^{18}$F(p,α)$^{15}$O, $^{25}$Al(p,γ)$^{26}$Si, and $^{30}$P(p,γ)$^{31}$S. Despite multiple efforts, $^{18}$F(p,α)$^{15}$O rate uncertainty is still dominated by the lack of spectroscopic information (i.e., spins and parities) of three broad and likely interfering resonances at $E_r$(c.m.) = –121 keV, 8 keV and 38 keV. New experiments aimed at improving this rate are being conducted by GANIL and IJCLab. Concerning $^{25}$Al(p,γ)$^{26}$Si and $^{30}$P(p,γ)$^{31}$S, more experiments are needed to improve spectroscopic information in the range Er(c.m.) = 163 – 965 keV in the former, and within ~ 600 keV of the $^{30}$P+p threshold in $^{31}$S in the latter. In particular, a new measurement for $^{30}$P(d,p) is planned with the new CARME array at CRYRING (GSI).

# Other nucleosynthesis sites

## Type I X-Ray Bursts and Related Phenomena

X-ray bursts (XRBs) are characterised by a sudden increase in brightness, reaching peak luminosities of ~ $10^4$ – $10^5$ $L_{sun}$ after a very fast rise (~ 1 – 10 s). The overall energy output in a typical XRB is ~ $10^{39}$ erg, released over 10 - 100 s. XRBs take place on neutron stars (NSs), with a stellar companion in the form of a main-sequence or a red giant star. XRBs are recurrent events but are characterised by much shorter recurrence periods, ranging from hours (or even minutes) to days.

An example of an XRB burst light curve is shown in Fig. 5.13. More than 100 stellar systems that exhibit XRBs have been discovered to date in the Milky Way.

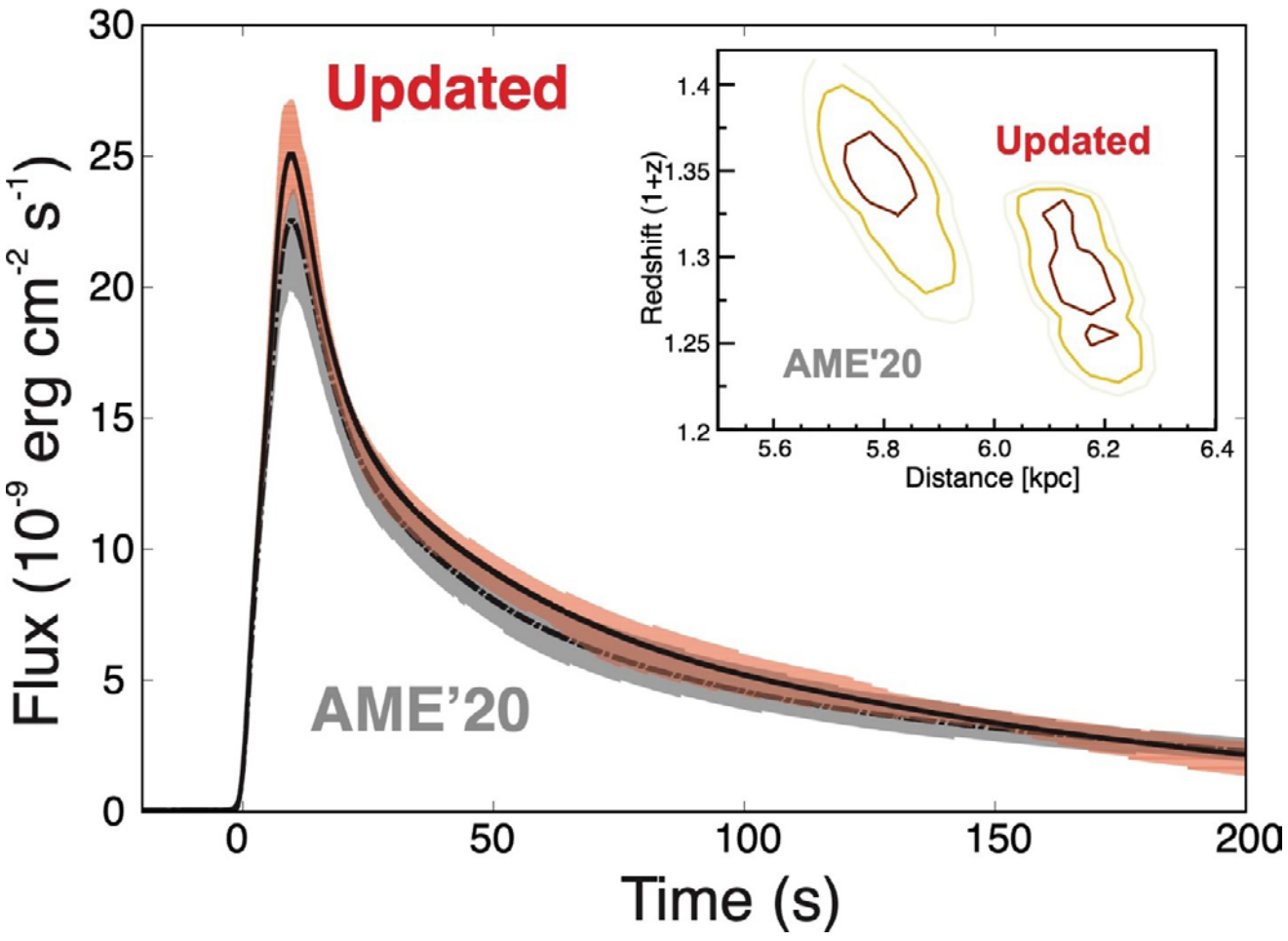

***Fig. 5.13: Impact of the new mass measurements near the $^{64}$Ge waiting point ("Updated") on the modelling of the XRB light curves, compared with the baseline expectation ("AME'20"). Insert: Surface gravitational redshift versus distance of the XRB GS 1826–24 obtained through a comparison of the modelled light curves and the observational data from the year 2007 bursting epoch. Figure adopted from Wang et al., Nat. Phys. 2023.***

Because of the strong surface gravity that characterises NSs, high temperatures and densities are reached at the base of the accreted envelope, resulting in complex nuclear activity that involves hundreds of nuclei extending all the way up to the SnSbTe mass region (or beyond). 1D simulations show that the main nuclear flow in XRB is driven by the rp-process (rapid proton captures and $\beta^+$-decays), the 3α-reaction and the αp-process (a sequence of (α,p) and (p,γ) reactions), and proceeds far away from the valley of stability, merging with the proton drip-line beyond A = 38 (see Box 5.1). Recent progress includes models with rotation which mostly affect the shape and duration of the light curves.

A major challenge is the lack of observational nucleosynthetic constraints. XRBs are unlikely to result in extensive mass ejection because of the extremely large escape velocities from an NS. The energy required to escape from the strong gravitational field of an NS is ~ 200 MeV/nucleon, whereas the nuclear energy released from thermonuclear fusion of solar-like matter into Fe-group elements is only ~ 5 MeV/nucleon. However, a large photospheric radius expansion is predicted for some models, such that the overall luminosity of the star can reach or exceed the Eddington limit, leading to the ejection of a tiny amount of material through a radiation-driven wind. A recent study that coupled hydrodynamic simulations and a radiation-driven wind model showed that about 0.1% of the envelope is ejected in a typical XRB, contributing to the Galactic abundances of $^{60}$Ni, $^{64}$Zn, and $^{58}$Ni.

Efforts aimed at providing observables from XRB spectra have been undertaken. Gravitationally-redshifted absorption lines in high-resolution spectra taken during 28 XRB detected in the source EXO 0748-676 revealed the presence of Fe XXVI lines (during the early phase of the bursts), Fe XXV and perhaps O VIII (at later stages). However, no evidence of such spectral features was found either during the analysis of 16 bursts observed in GS 1826-24, or from another series of bursting episodes in the original source. Another study identified strong absorption edges attributed to Fe-peak elements with abundances about 100 times solar in two XRBs exhibiting strong photospheric expansion. This suggests the presence of heavy-element ashes in the ejected wind, an issue that deserves further theoretical and observational work.

Most of the reaction rates used in XRB nucleosynthesis studies rely on theoretical estimates obtained from statistical models which may be affected by significant uncertainties. Thanks to large experimental campaigns, nuclear masses and half-lives are now experimentally known for most nuclides relevant to the rp-process. Sensitivity studies show that the masses around the $^{64}$Ge waiting point have the largest impact. Masses of $^{65}$As and $^{66}$Se have recently been measured with isochronous storage ring mass spectrometry at the CSRe in Lanzhou, thereby constraining the matter flow through $^{64}Ge$, making $^{68}Se$ and $^{72}Kr$ the next most uncertain waiting points for which mass measurements are needed. The impact of nuclear uncertainties has been assessed either through individual reaction-rate variations or simultaneously varying all reaction rates in a Monte Carlo approach. Main uncertainties include proton-capture reactions, such as $^{65}As(p,\gamma)^{66}Se$, $^{61}Ga(p,\gamma)^{62}Ge$, $^{96}Ag(p,\gamma)^{97}Cd$, $^{59}Cu(p,\gamma)^{60}Zn$, $^{86}Mo(p,\gamma)^{87}Tc$, $^{92}Ru(p,\gamma)^{93}Rh$, $^{102,103}In(p,\gamma)^{103,104}Sn$, and α-capture reactions like $^{12}C(\alpha,\gamma)^{16}O$, $^{15}O(\alpha,\gamma)^{19}Ne$, $^{18}Ne(\alpha,p)^{21}Na$, $^{30}S(\alpha,p)^{33}Cl$, and $^{56}Ni(\alpha,p)^{59}Cu$.

## Galactic Cosmic Rays

Galactic cosmic rays (GCR), which involve high-energy protons, α-particles and heavier nuclei, are considered an important production site of the light species Li, Be, and B, through high-energy nuclear collision (spallation) with interstellar nuclei (mostly CNO nuclei). The spallation cross sections of protons and α-particles on C, N and O nuclei at energies greater than 100 MeV/nucleon favour the synthesis of B, Li, and Be isotopes (in particular, $^{10}B$, $^{9}Be$, and $^{6}Li$), in the same decreasing order as observed in GCR. Production of light antinuclei is also expected in the interaction between GCR and the interstellar medium. Measurements of production and annihilation cross sections for antideuterium and antihelium at LHC and SPES are key to controlling the contribution of GCR to antimatter production in our Galaxy with high precision.

Galactic cosmic rays are expected to be associated with the most energetic astronomical objects and have traditionally been linked to supernovae and their remnants. However, the true origin and acceleration mechanism are still a matter of debate, partially because during propagation GCR may be deflected by Galactic magnetic fields, and most information regarding its original direction of motion is expected to get lost. Observationally, Be and B abundances reveal a linear dependence on metallicity. It is unlikely that GCR could originate from supernova-induced energetic protons or α-particles colliding with CNO nuclei of the interstellar medium since that would give rise to a Be or B abundance proportional to the metallicity squared. Neither can these observations be explained by a CNO abundance in Galactic cosmic rays that increase with time, since the relationship between the Be or B abundance and metallicity would not be linear. Accordingly, it seems unlikely that individual supernovae accelerate their own ejecta, contributing directly to the GCR flux. The leading model nowadays links the origin of GCR to the propagation of the supernova shock wave through material ejected as wind in the early pre-supernova evolution of massive stars.

# Perspectives

Nuclear astrophysics is an interdisciplinary field often motivated by observations of cosmic events that require a scientific explanation. These observations have to be carefully modelled to obtain a meaningful understanding and interpretation. The models typically require many nuclear physics inputs, both from nuclear theory and experiments. Thus, perspectives in nuclear astrophysics depend on the developments and expected future outcomes in observations, modelling, nuclear theory and experiments, as detailed below.

## Observations

The first measurement of the electromagnetic counterpart of a gravitational-wave source GW170817 (kilonova AT2017gfo) has had a profound impact on the field. It provided the first unambiguous identification of an r-process site and has notably enriched our knowledge of nuclear physics in extreme environments. In the future, the number of gravitational-wave observations, currently done by LIGO, Virgo and Kagra, could be further increased with the Einstein Telescope and the space-borne LISA observatory. Combined with optical-counterpart detections, the knowledge of such compact mergers will increase dramatically.

The number of stars with detailed chemical information has not only increased significantly but rarer objects and chemical patterns have also been discovered via recent spectroscopic surveys within the Milky Way, as well as via asteroseismology and astrometry, such as the Gaia mission. The data are now enriched with kinematic and age information, showing the distinction between stars that formed *in situ* and those that were accreted by the Milky Way throughout its evolution, as well as remnants from globular clusters. With the multi-object spectrographs currently evolving to accommodate thousands of fibres, the observational data is expected to increase substantially. Artificial intelligence techniques will certainly play a key role in the analysis, classification and interpretation of observations in the forthcoming years.

Very high-precision measurements of meteorites and other Solar System samples allow us to clearly identify the signatures of stellar nucleosynthesis and help us to understand the circumstances of the birth of the Solar System, including planet formation. In addition, live radioactive nuclei of astrophysical origin have been measured from Earth samples using accelerator mass spectrometry. These provide us with unique indicators of nearby astrophysical explosions.

Further discoveries are expected from future European facilities, such as the Extremely Large Telescope, the Wide Field Survey Telescope and the James Webb Space Telescope (JWST), which has already provided unprecedented data on very old galaxies, and indications of first-generation stars. These developments will revolutionise our understanding of stars in extremely metal-poor conditions, leading to more precise constraints on the nature of stars in the earliest stages of the Universe, and of kilonovae, potentially enabling the spectroscopic dissection of these explosions and their nucleosynthetic yields.

## Modelling

The dawn of supercomputing and the development of faster processors have provided new opportunities to study scenarios requiring a truly multidimensional approach. Current (super)computers have not yet reached the required performance to achieve such goals for many astrophysical scenarios, and stellar models are still often performed in 1D or rely on post-processing calculations based on temperature and density versus time profiles extracted from models.

Certain aspects are still poorly understood or very difficult to implement in existing stellar evolution codes. For instance, rotation is essential to characterising the chemical profiles inside stars driven by rotational-induced mixing, but it tends to destabilise the codes. Accurate descriptions of convective transport mixing and mass loss are particularly lacking for explosive scenarios, and the role of magnetic fields has also been frequently omitted. Evolution in binary stellar systems suffers from the very uncertain common envelope phases that strongly affect the dynamics of those systems.

Sensitivity studies guiding nuclear physics experiments should be carried out, in particular when more advanced experiments, e.g. multizone or 3D models, become computationally available. Nuclear reaction networks could be further developed to also include isomers.

## Nuclear theory

A common challenge both for nuclear theory and astrophysics modelling is the proper treatment of theoretical uncertainties in model parameters, chosen method and extrapolations. Theoretical errors must be carefully propagated for a trustable prediction of astrophysical observables, and different approaches should be further developed and systematically applied (Backward-Forward Monte Carlo approach, Bayesian methods, etc.). The use of machine-learning techniques (Bayesian Neural Networks, Kernel Ridge Regression, Gaussian Processes, Radial Basis Functions, etc.) is extremely promising in speeding up computer -expensive calculations in nuclear astrophysics and in scanning the extremely large space of model parameters, even if special care must be taken in the use of these methods for (uncontrolled) extrapolations. Although phenomenological models give the best accuracy in the reproduction of available experimental data, their extrapolations to temperature, density, and isospin conditions, where data do not exist, are often largely uncontrolled. The *global, universal* and *ab-initio* models are therefore often better suited for such problems.

Dense matter applications, such as the interpretation of the gravitational waves and coincident multi-messenger signals from neutron-star mergers, challenge nuclear theory as the static nuclear properties and reaction rates are in-medium modified. This requires a theoretical description of the nuclear interaction in dense matter. To construct the nuclear EoS, the bulk properties of nuclear matter must be known. These have to be calculated in a very large domain of temperatures and electron fractions which measure the relative neutron richness of the system to give inputs for the related hydrodynamical simulations and reliable predictions for the post-merger signals. The chiral effective field theory (ChEFT) has opened a new avenue to describe nuclear interactions and nuclear systems consistent with QCD, the fundamental theory of strong interaction. In ChEFT, many-body nuclear interactions can be calculated systematically and extended to the strange-baryon sector to model two-body and three-body interactions involving nucleons and hyperons. The ChEFT interactions have been used to calculate the EoS of nuclear matter both at zero and at finite temperatures using different many-body approaches. Crucially, this approach provides a quantitative understanding of systematic uncertainties coming from nuclear theory up to its breakdown density, and helps to select which EoSs are extrapolated to higher densities. Ab-initio methods are also very rapidly progressing in the description of ground and excited state properties of finite nuclei, thanks to the increased computing power and the great progress of numerical solution methods of the many-body problem.

Nuclear masses are perhaps the most fundamental nuclear properties for astrophysics applications. Mass models can typically predict measured masses with a root-mean-square (rms) deviation smaller than 800 keV. Ab-initio methods are not yet sophisticated enough to comply with this level of accuracy, but microscopic phenomenological models grounded in the density functional theory (DFT) could replace the phenomenological inputs in astrophysical simulations. Approaches based on the DFT can also be used for the EoS modelling in the range of densities and electron fractions where ab-initio approaches are difficult to apply. The field is witnessing impressive progress with the development of beyond-mean-field techniques, such as the proper account of symmetry breaking and symmetry restoration via projection techniques, configuration mixing through the generator coordinator method, and high-order terms in the nuclear response (QRPA, SRPA and beyond).

## Nuclear astrophysics experiments

Nuclear astrophysics is an active research field. For ground-state properties, novel mass-measurement techniques combined with new radioactive beam facilities can further extend the region of precisely known masses. Moreover, long-lived isomeric states potentially playing a role in astrophysical processes can be resolved with unprecedented precision. With high-quality radioactive beams and powerful detector setups, beta-decay studies can provide more accurate data on the decay of exotic nuclei, including the ones in highly charged ions. Level density and gamma-strength functions relevant for reaction rate calculations can be explored for new nuclei, for example, via the so-called beta-Oslo method. High-precision data on excited states are also needed to pin down the resonance states and their properties, e.g. for explosive hydrogen burning.

Atomic-level phenomena, such as ionisation, electron screening and opacities, should also be better constrained for nuclear astrophysics calculations; for example, opacities of (heavy) r-process nuclei are needed for the interpretation of kilonova light curves. To achieve this goal, a huge amount of atomic data is required. Available and ongoing theoretical calculations need to be benchmarked by dedicated experiments (e.g. PANDORA).

Many of the outstanding reactions for nuclear astrophysics require rather low beam energies that can also be achieved at smaller-scale facilities. Often, background suppression plays a central role in extending the measurements to the astrophysically relevant energy region where the cross sections are typically vanishingly low. The newly commissioned Bellotti Ion Beam Facility and Felsenkeller underground laboratories will provide new opportunities for such studies. The low astrophysical energies are also challenging for explosive hydrogen burning, where the low-energy resonances often require state-of-the-art detector setups such as time projection chambers or their combinations with an active target.

Reactions with radioactive targets or beams are challenging but pivotal to extend the knowledge of astrophysically relevant reactions beyond the stable nuclei. Developments in the target preparation, for example with samples created using radioactive waste, are essential in this respect. Combined with intensive neutron sources, these targets will help to constrain many neutron-capture cross sections. Using radioactive ion beams in storage rings would open new opportunities by enabling a broad range of reactions to be studied, even with short-lived species. In addition, photodissociation reactions with gamma beams and reactions in laser-induced plasma can yield relevant data for nuclear astrophysics.

# Recommendations: Nuclear Astrophysics

Nuclear astrophysics has been boosted by trailblazing observations since the previous Long Range Plan of 2017. To respond to the forthcoming observational data and the new multimessenger era in nuclear astrophysics, recommendations for networking and databases, nuclear theory and modelling, nuclear astrophysics experiments as well as facilities and infrastructures are in place.

### Networking and databases

- Joint research activities and networks in nuclear astrophysics in Europe should be further continued and strengthened and links to international nuclear astrophysics collaborations should be maintained.

- The connections to other research areas and communities, e.g. gravitational-waves, astronomy and atomic physics, should be maintained or established to fully benefit from future observations.

- We highly recommend the maintenance and updating of evaluations and databases important for nuclear astrophysics.

- We highly recommend continuing the training and networking activities at ECT*.

### Nuclear theory and modelling for nuclear astrophysics

- Uncertainty estimates for theoretical models and machine-learning methods for data processing and analysis should be further utilised and developed.

- To achieve a better understanding of the heavy-element synthesis and chemical evolution in the Cosmos, a theoretical description of fission, nuclear properties far from stability and neutron-capture rates are needed.

- Sensitivity studies are needed to guide the experimental efforts for various astrophysical scenarios.

## Nuclear astrophysics experiments

● Direct method experiments pursuing the reactions to the astrophysically relevant low energies are still badly needed for several reactions. Indirect methods, e.g. surrogate reactions, need to be explored when direct methods are not feasible.

● Experiments further from stability and more sensitive measurement techniques are strongly recommended, as these serve as important inputs for the astrophysical calculations and benchmark the nuclear models.

● The development of intense radioactive beams and related methodology (e.g. storage rings) should be supported to make the next step for reaction-rate measurements with radioactive beams.

● The expertise and availability of high-quality and also radioactive targets for nuclear astrophysics experiments will be needed and efforts to steer the activities in this respect should be continued/initiated.

## Facilities and infrastructures

● Physicists in Europe should have access to radioactive beam facilities at the frontier of exotic nuclei research. We strongly recommend the completion of the NUSTAR experimental facilities at FAIR, and the planned and ongoing upgrades for more intense and new radioactive beams, e.g. at ISOLDE (including the low-energy storage ring), SPES and GANIL-SPIRAL2.

● Smaller-scale facilities, and in particular the European low-background laboratories, are important for nuclear astrophysics research; activities connecting their work should be further supported.

● A high-energy AMS system would help in the isotopic abundance measurements and would be beneficial not only for astrophysics but also for fundamental physics and environmental applications.

● Access to large and fast HPC facilities in Europe should be guaranteed.

**References**

*A. Arcones & F.-K. Thielemann, "Origin of the elements", Astron. Astrophys. Rev. 31 (2023) 1*

*M. Arnould & S. Goriely, "Astronuclear Physics: A tale of the atomic nuclei in the skies", Progr. Part. Nucl. Phys. 112 (2020) 103766*

*M. Aliotta, A. Boeltzig, R. Depalo, & G. Gyürky, "Exploring Stars in Underground Laboratories: Challenges and Solutions", Annu. Rev. Nucl. Part. Sci. 72 (2022) 177-204*

*M. Lugaro, M. Pignatari, R. Reifarth, & M. Wiescher, "The s-Process and Beyond", Annu. Rev. Nucl. Part. Sci. 73 (2023) 315–40*

*J.J. Cowan, C. Sneden, J.E. Lawler, et al., "Origin of the heaviest elements: The rapid neutron-capture process", Rev. Mod. Phys. 93 (2021) 015002*

# Symmetries and Fundamental Interactions

**Convenors:**
**Paolo Crivelli** (ETH Zürich, Switzerland)
**Pierre Delahaye** (GANIL, Caen, France)

**NuPECC Liaisons:**
**Klaus Kirch** (ETH Zürich and PSI Villigen, Switzerland)
**Eberhard Widmann** (SMI Vienna, Austria)

**WG Members:**
- **Michail Athanasikis** (KU Leuven, Belgium)
- **Sebastian Baunack** (JGU Mainz, Germany)
- **Matteo Biassoni** (Università degli Studi di Milano-Bicocca, INFN Milano-Bicocca, Italy)
- **Bertram Blank** (LP2I Bordeaux, France)
- **Luigi Coraggio** (Università della Campania "Luigi Vanvitelli" and INFN Napoli, Italy)
- **Jacek Dobaczewski** (University of York, UK and University of Warsaw, Poland)
- **Martín González-Alonso** (IFIC, Univ. Valencia – CSIC, Spain)
- **Tobias Jenke** (ILL, Grenoble, France)
- **Andreas Knecht** (PSI, Villigen, Switzerland)
- **Magdalena Kowalska** (CERN, Geneva, Switzerland)
- **Bastian Märkisch** (TU München, Germany)
- **Natalia Oreshkina** (MPIK Heidelberg, Germany)
- **Nancy Paul** (LKB, Paris, France)
- **Guillaume Pignol** (LPSC Grenoble, France)
- **Martino Trassinelli** (INSP, Sorbonne Université, Paris, France)
- **Stefan Ulmer** (HHU Düsseldorf, Germany and RIKEN, Wako, Saitama, Japan)
- **Sven Sturm** (MPIK Heidelberg, Germany)

**Contributors:**
- **Barbara Maria Latacz** (CERN, Geneva, Switzerland)
- **Ekkehard Peik** (PTB Braunschweig, Germany)
- **Ubirajara Van Kolck** (IJCLab, Orsay, France, and University of Arizona, Tucson, USA)

# Introduction

Symmetries, referred to as the invariance of the laws of physics under a given transformation, play a fundamental role in physics. More than a century ago, Emmy Noether proved mathematically that for every continuous symmetry there exists a conserved quantity. An exemplary illustration of Noether's theorem is that the invariance of a physical system under time translation (i.e. the laws of physics remain the same if we repeat the experiment at different times) results in energy conservation, thus forming a fundamental principle in our understanding of the Universe. Lorentz and local gauge invariance are guiding principles upon which the Standard Model rests. However, not all symmetries perceived as natural in our everyday life - such as the absence of a preferential direction in empty space (vacuum) - are conserved in the Standard Model. The discovery in 1957 that spatial inversion (parity, P) is violated in weak interactions was a shocking revelation for the physics community, as evidenced by Pauli's letter to MIT director Weisskopf at that time: "Now, after the first shock is over, I begin to collect myself. Yes, it was very dramatic." A few years later, CP violation, the combination of parity with charge conjugation (C), was also discovered in the kaon system. Over the years, accumulating evidence has shown that all discrete symmetries, including time reversal (T), are violated, with just their combination CPT still standing up against the most accurate tests. CPT is considered to be truly fundamental because Lueders' theorem states that if a quantum field theory is invariant under Lorentz transformations and satisfies the condition of locality, then the combined CPT symmetry is conserved. Investigating fundamental symmetries is therefore a task of utmost importance, vigorously pursued to reveal new phenomena that could help answer some of the open questions of the Standard Model, such as the origin of the baryon-antibaryon asymmetry. In the next decade, various nuclear physics experiments, including searches for electric dipole moments (EDMs), precision studies of radioactive molecules and beta decay, will push CP violation testing to the next level. Simultaneously, the accuracy of tests for Lorentz/CPT symmetry will be boosted by more precise studies of matter-antimatter systems, such as comparing hydrogen and anti-hydrogen or studying the properties of neutrinos and anti-neutrinos. These precision tests explore facets of the four fundamental interactions that are not always accessible through colliders. They are conducted in smaller-scale facilities, including laboratories, reactors, and accelerators producing and manipulating highly charged ions, neutrons, muons, antiprotons, and radioactive ion beams. These experiments provide unique access to precision measurements of fundamental constants, such as masses or radii of particles. Teams of physicists and engineers compete with ingenuity to develop a varied range of techniques and probes, which drive this research forward. Quantum Electrodynamics (QED) is tested to high precision in highly charged ions and exotic atoms. The fundamental nature of the weak interaction is probed by scrutinising the beta decay of neutrons and radioactive ions. On the theoretical front, continuous progress in Effective Field Theories (EFT) makes possible a meaningful comparison between colliders and low-energy searches in terms of their sensitivity to New Physics.

The behaviour of a system under a given transformation can reveal the properties of its interaction, as demonstrated by parity violation showing that weak interaction is a chiral theory. Nuclear physics has played a major role in these advancements and it will continue to do so, offering new, unique, and complementary insights into all known interactions—electromagnetic, weak and strong—and also into the gravitational behaviour of neutrons and antimatter, as presented in this chapter.

# Fundamental symmetries

## CP violation

### EDMs

Because the existence of a non-zero electric dipole moment (EDM) for any fundamental particle with spin would violate both parity (P) and time reversal (T) symmetry, the search for EDMs provides a promising avenue for discovering new CP-violating interactions that could potentially explain the matter-antimatter asymmetry of the universe. Due to their high discovery potential, numerous experiments have been conducted or are planned on different systems, including neutrons, stable and radioactive atoms, molecules, and charged particles.

While the search for the neutron EDM is a well-established field, future experiments hold the promise of improving sensitivity by up to two orders of magnitude beyond the current upper limit of $2 \times 10^{-26}$ e cm. Over the next decade, several dedicated instruments are expected to become operational at various facilities in Europe – the most advanced experiments being n2EDM at PSI and PanEDM at ILL – and in North America at TRIUMF and Los Alamos. In addition, two new concepts are being developed in long-term R&D programmes in view of a future experiment at the ESS: Beam EDM based on a cold neutron pulsed beam, and EDM$^n$ which takes a cryogenic approach. All experiments involve exposing spin-polarised free neutrons to an electric field and then measuring the resulting (tiny) effect on the neutron spin. The progress in this field critically relies on the availability, performance and reliability of ultracold neutron sources. Ultracold neutrons are low-energy neutrons that can be stored in a measurement chamber for durations comparable to the beta decay lifetime of the neutron. In Europe, significant efforts are being made to develop such sources, particularly at PSI, ILL, FRM-II and possibly at the ESS. Additionally, achieving precise control of the magnetic field within the measurement chamber is crucial, necessitating advancements in dedicated techniques for magnetic shielding and quantum magnetometry. This aspect is also relevant to the search for other EDMs in paramagnetic molecules (such as YbF, ThO, BaF, HfF) or diamagnetic atoms (such as $^{199}$Hg, $^{129}$Xe).

Another approach to discovering fundamental EDMs involves charged particles in storage rings. In this class of experiments, an EDM could be observed as an out-of-plane precession of the particle spin. Currently, experiments in this area are in the development stage, with a dedicated setup being built for measuring the muon EDM at PSI. Furthermore, more ambitious plans involve constructing storage rings for protons and deuterons, which require substantial research and development efforts. In Europe, this effort is led by the JEDI and CPEDM collaborations.

Figure 6.1 shows the overview of past and future EDM searches of various systems. They are all complementary in the search for new physics. To establish connections between low-energy observables, particularly the EDMs of various simple or composite systems, and the diverse possible sources of CP-violation in new physics models, a comprehensive theoretical framework is necessary. This framework involves a complex ladder of effective field theories encompassing atomic physics, nuclear physics, QCD, and the so-called SMEFT. However, there are still important gaps in this theoretical framework, with some of the connecting factors carrying theoretical uncertainties greater than 100%. Therefore, improvements on the theoretical side are also relevant and necessary.

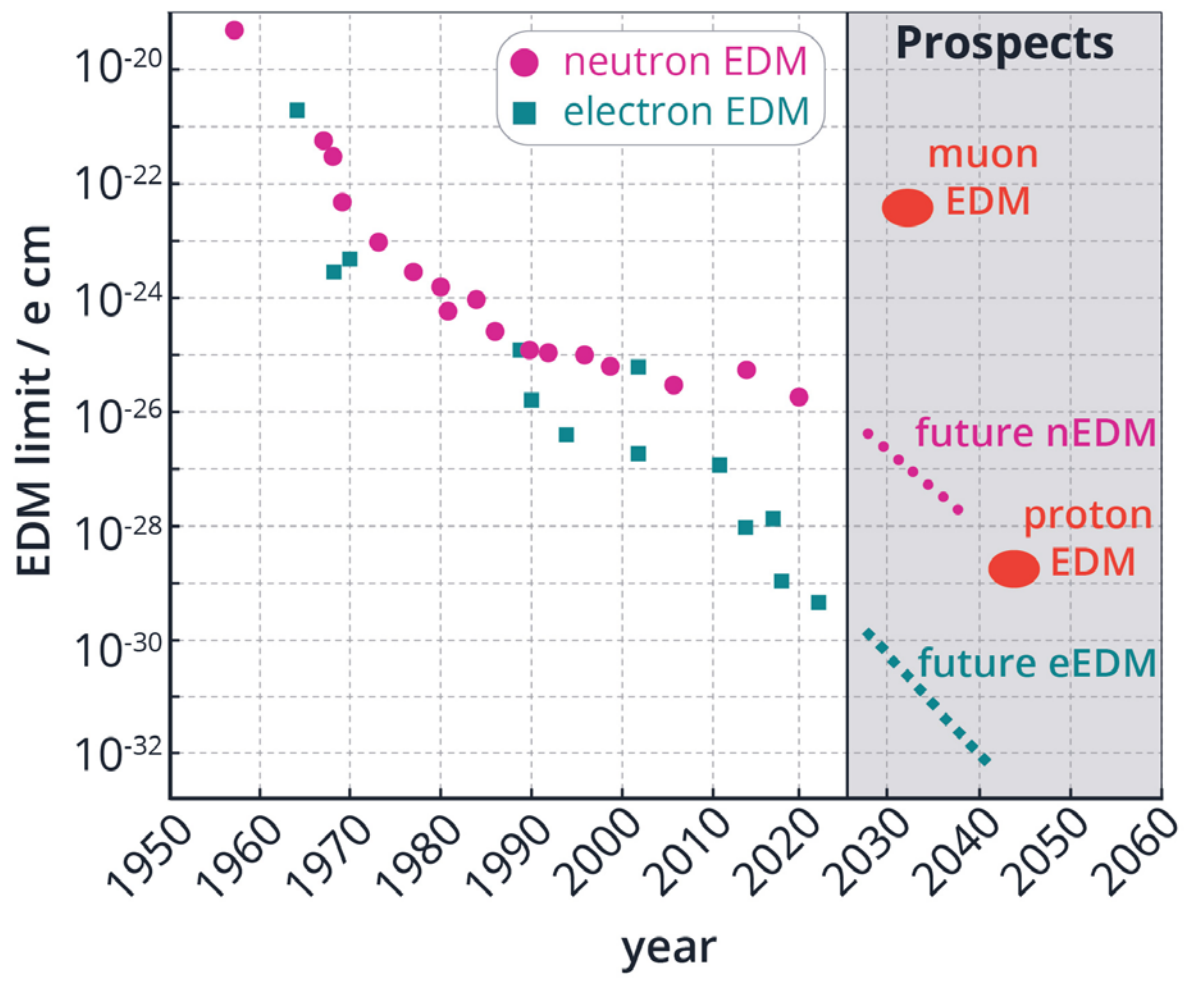

*Fig. 6.1: History of the upper limit on the EDM of the neutron and electron, and prospects of future experiments with neutrons, electrons, protons and muons.*

## Radioactive molecules

### Box 6.1: Radioactive molecules: powerful tool and unique laboratory

**The production and study of radioactive molecules is fast acquiring momentum at radioactive ion beam facilities across Europe and beyond. The motivation for studying the structure and dynamics of molecules containing short-lived radioactive nuclei is multi-faceted and covers areas of both fundamental and applied science in regions of the nuclear chart where molecular studies have so far been too challenging.**
**For heavy species, gas-phase spectroscopy provides powerful benchmarks of the predictions of ab initio quantum chemistry in regions where relativistic effects are crucial, the chemistry of 5*f*-electrons is not fully understood, and experimental data is scarce. Meanwhile, producing isotopically pure compounds of the early actinides is important for understanding the isolated molecular dynamics of relevance to nuclear engineering and radioactive waste management. Simultaneously, the optimisation of the ISOL production of molecular beams that are purer and more intense than the constituent atomic beams is also of direct importance for the future of ISOL as a production plan for medical radioisotopes. Finally, some of those radioactive molecules may prove ideal laboratories for searches of physics beyond the Standard Model.**

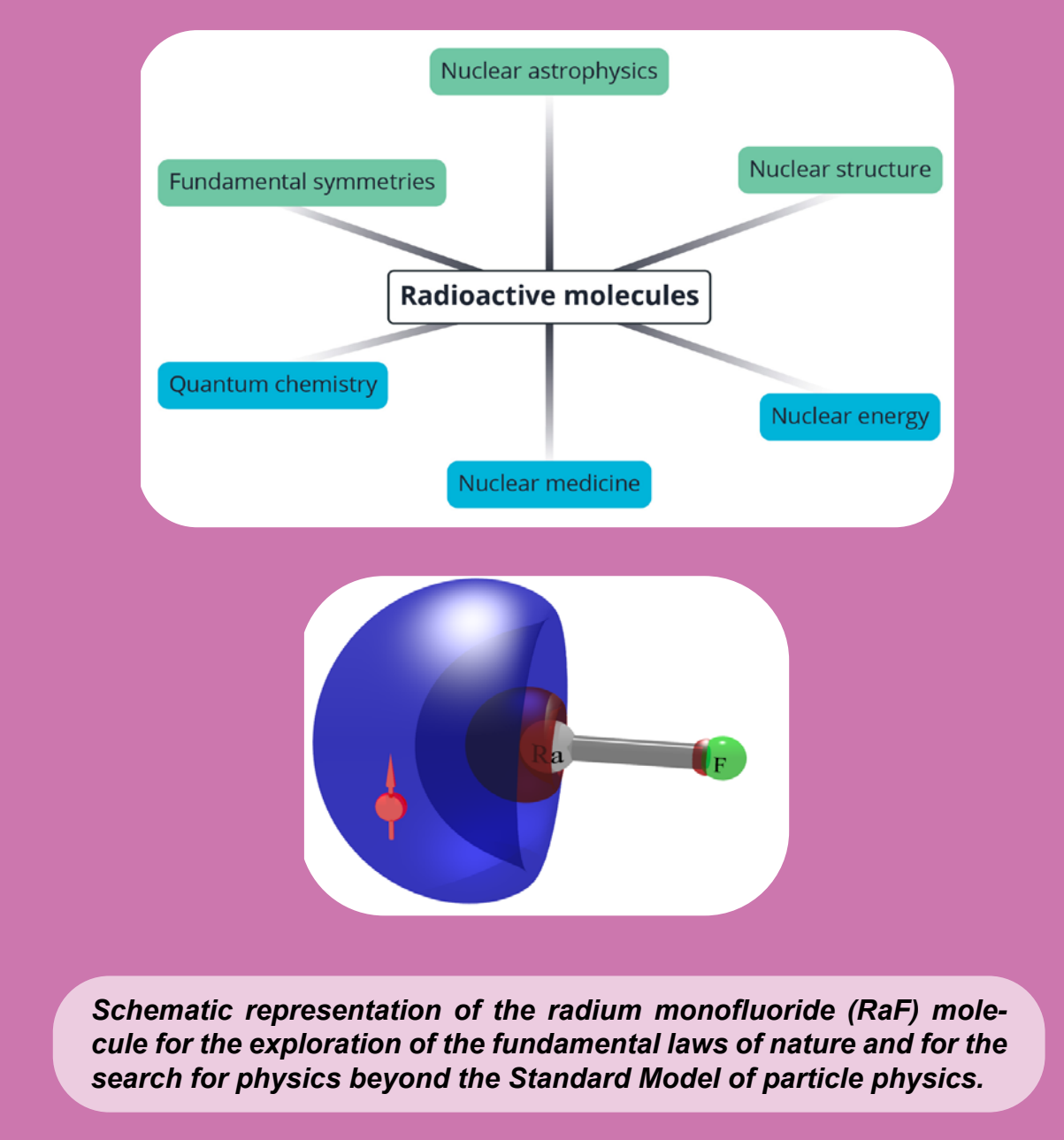

*Schematic representation of the radium monofluoride (RaF) molecule for the exploration of the fundamental laws of nature and for the search for physics beyond the Standard Model of particle physics.*

Radioactive molecules (Box 6.1) are a promising new area of study in which the signatures of symmetry breaking can be amplified by many orders of magnitude relative to nuclear and atomic systems. Though studies of short-lived radioactive molecules are in their early stages, the rapid advances made in the ability to produce, trap and perform spectroscopy upon them have shown great promise. Furthermore, future radioactive beam facilities will be able to produce larger samples of the radioisotopes needed for such studies, allowing for higher precision measurements to be made.

The rich electronic structure in molecules can also be employed for nuclear-structure studies. Analogous to the atomic counterpart, the molecular structure can be highly sensitive to nuclear size, electromagnetic moments and nuclear observables that are not routinely accessible in the atomic systems. For example, due to the presence of stronger electromagnetic field gradients within certain molecules compared to their constituent atomic systems, the extraction of electric quadrupole or magnetic octupole moments of isotopes not accessible via atomic spectroscopy or other methods could become possible, or the precision of existing data could be significantly improved.

To explore the limits of the Standard Model and its extension theories, precision searches for the CP-violating electric dipole moment of the electron (eEDM) and nuclear Schiff moments are a promising approach, because the magnitude of the moments scales with the level of T breaking in the fundamental forces. Heavy polar molecules are a highly sensitive laboratory for these searches, thanks to the strong electric field (in the order of GV $cm^{-1}$) that is formed along their internuclear axis and the high polarisability of linear molecules which enables aligning the molecular field along the laboratory axes.

The most stringent upper bounds to the eEDM, down to the $10^{-30}$ e cm level, have been set from the spectroscopy of ThO and $HfF^+$, while ongoing global campaigns for nuclear Schiff moments expect an unprecedented sensitivity in molecule-based searches. The sensitivity of these searches scales with the strength of the internal electric field and thus with the atomic number of the heavy nucleus. Radioactive molecules, such as FrAg, RaF, or the polyatomic $RaOCH_3$ have thus been highlighted by theory for their expected sensitivity. In nuclear Schiff moment searches, the octupole deformation of heavy nuclei leads to an additional enhancement in sensitivity.

In the near term, laser spectroscopy of radioactive molecules proposed for precision tests is necessary to benchmark the quantum-chemistry methods that have indicated their sensitivity to CP-violating properties and to confirm their suitability for precision experiments. At the same time, progress in the delivery of intense and pure molecular beams with existing and novel sources is pivotal for improvement in the achievable experimental precision. In the long term, progress in coupling state-of-the-art precision setups to facilities optimised for the production of radioactive molecules should be followed by proof-of-concept experiments with intensely produced beams. ISOLDE can be expected to host a large number of short- and long-term experimental studies as part of its ongoing programme in molecular production and spectroscopy. Other European facilities that utilise gas-cell sources, such as IGISOL and next-generation installations, should be supported in the development of similar programmes, as the gas-cell environment is ideal for efficient molecular formation.

In parallel, advances in nuclear modelling will be crucial for interpreting results from measurements performed on molecular systems. An accurate description of nuclear magnetisation will be essential for linking the measured hyperfine structures and isotope shifts to nuclear properties. Cross-community efforts by nuclear physicists and quantum chemists will be key to tackling the interdisciplinary nature of the challenges posed in this exciting new field. Close collaboration will also be needed to link our descriptions of fundamental interactions within the nucleus to those used in the domain of high-energy particle physics. Our complementary studies of nuclear moments in the low-energy regime could play a critical role in shaping future descriptions of the fundamentals of nature and directly contribute to ground-breaking discoveries.

### Beta decay

Complementary to the search for permanent EDMs is the measurement of T (or equivalently CP) violating correlations in nuclear and neutron beta decays (see Box 6.3). As to nuclei, the measurement of the D triple correlation in the beta decay of spin-polarised mirror nuclei appears of particular interest. The CP violation would then come from a phase between vector and axial-vector couplings. This correlation will be precisely measured in the decay of trapped and laser-polarised $^{23}Mg$ and $^{39}Ca$ mirror nuclei by the MORA experiment. MORA has just started taking data at the University of Jyväskylä. A degree of polarisation exceeding 99% can be expected, using collinear laser spectroscopy on the cooled and trapped ions. Beam intensity and purity are assets for these kinds of measurements. Dedicated developments are undertaken for the upcoming DESIR facility at GANIL-SPIRAL2, which should eventually host the setup. Correlations depending on the transverse electron polarisation, a quantity that vanishes for the SM weak interaction, are a powerful probe of both CP-violating and CP-conserving BSM physics. With advanced neutron decay spectrometers using Mott scattering and particle tracking techniques and correlating the neutron spin with electron and recoil proton momenta, as is being developed in the framework of the BRAND project, seven such correlation coefficients (H, L, N, R, S, U, V) become accessible, five of which have never before been attempted experimentally (H, L, S, U, V). Such measurements can reach an absolute accuracy of about $5 \times 10^{-4}$ and offer completely different systematics to measurements of the more 'traditional' correlation coefficients (a, A, B, D) and additional sensitivity to CP-violating parts of scalar and tensor couplings. Such experiments can be performed with cold neutron beams at ILL-Grenoble, FRM II-München and later at the ESS in Lund.

# Lorentz / CPT invariance and other

## Antimatter

The apparent imbalance of matter over antimatter in our Universe is an intriguing puzzle within the realms of contemporary particle physics and cosmology, beckoning for resolution. To resolve this problem, various theoretical scenarios have been proposed, while numerous ongoing experimental endeavours diligently strive to elucidate the genesis of this asymmetry. Among them are experiments at the antiproton decelerator/ELENA facility of CERN, which compare the fundamental properties of protons and antiprotons (BASE-collaboration) and the optical spectra of hydrogen and antihydrogen (ALPHA and ASACUSA collaboration). Additional efforts within ASACUSA perform spectroscopy on exotic antiprotonic helium atoms and access, in combination with advanced three-body quantum-electrodynamics calculations, the antiproton-to-electron mass ratio. All these experiments are performed at low energy, using techniques of atomic, molecular and optical physics and reach fractional accuracies on the parts per billion level and below. Any detected difference in these measurements would challenge the fundamental charge (C) / parity (P) / time (T) reversal invariance, which is deeply imbedded in the relativistic quantum-field-theories of the Standard Model and would thus hint at new physics that could potentially contribute to better understanding the matter/antimatter imbalance.

In recent years, CERN has upgraded the antimatter facility with the new 100 keV ELENA synchrotron, which allows for parallel user operation and for more efficient use of the antiprotons produced by CERN's accelerator infrastructure, implying a long-term perspective of the programme. In the meantime, collaborations at the AD have reported substantial progress in the development of their experiments and the high-precision measurements conducted with these, as shown in Fig. 6.2.

The ALPHA collaboration has successfully synthesised and trapped antihydrogen atoms and measured the optical transition frequency between the 1S and 2S states with a fractional resolution on the level of 2 parts in a trillion. Very recently, the collaboration reported on the laser-cooling of antihydrogen atoms heralds a bright future towards measurements at $\approx 10^{-15}$ accuracies similar to those achieved in hydrogen. The ASACUSA collaboration has performed two-photon laser-spectroscopy on antiprotonic helium and determined the antiproton-to-electron mass ratio with a fractional precision of $8.2\times10^{-10}$. Using the much-improved antiproton beam provided by ELENA, considerable enhancements in their spectroscopy seem to be within reach. Another effort within ASACUSA is currently preparing an experiment for the in-flight ground-state hyperfine splitting spectroscopy of antihydrogen at an accuracy at the parts-per-million level. The BASE collaboration has determined the magnetic moment of the antiproton with a fractional precision of 1.5 parts in a billion. With a recent substantial upgrade of their apparatus, ten to 100-fold improved measurements seem to be within reach. In addition, BASE compared the proton-to-antiproton charge-to-mass ratio with a fractional accuracy of 16 parts in a trillion. This measurement constitutes the most precise test of CPT-invariance in the baryon sector to date and also tests the weak equivalence principle by matter/antimatter clock comparisons. Future planned charge-to-mass ratio spectroscopy with co-trapped magnetron-locked matter/antimatter crystals has the potential to improve the precision by at least another order of magnitude. Using the newly constructed transportable antiproton trap BASE-STEP, the collaboration is planning to perform future antiproton experiments in dedicated offline laboratory space, insensitive to the noise imposed by CERN's accelerator infrastructure. In the meantime, a branch of the GBAR collaboration is preparing for 100-ppm-level measurements of the Lamb-shift in antihydrogen. In addition to the efforts at CERN's AD, an offline experiment with electrons and positrons (LSym) has been proposed, aimed at differential matter/antimatter magnetic moment measurements with sub-p.p.t. resolution.

While all reported measurements are so far consistent with CPT-invariance, such experiments are potentially sensitive to new physics, such as asymmetric interactions between dark matter and matter/antimatter, or the interactions discussed within the Standard Model Extension (SME). The conceptual framework of the SME, an effective low-energy field theory that adheres to micro-causality and renormalisability, preserves translational invariance and covariance under shifts in the observer's inertial frame. However, it introduces violations of CPT symmetry and partially disrupts covariance under particle boosts. This framework offers a means to assess the sensitivity of various experiments to CPT and Lorentz-violating extensions to the Standard Model. The inclusion of interactions that break Lorentz and CPT symmetries leads to alterations in the Lagrange densities, extending them with interactions of dimensions of mass or energy. Consequently, within the framework of the SME, those experiments achieving the highest absolute energy resolution are deemed to yield the most pronounced sensitivities in Lorentz and CPT violations. In Figure 6.2, the current CPT-violating sensitivity of measurements across diverse systems is illustrated.

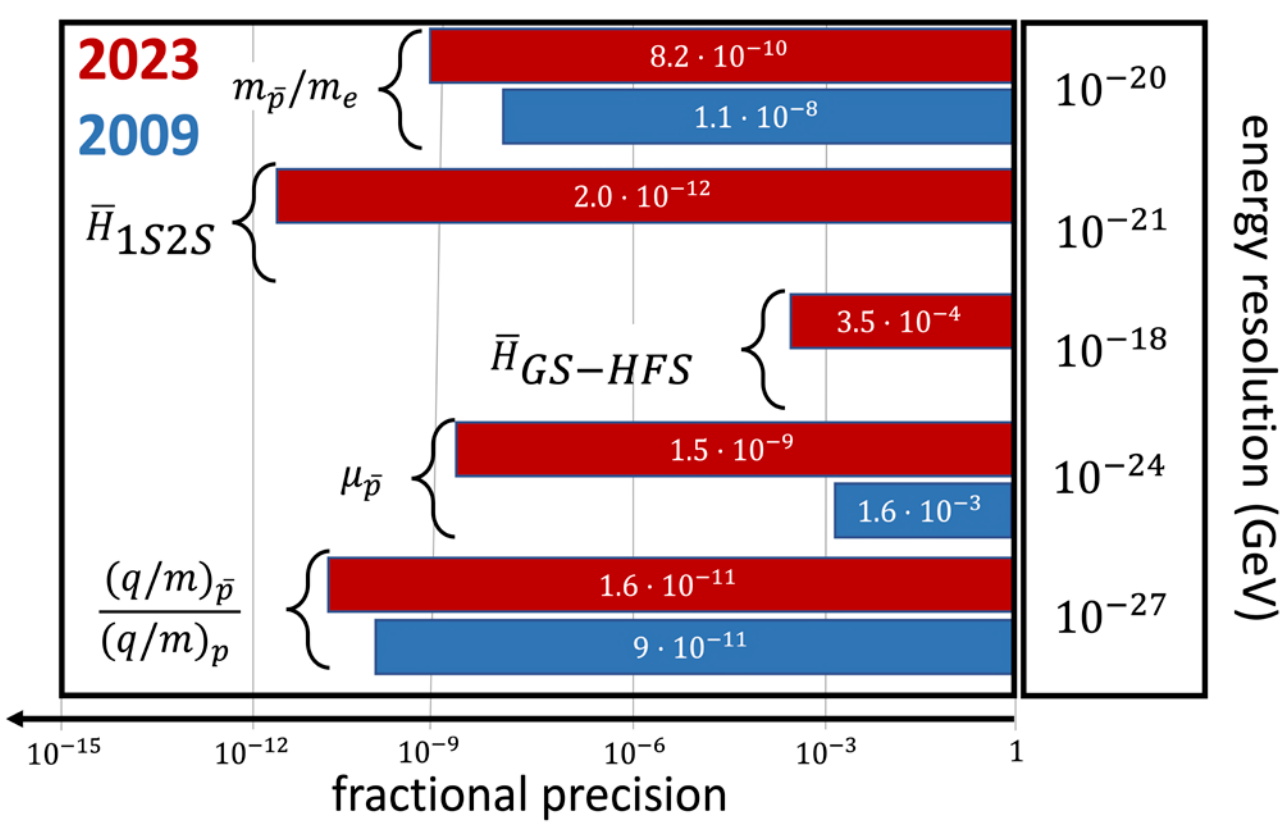

*Fig. 6.2: Comparisons of fractional precision of quantities measured within the AD/ELENA Programme of CERN, status 2009 (blue) and 2023 (red). Antiproton-to-electron mass ratio (ASACUSA), spectroscopy of antihydrogen (ALPHA), fundamental properties of antiprotons (BASE). The numbers on the right-hand side indicate the absolute energy resolution of the respective measurements. In the next 10 years, an improvement of at least one order of magnitude in the precision of those quantities is expected (see the text for more details).*

## Neutrinos

Being very abundant in the universe and participating in many processes across a wide range of energies and environments, neutrinos can be considered sensitive probes for CPT violation through a test of Lorentz invariance. For some of the most sensitive experimental contexts for this phenomenon nuclear physics represents a critical input to extract useful information. Direction-dependent effects of deviations from perfect Lorentz invariance can show up in kinematic measurements of the electrons from beta decays. In MAC-E spectrometers "KATRIN", for example, a time-dependent analysis of the Tritium decay end-point could show a correlation of the decay kinematics on the absolute orientation of the guiding magnetic field. Double beta decay is another context in which Lorentz violating mechanisms can generate measurable effects, in particular in the form of deformation of the kinetic energy spectrum of the two electrons emitted in the two-neutrino mode. Next-generation experiments, with high-resolution detectors and negligible background in a wide energy range overlapping with the electron energy spectrum, can provide high precision and high accuracy measurements of the spectral shape. Furthermore, CPT-violating Majorana couplings in SME can trigger and enhance neutrino-less decay branch even for negligible values of the Majorana mass.

### Thorium Clock

The nucleus $^{229}$Th is a unique system for precision spectroscopy, featuring a low-energy (8.3 eV), long-lived (about 2200 s in vacuum) isomer connected to the ground state by a magnetic dipole transition (see also nuclear clock in chapter 7 - Application and Societal Benefits). With a wavelength of approximately 148 nm in the vacuum ultraviolet, the radiative transition allows for precision laser spectroscopy techniques developed in atomic physics. This property makes the system highly attractive for tests of fundamental principles and for frequency metrology. Considerations of the nuclear level structure indicate that this case of a nearly degenerate nuclear ground state and isomer must be due to a fortuitous coincidence where the changes in the contributions of the strong and the electromagnetic interactions to the energy levels cancel each other nearly perfectly. Theoretical determination of both components calls for performing advanced modelling within the nuclear DFT. Although the near degeneracy of the two levels is certainly beyond the reach of theory, their electromagnetic moments, radiative transition rate, and Coulomb energies can, in principle, be estimated. This will require simultaneously including the effects of the quadrupole and octupole deformations, spin polarisations, and configuration mixing along with full symmetry restoration and uncertainty analysis.

Such a finely tuned balance of different interactions will be highly sensitive to hypothetical effects of physics beyond the Standard Model, like for example variations in the fundamental coupling constants. While the precise information required on the $^{229}$Th nuclear properties is still incomplete, it can be safely stated that the sensitivity of the $^{229}$Th nuclear transition frequency to the value of the fine structure constant is significantly higher (possibly by 3 orders of magnitude) than in the most sensitive presently operated atomic clocks. At the same time, the nuclear resonance frequency can be largely insensitive to the common systematic frequency shifts that are of concern in atomic clocks. In consequence, a nuclear clock could be both highly precise and highly sensitive in fundamental tests like searches for violations of Einstein's equivalence principle and new particles and interactions. These tests could be performed by comparing the nuclear clock to atomic clocks based on resonant frequencies of the electron shell. Two different experimental approaches towards a $^{229}$Th nuclear clock are investigated: laser-cooled trapped $^{229}$Th ions that allow experiments with control on the nucleus-electron interaction and minimal systematic frequency shifts, and Th-doped transparent solids (for example fluoride crystals) enabling experiments at much higher particle number. The recent demonstration of the first laser excitation of the $^{229}$Th low-energy nuclear transition is a major breakthrough, marking a fundamental step towards the narrowband excitation required for clock operation.

An experiment is planned on the spectroscopy of one-electron ion $^{229}Th^{89+}$ at the ESR storage ring of GSI-FAIR. The expected lifetime of the ion state is estimated to be 5–6 orders of magnitude smaller than the neutral atoms, with a consequent gain in laser excitation probability.

### Lorentz invariance from fast ions

Lorentz invariance has recently been tested to the $10^{-9}$ accuracy level by measuring the time dilation in $Li^{+}$ ions with a velocity of 30% of the light speed at the ESR storage ring at GSI-FAIR. New experiments are planned in the future HESR ring of the same facility for ions in the ultra-relativistic regime.

### Pauli exclusion principle and collapse models

The Pauli exclusion principle is at the heart of quantum mechanics and the Standard Model. Its possible violations are deeply connected to CPT and Lorentz symmetry, but also quantum gravity.
World records have been achieved by the VIP collaboration at the INFN-LNGS laboratory in both Pauli exclusion principle violation probability and collapse models, through the observation of spontaneous X-ray emission from copper conductors. New developments in this area are foreseen. The activity is divided into Open system and Closed system experiments. In the former, the existing apparatus will be upgraded with a focus on new detectors and target designs for increasing the sensitivity. In the latter, dedicated data-taking will investigate new aspects and effects in Pauli exclusion principle violation predicted by quantum gravity, such as anisotropy of a possible signal, and explore the connection with CPT and Lorentz symmetries.

## Lepton-flavour physics

### Neutrinoless double beta decay and Lepton number violation

Neutrino-less double beta decay is currently considered the most experimentally accessible lepton-number-violating process to be observed in a laboratory with the currently available experimental techniques. Double beta decay is the preferred decay mode for some even-even nuclei that cannot undergo single beta decay due to energy conservation or angular momentum suppression. Two neutrons are converted into two protons within the nuclear medium in a fourth-order weak decay with the production of two electrons and two anti-neutrinos in the final state. Neutrino-less double beta decay (NDBD) is an SM-forbidden alternative decay channel. Resulting in a final state where two electrons appear but no anti-neutrino is produced (as is the case for the two-neutrino decay mode), the total lepton number is increased by two units by this process. As a consequence, matter is effectively created in decay, making it among the most experimentally accessible candidates to explain the matter-antimatter asymmetry that characterises our universe.

Despite the unprecedented impact of its discovery on our understanding of Nature, the search for NDBD comes with exceptional experimental challenges. Candidate nuclei have half-lives in the $10^{19}$ - $10^{24}$ years while current limits on the neutrino-less mode correspond to half-lives over $10^{25}$ years. This translates into several expected events of a few per ton of observed isotope per year of measurements. Therefore, the scale of current generation competitive experiments is automatically defined to be in the hundreds to thousands of kilograms, extremely low background levels are needed as well as deep underground hosting laboratories (Laboratori Nazionali del Gran Sasso and Laboratorio Subterraneo Canfranc in Europe, SNOLAB in Canada, Kamioka Observatory in Japan, Jinping Underground Laboratory in China). Few competitive technologies are currently available and demonstrated, as described in chapter 9: cryogenic calorimeters (CUORE, CUPID, AMORE), high purity germanium diodes (GERDA, MAJORANA, LEGEND), liquid xenon time projection chambers (EXO) and scintillator-based setups (SNO+, KAMLAN-ZEN). With every new generation, background reduction and mass increase become more challenging. A worldwide effort is needed to make even the procurement of the required amount of stable isotopes feasible, as the cost and time constraints associated with the required enrichment process are quickly becoming a limiting factor. At the same time, R&D on new technologies able to increase the purity of materials (contamination levels are already at the limits to which currently available screening techniques are sensitive) and the capability of rejecting background (quantum technology-based sensors, for example) require a constant effort and should take advantage of any available synergy within and outside of the field.

### Lepton flavour violation

Lepton flavour violation has been established in neutrino oscillations, but has so far not been observed in the charged sector. With a large suppression, lepton flavour violation in the neutrino sector also induces lepton flavour violation in the charged sector. However, this violation should appear only many orders of magnitude below experimental sensitivities. This leaves room for a large variety of Beyond Standard Model physics where charged lepton flavour violation (cLFV) arises quite naturally. Searches for cLFV in muon decays have a long history and currently provide the most stringent limits ($\tau$ decays are also being used, albeit with less sensitivity) such as, for example, the MEG experiment setting the upper limit for the branching ratio of the $\mu \rightarrow e\gamma$ decay to $3.1\times10^{-13}$ at 90% C.L. Several international collaborations aim at considerably improved searches: MEG II at PSI aims at $5\times10^{-14}$, Mu2e at FNAL and COMET at J-PARC search for the conversion of negative muons $\mu \rightarrow e$ in the field of a nucleus aiming at 10−16 and Mu3e at PSI will search for neutrinoless $\mu \rightarrow eee$ decays in two steps pushing to $10^{-15}$ and $10^{-16}$, requiring for the second step the completion of the HIMB project at PSI to boost the currently available muon beam intensities. All the different searches are complementary, probing the underlying new physics in different ways and will be crucial if cLFV is discovered.

### Lepton flavour universality

A measurement of the charged-pion branching ratio to electrons vs. muons – $R_{e/\mu}$ is a sensitive test of lepton flavour universality (LFU). PIONEER, a recently approved experiment at PSI, aims to improve the current knowledge of this quantity by an order of magnitude, thus reaching theoretical accuracy. Data-taking will start in 2029. In addition, PIONEER will contribute in a later phase to tests of the weak interaction by precise measurement of the pion beta decay — see **3.2.1**.

### Parity violation and weak mixing angle

In the past decades, parity-violating electron scattering (PVES) has been established as a tool for precision measurements in the field of hadron physics. In the future, low energy PVES measurements will test the Standard Model and contribute to the search for new physics beyond the Standard Model with sensitivity to the TeV range. Such an experiment is the P2 experiment, which is currently under construction at the upcoming MESA accelerator facility in Mainz. It will measure the asymmetry $A_{PV}$ in the cross sections of polarised electrons off nuclei

$$A_{PV} = \frac{\sigma + - \sigma -}{\sigma + + \sigma -}$$

where $\sigma^+$ and $\sigma^-$ denote the cross section for positive and negative electron helicity. In the case of a proton target, $A_{PV}$ is related to the weak charge of the proton $Q_w(p) = 1 - 4 \sin^2 \theta_W$ and hence to the weak mixing angle $\sin^2 \theta_W$. Since in the proton case the asymmetry is small ($A_{PV} \approx 30 \cdot 10^{-9}$) and a relative uncertainty of $\sigma_{APV}/A_{PV} = 1.6\%$ is targeted, the luminosity in the experiment will be high ($L = 2.4 \cdot 10^{34}$ cm$^{-2}$s$^{-1}$). The MESA accelerator at Mainz is designed to fulfil the needs of the P2 experiment. Its uniqueness is the energy-recovering mode in a multi-turn, high-intensity operation which will be used for the first time for physics experiments. It will deliver a polarised electron beam with a beam current of $I = 150$ $\mu$A and a remarkable stability that will contribute to the uncertainty with only $\sigma_{Abeam} = 0.1 \cdot 10^{-9}$. The beam energy will be $E = 155$ MeV and the scattered electrons will be detected by a large solid angle detector with full azimuthal coverage and polar scattering angle acceptance of $25^{\circ} \leq \theta \leq 45^{\circ}$. This detector will consist of fused silica bars together with a photomultiplier tube readout. The averaged momentum transfer will be $Q^2 = 4.5 \cdot 10^{-3}$ GeV$^2$. The momentum transfer will be determined experimentally by a tracking detector based on HV-MAPS.

10000 h of data taking with a hydrogen target will provide a precision measurement of the weak mixing angle with $\Delta \sin^2 \theta_w / \sin^2 \theta_w = 0.15\%$, see Fig. 2.13. This precision is comparable with the most precise high-energy measurements so far at the Z-pole. Any deviation of the measurement result from the Standard Model prediction points to new physics. Examples are hidden weak scale scenarios such as compressed supersymmetry, lepton number violating amplitudes such as those mediated by doubly charged scalars, and light MeV-scale dark matter mediators such as the “dark” Z. The mass range for new physics in a four fermion contact interaction is up to 49 TeV. Measurements with other targets such as carbon or lead will complement the physics programme of the P2 experiment.

Alternative to parity-violation in neutral atoms, a measurement concept based on laser spectroscopy of $n = 2 \rightarrow 2$ atomic transitions in heavy He-like ions was proposed in the 80s, with a gain of several orders of magnitude in accuracy. At present, the limitation on theoretical prediction accuracy, which includes the nuclear size uncertainty, prevents the proposition of realistic experimental proposals. The goal of the new planned experiments on He-like uranium spectroscopy at the GSI-FAIR facility is to drastically reduce the uncertainty of $n = 2 \rightarrow 2$ intrashell transition energies, providing valuable new benchmark tests for the theory and thus making possible future parity-violation measurements in some electron heavy ions.
Parity violation can also be studied with high precision in nuclear beta decay, where it was first observed.

# Interactions

## QED tests and radii

### Highly Charged Ions

In highly charged ions (HCI), all but a single or few atomic electrons are removed. The remaining electrons are then tightly bound by the Coulomb field of the nucleus, orbiting very close to it, and are thus exposed to strong electromagnetic fields. In such few-electron systems, quantum electrodynamic (QED) effects are enhanced and extraordinarily precise ab-initio calculations of fundamental observables can be performed, such as transition energies, magnetic moments, Lamb shifts or hyperfine structure up to 11 digits of precision. By comparing complementary HCI systems, different aspects of the fundamental theories can be tested, notably strong-field QED. As a bonus, the strong binding of the remaining electrons makes HCI far less sensitive to external perturbations, such as blackbody radiation or Stark shifts, further enhancing the sensitivity of the measurements.

From the theory side, perturbation theory allows for separate contributions of different origins: leading-order QED, two- (three-) loop QED, and nuclear effects: finite size, deformation, magnetic distribution and polarisation. With increasing experimental accuracy, all-order theoretical methods are needed, an active area of development. Whereas QED effects can in principle be calculated to an unlimited accuracy, uncertainties of nuclear effects are determined and limited by our knowledge of nuclear physics. Especially in the high-$Z$ regime the nuclear charge radius is a major source of uncertainty for QED tests. Consequently, it is of utmost importance to have reliable independent measurements of charge radii, which might imply the re-measurement of already tabulated values with muonic atom spectroscopy (see next section). The bottleneck and most demanding aspect is the nuclear polarisation effect, which describes the interaction between electrons and internal degrees of freedom of the nucleus and requires detailed knowledge of the complete electronic and nuclear spectra including wavefunctions. Nuclear effects can be minimised by either considering less-affected quantities or (highly-excited) states or by combining a few observables to cancel nuclear contributions (specific difference approach or reduced g-factor). While this approach shows great potential to push strong field QED tests beyond the current limits, it requires significant development of the QED prediction for few-electron systems.

Until now, precision experiments have been performed mostly in low- and medium-Z ions. In the case of heavy ions, measurements of the g-factor of hydrogen-like ions up to $^{118}Sn^{49+}$ yielded a stringent test of strong-field QED up to 2-loop level. In the case of atomic transition spectroscopy, state-of-the-art experiments with H-like heavy ions were sensitive only to (all-order) one-loop terms. Measurements with Li-like heavy ions can be sensitive to isotopic effects and to two-loop effects. However, the latter cannot be disentangled from many-electron effects. Only recently, two-loop terms have been cleanly tested in He-like uranium by Bragg spectroscopy.

New ion sources for cold, heavy HCI are currently under development and will be available in the coming years. This will enable the determination of the bound-electron g-factor and laser spectroscopy of the hyperfine structure in trapped, heavy HCI at MPIK and at HITRAP at the GSI-FAIR facility. The Lamb shift will be measured to sub-eV precision at the ESR and CRYRING storage rings at GSI-FAIR, with new cutting-edge X-ray detectors and state-of-the-art crystal diffractometers. Furthermore, in the new planned HESR ring at GSI-FAIR ultra-relativistic collisions of heavy ions supercritical fields can be reached, resulting in the creation of on-shell electron-positron pairs and providing a unique window into the QED vacuum polarisation.

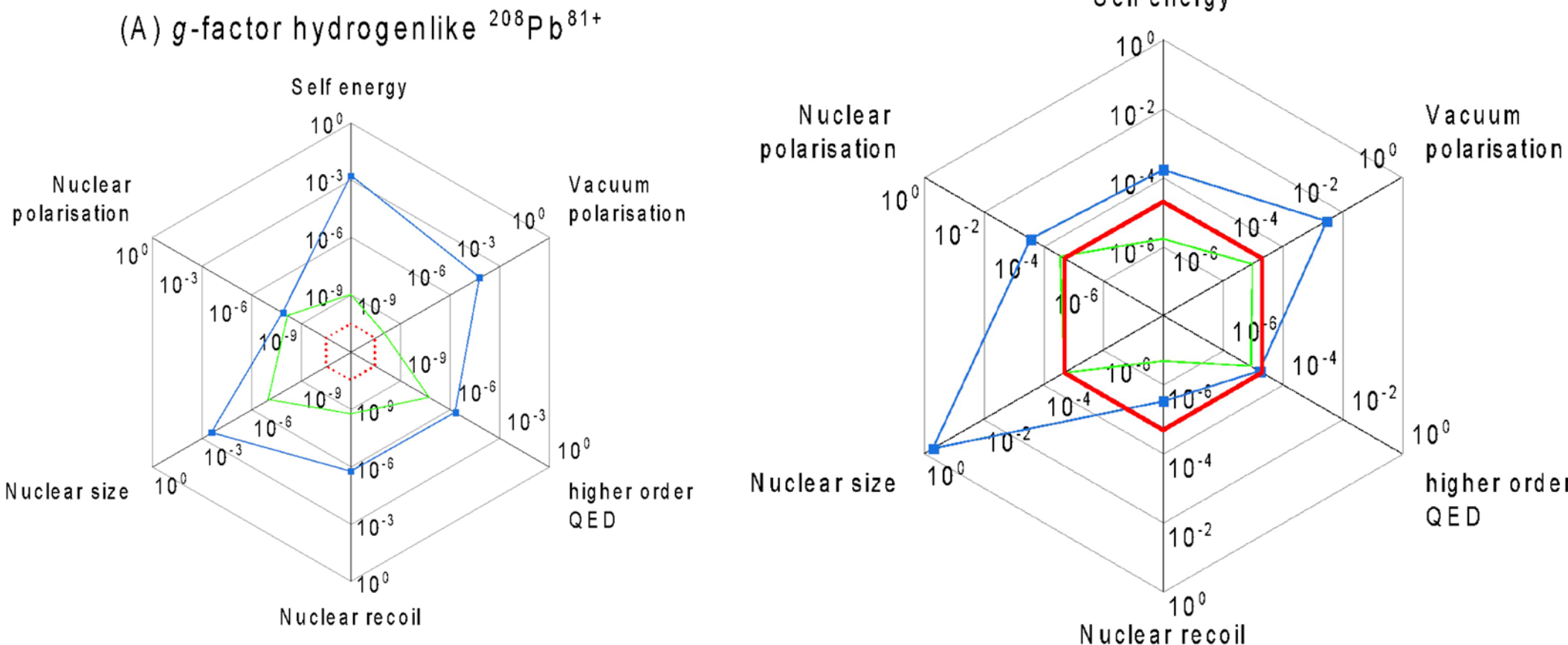

*Fig. 6.3: Fractional value (blue), theoretical (green) and experimental (red) uncertainties of selected contributions to different systems and observables. For (A), where a measurement has not yet been performed, the projected uncertainty for the experimental result and the yet uncompleted non-perturbative (binding) two-loop QED contributions are shown. The different systems are complementary so measurements should be done on several observables.*

With novel techniques such as quantum logic spectroscopy, microcalorimeter x-ray detectors and the coherent determination of g-factor differences with co-trapped ions, but also old techniques with new implementations like in-ring COLTRIM and laser pump/probe experiments, the accessible precision and consequently the bounds on new physics can be boosted by orders of magnitude, while complementary measurements in different HCI systems make it possible to disentangle the various facets of the theory. This complementarity is highlighted in Figure 6.3. Furthermore, if experiment and theory agree, stringent bounds can be placed on the existence of physics beyond the standard model, such as a fifth force.

## Muonic/antiprotonic atoms

A complementary type of highly-charged ion can be created with exotic atoms (Box 6.2) where the electrons are replaced by another heavy, negatively charged particle like a muon or antiproton. These systems are unique as the heavy muons (antiprotons) are about 200 (2000) times closer to the nucleus compared with electrons. This means that for low-lying atomic states, the sensitivity to the nuclear structure is enhanced by orders of magnitude, and short-range interactions like nuclear polarisation become prominent. For this reason, muonic atoms have been extensively used to study nuclear properties and extract nuclear charge radii, advancing in tandem with measurements of electronic species. For example, comparing the 1S-2S transition measurement in H with its theory prediction that makes use of the precise proton radius value from μH, leads to a precise determination of the Rydberg constant with a relative accuracy of $8 \times 10^{-13}$. Similarly, the alpha particle charge radius from μHe spectroscopy is needed to interpret the future 1S-2S measurement in He. With the Rydberg constant obtained by combining μH and H measurements, the confrontation theory-experiment in He provides a bound-state QED test particularly sensitive to challenging higher-order corrections scaling with $Z^{5..7}$. Future experiments at J-PARC, RAL, and PSI aim at measuring the ground-state hyperfine structure (HFS) in μH via laser spectroscopy, which in combination with the HFS in H will allow disentangling the Zemach radius and proton polarisability contributions from two-photon exchange. The laser advances required for these measurements will open the door to improving 2S-2P spectroscopy in μH, μD, and μHe by a factor of 5. When moving to heavier muonic atoms, new cryogenic x-ray microcalorimeter detectors will enable improvements by factors of 3-10 of the charge radii of Z = 3 − 8 nuclei from muonic atom spectroscopy, which will be compared with state-of-the-art *ab initio* nuclear theory calculations from chiral EFT. For medium and high-Z systems, large germanium detector arrays allow us to obtain the charge radii and nuclear ground-state properties from the measured muonic x-ray spectra. Current efforts are extending this method to measurements of scarce and/or radioactive elements available only in microgram quantities - with concurrent developments underway in the preparation of suitable targets and a novel method to transfer the muons to the nuclei of interest.

### Box 6.2: Exotic Atoms: unique probes of the Standard Model and Beyond

**Exotic atoms offer a unique and complementary approach to extracting fundamental constants, testing all known interactions including the validity of the weak equivalence principle for antimatter and searching for new physics while probing fundamental symmetries. Recent years have witnessed impressive progress in the field of exotic atoms driven by the development of improved beamlines and trapping techniques, manipulation of the constituent particles, quantum logic spectroscopy and tremendous advancement in technology (e.g. lasers, microcalorimeters, etc.). The next decade offers great prospects for this multidisciplinary research area, which merges different fields such as nuclear, atomic, particle, laser, quantum information and plasma physics.**

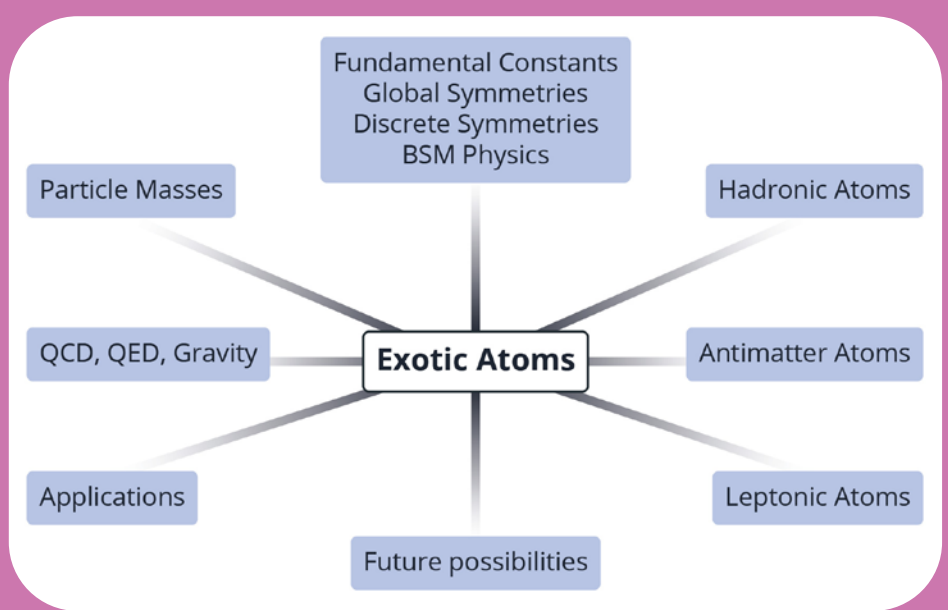

**What are exotic atoms? Ordinary atoms: positive nucleus which interacts electromagnetically with e⁻**

**Exotic Atoms: replace at least one of the two:**

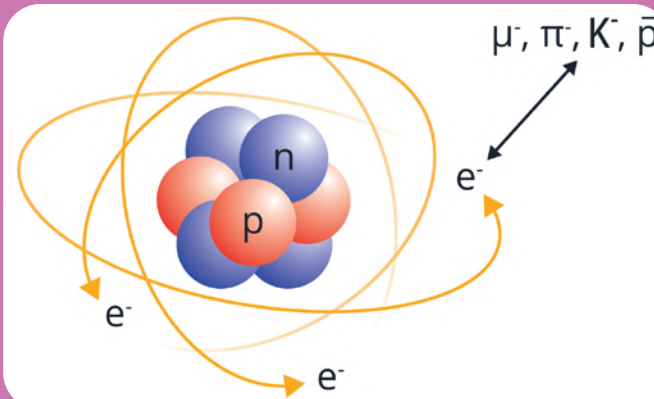

- **an e⁻ replaced by any negatively charged particle (muonic, pionic, kaonic and anti- proton atoms)**

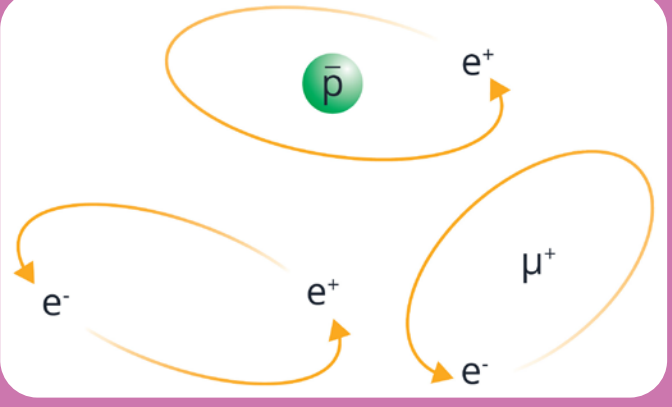

- **negative nucleus and positive orbiting particle (anti-hydrogen) or nucleus replaced by a positive particle (e⁺e⁻ positronium (Ps), μ⁺e⁻ muonium**

Antiprotonic atoms may also be used to probe nuclear properties and radii, through either historical x-ray spectroscopy, or via analysis of the annihilation products after antiproton nucleus collisions. The PUMA project at the CERN/ELENA facility will use the annihilation of antiprotons on nuclei to probe the so-far unexplored isospin composition of the nuclear-radial-density tail of radioactive nuclei. The observable PUMA pion annihilation will be complementary to the new HCI and muonic atom experiments planned for the next 5+ years to extract rms charge radii and isotope shifts.

The second global feature of exotic atoms is that the small radius of the exotic particle means that it also lies in orders-of-magnitude stronger Coulomb fields than traditional HCI. These systems thus allow for exploration of the strongest-field QED, notably by focusing on transitions between Rydberg states where nuclear uncertainties are negligible. Next-generation strong-field QED tests using quantum sensing x-ray microcalorimeter detectors and exotic atoms will be pursued and further developed with muonic and antiprotonic atoms accessing second-order QED across a vast range of Z, using the same technologies as to measure nuclear properties. As the leading order QED contribution in exotic atoms is vacuum polarisation and self-energy dominates in electronic species, these QED tests will complete and complement those with HCI.
Concurrently with these experimental goals, corresponding theory developments are essential to correctly describe exotic atoms. For muonic atoms, the limiting theory uncertainty for low angular momentum states derives from nuclear- and nucleon-structure effects. Improved theory predictions, especially for polarisability contributions, are needed to match up with the present experimental precision. Higher-order QED corrections including mixed vacuum polarisation self-energy diagrammes, diagrammes with mixed muon and electron loops and hadronic vacuum polarisation may be needed for high-accuracy measurements for all states. For antiprotonic atoms, the theory must be able to encompass different energy scales, starting from pure QED in the Rydberg regime, in the same way as for muonic atoms, to matter-antimatter interactions via the strong force leading to annihilation and meson production. Progress will be made by matching ab initio nuclear calculations with halo effective field theory in intermediate-mass systems. In heavy deformed muonic atoms, where the exact determination of nuclear polarisability is a challenge, nuclear DFT may offer an alternative approach. By treating the muon-nucleus Coulomb interaction self-consistently, where the muon moves in a deformed nuclear Coulomb field, and by performing the symmetry restoration for the entire atom an improved insight is obtained into the link between atomic measurements and nuclear properties. A complete theory is needed, starting from exotic particle cascade and including Auger and radiative decay until annihilation, and will be developed in the coming years. For PUMA, additional development is needed for a comprehensive treatment of final state interactions between annihilation products and the residual nucleus. For both muonic and antiprotonic atoms, developments at the interface between the nuclear and atomic theory communities are needed to support the new experimental efforts.

# Weak interaction

## CKM matrix

Beta decays (Box 6.3) give precise and unique access to a fundamental SM quantity: the $V_{ud}$ element, which parametrises the coupling of the W boson to first-generation quarks. In combination with extractions of $V_{us}$ from leptonic or semi-leptonic kaon decays and $V_{ub}$ from B decays, this permits checking the unitarity of the CKM matrix, a central feature of the SM.

The extraction of $V_{ud}$ (and the study of beta decays in general) represents a multidisciplinary effort, involving precise experiments with nuclei, neutrons and pions, as well as theory progress on nuclear calculations, QED and electroweak radiative corrections and New Physics implications.

Recent years have witnessed important progress on almost every front. Modern model-independent analyses based on EFT techniques have demonstrated the sensitivity of the CKM unitarity test to TeV scale new physics and its complementarity with searches at high-energy colliders such as the LHC. Interestingly, new techniques used to calculate the various theory corrections have revealed a tension with CKM unitarity. Additional progress on the kaon side has also generated tension between the various $V_{us}$ extractions. The overall intriguing picture ($\gtrsim 3\sigma$) goes by the name of Cabibbo anomaly. The progress on the experimental side is illustrated by the impressive precision of the latest measurements in neutron decay (lifetime and beta asymmetry), and in the decay of $^{37}$K, as well as by the development of new techniques. On the nuclear theory side, progress is needed in building consistent methodology based on unique approaches/interactions valid along the entire N=Z line that would incorporate varying nuclear configurations and analyse their impact on the beta decays.

Let us summarise the various methods used to extract the $V_{ud}$ element (see Fig. 6.4). Superallowed Fermi transitions currently provide the most precise determination, which is dominated by the error of the nuclear-structure and isospin-breaking corrections. With the expected increase in computational power and recent theory developments, ab initio nuclear methods may potentially produce precise and reliable results for these corrections, which would decrease the currently dominant error in the $V_{ud}$ extraction to subdominant levels. Low-mass nuclei (e.g. $^{10}$C) are particularly interesting because (i) they are amenable to nuclear ab-initio calculations with controlled uncertainties; and (ii) they play a crucial role in the extraction of the Fierz term (see next subsection). Precise measurements of higher-mass nuclei (A≥62) will allow the verification of the conserved vector current (CVC) hypothesis on a wider basis.

## Box 6.3: Beta decay

**Beta decay is a semi-leptonic process which has been used since the 1930's to probe the intrinsic symmetries of the weak interaction.**
**The beta decays of the pion, neutron and nuclei (see figure) grant access to a variety of observables in transitions which obey different spin selection rules. Complementary precision measurements enable the exploration of the large effective parameter space that characterises beta decay to look for new physics. $V_{ud}$, the first element of the Cabibbo-Kobayashi-Maskawa matrix, is inferred from Q-values, decay half-lives and branching ratios. Correlation terms in the beta decay Hamiltonian, which describe the decay probability as a function of the beta and recoil energies and directions, allow for the search of exotic currents or CP-violating interactions.**
**Each probe presents unique benefits and challenges: while nuclei are typically easier to manipulate than neutrons and pions, they require more extensive calculations for nuclear corrections. Progress for all probes is driven by collaborative experimental and theoretical efforts, with nuclei benefitting from advancements like ion traps and laser polarisation techniques, and neutrons and pions from enhanced beam intensities and detection methods.**

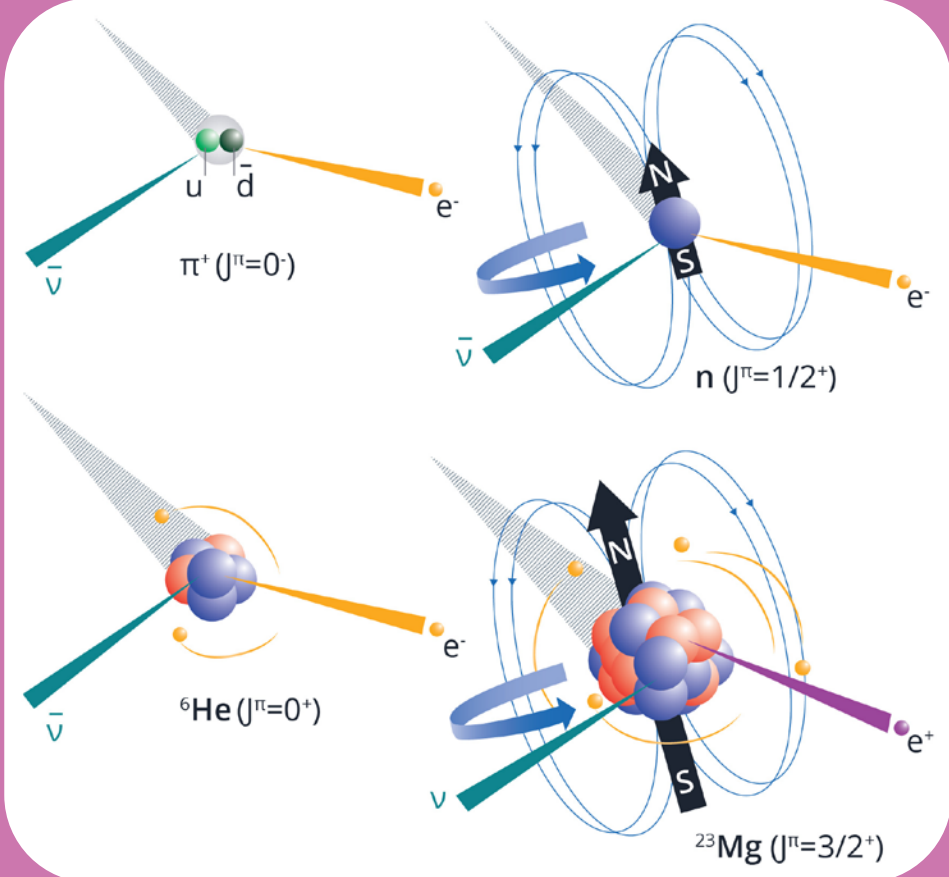

*Examples of beta decays used to probe the weak interaction. The dashed lines show the particle recoil. Not all electrons and nucleons are represented for $^{23}$Mg (Z=12).*

Secondly, we have mirror beta decays, mixed Fermi-Gamow-Teller $J^+ \rightarrow J^+$ transitions between members of an isospin doublet. The recent $^{37}$K measurement has shown that high-precision measurements of correlation coefficients are possible in these decays. Further, the corrected Ft-values for the transitions up to $^{39}$Ca are now at the level of 0.1% so they do not significantly contribute to the error on $V_{ud}$.
Thirdly, we have neutron decay, which is a particular mirror decay. Traditionally limited by experimental uncertainties, this could change in the next few years if recent experimental improvements continue and tensions between measurements are clarified. The aim of PERC at FRM II is the determination of the nucleon axial constant in neutron decay with a fractional precision of $10^{-4}$. Together with recent progress on the neutron lifetime, this could provide leading precision on $V_{ud}$ without nuclear corrections.

Lastly, we have pion beta decay, the cleanest channel from a theoretical point of view, but the hardest experimentally (due to the small branching ratio BR). A next-generation rare pion decay experiment (such as the PIONEER experiment at PSI) could increase the BR precision by a 3-10 factor, bringing this $V_{ud}$ extraction closer to the competitive region.
Recent developments have put the field in a position to make unprecedented progress if the necessary resources are available. In addition to the $V_{ud}$ extraction these improvements will have several other implications, such as the search for exotic interactions (see next subsection), QCD and nuclear-structure studies, and the precise study of sub-leading corrections (e.g. weak magnetism) which will open new research opportunities.

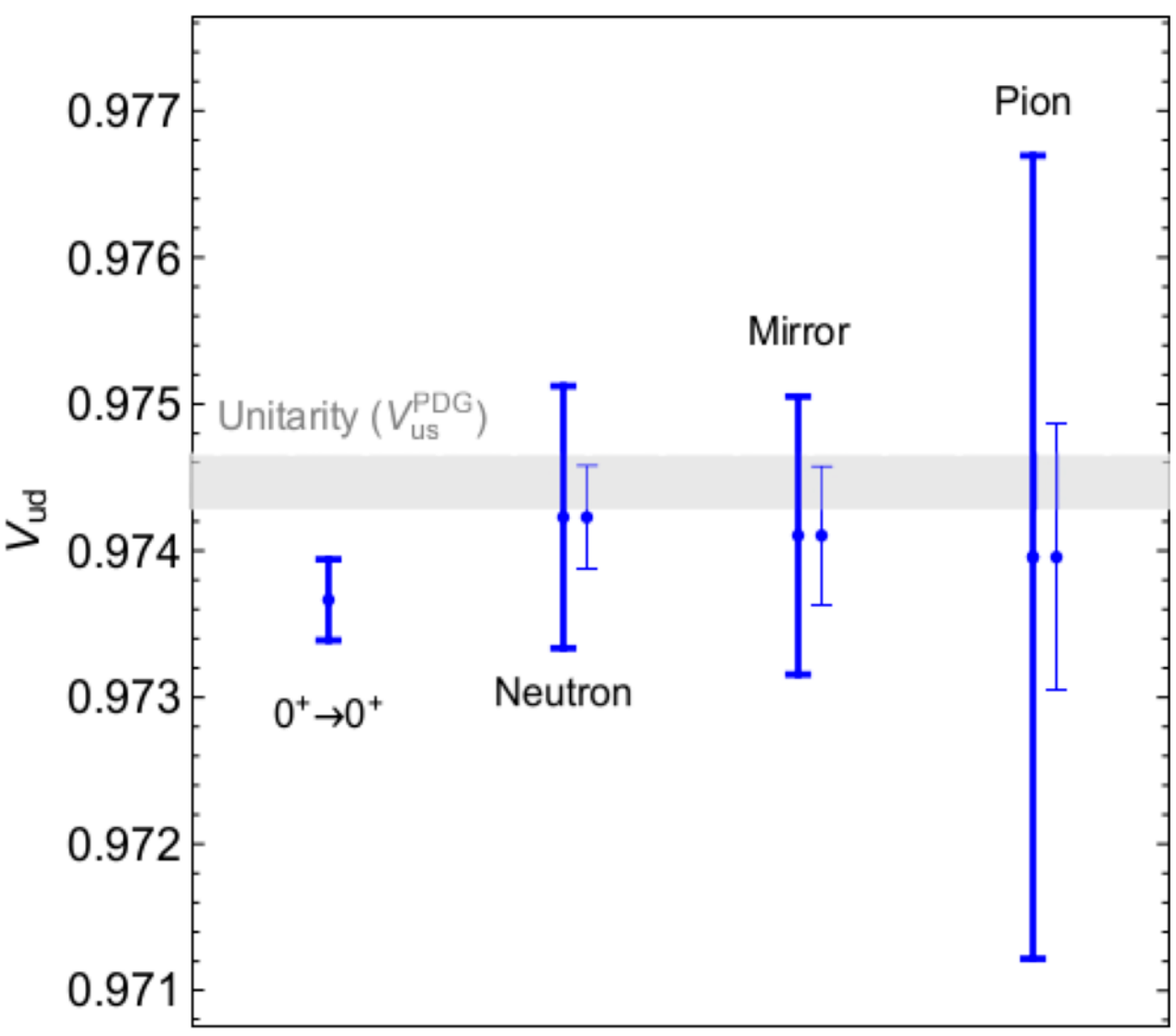

***Fig. 6.4: Summary of $V_{ud}$ extractions. Thick (thin) error bars show current results (projections). The neutron case projection is the nominal estimate obtained with current values for the lifetime and axial charge ($g_A$) without taking into account inconsistencies (i.e. without inflating errors by hand). This illustrates the expected reduction of the error if current internal tensions are resolved. In the mirror case we project a two-fold improvement in the final uncertainty, which is expected if more correlations and Ft values reach $10^{-3}$ - $10^{-4}$ sensitivities. The pion projection considers a factor of 3 improvement foreseen in the phase 2 of PIONEER. Improvements related to the $0^+ \rightarrow 0^+$ transitions are difficult to predict as they currently rely on intricate theoretical advances.***

## Exotic currents

Beta decays (Box 6.3) give precise and unique access to a fundamentIn addition to the determination of $V_{ud}$ , neutron and nuclear beta decay (Box 6.3) give information about hypothetical new particles and interactions that might be beyond the reach of current colliders. The study of such effects is carried out using an Effective Field Theory (EFT) framework. Non-standard effects are encoded in the so-called Wilson coefficients that can be constrained (or discovered!) using precision measurements, and whose sensitivity to New Physics can be compared to other low- and high-energy probes. A variety of measurements can be carried out at radioactive ion beam and neutron facilities. As meaningful examples, Wilson coefficients of non-standard (i.e. non V-A) scalar (S) or tensor (T) currents are sensitive to the existence of a first generation of Leptoquarks, while those related to right-handed currents are sensitive to the existence of $W_R$ bosons. For S or T currents, constraints competitive with high- and other low-energy searches are generally obtained from direct or indirect measurements of the so-called Fierz interference term, respectively in pure Fermi or GT transitions. This non-standard term affects the low-energy region of the beta spectrum shape, and appears in a normalisation factor for many observables in beta decay, whose sensitivity to the exotic currents depends on the type of transition and the experimentally accessible phase space. As such, spectroscopic measurements have been done for the precise determination of the Ft values of $0^+$ to $0^+$ transitions (see section above), for the test of CVC, and used for the determination of $V_{ud}$ , also providing stringent constraints on the existence of S currents. Experimental efforts are now focusing on light nuclei $^{10}$C and $^{14}$O to improve the limits on the Fierz interference term. To constrain T currents, several direct measurements of the beta spectrum shape are progressing: at GANIL with the b-STILED project (focusing on $^{6}$He), at ISOLDE with the InESS/WISArD setup ($^{114}$In) and at KU Leuven, in collaboration with the Jagellonian University of Krakow with the miniBETA detector (also $^{114}$In). The latter two experiments focus mainly on determining the weak magnetism recoil order term. Finally, the WISArD experiment at ISOLDE is looking simultaneously for S and T currents by measuring the beta–neutrino angular correlation in the decay of $^{32}$Ar for the two main GT and Fermi branches.

The above-mentioned experiments all aim at a precision of $10^{-3}$ on the Fierz interference term (or equivalently a sensitivity on the Wilson coefficient of $10^{-3}$ for these exotic currents) to compete with high-energy searches. They have long-term perspectives at upcoming radioactive ion beam facilities, like DESIR at GANIL and ISOL@MYRRHA, where dedicated instrumentation and enough beam time availability are expected. The use of new quantum detection techniques, like the superconducting tunnel junction developed within the frame of the SALER project, will be investigated for these searches. Such a detector will enable the measurement of the recoil energy spectrum of beta emitters with the highest precision. In neutron decay, the BRAND experiment will permit the simultaneous measurement of many correlations constraining different combinations of the Wilson coefficients at the same time. The instrument PERC at FRM II will employ a magnetic filter and a pulsed neutron beam to constrain the Fierz interference term from a measurement of the beta asymmetry or the beta spectrum on the $10^{-3}$ level.

## Muon capture

Nuclear models aimed at the description of the nuclear matrix elements of NDBD ($0\nu\beta\beta$) decays have traditionally been tested in connection with two-neutrino $\beta\beta$ ($2\nu\beta\beta$) decays and $\beta$ decays. However, some time ago it was proposed that the ordinary muon capture (OMC) could also be used for this purpose: $\mu^- + p \rightarrow n + \nu_\mu$. The $2\nu\beta\beta$ and $\beta$ decays are low-momentum-exchange processes ($q \sim$ a few MeV), whereas both $0\nu\beta\beta$ and OMC are high-momentum-exchange processes ($q \sim 100$ MeV). In this way the $0\nu\beta\beta$ and OMC are similar processes and possess similar features: they can excite high-lying nuclear states with multipolarities $J^\pi$ higher than $J^\pi = 1^+$. The $0\nu\beta\beta$ decay proceeds between the $0^+$ ground states of parent and daughter even-even nuclei through virtual states of the intermediate odd-odd nucleus. These same virtual states can be accessed by the OMC from either the daughter nucleus (electron-emitting $\beta\beta$ decays) or the parent nucleus (positron emitting/electron-capture (EC) $\beta\beta$ decays). Measurements on a variety of candidate isotopes are currently being pursued by the OMC4DBD/MONUMENT collaboration using the muon beams at the Paul Scherrer Institute. Additionally, theoretical calculations are being performed for the OMC processes to capitalise on the forthcoming results.

# Gravity

## Gravity of antimatter

The first proposal to measure directly the gravitational force acting on antiparticles using positrons was put forward 50 years ago by Witteborn and Fairbank. Such a measurement was later proposed at CERN with antiprotons. However, those experiments have not yet been carried out, since the use of charged particles to measure the effect of gravity requires the elimination of stray electromagnetic fields to a level that is extremely challenging. The solution to this problem is to use neutral systems composed entirely of antimatter, such as antihydrogen, or only partly, such as muonium and positronium. In 2022, the BASE collaboration reported a constraint on the differential matter–antimatter weak equivalence principle-violating coefficient to less than 0.03%. This measurement is extremely interesting; however, its results are partly model-dependent and not sensitive, for example, to gravi-photon coupling. A direct free fall experiment that investigates the ballistic properties of antimatter in the gravitational field of the earth would be model-independent and sensitive to a broader range of anomalous gravity/antimatter interactions. Three experiments with antihydrogen using different experimental schemes are currently pursuing this goal at CERN: ALPHA-g, AEGIS and GBAR. The experiment of the LEMING collaboration at PSI, and experiments at UCL and in Italy are being prepared to perform such measurements with muonium and positronium, respectively. The result obtained by ALPHA-g confirms that the behaviour of antimatter under gravity is the same as that of matter with 10% accuracy. In the next decade, those measurements will provide a test with an increasing degree of precision to a level of $10^{-3}$. With techniques such as laser interferometry and quantum gravitational states (developed for neutrons), uncertainty at a level of $10^{-6}$ should be achievable in the long term.

## Search for modified gravity with neutrons

Beyond-the-SM particles that may constitute the Dark Sector may modify known interactions and manifest themselves as exotic forces, e.g. the "fifth force" between two masses. In addition, the "new" particles may couple to charges and spins, producing a variety of possible exotic interaction potentials. Dark Energy, the nature of which is a prominent open question in Cosmology, could be caused by light scalar bosons with a coupling to ordinary matter and a screening mechanism. Possible realisations e.g. chameleon, symmetron, and dilaton fields, would manifest in precision fifth-force searches or equivalence-principle tests. Neutrons are powerful probes to precisely test theories of modified gravity, as they bypass the electromagnetic background induced by van der Waals and Casimir forces and other polarisability effects. Prominent examples are experiments with gravitationally bound quantum states of ultracold neutrons performed at ILL by the qBounce collaboration. In addition to probing for extra bosons and scalar fields as discussed above, searching for deviations from Newton's gravitational law also probes for other kinds of new physics, such as various forms of extra dimensions. Neutron interferometer experiments have contributed substantially to searches for Dark Energy scalar fields. In the next decade, the sensitivity of such studies will improve considerably due to several ongoing developments, i.e. split-crystal neutron interferometry at the instrument S18 (ILL), interferometry using very-cold neutrons and a setup of stored gravitational quantum states (qBounce), both implemented at the instrument PF2 (ILL).

# QCD

## Antiprotonic atoms

Antiprotonic systems offer unique opportunities to study strong-force interactions and annihilation, and eventually even to use the annihilation mechanism to create other exotic systems that would otherwise be difficult or impossible to study in the laboratory. The antiproton annihilation mechanism with respect to the participating nucleons and their constituents is poorly understood and models do not perform well in the sub-keV regime. The ASACUSA collaboration at AD/ELENA is developing a unique facility for slow extracted antiprotons that, when combined with a thin-foil method and a $4\pi$ detector, will permit determining total multiplicities after $\bar{p}$A annihilation for two nucleon systems, so-called Pontecorvo reactions, and three-nucleon systems such as $\bar{p}^3$He, for which no data exists.

The PUMA experiment at AD/ELENA and ISOLDE is also pursuing annihilation as a method to produce hypernuclei, where new symmetries and phenomena are produced by the strangeness degree of freedom and the in-medium hyperon-nucleon (YN) and hyperon-nucleon-nucleon (YNN) interactions may be explored. About 3% of antiproton-nucleon annihilations result in the production of strangeness, mostly kaons, which then re-interact with the residual nucleus and produce a Λ-hyperon which may then form a bound hypernucleus. Theoretical estimates show that 1% of the annihilations result in hypernucleus production, offering a new way to produce hypernuclei efficiently and perform high-resolution spectroscopy.

Antiproton annihilation can also be used as a novel route for producing radioisotopes. The AEgIS collaboration is developing a method based on charge exchange between co-trapped cold negative atomic ions and antiprotons, where annihilation may result in a broad range of trapped sympathetically-coolable HCIs with one to two fewer neutrons and up to three fewer protons, depending on the parent nucleus. In a second step, these cold HCIs can, through interaction with Rydberg positronium or Rydberg antiprotonic atoms produced in a subsequent burst, form highly-charged, hydrogen-like systems for a variety of applications like precision spectroscopy or EDM searches with non-symmetric heavy nuclei.

## Kaonic atoms

Kaonic atoms are an extremely efficient tool for the investigation of the strong interaction and relative symmetries at the low energy frontier in systems with strangeness, since they provide direct access to the kaon-nucleon/nuclei interaction at the threshold. The SIDDHARTA-2 experiment is currently being installed at DAΦNE aiming to perform, in the coming year, the first-ever measurement of the x-ray transitions to the fundamental level in kaonic deuterium, a crucial step towards determining the isospin-dependent antikaon-nucleon scattering lengths. In parallel, tests with newly developed detector systems are being carried out for future dedicated intermediate and heavy kaonic atom measurements. Beyond SIDDHARTA-2 the EXCALIBUR (EXtensive Kaonic Atoms research: from LIthium andBeryllium to URanium) proposal has been put forward for future dedicated high-precision kaonic atom measurements along the whole periodic table as a solid basis for all theories and models relying on kaonic atoms.

# Exotic physics

# Dark matter

Dark Matter (DM) is still a great puzzle, and the understanding of its origin is of massive importance for Cosmology and (Astro-)Particle Physics. A leading candidate for the constituent of DM is the Weakly Interacting Massive Particle (WIMP). However, so far, no WIMP has been found, despite the intensive searches at accelerators and direct detection experiments (see also chapter 9) A class of interesting theoretical models introduces the concept of an extended Dark Sector (DS), in which DM is not only composed of a single particle but in which a complete set of dark particles with their own phenomenology is conjectured. This is becoming an extremely fertile domain of exploration with many different techniques involved to address in a complementary way the very large parameter space of possible DM candidates (see Fig. 6.5).

Low energy experiments undertaken at Radioactive Ion Beam (RIB) facilities are capable of probing different portals to the Dark Sector. The recent quest for the X17 boson originates from an experiment carried out at ATOMKI. The boson's hypothetical existence was formulated to explain the observation of an excess of probability for pair creation at a certain $e^+$- $e^-$ angle after the de-excitation of high energy states of $^8$Be and $^4$He. Experiments such as New JEDI, currently taking data at Orsay, are testing the solidity of the experimental evidence of the excess. Others are looking for other types of evidence for this elusive particle in fixed target experiments at CERN, such as NA62 and NA64. A possible dark neutron decay was evoked to reconcile the differing measurements of neutron half-life, obtained in bottle and in-flight decay experiments. At GANIL, the possibility of a dark decay of one of the neutrons from the halo of $^6$He was investigated. The limit on the branching ratio has been enhanced by 5 orders of magnitude. Searches for a dark photon γ′ will be performed at the MESA accelerator by the MAGIX experiment via scattering of electrons on a heavy nucleus $e^-Z \rightarrow e^-Z\gamma'$ and the detection of its decay products. Light dark matter searches will also be performed at the beam dump experiment DarkMESA with an expected 3.5 · $10^{22}$ electrons on target. Antiprotonic atoms have also been proposed in the search for the sexaquark, a dark matter candidate, which may be produced through the formation of $\bar{p}$-$^3$He at rest. The AEgIS collaboration is exploring the possibilities of sexaquark detection via studies of antiprotonic atom annihilation products.

Possible DM candidates

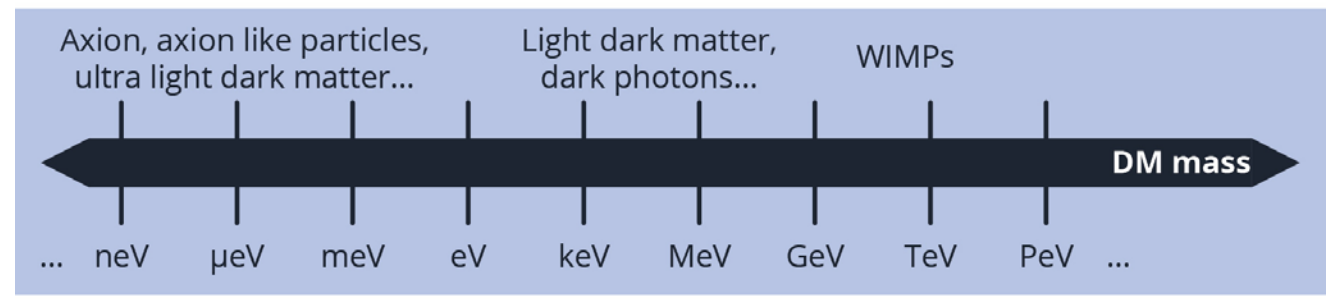

Atomic, molecular, nuclear clocks...

Accelerators, direct detection experiments...

Haloscopes, interferometers, mechanical resonators...

Non-exhaustive list of experimental techniques to search for DM

*Fig. 6.5: Mass range of DM candidates.*

# Fundamental constants and particle properties

## Neutrino properties

Among the many properties of neutrinos that still elude our insight, the absolute value of their masses, the mass ordering, their intrinsic nature (whether Dirac or Majorana) and the possibility of charge-parity violation stand as essential ingredients for constructing an accurate framework of physics beyond the Standard Model. Beta decay and double beta decay are the nuclear processes that permit the exploration of some of these properties.

### Mass

The non-vanishing mass of neutrinos, undeniably proven more than 20 years ago by flavour oscillations, is one of the obvious shortcomings of the Standard Model, where neutrinos are massless singlets with no coupling to the Higgs field. Given its deep implications on our understanding of Nature at the deepest level, an experimental determination of the absolute value of the mass (as opposed to mass differences that can be extracted with high precision by oscillation experiments) is paramount and the efforts in this direction have been growing at a steady pace in the last decades. There are currently three recognised ways of measuring neutrino mass: direct mass measurement, cosmology, and neutrinoless double beta decay. Direct mass measurement leverages the impact of neutrino mass on the kinematic properties of the final state of a weak process, in particular a nuclear beta decay or electron capture. A small but non-vanishing neutrino mass generates a deformation of the spectral shape close to the decay Q-value, de-facto moving the beta spectrum end-point at lower energy. Being the magnitude of the shift, a measure of the lightest mass eigenstate (or, more realistically, of an incoherent sum of mass eigenstates weighted by the corresponding PMNS mixing matrix elements), experiments aimed at observing this effect must implement an exceptional energy resolution on beta electrons (of the order of 1 eV or better) and a very deep and complete understanding of all instrumental effects potentially resulting in a sensitivity-spoiling systematic deformation of the spectrum. To maximise sensitivity, low Q-value nuclear transitions are preferred, like tritium or rhenium beta decays or holmium electron capture. Currently, the experiment leading the effort in this direction is KATRIN, hosted at KIT, Germany, which adopts a MAC-E filter to perform an integral measurement of the endpoint of Tritium beta decay. As more statistics are accumulated, KATRIN sensitivity will quickly become limited by systematic effects, and most of the community effort will be devoted to understanding and reducing their impact. In this respect, changing the operating conditions and replacing part of the system (for example the focal plane detector, as foreseen in the TRISTAN upgrade for the keV-scale sterile neutrino search) will provide valuable new information. The focus of the upcoming decade for the community working on this type of experiment should be to deepen the understanding and the modelling of all instrumental effects, as well as to develop new spectroscopic approaches that are more robust against them (like CRES Cyclotron Radiation Electron Spectroscopy, explored by the PROJECT-8 experiment). Experiments based on cryogenic calorimeters (HOLMES, ECHO) are currently limited to a relatively small number of detector channels. They can, potentially, quickly be scaled once the technology is mature. In this case, future efforts should focus on the multiplexing of the readout, enabling a cost-effective and technologically feasible scaling from tens to thousands of individual detector elements. Finally, the feasibility of combining two or more of the currently existing experimental approaches should be explored: a focal plane detector based on high-resolution cryogenic calorimeters coupled to a KATRIN-like spectrometer and high-activity gaseous source could, for example, overcome some of the sensitivity limitations currently imposed by background and systematic effects. The prospects regarding the sensitivity of experiments attempting to determine the neutrino mass in the next decade are shown in Fig. 6.6.

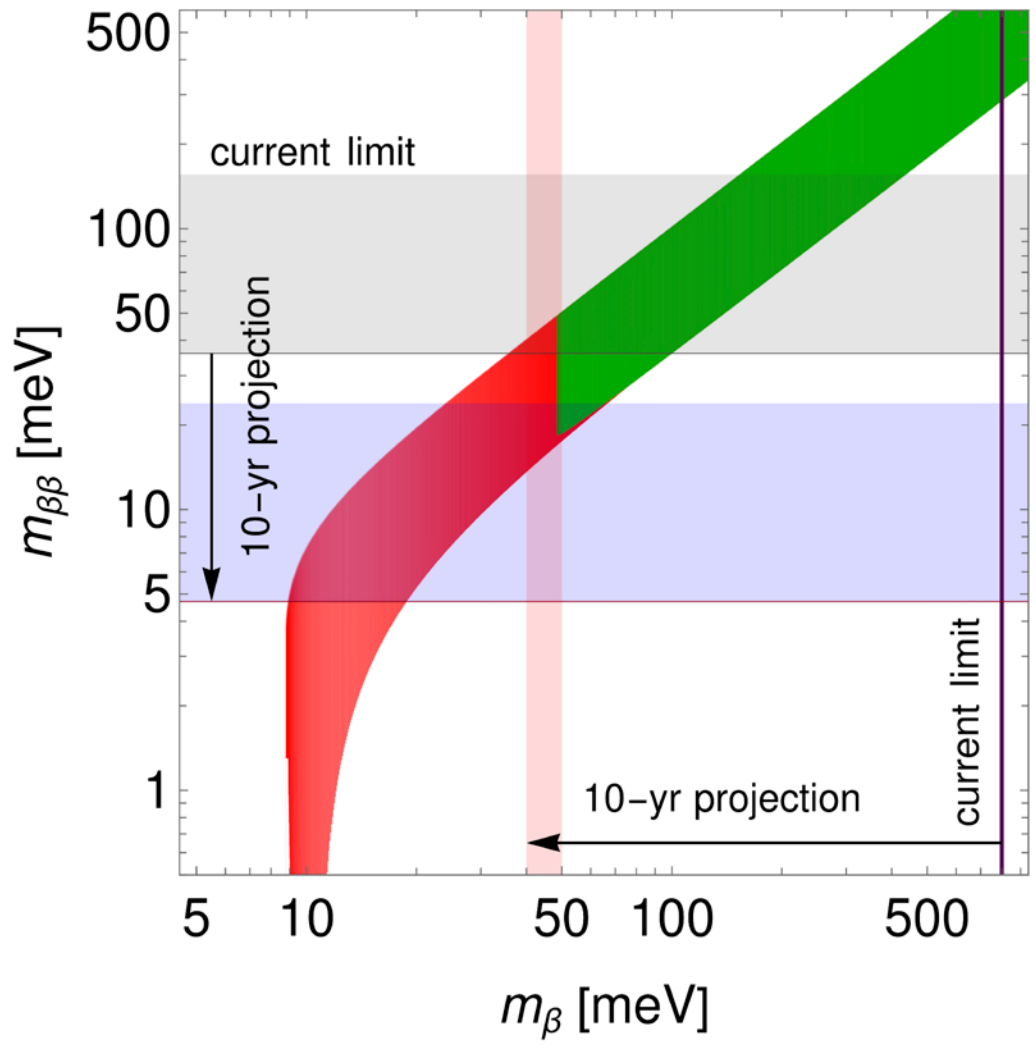

*Fig. 6.6: Combined parameter space for neutrino-less double beta decay (vertical axis) and direct neutrino mass measurements (horizontal axis); the regions currently explored are highlighted, as well as the regions that can be expected to be explored with the experimental developments in the next ~10 years. The grey band on $m_{\beta\beta}$ corresponds to the range of the currently published limits from the Kamland-ZEN experiment (depending on the choice of nuclear matrix element). The blue band spans the range of the projected sensitivity of the currently ongoing experimental effort 10 years from now, including nEXO, NEXT, Legend-1000 and CUPID with the available range of NMEs. On the $m_\beta$ axis, the purple line is the current KATRIN limit while the red band spans extrapolations from KATRIN++ and PROJECT-8 future roadmaps.*

### Nature: Majorana vs. Dirac and double beta decay

The origin of the non-vanishing mass of the neutrino raises an even deeper question about its nature, as almost all BSM theories require it to be a Majorana particle. At Standard Model energies, the effects of BSM theories are captured by higher-dimensional operators of an effective field theory, SMEFT, which are suppressed by powers of the large mass scale characteristic of BSM physics. The least suppressed operator, with dimension five, violates lepton number. It generates Majorana masses that allow neutrinos, the only fundamental neutral fermions in the SM, to be their own antiparticles. Consonant with SMEFT expectation, the first established deviation from the Standard Model was the measurement of non-vanishing neutrino masses. Since the same operator also leads, via Majorana neutrino exchange, to neutrinoless double beta decay, this most intriguing yet-unobserved nuclear decay process is a strong candidate for detection.

NDBD is for the foreseeable future the only way to test violation of lepton number. Should NDBD be observed and its magnitude (i.e. the branching ratio of this channel compared to the Standard-Model allowed two-neutrino double beta decay) measured, information about its absolute mass scale and the mass ordering could be extracted. For this to be accomplished, input from nuclear theory is essential. The formal expression of its inverse half-life $[T^{0\nu}_{1/2}]^{-1}$ relates this observable to the effective neutrino mass through the nuclear matrix element (NME) of the decay operator between the wave functions of the parent and daughter nuclei. This evidences the need to calculate accurate nuclear wave functions and, in particular, a sound knowledge of the mechanism of renormalising the decay operator (see Box 4.5 Nuclear Structure and Reaction Dynamics).

As has been discovered since the last NuPECC Long-Range Plan, the decay operator receives a leading contribution from the QCD mass scale. The strength of this contribution has been estimated from nuclear charge-independence breaking using constraints from chiral symmetry. However, a more precise determination, including its sign, is needed for a meaningful analysis of an eventual NDBD signal. Such a determination requires a combination of lattice QCD and chiral effective field theories. The new contribution, which represents correlations on the QCD scale, ensures the renormalisation of the decay operator in *ab initio* approaches. It has been incorporated in both *ab initio* calculations and more traditional many-body methods.

Ab initio solutions of the many-nucleon problem have progressed significantly in recent years and are starting to reach the NDBD of heavy nuclei. However, most of the nuclei of experimental interest lie in mass regions where exact calculations cannot be carried out. Truncations of the number of the nuclear wave-function components must be introduced and further renormalisation is needed for effective decay operators in reduced model spaces. The theory community focuses some of its efforts on the validation of nuclear models by reproducing observables that characterise the *β*-decay process. Since it is acknowledged that NDBD is dominated by the GT component of the decay operator, a main issue is the understanding of the so-called “quenching puzzle” of the axial coupling constant, namely the need by most nuclear structure models to resort to a reduction of $g_A$ in order to reproduce the observables directly linked to GT transitions. This empirical procedure affects the predictive power of nuclear models for reliable calculations of NMEs, being grounded on the availability of a large set of *β*-decay data — with neutrino emission – which may have very little connection with a decay that is a second-order electroweak process outside the framework of the Standard Model. During very recent years, the results of *ab initio* calculations for *β*-decay amplitudes have shown that the “quenching puzzle is an artefact of the truncation of the degrees of freedom in the nuclear Hamiltonian”. A controlled expansion of the components of the nuclear wave function and the inclusion of meson-exchange currents, accounting for the composite nature of the nucleons, may overcome such a long-standing problem. This paves the way toward a better understanding of the renormalisation mechanism of the GT decay operator, as well as to sounder predictions of the NDBD matrix elements.

# Particle properties

Fundamental constants, such as particle masses, cannot be predicted by theory and are thus an essential input to exploit the predictive power of the SM. In addition to the absolute value of neutrino masses, which are still unknown, improvements in our knowledge of the pion, kaon and muon masses are expected in the coming years.

## Mesons

At present, the most precise measurement of the negatively charged pion is obtained by X-ray crystal spectroscopy of pionic nitrogen with a final accuracy of 1.3 ppm. Laser spectroscopy of pionic helium was recently demonstrated at PSI. A possible gain of several orders of magnitude was obtained for the pion mass. The PiHe collaboration plans to search for other $\pi^4He^+$ transitions such as $(n, \ell)$ = (17, 16)→(16, 15) which are expected to be narrower by a factor of at least $10^{-3}$ compared to the resonance observed in the recent experiment. Helium gas targets in which the collisional shifts are small may be used, together with various laser spectroscopic techniques to increase the spectral resolution. The precision of the theoretical calculation of the transition frequencies is currently limited by the ppm-scale experimental uncertainty of the $\pi^-$ mass. The precision of the QED calculations themselves has recently been improved to a fractional precision of around $4 \times 10^{-9}$ for some transitions. This would lead to a determination of the pion mass, in principle, with a similar level of precision.

The most precise measurement of the negatively charged kaon is obtained by a weighted average value from several X-ray spectroscopy of transitions in kaonic atoms with a relatively high uncertainty of 0.013 MeV (two orders of magnitude worse than for pion) that is mainly caused by inconsistency between measurements. New experiments are planned at the DAΦNE facility with heavy atoms (Pb, W with solid targets) or light atoms (C, N with gaseous targets) to improve such accuracy.

## Muon

The measurement of the spectroscopic properties of Muonium, the bound state of the muon and the electron, allows extraction of fundamental constants such as the muon mass and its magnetic moment. This will become relevant for a robust comparison with theory of the g-2 muon projected uncertainty of the Fermilab experiment. At the Japan Proton Accelerator Research Complex (JPARC) the MuSEUM experiment has started commissioning, aiming to improve the muonium HFS and, thus, our knowledge of the muon magnetic moment. In Europe, the Mu-MASS experiment at PSI aims for a 1000-fold improvement in the determination of the 1S-2S transition frequency of Muonium which will provide the best determination of the muon mass at a level of 1 ppb. Combined with the results of MUSEUM, this will yield one of the most sensitive tests of bound state Quantum Electro-Dynamics (QED) with a relative precision of $1 \times 10^{-9}$.

The planned $\mu^+$ beam (HIMB) upgrade at PSI (see description of PSI in chapter 8) harbours tremendous opportunities for improving M spectroscopy experiments in the next decade. Higher statistics makes it possible to implement experimental techniques that are systematically more robust.

# Recommendations: Symmetries and Fundamental Interactions

### Theory

Precision tests of fundamental symmetries and interactions require theory to bridge the gap between very different energy scales. A diverse and comprehensive theoretical framework needs to be developed, where initiatives to build sustainable links between nuclear theory, quantum chemistry, atomic, molecular and particle physics are vigorously encouraged. Progress is driven by efforts in different fields like lattice calculations, ab initio methods, microscopic nuclear many-body models, EFT techniques, and dispersive approaches. In this context:

- Building links between these different approaches is of particular relevance for the emergence of new powerful probes, such as radioactive molecules for the search of CP violation, or the Thorium clock for fundamental tests.

- Continuous efforts must also be pursued for more traditional probes, which are reaching a level of precision where theoretical uncertainties dominate. This is, for instance, the case with precision measurements in nuclear beta decay and the determination of $V_{ud}$.

- The discovery of the striking process of neutrinoless double beta

decay will establish the existence of violation of lepton number and a Majorana mass for neutrinos. The quantitative interpretation of an eventual measurement directly relates to the nuclear matrix element, whose precise theoretical determination is required.

● Enhancing the sensitivity of QED tests in highly charged ions and exotic systems requires better knowledge of nuclear polarisation.

To enhance the discovery potential of these experiments, a precise and comprehensive global description of light and heavy open-shell even and odd nuclei is essential.

## Facilities, access to beam, and instrumentation

To have a substantial impact, precision experiments require dedicated efforts to develop tailored instrumentation at research facilities and secure access to beam delivery over extended time periods. In this endeavour:

● The specialisation of upcoming Radioactive Ion Beam facilities such as ISOL@MYRRHA and DESIR at GANIL-SPIRAL2 should be regarded as an opportunity not to be missed. In general, customised instrumentation and beam time availability should be guaranteed for fundamental tests at RIB facilities like ISOLDE, GANIL-SPIRAL2, and ACCLAB/IGISOL.

● Multidisciplinary research infrastructures like ILL, FRM-II and PSI should prioritise availability and access to their cold and ultracold neutron beamlines, and support infrastructure upgrades to strengthen their unique programmes in fundamental physics with neutrons. The long-term operation of ILL should be ensured beyond 2033 until the corresponding infrastructures for fundamental neutron physics at the upcoming ESS facility are established. A cold neutron beam line for fundamental physics also featuring a UCN source at ESS should be strongly supported.

● Continued support should be granted for the development and commissioning of facilities for the production and storage of heavy highly charged ions, such as HITRAP, HESR and high-energy EBITs. The operation of the CRYRING and ESR at GSI-FAIR should be maintained.

● The AD/ELENA physics programme at CERN should be strongly supported over the long-term, including running experiments, planned projects, and potential new proposals.

● The HIMB upgrade at PSI will provide for improved measurements and new experiments in both fundamental and applied physics. The project should be vigorously pursued and executed to allow for these exciting new possibilities, with the first beam expected in 2028.

To foster stronger links between research centres and universities, dedicated funding for joint positions should be allocated. Support for more permanent positions is essential for scientists involved in precision experiments, both at large facilities and universities, to ensure the continuous transfer of expertise to the younger generation of researchers.

Ensuring adequate funding for university groups engaged in precision experiments at large facilities is essential. Typically, the setups are developed and tested in university laboratories before being implemented in the facilities. It is worth noting that the facilities are primarily responsible for providing the necessary large-scale infrastructure and may not directly contribute to the setup's construction.

## Specific experimental developments

On the experimental side, we encourage specific developments, especially:

● Pursuing the development of quantum sensors, such as x-ray microcalorimeters, to advance precision studies of nuclear and atomic systems. It is also essential to actively pursue the necessary technical developments to ensure their compatibility with exotic beam facilities. In this context, we emphasise the importance of conducting crosschecks of old X-ray spectroscopy measurements on muonic atoms. It is highly recommended to provide support for remeasurement campaigns since underestimated systematic errors could potentially result in deviations from the Standard Model predictions. This becomes increasingly important as more precision experiments rely on precise charge radii as input.

● Addressing the present challenges of the direct neutrino mass measurement and neutrinoless double beta decay. Direct neutrino mass measurements complement model-dependent cosmological measurements. As sensitivity quickly becomes limited by systematic uncertainties in current experimental approaches, there is a critical need for cross-disciplinary efforts to integrate available technologies into experimental approaches that are more robust against systematic effects. Such an initiative can significantly enhance the sensitivity to allow cover inverted mass hierarchy. The search for a neutrino-less double beta decay must continue with a multi-isotope programme with at least two different technologies to mitigate the risk of false signal and reduce the impact of systematic uncertainties in observation. In this domain, a global and collaborative effort to increase the production rate and reduce the cost of isotopically enriched materials is needed. Also, relevant background sources are becoming harder to identify and mitigate before the full experiment is built and run. Interdisciplinary efforts to develop new materials, technologies and screening methods to reduce background sources are encouraged.

● Undertaking dedicated beam development at RIB facilities for precision measurements in nuclear beta decay. Several high-intensity and high-purity beams for nuclei close to the $N = Z$ line, from very light masses up to $^{100}$Sn would permit extension of the sets of $0^+$ to $0^+$ decays entering the determination of $V_{ud}$, enabling a critical test of theoretical corrections assuming CVC over a wider range of nuclei. Efforts should also be pursued for the determination of $V_{ud}$ from other systems, like mirror nuclei, neutron and pion decays at ad-hoc facilities. Other specific light nuclei must be developed to study recoil effects in nuclear beta decay and search for physics beyond the standard model via, for example, correlation measurements.

● Strengthening the development of radioactive molecular beams of interest for precision tests of the Standard Model and searches for new physics. The ongoing programme at ISOLDE for the production of actinide molecules provides a blueprint for implementation at other facilities, as a variety of production techniques and environments is important for the availability of all species of interest for precision spectroscopy. These efforts require ensuring adequate beam time and funding for technical developments in molecular beam production at radioactive ion beam facilities.

# Applications and Societal Benefits

**WG Conveners:**

**Thomas Elias Cocolios** (KU Leuven, Belgium)
**Charlot Vandevoorde** (GSI Darmstadt, Germany)

**NuPECC Liaisons:**

**Lucia Popescu** (SCK CEN, Mol, Belgium)
**Vladimir Wagner** (CAS, Řež, Czech Republic)

**WG Members:**

- **Hamid Aït Abderrahim** (SCK CEN, Mol, Belgium)
- **Michail Athanasakis-Kaklamanakis** (CERN/KU Leuven, Belgium)
- **Claude Bailat** (CHUV, Lausanne, Switzerland)
- **Sayani Biswas** (PSI, Villigen, Switzerland)
- **Iva Bogdanović Radović** (RBI, Zagreb, Croatia)
- **Daniel Cano-Ott** (CIEMAT, Madrid, Spain)
- **Seán Collins** (NPL, Teddington, UK)
- **Charlotte Duchemin** (CERN, Geneva, Switzerland)
- **Marco Durante** (GSI, Darmstadt, Germany)
- **Giles Edwards** (University of Manchester, UK)
- **Muriel Fallot** (Subatech, Nantes, France)
- **Anne-Marie Frelin** (GANIL, Caen, France)
- **Robin Golser** (TU Wien, Austria)
- **Angel Ibarra** (CIEMAT, Madrid, Spain)
- **Arnd Junghans** (HZDR, Dresden, Germany)
- **Zsolt Kasztovszky** (Budapest, Hungary)
- **Ulli Köster** (ILL, Grenoble, France)
- **Armadina Lima Lopes** (University of Porto, Portugal)
- **Renata Mikołajczak** (POLATOM, Świerk, Poland)
- **Enrique Nacher** (CSIC, Valencia, Spain)
- **Katia Parodi** (LMU München, Germany)
- **Vincenzo Patera** (Sapienza Rome, Italy)
- **Nikolas Patronis** (CERN/Univ. Ioannina, Greece)
- **Ekkehard Peik** (PTB, Braunschweig Germany)
- **Lino M C Pereira** (KU, Leuven, Belgium)
- **Arjan Plompen** (JRC, Geel, Belgium)
- **Marco Ripani** (INFN, Genova, Italy)
- **Nathal Severijns** (KU, Leuven, Belgium)
- **Zeynep Talip** (PSI, Villigen, Switzerland)
- **Eugenia Toimil-Molares** (GSI, Darmstadt, Germany)
- **Ana Vaniqui** (SCK CEN, Mol, Belgium)
- **Clemens Walther** (University of Hannover, Germany)

# Introduction

We are living in a world where technology and digitisation are evolving at a tremendous pace, causing a significant global impact on both individual persons and society as a whole. It puts so much pressure on our economy, politics, the social organisation of society and the environment that it has become clear that the present global organisation is not sustainable any more. In 1987, the United Nations (UN) declared in the report 'Our Common Future' (the so-called 'Brundtland Report') that humanity is capable of making development sustainable to ensure that it meets the basic needs of all to fulfil their aspirations for a better life without compromising the ability of future generations to meet their own needs. The concept of sustainable development does imply limitations imposed by the present state of technology and social organisation on environmental resources and by the ability of the biosphere to absorb the effects of human activities. To further manage and improve both technology and social organisation, the concept of Planetary Boundaries was developed. These show nine systems important for humans on planet Earth, which our own activities put under severe pressure, e.g. climate and biodiversity. These were later combined with basic human needs to yield the Sustainability Doughnut, the safe space for humanity providing these basic needs for each person without compromising the planetary boundaries.
The United Nations adopted the 2030 Agenda for Sustainable Development in 2015 to secure peace and prosperity for people and the planet. The agenda comprises 17 Sustainable Development Goals (SDGs), see Fig. 7.1. These are meant as a call for action to all governments across the globe, but research communities can also make a significant contribution. Nuclear physics and its applications in Europe affect all 17 of the SDGs. Nuclear physics and its applications in Europe can play a major role in the domains of **energy**, **health**, and **space**.

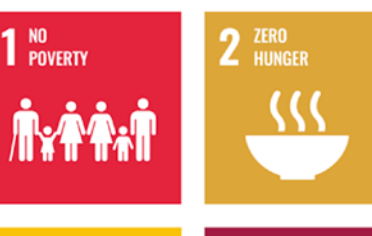
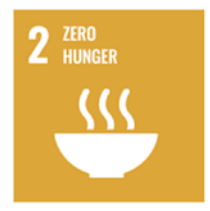
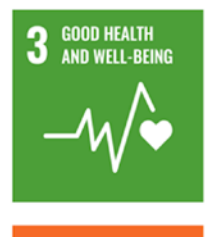
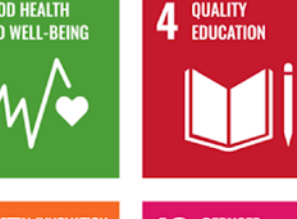
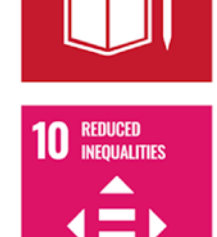
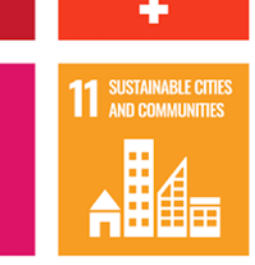
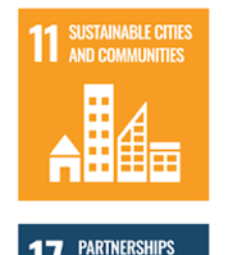
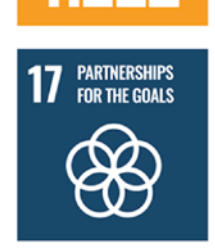
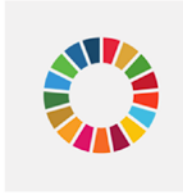

*Fig 7.1: The nuclear science research community contributes to all of the 17 sustainable development goals of the United Nations.*

In addition to SDG7 (energy), clean and affordable nuclear energy that is available anywhere contributes to SDG1 (no poverty), SDG8 (economy), SDG9 (industry) and SDG10 (reduced inequalities). Imaging and therapies arising from nuclear physics are commonly used in clinical practice around the world, contributing to SDG3 (health). Nuclear physics techniques such as isotopic markers to study plants and the water cycle have strong effects on SDG2 (zero hunger), SDG6 (clean water), SDG13 (climate action), SDG14 (life below water), SDG15 (life on land). Nuclear physics applications require a highly educated and inclusive workforce, contributing to SDG4 (education) and SDG5 (gender equality). The responsible treatment of nuclear waste from medical and energy applications addresses SDG11 (sustainable cities) and SDG12 (responsible consumption). The nuclear physics-based monitoring of non-proliferation aims to address SDG16 (peace). Finally, the strong collaborative nature of nuclear physics in particular in Europe supports SDG17 (partnership).

In this chapter, we shall in particular address the following SDGs.

**#3 Good health and well-being:** With techniques such as external beam radiation therapy (EBRT) or targeted radionuclide therapy (TRT), nuclear science applications can contribute to the treatment of cancers that are not yet curable. These approaches may also be used to improve not only patient survival but also quality of life, which is an increasing concern in good health and well-being.

Besides cancer, many developments are ongoing in support of our understanding of ionizing radiation, as is developing novel techniques for the in vitro or in vivo study of biochemical systems. With a better understanding of the cause of biochemical dysfunction, we come one step closer to finding a cure, e.g. for Alzheimer's disease.

**#7 Affordable and clean energy:** It is undoubtedly true that nuclear power has a large part to play in stepping away from CO2-emitting energy sources. It is also undeniable that the existing nuclear park is ageing and that the nuclear waste produced in the last decades has become a growing question that the nuclear community must address.

The development of new Small Modular Reactors (SMR) can answer the need for flexibility in nuclear power generation, while the mass production of such reactors will drastically drive the costs down, making nuclear power more affordable.

Moreover, the use of different nuclear cycles will allow the transmutation of the minor actinides, which form the most concerning part of nuclear waste. Some of it could be burnt in SMRs, but the Accelerator Driven System (ADS) reactor could provide a more direct answer for the large-scale transmutation of this nuclear waste.

Finally, recent successes in the development of fusion power motivate our community to drive this research further. Fusion power could provide a large-scale clean source of energy to support the long-term development of Europe and the rest of the world.

**#13 Climate action:** Nuclear science and technologies are also used to study the biosphere and the climate, providing unique information that contributes to our understanding of our planet and its climate.

# Energy

Energy production is a very important field in which nuclear physics can have an impact. Decarbonisation goals and security of supply represent critical challenges for Europe, and nuclear energy is expected to continue to provide a significant contribution to energy production for decades to come[1]. On the other hand, innovation is becoming more and more important in the nuclear energy field, as indicated by the need to develop novel concepts like Small Modular Reactors (SMR) together with advanced nuclear systems as outlined in the Gen IV programme and SNETP/ESNII. At the same time, the fusion community continues to develop the main projects outlined in the global fusion roadmap, ITER and DEMO, while several private or public/private initiatives aim to pursue diversified avenues in an attempt to speed up the achievement of commercial fusion. The developments in nuclear reactor technology and new designs of SMR and larger systems also bring progress in passive and inherent safety. Specific research for the development of the next generation of fission reactors and the design of the first fusion reactors is directly linked to several basic research issues: precise cross section and decay measurements, understanding of nuclear reactions, evaluation of data and their verification, validation and development of reliable databases and irradiation tests of materials. All these initiatives will contribute to reaching a net zero-emission energy landscape in the next decades[2].

[1] *EU taxonomy: Complementary Climate Delegated Act to accelerate decarbonisation, Commission Delegated Regulation EU 2022/1214.* https://finance.ec.europa.eu/publications/eu-taxonomy-complementary-climate-delegated-act-accelerate-decarbonisation_en
*COP28: Nuclear energy makes history as final COP28 agreement calls for faster deployment, IAEA, 2023.* https://www.iaea.org/newscenter/news/nuclear-energy-makes-history-as-final-cop28-agreement-calls-for-faster-deployment
*Nuclear Energy Summit: Declaration on Nuclear Energy, Brussels, 2024.* https://nes2024.org/en/declaration/
[2] *UK Policy Paper Civil nuclear: roadmap to 2050, ISBN 978-1-5286-4560-7, 2024.* https://www.gov.uk/government/publications/civil-nuclear-roadmap-to-2050

## Fission reactors

Given the age of the worldwide nuclear reactor fleet, some countries are implementing lifetime extensions to be able to keep the plants operative for as long as 60 years. At the same time, to achieve improvements in the safety, security and efficiency of nuclear power, new reactor concepts are being developed. Here, the main aspects are minimisation of the typically long-lived Minor Actinides (MA) production, more efficient use of the fuel, better thermodynamic efficiency, opportunity to produce hydrogen, and finally stronger safety features to further minimise the risk of accidents. The so-called Gen IV initiative as well as the European Sustainable Nuclear Energy Technology Platform (SNETP) are dedicated to such innovative designs.

Among the innovative designs being studied, fast reactors use liquid metals or gas as coolants, which means that fission neutrons are much less moderated and exhibit an energetic spectrum with a higher fission-to-capture ratio which means lower production of transuranium elements and the possibility to burn these, for which fission occurs only above a certain energy threshold.

Recently, the concept of SMR, delivering up to about 300 MW of electric power, has become widespread, giving rise to several designs around the world. Apart from improved safety features and easier deployment due to high standardisation, SMRs are also considered interesting for non-electricity applications like district heating.

Reactor lifetime extensions with the higher exposure of structural materials to neutron bombardment, innovative fission reactor designs, e.g. with the much higher, more energetic neutron fluxes occurring in fast reactors, and nuclear fusion, need facilities for neutron irradiation and high precision neutron data for several reactions (also on actinides). This programme can benefit from several existing or planned experimental facilities.

## ADS

To reduce the amount of high-level, long-lived radioactive waste to be disposed of in geological repositories, Accelerator Driven Systems (ADS) have been proposed. In ADS, an accelerator (typically protons) is used to provide a neutron source which drives a subcritical fission core, with the advantages of better reactivity control and the possibility to load relatively high quantities of MA. To achieve this goal, it is necessary to separate Mas from the spent fuel for subsequent irradiation in the ADS, an approach called Partitioning & Transmutation (P&T). MYRRHA is a European endeavour to test this concept in a multipurpose facility, see Fig. 7.2. MYRRHA's phase 1 envisions the commissioning of a 100 MeV linear proton accelerator in 2027 which will also be used to provide Radioactive Ion Beams (RIB) enabling research in fundamental nuclear physics. Phase 2 anticipates the commissioning of a 600 MeV proton accelerator in 2033 and the commissioning of a coupling to a Lead-Bismuth-Eutectic-cooled fast subcritical core in 2036. **The completion of MYRRHA is crucial to achieving the ambition of reducing our current nuclear waste before its long-term geological disposal.**

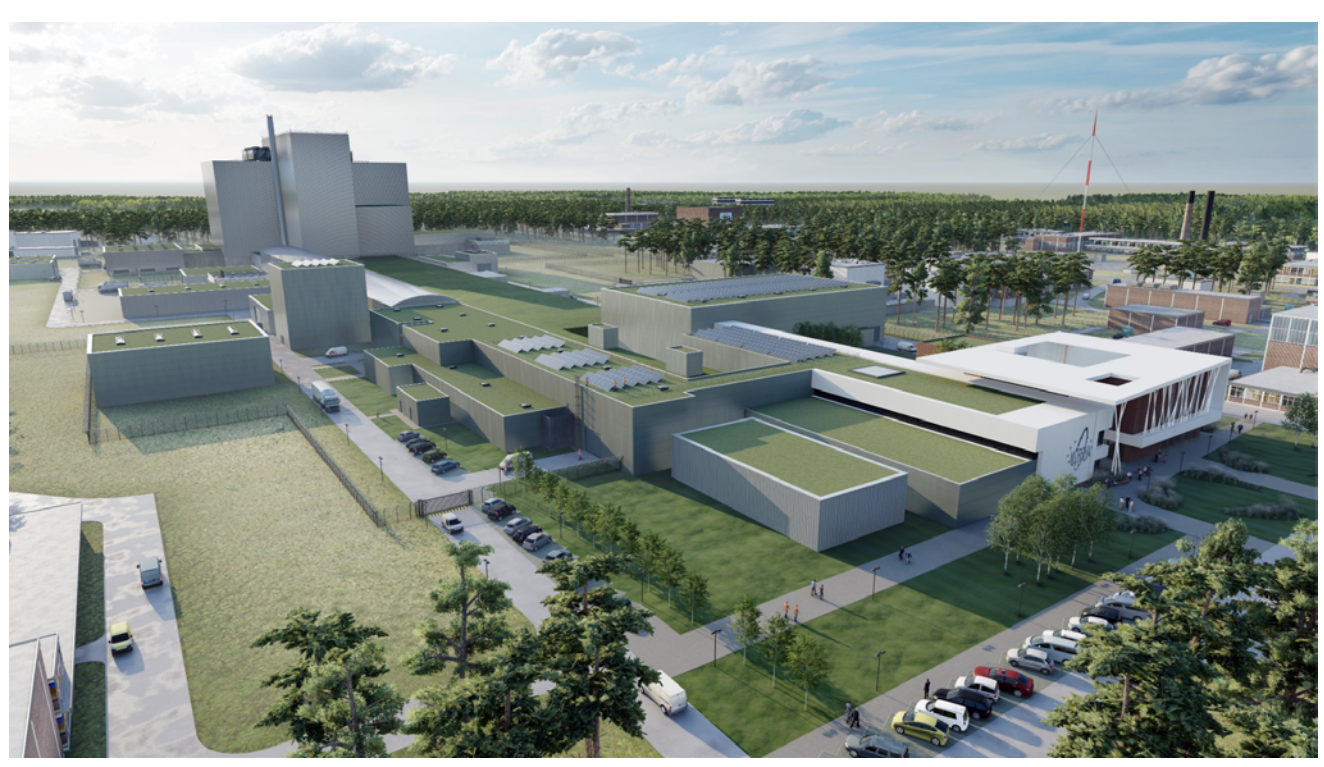

*Fig. 7.2: The future MYRRHA Accelerator Driven System complex in Belgium.*

## Fusion reactors

Fusion reactors are considered one of the few possible alternatives in the production of CO2-free energy in the long term. Over the last few years, overall interest has increased very significantly both in the public as well as in the private sector. This was triggered by milestones such as the world record of fusion energy production at JET on February 8, 2024 (69 MJ) for the case of magnetic fusion and the first demonstration of a significant net energy gain due to fusion reactions (around 50% gain for a production of around 3 MJ of energy) in the case of inertial fusion.

In the private sector, a significant number of startup companies have received significant investments (in the range of several billions of dollars up to now) in the EU, USA, and Japan to develop a wide portfolio of different approaches to fusion energy production besides the classical D-T reactions, and to focus on the fusion supply chain. Due to the very different approach in the private sector, much more tolerant to risk, this will probably accelerate the generation of new results coming from these companies in the coming decade. These important developments have induced a significant acceleration of the different national and international fusion programmes. In the next few years, the start of operation of many relevant plasma physics experimental machines is expected (JT-60 in Japan, DTT in the EU, CFETR in China, SPARC and NSTX-U in the US, STEP in the UK etc.) in parallel to the significant progress expected in the assembly of ITER and the future DEMO, see Fig. 7.3.

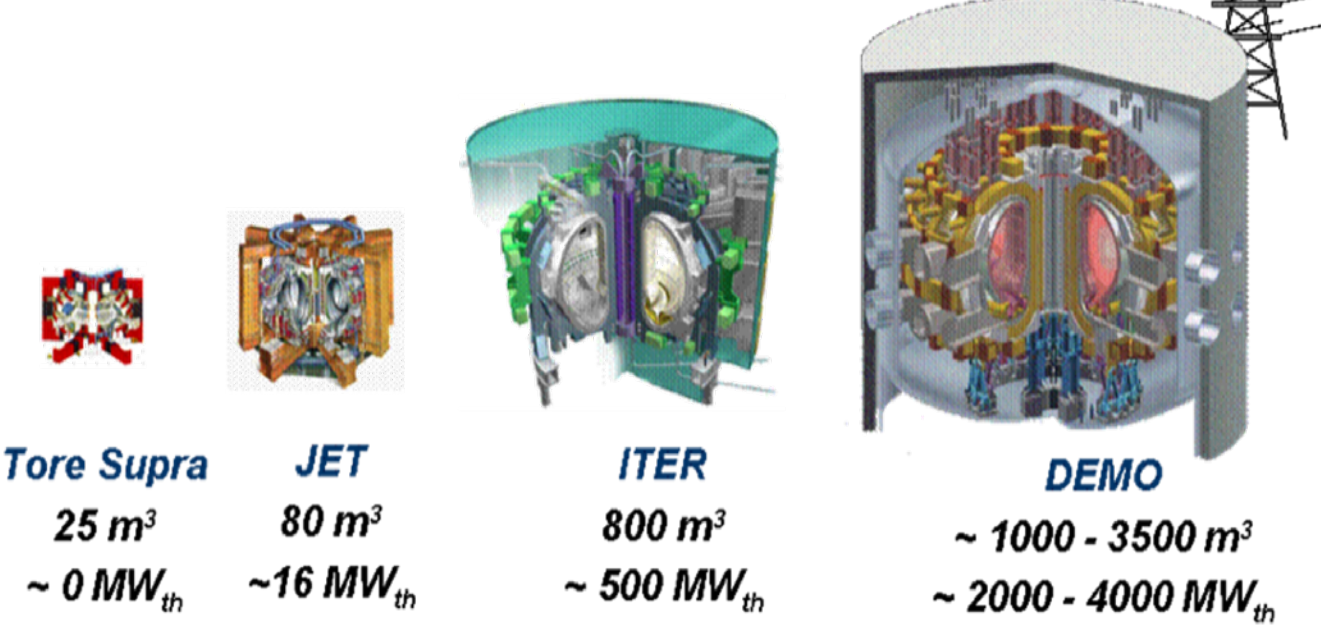

*Fig. 7.3: The fusion roadmap from the JET reactor today, to the ITER reactor under construction and the DEMO reactor in the future.*

As for inertial fusion, the ELI research infrastructure, the world's largest and most advanced high-power, high-intensity and short-pulsed laser Infrastructure, has strengthened European interest in this energy production approach. An initiative supported by the German Ministry of Education and Research has recently been launched by ELI-ERIC with the participation of ELI-NP, soliciting all relevant European stakeholders from Academia and Industry to identify practical ways for contributing to the development of methods and technologies relevant to laser-driven ignition.

The construction of a fusion reactor requires better understanding of radiation effects on material properties. This is one of the key challenges deriving from the extremely high flux of high-energy neutrons on reactor material and components. In this respect, the start of the construction of the IFMIF-DONES facility is notable. It will be an intense neutron source based on d-Li interaction able to produce up to $10^{18}$ n/s (fusion-like). This unique facility will become operational in the next decade and has the potential to significantly contribute to the production of improved cross-section data as required by the development of fusion reactor technology, but also by many other scientific and technology areas with societal impact. **Improvements in the nuclear physics information presently available, both in terms of data and theories, should be pursued with high priority in the next few years.**

## Nuclear waste

The decay time and number of radioactive products from nuclear reactor operation determines whether disposal of radioactive waste requires either short or long or very long-term storage. While for short or long-term storage, surface or near-surface disposal can be adopted, for the very long-term (in particular in the case of direct disposal of spent fuel) a possible solution envisaged is storage deep underground in so-called geological repositories. Surface or near-surface disposal facilities have been safely in operation for many years in several countries around the world. International recommendations have guided national policies, strategies, and programmes for the management of spent fuel and radioactive waste. Continued research on the optimal choices of sites and the prediction of the time evolution of the waste packages underground is performed at the international level. EURATOM has promoted the advancement of research on several aspects of decommissioning technologies, as well as on aspects and issues arising in interim storage, pre-disposal and final disposal where, in the latter point, the scientific aspects of deep geological disposal are of special interest. To reduce the amount of waste to be disposed of in geological repositories, it has been proposed to transmute at least part of the nuclear waste with the aforementioned ADS.

Critical aspects that remain to be addressed include the long-term stability of nuclear waste against chemical and physical reactions. Long-lived MA are of particular concern. Their radioactive nature and low availability have hindered our knowledge of their basic chemical properties. **Systematic radiochemical investigations remain critical to increase our knowledge of the MA and better predict their behaviour in the long term.**

## Security and safeguards

Enhancing security also means developing new technologies for the investigation of nuclear materials, be they stored in repositories or transported, and for the investigation of containers as a means to fight illicit traffic of nuclear materials and terrorism. For instance, high-brilliance γ-ray sources such as the ones developed at ELI-NP may be used to implement active interrogations based on nuclear resonance fluorescence (NRF). Also, some of the advanced detectors devised for nuclear installation decommissioning, radioactive waste management, heritage or material science, may be utilised for security applications, e.g. cargo inspections. The detection of antineutrinos has been proposed as a technique for assessing and monitoring fuel composition and possible fuel manipulations in nuclear reactors.

The many developments ongoing in nuclear science and its application to society will surely benefit security and safeguards. Adequate support should be given to facilitate the translation of the technology to this sector.

## Perspectives and needs

The underlying nuclear data libraries determine the precision with which nuclear systems can be simulated by modelling and particle transport simulations. These simulations have advanced to a detailed multiphysics level including neutron transport and thermal hydraulics with a realistic geometry, spatially resolving all technical components. More can be done in this respect, but on the other hand, nuclear data have still not reached sufficient levels of precision in many cases. The development and characterisation of novel radiation detectors are also beneficial in improving the reaction database. Finally, **maintaining nuclear application competencies in Europe requires the availability of laboratories and experimental facilities so that early-stage researchers may gain a fundamental understanding of the nuclear processes at work and the associated technologies.**

Several facilities can contribute to the goal of improving the nuclear database by offering accelerator-detector assemblies in which several types of beams, mostly ions and neutrons, can be used for this purpose. The development of neutron beams deserves particular attention as several laboratories are contributing to expanding the offer in terms of intensities and energy. These range from low-intensity beams performing measurements of fission, capture and other nuclear reactions on several nuclides including actinides, to high-intensity beams needed to test materials for use in the harsh environment of next-generation nuclear reactors, both fission and fusion. Gamma-ray spectra and multiplicities from capture and fission reactions are of interest for γ-ray heating and shielding in reactors. The experimental investigation of neutron inelastic scattering and (n,xn) reactions is of special interest for the design and safe operation of nuclear reactors, as they modify the neutron spectrum and the neutron population and produce radioactive species. The neutron-induced charged-particle-emitting (n,cp) reactions on isotopes frequently used on structural materials are of special interest due to possible material embrittlement from radiogenic light elements forming gas bubbles. Measurements of fission fragments (A and Z distributions) are interesting for decay heat in reactors and core poisoning. The neutron multiplicity distribution and fission neutron characteristics are also of particular interest in the design of fission reactors and ADS.

# Health

Ionising radiation and its interaction with organic matter still requires extensive investigation. While first principles are well understood, the impact on single molecules, cellular or organ level or full body remains difficult to fully comprehend. The impact on health is thus still a subject of both fundamental and practical nature. Moreover, ionising radiation has been used extensively for a wide variety of clinical applications and currently represents an indispensable pillar of modern contemporary medicine. In general, these applications can be divided into diagnostics and therapy, where we can further distinguish between administration through external (X-rays, γ-rays, or accelerated charged particles) and internal (decay of radionuclides) sources of radiation. In both areas, the nuclear physics community plays a central role, from the production of medical radionuclides to the development of new diagnostic and therapeutic techniques that make use of particle beams. Particularly in the fight against cancer, these technologies contribute significantly to all stages of patient care, ranging from early detection and diagnosis to curative treatment and palliative care. These developments are essential for meeting Europe's Beating Cancer Plan of the European Commission[3], as enshrined in the Strategic Agenda for Medical Ionising Radiation Applications (SAMIRA) action plan[4].

The last few years have seen many advancements in cancer treatment, such as the exponential growth of the particle therapy sector (118 proton and 14 carbon therapy centres worldwide today, see Fig. 7.4) or the approval of nuclear therapy drugs by the EMA (e.g., $^{177}$Lu-based drugs Lutathera® and Pluvicto®). However, despite this progress, some types of cancer still present poor prognosis and resistance to radiation, which calls for new therapeutic solutions. To address this need, technological developments are still ongoing, where the differentiated biological response between tumoral and healthy tissues has become a key aspect in the design of new treatment modalities.

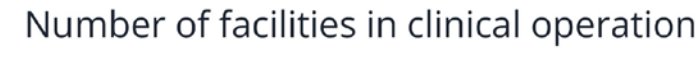

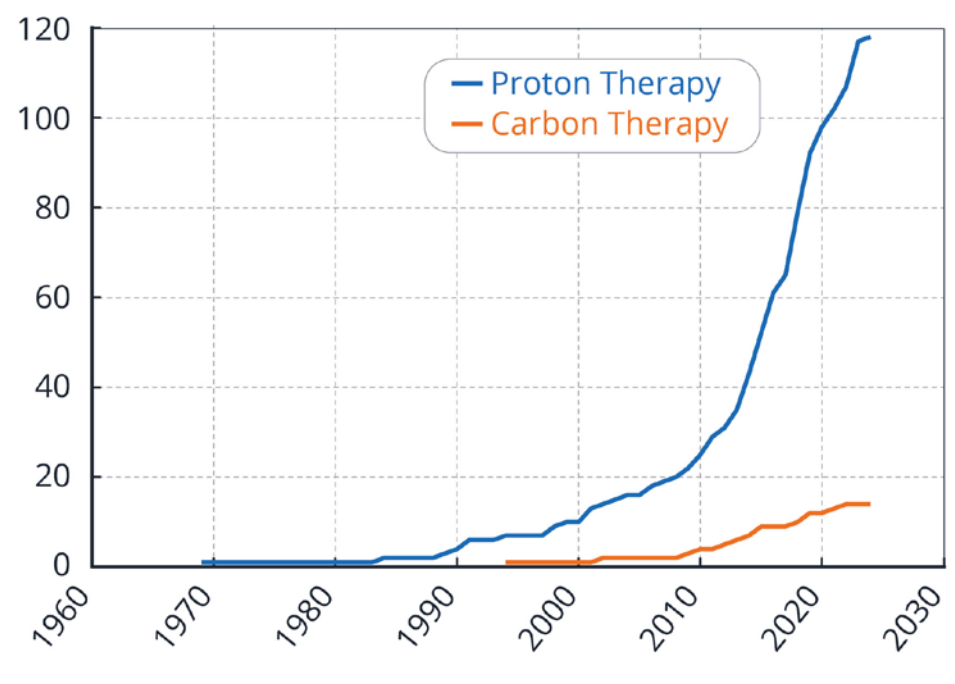

*Fig. 7.4: The increase of worldwide particle therapy facilities in clinical operation over the years*

[3]Europe's Beating Cancer Plan, 2021. https://ec.europa.eu/commission/presscorner/detail/en/ip_21_342
[4]Strategic Agenda for Medical Ionising Radiation Applications (SAMIRA), 2021. https://energy.ec.europa.eu/topics/nuclear-energy/radiological-and-nuclear-technology-health/samira-action-plan_en

## Imaging

Advanced medical imaging includes the following two approaches: external exposure with an X-ray beam; or internal emission of γ rays from radionuclides. While the former provides morphological information, and the latter provides molecular information, and the combination provides a global overview of the patient. Multiple projections coupled with advanced reconstruction algorithms permit the 3D reconstruction of the image.

In the case of **ion beam therapy**, patient representation is insufficient for planning treatment since ions and X-rays interact very differently with matter. A Computed Tomography (CT) image provides a map of X-ray attenuation coefficients, whereas for treatment planning one would need ion relative stopping powers (RSP). The conversion from photon attenuation to RSP results in uncertainties of 1-3% in the calculation of ion ranges in patients. This uncertainty can be reduced with advanced X-ray imaging techniques such as dual-energy or spectral CT. Alternatively, the use of proton or ion beams rather than X-rays could produce the primary tomographic images for the treatment plan.

The use of imaging radionuclides, such as $^{99m}$Tc and $^{18}$F for the diagnosis and staging of cancer through single photon emission computer tomography (SPECT) and positron emission tomography (PET) respectively, is well established in nuclear medicine, see Fig. 7.5. Recent international directives have prompted new developments in **quantitative imaging**, to establish **personalised medicine** which aims at improving outcomes for patients according to their features. **Both accurate nuclear data on the medical radionuclides and well-established protocols are needed to ensure the appropriate use of quantitative imaging for personalised medicine.**

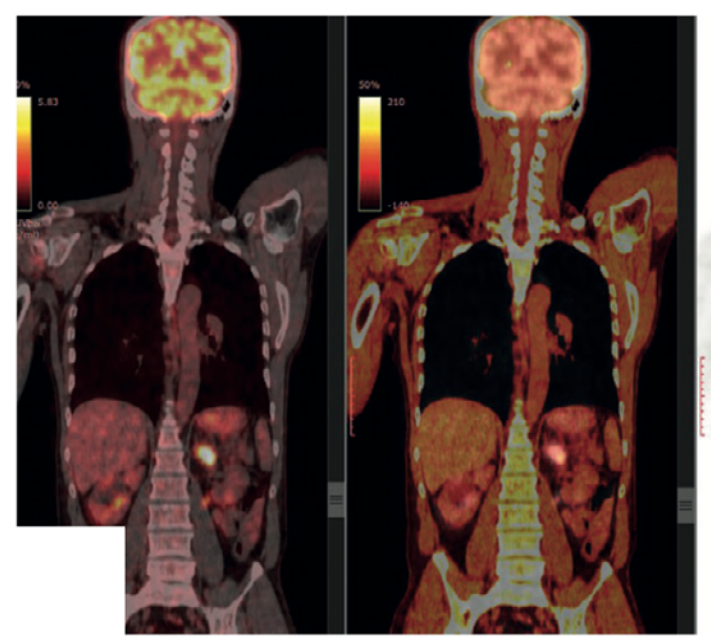

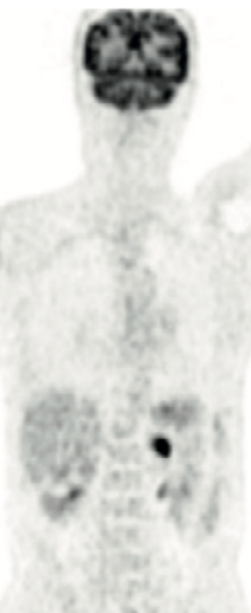

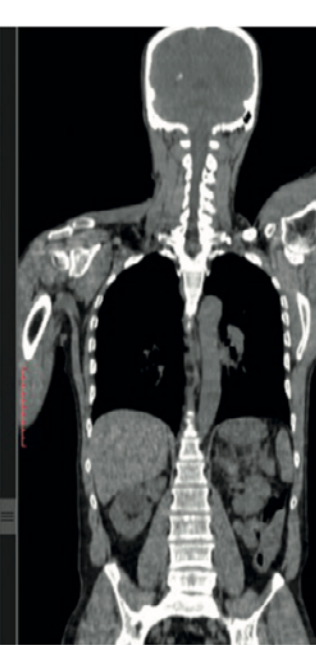

*Fig. 7.5: A PET-CT scan of the human body*

## External beam radiation therapy

Nowadays approximately two-thirds of all cancer patients receive radiation therapy during their disease trajectory, with external beam radiation therapy (EBRT) as the mainstay of radiation therapy (RT). The basic principles of RT have changed very little over the past century, where the goal is to deliver a sufficient radiation dose to kill the tumour cells while minimising the dose to the surrounding healthy tissues. EBRT has undergone important technological advances, resulting in more accurate beam delivery with image guidance and adaptive workflows, allowing modifications during the treatment course. This has permitted a significant increase in life expectancy and quality of life.

**Proton and heavy ion beam therapy** is an ideal tool to treat tumours that are located close to critical organs and are radioresistant (e.g. hypoxic tumours), because of the localised dose deposition (Bragg peak) and higher biological effectiveness compared to X-rays and electrons. Challenges especially related to target motion still prevent the optimal exploitation of ion beam therapy. A common approach to predict motion patterns during treatment planning uses 4D-CT and integral target volumes which account for the range of motion. In addition to the ballistic targeting accuracy, a better understanding of the radiobiological properties of particle beams could strongly impact their clinical implementation. In proton therapy, there are still unmet needs to fully understand the implications of varying radiobiological effec-

### Box 7.5: FLASH Radiotherapy

FLASH radiotherapy is an innovative EBRT technique for cancer treatment, where **ultra-high dose rates (> 40 Gy/s)** are used to deliver a radiation dose to the tumour in a very short time, typically in less than one second. This ultra-fast delivery distinguishes FLASH from conventional EBRT, where the radiation dose is delivered in fractions of several minutes (~ 1 Gy/min) over a longer period (typically 30 days). In addition, several preclinical studies illustrate that FLASH offers the possibility to **spare normal tissue** and reduce EBRT-related side effects while still providing tumour control. Despite the lack of clarity on the underlying biological mechanisms of this differential effect, the results arouse considerable enthusiasm in the international radiation oncology community. The first FLASH treatment was delivered to a patient with skin lymphoma at CHUV in Lausanne, where a single fraction of 15 Gy was delivered in 90 ms using 5.4 MeV electrons.

FLASH offers immense potential since it enables dose escalation, with the additional benefits of reducing treatment time and organ motion challenges. However, several hurdles make the clinical translation challenging, and this is where research at accelerator facilities can make a difference. To date, the majority of FLASH experiments are performed with electrons from modified linear accelerators or dedicated electron accelerators, characterised by a low penetration depth and short source-target distance. This restricts the technique to the treatment of small, superficial tumours. The use of very high energy electrons (VHEE) ( 0-200 MeV) could overcome this limitation. This has triggered several new initiatives, such as the CERN-CHUV project and laser-driven accelerator projects, see Fig. 7.15, including plasma accelerators or dielectric laser accelerators. The use of ultra high energy electrons (UHEE) (GeV) at synchrotron-based facilities is also being explored, such as at the ELISA facility in Bonn. Furthermore, the FLASH effect has also been illustrated with low-energy soft X-rays (<1 MeV) produced at very high intensity with synchrotron light sources.

Protons are by far the most mature technology for the clinical translation of FLASH, where a hybrid active-passive system is used with patient-specific 3D-range modulators. The intensity upgrade for FAIR enables the synchrotron at GSI to deliver the required beam currents to study FLASH effects with heavy ions in larger volumes than can be achieved at medical synchrotrons for animal studies. Recent advances in ultra-short high-power lasers enable ultra-high dose rate irradiation with very short duration ion pulses, opening new avenues for FLASH. Experiments using in vivo model systems, as well as human organoids, are crucial to gaining a better understanding of the underlying biological mechanisms since FLASH is a tissue effect, implying that *in vitro* experiments with cells have limited value.

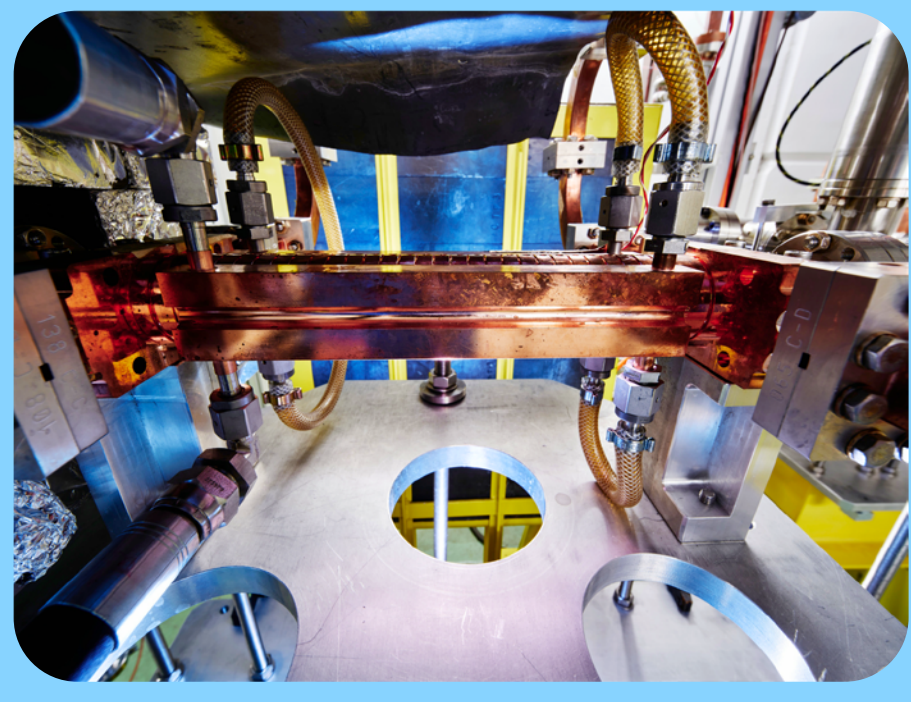

*Fig. 7.15: View of the Compact Linear Collider (CLIC) accelerator component to develop an accelerator for FLASH therapy by CERN and the Lausanne University Hospital (CHUV)*

tiveness (RBE). Carbon ions have recently been shown to generate a stronger immunological response, which offers scope to further improve the efficacy of immunotherapy.

**Technological developments** and increased clinical experience continue to foster exponential growth of particle therapy, with competitive and commercial turn-key solutions, especially for proton therapy. Nevertheless, the high cost of carbon facilities has long been a roadblock for their large-scale clinical implementation. Recent technological developments are aimed at reducing size and cost, for example through compact superconducting gantries and novel accelerators such as the C400 multi-ion cyclotron (ARCHADE). The resurgence of particle therapy with **upright beam delivery systems** arises from recent improvements in flexible scanning beam delivery and commercial availability of vertical patient CT imaging, along with advancements in robotics and image guidance systems to mitigate anatomical motion in seated treatments.
There is renewed interest in widening the spectrum of particles used clinically to **helium ($^{4}$He), oxygen ($^{16}$O) and neon ($^{20}$Ne) ions**, besides protons and carbon ions. The reduced lateral scattering promotes helium ions as an alternative to protons as shown in a first clinical treatment in Heidelberg. Despite the risk of increased normal tissue toxicity, heavier ions than carbon might also be interesting in the envisioned new delivery schemes combining multiple ions.

While conventional treatment schemes rely on homogeneous dose distribution delivered at a moderate dose rate and fractionated over several treatment days, some new developments break with these paradigms and show promising results in terms of healthy tissue response while maintaining tumour control. **FLASH radiotherapy** is an emerging method, using ultra-high dose rates to deliver the radiation dose to the tumour (see frame). In past years, preclinical studies have shown that FLASH radiotherapy can substantially widen the therapeutic window of EBRT. It is considered one of the major breakthroughs in radiation oncology, but to seize its full potential an understanding of the radiobiological mechanisms, optimisation of parameters and technological challenges related to beam delivery will have to be solved in the coming years.

**Spatially fractionated radiotherapy** (SFRT) also breaks with standards by using micro/mini beams to treat the total tumour volume with a non-uniform dose. SFRT divides treatment volumes into fractional subvolumes with alternating high and low doses. Preclinical studies have generated promising results regarding the reduction of normal tissue toxicity and induction of immunomodulation, which have resulted in renewed international interest in this technique. SFRT delivery techniques have mainly been explored using X-rays thus far. However, proton minibeam radiotherapy (pMBRT) is more promising, since it allows simultaneous homogenous tumour dose coverage and highly modulated spatial dose distributions in normal tissues. The use of heavier ions (e.g., $^{12}$C and $^{20}$Ne) for SFRT opens new possibilities for the generation of magnetically focused and scanned mini beams and will continue to receive growing attention in the next years.

For all these novel techniques, the **understanding of underlying biological mechanisms** is still limited and needs extensive study with in vivo models or alternative models such as zebrafish eggs and organoids. To pave the way for these promising new treatment modalities, preclinical radiobiological studies must also be accompanied by new technological developments in terms of accelerators, beam control and dosimetry instrumentation.

## Targeted radionuclide therapy

Targeted radionuclide therapy relies on administering to patients radiolabelled molecules (radiopharmaceuticals) targeting tumour biomarkers or their microenvironment. The use of radionuclides emitting charged particles (α, β, conversion or Auger electrons (AE)) ensures a localised dose deposition at the tumour site. This modality is adapted to the irradiation of diffuse cancers. For many radionuclides, decay is accompanied by γ-ray emission that can be used for imaging their biodistribution. The combined therapeutic and imaging use of radiopharmaceuticals, whether from a single radionuclide or a pair, constitutes the **theragnostic concept.** Complementary domains such as nuclear physics, radiochemistry, radiobiology and dosimetry must work together to fulfil this ambition.

### Box 7.2: Dosimetry: from ionizing radiation to health and safety

**The interaction of ionising radiation with atoms is well understood from first principles. Yet this understanding does not translate to predictability for the impact of ionising radiation on health, bulk material (e.g. shielding) or electronics. When considering systems globally, different approaches need to be considered.**

**While the activity, in Bq, is the physical quantity that can be determined for radioactivity, it has little relevance in dosimetry. Instead, the first quantity of reference is the absorbed dose, expressed in Gray (Gy=J/kg), which quantifies the energy deposited in a given volume.**

**This quantity, however, negates the interaction between the incoming radiation and the material. Since photons, neutrons and charged particles display very different forms of interaction with matter, a better understanding is obtained by considering the equivalent dose, expressed in Sievert (Sv). A radiation weighting factor is applied to the absorbed dose according to the type of particle (1 for photons and leptons, 5 to 20 for neutrons (depending on their energy), and 20 for α particles and heavy ions). This approach may already be satisfactory when describing the impact on electronics.**

**When considering a biological system, not all parts are as sensitive as others. For example, the skin would be much less impacted by radiation than a bone marrow cell. This has led to the concept of effective dose, also expressed in Sv, where a tissue weighting factor is considered for organ sensitivity, fractioning the weights across the different parts of the body.**

**While this approach offers a better approximation, it is still insufficient to fully answer the current developments in health applications (EBRT, TRT) as well as for astronaut health. Radiobiological understanding is required to build microscopic models of dosimetry that could translate to predictive capabilities for treatment planning or to support astronaut exposure prevention on long-term space missions.**

**Targeted alpha therapy** (TAT) is very promising in the treatment of disseminated diseases and chemo- and radiation-resistant cancers, or to be used as a complement to other therapies. The efficiency of such treatments relies strongly on the short range (comparable to the cancer cell diameter) and the high RBE of α particles. Currently, only $^{223}$Ra, under the name of Xofigo®, has been granted market authorisation and is used in clinical routine. Nevertheless, recent developments using α emitters like $^{211}$At, $^{212}$Pb, $^{213}$Bi, or $^{225}$Ac (see Fig. 7.6) have progressed from early in vitro studies through in vivo experiments to clinical trials. Proper facilities and training of personnel are needed to deploy TAT further to the clinic. Moreover, the isolation and labelling chemistry is challenging and radiolysis is more pronounced than with β-emitting radionuclides. Finally, imaging for biodistribution and dosimetry is limited without an analogue imaging isotope. TAT dosimetry needs multi-scale radiobiology and dosimetry studies, as the standardisation of clinical dosimetry practice is still heavily debated. Furthermore, all these multidisciplinary developments are strongly constrained by the limited accessibility to α-emitting isotopes.

Over the past fifty years, researchers have investigated the potential applications of **Auger electron**-emitting radionuclides in the field of nuclear medicine. Despite the scarcity of their clinical applications, radiopharmaceuticals radiolabelled with AE-emitting radionuclides continue to be an important topic due to their great potential. In general, the radiobiological effects of Auger electrons are still not well understood. **Further systematic investigations are urgently needed to understand their effects in therapeutic applications, which requires the availability of Auger electron-emitting radionuclides in appropriate activities and qualities.**

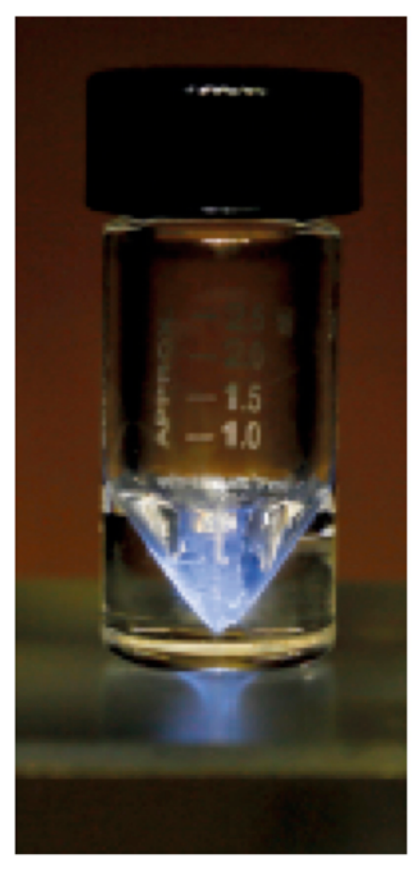

*Fig. 7.6: A capsule of actinium-225*

Conventional medical radionuclides can readily be produced, either in research reactors for neutron-rich species or with low-energy cyclotrons (<20MeV) for neutron-deficient species. Such cyclotrons are well distributed across Europe. Some of the innovative medical radionuclides investigated today require **medium-energy and high-intensity accelerators** with multiple particle beams (30-100 MeV) and sometimes even higher-energy protons (1-2 GeV). For some radionuclides, radiochemical separation is not sufficient and **physical mass separation** is currently only possible at the ISOLDE and MEDICIS facilities at CERN. MEDICIS has demonstrated the possibility of separating samples irradiated at any accelerator or high-flux nuclear reactor. However, with a growing interest in radionuclides with short half-lives (a few hours to a few days), the distribution radius from a single production facility becomes limited and a **better geographical distribution of new facilities is required to ensure full access to those opportunities across the European continent.** Several national initiatives contribute to this objective, such as the UK Medical Radionuclide Innovation Programme[5].

Following the radionuclide production, the **radiochemical separation process** is applied for two purposes: to isolate the desired material from residues of target material or co-produced elements and to recover the enriched target material for further usage. In this aspect, novel chelating agents and resins should be developed to facilitate efficient chemical separation processes. In particular, the radiation damage resistance of the ion exchange and extraction resins, which play a crucial role in large-scale productions and radionuclide generator concepts especially for α-emitting radionuclides, should be improved.

## Cross-over

EBRT and radionuclide therapies are complementary techniques that concentrate on different patient conditions. However, some special cases can benefit from combining the strengths of both approaches into a single modality.

**Boron neutron capture therapy** (BNCT) is based on a boronated vector targeting tumour cells followed by irradiation with epithermal neutrons. In principle, this technique has the advantage of reducing the effect of nonspecific targeting during administration and elimination of the active compound thanks to the selective activation by external beams, although the latter will also contribute to nonspecific irradiation. The interest in this modality is renewed by the development of compact accelerator sources, compatible with hospital practice.

**The use of radioactive ion beams (RIB) EBRT** can provide big improvements in image quality and signal-to-noise ratio compared to stable ions for range monitoring in ion therapy. While having a high potential for image-guided particle therapy, the main hindrance is the challenging production and low intensity of RIBs. Different paths are being explored, from the in-flight fragmentation of the ion beam to the injection and post-acceleration of mass-separated radionuclides. One could consider isotopes of C, N, and O as potential projectiles in RIB therapy, and pioneering work is ongoing in Japan. However, it was only until the recent intensity upgrades at existing facilities (e.g. SIS-18 at GSI/FAIR), that experiments on RIB characterisation and preclinical experiments could be fully explored in Europe (see frame). Several research projects are ongoing on the development of new detectors and to build sources of radioactive ions for medical synchrotrons.

### Box 7.6: $^{11}$C hadron therapy

**The uncertainty in the particle range within the human body is arguably the main physical problem of radiotherapy using protons or heavy ions.**

**Several range verification methods have been explored to mitigate this problem, such as positron emission tomography (PET), to monitor dose delivery by exploiting $\beta^+$-emitting isotopes produced from fragmentation induced by the impinging beam. Modern nuclear physics relies heavily on radioactive ion beams (RIB), used to explore the limit of the existence of nuclei and the precision of the nuclear models. Positron-emitting RIB in radiotherapy can provide significant improvement in image quality and signal-to-noise ratio compared to stable ions. By using the same beam for treatment and online PET imaging, the position of the beam can be corrected online, achieving real image-guided particle therapy. This idea was already proposed many years ago at the Lawrence Berkeley Laboratory, but only the recent intensity upgrades of the SIS-18 at GSI/FAIR made in vivo experiments feasible. As part of the ERC Advanced Grant BARB (www.gsi.de/BARB), $^{10,11}$C and $^{14,15}$O are tested for simultaneous treatment and online beam verification in an animal model. A major milestone was achieved during an experimental campaign at GSI in February 2024, when mice with a xenograft osteosarcoma tumour close to critical structures (e.g. spine and oesophagus) were treated with an $^{11}$C beam using online range verification with the in-beam detector developed at Ludwig-Maximilians-Universität (LMU) Munich (see Fig. 7.16).**

**Several research projects are ongoing in the development of new detectors and building sources of radioactive ions for medical synchrotrons, e.g. leveraging the developments in RIB production at CERN that can be used as injectors in running clinical facilities like MedAustron in Austria. These joint research efforts between several nuclear physics institutes, clinical centres and universities will improve the potential of RIB in cancer treatment in the coming decade.**

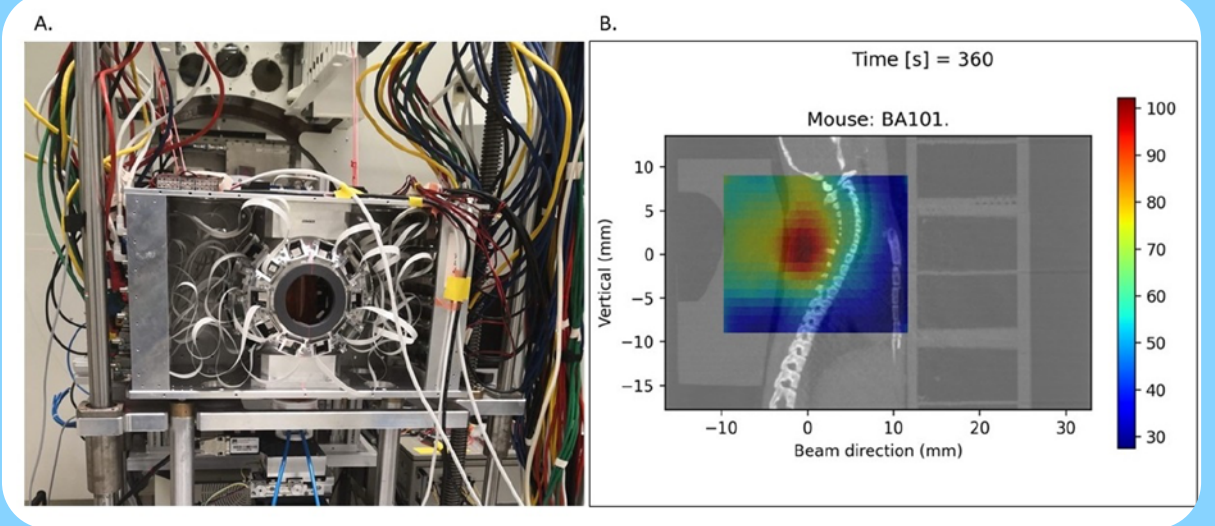

*Fig. 7.16: (A) The detector cage of the in-beam PET scanner developed at LMU within the ERC project SIRMIO (www.lmu.de/sirmio). (B) Online PET image superimposed onto a prior CT of a mouse with a tumour in the neck region, treated at GSI/FAIR with 11C in February 2024 as part of the BARB project. The colour palette indicates the radioactivity measured by PET in the target tumour, while the spinal cord is spared*

[5] *The Medical Radionuclide Innovation Programme (MRIP): selected projects, 2023.* https://www.gov.uk/government/publications/medical-radionuclide-innovation-programme/mrip-selected-projects

## Perspectives and needs

New treatment strategies, especially the ones based on differential biological mechanisms, require multiscale assays before clinical trials to identify their optimal implementation and effectiveness. These assays range from in vitro evaluation (with various models: 2D cell culture, organoids, ...) to *in vivo* experiments. The diversity of scales and geometries is thus challenging in terms of imaging and dosimetry. Even if progress has been made (particularly in preclinical radiotherapy), radiobiology experiments often lack optimal instrumentation and methods to provide accurate and reliable dose delivery and measurement. Developments in this domain are necessary to improve experimental reproducibility, allow intercomparison between experiments, enhance the valorisation of radiobiology assays and benchmark radiobiology numerical models.

It must be noted that beam access for EBRT and access to radionuclides for radionuclide therapy are essential to improve and develop treatments. New therapeutic strategies also rely strongly on various and complementary scientific domains (physics, biology chemistry, medicine); a truly interdisciplinary approach is required.

# Space

An exciting new space era lies ahead of us, with plans for a sustainable return to the Moon and long-term Mars missions. The space landscape is drastically changing, with the growing involvement of private companies next to the traditional space agencies. This fosters research interest and new funding opportunities for space-related radiation sciences. Research at nuclear physics accelerator facilities plays a central role in these funding programmes to mimic the complex space radiation environment on Earth. When astronauts leave the Earth's protective magnetosphere, they endure higher levels of radiation from galactic cosmic radiation (GCR) and the risk of a large solar particle event (SPE). Planned human exploratory missions can only be safely performed once we can substantially reduce the existing uncertainties on space radiation health risks, ensure reliable equipment and safe habitats in space, as well as develop effective countermeasures. **Ground-based research programmes and Monte Carlo-based approaches will be decisive in achieving these goals.**

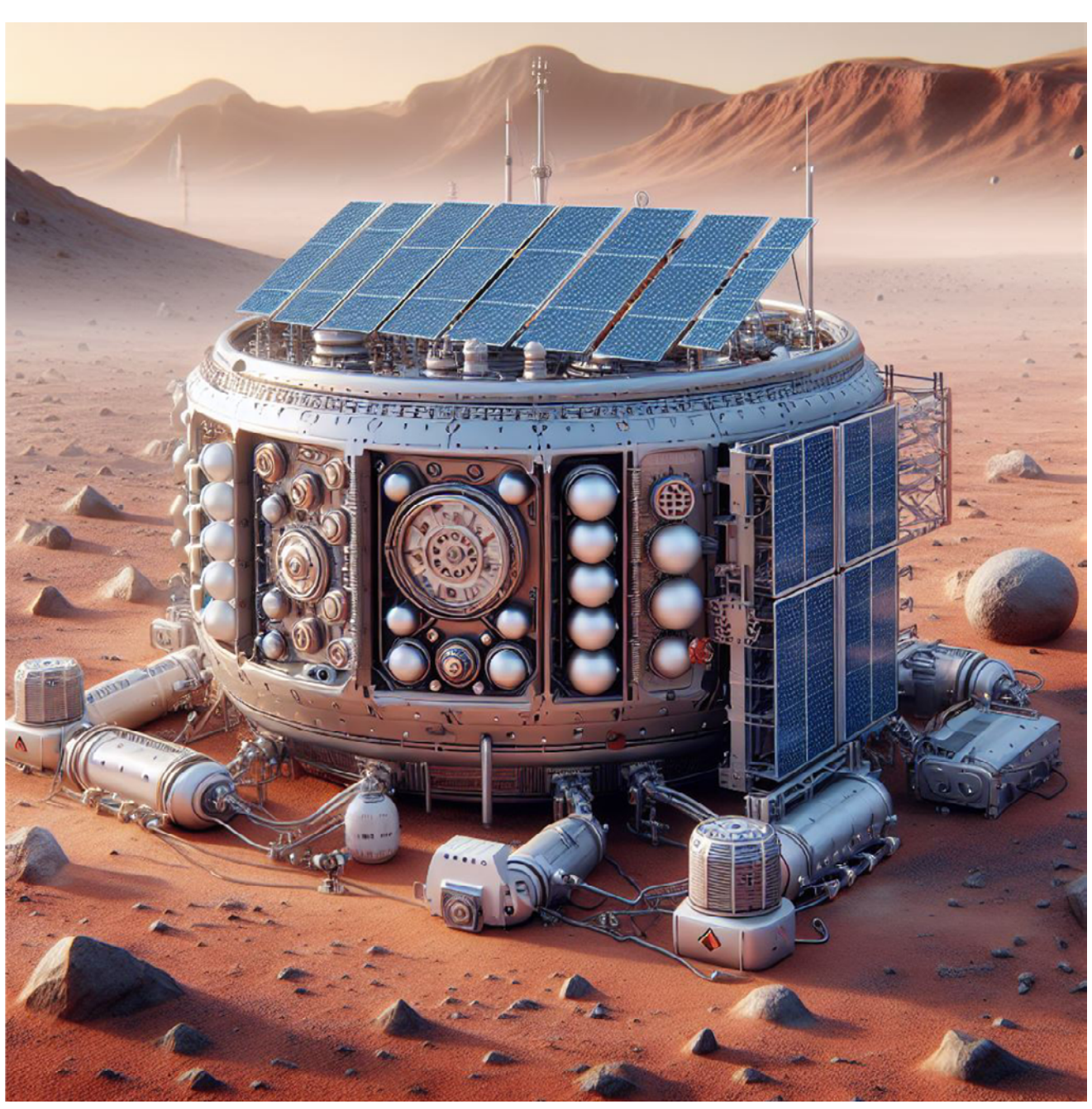

*Fig. 7.7: An SMR on Mars. AI-generated illustration*

## Energy sources for space applications

On the Moon, nights are almost 14 days long, during which solar panels cannot be used. On Mars, dust storms have proved to be detrimental to solar panels, which has forced NASA to equip the last Martian rovers with alternative energy sources. The same happens for far-away exploration missions (beyond Jupiter) where solar irradiance is too low to power spacecraft and their payload. Currently, the main alternative systems use Radioisotope Thermal Generators (RTG) to generate electricity (e.g. on Curiosity, Perseverance, and other Mars rovers). Several satellites are also based on these systems. A specific radioisotope (typically $^{238}$Pu) is decaying naturally, producing heat which can be transformed into electricity. The main advantage is its physical process independent of external factors so the heat production cannot be interrupted.

Establishing a habitat on the Moon or Mars requires a large uninterrupted power source, in the range of a few MW. Various studies are ongoing for small nuclear reactors with no or very reduced maintenance and long-term (several years) operation capabilities. NASA and DOE have, for instance, developed a system called Kilopower, which could be deployed on the Moon within a few years. China is also developing a similar system.

Developing mining systems on the Moon and Mars to extract the necessary raw materials for establishing a complete base on an external planet depends heavily on energy production, which can almost only be generated by nuclear power. The future of space exploration, and in particular planet exploration, thus relies on the use of nuclear power generation with SMR-type reactors.
ESA is calling on European nuclear research organisations and industries for developments in these fields to guarantee the EU's independence in this run for deep space exploration and settlement.

## Health of astronauts

The design of the space station orbiting the Moon (Gateway), lunar landers and surface habitats aims to protect the crew against SPEs in the best way possible by incorporating storm shelter concepts, optimisation of vehicle design and active dosimetry. Unfortunately, the penetrating GCR will continue to pose the most significant health risks to astronauts, see Fig. 7.8. Accurately predicting and modelling the complexity of the radiation environment inside a spacecraft is one of the greatest challenges in preparation for future space exploration. High charge and energy (HZE) ions, although only a small fraction of the overall GCR spectrum compared to high-energy protons, are more biologically damaging. The radiation field inside a spacecraft also comprises a non-negligible secondary radiation field, whose nature is highly variable, depending on factors like the ever-changing characteristics of the primary radiation field with time, or the type and thickness of spacecraft shielding. Secondary neutrons become a significant concern because of the significant uncertainties on their biological impact at higher energies (> 20 MeV). The experimental and calculated production cross sections of both neutrons and charged ions from the interaction of He, C, O, Si, and Fe ions on the spacecraft shielding material are crucial in assessing the dose contribution to astronauts.

Existing epidemiological data for radiation risk assessments in space are currently based on life span studies of atomic bomb survivors and a limited number of low Earth orbit (LEO) astronauts. These are not directly applicable to prolonged missions in deep space with GCR exposure. Ground-based high-energy particle accelerator facilities are required as analogues for space radiobiology experiments. These are currently performed with $^{56}$Fe at 1 GeV/nucleon, which is the most abundant heavy ion in the GCR spectrum. This simplification leads to inaccurate estimations of the harmful radiation effects of the full space radiation environment. At the moment, ESA and GSI/FAIR are building the first GCR simulator in Europe, exploiting fast energy switching of mono-energetic $^{56}$Fe beams on specially designed beam modulators. In addition, there is a lack of high-energy neutron sources, including Quasi-Monoenergetic Neutrons (QMN). Appropriate uniformity in the data collection and format is required in order to obtain a global conclusion on this distributed approach.

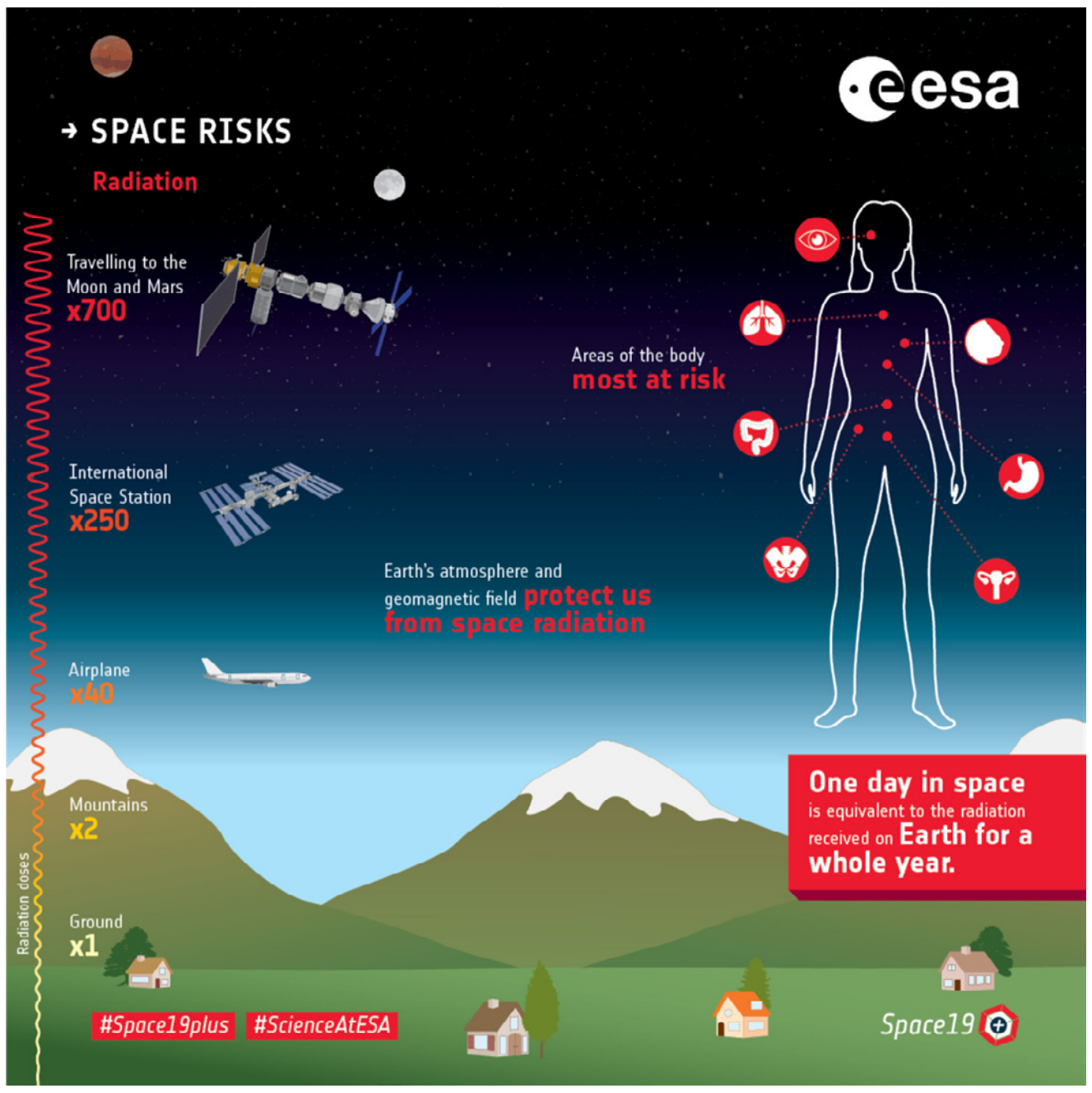

*Fig. 7.8: Radiation exposure incurred during space exploration and associated health risks*

Despite the new GCR investment and ESA's initiatives to support ground-based space research under the IBPER programme, **beam time opportunities for space radiation research** remain limited in Europe. **Monte Carlo-based approaches** might also be required, covering the broad spectrum of the space radiation field as well as the widely different scales of microscopic interactions to full-scale integration of space vehicles.

## Radiation hardness testing

Radiation hardness testing plays a critical role in the development and selection of detectors for space missions. However, as devices become more complex testing becomes increasingly challenging. **Modern ASICs** (Application-Specific Integrated Circuits) **or FPGAs** (Field-Programmable Gate Arrays) contain millions of transistors, making it difficult to identify all the potential disturbances that a single-event effect (SEE) might trigger.

Looking ahead to future missions, the **Galileo satellite navigation system**, which has to maintain a high level of accuracy throughout each satellite's 12-year lifespan while orbiting through the outermost Van Allen belt, is a top priority. The European Cooperation for Space Standards (ECSS) plays an important role in this field, by overseeing a comprehensive set of space manufacturing standards, among which are the Radiation Hardness Assurance standards.

Access to facilities for this type of testing has been facilitated through European-funded infrastructure projects (RADNEXT, HEARTS). The objective of these projects is to create **a network of facilities and related irradiation methodology** to address the emerging need for electronics components and system irradiations, as well as combining different irradiations and simulations. Another direction is the use of high-power laser-generated mixed particle beams at ELI-NP. High-power lasers offer an alternative for reproducing the interaction of cosmic rays with condensed matter since the energy spectrum of laser-accelerated particles is similar to the broad, multi-MeV-scale spectra of natural cosmic radiation. **To continue enhancing Europe's capacities and competitiveness in space-related radiation research in the coming years, sustained access to high-energy heavy-ion accelerators and high-energy neutron sources is needed, as well as long-term coordination of European infrastructure funding.**

# Climate and the Earth

Various techniques developed by the nuclear physics community are now routinely used for the measurement of elemental composition as well as high-sensitivity isotopic ratio spectrometry for a variety of environmental monitoring applications. These can reveal important information on the history of our planet's climate to help predict its future, as well as in exploring our biosphere. Climate change resulting from human interaction through changes in the environment poses a significant challenge to humanity and the biosphere in general. Gamma Ray Spectroscopy (GRS) and Liquid Scintillation Counting (LSC) techniques are particularly useful for monitoring radioactive species in the environment, with high sensitivity. Finally, Ion Beam Analysis (IBA) is frequently combined to provide comprehensive information. **As new technologies emerge from fundamental nuclear physics research, their application to climate and Earth science can provide new insights.**

## Environmental radiation monitoring

**The use of nuclear physics instrumentation and techniques for** the environmental measurement and characterisation of radioactive material can be split into the evaluation of normally occurring radioactive material (NORM) and anthropogenic radioactive material (e.g. waste streams from the nuclear power industry). GRS is used routinely to monitor a plethora of radionuclides. It may be used in conjunction with high-abundance sensitivity isotope-ratio mass-spectrometry techniques as signatures of uranium depletion or enrichment.

LSC remains the standard most sensitive analytical technique for the detection and quantitation of low-energy α-emitting and β-emitting radionuclides as well as radionuclides that decay by electron capture in the environment. It is used to monitor various chemical and biological cycles via radiotracers like $^{3}$H, $^{14}$C, $^{32}$P, $^{35}$S, and $^{131}$I. It is also used routinely to monitor NORM as well as waste stream discharges from nuclear facilities via radiotracers including $^{36}$Cl, $^{41}$Ca, $^{55}$Fe, $^{63}$Ni, $^{90}$Sr, $^{93}$Zr, $^{99}$Tc, and $^{129}$I. Recent advances in the technique include the use of plastic scintillators and improvements in the resolution of the detectors. **Development of the instrumentation and technology for exotic nuclei may be readily transferred to these applications.**

**Aerial radiation monitoring** systems may be deployed with manned or unmanned aerial vehicles to characterise, measure and track radioactive materials released into the environment or atmosphere, either before or during emergencies. See Fig. 7.9. Multidimensional detection systems are employed using γ-ray spectroscopy, neutron detectors, and infrared spectroscopy for online geospatial chemical and radiological data acquisition. These detection systems typically use $LaBr_3$(Ce), NaI(Tl), or plastic scintillation detectors. **Continued improvements in instrumentation and data processing within the nuclear science community will result in more compact solutions providing greater diversity in information and higher spectral resolution.**

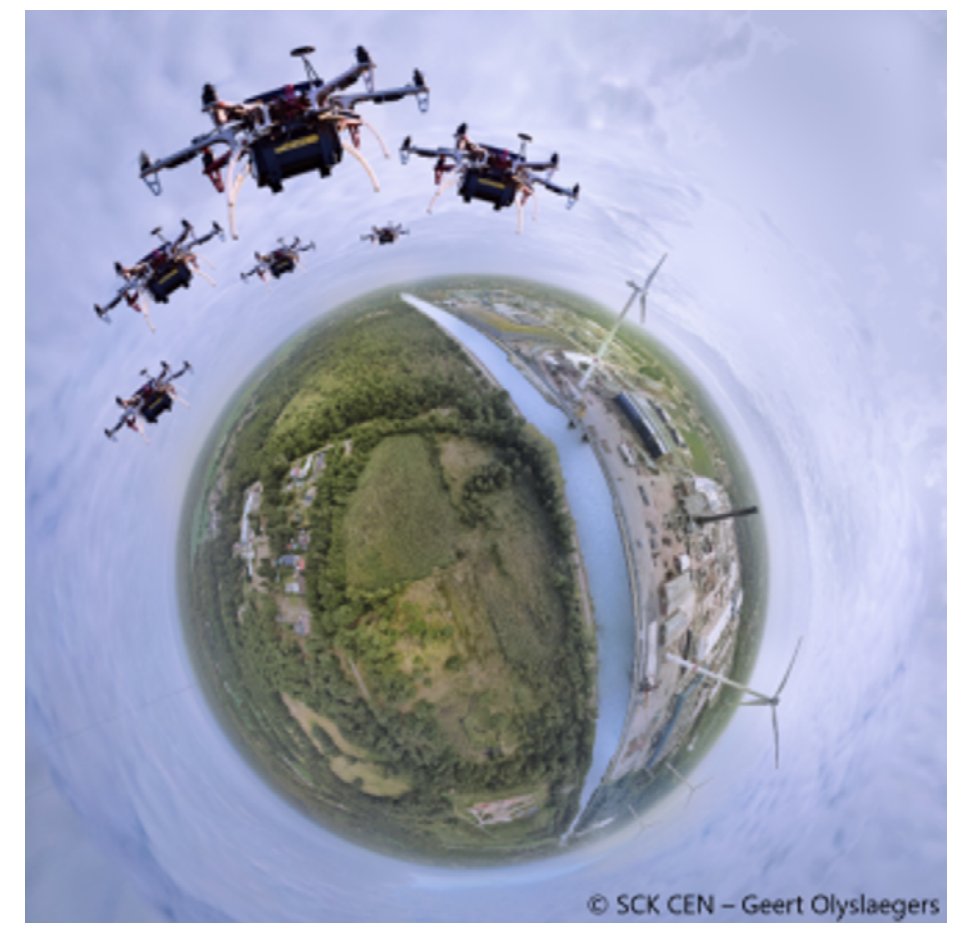

*Fig. 7.9: ARMAGED-drones for radiation monitoring*

## Box 7.3: Exploring our biosphere

**Monitoring changes in the environment through the migration of legacy waste streams within the biosphere is paramount. Ion beam analysis (IBA) delivered by relatively small accelerators is used to investigate the response of materials to radiation environments. In particular, PIXE techniques are useful, typically using protons (H-PIXE). This approach is limited to light elements (Z<10). Detection limits for the lighter elements are improved when using α particles (He-PIXE), which also have the advantage of being able to detect within a high-Z-element matrix (Z>10). PIXE is fast and, when combined with other IBA techniques such as Particle-Induced Gamma-ray Emission (PIGE), Elastic Backscattering Spectrometry (EBS) and Particle Elastic Scattering Analysis (PESA), can analyse elements from H to U. The main caveat when using IBA techniques is the lack of isotopic resolution which is standard with mass spectrometric-based techniques.**

**PIXE is particularly suited, due to its inherent sensitivity and simultaneous elemental measurement capabilities, to the offline monitoring of particulate matter (PM) with small particle sizes even down to 2.5 µm in size ($PM_{2.5}$). The PIXE technique is also non-destructive, so it can be used as a complementary technique with new state-of-the-art Inductively Coupled Plasma tandem quadrupole Mass Spectrometry (ICP-MS/MS). This presents many advantages over the less sensitive (but more robust) Optical Emission Spectroscopy (ICP-OES) technique. ICP-MS/MS generally offers lower method detection limits for crustal elements than PIXE. However, this is dependent on the solubility and hence digestion/sample preparation methods employed. The technique is also destructive.**

**The use of multi-collector sector instrument-based inductively coupled plasma mass spectrometry (MC-ICP-MS) is more sensitive than quadrupole-based systems and offers the additional benefit of being able to accurately measure isotopic ratios with high abundance sensitivity ($10^8$) necessary to measure anthropogenic radionuclides. The development and increased adoption of tandem mass spectrometry techniques including ICP-MS/MS are now becoming mainstream and mean that analytes can be measured in complex sample matrices with increased precision. They are more cost-effective than PIXE systems. If sample size is limited, the non-destructive PIXE technique is preferred.**

**3D elemental mapping combined with the study of internal morphology using PIXE, PIGE, RBS and Scanning Transmission Ion Microscopy (STIM) is utilised for the comprehensive characterisation of complex polymer/graphene oxide/nanoparticle composites and the imaging or irradiation of living cells and small animals for environmental studies, as already introduced for cultural-heritage samples. Radiation-resistance testing and studies can be performed under well-controlled and monitored conditions for electron and photon beams produced by a LINAC or Rhodotron®. Photon beams are also used for the Photon-Activation Analysis (PAA) of geological, biological, environmental and other samples.**

**Magneto-Optical Trapping (MOT) instrumentation has been successfully developed for the monitoring of rock-type change mapping when performing survey scans down boreholes for reservoir porosity determinations. The quantum technology based on atom interferometry MOT gravity sensors uses light that is tuned slightly below an atomic resonance of $^{87}$Rb in conjunction with a quadrupolar magnetic field. Scattered light from the atom cloud is measured through fluorescence detection.**

## Nuclear decommissioning

There is a variety of signature fission or activation products present within the atmosphere that require monitoring, including $^{111}$Ag, $^{125}$Sb/$^{125}$Sn, $^{131}$I, and $^{140}$La/$^{140}$Ba. $^{85}$Kr is a radioactive noble gas present in the atmosphere in small amounts through the capture of moderated cosmic neutrons with the neighbouring stable isotope $^{84}$Kr, yet the majority originates from anthropogenic sources, as a fission product formed in nuclear reactors and held within the fuel cladding. An increase in the atmospheric concentration of $^{85}$Kr in the vicinity of a storage facility can indicate a leak in fuel containers. When spent fuel rods are reprocessed, most of the $^{85}$Kr is released into the atmosphere and is difficult to contain. Considering these implications, accurate $^{85}$Kr measurement also has applications in nuclear security. The standard method involves sampling the air for approximately 1 week by 2 stages of gas chromatography with methane as the carrier and counting gas. **A more sensitive approach would lead to faster sampling and monitoring of this crucial isotope.**

**Muon tomography** or 'muography' is a technique for imaging objects which is impenetrable by more conventional X-ray imaging techniques, see Fig. 7.10. It has evolved from archaeological and geophysical interests to practical applications such as determining the density of the contents of a barrel and from that data inferring the identity of the contents. This technology uses muons from naturally occurring high-energy cosmic rays. After detection, their path through the object is reconstructed to give a 3D density map.

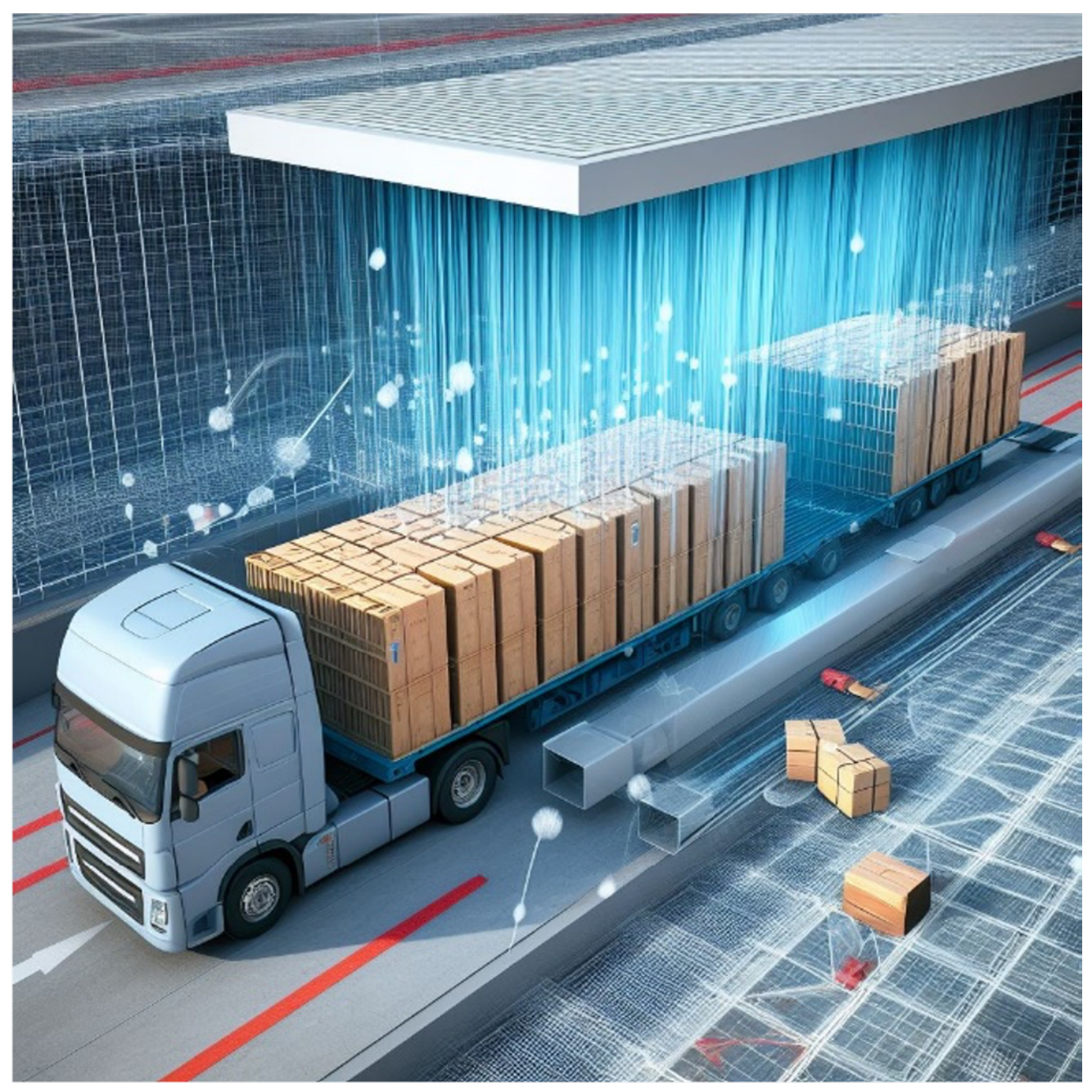

*Fig. 7.10: A muon scanner to inspect the content of cargo containers*

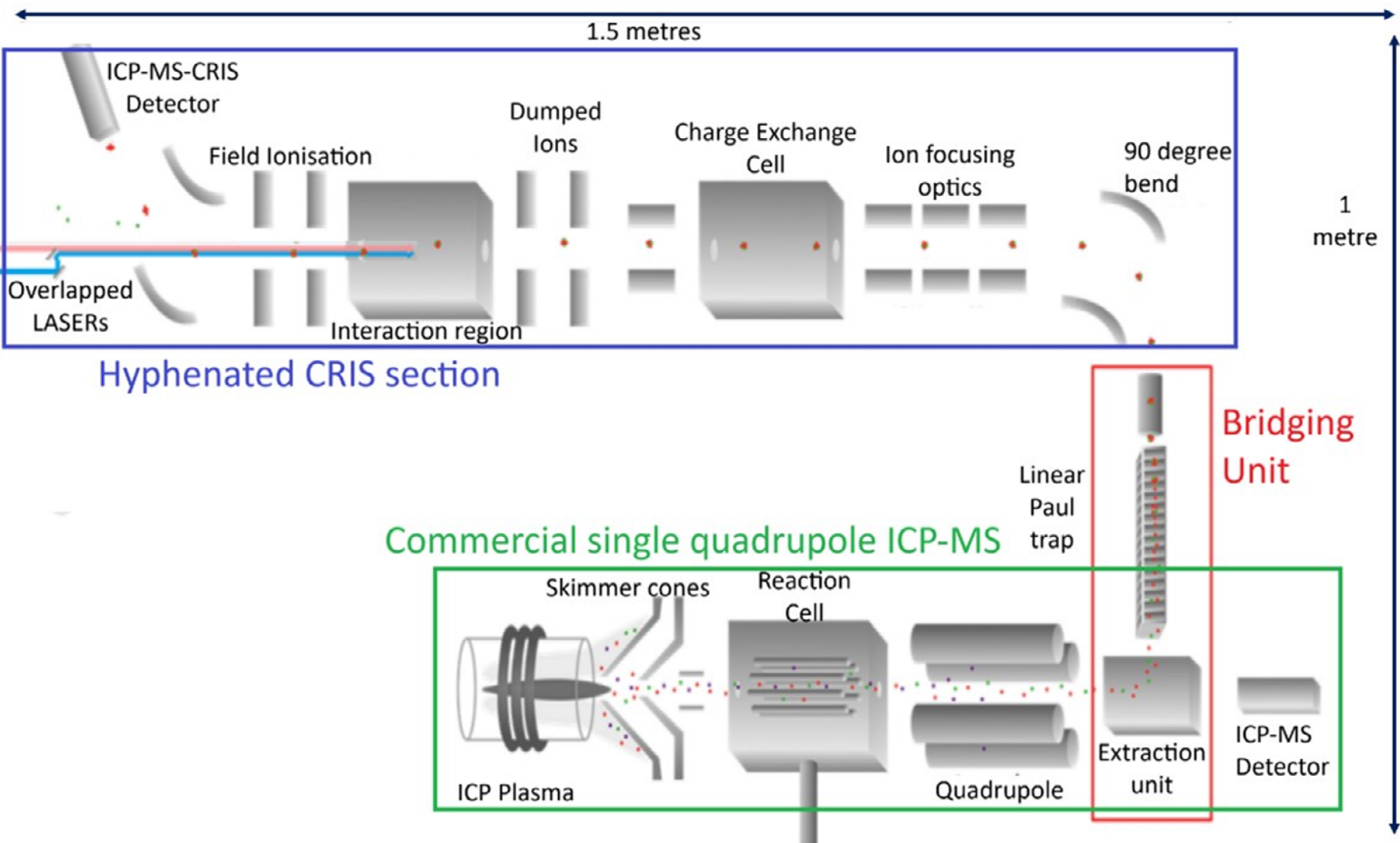

*Fig. 7.11: Layout of the ICP-MS-CRIS combined analytical technique*

## Perspective and needs

As the nuclear decommissioning programmes across Europe accelerate, so the ability to characterise larger volumes and sample numbers requires cost-effective solutions. Additionally, the practical requirements of site characterisation, remediation and verification/validation demand quantification techniques with a rapid capability, ideally with a turnaround time of a few hours.

An alternative method proposed for monitoring exotic radionuclides in environmental samples involves the use of the Collinear Resonance Ionisation Spectroscopy (CRIS) technique, see Fig 7.11. Further development of the technique, including the feasibility of hyphenating the CRIS technique with existing ICP-MS instrumentation for ICP-MS-CRIS, can provide new opportunities to explore our planet.

# Cross-disciplinary applications

## Heritage

Heritage science is the interdisciplinary domain of the scientific study of cultural or natural heritage. It draws on diverse humanities, science and engineering disciplines and focuses on enhancing the understanding, care and sustainable use of heritage to enrich people's lives, both today and in the future.

Cultural heritage refers to the legacy of tangible artefacts inherited from past generations, maintained in the present and bestowed for the benefit of future generations. This includes everything from buildings, monuments, landscapes, books, works of art and artefacts. Cultural heritage is focused on studying and preserving the practices, knowledge and physical remnants of human cultures.

Natural heritage refers to physical, biological and geological formations, as well as the habitats of threatened species of animals and plants and areas with scientific, conservation or aesthetic value. This encompasses ecosystems, biodiversity and geological structures such as caves, volcanoes or fossils with significant scientific or aesthetic value. Natural heritage is focused on studying and preserving the natural environment and its many species.

Nuclear physics offers a diverse array of **nuclear analytical techniques (NAT)**, from IBA to muon-induced X-ray emission (MIXE) and neutron techniques, to ascertain the composition and age of tangible heritage. It plays a pivotal role in the conservation of art and archaeological artefacts. These techniques can detect trace elements without requiring physical sampling. They provide elemental distribution through mapping, depth data, and even isotopic ratios.

**MIXE** is a new development that provides depth-dependent information up to a few cm (sample dependent), is sensitive to all elements in the periodic table (from Li to U), and can even be used to determine the isotopic composition for elements with $Z > 25$. However, this approach is only available at PSI and access is thus highly limited.

Unearthing the timelines and preserving the stories encapsulated within our shared heritage is complementary to the composition analysis. $^{14}C$ dating, facilitated by **Accelerator Mass Spectrometry** (AMS), is a pivotal tool in this context. Europe is witnessing a surge in the number of AMS facilities, and the advent of compact versions has enhanced accessibility. However, when applying these tools to heritage science, the preparation of samples holds equal, if not greater, significance than the actual measurement process. Incorporating standardised systems for graphitisation and carbonate sampling setups is indispensable to achieving the highest accuracy and quality of results. Besides AMS, other (quantum) optical methods are promising and may soon find their field of applications (e.g. ICP-MS-CRIS).

While some analytical methods are deemed non-destructive, irradiation can still activate the sample or induce changes based on the material and experimental variables. Reducing both the beam current and acquisition duration can mitigate this impact but requires efficient detector systems. Progress aimed at material analysis can also improve the capabilities of heritage science, e.g. improve lateral resolution in the micrometre range.

Cultural heritage artefacts made of wood, paper or textiles can be attacked by different biological species. **High-intensity $\gamma$ rays are a valuable tool** for conservators in destroying bioactivity (disinfection) or for consolidation by radio-polymerisation. Irradiation acts on all biological aggressors and large numbers of objects can be treated simultaneously. Considerable international efforts are ongoing to develop this approach for cultural heritage, under the auspices of the IAEA.

Access to these techniques by the heritage community remains exceptional. The lack of dedicated facilities and the limited access to muon and neutron sources is a challenge to the widespread use of them. Continuous cooperation between human and natural science researchers and access to archaeologists, conservators and curators remain significant challenges. Portable machines, e.g. MACHINA - the Movable Accelerator for Cultural Heritage In-Situ Non-destructive

Analysis - will bring IBA techniques directly to museums and therefore improve both access and visibility towards that community, see Fig. 7.12.

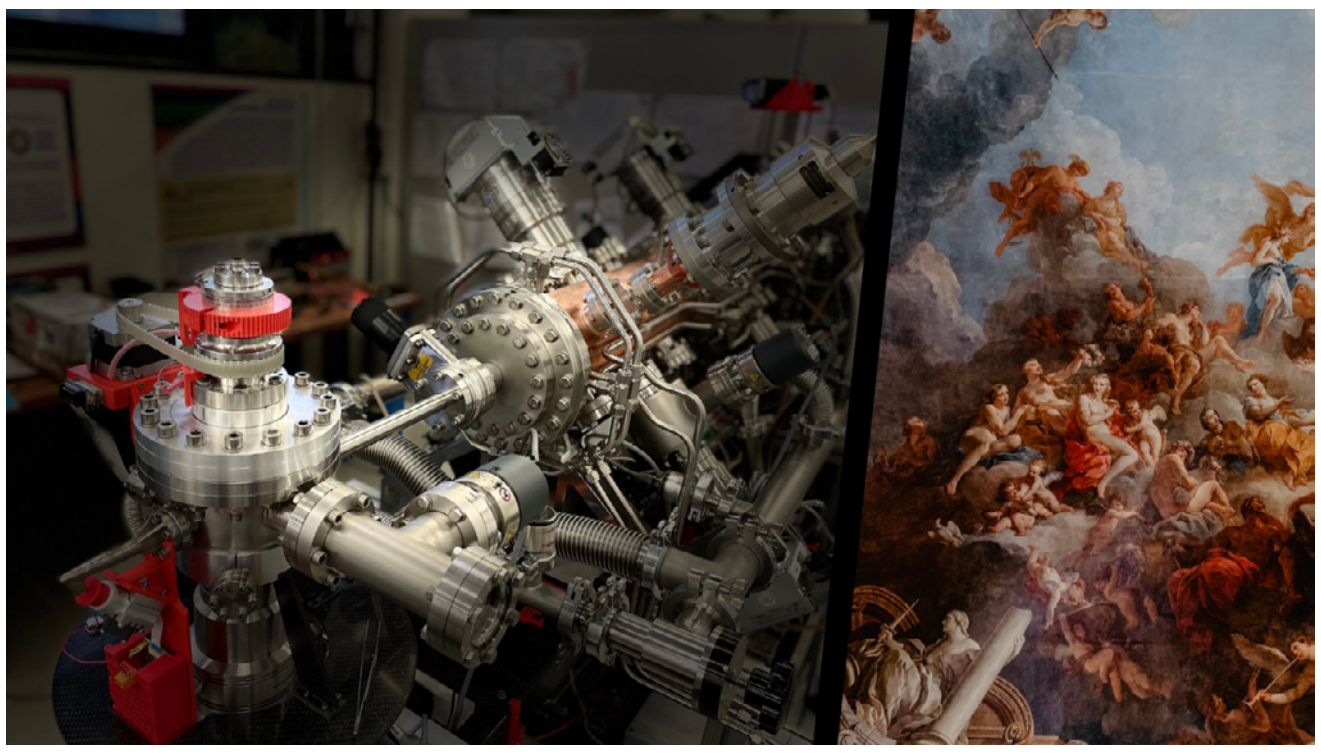

*Fig. 7.12: CERN and INFN have developed a new transportable accelerator, MACHINA, for analysis of works of art*

An important step forward is the European Research Infrastructure for Heritage Science (E-RIHS) initiative on the ESFRI Roadmap. E-RIHS aims to supply state-of-the-art tools and services to a diverse community of researchers dedicated to expanding our understanding of heritage and refining preservation strategies. Within E-RIHS, nuclear physics techniques are recognised as invaluable tools that shed light on historical methodologies, materials, timelines and degradation processes.

## Materials Research

Nuclear techniques provide unique tools for materials research, including the synthesis, modification and characterisation of materials. These methods are applied experimentally with enormous success, yielding novel nanostructures, new phases and/or unique morphologies hardly accessible via other techniques. Great progress is also being made in modelling experimental results and thus enhancing the impact of the measurements. Around Europe, research facilities provide different types of particle beams from protons to U and even radioactive ions to probe/modify a large variety of materials.

**Swift Heavy Ions** (SHI) of MeV-GeV energy are delivered at large-scale infrastructures such as GSI/FAIR and GANIL, whereas ions of lower energies (keV-MeV) are available at numerous smaller accelerators across Europe. The future linear accelerator HELIAC at GSI will deliver a continuous wave beam with improved conditions for materials science. At GANIL, online Electron Paramagnetic Resonance (EPR) is in preparation, which will mitigate the need for multiple irradiations in the study of the effect of fluence. Examples of current and prospective materials research activities with SHIs include radiation hardness tests of functional materials in high-dose environments, tests of space electronics, beam-induced sputtering and desorption, defect engineering, astrochemistry, radiation effects on radiolysable and organic materials and ion-track nanotechnology, including the implementation of nanochannels and nanowires produced by ion-track nanotechnology in health- and energy-related applications. Ion beams of various energies are also applied to investigate the response of materials to fission/fusion-related environments.

Many ion accelerator laboratories providing keV-MeV ion beams are equipped with a range of specialised techniques for IBA (e.g., RBS, ERDA, PIXE, PIGE). This broad set of instrumentation plays a crucial role in the analysis of materials, but keV-MeV ions can also be applied for synthesis, processing and near-surface modification of materials. Ion-beam implantation is also valuable to dope a variety of crystalline materials like GaN, ZnO, diamond, sapphire and $LiNbO_3$, resulting in the formation of structures with carefully tuned optical, luminescent, and ferromagnetic properties. By adjusting the beam parameters, these techniques offer the capability to precisely profile the depth and accurately position dopants within the lattice structure. Research examples include MeV ion irradiation for nanopatterning of semiconductor surfaces and the modification of 2D material through ion irradiation, for example to enhance its hydrogen evolution reaction capabilities.

**Radioactive ion beams (RIB)** are applied to probe local environments for materials, using techniques like emission channelling, perturbed angular correlation spectroscopy and Mössbauer spectroscopy. These methods are experiencing renewed interest as the role of individual distortions and defects in materials becomes more important in nanoscale devices. This is particularly evident in semiconductor and quantum technologies where the control of defects and single-defect centres is crucial. A new generation of spectrometers allows *in situ* probes of materials while strong external magnetic fields are applied, creating opportunities for studies of quantum optical materials, multiferroics, novel green-energy materials and materials relevant to constructing q bits for quantum information. The precise localisation provided by implanted radioactive probes also provides opportunities for studying surfaces, particularly in topological insulators and low-dimensional 2D materials and heterostructures. ISOLDE combines a large diversity of radioisotopes and equipment to address the needs of these research domains. Soon, ISOL@MYRRHA will offer a second facility where these applications will become possible.

### Box 7.4: Mössbauer spectroscopy

**Mössbauer spectroscopy (MS) is one of the key techniques in nuclear solid-state physics. More than 90% of MS applications use the stable isotope $^{57}$Fe due to the eminent role of iron in the animate and inanimate world, ranging from modern materials science to essential functions in biological systems. Laboratory-based $^{57}$Fe MS relies on the radioisotope $^{57}$Co as a source of 14.4 keV radiation to analyse the hyperfine interactions of the $^{57}$Fe nucleus in the respective samples, thereby providing a powerful local probe for electronic, magnetic and dynamical properties of condensed matter. MS is currently being applied in nearly 200 laboratories with 1300 active researchers within the European Economic Area (EEA).**

**The synchrotron-based version of MS comprises nuclear resonant forward scattering (both in time and energy domains), nuclear inelastic scattering and synchrotron-radiation perturbed angular correlation. The latter methods are extensively used for studying the dynamical properties of materials, particularly in extreme conditions, with an important impact on material science and earth sciences as well as biomolecular dynamics. The ongoing increase in the source brilliance, reflected in the upgrades of facilities like the ESRF and PETRA IV in DESY will lead to further decreasing length and time scales.**

**With X-ray lasers like the European XFEL, the application of nuclear resonant methods has just started with fascinating perspectives for fundamental research, e.g. in the field of nuclear quantum and nonlinear optics. It is thus anticipated that accelerator-based MS will continue to shape the research portfolio at these facilities in the coming 10 years.**

**Besides $^{57}$Fe, other relevant isotopes in this field are $^{119}$Sn, $^{161}$Dy, $^{149}$Sm, $^{61}$Ni, and $^{40}$K. In many important applications, the natural abundance of these isotopes is not sufficient, therefore enriched material is needed. This applies in particular to samples which contain these atoms in dilute quantities only, such as experiments with very small sample volumes, like in nanoscience or high-pressure physics. Dilute systems can also be investigated using emission Mossbauer spectroscopy, in which a radioactive isotope is incorporated in the material under study and then decays to the excited Mössbauer state; typically, ion implantation in ISOL facilities is used for the incorporation, making it possible to use short-lived isotopes (e.g. $^{57}$Mn/$^{57}$Fe at ISOLDE).**

### Box 7.1: The nuclear clock: a new tool in the nuclear toolbox

**The atomic clock is a prominent example of quantum technology, now sometimes referred to as a first generation. Its development has advanced the accuracy of time and frequency measurements by about ten orders of magnitude, into the $10^{-18}$ range, making it the world's most precise timekeeper. Much of the progress with atomic clocks over the past 30 years has been achieved by the application of methods of laser cooling and trapping of atoms and ions. The idea of a nuclear clock was first proposed in 2003 and is now intensely discussed, since it would be more precise than today's most precise atomic clock due to the difference in size and constituents of a nucleus compared to an atom.**

**The conceptual shift from an atomic to a nuclear clock is straightforward: instead of a transition in the electron shell, a periodic transition between two states of an atomic nucleus are used as the frequency reference and a laser oscillator is stabilised to the resonance. The small size of the nucleus and tight binding of its constituents make resonant frequencies associated with internal excitations highly insensitive to external fields, providing strong suppression of some systematic frequency shifts that are of concern in atomic clocks. Two additional aspects contribute to the interest in a nuclear clock: first, Mössbauer γ-ray spectroscopy has shown that extremely high resolution may be obtained for nuclear resonances in solids, potentially providing a much stronger signal. Second, in contrast to the transition frequency of a valence electron, the nuclear resonance frequency is not purely determined by electromagnetic interactions but also by strong interaction. The comparison of atomic and nuclear clocks could therefore make unique contributions to tests of fundamental physics and searches for new physics beyond the Standard Model based on precision clock comparisons.**

**Since 2003, the transition between the ground state of the thorium-229 ($^{229}$Th) nucleus and the first isomer state has been proposed as the basis for the nuclear clock. This nucleus is unique with an isomer energy of about 8.3 eV, in the range of outer-shell electronic transitions. Recent experiments have provided essential information on the nuclear properties of $^{229}$Th, such as nuclear moments, decay modes of the isomer and the first observation of the ultraviolet gamma-ray photon. The latter was achieved at ISOLDE, by producing the isomeric $^{229}$Th via the β decay of $^{229}$Ac within an ultraviolet-transparent crystal. A major breakthrough was achieved with the laser-excitation of that transition using a broadband laser. A large gap remains between the current knowledge of the transition frequency (GHz range) and the natural linewidth (in the mHz range). A few laboratories have developed dedicated tunable laser systems in the ultraviolet vacuum at about 150 nm wavelength to search for the $^{229}$Th resonance. The success of these efforts will open a new field of experimental work with novel opportunities in coherent state preparation and precision frequency measurements in a nucleus.**

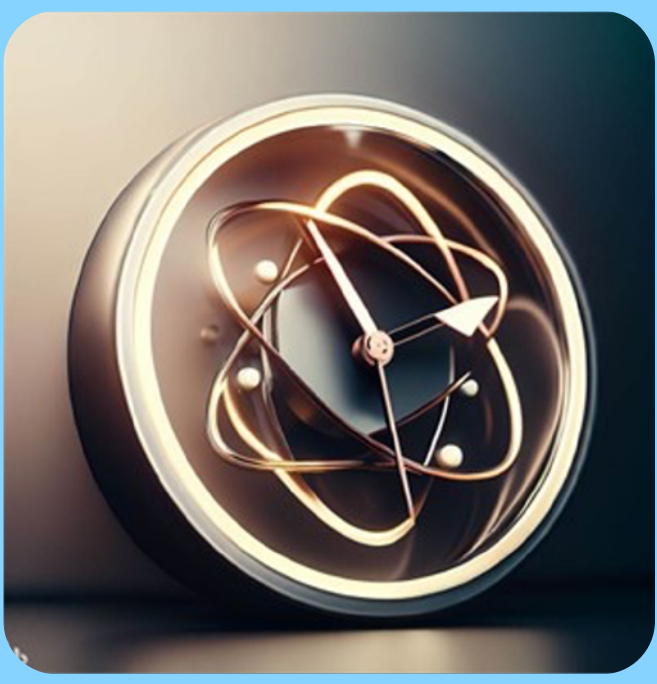

*Fig. 7.14: AI-generated illustration of a nuclear clock*

**Neutron beams** are also pivotal for materials science, with activities spanning investigations of atom arrangement in crystals to nano and microscale heterogeneities. The high penetrability of neutrons allows for bulk and *in situ* analyses under different conditions such as mechanical strain and high temperature. Advanced neutron activation methods (e.g. NAA, PGAA, NDP) also contribute to the field of materials research. Supporting small-scale European neutron facilities is deemed crucial to enable science programmes to continue and expand high-impact applications.

**Muons** are used for the study of magnetism and other bulk properties using the muon spin rotation, relaxation or resonance (**μSR**) technique with high-energy positive muons, as well as the characterisation of semiconductor detectors with μSR using low-energy positive muons. **MIXE** can be used to investigate sensitive structures in a non-destructive way, such as operando studies of Li-ion batteries. PSI has plans to increase the available muon rates by up to a factor of 100 for particle physics research as well as materials science at the High-Intensity Muon beams facility envisaged at HIPA.

Testing of innovative accelerator materials under the extreme conditions they may encounter in the case of accidental beam impact takes place at ELI-NP and GSI/FAIR. This includes testing, optimising and developing materials for e.g. collimators, beam-dumps, and secondary particle production targets, as well as quantifying material damage for operating scenarios.

## Forensics

Although in daily forensic work, commercially available, high-throughput table-top instruments dominate, NAT has some unique features that make them increasingly attractive when standard forensic techniques fail. One of the main obstacles to more widespread use of NAT in forensics is the necessity to use external accelerator or reactor facilities outside the forensic laboratory. To bridge the gap, access to NAT facilities is provided through European projects, the IAEA, and the CERIC-ERIC, which define a new model that may be relevant to many other distributed infrastructures.

NAT applied to heritage helps to identify archaeological objects from forgeries. NAT and IBA are also successfully applied in bulk analysis of food, drugs, counterfeited medicine, automotive glass, paints, gunshot residues and many more. In the case of AMS, there are many examples of the use of $^{14}$C dating in solving criminal cases by analysing small samples of human tissue, bone, teeth or hair. This technique is also applied in solving cases of smuggling, trading of endangered species or counterfeiting food, drugs and wine, or to screening forged objects of art.

## Quantum technology

The second quantum revolution sees the development of quantum systems not only for their fundamental interest but also for their industrial and large-scale applications. Nuclear science can still offer new opportunities for quantum technology research and developments that should be fully explored.

One of the main approaches to quantum information is the use of q-bits in the form of ions. Radioactive ions may provide additional degrees of freedom for manipulation with respect to stable ions. For example, the atomic structure of Ba+ ions has been identified as being particularly suited for optical quantum manipulation. Its naturally occurring odd-mass isotopes $^{135,137}$Ba both feature a ground-state nuclear spin of $3/2^+$, which complicates their manipulation. However, the radioactive $^{133}$Ba, with a spin of $1/2^+$, would be more appropriate. **New openings to the quantum information community should be actively sought out to promote those opportunities.**

A new development in nuclear science is the nuclear clock (see frame). Whether based on trapped ions or on solid-state devices, the nuclear clock will offer new opportunities for high-precision time measurement. Many developments are ongoing across Europe to achieve the ability to manipulate the nuclear states in $^{229}$Th, and, once achieved, many new opportunities will arise in the decade to come.

## Quantum chemistry with radioactive molecules

Across different societal applications of nuclear radioactivity, such as monitoring nuclear safety, nuclear waste management and nuclear medicine, there is a shared need to develop our understanding of the chemical compounds formed by the radioactive isotopes of interest.

Nuclear chemistry research in the gas, liquid, and solid phases has been ongoing for several decades using long-lived radioisotopes found in natural abundance on Earth or separated from spent nuclear fuel at specialised facilities. In certain cases, however, the radioactive elements of interest have no isotope with a half-life longer than a few hours or days. Examples are astatine and francium, where the longest-lived isotopes are $^{210}$At ($T_{1/2}$=8.1 hours), $^{211}$At ($T_{1/2}$=7.2 hours) and $^{223}$Fr ($T_{1/2}$=22 min). This also applies to compounds of interest that contain specific short-lived radioisotopes, such as biomolecules of $^{211}$At and $^{225}$Ac for targeted α therapy. In such cases, radioactive ion beam facilities (RIB) can offer purified beams of short-lived radioisotopes for research in a variety of fundamental and applied areas.

The first spectroscopic experiments with radioactive molecules at ISOLDE have demonstrated that gas-phase molecular spectroscopy at RIB facilities can provide a powerful benchmark of ab initio quantum chemistry. **Further developments in molecular production and study at European RIB facilities should expand this programme to a region of the periodic table where relativistic effects cannot be ignored.** A large number of electrons have to be considered, and the interplay of 5*f* and 6*d* electrons in bonding is not fully understood

## Perspective and needs

Besides $^{14}$C, other isotopes of relevance to heritage science ($^{10}$Be, $^{26}$Al), and the actinides benefit from smaller AMS facilities. However, $^{53}$Mn and $^{59}$Ni, important for natural heritage, require electrostatic accelerators dedicated to AMS with more than 10 MV terminal voltage. These are currently not readily available in Europe. The addition of one such machine in Europe is essential to complement the existing infrastructure.

To maintain and further develop research programmes, the materials science community needs a different beam access model compared to the nuclear science community. This model should reflect the fast-paced domain of materials science and quantum solid-state physics, where novel ideas may grow and come to fruition in a shorter timeframe than the typical experimental cycle of large-scale accelerator facilities.

National facilities play a crucial role in enabling research across various fields, filling a gap in the capabilities of large instruments and accelerators. Small-scale laboratories are part of European-funded infrastructure consortia. However, the short-term nature of those funding schemes brings a lot of pressure on their long-term reliability and in ensuring access for the various communities.

# Nuclear data

Nuclear theory lacks the ability to predict fundamental physical quantities, such as nuclear reaction cross sections, nuclear structure and nuclear decay parameters, with the level of precision necessary to assess the performance, safety and security aspects of various nuclear research, technological and health applications. Therefore, the precise determination of these nuclear data must be obtained through experiments. Experimental data, where needed complemented by theoretical models, form the basis of evaluated nuclear data, consisting of our best knowledge of the physical quantities. Evaluated nuclear data need validation with benchmarks sensitive to the quantities of interest. These benchmark experiments require well-defined radiation fields generated in dedicated experimental facilities.

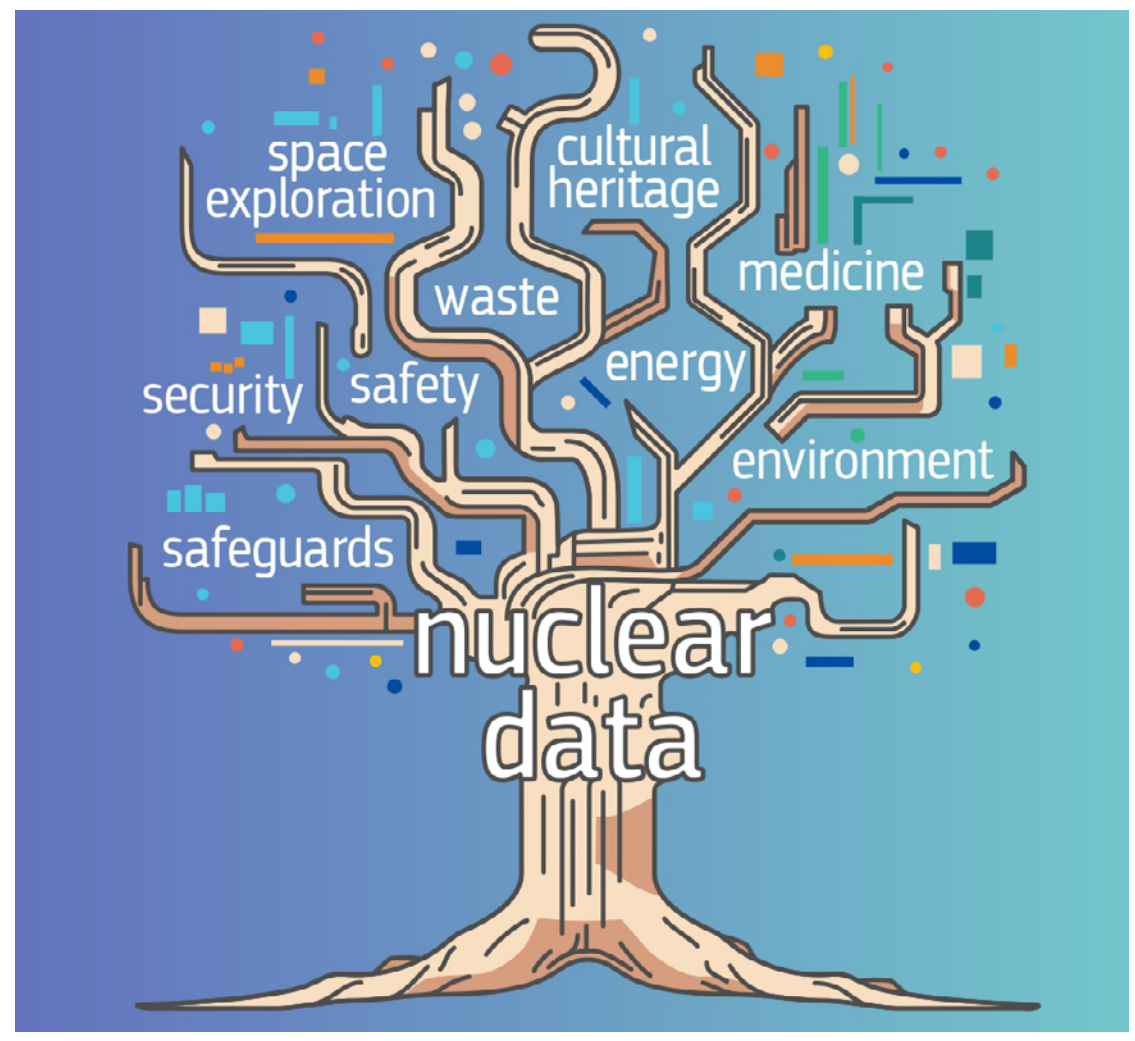

*Fig. 7.13: Nuclear data for science and applications © European Atomic Energy Community, 2023 - C.L. Fontana (JRC-Geel)*

**Final users of nuclear data** cover a wide array of nuclear engineers and scientists working for nuclear safety authorities, radioactive waste management institutions, operators of nuclear and radioactive facilities, engineering companies developing innovative nuclear devices and applications, hospital health physics and nuclear medicine units, institutes involved in nuclear technology research, basic research, environmental nuclear applications, heritage science and other non-energy nuclear applications. Their needs, particularly those related to safety and safety margins associated with reactors and other nuclear facilities, require exceptional precision. In most cases, nuclear data with a precision better than a few per cent is imperative.

There are different types of data and classifications depending on the observables and the purpose of the measurements. In all cases, it must be underlined that the path of experimental data reduction, analysis, evaluation and incorporation into theoretical models and simulation tools is a **complete cycle** as discussed in the Chapter on Open Science and Data.

## Nuclear data needs and priorities

### Fission technology

To enhance the operational conditions and safety of existing reactors and, notably, to advance ongoing research for the development of innovative fission reactor models like Gen IV fast neutron reactors and sub-critical ADS, **high-accuracy cross-section data for neutron-induced fission reactions and capture reactions for thorium, uranium, plutonium, and minor actinides are imperative**. Given that future nuclear reactor designs predominantly feature fast neutron reactors, the necessity for high-precision cross-sectional data extends from the thermal neutron energy range to the MeV energy domain. Recent examples of key measurements carried out for the U/Pu and Th/U fuel cycles are the fission and capture experiments on $^{233,235}$U, $^{239}$Pu, and $^{230}$Th measured at the n_TOF facility at CERN.

Cross-sectional data on bismuth and lead assume paramount significance for the advancement and operation of ADS and fast critical liquid-metal reactors, as lead-bismuth eutectic has been proposed as a coolant. For other structural materials such as Fe, Mn, Cr, Ni, Al, or alternative coolant candidates, e.g. Na, comprehensive data encompassing neutron capture reaction cross-sections, γ-ray yields, and neutron elastic and inelastic scattering cross-sections are of vital importance. Neutron-induced reactions where one of the reaction products consists of light-charged particles, such as protons, deuterons, tritons, or α particles, significantly influence the radiation damage sustained by structural materials.

Despite the diligent efforts of numerous national and international collaborations and facilities, the precision requirements set forth by organisations like the IAEA, NEA, and other stakeholders in future fission reactor technologies remain unmet. **The accuracy of these data directly impacts the precision of any neutron economy calculations and studies concerning the radiation resilience of reactor structural materials.**

## Fusion technology

The design, construction, and safety operation of fusion reactors require a large set of nuclear data. One of the most important issues is the effect of radiation damage on structural materials in the inner part of the fusion reactor, subject to an extremely large flux of 14 MeV neutrons (and lower). The interaction between neutrons and structural materials, specifically involving inelastic scattering, capture, and neutron-induced light-charged-particle reactions (n,cp), is responsible for displacement, transmutation and gas formation, ultimately limiting their operational lifespan. While the neutron damage in fusion reactors can be assessed experimentally at neutron irradiation facilities like IFMIF-DONES, the interpretation of the results and model calculations rely on neutron interaction cross sections, up to several tens of MeV.

Reaction data available in the Fusion Evaluated Nuclear Data Library (FENDL), and activation cross sections from the European Activation File (EAF) needed to address safety, licensing, decommissioning and waste management issues, are often based on scarce or discrepant measurements, or in a limited energy range. For this reason, **there exists a pressing need for new neutron data**, for a variety of reactions and isotopes, and for a wide energy range.

## Health applications

The enormous interest in targeted radionuclide therapy is resulting in the search for novel radionuclides with ideal characteristics for nuclear medicine applications, which is driving the need for new methods of production of these radionuclides. Optimisation of these production routes requires accurate determination of the irradiation cross-sections, for which accurate nuclear decay data is vital.

For many of the novel medical radionuclides, (e.g. Tb theragnostic quartet) the nuclear decay data are not of sufficient quality (sometimes last studied several decades ago). **There are ongoing efforts to determine new data for these radionuclides that should be further strengthened and supported.**

Concerning EBRT, there remains substantial uncertainty on the fragmentation of elemental constituents of organic matter (C, N, O) when irradiated with high-energy protons or light ions. Those fragments have a deeper range in the matter so they induce unwanted additional doses that should be accounted for. Detailed cross-section measurements covering the full range of beam energy are required to be able to fully model the irradiation condition for treatment planning.

## Materials research

IBA techniques today constitute one of the most prominent research fields in practically all the existing small accelerator facilities worldwide. However, the implementation of IBA techniques is quite often impeded by the lack of adequate and/or reliable charged particle differential cross-section data over a wide range of energies and backward detection angles. To deal with this limitation, a new, comprehensive library, IBANDL (Ion Beam Analysis Nuclear Data Library) has been created under the auspices of the IAEA. I**BANDL requires considerable research efforts over the next decades before it becomes finalised**. For example, important differential cross sections for $^{3}$He-induced reactions, particularly useful for fusion materials applications, are still missing; proton, deuteron and α-particle data at relatively high energies (namely above 3 MeV/nucleon) are scarce; theoretical evaluations for several key isotopes are still pending, while the validation of all the widely used cross sections is clearly needed to create a firm basis for analytical work.

## Space applications

Nuclear data and reaction rate calculations are integral to evaluating the impact of GCR and SPE on the structural materials and electronic systems of aircraft and spacecraft, as well as on the health of astronauts for long-term space missions to the Moon and Mars. **Continued refinement of the accuracy of nuclear data, especially at higher energies, is paramount.** Even for the most common particles like protons, neutrons, and α particles, data on reaction cross sections at high energy (GeV) are scarce and may exhibit disparities. Consequently, existing simulation packages predominantly rely on general reaction models, which may lead to significant inaccuracies.

Lastly, we must underscore the significance of high-precision nuclear data in the development of advanced nuclear detection technologies for security purposes, as well as in dosimetry and metrology studies. Despite the concerted efforts of the scientific community to enhance nuclear structure, decay and reaction cross-section data, there remains room for improvement in this critical domain.

# Facilities and future experiments

Existing and future accelerator and research reactor facilities serve a dual purpose: they provide well-defined particle beams that enable high-quality nuclear research activities, and they also serve as invaluable training grounds for young researchers and as hubs for interdisciplinary collaboration among researchers with varying levels of experience and expertise. These facilities play a major role in advancing nuclear data quality. Each facility features unique characteristics and operates in a complementary manner to reach more comprehensive and accurate nuclear data. This becomes particularly critical in cases where extreme precision and minimal uncertainties are imperative, such as in reactor technology and medical applications.

Europe is home to a diverse park of large-scale facilities that offer stable/radioactive ion and neutron beams. However, the pivotal role played by existing infrastructure in advancing more precise, comprehensive and robust nuclear data for nuclear physics applications can only be sustained if continuous efforts are made to upgrade and improve these facilities. Finally, smaller-scale facilities allow for a more hands-on approach, facilitating greater involvement in experiments. Moreover, these facilities can serve as testing grounds for the development of innovative detection systems, which can subsequently be applied in larger-scale facilities to facilitate more inclusive measurements.

Measurements of nuclear data require **special samples and targets** that can be radioactive or require high isotopic enrichment. Cf. Chapter on Nuclear Physics Tools Detectors and Experimental Techniques, Sections on material enrichment and target preparation for further detail. Furthermore, accurate and precise nuclear data require **advanced measurement protocols and detector systems**, such as those described in Chapter on Nuclear Structure and Reactions as well as in Chapter on Nuclear Physics Tools Detectors and Experimental Techniques. Finally, a wide variety of experimental data must be compiled, assessed, evaluated and validated following a **complete nuclear data cycle** before its incorporation into the nuclear databases, as described in Chapter on Open Science and Data.

## Metrology standards

The determination of all nuclear decay data comes through the application of measurement techniques. It is therefore vital that all measurements be carried out with best **metrology** practices, supported using standards and application of rigorous uncertainty analysis. Many countries across Europe have national metrology institutes (NMIs) or designated institutes (Dis) that are responsible for maintaining standards for the SI units and disseminating standards to end-users. The use of these standards and the application of metrology to nuclear decay data measurements has been shown to ensure that the data being generated is consistent across continents and provides confidence in the data being generated. This is of special importance when new evaluations of nuclear decay data are undertaken.

Nuclear data measurements are typically simpler to execute when performed relative to reference standards. In most neutron reaction cross-section measurements, for example, the direct measurement of the incident neutron fluence is replaced by the measurement of a standard reaction cross-section. This also facilitates the evaluation procedure, as the results of individual experiments can easily be rescaled if the standards are re-evaluated. With increasingly stringent requirements on the target accuracy of the general-purpose libraries, **it is, therefore, crucial to work continuously on the progress of the standards,** as any improvement in the standards will cause all measurements relative to that standard to be improved.

In past years, advances in nuclear technologies, medical applications and air and space travel have highlighted the need for high-energy neutron standards ranging beyond the 20-MeV limit usually considered for fission applications. Great effort is being invested in extending the range of validity of certain standards; at present, two standard cross sections ($^{235}$U(n,f) and $^{238}$U(n,f)) have been extended up to 200 MeV, to ideally reach 1 GeV. The main limitation is represented by the experimental database, as measurements extending above 100 MeV are rare. **For this reason, there is an increasing interest in the measurement of (n,f) and (p,f) reactions at high energy up to 1 GeV**, in which the main challenge is to produce absolute results with accuracy suitable for a standard. In Europe, the n_TOF facility at CERN is the only one with such capacity and its uniqueness should be fully exploited.

## Box 7.7: Detector technology: from the nuclear lab to applications

**Detector developments are essential to leverage the potential of new facilities, and special effort is devoted to this field of research across all layers of nuclear science. When we refer to nuclear information for energy, medical or other kinds of applications, precision and accuracy are key factors. To fulfil these requirements, dedicated experimental setups must be developed with advanced characteristics.**

**The full exploitation of future high-intensity beams can be achieved through highly segmented detection setups that provide wide solid angle coverage along with advanced spatial and energy resolution. Recording multiple observables per event can yield more inclusive measurements. Typical examples are fission reaction cross-section studies that have so far been performed by recording counting rates. However, a more complete understanding of the fission mechanism requires the comparison of the predictions of fission models with many observables. These include the angular distribution, charge and mass distribution of the fission fragments, or even the recording of prompt and delayed (isomeric) γ rays. Such complete measurements are scarce and as a result, the fission process remains one of the less understood nuclear phenomena. The quality of the recorded data can be ensured only when advanced front-end electronics that allow for high sampling rates and high-resolution data are employed.**

**In the health sector, imaging is at the heart of patient care, be it in theragnostics applications or image-guided radiotherapy. The latest developments are found in PET imaging, where many new avenues are being explored to increase sensitivity, thereby reducing the required radioactivity for the patient, as well as increasing the speed of data collection and reducing cost. Several approaches are being explored, such as:**

**Time-of-flight (TOF) PET, which requires detectors with the highest time resolution;**

**3γ-PET, whereby not only the two 511-keV γ rays from electron-positron annihilation are being detected, but also a third γ-ray photon emitted at the same time;**

**Total-body PET, where the PET scanner is increased in length to cover the full body of the patient and thus takes the picture in a single shot without having to scan through slices of the patient (see Fig. 7.17);**

**CZT-PET detectors, which follow the patient's morphology, go closer to the head or neck of the patient to take higher-quality pictures while being able to retract further out for the rest of the body (see Fig. 7.18).**

**In heritage research, the techniques developed by the nuclear science community are being leveraged to offer a broad spectrum of analysis techniques, such as MACHINA (see Fig. 7.12) or MIXE (see section on cross-disciplinary research). These novel approaches require sensitive detectors that are adapted to the new opportunities offered by these techniques, like an adapted spectrometer to fully benefit from the depth sensitivity offered by the muon's penetration with MIXE. A very synergistic approach with the nuclear science community, e.g. between MIXE and other muonic x-ray spectroscopy collaborations at PSI, ensures an efficient knowledge transfer that should be fostered and exploited.**

**All these developments are enabled by the many advances in detector technology reported in Chapter Nuclear Physics Tools, which have an immediate impact and benefit for society.**

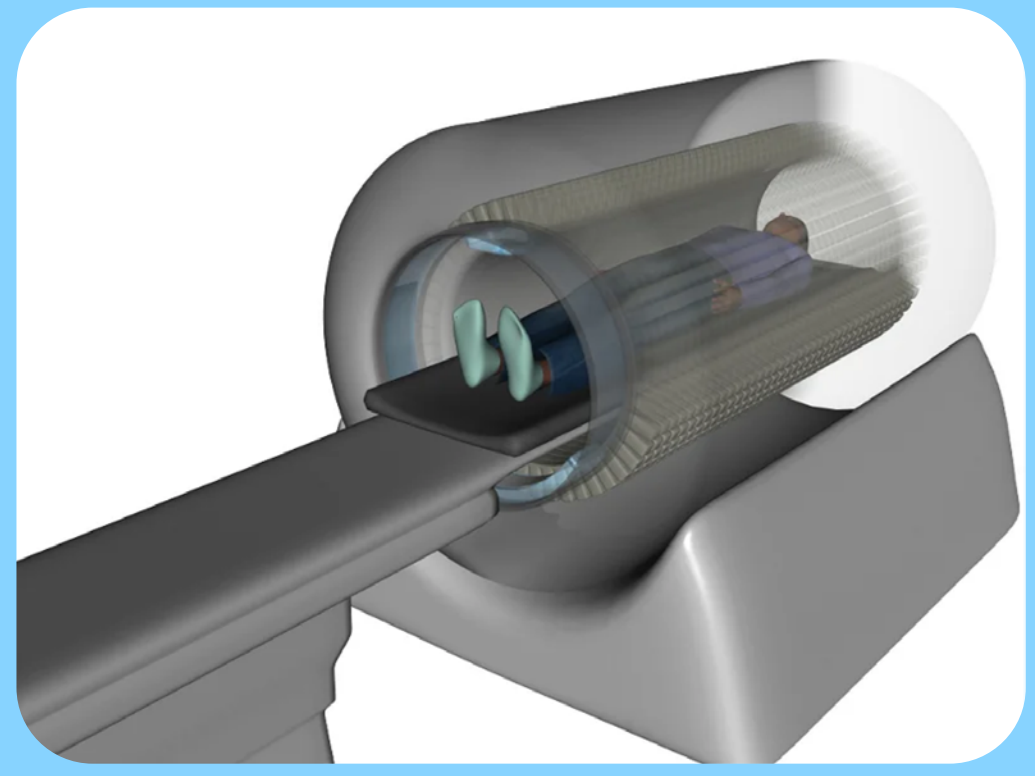

*Fig. 7.17: The EXPLORER scanner is designed to produce 3D pictures of the entire human body*

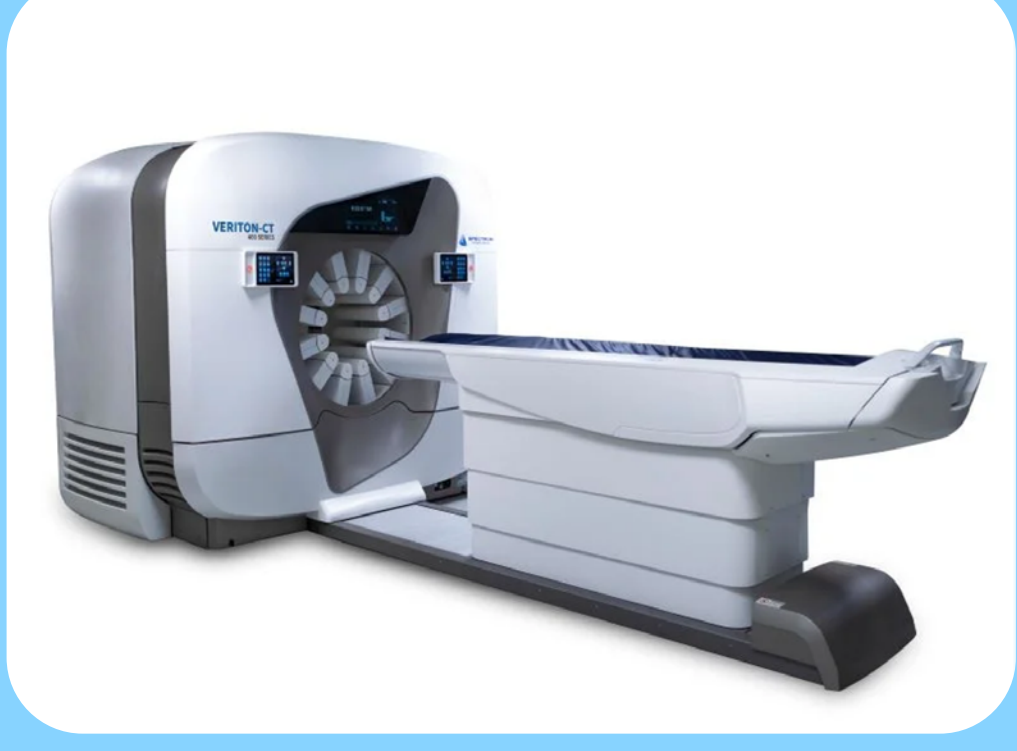

*Fig. 7.18: VERITON-CT 400 Series digital SPECT/CT scanner*

## Perspectives and needs

For nuclear applications that will have societal relevance for the foreseeable future, it is indispensable to continue to work on improved nuclear data. The next generation of applied nuclear physicists will need research infrastructures for the development of nuclear power as a future energy source, as well as for nuclear waste management and the choice of a final repository for safety, radiation protection and cross-cutting activities like medical radiation therapy.

The **maintenance of competencies in the nuclear data sector**, especially in the regime of nuclear applications where the need for accuracy is enhanced, requires the availability of large and small-scale facilities of stable and unstable ion beams. Neutron beam facilities should also be expanded to allow early-stage researchers to gain a fundamental understanding of the interaction probabilities of neutrons with nuclei. For integral validation experiments, benchmark and reference neutron and γ-ray field experimental facilities in Europe are becoming very rare. **The few of them still operating should be maintained through future European support programmes**.

Ultimately, the increasing demand for more precise nuclear data can surely be met through coordinated actions within the European nuclear community provided all stages depicted in Fig. 11.4 are accomplished.

# Research Infrastructures

**Convener:**
**W. Korten** (IRFU, CEA, Université Paris-Saclay, France)

**NuPECC Liaisons:**
**J. Goméz-Camacho** (University of Sevilla, Sevilla, Spain)
**P. Roussel-Chomaz** (GANIL, Caen, France)

**WG Members:**

- **Navin Alahari** (GANIL, Caen, France)
- **Angela Bracco** (Università degli studi di Milano and INFN Milano, Italy)
- **Francesca Cavanna** (INFN Sezione di Torino, Italy)
- **Letitia Cunqueiro-Mendez** (Sapienza, Rome, Italy)
- **Bogdan Fornal** (IFJ PAN, Kraków, Poland)
- **Sean J. Freeman** (CERN, Geneva, Switzerland, and University of Manchester, UK)
- **Zsolt Fülöp** (HUN-REN Atomki, Debrecen, Hungary)
- **Tetyana Galatyuk** (GSI and Technische Universität, Darmstadt, Germany)
- **Frank Gunsing** (IRFU, CEA, Université Paris-Saclay, France)
- **Ari Jokinen** (JYFL-ACCLAB, University of Jyväskylä, Finland)
- **Alexander P. Kalweit** (CERN, Geneva, Switzerland)
- **Andreas Knecht** (Paul Scherrer Institut, Villigen, Switzerland)
- **Ulli Köster** (Institut Laue-Langevin, Grenoble, France)
- **Razvan Lica** ("Horia Hulubei" National Institute for R&D in Physics and Nuclear Engineering, Bucharest, Romania)
- **Frank Maas** (Helmholtz-Institut Mainz, Johannes Gutenberg Universität, Mainz, Germany)
- **Carlos Munoz Camacho** (Université Paris-Saclay, CNRS, IJCLab, Orsay, France)
- **Catarina Quintans** (LIP, Lisbon, Portugal)
- **Marco Radici** (INFN Sezione di Pavia, Italy)
- **Berta Rubio** (Instituto de Física Corpuscular, CSIC - Universitat de València, Spain)
- **Konrad Schmidt** (Helmholtz Zentrum Dresden Rossendorf, Germany)
- **Paul Schuurmans** (TU Hasselt, The Netherlands)
- **Nathal Severijns** (KU Leuven, Belgium)
- **Thomas Stöhlker** (Helmholtz-Institut Jena and GSI Darmstadt, Germany)

The ambitious science programme of the European Nuclear Physics community is based on a wide variety of research infrastructures and their supporting integrating activities. These are succinctly presented in this chapter. This ecosystem of facilities provides a wide range of capabilities, necessary to cover the large breadth of the nuclear sciences, but also sufficient capacity to support a large and vibrant European community. A comprehensive description of worldwide research infrastructures in nuclear physics can be found in the latest IUPAP report[6].

# Hadron and Heavy-Ion Beam Facilities

Europe has a well-established infrastructure delivering hadronic and heavy-ion beams. These facilities have evolved to support a breadth of science opportunities from fundamental nuclear physics through to applications of nuclear techniques. Stable light and heavy ions are used at different energies, ranging from several TeV/u to probe new forms of matter, down to intense MeV beams used in isotope production for medical research and ions at tens of keV, to measure reactions that occur in stellar environments. Several facilities deploy accelerator systems using hadron beams to produce secondary beams that are used as tools for research. These include a very strong portfolio of radioactive beam facilities, together forming a formidable combined research capacity. High-energy proton drivers facilitate the production of secondary particles, such as muons and antiprotons, used as probes for material science and antimatter research respectively. New facilities or upgrades at many facilities maintain the world-leading position of European scientists exploiting hadronic and heavy-ion beams for their research.

## CERN

CERN (http://www.cern.ch/) is a global centre for high-energy physics focused on the exploitation of the Large Hadron Collider (LHC). At the same time, the LHC and the chain of accelerators that form its proton injector provide unique opportunities for nuclear physics. The LHC accelerates heavy ions to energies sufficient to create new states of QCD matter. ISOLDE is the only radioactive ion beam facility that uses GeV protons as a driver for nuclear physics and its applications. Protons at 20 GeV are used to produce high intensities of neutrons with a wide energy spectrum at n_TOF. World-unique low-energy bunched antiproton beams are produced at the Antiproton Decelerator (AD). Protons with up to 450 GeV are used to produce muon, pion, and kaon beams at 50-280 GeV for fixed-target experiments.

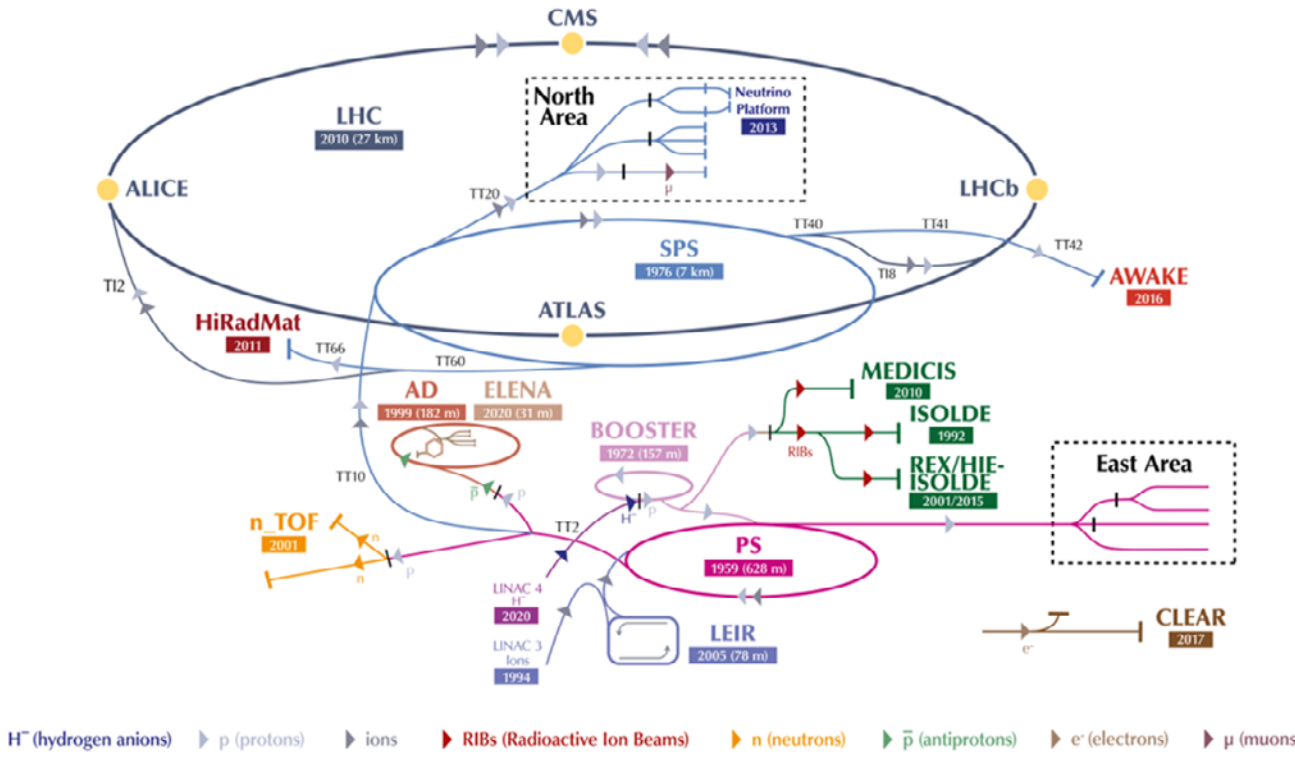

*Fig. 8.1: The CERN accelerator complex and experimental facilities (© CERN)*

Each of these nuclear-physics infrastructures relies on the unique capabilities of the CERN accelerators and have provided many high-profile exciting scientific results including, for example, studies of the quark-gluon plasma (QGP), characterisation of isomers for nuclear clocks, critical tests of matter-antimatter symmetry and rates of astrophysical nuclear reactions. In the medium term, each of these facilities has upgrades to maintain their strong positions.

CERN accelerators operate a sequence of running periods followed by long shutdowns[7]. After the current Run 3, Long Shutdown 3 (LS3) will begin in late 2025. While LHC Run 4 runs from 2029 to 2032, the proton injector chain will be available earlier from mid-2027.

### Nuclear Physics at the Large Hadron Collider

The LHC is also used to accelerate heavy ions to study the properties of the QGP, a deconfined state of matter at high temperatures and densities. ALICE (alice.cern) is the dedicated detector for this work, but all major LHC experiments (ALICE, ATLAS, CMS, LHCb) run successful and rich hadron and heavy-ion programmes. More recent developments on the synergies of nuclear, hadronic, particle, and astroparticle physics make successful use of the proton-proton collisions, providing unique insights into hyper-nuclei properties, anti-nuclei, hyperon-nucleon potentials and also nuclear structure via particle correlation measurements.

Each year, the LHC operates for one month with ions, where ion-ion or proton-ion collisions are studied. Major upgrades have been carried out for the current Run 3 at ALICE and LHCb, with intermediate Phase 1 upgrades at ATLAS and CMS. The ALICE upgrade was motivated by higher ion luminosities available in Run 3. From 2029, significantly higher proton luminosities will be available, fitted with major upgrades at ATLAS and CMS. In later run periods, the ion luminosities will be improved further and an upgrade, ALICE 3, will be needed. An upgrade to LHCb is also planned in this period.

The recent ALICE upgrade with a completely new inner tracking system and a GEM-based Time Projection Chamber running for the past year has enabled a continuous data mode, rather than triggered operation, facilitating an increase from 1 kHz to 50 kHz in interaction rate. ALICE 3 will involve the construction and installation of a completely new dedicated heavy-ion experiment to operate in the period 2035-41. It is based on a compact low-material budget, all-silicon tracker with excellent vertex reconstruction and particle identification capabilities. A Letter of Intent was reviewed by the LHC programme committee in 2022 and a funding strategy is being developed. The main physics goals are the observation of chiral symmetry restoration and thermal radiation in the QGP, as well as precision measurement of its properties using charm and beauty hadrons.

Whilst upgrades to the other LHC experiments aim for improved performance in proton-proton collisions, the heavy-ion programme at the LHC will nevertheless benefit significantly. In particular, the new CMS and ATLAS Inner Tracking System will increase coverage in rigidity and new timing layers will enable time-of-flight particle identification. During the present Run 3, the System for Measuring Overlap With Gas (SMOG-II) was installed, allowing fixed-target collisions with various ion species at previously unstudied energies; future LHCb upgrades will improve performance in pp and for all centralities in heavy-ion collisions.

### Antiproton Decelerator (AD)

The Antiproton Decelerator (https://www.home.cern/science/accelerators/antiproton-decelerator) is a CERN facility delivering low-energy antiprotons for antimatter studies. Antiprotons are created using 26-GeV protons on a metal target. The AD ring cools the antiproton beam and reduces the energy to 5.3 MeV. This feeds the ELENA ring, which was commissioned in 2017 as a priority in the last NuPECC long-range plan, and now provides exceptionally low-energy (100keV) and well-bunched (100ns FWHM) antiproton beams suitable for precision experiments based on traps and cold plasmas. A focus of the ELENA programme is the production and study of antihydrogen for matter/antimatter symmetry tests, probing differences in hyperfine structure, magnetic moments and gravitational effects between hydrogen and antihydrogen.

[6] https://wg9.triumf.ca/report41.html

[7] A sequence of runs and shutdowns is planned during the lifetime of the LHC (currently to 2041) with more detail at: lhc-commissioning.web.cern.ch/schedule/images/2023/Complex-long-term-Apr23.png.

ELENA (see description of ELENA in the chapter 6 Symmetries and Fundamental Interactions) also offers new opportunities for nuclear physics. A major project under construction and commissioning is PUMA, to study the interaction of antiprotons with nuclei. Trapped antiprotons will be transported to ISOLDE to study the capture of antiprotons by radioactive nuclei probing the proton-neutron asymmetry of nuclear-matter distributions. Other topics, such as hypernuclei, short-range correlations in low-density nuclear matter and the formation of antiprotonic atoms are other opportunities arising from such measurements.

Developments at the AEgIS, ASACUSA and GBAR experiments will provide opportunities to study antiproton annihilation at very low energies, strong-field QCD effects in antiprotonic atoms, and production of highly charged, trapped ions and hypernuclei for precision studies. Several parts of the AD facility were commissioned in the early 2000s and a programme of consolidation and updating is currently being developed for the facility beyond 2028.

## Isotope Mass Separator OnLine (ISOLDE) facility

**ISOLDE** (isolde.cern), a world-class ISOL facility at CERN, provides radioactive ions with high intensity and excellent emittance over a wide range of energies. It is unique worldwide in using 1.4-GeV protons on thick targets, resulting in the widest range of isotopes (>1300) and elements (>75) from He to Pu. HIE-ISOLDE, a system to bunch, charge breed and post-accelerate ions, was completed in 2018 and routinely delivers beams with energies up to 9.2 MeV/u for A/Q=4.5 and 11 MeV/u for A/Q=3.5.

A series of detectors and beam-line installations is maintained by users, exploiting radioisotope ions for a wide range of science. There is a focus on nuclear physics with precision measurements of nuclear properties (masses, moments, radii and spins), radioactive decay and reaction studies. These include nuclear astrophysics measurements and searches for physics beyond the Standard Model. Hyperfine effects are used to address aspects of atomic and molecular physics. Radioactive ions are also used as probes of the environment within hard, soft and biological materials, addressing aspects of condensed-matter physics and life sciences. The MEDICIS Facility has provided mass-separated radioisotopes for medical R&D projects since 2018 and developed its first therapeutic clinical translation medical project in 2023. Emerging science opportunities for the LRP period include radioactive molecules for fundamental physics; binding in negative ions; nuclear magnetisation distributions; high-sensitivity β-NMR for biology and medicine; antiproton interactions with radioactive ions; transfer-induced fission; and a new generation of spectrometers for condensed-matter physics.

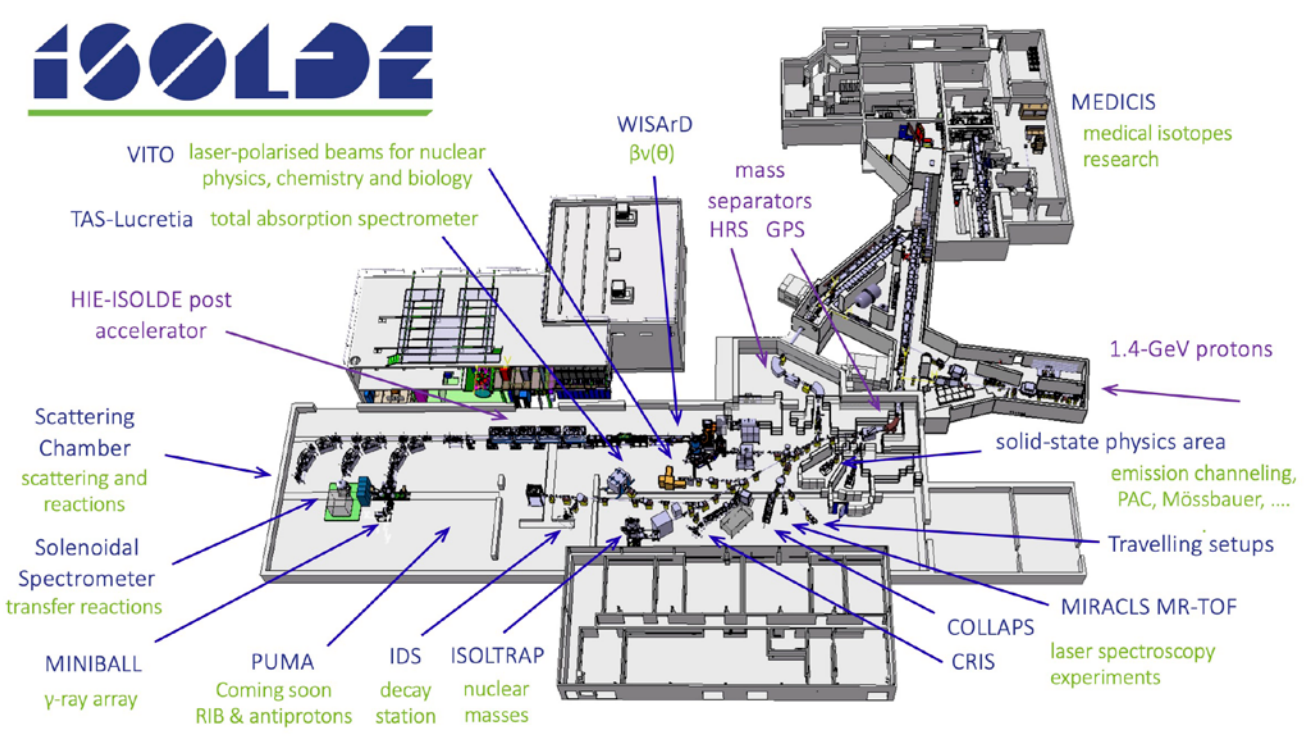

*Fig. 8.2: Layout of the experimental facilities at ISOLDE (© CERN)*

A series of improvements is planned up to 2027:

● Replacement of primary beam dumps, upgrade to delivery line and other infrastructure developments to receive higher energy and intensity protons. Operation with 2-GeV protons will improve both the capacity and capability of the facility.

● Consolidation of the post accelerator system will improve performance and operational capacity for accelerated beams.

● Improvements to beam purity by employing high-voltage mass-resolving time-of-flight techniques are being developed, especially for the future PUMA experiment which will bring antiprotons from CERN-AD.

● Beam switching for simultaneous delivery of different beams from the two separators.

● R&D activities have begun which underpin future construction of a Superconducting Recoil Separator. The development of novel magnets may prove to be important in creating a future ion storage ring.

After several focused workshops, ISOLDE users proposed an additional hall for receiving beams from new target stations in parallel with the current area, increasing the capability of the facility for new science, facilitating the installation of new experiments (including a storage ring) and beam purification devices for demanding precision experiments. On similar timescales, CERN will develop a wider strategy for the post-LHC era (2041+) to present ISOLDE with significant new opportunities. Future strategies for ISOLDE will be developed during the LRP period to encompass the aspirations of users and future directions of CERN.

## Neutron Time-of-Flight (n_TOF) facility

The pulsed broad-range neutron source (n_TOF https://home.cern/science/experiments/n_tof) at CERN is a world-leading research facility for studying neutron-induced reactions. It has been operational for over twenty years while being continually upgraded. Experiments focus on measuring neutron-induced reaction cross sections for stellar nucleosynthesis, nuclear fission and fusion technology, dosimetry, medical applications, and in general for nuclear data.

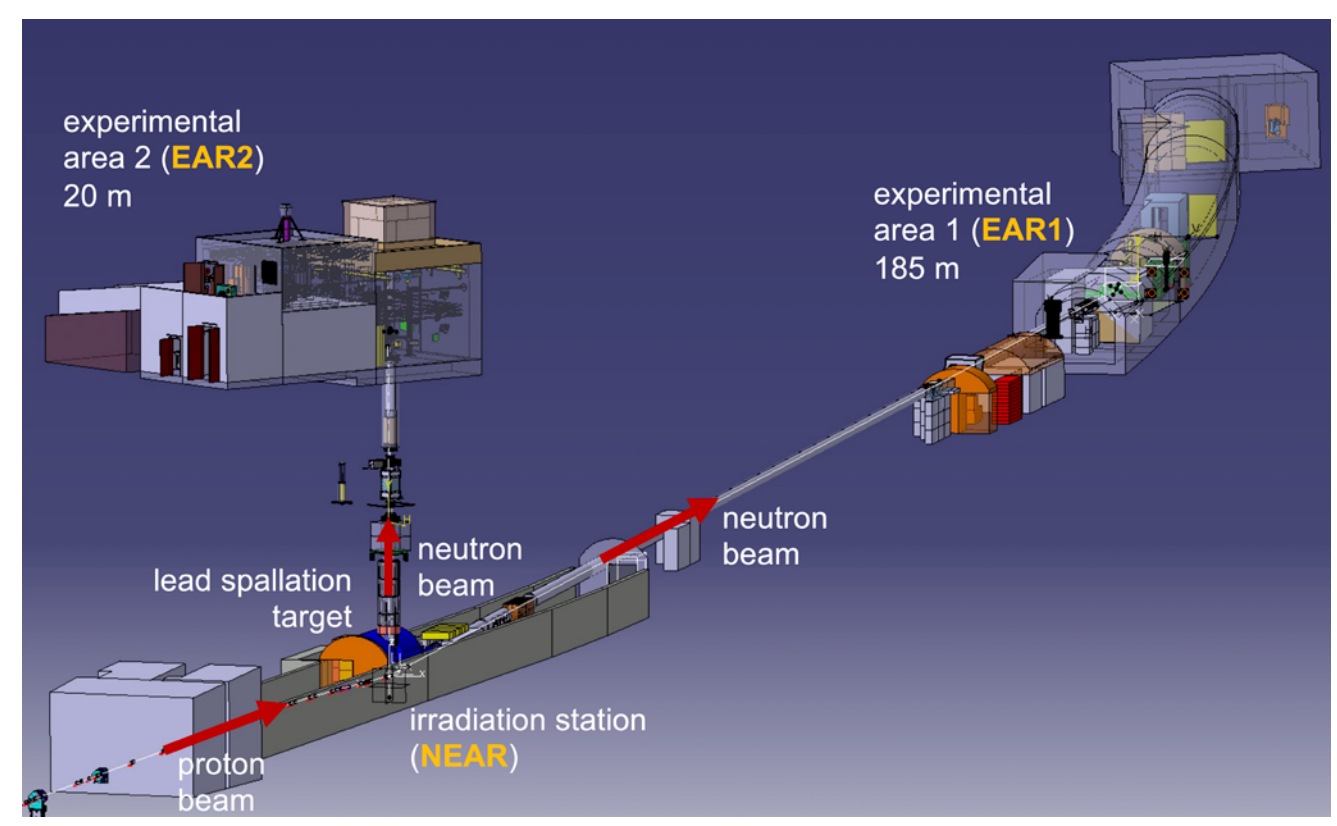

*Fig. 8.3: Overview of the n_TOF facility at CERN showing the neutron spallation target and the three experimental areas (© CERN).*

The facility is driven by 20-GeV protons from CERN's Proton Synchrotron accelerator. For n_TOF the beam consists of 7-ns pulses containing $9x10^{12}$ protons, repeated by multiples of 1.2 s and creating around $3x10^{15}$ neutrons per pulse in a lead spallation target. A novel target using nitrogen gas cooling was commissioned in 2021. Neutron moderators shape the neutron-energy spectrum, covering energies from the sub-thermal meV region up to GeV. Neutrons are guided to two experimental areas for high-resolution neutron-energy measurements, EAR1 and EAR2 (located at distances of 185 m and 20 m, respectively) and a third station, NEAR (at 3 distance and operational since 2021), focusing on neutron-activation measurements.

The facility's uniqueness stems from several factors. The very short primary proton pulse width is crucial for time-of-flight measurements vital to obtain high-resolution reaction cross sections. The high number of neutrons produced by each pulse constitutes a very high instantaneous flux and provides a signal-to-noise ratio favourable for measurements of low-mass or radioactive samples. The low repetition rate of pulses permits the use of a moderator, yielding a wide energy range of more than eleven decades without overlap between pulses. Future plans for n_TOF include the development of detectors for reaction channels completing neutron-induced capture, fission and light charged-particle reactions. This involves in particular (n,xn) and total cross section measurements in EAR1 and EAR2, along with neutron-activation measurements at the NEAR station. Plans for facility developments during LS3 include a dedicated moderator for NEAR, an off-beam spectroscopy station with a pneumatic rabbit system, a dedicated transmission station, and an upgraded beam-collimator system for EAR1 to accommodate neutron capture on small-diameter samples. For LS4, the current spallation target will come to its normal end-of-life cycle and will need to be replaced. Consideration is being given to a new design with improved performance.

## Super Proton Synchrotron (SPS)

The SPS accelerator (https://home.cern/science/accelerators/super-proton-synchrotron) at CERN delivers protons at 400 GeV, to produce several secondary beams for fixed-target experiments. M2 is a beamline providing secondary hadron beams and tertiary muon beams of both charges, with tunable energy (60-280 GeV) and high intensity ($10^{5\text{-}8}$ particles per s). Beams are debunched, giving constant intensity lasting 5s, and a repetition cycle of 20-60s. The muon beams are discussed in the section on Lepton beams. The hadron beams available at M2 will be used by the AMBER collaboration for hadron spectroscopy. The intense negative beam consists mostly of pions, with a small contamination from kaons and antiprotons while the positive one contains mostly protons and pions. Unique pion- and kaon-induced Drell-Yan measurements at AMBER (Apparatus for Meson and Baryon Experimental Research) will provide insights into meson structure. Hadron spectroscopy addressing the strange sector and kaon polarisabilities are other measurements proposed for AMBER. These physics programmes rely on the availability and quality of the kaon beam. The kaon identification is provided by Cherenkov-type detectors and a good beam parallelism is critical. Optimisation studies are underway for the M2 beam optics and to reduce beam scattering for parts of the beam path in air.

For the SPS heavy-ion programme, there are also new proposals to measure heavy quarks and electromagnetic processes in collisions of heavy nuclei between 5 and 20 GeV (the operating NA61/SHINE detector and the proposed new NA60+ experiment). In addition, NA61/SHINE plans to explore the onset of fireballs in nucleus-nucleus collisions, thereby extending the scan in the momentum/ion space of lighter ion beams.

# GANIL & SPIRAL2

The Grand Accélérateur National d'Ions Lourds (GANIL) (https://www.ganil-spiral2.eu) is primarily focused on cutting-edge research in basic nuclear physics and nuclear astrophysics, but also includes strong programmes in astrochemistry, material irradiation, nano-structuration, radiobiology and industrial and medical applications.

GANIL hosts a heavy-ion cyclotron complex delivering stable beams from $^{12}C$ to $^{238}U$ from 1 to 95 MeV/nucleon. Radioactive ion beams are produced both by in-flight and ISOL methods. For the former, the LISE spectrometer selects fragments with energy up to ~65 MeV/u and half-lives down to μs. The SPIRAL1 ISOL facility reaccelerates isotopes with half-lives of more than 1ms to energies from 1.2 to 25 MeV/u with the CIME cyclotron. Intensities for stable beams are around 5 μA and for radioactive beams range from 1 to $10^7$ per second. SPIRAL2 is a major upgrade of GANIL based on a very high-intensity LINAC, which provides high-intensity ions and neutron beams.

The cyclotron's experimental halls are equipped with a wide range of versatile and state-of-the-art instrumentation including two upgraded magnetic spectrometers and a new highly granulated charged-particle detector. In addition to these devices, GANIL regularly hosts other experimental set-ups brought by international users.

VAMOS++ is a large acceptance spectrometer, recently upgraded for increased isotopic resolution and higher data rates. It is regularly coupled with other devices, such as γ-ray arrays (FATIMA, PARIS, EXOGAM) and particle detectors (DIAMANT, MUGAST, NEDA), and has hosted the AGATA γ-ray tracking array. Emerging science topics include the study of fission processes in inverse kinematics for exotic systems and in a large range of excitation energies.

The LISE spectrometer is used to separate, focus and identify exotic projectile-like reaction fragments with a Wien filter to provide an additional velocity selection. A recent upgrade (LISE2022) now allows the use of the unreacted beam to induce further reactions. This allows for a secondary new detection system at 0° to make simultaneous, complementary measurements.

The recently combined INDRA-FAZIA highly-segmented $4\pi$ charged-particle multi-detector array is a powerful system for probing various reaction mechanisms. It operates in a stand-alone mode, but can also be coupled to magnetic spectrometers.

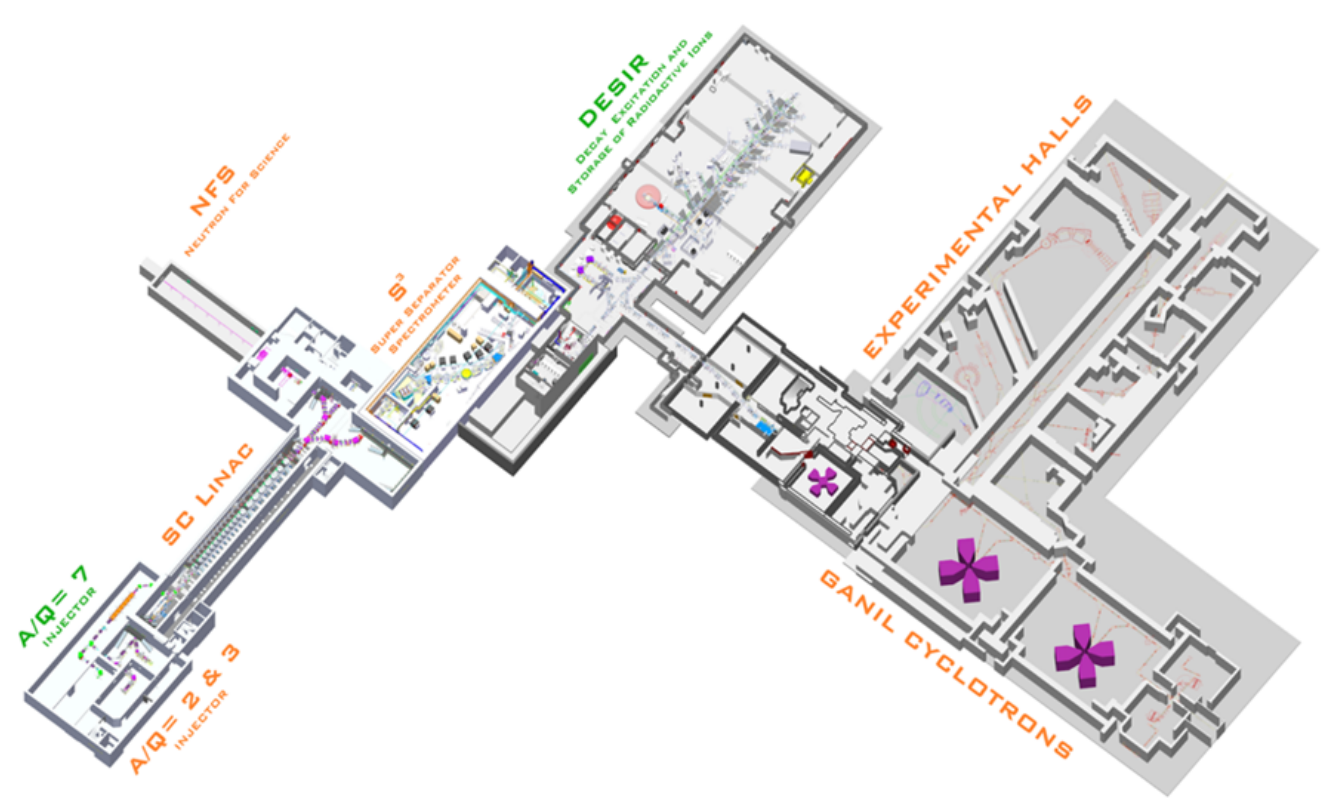

*Fig. 8.4: Schematic view of GANIL-SPIRAL2 facility (DESIR and A/Q=7 injector under construction). (© GANIL)*

Around half of the GANIL beam time (including parallel operation) is allocated to interdisciplinary research and societal applications such as nuclear waste management, ageing of materials in nuclear power plants and radiobiology with heavy ions. Material irradiation allows access to the region within a sample corresponding to the maximum energy deposition using a wide choice of ions and energies, and accessing linear energy transfers over a broad range of energies ranging from electronic to nuclear stopping powers. This, combined with the development of much in-situ experimental equipment, makes GANIL a leading facility for using swift heavy ions in a broad area of research and applications. A beamline is also devoted to industrial beam applications and radiobiology. A new laboratory dedicated to radiobiology handles the increased activity in this area. Requests for radiation hardness tests of electronic components have soared recently and various strategies are being considered to increase beam time for these applications.

GANIL cyclotrons have been operated successfully for more than 40 years, supported by refurbishment and upgrades to improve their reliability and performance. It is possible to deliver the beam simultaneously to four experiments and two test setups, maximising the available beam time of the facility. Stable beam development has led to increased intensities (e.g. U) and unique species (e.g. $^{186}W$, $^{232}Th$ and $^{130}Te$). The development of target/ion-source systems has broadened the range and intensity of radioisotope beams (e.g. $^{47}K$). In order to maintain operational efficiency, a major project for the refurbishment of the complete cyclotron facility is underway. These activities will be carried out in a phased manner in the period 2024-2030, minimising the impact on the available beam time.

## SPIRAL2

SPIRAL2 is an ESFRI landmark facility for nuclear physics, addressing topics such as the evolution of shell closures, exotic N=Z nuclei and super-heavy systems, as well as interdisciplinary research in radiobiology, radiotherapy, material science and nuclear data. Since the last LRP, several milestones have been completed.

After commissioning the superconducting LINAC, protons and deuterons are now routinely accelerated for experiments in the Neutrons for Science (NFS) hall. The nominal intensity of 5 mA for light beams has been demonstrated with protons, although only 10% of the maximum intensity was accelerated due to nuclear safety regulations. Alpha beams have also been used, for example, to optimise the production of innovative α-emitting medical isotopes. Intense beams of neutrons produced between 1 and 40 MeV are being used for a rich programme in basic and applied science including probing the fission process, resonances and neutron-induced, charged-particle emission related to nuclear-reactor studies. A variety of new instrumentation for neutrons, charged fragments and γ rays is also available.

The Super Separator Spectrometer ($S^3$), dedicated to the study of neutron-deficient and super-heavy nuclei produced by fusion-evaporation reactions, is in an advanced stage of construction. Step-by-step commissioning of the high-intensity, heavy-ion beams for $S^3$ is in progress and routine exploitation is expected from 2025. To further increase beam intensity a new LINAC injector, optimised for lower charge states (A/Q=7) and based on a superconducting ions source, is under construction.

DESIR (Decay Excitation and Storage of Radioactive Ions) will be a low-energy facility dedicated to the precision study of nuclear properties using trapping, decay- and laser-spectroscopy techniques. DESIR will receive radioisotope beams from both SPIRAL1 and $S^3$. The construction of the associated building started in 2023 and is expected to be ready for exploitation in 2027.

Longer-term recommendations by a recent international expert committee involve building a high-intensity facility for radioisotope production using fission and multi-nucleon transfer reactions, followed by post-acceleration up to 100 MeV/u. An electron-ion scattering facility is planned, benefitting from ion traps developed for DESIR. Several different scenarios to meet identified scientific objectives are being considered.

## GSI & FAIR

GSI Darmstadt hosts a suite of accelerators and storage rings enabling acceleration, storage and cooling of all ion species up to uranium that fosters a broad multidisciplinary research programme from nuclear structure and nuclear matter studies to atomic physics and materials science.

The international Facility for Antiproton and Ion Research (FAIR) (https://fair-center.eu), an ESFRI landmark for nuclear and hadron physics, is under construction on the GSI campus. The GSI accelerators will serve as injectors for FAIR and a major upgrade programme has been initiated to meet the requirements in beam intensities and qualities. Upgrades to the SIS18 synchrotron and construction work to connect it to FAIR were finished in 2018. To exploit the upgraded GSI accelerators and the available novel FAIR instrumentation for science and testing, the FAIR Phase-0 programme was initiated. This provides regular beam delivery periods of around three months per year, in line with the FAIR construction.

### GSI overview

The large, unique heavy-ion accelerator complex at GSI (https://www.gsi.de) is comprised of the UNILAC linear accelerator that accelerates ions to 11 MeV/u, the SIS18 synchrotron for further acceleration up to 2 GeV/u and an experimental storage cooler ring (ESR) for highly-charged ions at energies from 4 MeV/u to 0.5 GeV/u. A further storage cooler ring (CRYRING) has been installed behind the ESR for atomic and nuclear physics experiments from 15 MeV/u down to a few 100 keV/u. The HITRAP facility currently being commissioned decelerates and traps highly charged ions. Pion beams can also be provided in the momentum range of 0.5 to 2.5 GeV/c. PHELIX is a unique laser facility delivering short (0.5-20 ps) or long laser pulses (1-10 ns) between 0.3-2 kJ and up to an intensity of $2x10^{21}$ W/cm$^2$ for experiments with combined laser and ion beams.

UNILAC accelerates ions for both experiments and injection into the SIS18. It was upgraded recently to match the requirements for FAIR. The combination of a novel pulsed gas stripper and an upgraded post-stripper section will fill SIS18 up to the space-charge limit. UNILAC beams are used for super-heavy element research with the velocity filter SHIP, the gas-filled separator TASCA and a Penning trap. Irradiation studies are also performed for materials science and biophysics with a multitude of diagnostic devices to probe radiation effects in various materials.

A new super-conducting continuous-wave high-intensity linear accelerator (HELIAC) is being developed. It will provide ions at Coulomb barrier energies for the super-heavy element programme and could serve as an alternative injector for SIS18 while the UNILAC is being upgraded for FAIR operation. It will also provide beams for ion-based pre-screening of materials for fusion energy generation preceding neutron-based studies. It comprises a normal-conducting injector followed by a chain of four cryo-modules, providing continuous beams with an intensity of 1 mA and energies from 3.5 to 7.5 MeV/u. HELIAC will be crucial in providing long-duty cycle and continuous-wave beams after the upgrade of UNILAC as an optimised SIS-injector.

SIS18 is a highly flexible heavy-ion synchrotron providing highly energetic beams of protons to U ions with energies up to 2 GeV/u, delivered either to experiments or to production targets for secondary beams. The precise control system, the cycle-to-cycle flexibility of the magnets and the RF ramps allow multiple user operation. The FRS, a 72-m fragment separator, delivers unstable isotopes by in-flight production; these are separated and identified event-by-event. Radioactive beams can be used for fixed-target experiments or for injection into the storage rings.

The High-Acceptance Di-Electron Spectrometer HADES uses SIS18 beams for studying the properties of dense hadronic matter via di-lepton production in relativistic heavy-ion collisions. The set-up for Reactions with Relativistic Radioactive Beams R3B as well as secondary target stations in the FRS are dedicated to nuclear reaction and structure studies. Multi-purpose stations for atomic physics, material science and radiation biology research complete the suite of facilities at SIS18. The storage rings facilitate a large variety of different projects from astrophysically relevant reactions with radioactive beams near the Gamow window to laser-based investigation of the hyperfine structure of highly charged ions.

## FAIR

The FAIR accelerator complex will consist of SIS100, a fast-ramping superconducting synchrotron to provide intense primary beams; the Super-FRS, a large-aperture superconducting fragment separator to select exotic ions; the CR cooler-storage ring; and a high-energy storage ring, HESR. (see Fig. 8.5). The upgraded GSI accelerators will serve as pre-accelerators for SIS100, which will provide uranium beam intensities ten times higher than the GSI accelerators before their upgrades. Moreover, the magnetic rigidity of SIS100 allows acceleration of fully-stripped heavy ions (e.g. $U^{92+}$) at energies up to 10 GeV/u. The SIS100 beams can be used directly for experiments or be directed to production targets for rare-isotope and antiproton beams.

The unique research opportunities at FAIR are organised into four pillars:

- APPA is devoted to precision studies on fundamental interactions and symmetries with highly charged ions, high-density plasmas, atomic and material science, radio-biological investigations and other applications.
- CBM studies QCD matter and its phase diagram at the highest baryon densities and is complementary to LHC experiments that focus on low densities. The detector is designed to study rare probes in heavy-ion collisions with the highest beam intensities.
- NUSTAR makes use of a versatile suite of experimental set-ups for NUclear STructure, Astrophysics and Reaction studies of exotic nuclei produced by the Super-FRS. The low-energy stable-beam programme will use HELIAC.
- PANDA, with its beam of stored antiprotons, will probe a broad range of aspects of QCD at the scale where quarks and gluons form hadrons. In particular, the antiproton beam enables investigations of exotic hadrons.

Wide acceptance of the Super-FRS and the gain in primary uranium intensities translates into an increase of more than 1000 for secondary intensities of exotic isotopes as compared to the FRS. After their production, separation and identification the ions can be (i) stopped to study ground-state and decay properties, (ii) used in-flight to produce secondary reactions generating even more exotic species, or (iii) stored and pre-cooled in the CR. The pre-cooled secondaries – rare isotope or antiproton beams – will then be transferred from the CR to the HESR where they can be accumulated and accelerated up to energies of 15 GeV for antiprotons and about 5-6 GeV/u for very heavy ions. For experiments in the HESR, stochastic and electron cooling can compensate for momentum loss and emittance growth due to the interaction with internal targets of PANDA, NUSTAR, or APPA. The HESR can also store and cool highly charged heavy ions directly injected from the SIS100 via the CR.

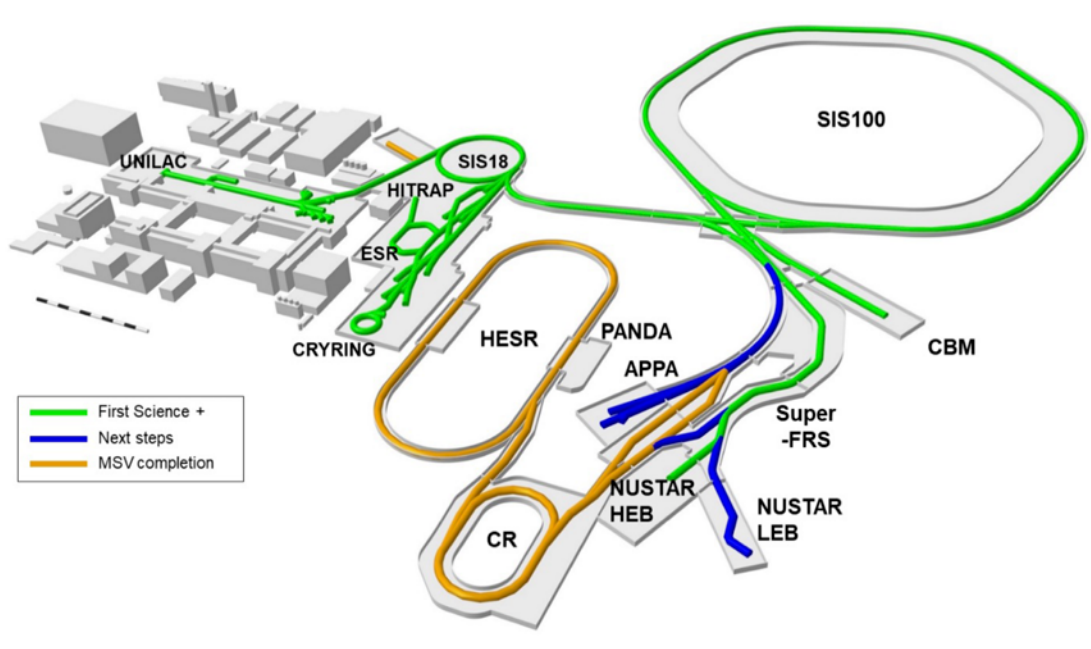

*Fig. 8.5: Layout of the FAIR facilities, including the different steps towards completion. (© GSI-FAIR)*

This scenario is known as the FAIR Modularised Start Version (MSV). Several events had affected progress towards the MSV and resulted in the need for substantial additional resources. An international review in 2022 confirmed that the planned scientific programmes of all four pillars were outstanding and in many cases world-leading. It recommended the completion of FAIR in a staged approach. The first step takes the Super-FRS into operation with SIS18 beams, facilitating *Early Science Experiments* with the NUSTAR High-Energy Branch (HEB). *First Science* will follow the completion of the SIS100 and use of its beams with the Super-FRS. Taking CBM into operation is dubbed *First Science+*. The Next Steps include the APPA cave and the NUSTAR Low-Energy Branch (LEB). The FAIR partner countries, with Germany contributing the largest share, have secured sufficient funding to achieve *First Science*. The final steps will be implemented as additional funding becomes available. The completion of the full MSV, adding the Collector Ring and High-Energy Storage Ring, is the declared goal of all FAIR shareholders.

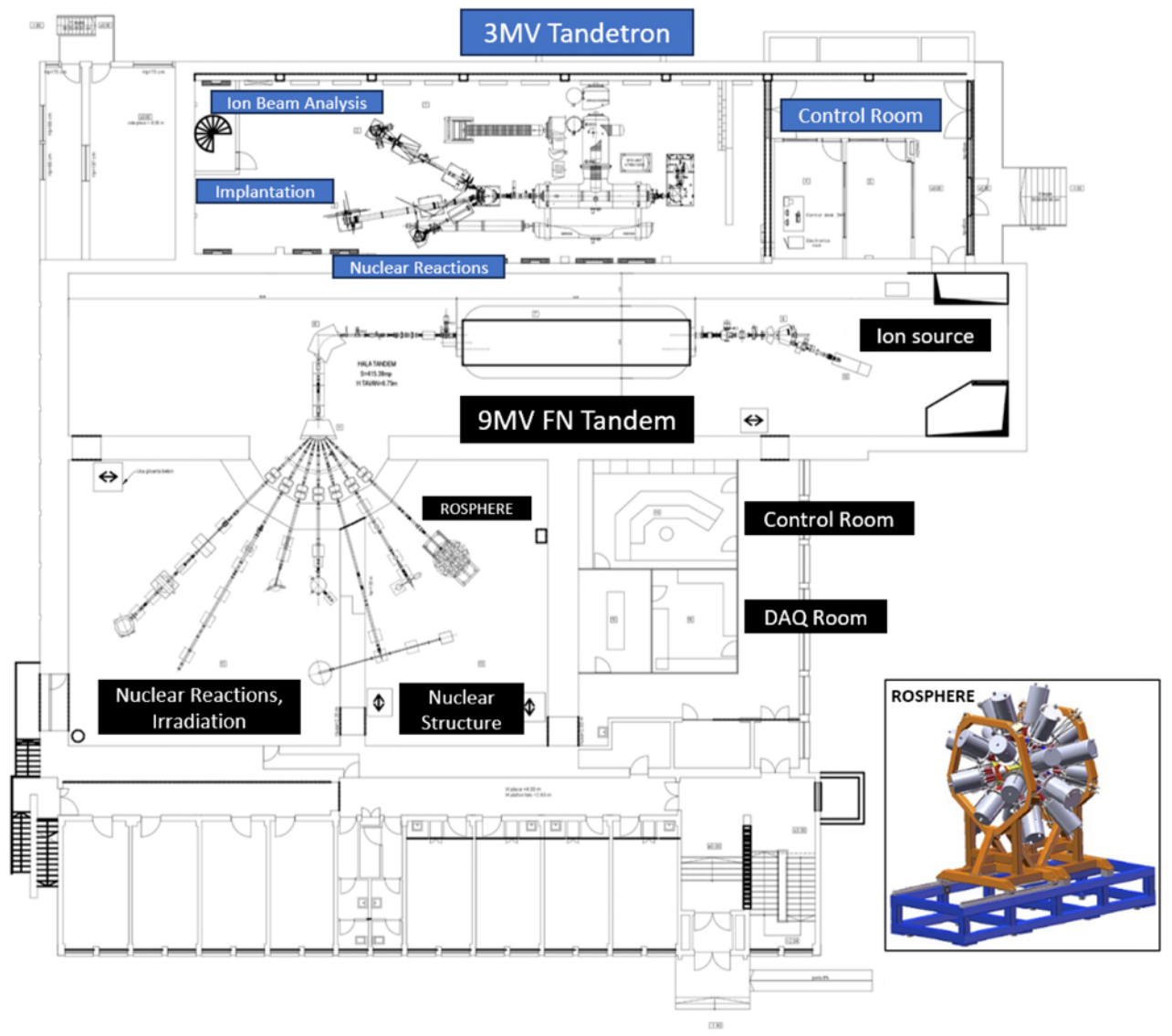

*Fig. 8.6: The 9 MV and 3 MV Accelerator layouts including the ROSPHERE (ROmanian array for SPectroscopy in HEavy ion REactions), dedicated to γ-ray spectroscopy, consisting of 25 detectors of two types: Compton-suppressed HPGe detectors and fast $LaBr_3$(Ce) scintillator detectors. (© IFIN-HH)*

## Horia Hulubei National Institute for Physics and Nuclear Engineering

The Horia Hulubei National Institute for Physics and Nuclear Engineering (http://www.ifin.ro) (IFIN-HH) is dedicated to research and development in physical and natural sciences. These include nuclear physics, engineering and related areas including nuclear photonics, astrophysics and particle physics, field theory, mathematical and computational physics, atomic physics and physics of condensed matter, and life and environmental physics. In all these fields, IFIN-HH conducts both experimental and theoretical research.

The accelerator infrastructure includes three tandem accelerators with terminal voltages of 1, 3 and 9 MV and a cyclotron delivering proton beams with a maximum energy of 19 MeV, dedicated to applications. The 1-MV and 3-MV tandems and the cyclotron were installed in 2012. The 9-MV Pelletron Tandem was recently upgraded and is used for basic nuclear physics studies. Recent highlights include the measurement of charged-particle-induced reaction cross sections and of nuclear state lifetimes, from tens of femtoseconds using the Doppler-Shift Attenuation method, to picoseconds using the Recoil-Distance Doppler Shift technique and to nanosecond range using in-beam fast timing. The applications developed at the 3-MV Tandetron include the physics of materials and the characterisation of cultural heritage artefacts. Additionally, cross-section measurements for nuclear astrophysics are performed using prompt spectroscopy or activation techniques. The sensitivity of the latter is facilitated by IFIN-HH's underground ultra-low background detector laboratory in the Slanic salt mine. The 1-MV Tandetron is dedicated to accelerator mass spectrometry, forming the internationally accredited RoAMS radiocarbon-dating laboratory. The TR19 cyclotron is the source for the Centre of Radiopharmaceutical Research. The industrial-sized irradiator IRASM is used for the irradiation of medical devices, materials and cultural artefacts.

The Radioactive Ion Facility at IFIN-HH (RIF@IFIN) is a newly-proposed facility for the production of radioactive ion beams using the Ion-Guided Isotope Separation On-Line method. It will focus on fundamental nuclear physics and astrophysics. The production of radioisotopes for medical research using photonuclear reactions and other applications, such as solid-state physics, is also envisaged. The radioisotopes are planned to be produced through photofission induced in an array of thin foils placed in a cryogenic gas cell by bremsstrahlung γ-rays. They are generated in a convertor using an electron beam from a linear accelerator with energy in the range of 200-300 MeV and intensity up to 10 μA limited by space-charge effects in the gas cell. The uniqueness of the RIF@IFIN facility lies in the use of state-of-the-art technologies in the extraction and separation of fission products, optimised for efficient production of short-lived neutron-rich isotopes, especially refractory elements that cannot be extracted in current ISOL facilities using thick hot targets.

## IJCLAB-ALTO

IJCLAB hosts two accelerators at the ALTO Facility (https://www.ijclab.in2p3.fr/en/platforms/alto/). The first is a 15-MV tandem accelerator that produces a wide range of heavy-ion beams from protons to gold. It can also produce directional neutron beams in inverse kinematics. The second machine is a linear accelerator for 50-MeV electrons at 10μA used to bombard a uranium-carbide target and produce neutron-rich radioactive beams via the photo-fission process. It is the first facility in the world to use this pioneering technique. After ~10 years of development, many physics experiments have been performed over the past few years.

With the delivery of a broad range of stable and radioactive beams, the six beamlines and experimental halls shown in Fig. 8.7 are equipped with diverse instrumentation, spectrometers and detectors: the high-energy branch ALTO-HEB, consisting of a γ-spectroscopy area, the LICORNE neutron source, the Radiograff precision dosimetry beamline for radiobiology, SPACE-ALTO, an irradiation station for space applications and the SPLIT-POLE spectrometer for nuclear reactions. The low-energy branch ALTO-LEB for ion source investi-gation houses a decay station with electron, beta-gamma and neutron detectors, LINO for Laser-Induced Nuclear-Orientation, POLAREX for low-temperature nuclear orientation measurements, and MILL-TRAP for precision mass measurements. Measurements encompass ground-state properties such as masses, magnetic and quadrupole moments and radii; decay properties of nuclei including beta-delayed γ rays; neutron decay and conversion electrons; nuclear structure and nuclear astrophysics studies, including neutron-induced reactions; and the interaction of ions with matter for biological and space applications.

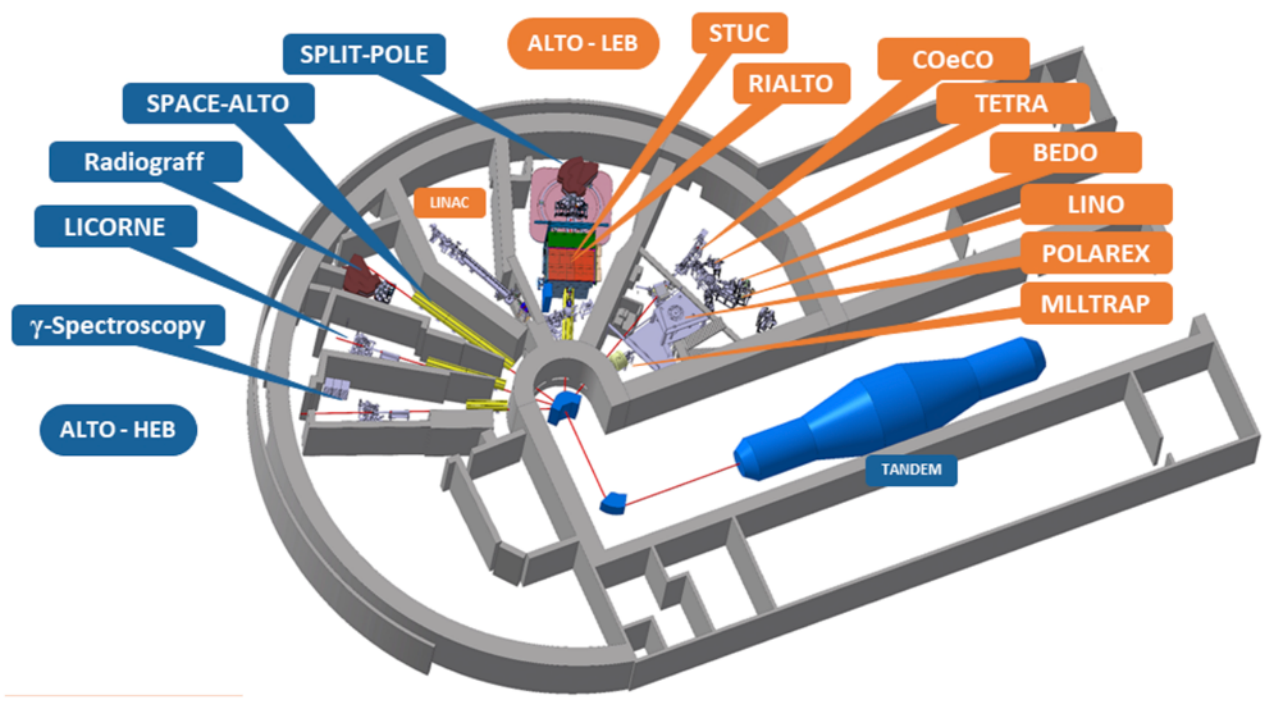

*Fig. 8.7: The installations at the ALTO facility (see text for details) (© IJCLAB)*

ALTO has developed a research strategy with strong synergies with GANIL; there is strong complementarity between the ALTO programme and future opportunities at DESIR, such as laser spectroscopy, ion trapping and online nuclei polarisation techniques. The platform has been opened to industrial users to reap potential societal benefits, with a quarter of the beamtime dedicated to R&D and applications. A growing number of industrial partners have interests in material irradiation, particularly in radiobiology and space science. The overall aim is to create synergies between the academic and industrial worlds for the mutual benefit of research, education and economics, and to increase the visibility of all actors including CNRS, the University Paris-Saclay and regional industrial and societal partners.

## Institut Laue Langevin (ILL)

The Institute Laue-Langevin (ILL) (https://www.ill.eu) is a European facility with fourteen member states serving an international user community. The 58-MW high-flux reactor provides intense neutron beams with energies from a few neV to about 1 eV, distributed to over 40 instruments operated simultaneously. While many instruments are for neutron-scattering experiments, some are dedicated to nuclear physics and others can be shared for nuclear physics applications. Major reactor components were renewed in 2022, assuring safe reactor operation beyond 2045, which corresponds to the reactor vessel's lifetime.

Most of ILL's nuclear physics instruments are unique worldwide:

• LOHENGRIN is a high-resolution recoil separator for fission fragments produced by targets exposed to a thermal neutron flux of $5\times10^{14}$ $cm^{-2}s^{-1}$. Fission yields are measured as a function of mass and kinetic energy with excellent selectivity, reaching yields as low as $10^{-10}$ per fission. Ge, Si(Li), $LaBr_3$ and β detectors are used to measure energy-dependent isotopic and isomeric fission yields and for nuclear spectroscopy. In particular, micro-second isomers can be studied, including conversion electron and X-ray spectroscopy down to 10 keV.

• FIPPS (Fission-Product Prompt γ-ray Spectrometer) consists of an efficient array of eight HPGe Compton-suppressed clovers using a "pencil-like" thermal neutron beam to study (n,γ) and (n,f) reactions on different targets including rare or radioactive materials. The detector array can be complemented with additional Ge detectors or ancillary detectors, such as $LaBr_3$ detectors for fast-timing measurements for sub-ns lifetimes.

• PF1b is a multipurpose beam port providing an intense beam of cold neutrons (<0.025 eV) with a capture flux of $2\times10^{10}$ $cm^{-2}s^{-1}$ over 20×6 $cm^2$. The intense cold neutron beam is used for measurements of fission cross-sections, prompt fission γ and neutron spectra, and ternary fission fragments. Using supermirror polarisers, PF1b can optionally provide neutron beams polarised up to 99.7%. The polarised PF1b beam is frequently used for fundamental physics in detailed studies of free neutron decay.

• The instruments PF2 and SuperSUN provide strong beams and high densities of ultracold neutrons (UCN) with energies below 250 neV. UCNs have the unique property of being storable and observable for extended periods (several hundreds of seconds). This makes them a valuable tool for nuclear and particle physics, cosmology, symmetries and fundamental interactions. Prominent examples are precise measurements of the free neutron's lifetime, limits for the neutron's electric dipole moment and searches for hypothetical gravity-like interactions using gravitationally bound neutron states.

• The V4 high flux irradiation position (the bright blue spot below the centre of Fig. 8.8) provides the highest neutron flux at ILL (up to $1.5\times10^{15}$ $cm^{-2}$ $s^{-1}$) used for producing radio nuclides for medicine and research applications. Unique samples for nuclear physics research can also be produced and have been used for experiments at FIPPS, at ILL, n_TOF at CERN, ISOLDE and MEDICIS at CERN and other facilities.

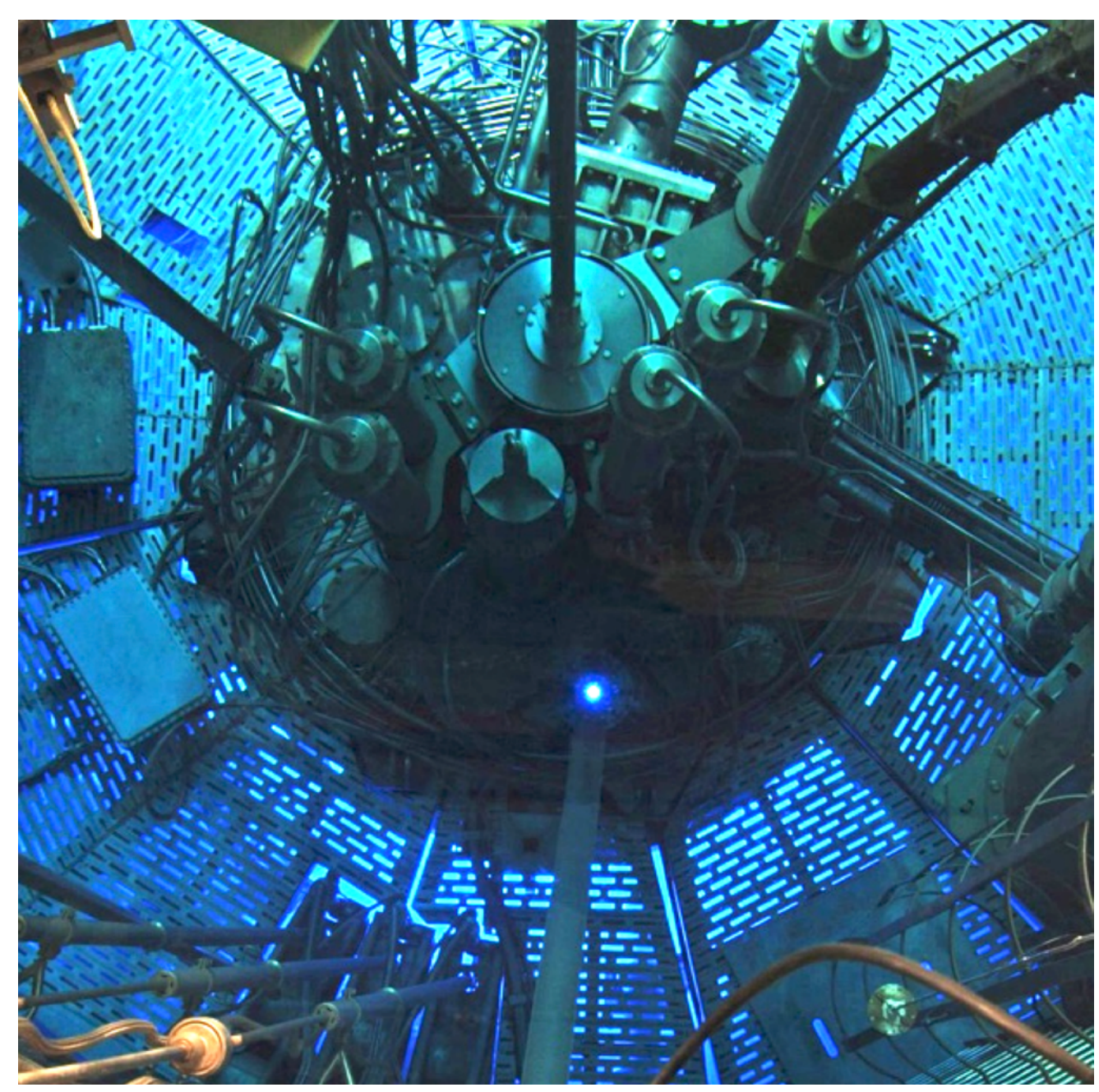

*Fig. 8.8: View into the pool of the reactor (Photo U. Köster, ILL)*

## JYFL Accelerator Laboratory

The Accelerator Laboratory of the University of Jyväskylä (JYFL) (http://www.jyu.fi/accelerator) hosts four accelerators for nuclear physics and applications, delivering a range of stable-ion beams (from protons to Au), electrons and photons, serving an international user community. The main K=130 heavy-ion cyclotron is served by three ECR ion sources and a multi-cusp ion source. The cyclotron provides a total beam time of around 6500 hours a year.

The major instrumentation at the laboratory includes (see Fig. 8.9):

• The gas-filled recoil separator RITU represents one of the most versatile and efficient systems for in-beam and decay spectroscopy of exotic nuclei in the world and is mainly used for studies of proton dripline and superheavy nuclei.

• The vacuum-mode separator MARA is complementary to RITU, and they share some detector arrays. MARA has been used to probe topics like isospin symmetry and pairing in N=Z nuclei.

• At the Ion-Guide Isotope Separation On-Line (IGISOL) facility, exotic beams of both neutron-rich and neutron-deficient nuclei are produced for comprehensive studies of nuclear ground (and isomeric) state properties and exotic decay modes. These studies are driven by questions in nuclear structure and astrophysics as well as neutrino and beyond-standard-model physics. IGISOL hosts a variety of ion traps, ion-manipulation devices and laser-spectroscopy instruments.

• The RADiation Effects Facility (RADEF) is the most important facility for applications and is specialised in the study of radiation effects in electronics and related materials. RADEF has been one of the three official test sites of ESA since 2005.

• A 1.7 MV Pelletron provides MeV ion beams to probe the elemental composition and structural properties of materials with state-of-the-art equipment.

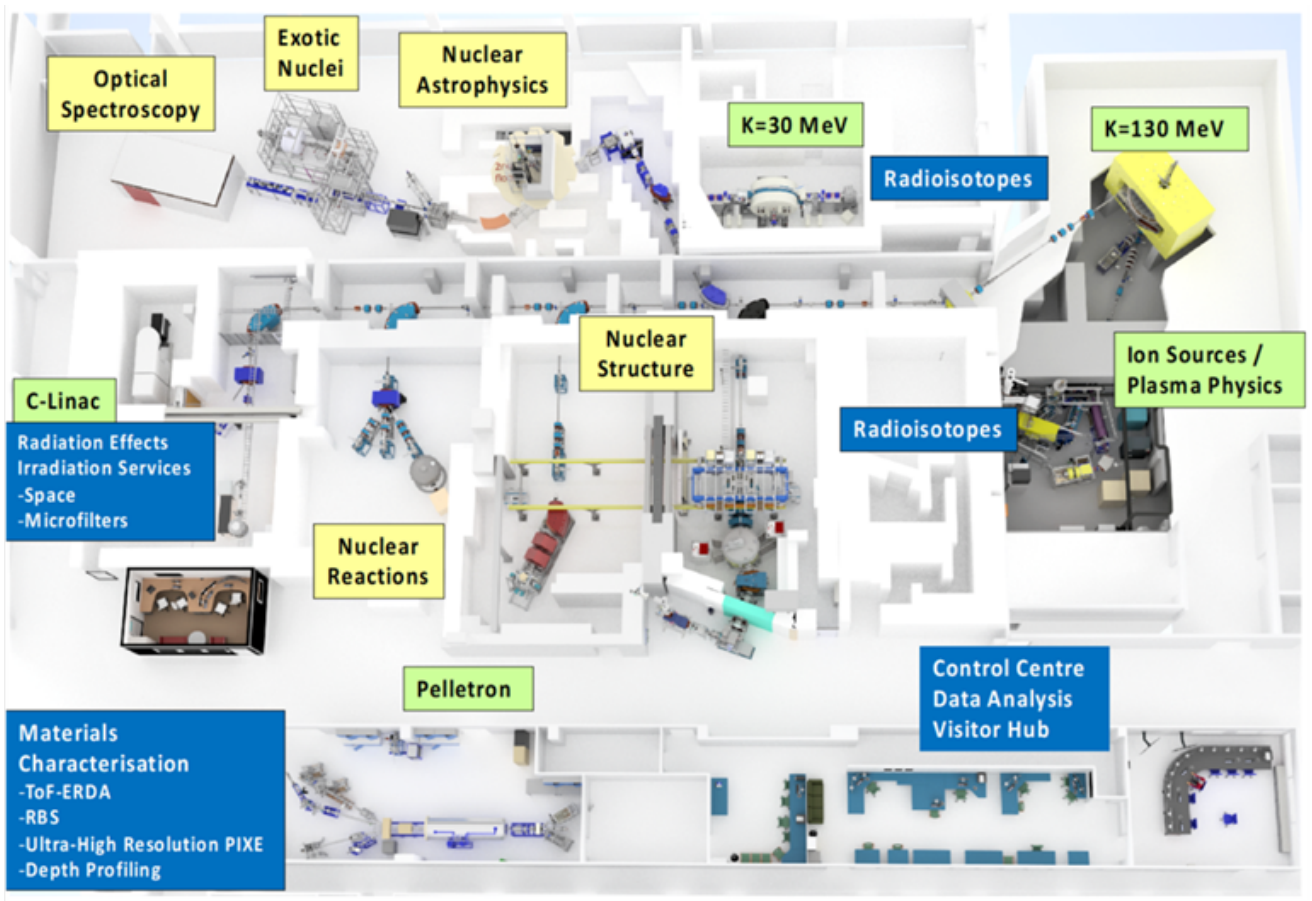

*Fig. 8.9: Laboratory layout of JYFL (see text for details) (© JYFL)*

Recent facility developments have included:

• The commissioning of an 18 GHz room-temperature ECR ion source has provided access to energetic (16.3 and 22 MeV/u) cocktail beams for radiation hardness tests.

• MCC30 light-ion cyclotron and the extension of the laboratory provide improved research conditions and additional capacity. More time can be released for beam development and for the use of heavy-ion beams from the K=130 accelerator in longer experiments and tests.

• The IGISOL-4 facility, served by both cyclotrons, offers considerable improvement in discovery potential for unexplored exotic nuclei with increased beam time using the K=30 cyclotron for the most challenging of cases. A recently commissioned Multi-Reflection Time-of-Flight (MR-ToF) Mass Spectrometer and the implementation of Phase-Imaging Ion Cyclotron Resonance measurements provide new possibilities for precision mass measurements and trap-assisted decay spectroscopy.

• Several laser systems are available for ionisation/spectroscopy applications and high-resolution collinear spectroscopy at IGISOL-4. New atom and ion-trapping systems are under development. Future developments will see radioactive beam research at the precision frontier.

• A new low-energy facility called MARA-LEB is under construction with a gas cell at the MARA focal plane, coupled with laser systems for selective ionisation and spectroscopy, as well as a versatile decay station. In the second phase, low-energy beams from MARA will be delivered to a cooler-buncher and MR-ToF mass spectrometer.

• A new 3 MV Tandem accelerator platform will mean that the capabilities for ion-beam modification and analysis of materials can be expanded and provide neutron beams, opening new lines of research.

## LEGNARO NATIONAL LABORATORIES (LNL)

The main research programmes at LNL (https://www.lnl.infn.it) centre on nuclear structure, reactions and astrophysics, and are carried out with the Tandem-ALPI-PIAVE accelerator complex (TAP), providing stable ion beams from p to Pb at energies up to approximately 10 MeV/u. Interdisciplinary activities are focused on the development of targets to produce novel radioisotopes for medicine and applications, their characterisation through ion-beam analysis techniques, and also on elemental microanalysis for material, earth and environmental sciences and cultural heritage. These activities are performed using two Van de Graaff accelerators: the CN (7 MV) and the AN2000 (2 MV). In the field of accelerator technologies, the main developments are for the European Spallation Source and for the IFMIF RFQ.

The SPES (Selective Production of Exotic Species) project (https://www.lnl.infn.it/en/spes-2) will construct Europe's first dedicated radioactive ion beam facility for fission fragments. The construction of the new complex is very advanced. It consists of a proton driver (a 70-MeV cyclotron) with two exit ports for a total maximum current of 750 μA. In one application, an ISOL facility will produce neutron-rich radioactive ion beams. An UCx ISOL target and ion source and a beam transport system with a high-resolution mass selection will deliver either non-reaccelerated beams for users or inject the produced species in the existing superconducting ALPI linear accelerator. In the other application, high-intensity proton beams will be exploited for the production of radioisotopes for applications (see Fig. 8.10).

Among the powerful detector systems available for the scientific programme, the large acceptance magnetic spectrometer PRISMA for heavy ions is used with different gamma-ray arrays, such as AGATA, which is hosted at LNL until at least 2026. Other set-ups include the GALILEO $\gamma$-ray spectrometer composed of 25 Compton-suppressed HPGe and 10 triple-cluster HPGe detectors, the GARFIELD large solid-angle multi-detector array for light charged particles and fragment identification for reaction studies, and an active target system.

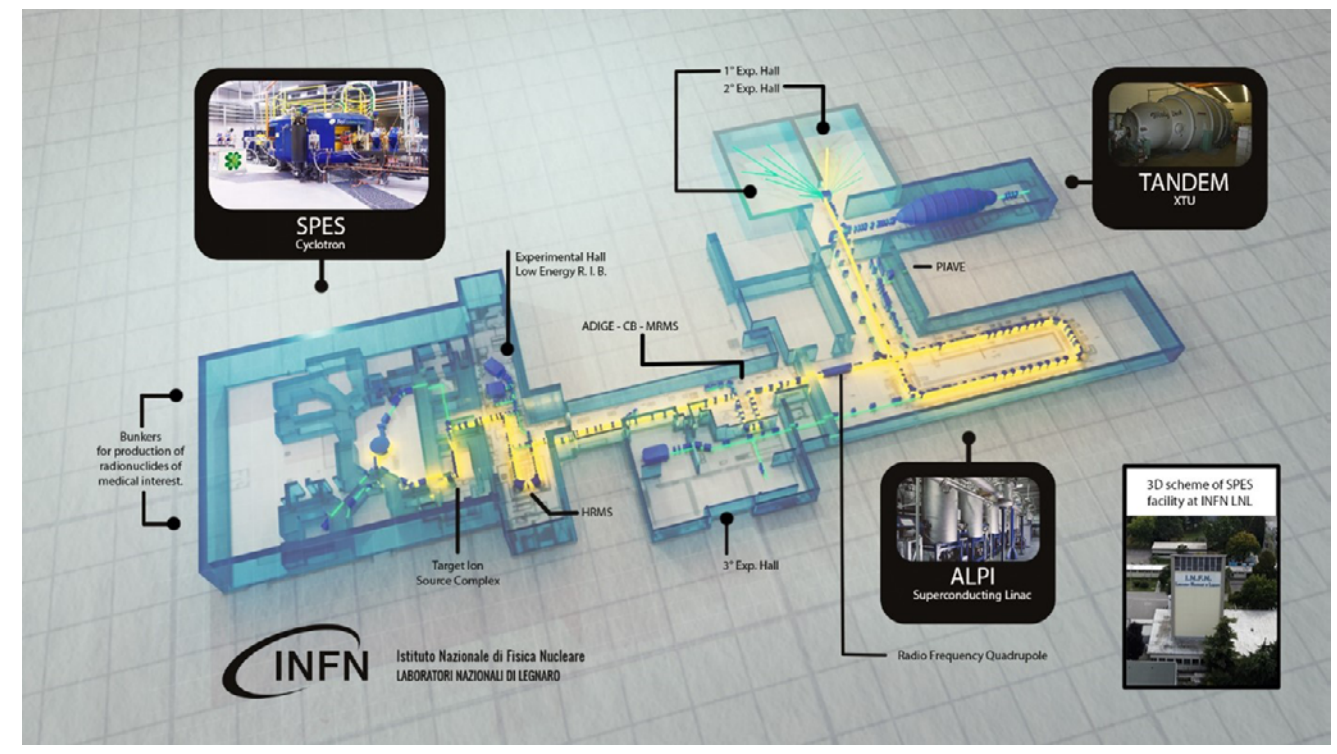

*Fig. 8.10: Layout of the SPES and Tandem-Piave-ALPI facilities at LNL. (© LNL)*

Additional facilities are available for in-flight production of light exotic nuclei, irradiations, ionisation-cluster-size distributions and neutron time-of-flight measurements, and elemental microanalysis of samples.

A series of medium-term upgrades is currently underway:

• A new phased approach for the SPES project will allow timely delivery of the successive milestones of the project.

• The exploitation of the facility, aimed at producing radionuclides, will start after the successful completion of the two first phases of SPES, namely the restarting of the cyclotron and the commissioning of the low-energy radioactive beams from SPES.

• A new Data Centre is being constructed to provide the upgraded AGATA DAQ infrastructure and the upgrade of the LNL Tier-2 Data Centre.

• The completion of ADIGE, the new injector RFQ to replace the

PIAVE injector and to improve reliability, beam intensities, beam energies and ion species of the ALPI accelerator which will also complete a large part of the post accelerator of SPES

● The refurbishment of the infrastructures at the AN2000 and CN accelerators.

## The Bellotti Ion Beam Facility at LNGS

The Gran Sasso National Laboratory (LNGS) (https://www.lngs.infn.it) is the largest underground laboratory in the world and is devoted to neutrino physics, astroparticle physics and nuclear astrophysics. It offers the most advanced underground infrastructure in terms of dimensions, complexity and completeness. For the past 30 years research in nuclear astrophysics has been carried out by the LUNA Collaboration (Laboratory for Underground Nuclear Astrophysics - https://luna.lngs.infn.it) at LNGS. Nuclear reactions of astrophysical interest are notoriously difficult to study in terrestrial laboratories because of the exponential drop in cross sections at the relevant stellar energies. The underground location assures a strong reduction of the cosmic radiation flux compared to a laboratory on the Earth's surface.

Two electrostatic accelerators, operating at 400 kV and 3.5 MV, are situated in different halls. Both machines deliver very intense proton and α beams (< 400 μA) and together cover a very broad energy range, from 50 keV to 3.5 MeV. The MV machine can also deliver intense carbon beams. Common features are the stability and high energy resolution of the delivered beams which make possible precision cross section measurements for nuclear astrophysics. The 400-kV accelerator is operated by the LUNA collaboration, while the 3.5-MV accelerator is operated as part of the LNGS Bellotti Ion Beam Facility, which has become a reality since the last NuPECC LRP. From 2024, calls for scientific proposals for beam time will be made annually and proposals will be evaluated by a dedicated Programme Advisory Committee.

Soon, the 400-kV accelerator will become part of the Bellotti facility by putting the two accelerators physically close together to fully exploit their capabilities. This creates the opportunity to upgrade the 400-kV accelerator, including the capability to produce doubly-charged $^{4}$He beams. The two accelerators will form an ion-beam facility permitting the study of several reactions over a broader energy range and reducing systematic uncertainties arising from normalisation between different data sets. For these reasons, the relocation and upgrade of the 400 kV accelerator inside the Bellotti Ion Beam Facility is of strategic importance.

The successful exploitation of the new facility requires the development of cutting-edge detectors and instrumentation. For example, total absorption spectroscopy would allow for total γ-ray energy measurements providing direct information on the total cross section with great efficiency. A notable case of long-term scientific ambition is the measurement of the $^{12}C(\alpha,\gamma)^{16}O$ reaction: despite extensive efforts in the last decades, this reaction still carries considerable uncertainties.

## National Cyclotron Laboratory Consortium (NLC)

The **National Cyclotron Laboratory (NLC)** is a consortium of two facilities in Poland, operating a double-site infrastructure: the Heavy Ion Laboratory (SLCJ) (https://www.slcj.uw.edu.pl) at the University of Warsaw, and the Cyclotron Centre Bronowice (CCB) (https://ccb.ifj.edu.pl) at the Institute of Nuclear Physics (IFJ PAN) in Kraków. The goal of the NLC consortium is to conduct basic and application research on 4 cyclotrons, two of which are installed at SLCJ and two at CCB. The SLCJ runs the K=160 isochronous heavy-ion cyclotron U-200P accelerating beams from He to Ar up to an energy of 10 MeV/u, and the PETrace K=16.5 cyclotron, delivering intense beams of protons and deuterons. CCB is equipped with the 60 MeV light-ion cyclotron AIC-144 and the new IBA Proteus-235 proton cyclotron, which delivers a proton beam at energy 70-230 MeV – it is dedicated mainly to proton radiotherapy, while nuclear physics research is conducted at weekends.

In recent years the two laboratories, CCB and SLCJ, have modernised several of their detection systems and developed new ones. The main instrumentation at SLCJ consists of discrete gamma-ray and charged-particle detection systems (see Fig. 8.11), while at CCB a high-energy gamma-ray detection set-up and various detectors of high- and low-energy charged particles are available (see Fig. 8.12). In addition, two irradiation stations at the AIC-144 cyclotron, which allow for proton exposure over a wide range of doses and dose rates are installed. The NLC facility is also equipped with state-of-the-art instrumentation for dosimetry.

The nuclear physics research programme of NLC aims at obtaining high precision data on properties of nuclei at and around the valley of stability, which makes it in many aspects complementary to the programmes of large-scale European RIs which often explore the exotic parts of the nuclear chart. The main line of research at SLCJ encompasses nuclear structure studies using the Coulomb excitation technique as well as investigations of nuclear reactions focused on the Coulomb barrier distributions. Research in radiochemistry, radiobiology and material science is also carried out, together with particle detector development and testing.

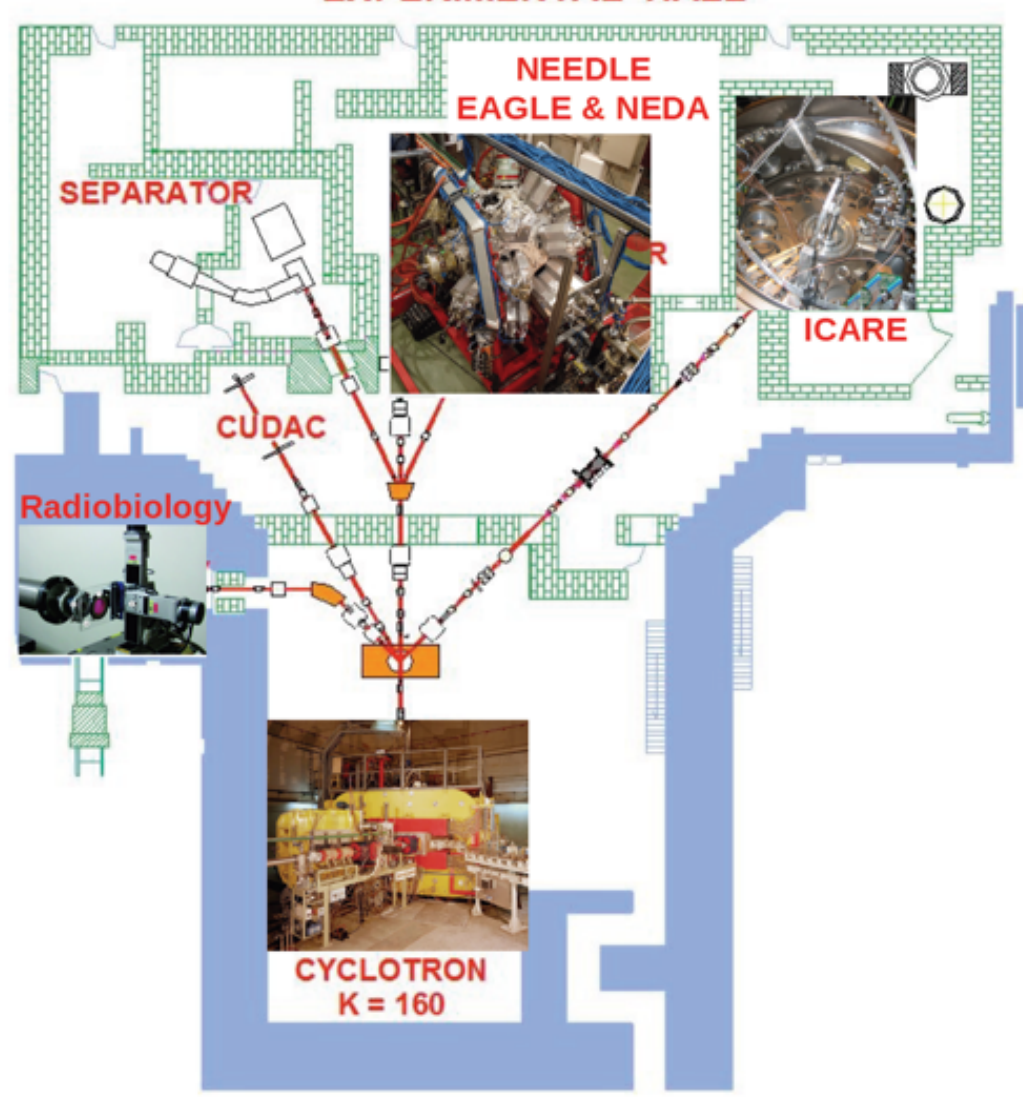

*Fig. 8.11: Experimental hall at the SLCJ facility, showing NEEDLE, the Ge multidetector array EAGLE coupled to DIAMAND and NEDA neutron detector (coupling to FATIMA, PARIS or RFD systems are also possible), the charged particle array ICARE, the CUDAC scattering chamber, the on-line mass separator IGISOL and a setup for radiobiology.*

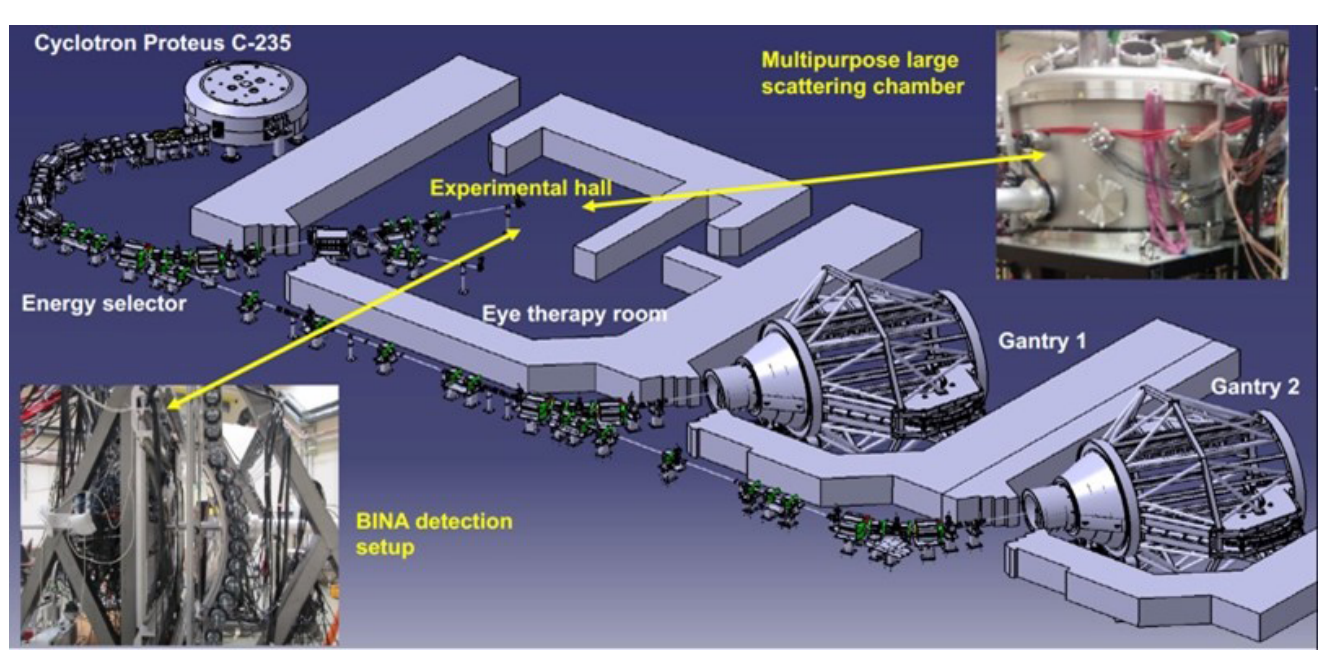

*Fig. 8.12: Layout of the CCB facility, displaying the experimental hall with the multipurpose large scattering chamber (which houses the high-energy gamma-ray detection system consisting of a set of $LaBr_3$ scintillators - occasionally complemented with the PARIS array clusters, the Kraków Triple Telescope Array (KRATTA) and a set of DSSSD detectors), the Big Instrument for Nuclear Data Analysis (BINA), the eye therapy room and two gantries for proton therapy of cancer sites at all body locations.*

At CCB, investigations of the decay of giant resonances and other specific states in the continuum, as well as dynamics of few-nucleon systems, are the main research projects. A large part of the research is also devoted to in-beam testing of detectors constructed for nuclear physics experiments in European facilities, to irradiations of different samples with high-intensity protons, and investigations of the clinical efficacy of the state-of-the-art scanned proton beam technique in the treatment of selected tumours.

In the coming years, several upgrades are planned at NLC. For example, at SLCJ, the development of new, heavier beams towards nickel and the construction of a vertical beam facility for radiobiology and material science studies are envisaged. Installation of a capillary line connecting the PETrace cyclotron with the ECR ion source of the K=160 cyclotron to accelerate short-lived isotopes is another expected development.

At CCB, the installation of polarised He-3 targets with a vertex detection system and an upgrade of the BINA detector enabling proton polarisation measurements is planned. Also, a new DAQ system, compatible with fully digital front-end electronics should become available. Other expected modernisations include the construction of the fully automatised scanning table for irradiation purposes and upgrading the irradiation lines and the relevant dosimetry methods for implementation of the high-dose-rate (FLASH) irradiation technique.

## NATIONAL LABORATORIES of SOUTH (LNS)

LNS (www.lns.infn.it) hosts two accelerators, the MP Tandem and the Superconducting Cyclotron delivering light and heavy-ion beams at low and medium energy. The LNS scientific mission is mainly the study of nuclear reactions at low and intermediate energies. The focus is on the equation of state of nuclear matter and the role of symmetry energy through multifragmentation, charge-exchange reactions to determine the NME of the double beta decay, the study of clustering in light nuclei, as well as the stellar and primordial nucleosynthesis with the Trojan-Horse method.

There are strategic activities in acceleration systems, in ion sources and in several multi-disciplinary fields including atomic physics, solid-state physics, single-event effects, biology and medicine, cultural heritage, nuclear waste monitoring and dosimetry. The operating CATANA proton-therapy programme using 62-MeV proton beams is dedicated to the treatment of eye melanoma. The scientific programme uses ion beams from protons to Pb (up to 25 MeV/u) and several detector systems and installations (see Fig. 8.13). The major instrumentation includes CHIMERA, a $4\pi$ charged-particle detector and the FARCOS array; MAGNEX, a large acceptance spectrometer with ancillary detectors; CLAD and LHASA, arrays of double-sided silicon strip detectors for nuclear astrophysics; POLYFEMO, a $4\pi$ thermalisation neutron counter; and several multipurpose scattering chambers to host complex detection systems for experiments with cyclotron and fragmentation beams.

The decommissioning of the Medea Hall will make available a large free space for new experimental halls, one of which will have laser pulses plus intense ion beams accelerated by conventional accelerators. There are two irradiation beam lines dedicated to multidisciplinary applications equipped with the necessary detectors and experimental setups that can easily be shared between different experiments.

A series of medium-term upgrades is underway and expected to be completed by 2027:

- The upgrade of the superconducting cyclotron.

- The installation of the new fragment separator FRAISE for production of high-quality, in-flight radioactive beams.

- Two new ion sources, one for noble-gas elements with the tandem and another for high-intensity injection into the cyclotron.

- The installation and commissioning of an ultra-short pulse power laser (up to 1 PW, down to 23 fs) for the generation of ion and electron beams aimed at basic nuclear physics, astrophysics, biophysics and material science studies.

- A research activity (PANDORA) using a superconducting plasma trap for the experimental study of transmutations in plasma to constrain β-decay rates in highly ionised systems and to investigate opacities relevant for multi-messenger astrophysics and the modelling of kilonovae ejecta.

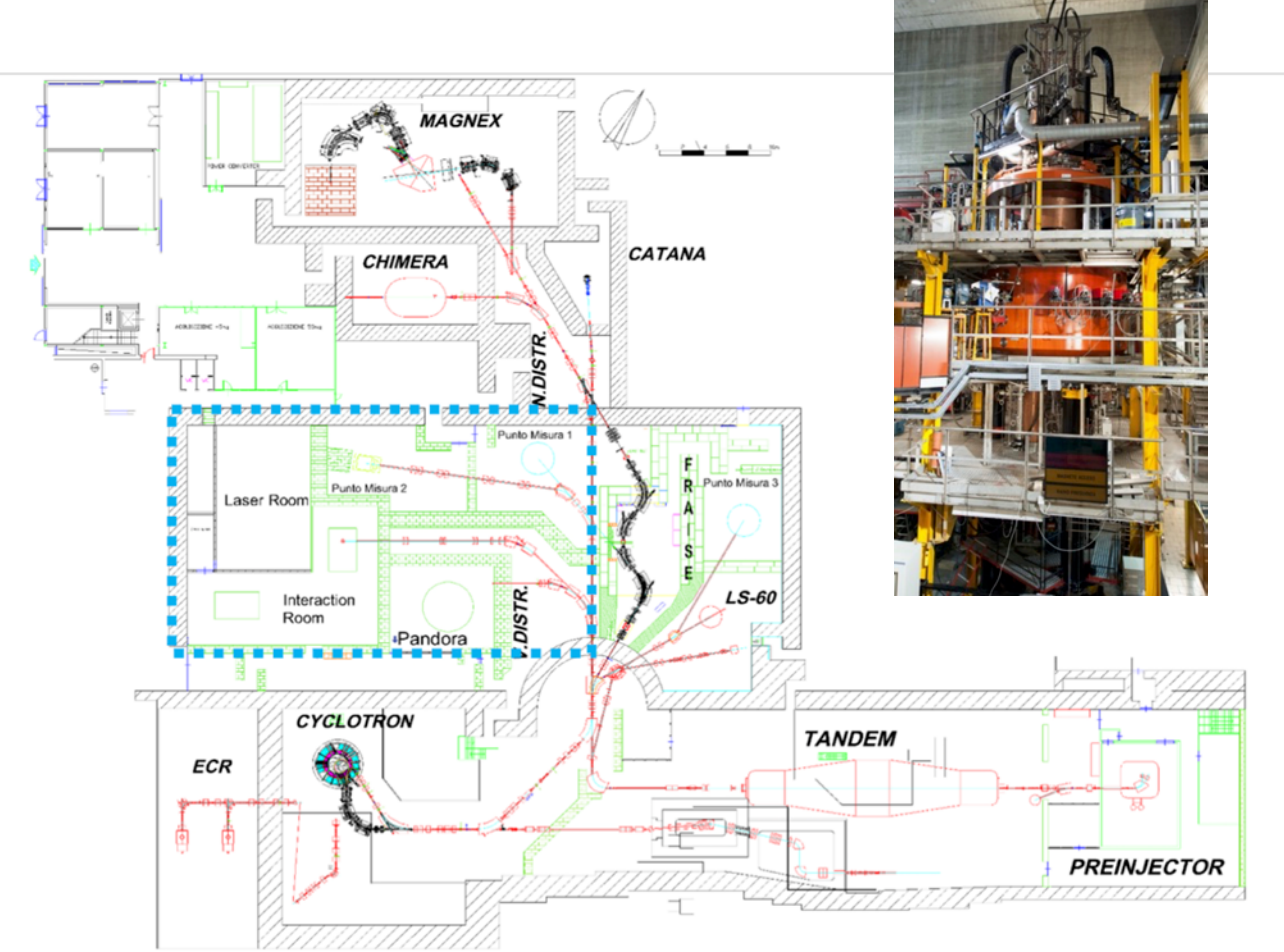

*Fig. 8.13: New layout of the LNS experimental halls after 2023. The area framed with dots hosts new experiments and a line for high intensity accelerated beams the Superconductive Cyclotron (photo).*

## Paul Scherrer Institute (PSI)

The Paul Scherrer Institute (PSI) (www.psi.ch) offers several user facilities covering a wide variety of fields. The main driver for the facilities relevant to nuclear physics is the High-Intensity Proton Accelerator (HIPA) with an energy of 590 MeV and a beam power of 1.4 MW. Since the first beam in 1974, HIPA has undergone a continuous upgrade programme targeted at improving performance and reliability. HIPA delivers protons to two target stations for the production of muons, pions and electrons and two neutron spallation sources for ultracold, cold and thermal neutrons. The beam from the injector cyclotron of HIPA is used to produce various radioisotopes for radiopharmaceutical applications at the IP2 facility. A dedicated small cyclotron COMET feeds three gantries for tumour treatments and is also used

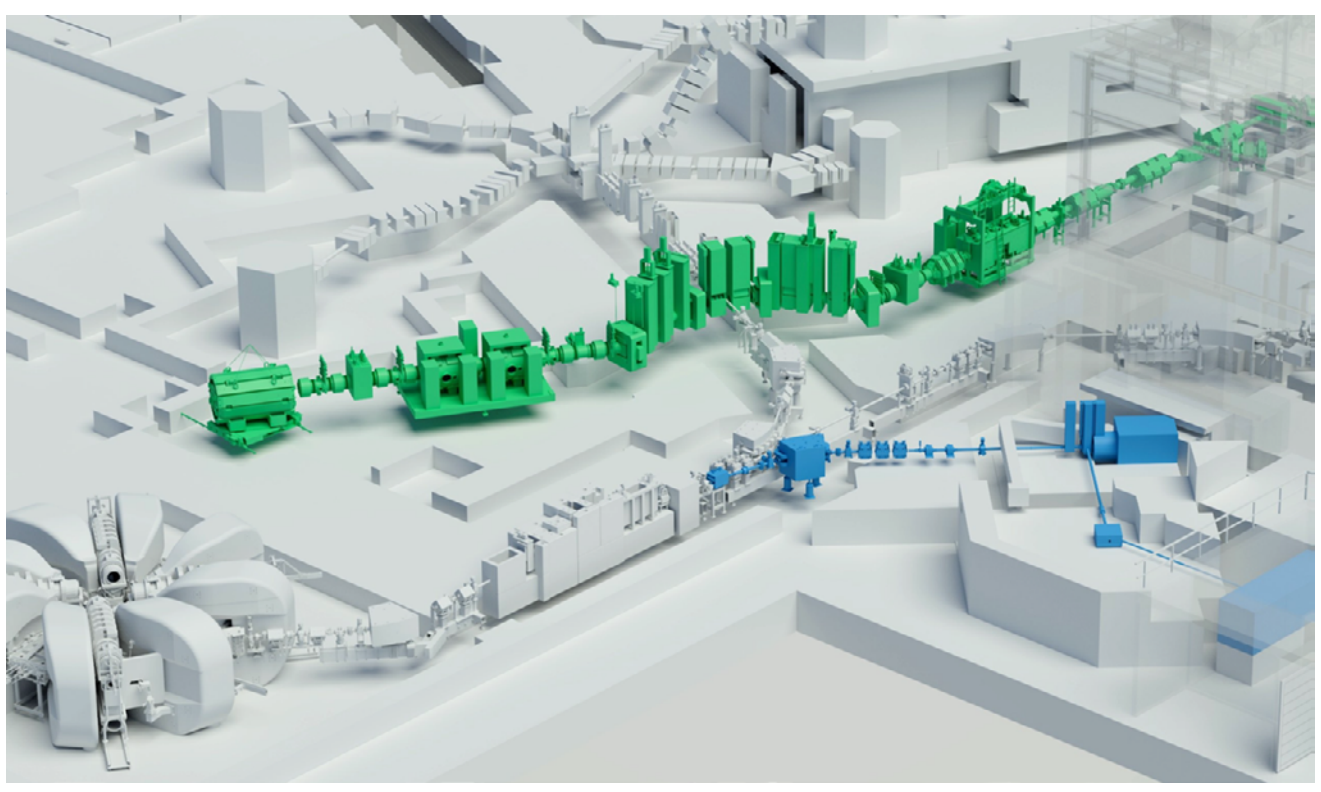

*Fig. 8.14: Overview showing the HIPA ring cyclotron of PSI together with the new HIMB and TATTOOS installations in green and blue, respectively, which are part of the IMPACT project. HIMB aims at increasing the muon intensities by up to two orders of magnitude, while TATTOOS will provide unrivalled quantities of a wide range of previously unobtainable radionuclides. (Graphics: Paul Scherrer Institut/Mahir Dzambegovic)*

by the proton irradiation facility PIF.
Muon beams have been used to obtain the strongest limits on all three important charged-lepton flavour-violating channels, ($\mu \to e\gamma$, $\mu \to eee$ and $\mu \to$ e-conversion) with improved experiments for the first two channels still being pursued. The physics of exotic atoms is another key component of the programme including muonium, pionic hydrogen and light muonic atom laser spectroscopy. The determination of charge radii of heavy isotopes through muonic atom x-ray spectroscopy has regained focus. Muons and muonic atoms are also being used for weak interaction studies, such as muon capture on the proton and on other elements to test low-energy QCD and provide input relevant for neutrinoless double beta decay. New experiments, such as a measurement of the muon electric dipole moment and a test of lepton universality in pion decays, are being prepared. Additionally, an instrument is being commissioned to use negative muons for non-destructive, depth-dependent elemental analysis.

The high-intensity source for ultracold neutrons serves the world's leading search for the neutron electric dipole moment and offers a total of three beam ports for experiments. A replacement of its central solid-deuterium unit to provide reliable and improved operation is planned.

The proton, pion and electron beams are also used for irradiation applications. At the spallation neutron source SINQ, several irradiation options exist, and a leading neutron radiography and tomography is in place. SINQ has recently undergone an extensive upgrade of its guide system and is exploring a further extension of the guide hall.

The IMPACT project (https://www.psi.ch/impact) is to produce much-improved quantities of muons and radionuclides at HIPA, benefitting a wide research field. Two new target stations, one for the High-Intensity Muon Beams (HIMB) and one for the Targeted Alpha Tumour Therapy and Other Oncological Solutions (TATTOOS) online isotope separation facility will be constructed, thereby significantly extending the existing infrastructure (see Fig. 8.14). HIMB will provide two orders of magnitude higher muon intensities and TATTOOS unrivalled quantities of a large range of previously unobtainable radionuclides. The first beam for HIMB is expected in 2028 with TATTOOS following in 2030.

# Lepton and Photon Beam Facilities

Electromagnetic probes complement hadronic probes and the combination offers deeper insight into hadrons and nuclei due to the different interaction mechanisms. Highly energetic real photons are produced from electron beams and used to study hadrons and nuclei with zero four-momentum transfer ($Q^2$). The interaction of leptons with hadrons and nuclei is described in terms of virtual photon exchange, where a positive or negative four-momentum transfer ($Q^2 \neq 0$ GeV$^2$) occurs in the kinematics of $e^+$-$e^-$ annihilation or scattering, respectively. The combination of real and virtual photons enables the complete study of the exchange amplitude. The size of the electromagnetic coupling ($\alpha_{EM}$=1/137) renders a clean separation: the emergence of strong interaction phenomena in the nonperturbative regime can be separated from the electromagnetic amplitude. QED allows for a perturbative treatment of higher-order corrections beyond the one photon exchange in a controlled way. The Muon beams at PSI have been discussed in the preceding section.

## CERN-SPS

The M2 beamline at the CERN/SPS (see Fig. 8.1 and the CERN section) provides an internationally unique high-energy and high-intensity muon beam, longitudinally polarised with respect to the beam direction. Currently, three experiments share the infrastructure: AMBER, NA64-mu and MUonE: the muon beam will be used by the AMBER collaboration for the measurement of the proton charge radius. The same facility will be used by the NA64-mu collaboration for the search of light dark bosons. MUonE proposes to use the muon beam for measuring the leading hadronic contribution to the muon anomalous magnetic moment. Both NA64-mu and MUonE experiments require a low divergence of the muon beam and an intensity of up to $10^6$ muons per second.

## Extreme Light Infrastructure ELI-NP

The Extreme Light Infrastructure, an ESFRI facility, is a large-scale distributed research infrastructure based on high-power lasers. It consists of three sites, each with a different research focus: ELI-Beamlines in the Czech Republic, ELI-ALPS in Hungary and ELI-Nuclear Physics. The latter, ELI-NP (https://www.eli-np.ro/), was set up in Romania by IFIN-HH and is dedicated to nuclear photonics, i.e. nuclear physics using extreme photon beams or their secondary radiation. These beams will be used for fundamental research studies as well as for developing high-impact applications. The nuclear programme at ELI-NP is particularly rich, the ELIMED facility at ELI-BEAMLINES has recently been completed and developments for muon beams and laser-induced fusion research are presently ongoing.

ELI–NP hosts a 2 x 10 PW laser system, based on Ti:Sapphire technology and delivering ultra-short laser pulses. It is the most powerful laser system worldwide, reaching intensities as high as $10^{23}$ W/cm$^2$, and has been operational at nominal parameters since 2020. High-intensity γ-ray beams will be provided by a system based on Laser Compton Backscattering (LCB) of laser light from relativistic electrons. The system will consist of a linear accelerator delivering electrons with an energy of up to 800 MeV, a storage ring in which electrons revolve at high frequency, an optical cavity for the recirculation of laser pulses, a frequency matched with the storage ring, and a collimator to select the bandwidth of the LCB γ-ray beam. The system will provide quasi-monochromatic γ beams up to 19.5 MeV, with a relative bandwidth of ≤0.5% and spectral density of ≥5000 photons/s/eV. The construction of the γ-beam system is underway and completion is expected in 2026.

Systems generating light at ELI–NP are complemented with instrumentation optimised to fully utilise the characteristics of available beams (see Fig. 8.15). Laser-driven setups have been developed for ion acceleration with solid targets and electron acceleration with gas targets, search for dark matter candidates, pair production in high energy photon collisions (Breit-Wheeler effect), radiation reaction and non-linear Compton scattering. Gamma-beam-driven experiments employ ELIADE (HPGe segmented clover array) for Nuclear Resonance Fluorescence studies, ELIGANT-GN (an array of $LaBr_3$(Ce), $CeBr_3$ and neutron detectors) for Pygmy and Giant Dipole Resonance studies, ELIGANT-TN (an array of $^3$He detectors) for (γ,xn) cross section measurements, ELISSA (DSSSD array) and e-TPC for charged particles detection, ELI-BIC (an array of ionisation chambers), ELI-THGEM (an array of THGEM detectors), and IGISOL for photofission experiments.

ELI-NP has been operating since 2022, serving an international user community. Laser-driven physics cases will be pursued in the short term to develop ion acceleration to several hundred MeV and genera-tion of high-power (Petawatt) γ-ray sources. Nuclear studies will aim at measuring cross sections of reactions of astrophysical interest under plasma conditions relevant to the stellar environment, studying nuclear lifetime behaviour in plasma and developing intense short-pulsed laser-driven neutron sources. The research on laser-driven medical applications will follow in the medium term. When the γ-beam system becomes available, the full physics case of ELI-NP will be pursued, including photonuclear reactions with monochromatic beams.

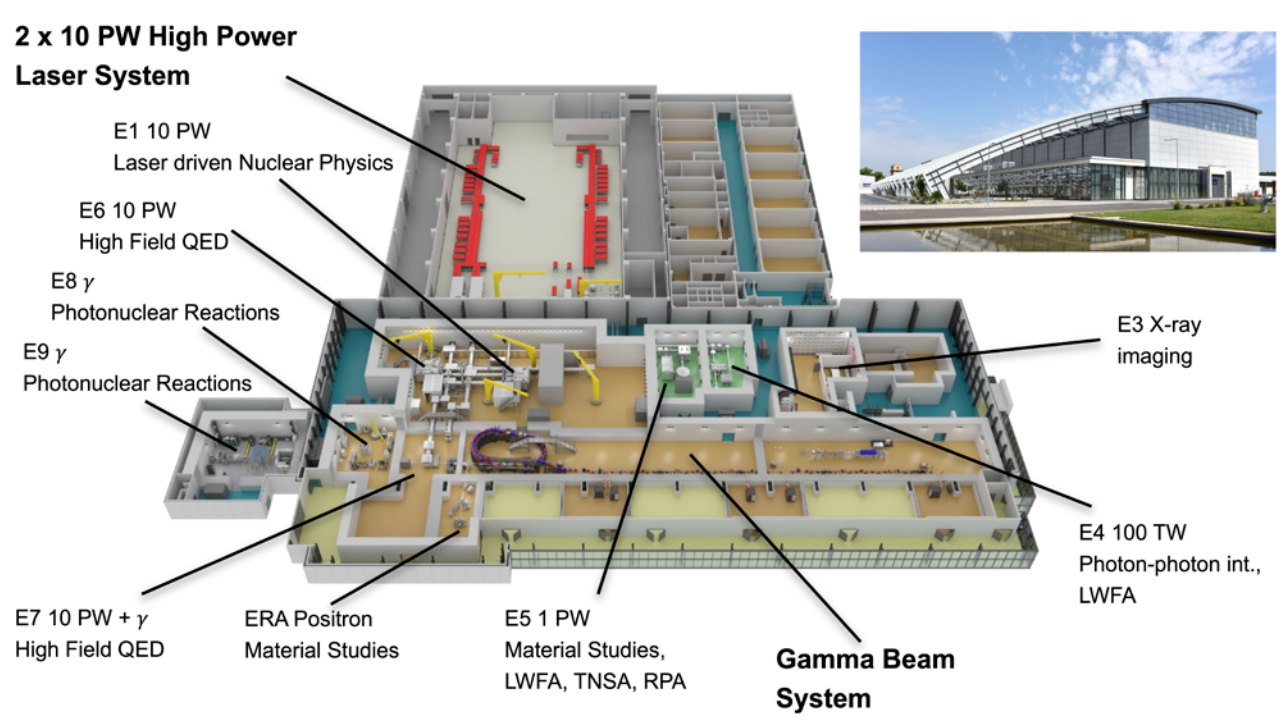

*Fig. 8.15: Schematic view of ELI-NP including its main systems and experimental facilities. The commissioning of the 10 PW laser system, which took place on August 19th, 2020, involved a successful laser endurance operation utilising both the pulse compressor and the laser beam transport system. (©IFIN-HH/ELI-NP)*

## ELSA

The Electron Stretcher Accelerator (ELSA) (http://www-elsa.physik.uni-bonn.de/) is operated by the Institute of Physics of the University of Bonn. It consists of two electron LINACs, a booster synchrotron and an electron stretcher ring (see Fig. 8.16). Unpolarised or polarised electron beams are injected at energies around 20 MeV into the synchrotron by either LINAC, accelerated to 1.2 GeV and transferred into the stretcher ring. The latter can be operated in booster, stretcher and storage mode. In the booster mode, used for hadron-physics experiments, several pulses from the synchrotron are accumulated with an internal current of typically 20 mA. The electrons are further accelerated to a maximum of 3.5 GeV, extracted slowly via resonance extraction and delivered to experiments with a typical spill time of 4–9 sec. A new beamline connected to the stretcher ring is in operation providing electron beams for detector tests and characterisation up to full energy with currents of 1 fA to 100 pA and a duty factor of 80%. The focus at ELSA is on hadron physics and irradiation for medical physics and other applications. The hadron-physics programme is devoted to baryon spectroscopy via meson photoproduction with an emphasis on polarisation experiments. ELSA will continue to provide beams for the existing hadron-physics experiments to allow for the best exploitation of their physics potential. Two experimental areas exist, each equipped with photon tagging systems including diamond radiators to provide polarised photon beams.

The CBELSA/TAPS experiment (http://www.cb.uni-bonn.de/) combines Crystal Barrel and TAPS electromagnetic calorimeters and is ideally suited to investigate photoproduction of neutral mesons decaying into photons with a nearly complete solid-angle coverage. Together with the polarised frozen-spin target and a circularly or linearly polarised photon beam, single and double polarisation experiments have been successfully performed. A major upgrade to trigger on final states including neutrons has been performed. For the future, an upgrade is planned to include a forward spectrometer to access final states exhibiting strangeness. A dipole magnet and tracking detectors with a segmented electromagnetic calorimeter will cover the forward acceptance.

Detector tests are pursued in two areas. At LINAC I, a pulsed high-current 20-MeV beam can be used for material irradiation. The test beamline provides beam for detector development. Both test areas serve as a facility for the Centre for Detector Physics FTD at the University of Bonn campus which recently went into operation (https://www.ftd.uni-bonn.de/).

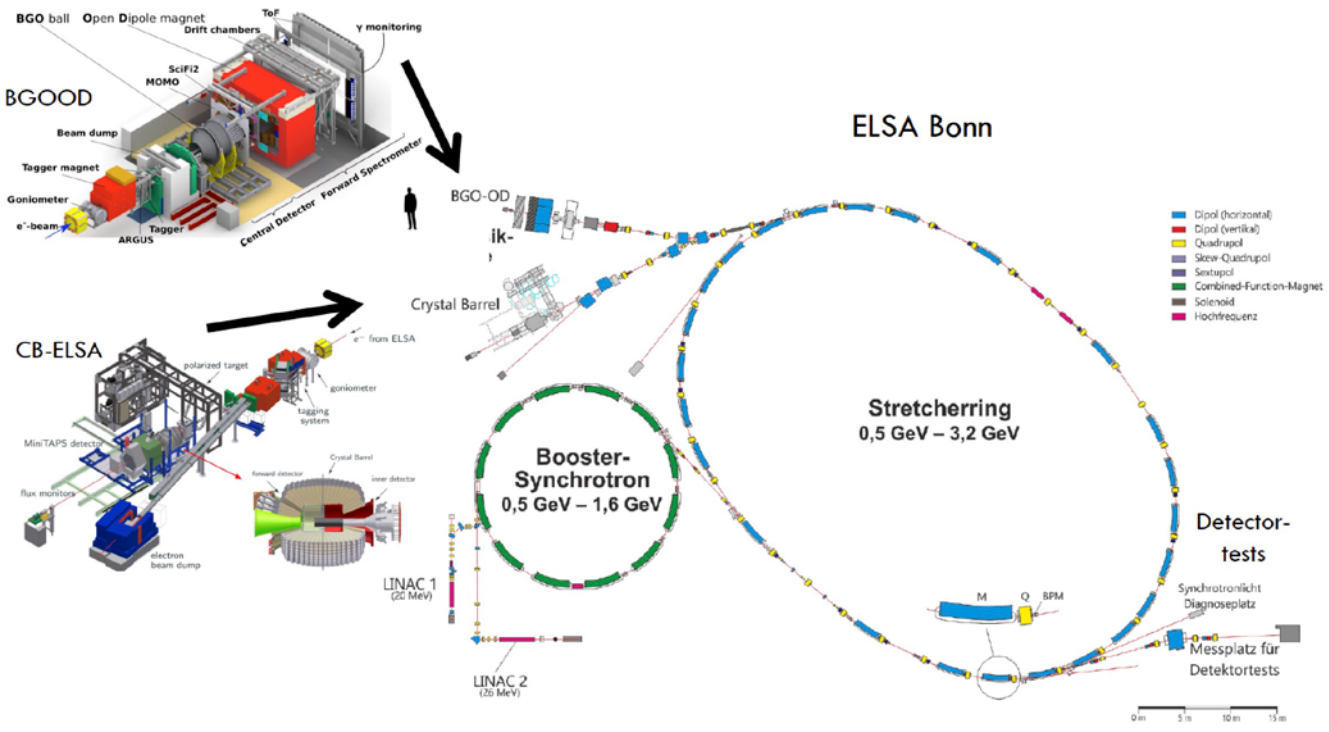

*Fig. 8.16: Floor plan of the ELSA electron accelerator and the experimental facilities. (© Univ. Bonn)*

## Frascati National Laboratories (LNF)

The INFN Frascati National Laboratories (LNF) (http://www.lnf.infn.it) host the e+e- meson production factory DAΦNE. Its accelerator complex consists of a double-ring electron-positron collider working at the c.m. energy of the Φ-resonance (1.02 GeV) and an injection system. The infrastructure includes two independent rings, each about 97 m long. The two main rings cross in two symmetrical sections: the Interaction Region (IR), specifically designed for hosting the experiment taking data, and the Ring Crossing Region, where the beams travel in two vertically separated beam pipes intersecting with a 50 mrad horizontal crossing angle, as in the IR. A full energy injection system, including an S-band LINAC, 180-m long transfer lines, and an accumulator/damping ring, provides fast and high-efficiency electron–positron injection.

DAΦNE also supplies four synchrotron light lines, and the accelerator complex contains a beam test facility with two beamlines. DAΦNE is still a unique machine for studies using low-energy, charged kaons with momenta below 140 MeV/c. In addition, a new approach to collisions, the Crab-Waist collision scheme, has been developed giving the machine luminosity increases of up to a factor of three. The achieved luminosity is one order of magnitude higher than that of other colliders working at the same energy, and the Crab-Waist scheme has become the main approach for present and future lepton colliders.

For many years experiments using kaons delivered by the phi-decay have produced unprecedented results in rare decays and symmetry studies (C, P, and CPT) and hadron and strangeness physics in the phi energy range. Currently, the SIDDHARTA-2 experiment is running at DAΦNE, which aims at measuring a series of kaonic atoms, in particular the first kaonic deuterium X-ray transitions to the fundamental level, from which the antikaon-nucleon isospin-dependent scattering lengths can be deduced. Investigating kaonic atoms provides an exceptional tool to comprehend the strong interaction among particles in the non-perturbative regime involving strangeness. This research has far-reaching implications, ranging from nuclear and particle physics to astrophysics.

At the same time, LNF pursues goals in different fields of research: elementary and astroparticle physics, theoretical physics, multidisciplinary activities with synchrotron radiation beams and detector and accelerator developments with the Beam Test Facility. The Positron Annihilation to Dark Matter Experiment (PADME) was designed and constructed to search for dark photons in positron-electron annihilation. The PADME detector has been modified and the latest data-taking run has been undertaken to probe the existence of the X17 particle detected at ATOMKI.

A new laboratory (SPARC_LAB) has been set up with a high-brightness electron beam driving a free electron laser in the green light. The SPARC injector, in conjunction with the very powerful infrared laser FLAME (300TW and 25-fs width), will permit investigations into the physics of plasma-wave-based acceleration and the production of X-rays via Compton scattering.

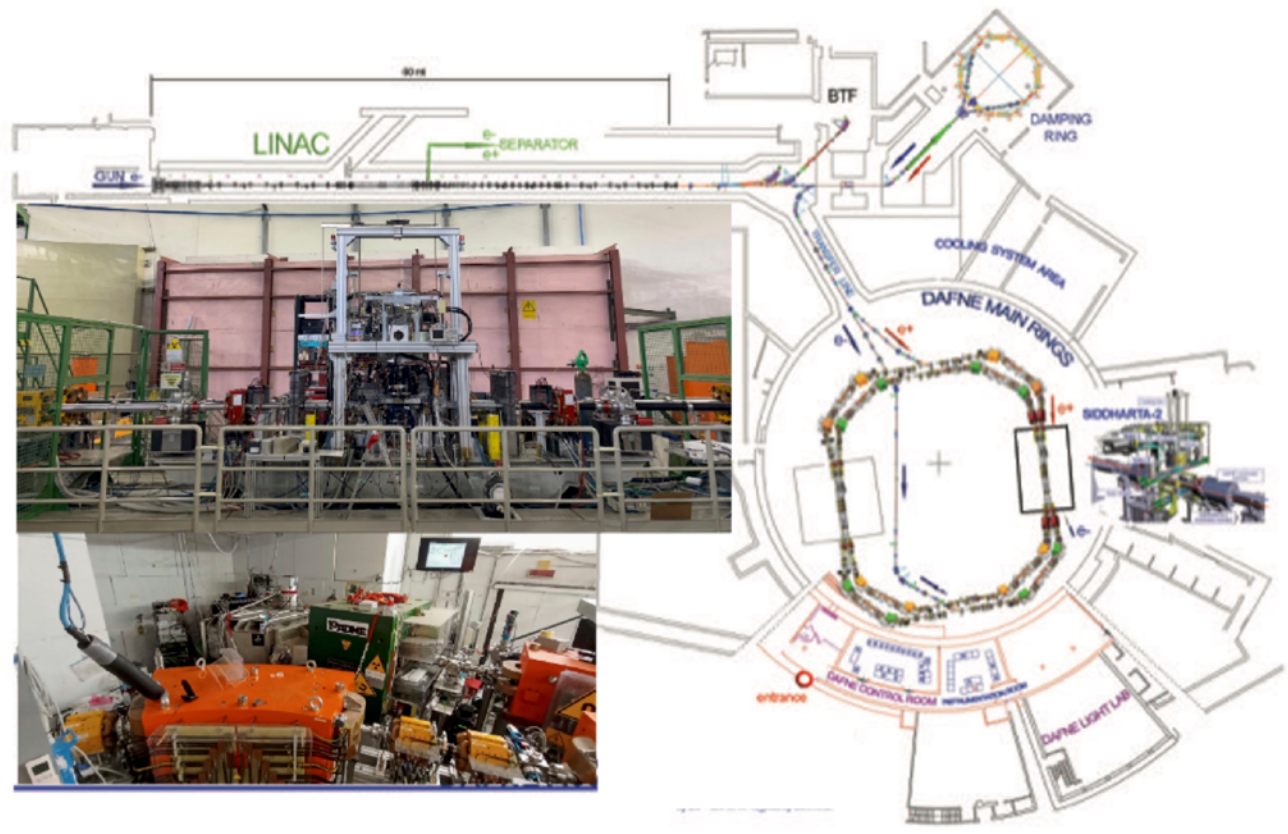

*Fig. 8.17: The DAFNE Accelerator Complex, with the SIDDHARTA-2 and PADME setups (© LNF).*

# Mainz Microton (MAMI) and Mainz Energy-Recovering Superconducting Accelerator (MESA)

The Mainz Microtron (MAMI) electron accelerator is operated by the Institute of Nuclear Physics of the Johannes Gutenberg University of Mainz. During the past ten years, the institute has worked on the preparation of a new, high-intensity electron accelerator, the Mainz Energy-Recovering Superconducting Accelerator (MESA).

The MAMI continuous-wave accelerator consists of sources for unpolarised and polarised electrons, followed by an injection linac (see Fig. 8.18). Three consecutive race-track microtrons and a harmonic double-sided microtron provide a beam energy for experiments ranging from 180 MeV up to a maximum beam energy of 1.6 GeV. Hallmarks of the MAMI accelerator are the excellent intensity of up to 100 µA, a high degree of polarisation of up to 85%, and an absolute energy resolution of $10^{-4}$. With these parameters, MAMI and its experiments are ideally positioned for highly competitive investigations in the field of hadron and nuclear physics. Currently, two major experimental setups are operated at MAMI: the A1 high-resolution five-spectrometer setup, and the A2 experiment at the tagged-photon beam line.

The core A1 components are three big spectrometers with momentum resolutions of $10^{-4}$, a short-orbit spectrometer for momenta of up to 200 MeV/c, and a compact spectrometer (KAOS) for kaons up to 1900 MeV/c. A highly segmented, large solid-angle neutron detector has been completed recently. High-power cryotargets are available for liquid hydrogen and deuterium, pressurised $^{3,4}$He and polarised $^{3}$He. Recently, a cryogenic cluster jet target of noble gas targets has been added. The setup has been updated to enable accurate measurements of small ($10^{-6}$) single-spin asymmetries with normal beam polarisation. The A2 Collaboration runs a facility for energy tagging of bremsstrahlung photons designed by Glasgow and Edinburgh Universities. The energy of the outgoing electrons is momentum analysed to tag the photon energy. The primary detector arrangement consists of the large acceptance Crystal Ball together with the TAPS calorimeter wall. This setup is particularly suitable for the detection of final-state photons from neutral meson decay with a solid angle of almost $4\pi$ with high resolution and count rate capability. A polarised frozen-spin target for protons and deuterons with longitudinal and transverse polarisation is operating successfully. Additional targets and sub-detectors are available. Measurement of the proton charge radius employing a time projection chamber is envisaged.

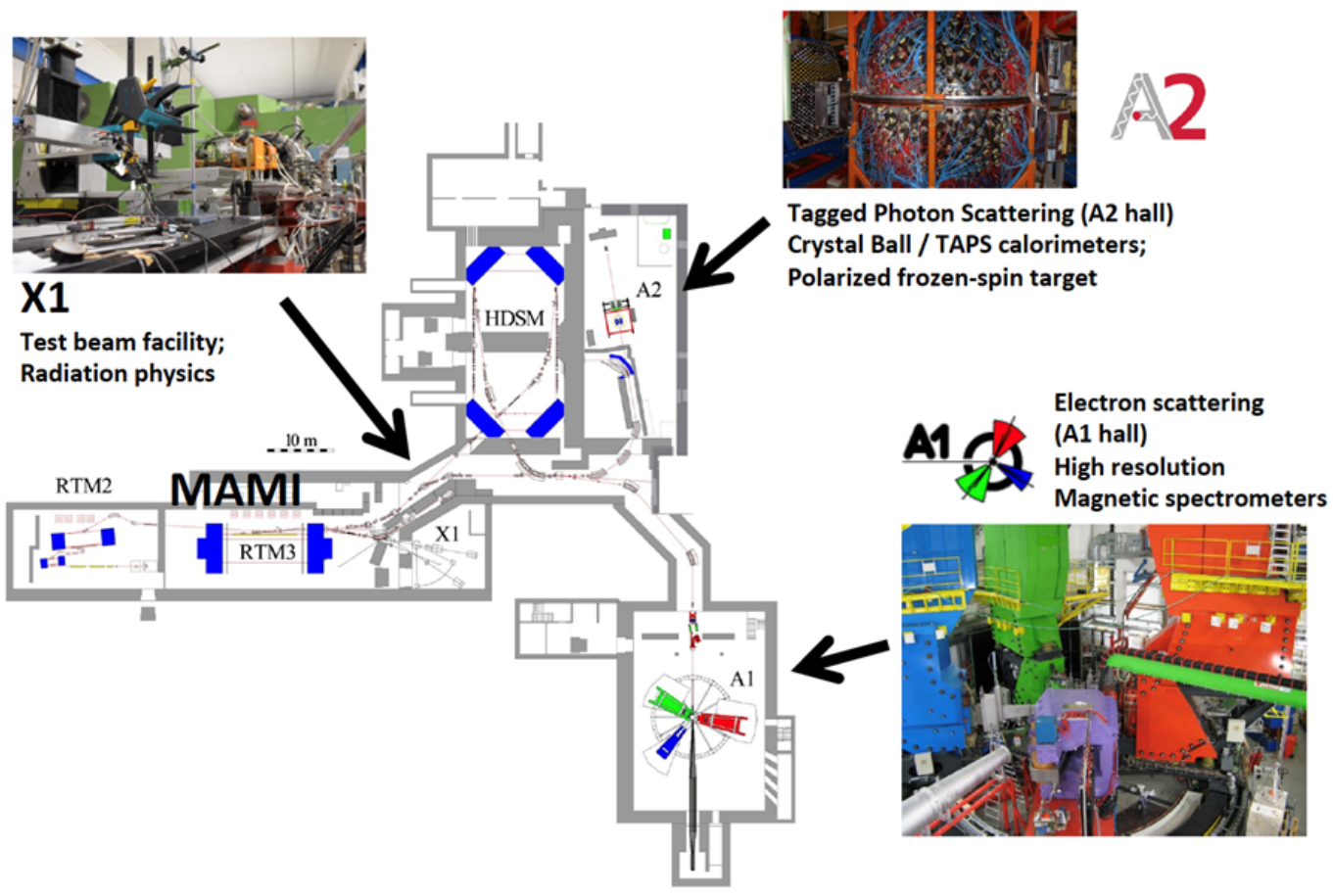

*Fig. 8.18: Floor plan of the MAMI electron accelerator and the experimental facilities. (© Univ. Mainz)*

The MESA facility (see Fig. 8.19) is currently under construction and will go into operation in 2025. The accelerator will be able to run in parallel with the Mainz Microtron MAMI, and its three experiments will provide the basis for a rich physics programme in nuclear, hadron, and particle physics for decades to come. Studies are being made of the possibility to produce coherent X-rays from the electron beam. A dedicated test beam area for detector tests and development of electron beam monitors is available and has been instrumental in the development of the MESA facility.

A new underground hall has been constructed to house the accelerator and experimental facilities. The use of superconducting accelerator cavities required the installation of a major cryogenics system. MESA will have two operation modes: an extracted-beam mode at an energy up to 155 MeV with a polarised electron beam of up to 150 µA in combination with various targets of considerable thickness in the P2 spectrometer. After interaction with the target, the beam will be dumped at full energy. Luminosities will be in the range of $10^{39}$ cm$^{-2}$ s$^{-1}$ and above; in addition, an innovative energy-recovering (ERL) mode will be used with a beam energy of up to 105 MeV and intensity of >1 mA (see the highlight Box 8.1 on ERLs). For the first time, this ERL mode will be used for hadron and nuclear physics experiments with the MAGIX spectrometer setup, using a very thin gas jet target with highly competitive luminosities of at least 1035 cm$^{-2}$ s$^{-1}$. After interacting with the target, the beam is decelerated down to injection energy by re-injecting it into the cryostats with a decelerating phase. The operation of the ERL mode in combination with a thin gas jet target will be new and unique. This will pave the way to extremely clean experimental conditions for precision experiments ranging from nuclear cross-section measurements of astrophysical relevance to measurements of the electromagnetic form factor of the nucleons and to the search for new particles.

In the extracted beam mode, the polarised electron beam at 155 MeV will interact with a liquid or solid-state hydrogen target. The P2 experiment will perform a world-leading measurement of the weak charge of the proton by parity-violating electron scattering with a spin-polarised beam. This will give the most precise determination of the weak mixing angle at low energies, thereby probing mass scales for particles beyond the Standard Model up to 50 TeV. Furthermore, a measurement of the parity-violating asymmetry using nuclear targets can probe the neutron-skin thickness of nuclei and thus permits extraction of the equation of the state of nuclear matter with impact on the understanding of neutron stars. Finally, the beam dump of the P2 experiment (see Fig. 8.19) can be used as a target to produce dark sector particles. A dedicated calorimeter will be placed behind the beam dump, providing world-leading sensitivity for light dark-matter particles. These particles carry the momentum of the accelerator and therefore produce highly energetic electron recoils inside the detector.

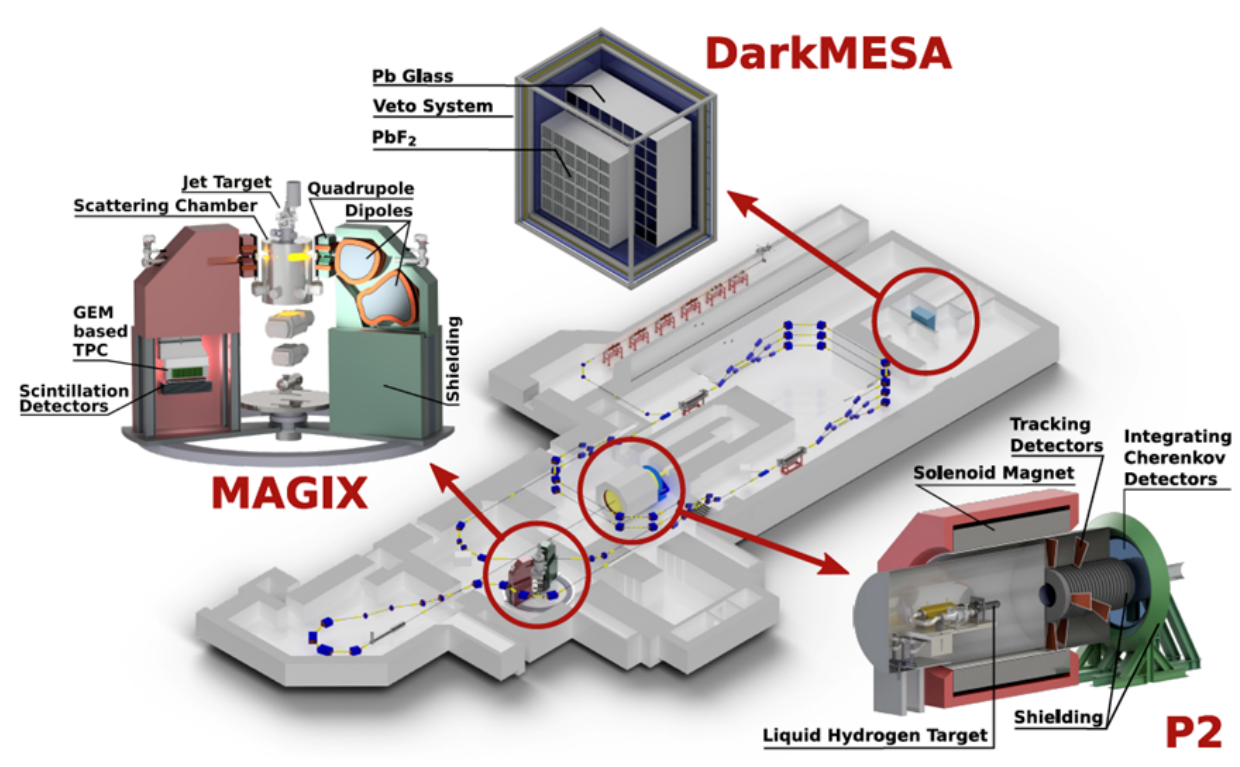

*Fig. 8.19: Floor plan of the upcoming new MESA facility with the three experiments MAGIX (energy-recovering beam mode), P2 and DarkMESA (both being operated in the extracted beam mode). (© Univ. Mainz)*

## Box 8.1: Future Energy-Recovering Linacs (ERL)

**Energy-recovery linear accelerators (ERLs) (see Fig. 8.24) are a new generation of emerging electron accelerators. They have the potential of high beam power in continuous-wave operation at moderate operational energy costs. In a conventional Linac, the beam is accelerated (green dots for electron bunches and blue arrows for the accelerating RF-field), utilised in an experiment and afterwards dumped. In a super-conducting Energy Recovering Linac (SRF-ERL) the accelerated beam enters the same accelerating cavity after the experiment, but with a 180° longitudinal phase shift, thus interacting with the decelerating RF-field (orange dots and red arrows). The beam bunches are decelerated and the beam energy is transferred back to the RF field with high efficiency. This process can be very efficient with small losses so that the operating energy cost is a small fraction of the power stored in the electron beam.**

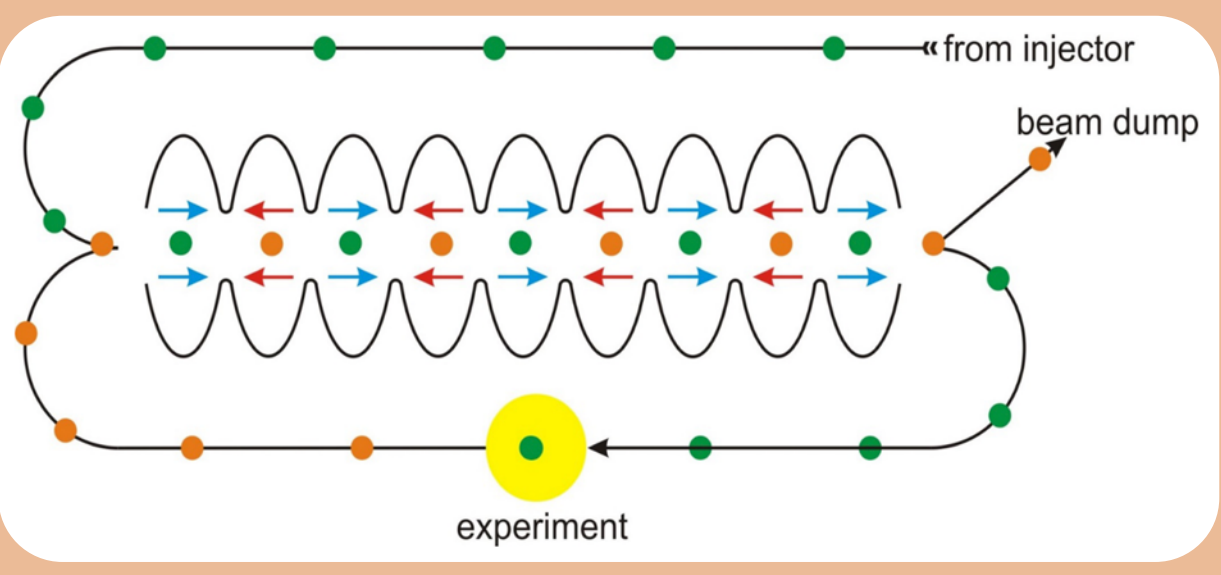

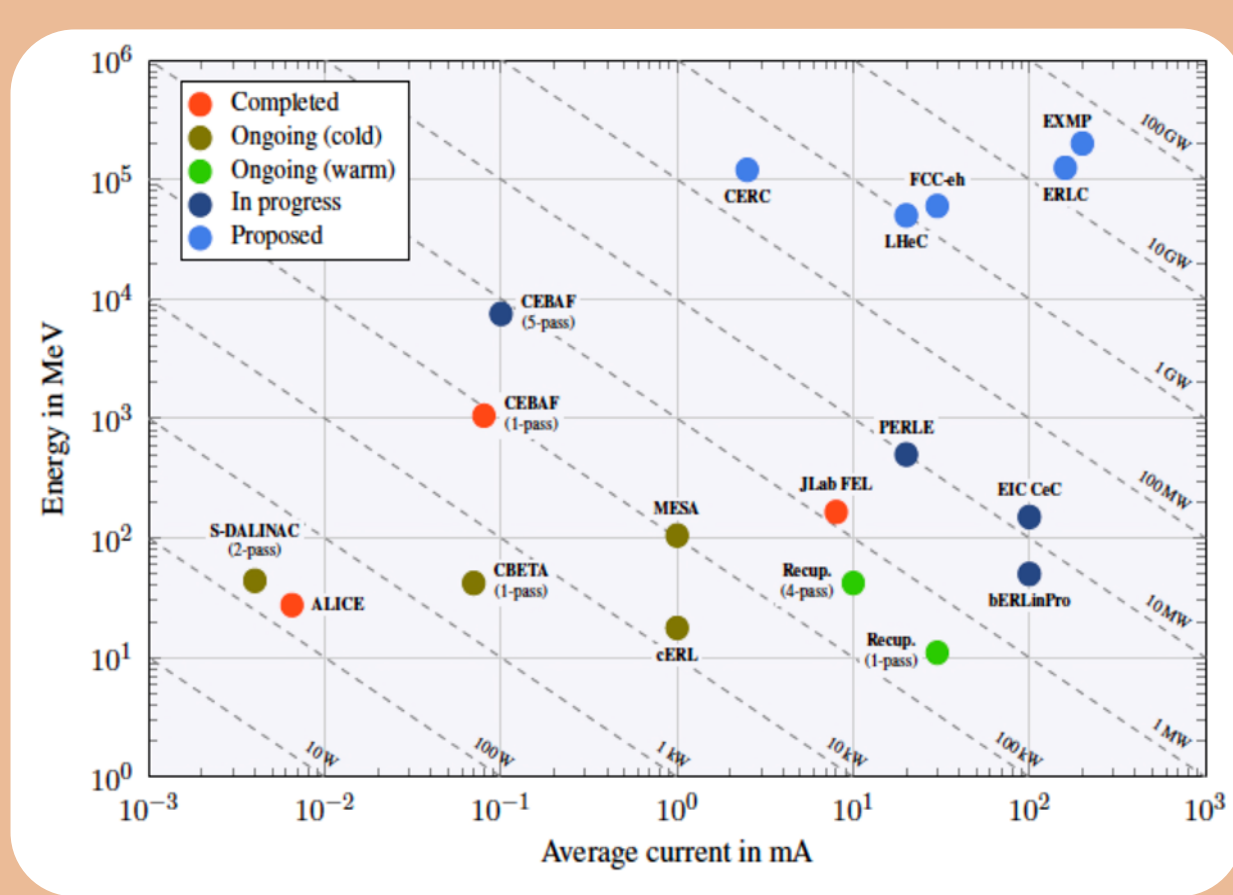

*Fig. 8.24: Principle of an Energy Recovering Linac (left, see text for details). Available and planned ERLs in the energy range up to 10 GeV.*

**ERLs can be set up as recirculating linear accelerators allowing for a multiple pass of the accelerating cavities. While single-turn SRF-ERL has been demonstrated at JLAB and Cornell, the S-DALINAC at Darmstadt has demonstrated the first performant multi-turn SRF-ERL[1]. The first usage of multi-turn SRF-ERLs for physics experiments is now in preparation at the S-DALINAC as an ERL-based light source and at MESA in Mainz including the interaction of the electron beam with a thin cluster jet target. High-current ERLs promise to advance the fields of precision measurements in nuclear and particle physics with low energy consumption. This high efficiency can be very interesting for collider projects or future light sources in condensed matter research.**

**A particularly interesting future application would be the combination of an ERL with a storage ring or an ion trap for exotic nuclei to perform electron-radioactive ion scattering experiments. Electron scattering on radioactive nuclei could provide nuclear observables with an unprecedented radial sensitivity for exotic systems. Developments in the accelerator design and electron-ion collision techniques should offer the increased luminosities required for measuring nuclear radii ($10^{26-28}$ $cm^{-2}s^{-1}$), expanding in the long-term to inelastic form factors ($10^{29-31}$ $cm^{-2}s^{-1}$).**

**[1] F. Schliessmann et al., Nature Phys. 19 (2023) 4, 597-602**

## Neutron facilities for nuclear physics

Various neutron beam facilities in Europe support a wide range of research and industrial application fields, including nuclear physics, solid-state physics, chemistry, biology, material science, cultural heritage, metrology, medical applications and others. While thermal neutrons (around 25 meV) are used for the investigation of the structure and dynamics of matter on the atomic and molecular scale, interactions of neutrons with atomic nuclei are studied in a wide range of energies (see Fig. 8.20). In the context of nuclear physics, this concerns mainly neutron-induced reactions and their reaction products respectively as well as nuclear structure aspects. Moreover, cold and ultracold neutrons are frequently used in experiments addressing properties of the neutron itself or other fundamental physics questions.

Both nuclear reactors and accelerator-induced reactions produce initially fast, typically MeV, neutrons. A moderator is required to slow down the neutrons to lower energies. Research reactors have historically been the main source of thermal and cold neutrons. In the domain of neutron time-of-flight experiments, primary electron, proton or deuteron beams with short (ns) pulse widths are used with neutron production targets. A moderator will extend the energy range, but puts a limit on the repetition rate to avoid the overlap of bunches. Compact accelerator-based neutron sources (CANS) have gained more interest in recent years, because of the easier construction and licensing compared to a reactor or spallation neutron source and can thus be implemented in a decentralised manner.

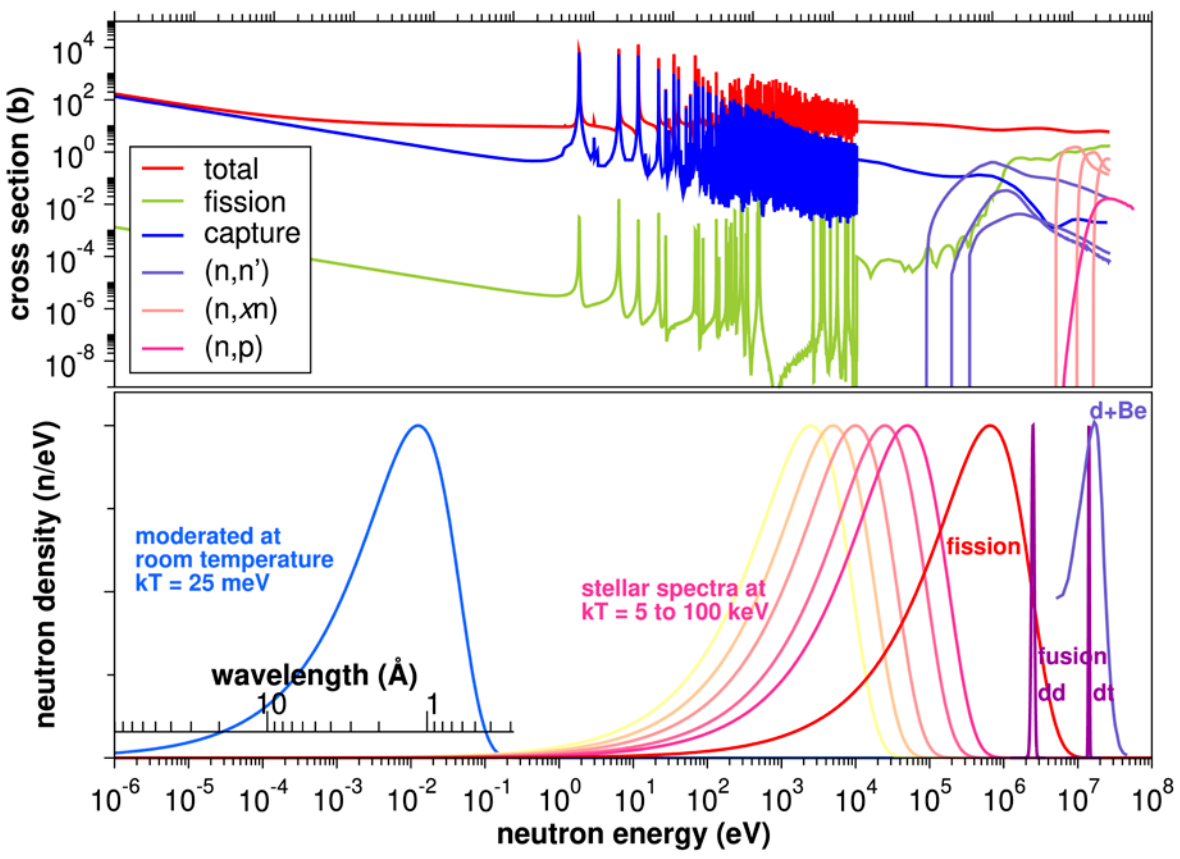

*Fig. 8.20: Neutron-induced nuclear reaction cross sections for a typical (non-fissile) heavy nucleus (upper panel), together with characteristic neutron density distributions from several reactions and in several environments (lower panel). All distributions have been normalised to their maximum value.*

Research reactors with a large fraction of beam time dedicated to nuclear physics include the high flux reactor RHF operated by ILL in Grenoble, providing worldwide the most intense beams with continuous neutron flux in the neV to 1 eV range, with a neutron flux up to $1.5x10^{15}$ neutrons $cm^{-2}$ $s^{-1}$. Several other research reactors also offer neutrons for nuclear physics experiments, like the FRM II reactor (MLZ at TU Munich), the TRIGA reactors (Mainz University and Atominstitut in Vienna), Budapest Neutron Centre (BNC) of KFKI, the LVR-15 of UJF at the Research Centre Řež, and the BR1 of the SCK-CEN in Mol.

Four operational dedicated neutron-time-of-flight facilities to measure energy-dependent neutron-induced reaction cross-sections are available in Europe, each with their own complementary set of characteristics. The pulsed white spectrum neutron source GELINA of JRC-Geel uses a linear electron accelerator with a post-accelerating beam compression system resulting in 1 ns short electron pulses. A rotating, mercury-cooled uranium target coupled to a water moderator serves 12 operating flight paths with distances from 10 to 400 m with a neutron spectrum ranging from thermal to about 20 MeV. The typical average electron beam energy is 100 MeV and repetition rates go from 50 to 800 Hz. The facility nELBE of HZDR is an unmoderated pulsed neutron source driven by the 40 MeV superconducting electron linear accelerator ELBE, producing neutrons in the 100 keV to 10 MeV range. The neutron facility NFS at GANIL in Caen produces pulsed neutron beams for time-of-flight, either continuous in the 1-40 MeV energy range or quasi mono-energetic, using primary p, d beams on a Be or C target [see section GANIL]. The neutron time-of-flight facility n_TOF at CERN uses a 7 ns short 20 GeV proton beam pulse on a lead spallation target, producing a high-intensity neutron beam in the energy range of meV to GeV [see section CERN]. The continuous spallation neutron source SINQ at PSI Villigen also has a dedicated intense ultracold neutron source, while the pulsed spallation neutron sources ISIS in Oxfordshire, and the future ESS in Lund, are at present not primarily planned for nuclear physics experiments.

Many facilities can be qualified as CANS, using light-ion accelerators to generate neutrons from proton, deuteron or helium beams incident on targets made of light elements. The neutron spectra can be chosen to be quasi-mono-energetic or with a broader distribution. The use of neutron beams is polyvalent and not exclusively related to nuclear physics. Here, we mention, for example, AIFIRA in Bordeaux, LICORNE at IJCLAB-ALTO Orsay, GENESIS with GENEPI2 in Grenoble, FNG in Frascati, PIAF at PTB Braunschweig, UJF in Řež, the new High-Flux Accelerator Driven Neutron Source at Birmingham, NPL in Middlesex, HISPANoS at CNA Sevilla, MONNET at JRC Geel, AMANDE and CARAT at Cadarache, and NCSR "Demokritos" in Athens.

Larger-scale facilities producing fast neutrons by breakup of 40 MeV deuteron beams, like NFS at GANIL, but at higher beam intensities, include the upcoming SARAF-II facility in Soreq (Israel), which will use proton and deuteron beams on a liquid gallium-indium jet to produce continuous or pulsed fast neutron beams that can also be moderated to thermal energies. An upgraded liquid lithium target (LiLiT2) is also planned. The future IFMIF-DONES accelerator in Granada [see section IFMIF-DONES] uses deuterons on a liquid lithium target, and a separate target to form a pulsed neutron beam (TOF-DONES).

Other upcoming CANS facilities include NESSA of Uppsala University, FRANZ of the Goethe University in Frankfurt, BELINA and NEPIR at the LNL Legnaro, and the ICONE project at Saclay. An epi-thermal neutron beam dedicated to BNCT is planned for construction in Caserta. New directions are being explored by the laser-based nuclear physics ELI-NP facility, which plans to exploit pulsed neutron beams from laser-induced fusion of deuterium and tritium gas mixtures.

## Smart and Salient: Small-Scale Facilities

Small Scale Facilities (SSF) in Europe constitute an ecosystem of nuclear physics-orientated accelerator facilities which collaborate intensely with the large-scale facilities detailed in this report, with each other, with universities and research centres, as well as with companies and society at large. They are the best examples of nuclear research infrastructures with high scientific impact, moderate investment, running costs and environmental footprint. While there are scientific topics, e.g. low-energy nuclear astrophysics, where SSFs are de facto instruments to solve problems, there is a very important synergy between SSFs and large-scale facilities as well, in a broad range of topics. There is also remarkable infiltration of nuclear science into other fields, e.g. cultural heritage studies, climate research and industrial applications, where SSFs provide the backbone technology.

For these reasons, it is not surprising that SSFs are well spread all over Europe and can be considered a leading facility in many countries where there is no large-scale nuclear research-related facility. Apart from the direct scientific impact, they play a key role in high-level hands-on training of the next generation of nuclear scientists and in outreach activities, extremely important for nuclear research.

Today, many SSFs are organised similarly to the large-scale facilities. They are also part of European networks and provide Trans-National Access for researchers. EU-funded networks such as CHETEC (Nuclear Astrophysics), RADIATE (Nuclear Applications) and CLEAR in EURO-LABS (synergies with large facilities), demonstrate the importance of the SSF in the European eco-system of Nuclear Physics facilities.

In this report not all Small Scale Facilities in Europe are presented. Instead, we highlight in particular those facilities which play a crucial role in national communities and have the character of national facilities or of those being primarily used for applications in their respective countries.

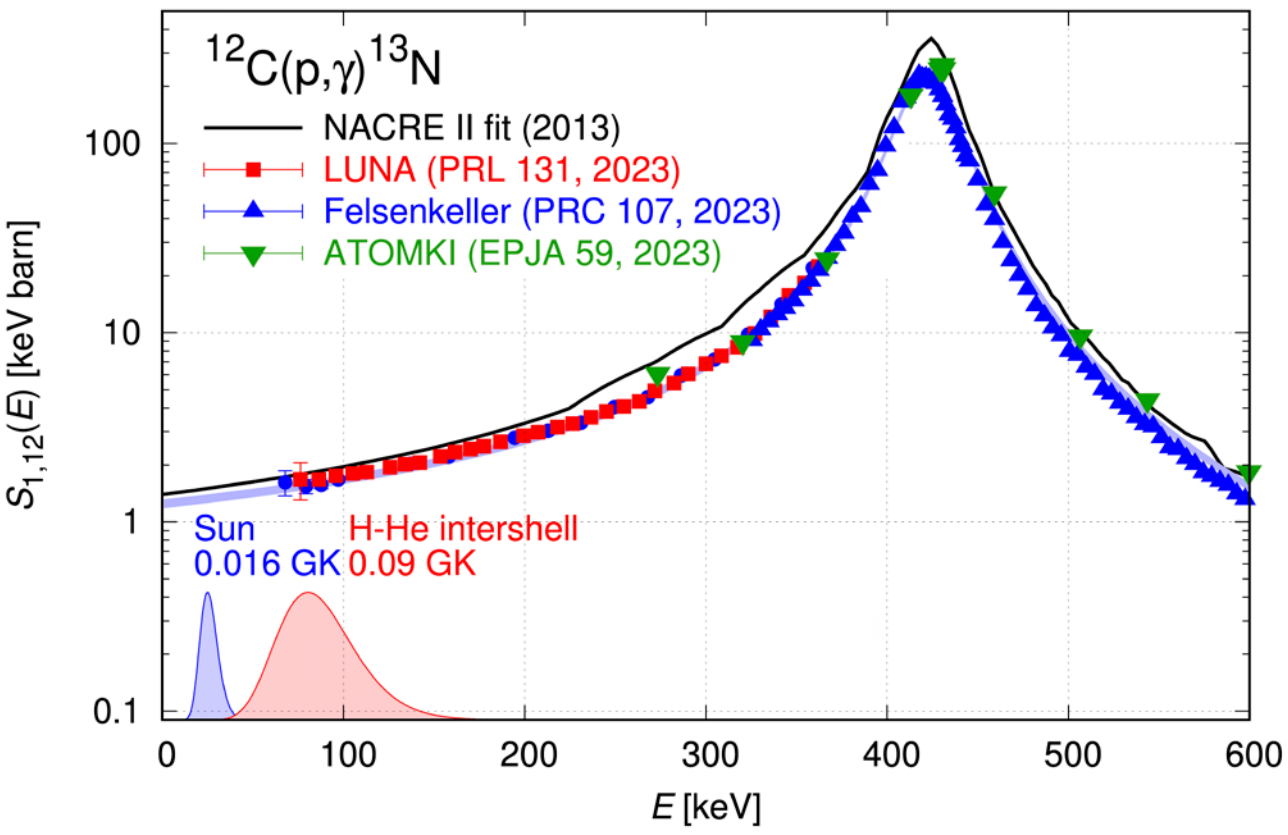

*Fig. 8.21: Comprehensive description of a nuclear reaction with astrophysical relevance at low energies using data from three small-scale facilities.*

**ATOMKI** (Debrecen, Hungary) is the Hungarian low-energy accelerator centre, which provides support both for scientific and technical projects including nuclear astrophysics, space chemistry, environment, heritage and material sciences, as well as for radiation biology and rad-hardness assurance and at the same time providing nuclear data for technological and medical applications. Main facilities include a cyclotron, a tandem accelerator, an ECR and an Accelerator Mass Spectrometer (AMS).

The **Centre of Accelerators and Nuclear Analytical Methods** is a research infrastructure belonging to the Nuclear Physics Institute (NPI) of the Czech Academy of Sciences (CAS) located in Husinec-Řež. It operates a Tandetron linear accelerator, two cyclotrons and a new AMS instrument, as well as a microtron external to the centre. These have multiple scientific applications.

The research highlights of the Tandem Accelerator Laboratory of **NCSR "Demokritos"** (Athens, Greece) include nuclear astrophysics related to stellar nucleosynthesis, studies of neutron-induced reactions and measurements of differential cross sections of reactions induced in light elements by ion beams of protons and deuterons.

**CNA** (Seville) and **CMAM** (Madrid) constitute the Spanish paired facility for accelerator-based research. The key research capabilities of CNA include the pulsed neutron beam line for astrophysics and neutron imaging and irradiation, and the AMS for precise measurements of Transuranides in the environment. Both CMAM and CNA have beam microprobes for detector studies and for proton therapy applications.

**The Jožef Stefan Institute** (JSI) in Slovenia operates a tandem accelerator supporting nuclear fusion research with broad and focused 3He beams, low-energy reaction studies for nuclear astrophysics, detector prototyping for FAIR and a broad spectrum of nuclear applications including medicine, biology and archaeometry.

**The Laboratory of Accelerators and Radiation Technologies** at IST of the University of Lisbon (Portugal) hosts two electrostatic accelerators, an ion implanter and a nuclear microprobe, among other techniques. They address topics such as materials science, energy and biomedical sciences as well as experimental nuclear physics in support of large-scale facilities.

The **Oslo Cyclotron Laboratory** (Norway) operates a K=35 cyclotron for light ions and the large $LaBr_3$(Ce) detector array OSCAR. Experiments focus on spectroscopy for nuclear structure and nuclear astrophysics, pioneering the extraction of nuclear level densities and gamma-ray strength functions. It also carries out research in radiation biology and R&D on the production of radionuclides for medical applications.

**RBI-Zagreb** (Croatia) carries out research on astrophysics and materials science. It has a unique dual-beam facility which will enable the development of new materials for fusion power plants as well as all other nuclear applications, such as hadron therapy.

**The Soreq Applied Research Accelerator Facility** (SARAF), under construction in Israel, is based on a proton/deuteron, 40 MeV, 5 mA accelerator. In combination with unique liquid metal irradiation targets it will be a high-flux source of continuous and pulsed neutrons from thermal to high energy, with radioactive nuclei from various areas of the nuclear chart. It will be used for basic and applied research in various fields of science and engineering.

At the **VERA** facility at the University of Vienna (Austria), special emphasis is placed on AMS applications within astrophysics, nuclear physics, atomic and molecular physics. The applications reach into many areas of our environment, from archaeology to climate research.

In France, in addition to GANIL/SPIRAL2 and ALTO, several small-scale facilities are available for interdisciplinary research through the network EMIR&A. **AIFIRA** in Bordeaux uses ion beams for irradiation of materials or ion-beam analysis, **CYRCe** in Strasbourg and **ARRONAX** in Nantes are dedicated to the production of radioisotopes and other nuclear applications for health. **GENESIS** in Grenoble is a neutron generator for detector developments and industrial applications, and **LSM** (Laboratoire Souterrain de Modane) is a platform for experiments related to rare events requiring very low background.

In Germany, **HZDR Dresden** is a research centre providing several installations: the Ion Beam Centre (IBC) is a key European facility for the application of ion beams in materials research. Plasma and ion sources generate ions of all species at energies between eV and 60 MeV. ELBE is a 40 MeV superconducting electron Linac producing secondary beams including neutrons and bremsstrahlung. The 5 MV Felsenkeller accelerator, located 45 m underground, provides ion beams for nuclear astrophysics. The 1 MV HAMSTER tandem accelerator is used for AMS and provides selective laser ionisation.

The Institute for Nuclear Physics at **TU Darmstadt** operates the superconducting thrice-recirculating electron linear accelerator **S-DALINAC**, providing a high-flux bremsstrahlung site, a photon tagger, and magnetic spectrometers with high momentum resolution or large acceptance that serve research and development in nuclear spectroscopy, nuclear photonics and accelerator science. In 2021, the S-DALINAC demonstrated world-wide the first energy recycling in a multi-turn energy-recovery LINAC.

The Institute for Nuclear Physics at the **University of Cologne** operates a 6 MV Tandetron for AMS, capable of measuring very small isotope ratios of down to $10^{-16}$, and a 10 MV FN Tandem for nuclear structure and astrophysics.

In Italy, several small-scale facilities are operated for interdisciplinary research based on accelerated ion beam techniques. The laboratory **CIRCE** in Caserta, with its 3MV Tandem-Pelletron accelerating H to U ions, has a programme focused on cultural heritage and on the study of environmental processes. Research and applications in these areas are also carried out at the 3 MV Tandetron of the **LABEC** laboratory (Florence) where Ion Beam Analysis and C-14 measurements with Accelerator Mass Spectrometry are made.

In the UK, the **Dalton Nuclear Facility** in Manchester houses a 5-MV tandem and a 2.5-MV Pelletron configured to provide a range of ion irradiation and analysis capabilities for radiation damage or radiation chemistry studies in nuclear energy, radiobiology, space missions and other applications. The University of **Birmingham Accelerator Facilities** operate a MC40 cyclotron for a wide range of commercial and research work. The new High Flux Accelerator-Driven Neutron Facility uses a >30mA 2.6 MeV proton beam to generate neutrons for materials science, nuclear data, medical physics and nuclear physics research.

# Future European Research Infrastructures

## MYRRHA

The future MYRRHA project (Multi-purpose hYbrid Research Reactor for High-tech Applications https://myrrha.be) at SCK CEN in Belgium has the ambitious goal of developing a liquid Lead-Bismuth Eutectic (LBE) spallation target and an LBE-cooled, sub-critical fast reactor core driven by a high-intensity proton beam. The maximum proton beam current expected is 4 mA at 600 MeV. MYRRHA is an innovative research installation, recognised as an ESFRI project, which will take nuclear waste treatment a significant step further. Using advanced partitioning processes, it will be feasible to isolate minor actinides as the most heat-emitting, long-lived elements from the spent nuclear fuel. One of the many benefits of an Accelerator Driven System with an LBE-cooled fast reactor is that it is capable of breaking down minor actinides in the transmutation process. MYRRHA will be built in three phases, starting with the construction of a 100-MeV linear accelerator, the proton-target facility hosting an ISOL facility and the full-power facility dedicated to material irradiation for fusion research. Phase 2 is focused on the construction of the 600-MeV linear accelerator, which will be coupled to the subcritical nuclear reactor in phase 3.

The ISOL facility at MYRRHA is being constructed in Phase 1 and scheduled for commissioning in 2027. The accelerator will deliver a high-intensity 100-MeV proton beam, up to 0.5 mA at 250 Hz. A wide variety of target materials from light (e.g. SiC) to heavy compounds (e.g. $ThC_x$), will be used for producing the radioactive ion beams. These have been developed over the years at established ISOL facilities, and together with the ion source, they will be further optimised to achieve the best performance under high-intensity irradiation. The delivery of a purified ion beam is envisioned by using a pre-separator followed by high-resolution mass separators (potentially, in combination with a gas-filled linear Paul trap). ISOL@MYRRHA will house one target station and an experimental area, as well as user laboratories (see Fig. 8.22).
ISOL@MYRRHA will provide a unique opportunity to perform measurements for extended periods of time, considerably longer than in other similar operational laboratories. Several user communities from different fields of nuclear, fundamental and solid-state physics have been identified with a growing demand for extended access to radioactive beams. For example, pushing the precision frontiers of nuclear structure research will greatly benefit from the long uninterrupted measurements by achieving high statistical significance and at the same time systematically investigating all sources of uncertainties. Moreover, various medical radioisotopes will be produced at ISOL@MYRRHA and exploited in research studies with radiopharmaceuticals or for searches for innovative medical radioisotopes.

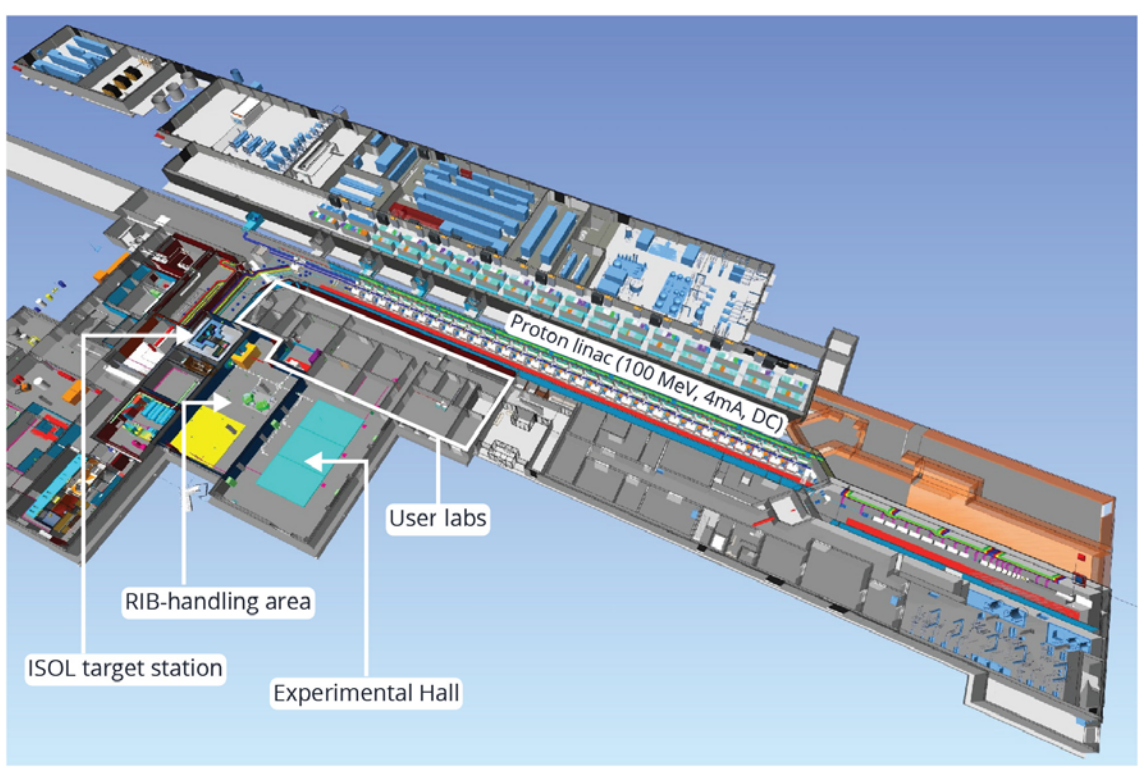

*Fig. 8.22: Layout of the MYRHHA phase1 facility, indicating the proton LINAC and the areas linked to the ISOL beams. (© MYRRHA)*

## IFMIF-DONES

The International Fusion Materials Irradiation Facility – DEMO Orientated Neutron Source (IFMIF-DONES, https://ifmif-dones.es) will be an accelerator-based neutron source that will produce a high neutron flux reproducing the conditions of a fusion reactor. IFMIF-DONES is recognised as an ESFRI project and will be built primarily for the irradiation and qualification of materials for the European fusion programme. DONES will utilise the Li(d,xn) nuclear reactions taking place in a liquid Li curtain bombarded by a 40 MeV deuteron beam. Apart from the infrastructure dedicated to the irradiation and qualification of materials, IFMIF-DONES will host several complementary experimental facilities. A versatile collimated neutron beam experimental area will make possible neutron scattering and neutron-induced reaction experiments, as well as biological and industrial applications of neutrons. A dedicated deuteron beamline using a fraction of the initial deuteron beam impinging on a secondary target is planned for a neutron Time-of-Flight facility (TOF-DONES). The production of radioisotopes for medical applications in neutron or deuteron-induced reactions is also being considered.

The extremely high intensity of the IFMIF-DONES deuteron accelerator will be used for driving one of the most intense neutron time-of-flight (TOF) facilities worldwide. A preliminary study of the performance of TOF-DONES has been carried out. It has considered the neutron flux, neutron energy distribution and energy resolution for various beamlines at different angles. A 175 kHz deuteron pulse repetition rate (i.e. 1/1000 beam extraction) with 5.3 ns pulse widths impinging on a graphite neutron production target was adopted as a realistic reference scenario for the calculations of the performance of the facility. The use of neutron moderators and degraders for extending the neutron beams to lower energies, from 10 eV to 40 MeV, was also investigated. The resulting neutron fluxes in the 10 keV to 30 MeV range are comparable with or larger than the currently most intense neutron TOF facilities in Europe: n_TOF Experimental Area 2 (EAR2) at CERN and Neutrons For Science (NFS) at SPIRAL-2. The TOF-DONES can cover a large variety of cross-section measurements and nuclear structure experiments required for nuclear technologies (fission and fusion), astrophysics and particle and astrophysics. Another experimental area at IFMIF-DONES will make direct use of the high continuous neutron flux produced for the fusion material sample irradiation. The neutrons, after passing through a materials irradiation module, have an energy spectrum ranging from 1 to 50 MeV. Such beams are interesting for isotope production and experiments where secondary particles have to be produced, e.g. for the production of unstable isotopes by (n,p) or $(n,\gamma)$ or even (n,f) reactions. These reactions are interesting in the context of studies on radioactive isotopes such as, for instance, the spectroscopy of fission fragments. The possibility of adding thermal or cold moderators is also being considered to allow for experiments with neutrons using scattering instruments or characterisation of samples with neutron activation techniques or neutron imagery. Production of radioisotopes for medical applications is also being planned.

## European Spallation Source (ESS)

The European Spallation Source (ESS) is currently under construction in Lund, Sweden (https://europeanspallationsource.se). It is organised as a European Research Infrastructure Consortium (ERIC) and currently has 13 member states. The ESS facility will be the world's first long-pulse neutron spallation source. The neutron pulse has a length of 2.86 ms and repeats at a rate of 14 Hz. The design target is to achieve an average power of 5 MW with a peak power of 125 MW. However, ESS has recently undergone a project re-baselining process, and it is now committed to delivering 2 MW of power by 2028, with the 5 MW capacity considered an upgrade[8].

There are currently 15 condensed-matter instruments under construction at ESS, which represent a subset of the full 22-instrument suite required for the facility to fully achieve its scientific objectives as defined in the ESS statutes. Furthermore, ESS has a mandate to include a fundamental physics programme, and the absence of a dedicated beamline for fundamental physics has been identified as a significant capability gap[9].

Two distinct beamlines have been proposed at ESS, the HIBEAM and the ANNI beamline. The HIBEAM beamline is optimised for conducting experiments related to neutron-antineutron oscillations and searches for sterile neutrons. It should be operational at the start of the ESS and also offers flexibility for other scientific investigations like searches for hadronic parity violation and electromagnetic properties of the neutron. The future ANNI beamline is specifically tailored for the study of neutron decay, investigations into hadronic parity violation, and the precise measurement of various neutron electromagnetic properties. Both beamlines would operate in modes similar to those of existing neutron facilities.

Other important tools for fundamental physics studies at neutron facilities are Ultra Cold Neutrons (UCNs). As part of the HORIZON 2020 INFRADEV project called HighNESS for the design of the ESS upgrade, several different possible UCN sources have been proposed. These concepts promise to deliver world-leading performance in this field.

Finally, one of the significant final deliverables of the HighNESS project is the development of the Conceptual Design for the neutron-to-antineutron search experiment, NNBAR. The NNBAR experiment aims to investigate neutron-to-antineutron oscillations, with the goal of achieving sensitivities greater than three orders of magnitude. This remarkable level of sensitivity has only been possible at the ESS due to the installation of a unique beam port that is three times larger than standard beam ports and is known as the "Large Beam Port." No other planned infrastructure matches this, making ESS the ideal location for conducting these searches.

# Theory and Computing Infrastructures

## ECT*

The European Centre for Theoretical Studies in Nuclear Physics and Related Areas (ECT*) is a globally recognised centre for research in theoretical nuclear physics and related areas in the broadest sense of the word. Since its inception in 1993, the international visibility of ECT* is largely driven by its annual programme of workshops and training schools attracting participants from around the world, as well as by a small but strong local research group.

In the next five years, ECT* should maintain and enhance its leadership role in the international community and consolidate its long-term financial position. This hinges on three pillars: the continuation of the delivery of its annual workshop and training programme; the development of its leadership role at the European level, in which ECT* should act as a natural place where ideas are developed; and extending its mission strategically, especially in related areas as an opportunity to increase the Centre's budget and reach.

The annual workshop and training programme is ECT*'s most visible activity. With about 20-24 workshops, an extended doctoral training programme and around 800 visitors per year, ECT* serves the nuclear physics community across all areas of research. As is currently the case, the programme should continue to support both "core" nuclear physics topics as well as developments in related areas, which may be relevant for research in nuclear physics or for which nuclear physicists can contribute interdisciplinary skills and expertise. For the coming years, it is expected that machine learning and quantum information will play an ever more important role in research in the physical sciences; ECT* should play a leading role in stimulating the community to engage with these developments. When large funding calls are made public, ECT* should lead on behalf of the theoretical nuclear physics community to ensure a coherent contribution is made at the European level. In parallel, initiatives at the local Trento level should be explored to firmly ground the Centre in the local research environment with the aim to secure additional and sustained funding via this route. The local research group is expected to engage with this, as a means to enhance their position.

[8] See https://europeanspallationsource.se/article/2021/12/10/ess-revises-project-plan-and-budget
[9] See https://europeanspallationsource.se/instruments/capability-gap-analysis

One of the aspects to be expanded upon is ECT*'s role as a strategic think tank, hosting small groups of researchers to stimulate new initiatives and act as a natural place where ideas are developed, from inception to grant submission and capture. Grant capture beyond the current mechanisms should play a more important role to alleviate present financial constraints and enable room for new initiatives.

## Computing Infrastructures

The forthcoming decade in frontier research will witness an unparalleled demand for data processing and analysis, where a substantial rise in data volume to be collected, shared, and processed is expected from nuclear and high-energy as well as from radio-astronomy experiments. Computing infrastructures in Europe will need to significantly enhance their performance and capacity to address the data storage, processing and analysis requirements. Several projects funded by the European Union, including EOSC projects like ESCAPE and OSCARS among others, are dedicated to achieving the adoption of novel data architectures, federation models, IT solutions such as Data Lake infrastructures and (supra)national Cloud infrastructures for research, to optimise the use of extensive computing and storage resources on a European scale. These initiatives aim to tackle the challenges posed by Open Science, particularly in collaboration with ESFRI facilities such as SKA, CTA, KM3Net, EST, ELT, HL-LHC, FAIR, as well as other pan-European research infrastructures like CERN, ESO, and JIVE. Moreover, the advent of HPC exa-scale infrastructure and QC offers European researchers an unprecedented processing capacity. This complements and integrates the capabilities provided by research High-Throughput Computing (HTC) and Cloud facilities at both national and European levels. However, exploiting HPC centres comes with several challenges, such as the diversity in access and usage policies and the heterogeneous and different computing architectures. To address these challenges and create a cohesive data processing system, the SPECTRUM project, funded by the EU, aims to integrate different European computing resources. This includes on-premises data centres, HPC clusters and Quantum nodes. The ultimate long-term objective is to establish a European exabyte-scale research data federation and a seamless computer continuum.

## European Integrating Activities

The European Commission is funding several Integrating Activities for the Nuclear Sciences to promote cooperation between the various research infrastructures. In particular, they grant research groups from less favoured countries access to all facilities, thus offering unique training opportunities for PhD students and young researchers. In the past ten years, important steps have been taken to foster the coherence of their multi-faceted research programmes, and transnational user access has been incorporated for many small-scale facilities.

**EURO-LABS** (EUROpean Laboratories for Accelerator Based Sciences) is a Horizon Europe project bringing together for the first time three research communities of nuclear physics, accelerator and detector technologies for high-energy physics in a pioneering super-community of sub-atomic scientists. It provides effective access to a network of 47 Research Infrastructures (including 3 RIs with Virtual Access) to conduct curiosity-based research, addressing fundamental questions and technological challenges and advancing projects with broad societal impact, fostering knowledge-sharing between scientific fields and enhancing Europe's potential for successfully facing future challenges. It also opens the way for a synergetic implementation of best practices for data management and activities relating to targeted service improvements at these RIs. Joint training activities are planned to develop the skills of the next generation of researchers to optimally use the RIs services for scientific and technological discoveries.

In the framework of the **STRONG-2020** project (H2020-INFRAIA), the hadron physics community in Europe leads a coherent effort from theoretical and experimental groups, complemented by challenging high-technology developments in instrumentation and industrial applications. It benefits from a number of selected leading European Research Infrastructures of excellent quality, ranging from test facilities for instrumentation to fully-fledged experimental facilities providing top-quality beam delivery and support to hadron physicists. They complement each other by the energy range and physics probes they provide. They operate accelerator facilities and experiments which are frequented by a large international community. The Transnational Access provides not only access to the infrastructures, but, in addition a platform for the experimental collaborations to meet and develop new ideas for future facilities like the international FAIR accelerator complex, the Electron Ion Collider in the USA or the low energy MESA accelerator. The Consortium, involving 2000 scientists from 16 European countries, plans to build a new infrastructure initiative benefitting from the Horizon Europe framework. The objective is to support transnational access to major world-class European facilities: two ESFRI landmarks (FAIR and CERN), emerging infrastructures (MAMI and MESA from the excellence cluster in Germany) and LNF Frascati and ECT* in Italy.

The H2020-INFRAIA project **ChETEC-INFRA** offers EU-supported access to three critical types of infrastructure that, together, provide the capabilities needed to study the origin of the chemical elements from the Big Bang to stellar burning and to neutron star mergers: astronuclear laboratories supply reaction data, supercomputer facilities perform stellar structure and nucleosynthesis computations, and telescopes and mass spectrometers collect elemental and isotopic abundance data. To address key questions on solar fusion, neutron capture nucleosynthesis and explosive stellar processes, ChETEC-INFRA develops improved nuclear reaction targets and detectors, open-source nucleosynthesis software tools and three-dimensional model atmospheres for stellar spectral analysis based on up-to-date physics. Data are archived and catalogued for long-term sustainability beyond the end of the project, ranging from evaluated nuclear reaction rates to detailed abundance data for a multitude of stars to tracer nucleosynthesis calculations. ChETEC-INFRA reaches out to PhD students, secondary school students and industry partners. The long-term aim is to grow, after completion of the four-year phase of a Starting Community of research infrastructures, from the present seed of a limited subset of nuclear astrophysics infrastructures to an Advanced Community by adding a few additional, excellent infrastructures and users.

**PRISMAP**, the European Medical Radionuclides programme, is a H2020-INFRAIA project delivering non-conventional radionuclides for medical research and is funded by the EC. MEDICIS at CERN, the mass separator facility for medical radionuclides, acts as the coordinator and core facility of this project. MEDICIS and additional production facilities at ILL, SCK-CEN, PSI, NCBJ, ARRONAX, DTU Risø and JRC Karlsruhe provide together a wide selection of non-conventional radionuclides for user projects. Non-conventional radionuclides are either "new" in their medical application or made available in new purity grades, e.g. through mass separation. PRISMAP provides either Remote Access, i.e. the radionuclides are shipped to the users' labs or via Transnational Access, i.e. users travel to PRISMAP biomedical facilities to perform their experiments. By supporting medical applications, in particular for targeted treatments of cancer, PRISMAP is implementing one of the key societal applications of nuclear physics. Thus, PRISMAP also acts as an important bridge between the nuclear physics and medical communities. Where required, PRISMAP generates and provides improved nuclear data for medical radionuclides and participates in harmonising the regulatory process to bring promising radiopharmaceuticals more quickly into clinical practice.

In the past decades, a succession of research and innovation funding programmes initiated by the European Commission has shaped the landscape of neutron beam facilities for nuclear physics by facilitating transnational access and institutional collaborations related to nuclear data initiatives. The recent ARIEL project, for example, has a rather comprehensive view of neutron beam facilities available for nuclear physics. Transnational access to research infrastructures delivering neutron beams is also part of the EURO-LABS programme.

# Research Infrastructures outside Europe

Beyond the European research infrastructures, the community is agile in using complementary and unique state-of-the-art facilities outside Europe. Several collaborations have invested in experiments and detectors sited at non-European facilities, drawing upon the technical skills and expertise of European scientists. Similarly, detector systems developed by European collaborations have been deployed at non-European facilities for dedicated campaigns, accessing beams that are complementary to those available in Europe. There is also the intellectual investment provided by European scientists as users in many different individual experiments at non-European facilities. All these activities that access infrastructures beyond Europe diversify and enhance nuclear physics, but also seed new initiatives and bring ideas, skills and knowledge to European facilities.

## Electron-Ion Collider (EIC) at BNL

The **Electron-Ion Collider (EIC)** (https://www.bnl.gov/eic) is a powerful and versatile new accelerator facility to be built in the United States at Brookhaven National Laboratory (BNL). It will be capable of colliding highly polarised electrons at high intensity with high-energy beams ranging from heavy ions (up to uranium) to polarised light ions and protons, at a variable centre-of-mass energy between 29 and 140 GeV and at a peak luminosity of around $10^{34}$ $cm^{-2}$ $sec^{-1}$, with a possible second interaction region.

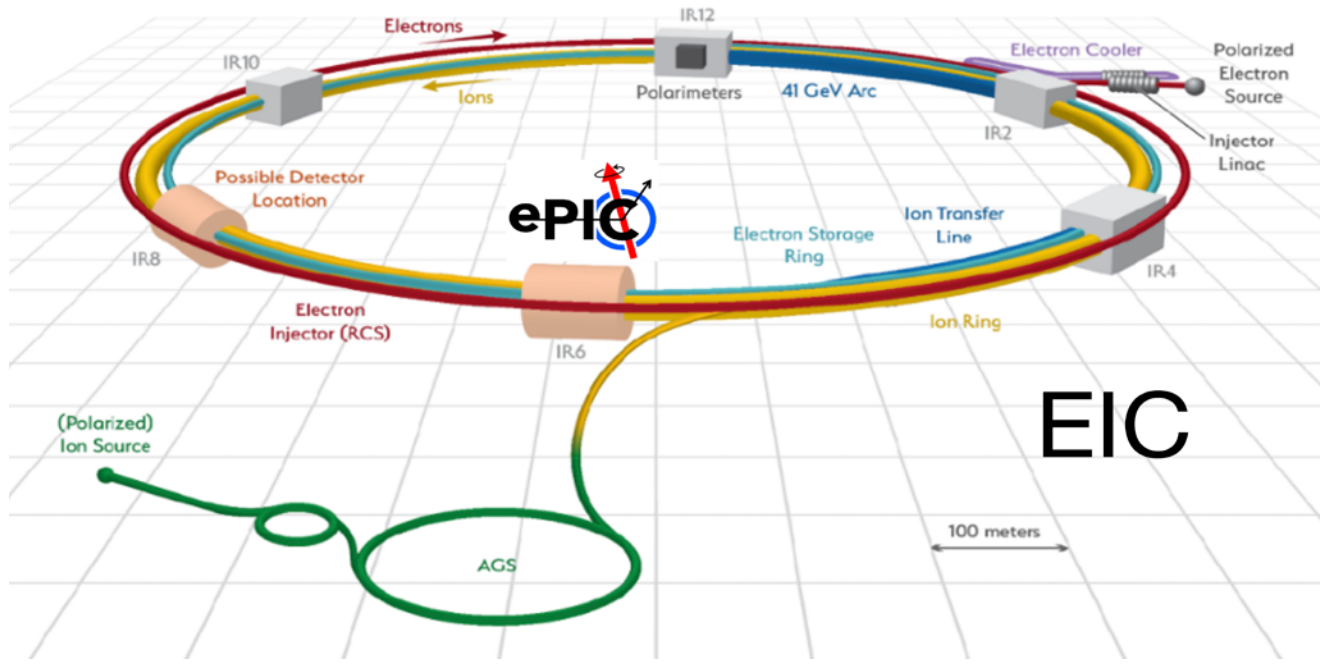

*Fig. 8.23: The EIC collider: red ring for 5-18 GeV electrons running clockwise, yellow ring for 41-275 GeV ions running anti-clockwise. IR6 indicates the location of the project detector ePIC (Courtesy of Brookhaven National Laboratory, USA).*

The EIC is designed to be a discovery machine addressing fundamental questions about how the most elementary building blocks (quarks and gluons) interact to build the structure and properties of all visible matter. The EIC will enable the exploration of new landscapes in QCD, permitting the construction of high-resolution multidimensional maps of the quark and gluon dynamics inside hadrons. To build such maps, different processes need to be analysed including inclusive and semi-inclusive deep-inelastic electron scattering and fully exclusive processes like deeply virtual Compton scattering. The EIC can analyse all these processes thanks to a project detector (ePIC) designed to provide hermiticity, excellent identification and complete kinematical coverage for particles emitted in the central, far-backward and far-forward directions (see Fig. 8.23). Its ability to collide a large variety of ion species will also permit the EIC to shed light on many open questions in heavy-ion physics (shadowing, EMC effect, initial stages of collisions, evidence of gluon saturation at high density). The high luminosity and cleaner environment (than in hadron colliders) will enable synergies between the EIC and other research fields like astrophysics (neutron stars, neutrino-nucleus interactions), electroweak physics (measuring $\sin^2 \theta_W$ at intermediate energy scales) and beyond-Standard-Model studies (constraining parameters of effective field theories). Following endorsement by the US Long Range Plan (2015 confirmed in 2023) and the National Academies of Sciences, Engineering and Medicine (2018), the EIC achieved U.S. Department of Energy CD1 status, which started the project execution phase. The science community released the EIC Yellow Report summarising the science case and detector concept before forming the ePIC collaboration around the project detector. The EIC is now in the R&D and design phase, and on track to start construction in 2025 and operation in the early 2030s.

The unique opportunities provided by the EIC facility have excited interest in scientists from all over the world, particularly in Europe, since 30% of ~180 institutions participating in ePIC are European. The current Deputy Spokesperson, an Analysis Coordinator and several working group conveners are European. The EIC Users Group also has a similar large European component (30% of ~280 institutions) and 2 of 8 members of the governing board, including the current Chair, are European, as is the co-Chair of the Resource Review Board. Europe is thus providing a substantial contribution to the development of the EIC science case and of the related detector technologies, thanks to its recognised leadership in many fields. Several European groups collaborate with U.S. groups in R&D projects. These strong contributions will put European physicists at the forefront in certain programmes in the coming years and enrich related programmes in Europe.

## Facility for Radioactive Isotope Beams (FRIB)

**FRIB** (https://frib.msu.edu), located on the campus of Michigan State University (MSU), is the newest scientific user facility for the Department of Energy, Office of Science (DOE-SC), with more than 1800 registered users. FRIB is open to users from around the world with machine time recommended based on scientific merit by an international Program Advisory Committee. The FRIB project was completed in 2022, with the first science results already published. FRIB enables scientists to make discoveries about the properties of rare isotopes, nuclear astrophysics, fundamental symmetries and applications for society. Over the coming years, FRIB is ramping up to the design beam power of 400 kW. Ions of any stable element can be accelerated in FRIB's superconducting radiofrequency linear accelerator to at least 200 MeV/nucleon. The rare isotopes produced at 50% of the speed of light via projectile fragmentation or fission are guided to experimental areas where the short-lived nuclei can be either used directly as fast beams for reactions or for implantation-decay experiments, or they can be slowed down in a gas cell and used in precision experiments after extraction or made into reaccelerated beams of pristine quality and energies ranging from hundreds of keV to well above the Coulomb barrier.

**International context** – Over the first year of operations, FRIB has supported the research of users from more than 50 countries. In July of 2023, MSU and the French CNRS established the International Research Laboratory on Nuclear Physics and Astrophysics (IRL NPA), located at FRIB, to jointly answer fundamental nuclear physics and astrophysics research questions. The DOE-funded FRIB Theory Alliance established the Europe-U.S. Theory Institute for Physics with Exotic Nuclei (EUSTIPEN), an initiative that provides travel grants and subsistence grants to ECT* for US-based scientists interested in collaborating with European colleagues. The International Research Network for Nuclear Astrophysics (IReNA), supported by the US National Science Foundation and headquartered at MSU, brings together nuclear physicists, astronomers and computational scientists from four continents, including Europe. Furthermore, research groups from the UK pursue instrument developments for FRIB-based research programmes. FRIB is member of EURO-LABS, and holds collaboration MoUs with European facilities such as LNL and CERN.

**FRIB400** – The discovery potential of FRIB can be further expanded with an energy upgrade (FRIB400) of the FRIB linear accelerator to 400 MeV/nucleon for uranium and to higher energies for lighter ions. FRIB400 will double FRIB's reach along the neutron dripline, generally bringing more neutron-rich nuclei into reach for nuclear structure, nucleosynthesis and neutron-star studies, and compressing asymmetric nuclear matter to densities required for experiments relevant to multi-messenger astronomy. In anticipation of this science potential, space was provided in the FRIB tunnel for an upgrade of the accelerator to 400 MeV/nucleon for uranium. The required accelerator technology has been proven. Once funded, the upgrade can be carried out in a staged approach with no major interruption of the FRIB science programme.

## Radioactive Isotope Beam Factory (RIBF) at RIKEN

**RIBF (**https://www.riken.jp/en/collab/resources/ribf/) at RIKEN Nishina Centre in Japan, delivers the world's highest intensities for rare isotope beams at intermediate energies between 100 and 300 MeV/nucleon. The European community has been heavily engaged in the RIBF's experimental programme from its beginning, forming various individual collaborations that have brought crucial equipment to RIKEN. Temporary deployment of equipment destined for FAIR (e.g. FATIMA, NeuLAND, DTAS, AIDA, BELEN) has offered European scientists early exploitation of NUSTAR@FAIR equipment in exotic beam experiments. Also, novel instrumentation has been specifically designed for the RIBF, offering unprecedented physics opportunities. A prominent example is the liquid hydrogen target system MINOS, which was used to establish the existence of a correlated free four-neutron system, the doubly magic isotope $^{78}$Ni, as well as the ground-state energy of $^{28}$O beyond the neutron-dripline. Together with an upgraded ion source and BigRIPS fragment separator, the available beam intensities for heavy beams will increase by a factor of 20 from 2031.

Active European commitment manifests itself in the design and delivery of dedicated detector systems, such as the silicon tracker system STRASSE, the neutron detector NEBULA+, and the novel scintillation array for in-beam γ-ray, invariant- and missing-mass spectroscopy. Involvement at the RIBF facility has enabled the European community to lead cutting-edge programmes at a second-generation fragmentation facility, training students and postdocs and preparing the community for the FAIR facility. RIBF will remain largely complementary to FAIR, providing mainly lighter nuclei at lower beam energies, but often with higher secondary beam intensities.

## Thomas Jefferson National Accelerator Facility (TJNAF)

**CEBAF**, the Continuous Electron Beam Accelerator Facility (https://www.jlab.org/accelerator) at Thomas Jefferson National Accelerator Facility in Newport News, U.S.A. delivers the world's highest intensity and highest precision multi-GeV electron beams. In 2017, a 12-GeV upgrade was completed and its science programme began. CEBAF operates 11-GeV electron beams simultaneously in 3 experimental halls and a 12-GeV beam in a dedicated hall for research on photoproduction of exotic mesons.

The CEBAF community is looking at future upgrades of the current facility, including the addition of polarised positron beams. A novel positron injector is being designed based on the production of polarised e+/e- pairs from the circularly polarised bremsstrahlung from a low-energy, highly polarised electron beam. A high-duty-cycle polarised positron beam will enable a unique science programme at the luminosity and precision frontier, mapping two-photon exchange effects as well as essential measurements of the 3D structure of hadrons. It will also offer new opportunities to investigate electroweak physics and physics beyond the Standard Model. A second upgrade possibility is to increase the electron energy to 22 GeV. This energy upgrade is based on the multi-pass acceleration of electrons in a single Fixed Field Alternating Gradient beam line. The potential of higher energies opens a rich and unique experimental programme, extending the life of the facility well into the 2030s and beyond.

The European community has played a major role in the Jefferson Lab's scientific programme over many years, contributing to the design, construction and operation of major experimental equipment and detectors. Several of the flagship physics programmes, such as hadron spectroscopy or the structure of the nucleon and nuclei, have strong European leadership, both from theory and experiment. The European community is strongly committed to continuing its involvement at CEBAF by leading new experiments and delivering the detector systems that they require.

## ISAC-ARIEL at TRIUMF

Canada's particle accelerator centre, **TRIUMF**, with its Isotope Separator and Accelerator (ISAC) rare isotope facility is the highest beam power ISOL facility in the world. In addition, the Advanced Rare Isotope Laboratory (ARIEL) upgrade will augment the capability with a new method of rare isotope production via photofission. ARIEL will also be the only facility in the world with the capability to deliver three simultaneous rare isotope beams. A diverse suite of experimental facilities brings scope for cutting-edge experiments with low-energy rare isotope beams. TRIUMF thus presents a multifaceted opportunity for the world to collaborate in discovering unknown secrets of nuclear structure, gaining knowledge on element synthesis in nature and providing stringent tests for some of the fundamental symmetries in nature.

Canada and Europe have extensive research collaborations in nuclear physics. The multi-reflection time-of-flight (MRTOF) mass separator at TITAN was produced through collaboration with the University of Giessen and GSI. The TRIUMF UK Detector Array (TUDA) and the TACTIC detector projects devoted to nuclear astrophysics are experimental infrastructures at TRIUMF with significant contributions from universities in the UK. The highly segmented silicon detector array SHARC was developed by the University of York and coupled to the TIGRESS Germanium detector array for studying reactions with particle-gamma coincidence. The ACTAR collaboration with GANIL in France, Spain and Belgium aims to operate a dedicated campaign of experiments using the complementary beam capabilities at TRIUMF. Researchers from European institutions lead a large fraction of the experiments. Moreover, the on-site collaborations host many European students and postdocs for training and data taking, leading to additional research and networking opportunities for junior scientists.

ISAC and ARIEL may also position themselves as world-leading facilities for future high-precision tests of physics using rare isotopes. The search for parity and time-reversal violation effects with radioactive molecules, RadMol, is a new project under planning at TRIUMF in collaboration with Germany. The study of the unitarity of the CKM matrix through precision measurements of nuclear-superallowed beta decay is a long-standing project that involves European collaboration. This is complemented by high-precision mass measurements at TITAN.

The accelerator physics collaborations are with CERN-ISOLDE for rare isotope production, target and ion-source developments. The future ISOL@MYRRHA facility in Belgium and PSI are collaborating closely with TRIUMF to adopt design expertise from the ARIEL development. In addition, the CANREB EBIS is the product of collaboration with MPIK in Germany.

TRIUMF is also pioneering the development of a "neutron-capture storage ring" for the direct measurement of astrophysical neutron capture reaction cross sections on short-lived nuclei down to seconds of half-life. This project is a collaboration involving scientists from Spain (IFIC Valencia) and Germany (GSI Darmstadt). The TRIUMF Storage Ring (TRISR) aims not only to be the first heavy ion storage ring coupled to an ISOL facility but also includes novel detection techniques via extraction of neutron capture reaction products into a recoil separator. A Canadian funding proposal for a design study has been submitted, aiming at the start of the construction of such a facility by 2029.

# Recommendations for research infrastructures

## GSI/FAIR

The Facility for Antiproton and Ion Research (**FAIR**), an ESFRI landmark facility, is under construction as an international facility on the campus of the GSI Helmholtzzentrum for Heavy-Ion Research in Darmstadt, Germany. It will open up unprecedented research opportunities with a focus on hadron and nuclear physics. However, it will also encompass atomic physics and nuclear astrophysics and applied sciences like materials research, plasma physics and radiation biophysics with applications towards novel medical treatments and space science.

The first phase of FAIR is expected to be operational by 2028 and will facilitate experiments with **SIS100** using the High-Energy Branch of the **Super-FRS**, the **CBM** cave and current **GSI** facilities. The next steps include the **APPA** cave and the Low-Energy Branch of the Super-FRS. The **CR** and **HESR** storage rings will be unique tools for precision measurements for atomic, nuclear and hadron physics. Long-duty-cycle and continuous-wave beams for the low-energy programme will be provided by HELIAC.

The completion of the full FAIR facility is the declared goal of all FAIR shareholders and should be vigorously pursued, as it will provide European science with world-class facilities for many decades. In particular, NuPECC recommends:

● Full exploitation of the novel research opportunities as provided by the FAIR facility for the **APPA, CBM, NUSTAR** and **PANDA** collaborations.

● Accomplishment of First Science+ by 2028, followed by the expedited completion of the **APPA** cave and the **Super-FRS** low-energy branch.

● Construction of the **CR** and **HESR** storage rings, together with the PANDA detector and other experimental installations to be concluded by 2032.

## GANIL/SPIRAL2

**SPIRAL2** at GANIL is an ESFRI landmark facility. The new Super-Separator Spectrometer **S³** is in an advanced stage of completion. The ongoing projects, the low-energy DESIR facility and the new heavy-ion injector **NEWGAIN**, will be operational in 2027 and 2028 respectively. The LINAC and the NFS experimental hall are now fully operational and have opened new research fields for GANIL/SPIRAL. Recent funding was received by GANIL from the French ministry of research to ensure the completion of DESIR and NEWGAIN and to increase the reliability and capabilities of the cyclotrons (**CYREN** project).

NUPECC recommendations for the future development and exploitation of the SPIRAL2 facility are as follows:

● Timely completion of ongoing projects (**S3, DESIR, NEWGAIN, CYREN**) and full completion of their novel research opportunities.

● Progress in the construction of a very high intensity **reaccelerated RIB facility** (using fission and multi-nucleon transfer reactions) up to 100 MeV/u, as recommended by the international committee for the future of the GANIL/SPIRAL2 facility.

● In the longer term, **electron-radioactive ion scattering** with the trapped ions in DESIR will give a unique view into the structure of nuclei far from stability and was also recommended by the international committee for the future of GANIL/SPIRAL2. It will open unique avenues for testing nuclear interactions and models.

## CERN

At **CERN**, the LHC and its proton injectors provide unique opportunities for nuclear physics exploited by a series of programmes encompassing relativistic heavy-ion physics, antimatter research, hadron spectroscopy and science with neutrons and radioactive isotopes. These constitute world-leading nuclear physics programmes that have successfully diversified the research output of CERN. A phase of decision-making will begin soon concerning the machine that will replace the LHC in the 2040s.

● NuPECC recommends that a strategy be developed to secure future opportunities for continuing world-leading nuclear-physics programmes unique to CERN, following the success of the diversity programme using the LHC and Proton Injectors.

With regard to individual programmes, NuPECC makes the following recommendations:

● The construction of **ALICE 3** as part of the **HL-LHC** plans is strongly supported. ATLAS, CMS and LHCb all have successful heavy-ion programmes which should remain an integral part of the lifetime of the LHC. Upgrades to these experiments should also be exploited for heavy-ion physics.

● Infrastructure in the CERN North Area should be developed to supply the unique high-energy and high-intensity hadron and lepton beams for the proposed experiments in hadron structure (**AMBER**) and heavy-ion collisions (**NA60+** and **NA61/SHINE**).

● The proposed programme of improvements and consolidation that will lead to increased capacity and capability at **ISOLDE** should be completed. Developments should be sought to maximise the scientific exploitation of this unique ISOL facility, including increases to operational capacity and space.

● Improvements should be made to the **n_TOF** facility during the next long shutdown and new designs investigated for the necessary target replacement in 2033 (during CERN LS4). The facility should fully exploit its unique potential to address key nuclear data needs using neutron-induced reactions in the coming years.

● A programme of consolidation and updating should be developed to ensure the operation of the unique antiproton facility, the **Anti-Proton Decelerator**, beyond 2028.

## ELI-NP

The Extreme Light Infrastructure - Nuclear Physics (**ELI-NP**) in Romania is an advanced nuclear photonics facility, currently harnessing the most powerful laser system in the world. It was set up as part of the ESFRI Extreme Light Infrastructure (ELI) project and funded by the European Commission and the Romanian government. Unique high intensity **gamma-ray beams** produced by Laser Compton Backscattering (LCB) will become available in the future. The main research directions of ELI-NP are related to nuclear physics driven directly by extreme photon beams or by the secondary radiations generated by them.

NuPECC strongly supports the future development and exploitation of the ELI-NP facility. With regard to the individual aspects NuPECC makes the following recommendations:

● Prioritise nuclear physics studies related to laser-matter interaction under extreme conditions, leading to relevant issues such as laser-driven particle beams, novel nuclear and photonuclear studies and the fostering of industrial and medical applications for important societal benefits.

● Very strong support for finalising the implementation of the gamma beam system, achieving full completion of the facility with the potential of providing breakthrough results in the field of nuclear photonics.

● Strong support for developing technologies for high-impact societal applications based on high-power lasers, such as interferometric phase contrast X-ray imaging and laser-driven protons and carbon ions for hadron therapy.

## Facilities for Hadron and Lepton Beams

A range of complementary facilities with a variety of hadron and lepton species is available to European researchers within and also outside Europe. They cover the range of beam energies needed for nuclear physics, hadron physics and particle physics and are supported by large EU-integrated infrastructure initiatives.
The construction of future flagship facilities, the antiproton beam experiment **PANDA** at FAIR and **ePIC** at the Electron-Ion Collider (EIC), which are of crucial importance for the hadron-physics community is strongly recommended.

In addition, we recommend:

● Vigorous exploitation of the full physics potential offered by the recent upgrades of the electron facilities available in Europe at different energy scales (including **ELSA, LNF, MAMI, S-DALINAC**) and the **completion** of the **MESA** facility.

● Achievement of the **High-Intensity Muon Beams (HIMB)** project at **PSI**.

## Facilities for Radioactive Ion Beams and Isotopes

Europe is in a very strong position in terms of infrastructure for radioactive isotope science. Historically, many key techniques have been developed in Europe, where the ISOL method was pioneered and leading contributions to in-flight production were made.

A range of complementary facilities with a variety of production methods means European researchers can secure high-quality radioisotope beams from lightest to superheavy nuclides, both proton and neutron-rich, with or without post-acceleration, and unencumbered by the deficiencies of one specific technique. These features support a large and vibrant European community that develops state-of-the-art experimental techniques and instrumentation, securing a world-leading position in many areas of radioactive beam science.

In order to maintain and expand this world-leading position, we strongly recommend:

● Completion of the full **Super-FRS** facility at **FAIR**, including the low-energy and ring branches, providing radioactive ion beams by fragmentation, spanning the range from relativistic energies to being fully stopped and taking advantage of the exceptionally high energies and intensities of the primary beams.

● Completion of the **GANIL/SPIRAL2** and **SPES-LNL** facilities. At GANIL, new opportunities will be created by $S^3$ (for in-flight low-energy beams) and DESIR (for $S^3$ and SPIRAL1 ISOL beams), and in the longer term by a second-phase fission target. SPES-LNL, focusing mainly on neutron-rich nuclides from fission, is progressing towards operation in the coming years.

● Energetic exploitation of **ISOLDE** at CERN, **IGISOL** at JYFL, **ALTO** and **SPIRAL** in France to provide the community with a wide portfolio of (low-energy) ISOL beams. Continuing efforts in beam development will keep these facilities at a world-class level. Achievement of **ISOL@MYRRHA**, optimised for long-lasting experiments, and continuation of a vigorous R&D programme to define the **RIF@IFIN** facility.

● Continuation and expansion of the very successful **PRISMAP** initiative providing radionuclides for medical applications and thus putting in practice one of the key societal applications of nuclear physics. In particular, the PRISMAP core facility **MEDICIS** at CERN should be supported as well as the construction of **TATTOOS** at PSI.

## Facilities for Stable Ion Beams

Stable ion beams for nuclear physics, available both at European large-scale and small-scale facilities, provide a variety of beams from proton to uranium to cover a very wide energy range. At these facilities, innovative research programmes are carried out, including impactful studies of nuclear structure, reaction mechanism, nuclear astrophysics, fundamental interactions and applications of nuclear physics in cross-disciplinary research with high societal value. The small-scale facilities, which are well spread out over Europe with their variety of state-of-the-art instruments, are key players focusing on specific scientific and technology topics; they are also particularly adept in education and training.

The high beam intensities at these facilities are particularly useful in developing experimental techniques or production methods relevant to radioactive beam facilities. Stable ion beams are thus critical for creating new avenues for exploring unstable hadrons and nuclei far from stability and to enrich the portfolio of isotopes for medicine or other applications.

Therefore, to keep and enhance the worldwide competitiveness of stable ion beam facilities, we recommend:

● Providing full support to the ongoing exploitation of all **stable-ion beam facilities** (including **CERN, GANIL/SPIRAL2, GSI/FAIR, IFIN, IFJ-PAN, IJCLAB, JYFL, LNGS, LNL, LNS, PSI, SLCJ**) spanning the energy range from fully relativistic to those relevant for astro-physical processes.

● The travelling instrument **AGATA** for use at both stable and radioactive ion beam facilities must be completed to reach the $4\pi$ configuration.

● Vigorous efforts to augment the ion beam species and intensities and optimal access to their unique detection systems should be made to open new avenues; for example, to produce nuclei far from stability and for applications.

● Support for a new, dedicated 15 MV **accelerator mass spectrometry system** for astrophysics and applications.

● Further developments of underground ion accelerators should be pursued, such as for the **LUNA** 400-kV accelerator inside the Bellotti Ion Beam Facility at LNGS and a possible **Felsenkeller** upgrade at DZA.

● For **smaller-scale facilities**, it is recommended to further develop and equip them to strengthen their unique science programmes, reinforce the synergies among them and their large-scale partners, and develop a sustainable coordination model within the EURO-LABS and ChETEC-INFRA projects.

## Neutron facilities

The diverse range of neutron facilities operating concurrently supports having distinct user groups with their unique needs and objectives. Nuclear physics research using “slow” neutrons and the production of radionuclides for research and cancer treatment is strongly concentrated on the European flagship facility **ILL**.

We recommend:

● Fully exploiting this unique facility up to the technical lifetime of ILL’s reactor and to further develop the experimental possibilities at **MLZ** and **PSI**;

● supporting smaller neutron facilities, in particular the **TRIGA** reactors, playing an important role in developing and optimising techniques and experimental setups that are then used efficiently at the larger facilities.

The capacity of high flux neutron time-of-flight measurements will be augmented in the long term by complementing the present flagship facilities **n_TOF** at CERN and **NFS** at GANIL with the new facility **IFMIF-DONES.** It is therefore imperative to continue supporting these unique European facilities. The situation is similar for pulsed fast neutron beams and epithermal Maxwellian neutron beams, where existing facilities and new initiatives serve the demands, including those from the nuclear physics community, for this type of beams.

Therefore, we furthermore recommend strongly:
● providing support, on both national and European levels, to the facilities that exhibit distinctive characteristics or play a vital role in terms of their capacity to host experiments;

● establishing and enhancing a sustained strategic European initiative focused on nuclear data over the long term and including support for nuclear target production;

● developing higher quasi mono-energetic neutron beams well above 20 MeV currently not available in Europe, and strongly benefitting dosimetry and other applications up to several hundred MeV for space applications.

## Theory and computing infrastructures

**ECT*** is a globally recognised centre for research in theoretical nuclear physics and related areas in the broadest sense of the word. It is recommended to expand the role of **ECT*** as a strategic think tank to develop ideas for grant capture beyond the current mechanisms, to alleviate present financial constraints and make room for new initiatives.

Several virtual access (VA) initiatives have been established to reinforce the communication between researchers within the NuPECC communities and beyond. These VAs offer infrastructures facilitating the sharing of data, software and other digital objects in adherence to the F.A.I.R. principles. Consequently, they contribute to the broadening of the physics reach. Notable successes in this realm include **NLOAccess** and **3DPartons** within STRONG-2020 and **Theo4Exp** in the EURO-LABS project, designed specifically to optimise interoperability between theory and experiment. Additionally, the **ESCAPE** Open-Source Software and Service Repository stands out as a sustainable open-access platform, sharing scientific software and services with the wider scientific community. Furthermore, dedicated websites serve as expansive repositories for theoretical results, exemplified by the CEA Bruyères-le-Châtel platform.

● We strongly recommend that virtual access initiatives be supported in a sustainable way to preserve their unique impact in the future.

## Access to Research Infrastructures outside Europe

In addition to European research infrastructures, the nuclear physics community is agile in exploiting unique opportunities at facilities elsewhere in different ways. Several collaborations have invested in experiments and detectors sited at non-European facilities, drawing upon the technical skills and expertise of European scientists. Similarly, detector systems developed by European collaborations have been deployed at non-European facilities for dedicated campaigns, accessing beams that are complementary to those available in Europe. In addition, intellectual investment and theoretical support are provided by European scientists as users in many different individual experiments at non-European facilities. Prominent examples are **FRIB** and **JLab** (USA), **RIKEN** (Japan) and **TRIUMF** (Canada). These activities that access infrastructures beyond Europe diversify and enhance European nuclear physics, but also seed new initiatives and bring ideas, skills and knowledge to European facilities.

An important example for the future is the **EIC**, where the wide European interest calls for coordinated action across different communities to secure involvement in the facility of the ePIC and other future experiments.

More generally, we recommend:
● fostering collaboration with **non-European Research Infrastructures** in order to seize unique scientific opportunities and synergies that complement scientific programmes based in Europe.

# Nuclear Physics Tools - Detectors and Experimental Techniques

**Convener:**
**Silvia Dalla Torre** (INFN, Sezione di Trieste, Trieste, Italy)

**NuPECC Liaison:**
**Eugenio Nappi** (INFN, Sezione di Bari, Bari, Italy)

**WG Members:**
- **Dieter Ackermann** (GANIL, Caen, France)
- **Andrew Boston** (University of Liverpool, UK)
- **Paolo Finocchiaro** (INFN, Laboratori Nazionali del Sud, Catania, Italy)
- **Piotr Gasik** (GSI, Darmstadt, Germany)
- **Paola Gianotti** (INFN, Laboratori Nazionali di Frascati, Italy)
- **Andrea Gottardo** (INFN, Laboratori Nazionali di Legnaro, Italy)
- **Fritz-Herbert Heinsius** (Ruhr Universität, Bochum, Germany)
- **Silvia Masciocchi** (Heidelberg University, Germany)
- **Thomas Peitzmann** (Utrecht University, The Netherlands)
- **Tina Pollmann** (University of Amsterdam, The Netherlands)
- **Veronique Puill** (IJCLab, Orsay, France)
- **Jochen Schwiening** (GSI, Darmstadt, Germany)
- **Maria Dorothea Schumann** (PSI, Villingen, Switzerland)
- **Monica Sisti** (INFN, Sezione di Milano Bicocca, Italy)
- **Joachim Stroth** (Goethe University, Frankfurt, Germany)
- **José Javier Valiente Dob´on** (INFN, Laboratori Nazionali di Legnaro, Italy)

**Contributors:**
- **Pietro Antonioli** (INFN, Sezione di Bologna, Italy)
- **Marlène Assié** (IJCLab, Orsay, France)
- **Michael Block** (Johannes Gutenberg-Universität Mainz and GSI, Darmstadt, Germany)
- **Franco Camera** (Universit`a degli Studi di Milano and INFN, Sezione di Milano, Italy)
- **Giovanni Casini** (INFN, Sezione di Firenze, Italy)
- **Ruben de Groote** (Institute for Nuclear and Radiation Physics, Leuven, Belgium)
- **Pierre Delahaye** (GANIL, Caen, France)
- **Achim Denig** (Johannes Gutenberg Universität, Mainz, Germany)
- **Alfons Khoukaz** (Universität Münster, Germany)
- **Vladimir Manea** (IJCLab, Orsay, France)
- **Gerhard Reicherz** (Ruhr-Universität, Bochum, Germany)
- Thomas Roger (GANIL, Caen, France)
- **Riccardo Raabe** (Institute for Nuclear and Radiation Physics, Leuven, Belgium)
- **Haik Simon** (GSI Darmstadt, Germany)

# Introduction

Advancement in the understanding of fundamental Nuclear Physics has been deeply and intimately entwined with the development of new technologies and experimental methods which have often found their way into applications with high societal impact (see Chapter Applications and Societal Benefits). Often, tools are developed to answer specific theoretical questions and predictions which in turn lead to new discoveries. In the past, these developments have been made in a more spontaneous "bottom-up" manner, which can lead to overlaps and work being repeated in several projects. The need to include top-level strategic recommendations in a more coordinated manner in the NuPECC LRP is self-evident. The breadth of technologies required for Nuclear Physics is very wide, encompassing detectors and related electronics, separators, ion traps, laser technologies, etc.

A coordinated roadmap for technology developments in Nuclear Physics will help to avoid repetition of work and identify priorities for the community. Such an approach has been employed with great success by the Particle Physics community in their long-range planning (for example, the ECFA RD roadmap [1], following the European Particle Physics Strategy Upgrade process and resulting in the formation of DRD Collaborations at CERN, the 2021 Snowmass Community Planning Exercise [2] and CPAD effort in the USA, bringing to the formation RDC Collaborations). A systematic exercise has been performed in the USA in nuclear physics: the Generic R&D Programme for the EIC [3]. Indeed, in these roadmaps, there are synergies which can be exploited by both communities, particularly in detector developments.

This chapter is organised as follows:

**Experimental methods** for major coming and upgrading experiments and facilities are discussed first, in a set of sections where the experimental methods are grouped according to the physics domain of exploration, as each one of the research fields has specific requirements which shape the experimental approaches and the required tools.

**Technological tools and related R&D efforts** are discussed in the following sections, subdivided into research areas similar to those illustrated for the experimental methods. Here two more sections are added. One is dedicated to the supply and availability of stable and radioactive isotopes as well as target manufacturing, items which are singled out due to their relevance and specific requirements. The other deals with front-end ASICs, electronic read-out chains and data acquisition, fields requiring relevant steps forward in all the domains of nuclear physics studies.

The chapter summary is provided in the form of recommendations.

# Experimental methods - Low Energy Nuclear Physics

The experimental methods discussed in this section are those adopted for the Nuclear Structure and Reaction Dynamics and for Nuclear Astrophysics. The section also provides data needed for nuclear physics Application and Societal Benefits.

## High-granularity detectors

Over the next decades in the field of low-energy nuclear physics, major advances in our understanding of nuclear structure properties will demand simultaneous high detection efficiency, high counting rate capabilities and good detection position resolution. This has spurred the development of high-granularity detectors and new experimental methods for γ-ray as well as light-particle spectroscopy. These efforts have encouraged the forming of sizeable collaborations among several European groups to achieve the necessary amount of financing and workforce. Key to this effort is also the concept of travelling detector arrays, which allow the community to perform a rich range of physics campaigns, exploiting different beams and facilities around Europe.

## Gamma-ray detectors

### High-resolution spectroscopy by Germanium arrays

High-resolution γ-ray spectroscopy has traditionally been the realm of the Compton-suppressed large high-purity Germanium arrays. The need for larger angular coverage and better-resolving power has driven the development of 4π γ-ray tracking arrays. They rely on detector segmentation and pulse-shape analysis to reconstruct the photon path inside the Germanium crystal with a few mm precision.
To this end, the collaborative effort of 13 countries and over 40 research institutes started the AGATA project in Europe. Since the first physics campaign in 2010, AGATA has evolved from the demonstrator phase to the present ~1π system (Phase 1) and will aim at the full ~4π system of 180 detectors via the intermediate ~3π configuration (Phase 2). As a travelling detector, AGATA has been, and will be, used at all major current and near-future European research facilities included in the ESFRI list, delivering stable and radioactive ion beams coupled to a suite of detector systems including magnetic mass spectrometers (e.g., PRISMA at LNL, VAMOS++ at GANIL and FRS/SFRS at GSI/FAIR) and arrays for high-energy γ rays, neutrons and charged-particle detection (PARIS, NEDA and GRIT respectively), significantly enhancing sensitivity to extremely weak and rare phenomena. AGATA, with its unprecedented capabilities, opens a new high-precision era in nuclear structure studies moving far away from the valley of stability.

### Spectroscopy with scintillators

Scintillators made with heavy materials allow for γ-ray detection with great efficiency, also at high energies. High granularity is also a key requirement for these arrays, to withstand the high counting rates and considerable *γ*-ray multiplicities which are required in present and future facilities.

An international research project called PARIS has been started, to develop and build a novel 4π *γ*-ray calorimeter, benefitting from recent advances in scintillator technology. It is designed to play the role of an energy-spin spectrometer, a calorimeter for high-energy photons and a medium resolution *γ*-ray detector. Arrays like CALIFA for R³B allow for the detection of gamma rays up to recoil protons with energies of several 100 MeV using a 4π barrel geometry. The device is composed of two shells: the inner scintillators by one of the most advanced technologies ($LaBr_3$:Ce or $CeBr_3$) and a more conventional scintillator (NaI) for the outer shell.

## Particle and ion detectors

Charged particle detectors with wide angular coverage are being developed to measure reaction products in different energy ranges. A requirement common to all detectors is the capability to perform a pulse-shape analysis to identify different ion atomic numbers and even masses. Another key aspect is a high granularity to sustain high counting rates and achieve the angular detection resolution necessary for the desired precision of the kinematic reconstruction reaction.

### Silicon detectors for direct reactions

In direct-reaction studies, the main observables to be measured are the energy of the populated states and the corresponding differential cross-sections, namely the angular distributions. Inverse kinematic experiments consist of measuring the light recoiling particle produced by reactions between a radioactive beam particle and a light target (as p, D, T, $^{3,4}$He). The GRIT European collaboration has proposed a new generation portable Silicon detector array for the optimal study of direct reactions at present and future European Nuclear Physics facilities such as GANIL, SPES, ISOLDE and FAIR. It consists of a new type of compact, high granularity, ~4π acceptance Silicon array, with new digital electronics allowing seamless integration inside AGATA as well as in the PARIS scintillator detector in its final version. The timeline for the completion of GRIT is mid-2027. The use of flexible pixel detectors of ALPIDE type is being explored, e.g. for covering the closed geometry around the R$^3$B COCOTIER liquid hydrogen target at FAIR. In a longer-term view, the pursuit of an R&D programme to extend the capability of Pulse-Shape Analysis particle identification is desirable.

The exploitation of direct reactions in inverse kinematics also requires new light ion targets with high density to compensate for the limited intensity of the radioactive ion beams. The use of $^{3,4}$He cryogenic targets (as designed for the CTADIR project or HeCTOR), or of windowless $^{1,2}$H semi-solid targets (like CHyMENE), will permit reaching a luminosity great enough to fully exploit the high-granularity of the combined AGATA-GRIT setup.

## Charged-particle detectors for heavy-ion reactions

Several European collaborations have been set up to address the need for charged-particle detectors for heavy-ion fusion and fragmentation reactions to identify the charged reaction products in mass and atomic number. The goal is for composite detectors with improved performance for ions at energies well below the Fermi regime. A big step ahead in lowering identification thresholds could be achieved by using high-quality very thin sensors aiming at Z identification up to Z~20 with a threshold of around 2 MeV/u.

A key development is being carried out by the FAZIA collaboration, supported by the MOU signed by contributing parties (Italy, France, South Korea, Poland and Spain). FAZIA is an array of 192 solid-state silicon-scintillator telescopes devoted to the detection and identification of charged fragments produced in heavy ion fixed target reactions at beam energies from 15 to 100 MeV/u. Its pulse shape analysis (PSA) technique makes it possible to identify ions in charge and also in mass and it can operate when the conventional ΔE-E method is not applicable. It thus considerably lowers the energy thresholds. Since 2019 the 12 FAZIA blocks are coupled with the INDRA multidetector at GANIL. This proves the technology: in fact, the telescopes offer a good energy measurement in a range from a few MeV protons up to some GeV heavy ions.

Another example of the array for heavy ion reactions is CHIMERA. It is a 4π detector based on 1192 Si-CsI(Tl) telescopes to detect both charged particles and γ-rays. It is combined with FARCOS, a correlator with good energy and angular resolution, while a novel neutron detector is under development.

# Neutron production and detection

Many facilities and experiments all over Europe envisage the enticing possibility of detecting neutrons to investigate nuclear properties and reaction mechanisms involving neutron emission. On top of this, several facilities are already in operation and others are being built or planned for the production of neutron beams devoted to studies of fundamental physics and nuclear astrophysics, as well as to material science, non-destructive analysis, imaging and other industrial applications. Neutron beams can be produced via high-energy proton beams at spallation sources, like n TOF at CERN, SINQ at PSI and ESS in Lund, and can provide an almost white neutron spectrum. Other production mechanisms are the photoproduction, for instance at GELINA; specific neutron production reactions via low energy high-intensity ion beams impinging on special targets as already done or planned at IFMIF-DONES, GANIL, LNL, ALTO, SARAF; fission reactors as at ILL, FRM-II and SCK-CEN; and ultra-high-power lasers as at ELI.

In most ion accelerator facilities, many experiments with a large solid angle coverage are planned, requiring the complementing of studies

## Box 9.1 : Examples illustrating the increasing complexity of detection arrays for low energy nuclear physics studies

**AGATA** © AGATA Collaboration

**Rendering view (right) of the 2π and 4π AGATA spectrometer showing the cryostat dewars (blue) of the HPGe detectors (grey). On the left a current view of the AGATA HPGe detectors from the target position.**

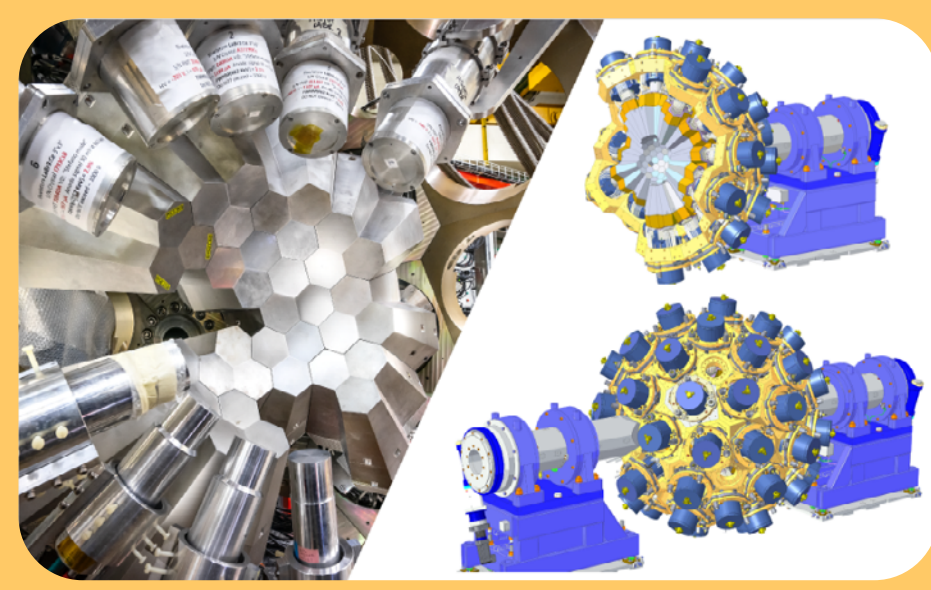

**GRIT** © GRIT Collaboration

**Rendering view of the GRIT array placed at the centre of the AGATA array. Some of the GRIT detectors are removed to allow the view of the internal face of the trapezoidal silicon detector and of the CTADIR cryogenic He target.**

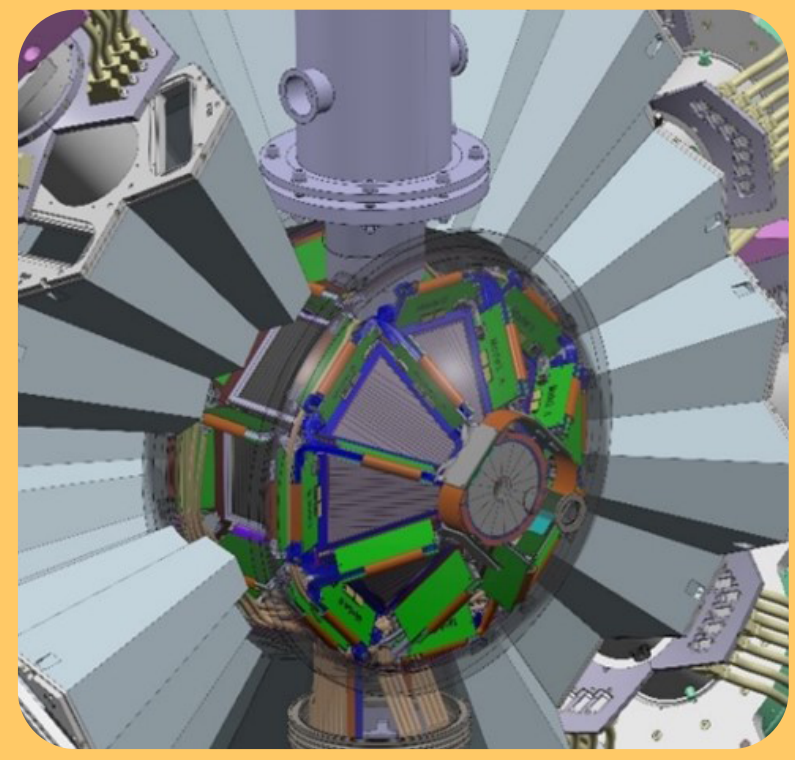

**INDRA-FAZIA** © GANIL

**Combined arrays for charged-particle detectors for heavy-ion reactions.**

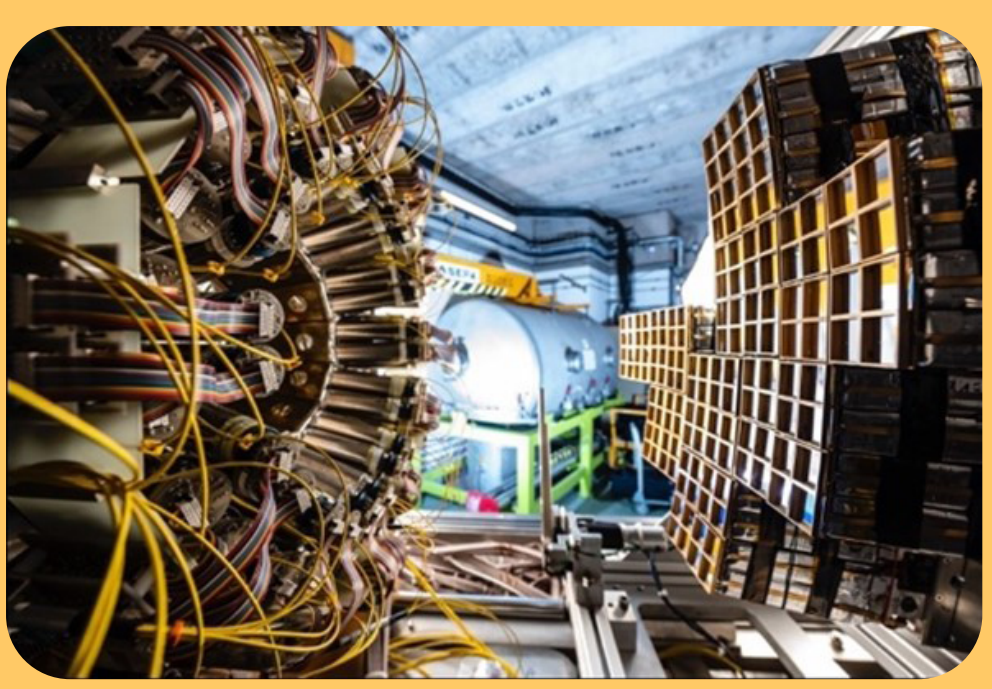

of gamma and charged particle emission with ever more precise information about neutron emission, like the TETRA, NEDA and NeuLAND large arrays of neutron detectors. Neutron time-of-flight spectrometers, like the NeuLAND array for FAIR, benefit from the obtainable high resolution using cost-effective FPGA-based TDC operations and combine high neutron detection efficiency, even for four neutron events, with a 4π coverage in the forward cone of reaction with relativistic velocity beams. The MOdular Neutron SpectrometeTER (MONSTER) was built for performing beta delayed neutron measurement at FAIR and at other European laboratories. Indeed, wherever neutrons are produced the need for improved detectors is compelling, as the existing neutron detector technologies still suffer from unsatisfactory intrinsic efficiency and/or poor energy resolution or range. Experiments aiming at the study of beta-delayed multi-neutron emission, neutron-neutron and neutron-proton correlation and neutron radioactivity will certainly benefit from new technological developments.

## Mass spectrometers and separators

### Heavy-ion mass spectrometers

Heavy-ion mass spectrometers for reaction recoils in the few MeV/nucleon energy range are well-established assets in several European laboratories. PRISMA at LNL and VAMOS++ at GANIL are mass spectrometers used mainly for deep-inelastic and fission reactions, while RITU, MARA and JYFL are mass separators mainly employed for fusion-evaporation reactions and superheavy detection.

The MAGNEX spectrometer at the INFN LNS is a heavy-ion spectrometer undergoing major upgrades in the framework of the NUMEN project to exploit the very intense Super-Conducting cyclotron beams. Target and detection systems will be completely rebuilt to make them compliant with the challenging radiation hardness required. The scattering chamber will be replaced by a thinner and smaller one to host a new calorimeter for γ-ray detection (G-NUMEN) based on ~100 $LaBr_3$(Ce) scintillators. The focal plane detector (FPD) will be replaced by a new one based on a gas-filled tracker and a wall of SiC-CsI telescopes, optimised to sustain the very intense flux of heavy ions reaching it. In addition, an array of $LaBr_3$(Ce) will be installed around the FPD to detect $e^+e^-$ pairs.

### Solenoid spectrometers for transfer reactions

The ISOLDE Solenoid Spectrometer ISS is situated at the ISOLDE facility, CERN. It is a solenoid spectrometer developed to measure direct nuclear reactions in inverse kinematics with reaccelerated beams from the HIE-ISOLDE linac. It consists of an on-axis position-sensitive silicon array within a solenoid magnet. ISS makes possible studies of direct reactions with radioactive beams bombarding deuterated-polyethylene targets to investigate the structure of atomic nuclei. At present, three such devices exist worldwide: ISS, HELIOS at Argonne National Laboratory and SOLARIS at FRIB.

### The Super Separator Spectrometer S3

The Super Separator Spectrometer (S3) takes advantage of the very high-intensity stable beams of the superconducting linear accelerator of SPIRAL2. The main requirement is to separate very rare events from intense backgrounds. The development of S3 will address three major technological challenges: the need for very intense heavy-ion beams to access very low cross-section reactions (picobarn and below), the need for a powerful recoil separator-spectrometer that can combine, thanks to its innovative superconducting multipole magnets, wide transmission with high selectivity and the capability to perform in-flight mass-number determination of short-lived nuclei.

The S3 project, considered a “radioactive nuclei production facility”, has motivated the development of a broad range of innovative instrumentation setups aimed at determining different observables of those nuclei, namely:

- The SIRIUS setup, an implantation decay station focused on the decay spectroscopy of very heavy and super-heavy elements.

- The Low Energy Branch - LEB setup couples a gas cell to stop the nuclei selected by S3 for laser ionisation/spectroscopy. The ions produced will be transported through a system of RFQs to a detection area combining a mass spectrometer, PILGRIM (MR-Tof-MS), and an α/conversion-electron/γ-ray detection station, SEASON. The major attribute of LEB is to use atomic physics techniques - more specifically, high-resolution spectral measurements of the atomic transitions - to provide fundamental and nuclear-model-independent data on the structure of ground and isomeric nuclear states.

### Ion traps and laser spectroscopy of nuclei

Ion trapping and laser spectroscopy methods have gained an important role in nuclear physics. Mass measurements performed with Penning traps and multi-reflection time-of-flight mass spectrometers at ISOLTRAP at ISOLDE, SHIPTRAP at GSI SHIP, JYFLTRAP at the IGISOL facility at Jyväskylä and the Fragment Separator Ion Catcher (FRS-IC) at GSI provide accurate binding energies. Many ion-manipulation techniques have been developed for cooling, stacking and bunching of radioactive ion beams (RIBs), for example with radiofrequency quadrupole cooler bunchers. Traps can also serve as high-resolution mass separators, removing nuclear isobars, and in certain cases even preparing isomerically purified beams. This is beneficial for trap-assisted nuclear spectroscopy. The use of buffer-gas cells (operating at the IGISOL, FRS-IC and SHIP facilities) has increased the reach of mass spectrometry and ion-manipulation techniques to essentially all production schemes, including fusion-evaporation reactions and fast-beam fragmentation. New ion-trap mass spectrometers are being commissioned for the SPIRAL2-GANIL facility (PILGRIM for S3-LEB and PIPERADE, MLLTRAP for DESIR).

Laser spectroscopy measurements provide the spin, the dipole moment, the quadrupole moment and changes in the mean-square charge radius of short-lived isotopes. A wide range of techniques has been developed to access them at RIB facilities. Collinear laser spectroscopy (CLS) methods provide the highest spectral resolution (COLLAPS and CRIS at ISOLDE and CLS at IGISOL), while in-source, laser ionisation and spectroscopy (LIS) techniques provide the highest sensitivities. The combination of LIS and particle identification, e.g. decay tagging or mass spectrometry, makes it possible to obtain measurements at the lowest production cross sections, as shown by experimental campaigns with RILIS at ISOLDE, RADRIS at SHIP and hot-cavity-catcher LIS at IGISOL. The use of buffer-gas cells and gas jets has has made possible studies of actinides at LISOL, IGISOL and SHIP, despite their challenging production and complex atomic structure, with new facilities being developed for SPIRAL2-GANIL (S3-LEB) and the University of Jyväskylä (MARA-LEB). Such techniques will also be required for nuclides produced by fast-beam fragmentation, in-flight fission, and multi-nucleon transfer reactions. Laser ion sources, nowadays available at most facilities, have improved their techniques for delivering pure RIBs and increased the range of accessible elements. Radioactive molecules open a new dimension for accessing elements not yet studied, or observables that might not be accessible in single atoms or ions. Methods which combine ion traps and laser spectroscopy have also been developed, e.g. using optical pumping for atomic state preparation, and may play an increasingly important role in coming years.
Cross-sections and decays in plasma environment could be measured at future facilities like PANDORA at INFN Laboratori Nazionali del SUD with laser plasma induced reactions.

## Experimental methods - High Energy Nuclear Physics

The experimental methods discussed in this section are those adopted for hadron physics (Chapter Hadron Physics) and for studies dedicated to strongly interacting matter under extreme conditions (Chapter Strongly Interacting Matter).

# Nuclear physics studies at high-energy colliders

## Strongly interacting matter at vanishing baryochemical potentials and high temperatures - collider experiments at LHC

In the period considered in the present LRP, LHC experiments will undertake phase II upgrades and then upgrades for the High Luminosity (HL) LHC.

The upgrades that have been installed in Long Shutdown 2 (2019-2021) in the LHC and ALICE will enable ALICE to collect Pb–Pb collisions at a higher interaction rate of 50 kHz, more than 50 times the rates achieved for minimum bias data taking in Runs 1 and 2. A key ingredient for the increased rate capability is the replacement of the sensor of the TPC, now equipped with GEMs. In addition, the new inner tracking system ITS making use of the MAPS-architecture ALPIDE tracking sensors improves the pointing resolution by about a factor of 2. A further upgrade of the innermost layers of the tracking system (ITS3) in 65 nm CMOS technology and the addition of the high granularity forward calorimeter FoCal, equipped with Si sensors for the detection of direct photons, $\pi^0$s and jets are planned for Long Shutdown 3 (2026-2028).

The **ATLAS** apparatus is undergoing a series of general upgrades which enhance its capabilities to perform heavy ion physics. The new small wheel is a part of the muon trigger system that enhances the muon coverage and reduces fake rates. During the long shutdown 3 the new all-silicon inner tracking system with ten times more readout channels will extend ATLAS tracking coverage to ±4 units of pseudorapidity and will improve the precision of the vertexing. Improvements are planned in the calorimeter electronics and data acquisition system. In particular, the high-granularity timing detector will be used as a part of the trigger system. Besides general improvements of the ATLAS detector, upgrades are ongoing for the heavy-ion-specific studies. The zero-degree calorimeter that plays a crucial role in ultra-peripheral collision studies has been refurbished for Run 3 and equipped with a new reaction-plane detector. During the long shutdown-3, this will be replaced with a new-generation detector worked out in collaboration with ATLAS and CMS teams.

For **CMS (phase II upgrades)**, an increased charged particle tracking pseudorapidity ($\eta$) acceptance resulting from tracker upgrades will be a boon to bulk particle measurements, and the upgraded Zero Degree Calorimeters (ZDC) will improve the triggering and identification of ultraperipheral collisions (UPC). The addition of a time-of-flight particle identification capability, made possible by the Minimum Ionising Particle Timing Detector (MTD) and consisting of a central barrel region based on LYSO:Ce crystals read out with SiPMs and two end-caps instrumented with radiation-tolerant Low Gain Avalanche Detectors (LGAD), will open the way to identify low momentum charged hadrons, such as pions, kaons, and protons, which will improve measurements of heavy flavour particles and neutral strange hadrons such as $K^0$ and $\Lambda$.

On a longer time scale, both ALICE and LHCb collaborations plan further major upgrades to prepare for operation in **Run 5 and Run 6 of the HL-LHC**. The ALICE collaboration plans to build a new detector called **ALICE 3**. The detector consists of a large pixel-based tracking system covering eight units of pseudorapidity, complemented by multiple systems for particle identification including silicon time-of-flight layers, a ring-imaging Cherenkov detector, a muon identification system and an electromagnetic calorimeter. Track pointing resolution of better than 10 μm for $p_t > 200$ MeV/$c$ is achieved by placing the vertex detector on a retractable structure inside the beam pipe. The **upgraded LHCb** detector will need to meet the beam conditions of the HL-LHC phase with around forty simultaneous pp collisions per bunch crossing. The aim is to preserve the same performance as in Run 3 and Run 4 with a much larger charged particle multiplicity and higher detector occupancy. The next upstream tracker is designed to reduce the number of fake reconstructed tracks in high-occupancy collisions thanks to a pixel-based design. Similarly, the MIGHTY tracker located downstream of the magnet will also use pixels in the innermost region of the detector close to the beam pipe. The PID systems will be upgraded to increase the resolution. The upgraded LHCb will grant access to the most central ion-ion collisions at the LHC where droplets of QGP are expected. Finally, the addition of magnet stations will improve the efficiency of low-momentum track reconstruction.

It is relevant to note here the adoption at LHCb and ALICE of trigger-less streaming readout data acquisition systems for improved data selection flexibility and data acquisition/reconstruction integration.

## Exploring nuclear matter in a wide range of nuclear species and with polarised proton beams - collider experiments at RHIC and at EIC

During the long RHIC collider lifetime the experiments at RHIC, in particular STAR and PHENIX/sPHENIX, have largely contributed - also thanks to a series of upgrades - to establishing and improving the methods and technologies required for experiments at hadron colliders. A specific uniqueness of RHIC is the collision of polarised proton beams with longitudinal and transversal spin orientation. Together with accelerator techniques to handle proton-polarised beams, adequate polarimeter methods have been developed to support the experimental studies at the polarised collider. Absolute polarisation is extracted from double-spin p-p asymmetry measurements and polarisation monitoring bunch by bunch is obtained from single-spin asymmetry in p-C scattering. The unique development and expertise gained at RHIC provide a key ingredient for the coming EIC. QCD investigations with lepton probes - the ePIC experiment at EIC.
The ePIC experiment at EIC, with initial data-taking at the beginning of the 30's, offers a future to Deep Inelastic Scattering (DIS) physics studies with up-to-date technologies and tools, taking advantage of the collider's high luminosity and beam polarisation. Its design combines previous experience in DIS-dedicated detectors at the HERA Collider and fixed target experiments at SLAC, CERN, FNAL and JLab with innovative frontier technologies, some specifically developed for ePIC. The EIC physics scope, pursued by ePIC in its globality, requires DIS, Semi-Inclusive DIS (SIDIS) and selected exclusive channel measurements. The ePIC detector thus aims at as complete an acceptance coverage as technically possible by the 9.5 m long Central Detector (CD), sitting at the interaction region and far from detectors along the outgoing beam lines instrumenting a total length of almost 80 m. This scientific mission dictates the CD requirements, namely high-resolution electromagnetic calorimetry, accurate momentum resolution and vertex reconstruction by the combination of a new 1.7 T superconducting solenoid and low-material-budget trackers, where Si MAPS provide very fine space resolution and MPGDs offer timing information and the extended usage of particle identification counters to improve electron identification in synergy with the electromagnetic calorimeters and hadron identification for flavour-dependent studies. External hadron calorimeters support the high energy jet reconstruction including their neutral component. The forward counters impose operation in high-rate regions, where both rate and radiation hardness represent challenges. Examples are the far forward roman pots by pixelised AC-LGADs and the zero-degree calorimeter, including an imaging electromagnetic section and a sampling hadronic section, the precision measurement of the luminosity via the Bethe-Heitler process by a high rate zero-degree electromagnetic calorimeter, and high-precision pair spectrometer in the backward. Some counter highlights include: the MAPS in 65 nm CMOS technology allowing for support-less arrangements in the vertex layers; the extended use of MPGD technologies; the hybrid imaging electromagnetic calorimetry in the CD barrel; the duet of tungsten with scintillating fibres followed by SiPM on tile approach for the forward CD calorimetry, providing the electromagnetic and hadronic response respectively; and the extended usage of Cherenkov imaging devices with ePIC-guided validation of HRPPDs (large area MCP-PMTs by INCOM) and SiPMs as single-photon detectors for imaging applications. The detector read-out and data acquisition system are based on a fully triggerless principle, with hits identified by time stamp rather than by event label for full data selection flexibility and data acquisition/reconstruction integration.

Polarimetry, a complement needed in the experimental approach, takes advantage of the expertise gained at RHIC for hadron beams, and at HERA and JLab for electron beams. p beam polarimetry requires internal targets: polarised H jet target for absolute measurement, and thin C target for fast (burst by burst) monitoring. The electron beam polarisation is obtained by measuring backward Compton scattering. A radio-frequency pulsed laser system is under development for the high-rate operation at EIC.

# Strongly interacting matter at large net-baryon densities and moderate temperatures - fixed target experiments

## HADES at GSI SIS18

HADES is a large acceptance particle detector with a low-mass tracking system, time-of-flight detectors, a RICH and an electromagnetic calorimeter optimised to detect electron pairs emitted in elementary and heavy-ion reactions. It has established di-electron spectroscopy of baryon-rich matter. A unique worldwide facility, it is the combination of a pion beam in the few GeV/c momentum range and the HADES dilepton spectrometer. In the future, the HADES spectrometer will be transferred to the SIS100 nuclear reaction cave and complement the physics programme of CBM.

## CBM at GSI SIS100

The Compressed Baryonic Matter (CBM) pillar of FAIR aims to explore the phase structure of QCD matter in the region of the highest baryon density reachable in heavy-ion experiments, thanks to its large solid angle coverage. Crucial to the successful operation of CBM are developments in various technological areas, including i) a full free-streaming data acquisition system without hardware trigger even at the peak collision rates of 10 MHz, ii) self-triggered readout electronics, high-speed data processing, fast algorithms for online selection and iii) high-rate radiation-hard detectors, including several key technologies for which intense R&D is ongoing, such as timing detectors (MRPCs, LGADs), novel MAPS silicon detectors for tracking, and MPGD-based muon systems by GEM chambers.

# Hadron Spectroscopy, Structure and Interactions

## PANDA at FAIR

The future fixed-target PANDA experiment at FAIR will make use of a high-density cluster jet target, which requires thin nozzles to produce a pencil-like supersonic gas jet in the antiproton storage ring. To achieve the goals of this outstanding physics programme, the PANDA detector design features high-resolution tracking, calorimetry and particle identification, all covering nearly the full solid angle. This requires the development of several new types of detectors: low-mass tracking detectors with large-area GEM foils, a compact, fast and radiation-hard, high-resolution lead-tungstate crystal calorimeter, and fast focusing DIRC detectors. Some of these detectors are near completion or under construction and will participate in experiments at FAIR, ELSA, MAMI and JLab as part of the intermediate PANDA physics programme to evaluate the detector performance before installation in PANDA in 2030. A streaming DAQ system with internal rates above 106/s requires fast and highly integrated digital electronics and new algorithms, including machine learning, for online feature extraction.

## TJNAF

An upgrade of the detectors in all the TJNAFCEBAF halls (A, B, C and D) is ongoing to cope with the increased opportunities offered by the CEBAF energy upgrade to 12 GeV for a deeper probing of the nucleon and nuclear structure. A key option is offered by the CEBAF beam polarisation, which is complemented by polarised targets with substantial progress in the technique of transportable frozen spin targets. The instrumental upgrade of the hall equipment is diversified with a common effort to equip them all with PID-dedicated devices. The more complete one is the CLAS12PID system, comprising a fine-resolution ToF setup, two threshold Cherenkov counters and an aerogel proximity focusing RICH. Another remarkable piece of equipment at CLAS12 is the tracking system, where the vertex is equipped with cylindrical MicroMegas.

## ELSA

A new upgraded experiment will extend the present capabilities for the study of the emergence of complex QCD bound states made of up, down and strange quarks, making use of ELSA polarised electron beams with energies up to 3.2 GeV used in combination with polarised targets. The successful photoproduction experiments off the proton will be further extended to the neutron, where polarisation data is especially scarce. The experiment will make use of the recently upgraded Crystal Barrel calorimeter as well as the forward-endcap of the PANDA electromagnetic calorimeter for high-resolution photon measurement. In addition, a new vertex detector for charged particles will be installed in the space between the target and the calorimeter (several layers of silicon detectors based on depleted monolithic active pixel sensors). The momenta of forward-going charged particles will be measured by a magnetic spectrometer, consisting of several planes of high-resolution tracking detectors by GEM technology up and downstream of a new dipole magnet. Particle identification will be provided by a wall of scintillator bars measuring the time of flight of charged particles.

A new experiment at ELSA, Lohengrin, to search for dark photons exploiting the Dark Bremsstrahlung process is in its conceptional design phase. The experiment will feature ultra-high-rate electron tracking using thin pixel detectors in a permanent magnetic field together with fast photon and hadron veto calorimetry.

## MAMI and MESA

At the Institute for Nuclear Physics of the Johannes Gutenberg-University of Mainz, the Mainz Microtron (MAMI), a continuous wave (CW) electron accelerator with beam energies of up to 1.6 GeV, has been operating for many years. The beam intensities the degrees of beam polarisation, and the reliability of the machine are hallmarks of MAMI. Two experiments, A1 and A2, provide the basis for the experimental programme at MAMI, where the latter experiment is located at the tagged photon beam line of the accelerator. In addition to MAMI, the new MESA accelerator, a low-energy electron accelerator for electron energies of up to 155 MeV, will become operational in 2025. MESA will be a multi-turn superconducting electron accelerator which can be operated with the innovative concept of energy-recovery, providing in this way a CW beam with intensities of more than 1 mA. While the energy of MESA is therefore below the available energy range of MAMI, the beam intensity will be further increased by more than an order of magnitude, which opens the avenue for a programme of high-precision experiments at low energies. Three new experimental setups are prepared for MESA: the P2 and DarkMESA experiments to be operated in the conventional extracted beam mode, as well as the high-resolution spectrometer setup MAGIX for the energy-recovering beam mode of MESA.

The operation of a high-intensity energy-recovery beam in conjunction with a light window-less gas target is a novel technique in nuclear and particle physics and will reduce the effects of multiple scattering to a minimum. A high-density supersonic gas jet target for MAGIX has been developed and commissioned at MAMI and is ready to be used at MAGIX. The two identical spectrometers of MAGIX will have time projection chambers with GEM readout as focal plane detectors, which is also innovative. The physics programme of MAGIX ranges from measurements of electromagnetic form factors of nucleons to measurements of nuclear reactions of importance for nuclear astrophysics. Furthermore, as already proven at A1/MAMI, the high-intensity environment is also ideally suited for searches for particles of a dark sector, such as dark photons. The parameter reach will be extended at MESA towards lower energies.

The P2 experiment addresses the weak mixing angle as well as a novel method to assess the neutron skins of nuclei using parity-violating processes. To achieve this purpose, the spectrometer will consist of a solenoidal magnetic field around the target (liquid hydrogen in the case of the weak mixing angle) and an arrangement of fused silica bars and HV-MAPS tracking detectors. As the main observable will be a measurement of left-right asymmetry of spin-polarised electrons, a series of polarimeters needs to be constructed, for which the targeted precision will go beyond state of the art. By measuring the weak mixing angle at low momentum transfer with a precision of 0.15%, the comparison with the Standard Model prediction makes it possible to test New Physics mass scales of up to 49 TeV.

## AMBER at CERN

AMBER at CERN SPS on the M2 beam line capitalises on the long tradition of hadron-dedicated experiments, making use of a secondary beam line capable of providing high-intensity muon beams and hadron beams. The backbone of AMBER is the COMPASS double-folded open spectrometer, complemented by dedicated components for specific measurements: for the proton radius measurement: a high-pressure TPC acting as active target and telescopes of Si trackers in MAPS technology (ALPIDE); for the Drell-Yan studies, a segmented target interlined with Si-strip trackers to reduce the systematic effect due to energy loss and multiple scattering of the produced muons when crossing the target layers.

The second phase, after the coming LHC shutdown period, will benefit from an upgrade of the M2 beam line aimed at enriched and more intense kaon beams. From the instrumentation point of view, an essential ingredient will be the upgrade of COMPASS RICH-1 to make it compliant with restrictions in the usage of fluorocarbons due to their high Global Warming Power (GWP), and the need for a RICH-0 dedicated to hadron identification in the few GeV/c range.

### Box 9.2 : Novel projects in high energy nuclear physics require method and technology progress in detection tools

**ALICE 3** © ALICE Collaboration
**A novel detector for the high luminosity era at LHC**

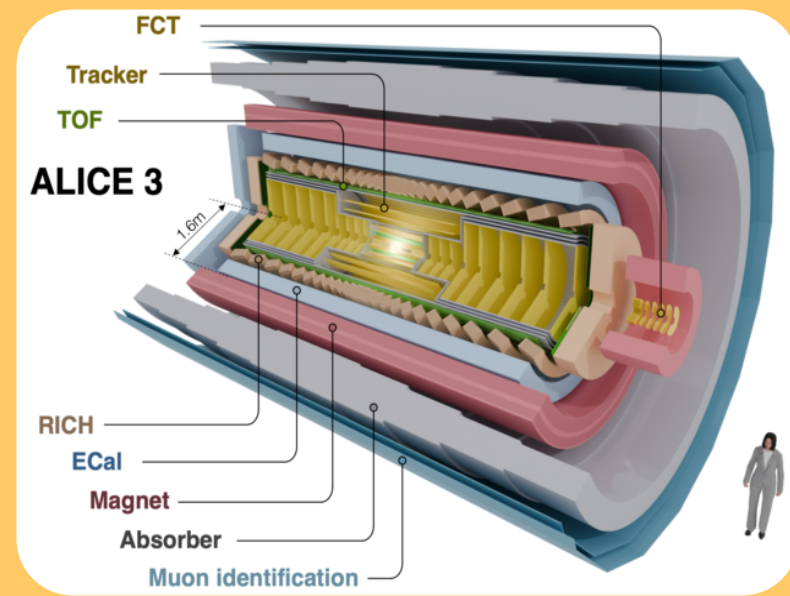

**ePIC** © EIC Project
**The project detector at the novel collider EIC**

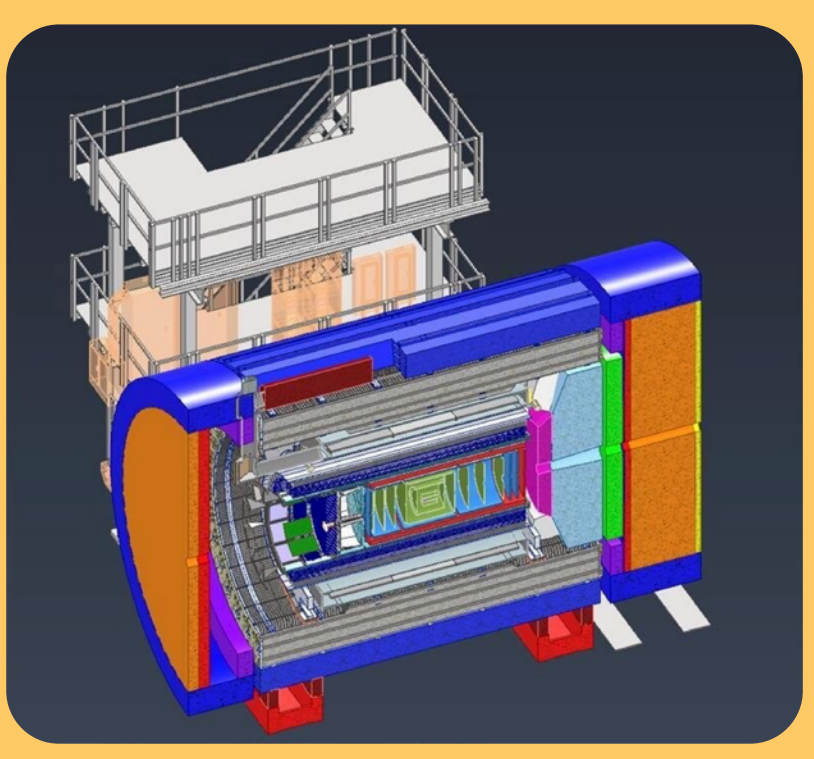

# Experimental methods - Symmetries and Fundamental Interactions

The experimental methods discussed in this section are those adopted for Symmetries and Fundamental Interaction studies and aspects of Nuclear Astrophysics.

Experiments aimed at the measurement of new physics phenomena beyond the Standard Model predictions need complex apparatus, high-dimensional statistics, long exposures, exceptional stability of all experimental parameters, reliable control of systematic uncertainties, and verification of results by independent methods. For most of them, radioactive background is a major limitation so, whenever feasible, experiments are located in deep underground laboratories.

## Neutrino properties

The energies of neutrinos produced by natural and man-made sources span several orders of magnitude, ranging from a few meV for relic neutrinos to several PeV for extra-terrestrial neutrinos. This diversity in energies and fluxes necessitates a wide range of detector technologies. As neutrinos travel from their source to the detector they undergo flavour oscillations. A range of neutrino parameters can be extracted by comparing fluxes of neutrinos of a specific flavour at the source and the detector. Current and upcoming oscillation experiments rely on the detection of the Cherenkov light generated by neutrino interactions in water or ice (KM3NeT, ICECube, Hyper-Kamiokande), of the scintillation light emitted by appropriate scintillation detector materials (SNO+, JUNO, T2K, NoVA), or the use of liquid argon time projection chambers (DUNE, ICARUS/MicroBooNE/SBND). Given their huge dimensions and anticipated performance, these experiments can simultaneously measure several neutrino sources and thus offer a broad physics programme, primarily devoted to shedding light on the fundamental neutrino properties. The study of beta decay, on the other hand, is the sole model-independent method for directly assessing the absolute value of the neutrino mass. The KATRIN experiment currently leads the effort, aiming to achieve a sensitivity of 0.2 eV on the neutrino mass by measuring the end-point of the $^{3}$H beta spectrum with unprecedented precision using a huge spectrometer. However, as statistics accumulate, KATRIN sensitivity will be quickly limited by systematics: the focus for the upcoming decade should, therefore, be on deepening the understanding of all instrumental effects as well as on developing more robust spectroscopic approaches (like Cyclotron Radiation Electron Spectroscopy, explored by PROJECT-8). Promising complementary efforts involve the use of cryogenic microcalorimeters for studying the end-point of the electron capture decay of $^{163}$Ho with high sensitivity, as pursued by HOLMES and ECHO. They can potentially quickly ramp up the currently limited number of detector channels, once the technology maturity is reached. Finally, the combination of existing experimental approaches such as a high-resolution cryogenic calorimeter coupled to a KATRIN-like spectrometer and a high-activity $^{3}$H gaseous source, holds promise in overcoming current sensitivity limitations posed by background and systematic effects.

## Neutrinoless double beta decay

Experimental methods for neutrinoless double beta decay ($0\nu\beta\beta$) focus on the precise measurement of the energy of the emitted electrons. However, due to the rarity of the process, the expected signal – a distinct monochromatic peak – is overshadowed by background events, mainly natural radioactivity and $2\nu\beta\beta$-decay. Consequently, it becomes crucial to increase the number of candidate $0\nu\beta\beta$-decaying nuclei under observation. Background reduction is the other significant challenge: it involves not only locating the experiment in a deep underground laboratory and ensuring meticulous material purification and selection, but also necessitates the use of state-of-the-art detectors with high energy resolution to enhance discovery potential and fast time response and effectively minimise pile-up resulting from the unavoidable 2νββ-decay events. The use of different candidate isotopes and technologies is essential in these searches, both for minimising uncertainties related to nuclear matrix element calculations and for maximising the potential for rejecting radioactive background. Cryogenic calorimeters are the only intrinsically multi-isotope approach, opening the way to study double beta decay simultaneously in different nuclei; experiments based on high purity germanium semiconductors, although limited to the study of Ge isotopes, are currently achieving the lowest background levels in the field; liquid xenon time projection chambers and scintillator-based setups are the most easily scalable in terms of target mass. Among the 35 isotopes that could undergo 0νββ-decay, the following are directly suited as the active medium in different ongoing and upcoming experiments: enriched germanium semiconductor detectors are used by LEGEND; natural $^{130}$Te is present in the $TeO_2$ cryogenic calorimeters currently taking data for CUORE, while $Li_2MoO_4$ enriched scintillating cryogenic calorimeters in $^{100}$Mo are planned for CUPID; enriched $^{136}$Xe in liquid form is chosen as target of a time projection chamber by nEXO, DARWIN/XLZD, while it is used in gaseous form in the high-pressure time projection chamber of NEXT. Alternatively, active detector materials can be doped or interleaved with the 0νββ-decaying isotope, like enriched $^{136}$Xe and natural $^{130}$Te as dopants in organic scintillator calorimeters (KamLAND-Zen, SNO+), and $^{82}$Se in solid-state tracking calorimeters (SuperNEMO).

## Dark-sector particles

Among the multitude of complementary detection approaches for dark-sector particles, we focus here on the ones with links to nuclear physics.

Searching for heavy galactic dark matter (WIMPs) with nuclear scattering is a well-established strategy with currently running (e.g. XENONnT, LZ, PandaX-4t, DEAP-3600, CRESST, DAMIC, SENSEI, PICO, NEWS-G) and future detectors (e.g. DARWIN/XLZD, DarkSide-20k, PandaX-xT, SBC, SuperCDMS@SNOLAB). Due to the persistent non-detection of WIMPs, the strategy has evolved from focusing mainly on WIMP dark matter in the 100 GeV region to covering as much mass range as possible to as low an interaction cross section as possible.

The experimental technique boils down to looking for nuclear recoils at energies below 100 keV in large ($O$(10 kg - 100 t)) target volumes. The main challenge is reducing backgrounds while increasing the target volume. After material selection, purification and shielding, important remaining backgrounds are neutrons from (α,n) and (μ,n) interactions in and near the target volume, as well as solar and atmospheric neutrinos interacting in the target. R&D has started, and should be further encouraged, in developing detectors with sensitivity to the direction of the incoming particle, to suppress solar neutrino background.

Indirect detection of galactic dark matter through its decay or annihilation into standard-model particles is another well-established technique. Cosmic ray experiments (charged particles, gamma rays, neutrinos) all look for such excesses. An important challenge and main link to nuclear physics is that interaction cross-sections with nuclei at the high energies relevant to cosmic accelerators are often not well known, so calculations of expected background levels have significant systematic uncertainties.

Dark Matter particles could be part of a 'dark sector' in particle physics, which may consist of several new particles and forces. Such particles could be created at accelerators. Detectors located near accelerator collision points are not suitable for detecting these particles directly. The particles would use up some of the collision energy and fly away unnoticed. However, their production can be inferred by looking for a larger-than-expected amount of undetected (missing) energy in the collision products. The dark particles can traverse shielding that stops any other particle (besides the neutrino). Beyond the shielding, far away from the collision region, some of them may decay into standard model particles that can be detected. Beam dump experiments make use of this to infer that dark-sector particles were created in the original collision.

## Muon-induced reactions

Complementary to existing neutrino and dark matter physics programmes are the searches for lepton number violating processes using muon-induced reactions.

The Mu3e experiment will search for the decay of a positive muon into two positrons and one electron. Using a two-phase approach and an innovative design, this requires a high-intensity muon beam, which will be provided in the second phase at PSI. Moreover, strong suppression of the accidental background will be achieved through excellent vertex and timing resolutions. These requirements will be met by a novel HV-MAPS silicon pixel detector in a 1 T magnetic field built around a double-cone target, plus two timing detector systems.

The MEG II experiment at PSI is focused on detecting the lepton flavour-violating decay of a muon into an electron and a photon. This is achieved by halting muons in an extremely thin target and subsequently measuring the emitted photons and positrons with the utmost precision. MEGII aims for sensitivity enhancement of one order of magnitude by exploiting the largest possible beam intensity available at PSI and by using improved detectors.

## The key role of nuclear physics

### Nuclear matrix elements

A significant source of uncertainty in the interpretation of 0νββ experiments is the value of the nuclear matrix elements (chapter Symmetries and Fundamental Interactions). Double charge exchange reactions (DCE) induced by heavy ions are a pivotal experimental method to support the extensive theoretical efforts in determining nuclear matrix elements for 0νββ decay since they probe the same nuclear states with $\sim 10^7$ times larger coupling constants. Nowadays, a deep investigation of DCE is being pursued at different laboratories worldwide, with a central role played by the NUMEN project at the INFN-LNS facility of Catania.

### Beta decay studies

Beta decays are used to generate the signal in some neutrino experiments and are a background in many dark matter experiments. Better understanding the shape of beta spectra would reduce the systematic uncertainties in these experiments.

### Scattering on nuclei

Incomplete knowledge of the cross sections and nuclear form factors for interactions between particles and nuclei is a source of systematic uncertainty in many experiments described here. Neutrino and dark matter experiments both need to know the cross sections of neutrinos interacting with the nuclei in the target materials. Dark matter experiments will eventually need to know nuclear form factors to higher precision.

### Rare isotope production

The experimental methods outlined in this section rely heavily on the availability of large amounts of enriched stable isotopes. Unfortunately, the scarcity of such materials, exacerbated by the geopolitical situation, poses a pressing challenge that urgently demands a solution.

Similar to the case of low-energy nuclear physics, **ion traps and laser spectroscopy** play a key role in experiments studying fundamental interactions and symmetries. Penning traps allow the highest resolution measurements of properties of elementary particles like the charge-to-mass ratio and magnetic moment of proton and antiproton performed at CERN-AD (see chapter Symmetries and Fundamental Interactions). Nested Penning traps are used to produce antihydrogen atoms from their constituents, antiproton and positron, which are then studied using microwave and laser spectroscopy for tests of CPT symmetry. Antihydrogen is trapped and laser-cooled in neutral atom traps to achieve the highest precision for laser spectroscopy. Similarly, the emerging field of spectroscopy of **radioactive molecules** requires trapping techniques to achieve the highest resolution, and beta decay studies as well as spectroscopy of highly charged ions are performed using trapped ions.

# Technological tools and related R&D - Low Energy Nuclear Physics

A recent overview of the technological tools in this domain of nuclear physics was provided by the conference Animma2023 [4].

## Neutron detectors

Current nuclear physics experiments and applications often require the detection of neutrons according to their specific needs, in a wide energy range spanning about fourteen orders of magnitude. Apart from passive detection methods like emulsions, plastic etchable sheets and activation foils, the three main technologies currently employed for neutron detectors are gas, scintillators and semiconductors. A gross sketch of the use of neutron detectors in several fields, along with the typical energy range of interest and the related technology, is shown in Fig. 9.1. The applicative fields, for instance radiation protection and medical and homeland security, basically cover the full energy range, whereas nuclear physics and fusion research target a more restricted high energy range. Neutron/material science is mainly focused on the (very) low energy range and requires detectors up to several square metres wide, typically based on gas technology, for instance at ESS.

The $^{3}$He crisis during the past decade has triggered research and development of alternative and better-performing neutron detectors which have been - and will be - beneficial to many new experiments and applications. The replacement of $^{3}$He with $BF_3$ has been strongly discouraged due to the toxic and polluting features of such a gas; the large gas detectors previously based on $^{3}$He are therefore currently being replaced by ordinary gas detectors featuring solid neutron converters like $^{10}$B or $^{6}$Li. Many concepts and geometries have been proposed, promising to significantly improve the detection efficiency and the spatial resolution of the new developments. To this end, arrangements of wire chambers, gas electron multipliers (GEM), straw tubes and others, in several geometries, are currently being studied in combination with solid neutron converters. New scintillating materials have become available and others are under development with quite interesting properties in terms of neutron detection and neutron/gamma discrimination. The discrimination, usually performed by analysing the pulse shape, is a peculiar feature required in particular for homeland security and nuclear physics. Scintillators are also being developed, sensitive to both thermal and fast neutrons, while other materials like stilbene have recently been rediscovered. Inorganic crystals like CLYC, CLLC, CLLB and several new plastic scintillators, most of them loaded with $^{10}$B or $^{6}$Li, are a few examples. Another very promising development line is on the solid-state detectors that exploit thin solid converters made of $^{10}$B, $^{6}$Li or $^{6}$LiF placed in front of silicon diodes to detect the charged particles produced by the neutron capture. This technology, best suited for small detectors, has already produced remarkable results and some commercial solutions are already available. The same detection principle can be applied to configurations where the silicon diode is replaced by a silicon carbide diode or a diamond to increase radiation hardness and speed.

Research and development on neutron detection technologies are currently quite active, as testified by an increasing number of thematic conferences and publications.

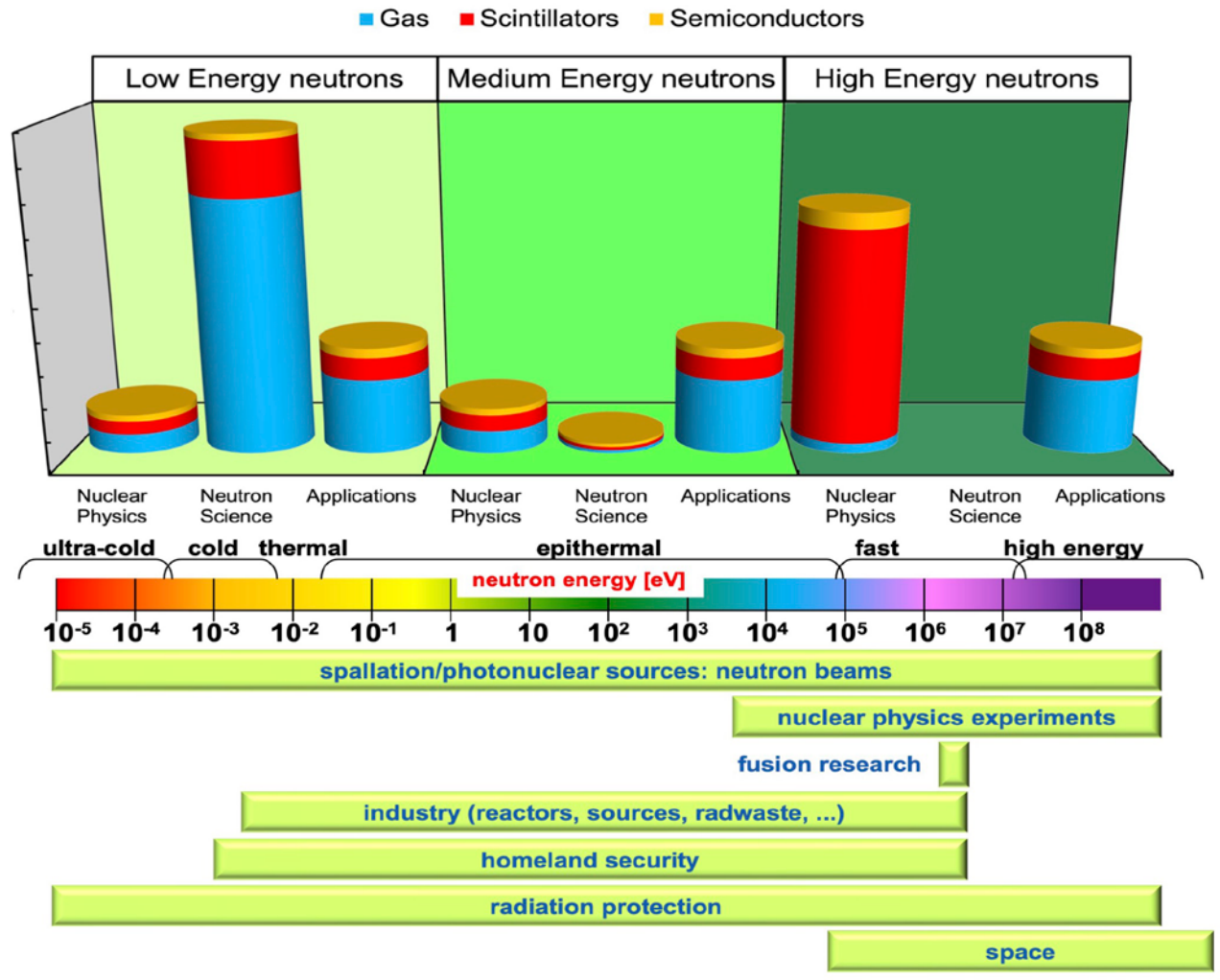

*Fig. 9.1: A gross sketch of the use of neutron detectors in several fields, along with the typical energy range of interest.*

## Technological developments for γ-ray spectroscopy

### Segmented Germanium detectors

New accelerator facilities require HPGe detectors able to cope with highly damaging radiation backgrounds and high counting rates. Recent advancements in HPGe contacts technology are paving the way for fabricating stable electron-collecting segments. These segments have the potential to offer better energy resolution for each γ-ray interaction in the crystal, achieving higher counting rates and decreasing sensitivity to particle radiation damage. The long-term objective is to develop this technology at an industrial level to produce complex coaxial HPGe segmented detectors. R&D has already been initiated at the laboratory level, thanks to the INFN's financial support for explorative research. Additionally, a research agreement with MIRION Technologies has been established to investigate the industrial feasibility of this technology. An important outcome of this R&D effort will be the acquisition of know-how for effectively repairing existing detectors, which are vital and costly tools in numerous experiments.

#### Developments for scintillators

Technological developments for scintillators encompass both scintillation material developments and signal readout.
Concerning the former, new highly performing scintillating materials (e.g. Elpasolite which can detect and identify both gamma radiation and neutrons) and devices for the readout of the scintillating light (e.g. SiPMs which, contrary to PMts, are insensitive to magnetic fields and allow the use of scintillators in the proximity of magnetic spectrometers) have been developed in recent years. The technological challenge is producing spectroscopic detectors that are also capable of providing additional information (position sensitivity, particle identification, etc.) using deterministic or AI techniques. In particular, Elpasolite scintillators like CLYC or CLLBC can measure the kinetic energy of an incident-monochromatic neutron from the signal amplitude. In the case of non-monochromatic neutrons, this is possible only in a restricted neutron-energy range. The increase in the energy range is a challenge for the future. On the other hand, technological developments in scintillator readout are also needed. Alternative readout systems, which may improve the position determination of the γ photon interaction, based on Silicon Photomultipliers (SiPMs), are under development for large-scale scintillator arrays like PARIS.

**Box 9.4: New accelerator facilities (ALPI-SPES, FAIR, HIE-ISOLDE, or SPIRAL2) require HPGe detectors able to cope with high damaging radiation background and high counting rates**

**High performance germanium detectors**

**© On the left, planar segmented HPGe detectors with PLM (Pulsed Laser Melting) contacts. On the right, coaxial segmented HPGe detector with PLM contacts and a flexible PCB board.**

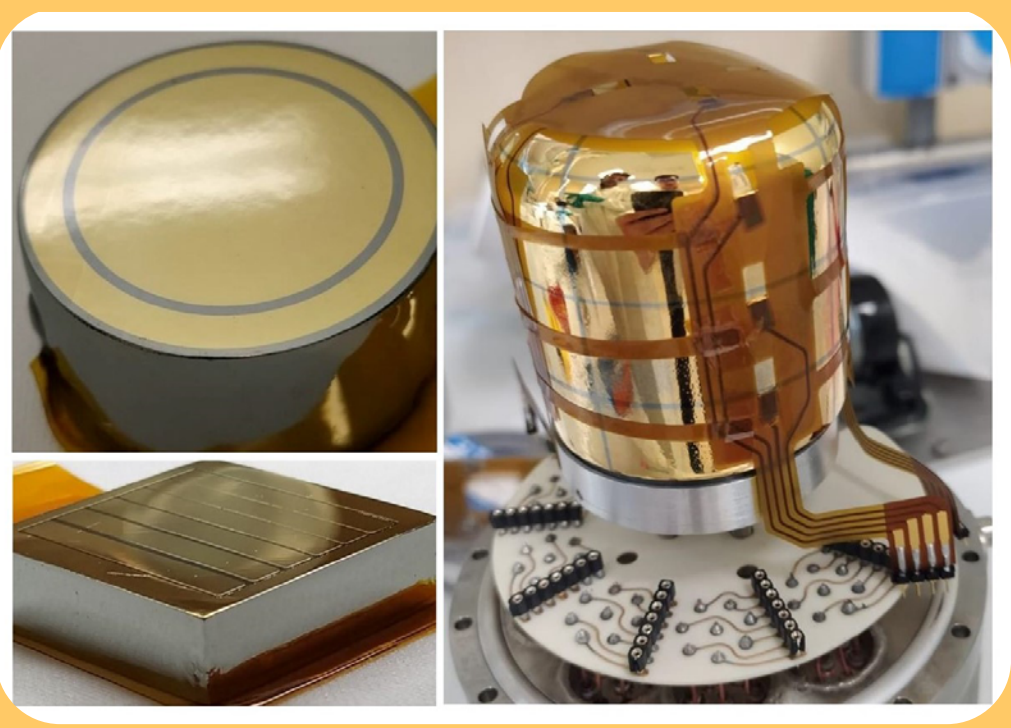

# Technologies for active targets

## Gaseous TPC

In low-energy nuclear physics, several initiatives have been set up for the design and operation of active gaseous targets, where the atoms of the detection gas are used as targets to study nuclear reactions. The main advantages of such devices are a solid angle coverage of nearly 4π and the possibility to increase the target thickness without suffering from the energy loss of the outgoing particles in the target (vertexing). A last-generation device is ACTAR TPC. Based on the principle of Time Projection Chambers and equipped with digital GET electronics ACTAR TPC permits sampling of the reaction volume into 128 × 102 × 512 voxels, hence imaging nuclear reactions in 3 dimensions. The amplification of the ionisation signal in ACTAR TPC is performed with micromegas. Such a device permits local increase in the detection dynamics by polarising the pads of the detection plane. Micromegas permits working with various gas mixtures over a wide range of pressures. The possibility of adding a Thick-GEM amplifier makes it possible to run with pure monoatomic gases. Operational since 2019, ACTAR TPC is currently installed in GANIL. Other detectors are being developed by the European community, optimised for different facilities: the ATS project at LNL will provide a reduced-scale ACTAR TPC detector with optimised performance for the post-accelerated exotic beams provided by the SPES facility, while the SPECMAT project, installed at ISOLDE, exploits the magnetic field provided by the ISS facility and a cylindrical configuration for high-efficiency studies of inverse-kinematics direct-reactions.

## Solid Active Targets

The resolution for γ-ray spectroscopy experiments with intermediate energy beams, i.e. velocities around β = 0.5, is affected by the knowledge of the velocity at which the emission took place. A significant contribution to this effect is given by the energy loss of the beam and ejectile particles in the target. To maximise the luminosity of the experiment, thick targets in the order of a few mm or g/cm$^2$ have to be used to access the most exotic nuclei. The ensuing γ-ray energy resolution degradation can be mitigated by replacing the passive target with an active one that records the position of the reaction. This is the goal of the LISA project. The fast ion beam from the Super-FRS impinges a stack of diamond detectors which induce direct proton knockout and nucleon removal reactions. The active target will detect the location of the reaction and thus the proper velocity β can be used in the Doppler correction. This will also make possible extraction of excited state lifetimes from their effect on the Doppler-corrected γ-ray energy spectrum.

## Ions traps and laser spectroscopy

Several ongoing developments aim to boost the use of ion traps and laser spectroscopy in nuclear physics studies. High-precision voltage control and measurement (the PENTATRAP and COALA experiments) and active voltage stabilisation systems (e.g. for MR-TOF MS setups) are needed to increase the measurement resolution and precision. For ion traps, this in turn brings a demand for high-resolution TOF detectors/acquisition systems and high-precision timing control, reaching into the sub-nanosecond range. Laser ionisation and spectroscopy require expansion of the range of achievable wavelengths, especially using solid-state lasers (with the notable development of a tuneable diamond Raman laser at ISOLDE). The use of gas jets demands high power/repetition-rate pulsed lasers, while the pulsed amplification of narrow-band light and the use of frequency-mixing techniques motivates the development of single-longitudinal-mode pump lasers. To meet the increased demand for low-energy beams, new gas and solid catcher technologies are being developed at GSI-FAIR, the Accelerator Laboratory in Jyväskylä and SPIRAL2-GANIL, aiming for high efficiency, low extraction time and the production of neutral radioactive beams for laser spectroscopy.

# Technological tools and related R&D - Isotope and target supply, unconventional targets

The continuously increasing demand for precise, accurate and reliable experimental nuclear data in scientific research fields like nuclear astrophysics, fundamental nuclear physics, nuclear medicine, geoscience, superheavy element research and others, results in an equally increasing need for high-quality samples (mainly for ion beams) and target material in sufficient amounts of either high enrichment in the desired isotope or of well-known isotopic composition, often custom-manufactured for the special envisaged application (Fig. 9.2). The desired samples very often represent or contain rare enriched stable or radioactive isotopes. Development efforts and funding for isotope provision and target fabrication need to be adapted to this demand as for other tools of nuclear research, like beam facilities, detectors, data acquisition and analysis hard- and software.

## Production and availability of enriched stable isotopes

The supply shortage of enriched stable isotopes (ESI), severely aggravated by the Russian aggression against Ukraine and its consequences, calls for a strategy to guarantee secure provision to European research institutions, without which fundamental research activities will come to a halt once the limited still-available reserves are consumed. The EURASIS [5] initiative by scientists from European research institutions, (GANIL/SPIRAL2; GSI/FAIR; HIL; ILL; IJCLab; JYFL; JGU Mainz, LNL, HI Jena/DESY; Van Swinderen Inst. f. Particle Physics & Gravity, University of Groningen, The Netherlands; Wigner Research Centre for Physics, Budapest, Hungary) aims for a secure supply to European research institutions, ideally as an international and discipline-overarching effort to solve this global problem, in synergy with all disciplines and communities concerned. This also applies to nuclear physics fields like, for example, research in nuclear medicine applications (lanthanides, e.g. $^{152}$Sm, $^{160}$Gd, $^{176}$Yb), M¨ossbauer spectroscopy ($^{57}$Fe/$^{57}$Co) and neutrino-less double-β decay ($^{48}$Ca;). The situation is particularly alarming for isotopes like $^{48}$Ca or the rare earth elements which can be enriched only by electromagnetic isotope separation (EMIS), with no installation of this kind existing in Europe. European research facilities have been receiving these materials hitherto mainly from Russian sources. Measures to mitigate this dilemma must include the setting up of an EMIS facility in Europe.

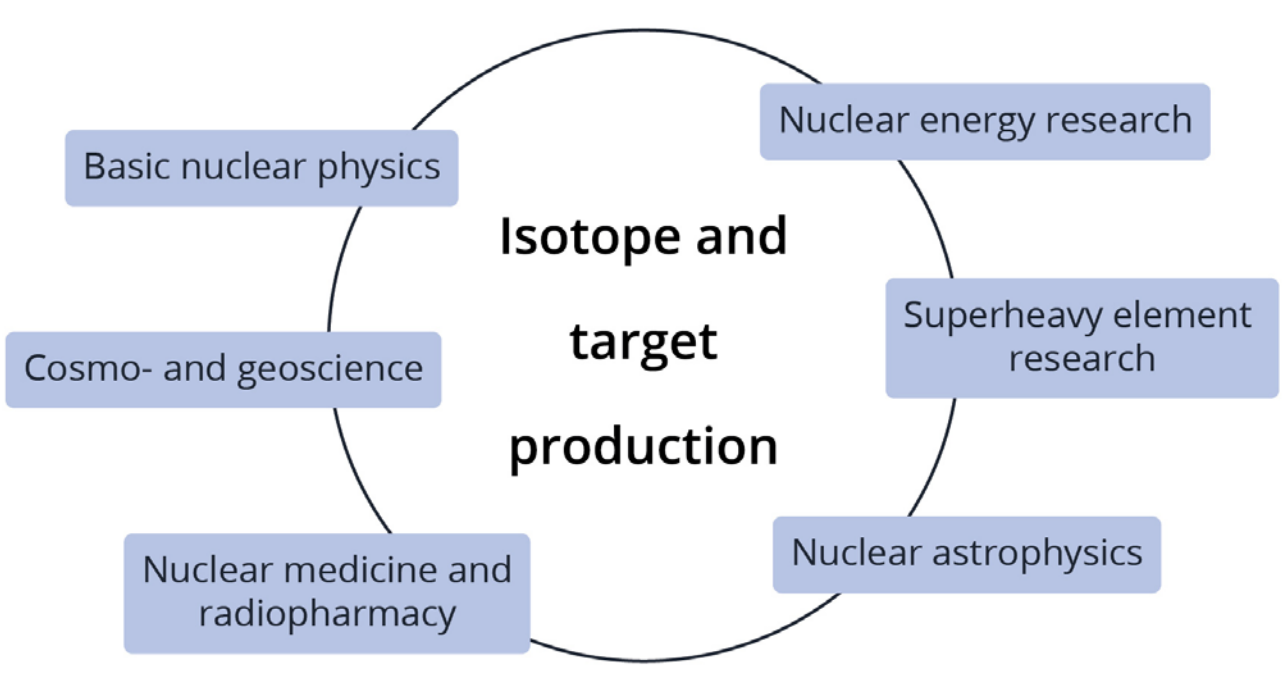

*Fig. 9.2: Isotope and target production and their fields of application*

To run their accelerators and serve their experimental installations of separators and complex particle and photon detection set-ups, European heavy ion accelerator laboratories operating cyclotrons (Univ. Jyväskylä, GANIL Caen, LNS Catania, HIL Warsaw, Univ. Groningen, LNL Legnaro) as well as linear accelerators and electrostatic machines (GSI Darmstadt, LNL Legnaro, GANIL Caen, IJCLAB/ALTO, Cologne Univ.) depend on enriched material. High-current stable beams are essential in the study of low cross-section processes. Hence, stable isotopes, often enriched material of low natural abundance for ion sources (up to tens of grams) as well as target material (several mg), are essential for the operation of heavy ion accelerators. Given the advent of high-intensity accelerators presently coming online and being planned for the mid to long-term future, for example the LINAC of GANIL/SPIRAL2 or the HELIAC project of GSI/FAIR, a secured ESI supply will remain important for many years to come.

Only light elements and mercury can be efficiently mass-separated by exploiting isotope effects via fractional distillation or chemical methods. Today, all other elements are either separated by EMIS or by the more cost-efficient centrifuges which can only be used for gases and volatile chemical compounds. Worldwide, the main supply of isotopes enriched by EMIS stems from two facilities in Russia (EHP Lesnoy and IPPE Obninsk). In the USA, DOE is now in the process of replacing the historic calutrons in Oak Ridge with newly developed EMIS machines. For historical reasons, Europe does not have high-current electromagnetic separators (only low-current but high-resolution variants for specific research applications such as SIDONIE at IJCLab, RISIKO in Mainz or off-line use of different ISOL facilities). All these isotopes have to be imported, like $^{176}$Yb (required for the production of $^{177}$Lu for cancer treatment), $^{48}$Ca (for nuclear physics research and ν-less double beta-decay experiments), but many other isotopes as well, namely all lanthanides, as well as other calcium isotopes ($^{43,44,46}$Ca), most platinum group metals (e.g. $^{96}$Ru, $^{102,110}$Pd, $^{190,192,194}$Pt) etc. The consequence is a highly competitive situation, where, for example, the exploding demand for “medical” $^{176}$Yb puts a heavy strain on nuclear physics supplies like heavily demanded $^{48}$Ca, and vice versa. The mismatch between steadily rising demand due to novel medical applications and ramp-up of new nuclear physics facilities on the one hand, and very insufficient supply on the other, has already led to a severe bottleneck in the past years. This has just been exacerbated by dwindling supplies from Russia since February 2022.

In the future, resonant laser ionisation and other laser-based techniques could complement electromagnetic enrichment for certain elements. Together with other enrichment schemes, they could be used in a combined, staged approach, using two (or more) methods in sequence to efficiently reach the highest final enrichment.

## Production and separation of radioactive isotopes

In general, radioactive isotopes can be produced at high-flux facilities like the research reactor at ILL Grenoble, the proton accelerator HIPA or the neutron spallation source SINQ at PSI. To obtain end-products with high specific activity, often enriched stable isotopes are needed as seed material. The desired radioactive products are extracted by radio-chemical separation processes from the bulk of irradiated material. Only a few institutions in Europe, such as JRC-Geel (Belgium), JGU Mainz and GSI Darmstadt (Germany), CEA (France) or PSI (Switzerland) can provide the complex equipment to handle those radioactive samples.

Isotopically pure samples also require, besides the prior chemical separation, mass separation. In Europe there are only a few installations like the RILIS installations at CERN-ISOLDE and MEDICIS-ISOLDE facilities at CERN or the RISIKO separator of JGU Mainz providing high mass resolution, and very limited capabilities regarding larger material quantities. The increasing demand for isotopically pure radioactive samples urgently calls for new innovative techniques and, in particular, for the development of a dedicated facility for the offline separation of radioactive samples. Both online and offline isotope separation will be applied at ISOL@MYRRHA, where the offline separator will be completed and commissioned in 2024. The ISOLPHARM facility at SPES will also provide mass-separated radionuclide samples for medical research. PSI plans to develop an installation similar to MEDICIS-ISOLDE, called TATTOOs, within the planned upgrade of the accelerator facilities (project IMPACT) included in the Swiss Roadmap for Research Infrastructures for the years 2025-2028, for the separation of radionuclides for medical research. In addition, a complementary off-line isotope separator also to be built at PSI is currently envisaged. This technique has the additional advantage that the desired isotope can be implanted directly into a suitable backing, thus avoiding the production of salts or oxides, thick layers or inhomogeneities. The EU-funded project SANDA, as well as its hopefully coming follow-up project APRENDE, should obtain part of the cost of the design, development and construction of a highly-efficient mass separator.

## Target manufacturing

The specification of a target is determined by the requirements of the underlying physics (nuclear reactions, unwanted by-products), signal detection capabilities and the specific configuration of the experimental setup. Front-end experiments in the field of nuclear data are not uniform. Every single experiment is unique with varying specific parameter sets. The applied techniques are by no means standard, but specially designed for a given application. Targets are to a large extent unique, requiring often complex custom production procedures. The standard manufacturing techniques, already available and used in the field (e. g. electrodeposition, evaporation, electrophoretic, conventional sintering and brazing, mechanical rolling, laser melting etc.) meet requirements like uniform layer thickness, high heat transfer effectiveness, good mechanical properties and high chemical purity. Actinide targets are important, for example for SHE or fission and nuclear data studies, and require efficient and dedicated production methods due to their scarcity and radiochemical constraints. Moreover, when highly enriched actinide and other radioactive isotope targets (very costly and/or rare) are needed, it is essential to limit material losses to a minimum. Ideas to improve already available target production techniques are, for example: Vacuum Deposition –

PVD, Spark Plasma Sintering (SPS), High-Energy Vibrational Powder Plating (HIVIPP), ion implantation and drop-on-demand deposition.

The continued development of existing and new methods is essential. In the European framework, this is pursued within the SANDA EU-project (H2020-nfrp2018) which aims at maintaining and developing European facilities for target production and fostering their network, while target development is also a prominent part of the EU-funded EURO-LABS project.

## Box 9.3: Internal targets play a key role for a number of upcoming experiments

**PANDA at COSY © University Münster**
**Operation of the Münster PANDA Cluster-jet Target at the proton synchrotron COSY at FZ Jülich. The cluster generator is installed in PANDA geometry, i.e. at a distance of 2.2 metres above the COSY interaction point and provides hydrogen target thicknesses of > 2x10$^{15}$ atoms/cm$^{2}$.**

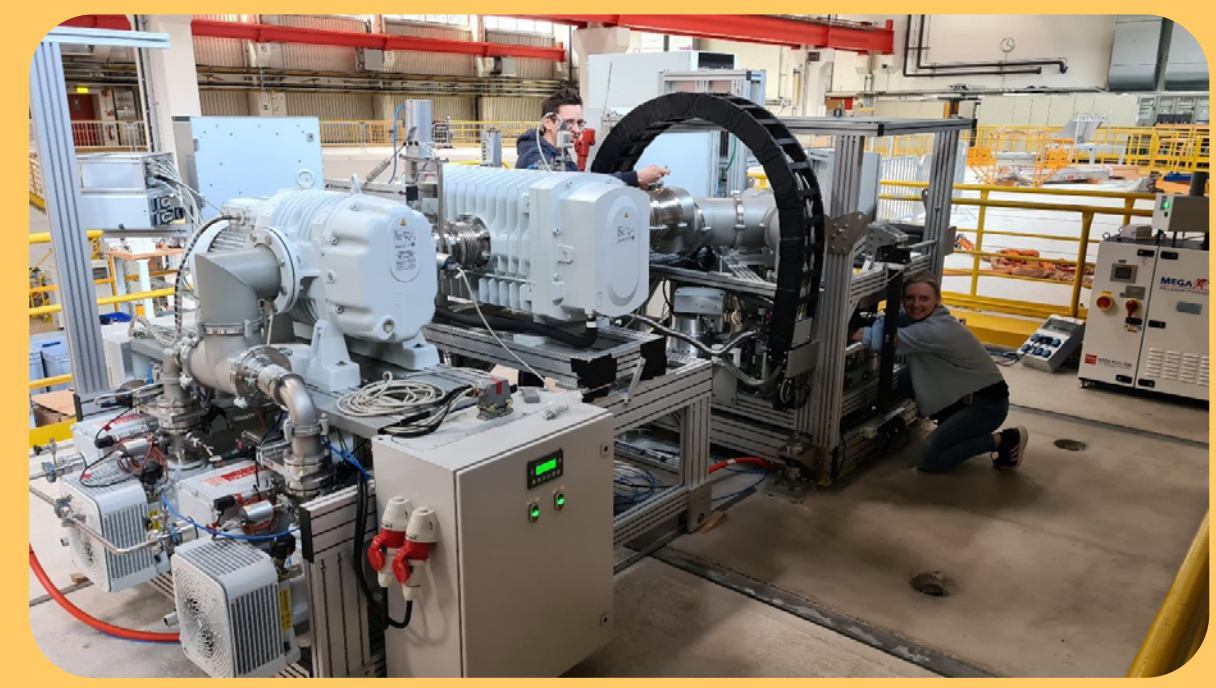

## Cryogenic targets

Light particle targets remain a major block for direct reaction measurements. One- and two-neutron transfer reactions like (d,p),(d,t) and (p,t) can be performed using (deuterated) plastic targets at the expense of excitation energy resolution and fusion-evaporation background. This difficulty has triggered the CHyMENE project of a windowless proton (deuteron) target. The counterpart reactions for proton transfer represent an experimental challenge and have been a severe obstacle to the study of proton shell evolution. The one-proton transfer reaction (d,n) suffers from difficulties in the detection of neutrons (low efficiency, low energy resolution, poor granularity for angular distributions). The (d,$^{3}$He), ($^{3}$He,d) and ($^{3}$He,p) transfer reactions represent the best alternatives. While the difficulties in identifying 3He particles for the pick-up reaction (d,$^{3}$He) have been overcome with the new generation of silicon arrays and the use of pulse shape analysis, the unavailability of sufficiently thick $^{3}$He targets is a clear showstopper for studying one-proton stripping reactions ($^{3}$He,d) and ($^{3}$He,p). The HeCTOr target has been designed for low-energy nuclear reactions and reaches an area density of 1020 atoms/cm$^{2}$. It was coupled to MUGAST, AGATA and VAMOS. In Europe, two follow-up projects (CIADIR in Italy and ATRACT in France) are being pursued to build thick $^{3}$He cryogenic targets based on a pulse tube technique.

## Polarised solid-state targets

Polarised solid-state targets used in European experiments at ELSA, Bonn, MAMI, Mainz and CERN explore spin-dependent phenomena. COMPASS at CERN has operated a polarised frozen spin target for nearly two decades. The AMBER collaboration plans polarised target measurements from 2028. European engagement extends to the Jefferson Lab, USA, studying nucleon spin structure. Targets comprise hydrogen (deuterium, $^{3}$He)-rich, highly polarised materials requiring advanced cryostats and superconducting magnet systems. Maintenance and adaptation of existing setups for future experiments are essential. The know-how should be preserved as well as the R&D for improved devices. Joint patents with the University of Bochum and General Electric exist. Bochum is developing a Q-Meter for polarisation determination to be deployed in the USA, Germany and Japan. To ensure continued operation and long-term storage, strengthening the groups involved and training a new generation of experts is recommended.

## Supersonic gas-jet targets

High-performance gas-jet, cluster-jet and micro-droplet targets are internal windowless fixed targets at accelerators and storage rings, or used at intense extracted beams.

They can be used with materials from hydrogen to heavy gases (like xenon) of the highest purity and they are key components for many experiments in hadron physics, enabling the recording of high-precision data with high luminosity at simultaneously low target density. This ensures that undesired secondary reactions of the ejectiles in the target itself are strongly suppressed and beam particles that do not interact with the target are again available for reactions during the next revolution in the storage ring. In particular, experiments with rare, expensive or radioactive reaction partners are sometimes only possible when these are shot as accelerator beams onto internal jet targets. Examples here are the planned experiments PANDA and KOALA at the future antiproton storage ring HESR (FAIR). Experiments of high luminosity can also be carried out at facilities using extracted accelerator beams, as the target densities can cover several orders of magnitude as foreseen for the future MAGIX experiment at MESA (Mainz). Prominent examples of precision experiments at storage rings with highly charged ion beams include the internal gas-jet target at the ESR (GSI) and the planned SPARC target for the HESR (FAIR). In addition, special storage cells in combination with molecular beams or polarised atomic beams offer the possibility of providing extended (un)polarised targets at the interaction point. Current developments with innovative storage cell technologies (SMOG2) are taking place, for example the LHCb experiment at the LHC (CERN). In addition to their use in conjunction with hadron and lepton beams, gas-jet, cluster-jet and droplet targets are also used at high-power lasers to develop innovative methods for generating high-energy electron or ion beams. For example, experiments at the Helmholtz Zentrum Dresden Rossendorf (HZDR) are performed at laser-driven plasma accelerators, in which micro-jet beams of liquid hydrogen are used as targets.

Extensive studies are required for gas, cluster and droplet jets on their optimised generation, on their properties as well as on their interchangeable operation through minimal modifications. This will enable universal and, in particular, more cost-efficient experimental setups that allow an even wider range of measurements.

# Technological tools and related R&D - High Energy Nuclear Physics

Up-to-date overviews of the technological tools in this domain of nuclear physics have been provided by recent dedicated conferences [6].

## Gaseous detectors

Despite increasing competition from solid-state detector developments, gaseous detectors are still indispensable in nuclear physics experiments. Low material budget, radiation hardness, operation in magnetic fields and the possibility to cost-effectively equip large areas make the use of gaseous detectors a very attractive solution. Future developments encompass a wide range of areas including miniaturisation, high-rate capabilities, radiation tolerance, industrialisation, advanced imaging techniques, eco-friendly gas mixtures or mechanical arrangements guaranteeing no dispersion of gasses with high global warming power in the atmosphere, and new detection concepts. These advancements promise to enable more precise and efficient detection capabilities including improvement of their spatial and time resolution. To fulfil the constantly increasing needs of nuclear physics experiments, all major gaseous detector technologies must be further developed in the coming years.

Over the next years, several **wire-based** detectors need to be constructed, including large area muon trackers for the NA60+ experiment, based on MWPC technology or a self-supporting, low-mass central straw tracker with 4D-tracking and PID for hadron physics (PANDA at FAIR) based on straw tubes. Developments of dedicated electronics will focus on leading and trailing time resolution for 4D measurements and for d*E*/d*x* with time-over-threshold and even cluster counting, with the measurement of the arrival time of each electron cluster generated in the gas ionisation process.

The introduction of **Micro-Pattern Gaseous Detectors (MPGDs)** has been made possible by the industrial development of modern photo-lithographic technology. They represent a natural development of gaseous detectors towards further improvements of the wire-based readout, partially overcoming its limitations. The ongoing, cross-disciplinary R&D on novel amplification structures, new coatings material (e.g., resistive layers, low outgassing materials) and gas mixtures make MPGD concepts adaptable to various operational conditions and performance requirements. One of the promising developments in this sector is represented by µ-RWELL chambers, consisting of a single-amplification stage resistive MPGD that combines, in a unique approach, several solutions and improvements achieved in past years in the consolidation of better-established technologies as GEMs and MicroMegas. From these R&D developments, improvements of stability in heavily irradiated environments and more simple construction procedures are expected, given an easy technology transfer to industry.

The outstanding capabilities of **Time Projection Chambers (TPC)** offering fine space resolution with multiple trajectory sampling and particle identification capabilities are also in the domain of gaseous detector technology, providing cost-effective solutions characterised by great performance figures. They are read out by gaseous sensors. MPGDs are especially interesting in this respect, as this technology can overcome the intrinsic rate limitations introduced by the gated readout, as already proven by the recent TPC upgrade at ALICE. Several new developments are ongoing in the area of TPCs for nuclear physics, including active-target TPCs, high-rate tracking and beam monitoring TPCs.

The main challenges addressed include i) high pile-up and space-charge fluctuations, ii) ultimate resolution, and iii) wide dynamic range and discharge stability. New structures (e.g. GridPix sensors, resistive micromegas), methods (e.g., hybrid MPGD stacks, gas mixture optimisation) and dedicated electronics (e.g. wide dynamic range ASICs) are proposed to further push the boundaries of the stable operation and ultimate resolution. Examples of ongoing and future developments include a high rate GEM-TPC in twin configuration for the Super-FRS at FAIR, HYDRA TPC for hyper nuclei studies with R3B at FAIR, ACTAR TPC development in GANIL, PUMA pion tracker at the antimatter factory of CERN, or a high-pressure TPC for AMBER at CERN SPS.

To cover large areas (~100 m$^2$) with precise time-of-flight detectors, a gas detector technology consisting of timing **(Multi-Gap) Resistive Plate Chambers (RPC)** is used in currently performed and planned experiments (e.g. CBM at FAIR). The main challenges of the future gas-based timing detectors include uniform response in terms of high-rate capability and timing precision over a large ToF detector area while operating with eco-friendly gases. R&D must continue towards an ultimate timing precision of ~20 ps in extended systems and a rate capability of up to 100-150 kHz/cm$^2$, necessary for systems in high radiation environments. RPCs by lower resistive material are being explored to these ends. A timing precision below 15-20 ps reaches the level of avalanche fluctuations and thus requires novel, very low-noise front-end electronics which can cope with high input capacitance and provide a large dynamic range. An interesting development based on novel surface resistivity electrodes is a Surface Resistive Plate Counter (sRPC). The electrodes of the sRPC exploit the well-established industrial Diamond-Like-Carbon (DLC) sputtering technology on thin polyimide foils, already introduced in the manufacturing of resistive MPGDs. Besides its use in particle and nuclear physics experiments as a timing detector, this new technology could be exploited as a thermal neutron device for homeland security applications.

## Silicon detectors

Silicon detectors provide the highest resolution and granularity at the same time. This characteristic is indispensable when extremely fine space resolution O(10) µm is required for the accurate reconstruction of short-leaving particle decays. The accuracy of the reconstruction is related to the pixel size. In addition, the particles are deflected by the material of the sensors, which in turn has a negative effect on the precision of the particle trajectory reconstruction, requiring an effort to limit the material budget of the detector stations. These considerations are the basis of one of the major development efforts in the silicon detector field (i). When occupancy and event rates are as high as at High Luminosity LHC, fine time information has to be associated with the fine space measurements developed in a second set of dedicated approaches (ii). To push the boundaries of silicon pixel tracker applications, the **CMOS Monolithic Active Pixel Sensors (MAPS)** are being developed continuously in international research consortia (R&D sector i). At 20 µm, the MAPS pixels are about five times smaller than those of the previous generation of sensors. The sensors, together with holding structures, supply cables and cooling systems and are integrated into a detector system that has an area of several square meters. At the same time, they can be thinned into a flexible silicon film with a thickness of 50 µm for extremely low material-budget solutions. The current construction technology limit for the size of individual MAPS is about 2×3 cm$^2$. Thanks to new capabilities in the CMOS industry it seems possible to overcome this limit, for example by exploiting the stitching process (65 nm TPSCo CIS process). Related research is aimed at bending future MAPS into self-supporting cylindrical structures and placing them particularly close to the interaction point. At the same time, the speed (particle rates up to about 100 MHz/cm$^2$) and life expectancy (tolerance to radiation doses above $10^{14}$ 1 MeV neq/cm$^2$) of the sensors should be further increased and their power consumption, thus cooling requirements, further reduced down to O(10) mW/cm$^2$. It is anticipated that the additional integration density provided by the 65 nm process may allow for improved granularity of CMOS sensors. Alternatively, it may serve to improve in-pixel functionalities including time stamping and data bandwidth, which may be exploited in high-rate tracking systems. Thanks to these features, MAPS are considered a leap innovation in the field of nuclear and particle physics detectors. They are used and planned in several nuclear and hadron physics experiments for, among others, heavy flavour tagging and low momentum tracking (e.g., ALICE, CBM, ALICE3). The development of wafer-thin sensors that become flexible and may be bent at radii in the order of 2 cm is currently a baseline for the replacement of the three innermost layers of the ALICE tracker in the ITS3 upgrade; the same approach is envisaged for the ePIC experiment at EIC. Using MAPS-based trackers in the target spectrometer of PANDA, or the R3B experiments, would allow for proton tracking and vertex reconstruction and presumably push the quality of the experiment to a level not reachable with the planar detectors.

Currently developed silicon time detectors (development sector ii) enable the planning of 4D trackers (3 spatial + 1-time coordinate). For example, the emerging technology of **low-gain avalanche diodes (LGADs)** has the potential to complement the position resolution of MAPS thanks to their excellent time resolution of O(10) (20-30 ps) and potential spatial resolution of up to 20 µm. The combination of MAPS and LGADs would enable unprecedented precision in both position and time measurements. To construct a large-scale LGAD-based 4D-tracking system, substantial R&D activities are required, including *i*) characterisation and usage of the next generation of LGAD strip sensors which feature a fill factor close to 100%, *ii*) dedicated module design to maximise the field-of-view while limiting the material budget, which is, in any case, affected by the sensor power requirement and related cooling system, and *iii*) development of a fast and compact

ASIC that can deal with the increased number of readout channels of a large-area particle detector system while still meeting the demanding time precision requirements. LGAD technology is being implemented in nuclear physics experiments (e.g. HADES start detector), beam monitor systems (e.g. at S-DALINAC in Darmstadt) or planned for future detectors (e.g. ePIC at EIC) and detector upgrades (one of the options for the future ALICE3 ToF system). In order to achieve a time resolution below 20 ps, further developments are needed to reduce the active volume thickness and optimise implantation parameters. One important effort specific to the time-of-flight is to develop long-strip AC-LGAD sensors to minimise the material budget in progress for the barrel detector of the ePIC experiment (EIC). Excellent rate capabilities and radiation hardness of LGADs make them a perfect candidate for the next generation of ion computed tomography scanners used in ion beam therapy.

## Ring Imaging Cherenkov Detectors

The identification of charged particles over a large momentum range is an important requirement for many current and planned nuclear physics experiments, in particular in the domain of nucleon/nuclear structure and spectroscopy. Cherenkov imaging detectors, namely RICH (Ring Imaging Cherenkov) and DIRC (Detection of Internally Reflected Cherenkov light) counters, utilise the prompt emission of the Cherenkov light in the radiator medium to provide charged particle identification. The momentum coverage can be tuned to experimental needs by selecting the refractive index of the radiator medium, where gaseous radiators (e.g., $CF_4$, $C_4F_{10}$, Ne) are available for high-momentum PID, silica aerogel or liquid fluorocarbons ($C_6F_{14}$) for the intermediate momentum range, and solid radiators (fused silica or $LiF_2$) for lower momenta. While the identification of charged hadrons is the primary goal of many RICH counters, they are capable of providing useful separation of electrons from charged pions at lower momentum and some are designed to separate charged pions from muons or electrons from hadrons. Experiments may include multiple RICH/DIRC counters with different refractive indices, sometimes in combination with time-of-flight techniques, to achieve clean PID over a momentum range of several orders of magnitude.

Examples of running or future experiments with Cherenkov imaging detectors include LHCb and ALICE3 at CERN LHC, AMBER at CERN SPS, CLAS12 and GlueX at TJNAF, Belle-II at KEK, HADES/CBM and PANDA at FAIR and ePIC at BNL EIC.

R&D for future RICH counters is focused on three areas: radiators, photon sensors and readout electronics. Despite their excellent match to the radiator requirement for gaseous RICHes, fluorocarbons will soon need to be replaced due to their high global warming impact. One possible alternative, for a similar momentum coverage, could be Argon, though the gas would need to be pressurised at a few bars. The availability of optical-quality aerogel is a concern. The commercial production of hydrophobic aerogel from Japan has been paused for an extended period, and of aerogel for Belle-II from a combined effort of industry and Japanese Academia, while the availability of aerogel from Novosibirsk remains questionable. The establishment of new commercial sources of aerogel for RICH counters would be highly desirable. Due to their potential ability to tune the refractive index to the exact requirements of an experiment, metamaterials, including photonic crystals, are an interesting candidate as radiators for future RICH counters. To increase tolerance to high event rates and deal with event pile-up, RICH counters at future high-luminosity facilities will require photosensors with small pixels and high-precision timing with single-photon sensitivity in strong magnetic fields, and an environment with high neutron fluxes and large doses of ionising radiation. SiPM technology has improved significantly in recent years, making them a promising candidate for future RICH and DIRC counters. The main challenge is the sensitivity to radiation damage, requiring additional technical infrastructure for operation at cryogenic temperatures and annealing at high temperatures, as well as precision timing to keep backgrounds at an acceptable level. Digital SiPM designs, based on CMOS technology, offer the possibility to mask individual noisy channels. Recent technological advances in the production of microchannel plates have increased the MCP-PMT lifetime to the level where they are suitable for RICH counters in high-luminosity and high-rate experiments. The main challenges are the development of MCP-PMTs with small pixels and high channel density, and reducing the cost per channel. Radiation-hard front-end readout electronics need to be developed in ASIC or FPGA to match the signal characteristics of the SiPM or MCP-PMT sensors. The fast timing for small single-photon signals at low gain and the high granularity of the sensor channels, combined with the need to keep the power consumption minimal, make the readout design challenging.

## Calorimeters

Calorimeters are widely used in high-energy nuclear physics to measure the energy and direction of particles. In electromagnetic calorimeters, photons, electrons, and positrons produce an electromagnetic shower which permits measuring the energy and, by segmentation of the calorimeter, measuring the direction of the particles to high precision. They also contribute to electron/hadron separation. Charged pions, kaons, $K_L$ as well as protons and neutrons typically interact in nuclear interaction and produce hadronic showers, where only a limited energy resolution is possible.

Excellent energy resolutions of about 2.5% at 1 GeV are achieved in homogeneous electromagnetic calorimeters consisting of scintillating crystals (which can have a long lifetime) like the Crystal Ball NaI(Tl) detector (constructed at SLAC, 1978, now at MAMI, Mainz) and the Crystal Barrel CsI(Tl) detector (CERN, 1988, now at ELSA, Bonn).

**Future developments** of scintillating material require the optimisation of mainly three parameters: **radiation hardness**, to reach the frontier of rare processes in high luminosity experiments, **light yield**, to achieve excellent energy resolution for low energy particles, **fast decay times** for high-rate experiments, as well as to achieve time resolutions in the picosecond region, e.g. for time-of-flight PET in medical applications. It is recommended to intensify systematic research on new materials for fast and cheap scintillators with high light output. Examples are new garnet materials, like GAGG and GYAGG, where their scintillation kinetics can be optimised by doping; glass ceramics, a cheap material suitable for large-volume detectors; and plastic scintillation materials based on polyethylene naphtholate (PEN) or polystyrene-based extruded plastic scintillator material doped with PTP and POPOP to provide a cost-effective alternative to previous plastic-scintillator calorimeters. Developments and studies are performed in cooperation between universities and European industrial crystal, glass and polymer-based scintillator experts like Crytur (Turnov, Czech Republic), Schott (Mainz, Germany) and ISMA (Kharkiv, Ukraine). Fast, compact and radiation-hard lead tungsten (PWO) scintillators are being developed in cooperation between university labs and a European company (Crytur in Turnov, Czech Republic). The company is the only producer worldwide and is essential in providing high-quality PWO-II crystals for the EIC and PANDA.

For certain physics cases, a more cost-effective sampling calorimeter type is chosen, where metallic absorbers producing the electromagnetic showers are interleaved with cheap organic scintillators to determine the energy of the initial particles. Due to less active volume, the energy resolution is reduced.

The ALICE forward calorimeter upgrade (FoCal, 2026-28), which requires only moderate energy resolution but unprecedented two-shower separation capabilities, consists of a high-granularity Si-W electromagnetic calorimeter and a spaghetti-style copper scintillator hadronic calorimeter. The high granularity of the EM calorimeter is achieved by a combination of Si-pad sensors (similar to the CMS HGCAL) and Si-pixel sensors (using the ALPIDE sensors developed for the ALICE ITS2). This pixel technology with its extreme granularity of ≈ 30 × 30 µm² has also been used in recent developments of a digital electromagnetic calorimeter (DECAL) which has shown very good performance. This kind of technology is a serious option for future high-energy experiments, both in nuclear and particle physics. For the EIC, a sampling calorimeter with scintillating fibres and tungsten powder will be adopted in the forward endcap. Scintillating glass blocks with projective geometry form the barrel calorimeter, whereas a hybrid calorimeter with imaging layers followed by a sampling section is considered a potential alternative.

Due to **fast** and **highly integrated electronics** like FPGAs (field programmable gate arrays) it has become possible to implement feature extraction algorithms directly at the front-end electronic, as planned for the PANDA experiment. The Belle-II experiment plans to facilitate fast digitisation and highly integrated FPGAs to apply pulse shape discrimination techniques. Advanced machine learning techniques will play a larger role in the future to extend the particle identification capabilities of calorimeters.

## Silicon photomultipliers

Silicon Photo Multipliers (SiPM), introduced towards the end of the 20th century for application in fundamental research, are nowadays largely used in a much wider domain including applications of social interest, such as medical imaging, and coupled with Light Detection And Ranging (LiDAR) in control systems and in industry, which largely absorb more than 50% of the production.

Nowadays, in fundamental research, SiPMs are ubiquitous. They are naturally attractive for nuclear and particle physics applications given their small size, high photon-detection efficiency (up to 60% in the visible range), relatively low price, high gain, fine time resolution (better than 100 ps per single photon and approaching 20 ps for multiple photon detection), and, contrary to PMTs or MCPs, relatively low operation voltage bias and insensitivity to magnetic field. The finite dynamic range, the temperature dependence of the breakdown voltage, a high dark count rate (10 kHz/$mm^2$ at room temperature) and a very moderate radiation tolerance are the main limitations: after a fluence of $10^9$-$10^{10}$ - MeV neutron equivalent per $cm^2$, all SiPM lose single photon sensitivity at room temperature. These features are the areas where research on these sensors is currently more active and where it must continue in the coming years.

In high energy and nuclear physics, the most common SiPM applications are for calorimeters and trackers and, increasingly, for particle identification detectors, especially via time measurements. The challenging application as single photoelectron sensors in Cherenkov imaging detectors, never used so far in experiments, will benefit from progress in controlling dark counting rates, which dramatically increase due to radiation damage. The recently demonstrated capability of restoring, after irradiation, the initial SiPM performance by thermal annealing cycles has resulted in adopting them as baseline sensors in the dual RICH detector at the ePIC experiment at EIC. In the coming years, several detectors will be equipped with SiPMs covering large areas, up to 10 $m^2$, in several large facilities like FAIR (time-of-flight detector at PANDA, calorimeters at CBM) and ePIC (calorimeters, RICH and time-of-flight). The time-of-flight-RICH approach is also considered for ALICE3. Cooling of the sensors remains a critical area in the design of the detectors (sPHENIX calorimeters and LHCb SciFi tracker).
Several promising sensor-dedicated R&Ds are ongoing: 3D-integration of the sensor coupled with a readout using CMOS technology, Back-Side Illuminated (BSI) SiPM simplifying the routing of the readout, the use of a layered anti-reflective coating, a textured surface of upright nano-micro pyramids with the potential to dramatically improve the photon detection efficiency, and other developments dedicated to the optical entrance making use of nanomaterials and metamaterials. These innovative developments must progress given the strategic potentialities that they will offer to nuclear physics.

### Box 9.5: Progress in a wide range of technologies is requested for the coming generation of high energy nuclear studies

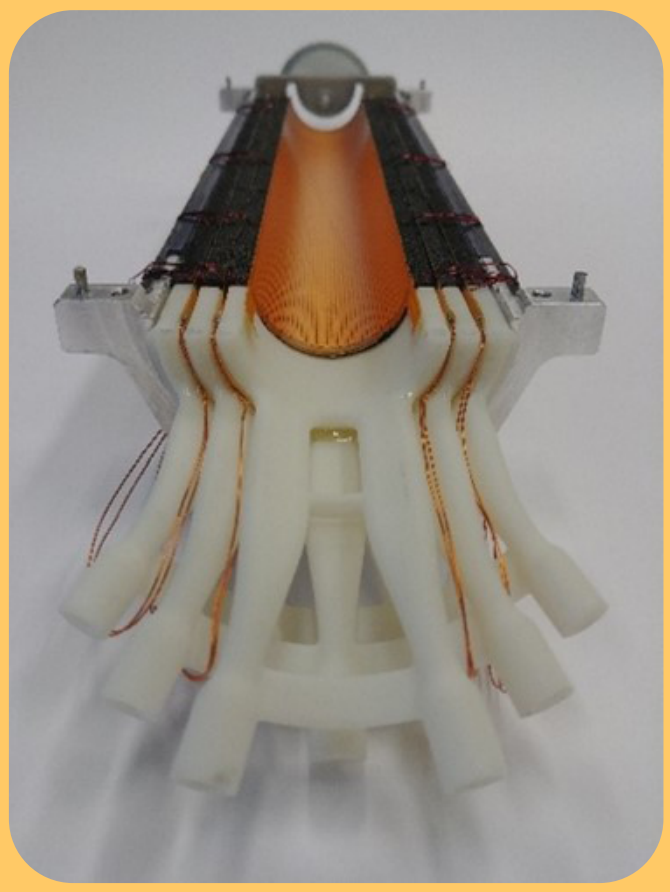

**ITS3 © ALICE Collaboration**
**Mockup of the novel inner tracker for the ALICE experiment. The sensors are MAPS in 65 nm CMOS technology.**

**HRPPD © INCOM Inc.**
**Exploded view of the HRPPD, a large-size (12 x 12 $cm^2$) MCP-PMT.**

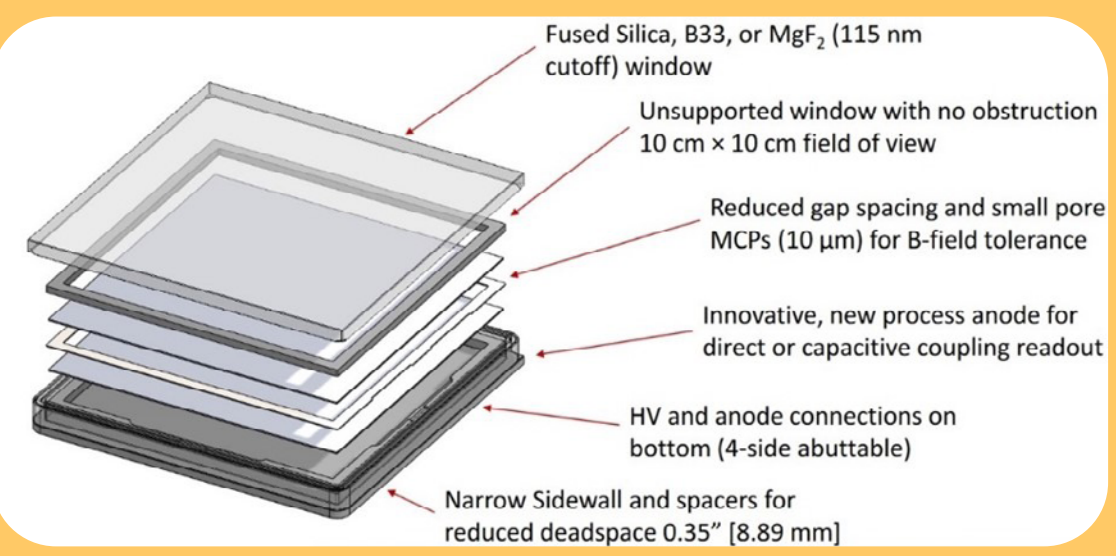

**PANDA endcap Calorimeter © PANDA Collaboration**
**Lead tungstate scintillators, thin carbon fibre alveoli, APD and VPTT photodetectors and electronics of the PANDA forward endcap calorimeter.$cm^2$) MCP-PMT.**

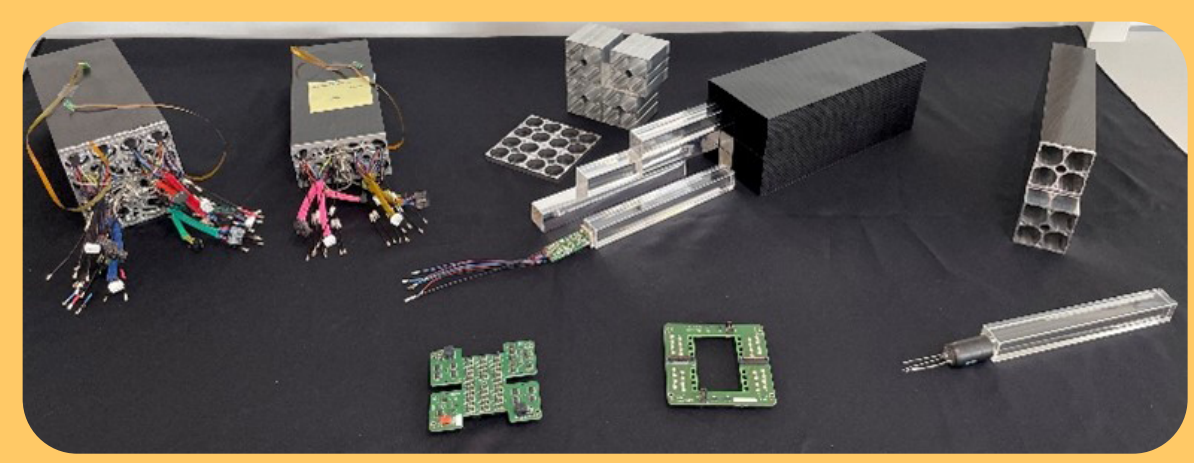

# Technological tools and related R&D - front-end ASICs, electronics read-out chain and data acquisition

## Data acquisition electronics

### High counting rate performance for high-energy nuclear physics experiments

The progressing challenges of modern experiments include continuously increasing rates of events to be detected to be able to perceive even the smallest deviations precisely enough. For this purpose, detectors have to be resistant to high particle rates on the one hand, and on the other, they have to be read out with ever-faster electronics. Modern experiments reach event rates in the range of several MHz.

Faster readout means that it is no longer possible to use individual signals to validate measurements (triggers) as in the past, but that it is increasingly necessary to switch to continuous readout. Here, the individual detector elements must be equipped with fast and selective electronics that are capable of separating background noise from physical signals at an early stage.

The pixel detectors described above incorporate readout electronics on the sensitive pixel elements to perform the initial processing of the signals in situ. For detectors with lower granularity, ASICs are suitable to either immediately digitise all signals or first test a signal threshold in analogue. FPGA-implemented feature extraction is one of the key components of modern detector front-ends.
For all types of newly developed chips, radiation hardness must be taken into account for stable operation in harsh conditions with continuously increasing luminosities of the currently operated and future machines.

### Granularity and PSA performance in low-energy nuclear physics experiments

High granularity germanium arrays like AGATA involve major upgrades of the data processing, using new data flow architecture and new algorithms combined with new hardware. The first objective is to develop innovative global data flow architectures to optimise the resources for the most interesting physics events. This could be achieved by adaptive event dispatchers able to push specific events in dedicated and optimised data pipelines. Advanced load balancing would be required to efficiently use the available resources. In parallel, global soft triggers could reduce the number of events processed by rejecting, as early as possible, the less interesting events. A significant gain is expected in the amount of data transferred through the local network and ultimately stored on the local disc before being moved to dedicated data centres. A second objective to making the whole system more efficient is to rely on heterogeneous hardware such as GPU or FPGA, which offer massively parallel treatments, or by using high bandwidth, low latency non-volatile memory (NVRAM) solutions. A third objective is to improve key algorithms (PSA/Tracking) to decrease the beam time required to reach a given statistic. Machine learning (Artificial Intelligence) techniques are probably the path to setting up such highly efficient data pipelines. For the period beyond 2030, the AGATA collaboration is preparing the next generation of front-end electronics based on cryogenic ASIC preamplifiers and very compact FADC close to the detector, and a new detector concept based on p-type crystal with new segmentation schemes. A further major upgrade of the detector infrastructure will be the change from LN2-based to electrical cooling. These upgrades are necessary to improve the reliability of the AGATA detectors and hopefully also reduce its environmental footprint.

Highly segmented Si detectors, such as GRIT, will need to handle approximately 104 channels. The overall triggering rate is expected to reach several thousand per second, especially with intense RIB at facilities like SPES. However, the acquisition of pulses from a large number of channels must be balanced against limitations in space and power consumption for the front-end electronics and cabling. These conditions require the selection of zero-suppression and serialisation in the front-end electronics, aiming to avoid free-running digitisers that consume significant power. A novel analogue memory chip called PLAS has been developed. It features self-triggering channels for sampling detector pulses and transmitting them at a slower pace. This advancement enables the utilisation of pulse shape analysis directly embedded in the readout system, resulting in significant improvements in particle discrimination.

# Technological tools and related R&D - Symmetries and Fundamental Interactions

The quest for a solution to open problems in fundamental physics requires an intense effort to innovate technology. Many detectors used in neutrino and dark matter experiments adhere to the 'calorimetric approach', where the observed nuclides or target elements are embedded within the detector's medium. This approach enables very high detector efficiency; however, it also necessitates unique detector research and development specific to each project. Despite that, some aspects covered here apply to all detectors relevant to this section.

#### Background reduction

Several aspects are used to mitigate backgrounds, each of which requires intense research and development and the construction of small-scale experiments to verify and validate the results.

#### Material purity

State-of-the-art technologies for material purification and highly sensitive methods for material selection are needed to ensure detector materials are free from undesired isotopes.

#### Particle identification

Signal channels are combined with each other and/or with sophisticated data analysis methods to enable particle identification, for example through pulse-shape-discrimination, and event topology reconstruction in a monolithic detector. Sometimes, a veto detector is added to the system to register the passage of unwanted particles.

#### Energy resolution

Especially in experiments where the signal is in a tight energy window, energy resolution is of paramount importance to make sure signal events are well-separated in energy from backgrounds.

### Time response

A fundamental characteristic when it is necessary to detect many closely spaced events, it is also very important in experiments searching for rare events. Poor temporal resolution does not permit discriminating distinct events that happen to coincide in time, thus contributing to the background in the regions of interest.

### Large exposure

Exposure is defined as the measurement duration multiplied by the mass of the isotope being observed. To enhance the sensitivity of rare event searches, increasingly larger numbers of the nuclei of interest must be put under observation for longer and longer times. Scaling-up challenges include procurement of the target nuclei in large enough quantities, engineering of increasingly larger detector infrastructure, and read-out techniques for large numbers of sensors.

### Signal extraction

To further enhance the sensitivity, all detectors in this category are meticulously designed to extract minute amounts of quanta, that is photons, phonons, or electrons emanating from the extensive detector volumes. Even marginal improvements in signal extraction efficiency can substantially bolster sensitivity to the rare events under investigation, not only because it leads to a higher event rate, but also because it enhances the discrimination power between signal and background. Technologies needed for this are efficient single-photon detectors covering a broad wavelength range from 120 to 550 nanometres; phonon sensors capable of detecting slight temperature variations; and efficient techniques for single-electron detection. In this context, the novel field of quantum calorimeter sensors deserves attention. Metallic Magnetic calorimeters (MMCs) are low-temperature detectors characterised by excellent energy resolution, fast response time and good linearity. These features, together with the possibility of building up arrays of detector pixels and their applicability across a broad energy range from X to gamma rays, make them highly solicited for both fundamental and applied science.

Very often all these items are closely interconnected. While continuous enhancement of existing technologies in these domains is crucial, fostering the development of novel solutions is equally imperative.

## Recommendations for Detectors and Experimental Techniques

### 1. Sensor technologies (Fig. 9.3)

(a) A new generation of segmented p-type coaxial **High-Performance Ge (HPGe) detectors** is fundamental in γ-ray spectroscopy to enhance the γ-ray counting rate as well as to reduce the effects of neutron damage. It must be accompanied by a corresponding development of dedicated front-end electronics.

(b) A crucial need for neutron detection and, therefore, a main path to follow is related to R&D for the **replacement of $^{3}$He-based detectors** with those using Boron, Lithium and their compounds as neutron converters.

(c) **Systematic research** on new materials for fast and **cheap scintillators** with high light output should be encouraged. These scintillators have multiple applications in future detectors for **gamma, electron and neutron detection and calorimetry** for nuclear and hadron physics experiments. Likewise, these also find multiple applications beyond fundamental research in medicine, industry and disarmament monitoring.

(d) **Gaseous detectors** are and will remain major ingredients in low- and high-energy NP experiments as well as in the studies dedicated to symmetries and fundamental interactions. Therefore, **R&D dedicated to gaseous-detector must continue** for key families of technologies: MicroPattern Gaseous Detector (MPGD), Resistive Plate Chambers (RPCs), Time Projection Chambers (TPC), Drift Chambers (DC) and Straw Tubes (ST). The target focus is progress in timing towards O(10ps) resolution, finer and finer space resolution, high rate capability in the control of the ion back-flow (for TPCs and photodetection) and in cluster counting for improved **PID capabilities**.

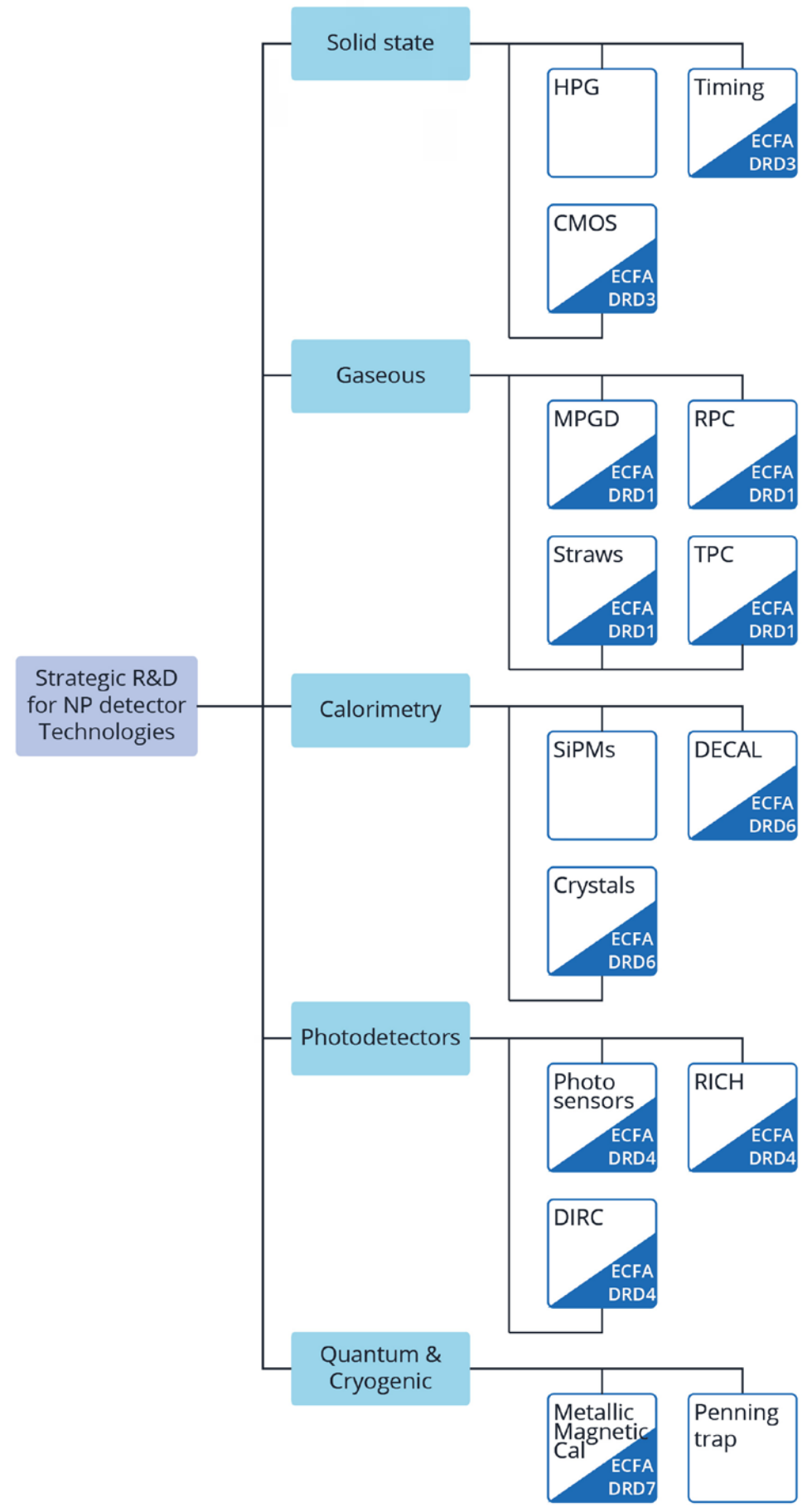

*Fig. 9.3: A gross sketch of the major needs of detector R&D envisaged for the period considered within the LRP. Where appropriate and synergistic, the novel R&D DRDs collaboration, promoted by ECFA and presently being formed at CERN, is indicated.*

(e) **Particle identification by Cherenkov imaging techniques** remains a fundamental need in hadron physics and spectroscopy and the concept of RICH and DIRC-based detectors need to evolve towards up-to-date and compact devices. Key elements are ensuring the availability of optical-quality aerogel as radiator material for future RICH counters as well as identifying and developing alternatives to fluorocarbons (gasses with high global warming power) as gas radiators in RICH detectors. The development of single photon detectors with improved time resolution and increased quantum efficiency (SiPMs, MCP-PMTs) must progress further.

(f) **SiPMs** are more and more present in high- and low-energy NP, symmetries and fundamental interaction studies. Even if today largely used outside fundamental research, still a large fraction of the production is dedicated to research. This imposes progress in the SiPM development, to overcome the present limitation related to dark counting rates, temperature-dependent performance and limited radiation hardness. More advanced concepts such as the 3D integration of the sensor coupled to CMOS technology, Back-Side Illuminated (BSI) SiPM, and the use of layered anti-reflective coating are promising paths to be pursued for SiPMs with improved characteristics.

(g) Further developments towards unprecedented precision in both position and time measurements with **silicon-based detectors** are required. These include support for an integrated effort to develop Monolithic Active Pixel Sensors (MAPS) based on commercial CMOS processes which target improved timing precision, radiation hardness and power management. These developments are complemented by those for depleted monolithic active pixel sensors (DMAPS), which are being developed to explore technological solutions for digital electromagnetic calorimeters to evolve this technology into a calorimeter tool with unprecedented capabilities and providing 5-D images - in particular if related to particle flow algorithms. To achieve a time resolution below 20 ps, further advancements are needed in silicon timing technology, including low-gain avalanche diodes (LGADs), Single Photon Avalanche Diodes (SPADs) or CMOS-integrated monolithic sensors with internal gain.

## 2. Strategic tools (Fig. 9.4)

(a) The European scientific community urgently needs a secure provision of **stable and radioactive samples and target resources, financial support and related infrastructure**, including essential improvements in material storage and transport concepts as well as expansion of networks and collaborations and a dedicated mass separator for radioactive isotopes.

(b) Develop a European strategy towards a guaranteed ESI supply for European research in nuclear physics and beyond, including sectors of societal interest; the **installation of a European EMIS facility to gain independence from extra-European support** is the essential ingredient.

(c) The increasing size, granularity and complexity of the experimental apparatus in all nuclear physics domains (low energy, high energy, symmetries and fundamental interactions) impose the pursuance of **progress in the overall read-out chain**, from front-end ASICs to electronics, to the use of heterogeneous computing resources and the development of advanced algorithms, also exploiting artificial intelligence techniques. These efforts must include the development of streaming read-out chains. The availability of front-end ASICs and read-out chains designed to match the ps-resolution capabilities of up-to-date and future sensors (SiPms, MCP-PMTs, LGAD Si trackers, specialised gaseous detectors and others) is a crucial ingredient.

(d) **Develop and exploit the potential of resonance laser ionisation of exotic radionuclides (RILIS)** in their hot-cavity and gas-cell variants beyond current capabilities in terms of spectral resolution and efficiency, both **for the selective production of low-energy radioactive ion beams (including isomeric selectivity) and as a laser-spectroscopy method.** Further improve the laser, gas-cell and ion-manipulation technologies required for laser ionisation close to the production or stopping point of the radioactive ion beams. Apply and consolidate in-gas-jet laser spectroscopy for the study of exotic nuclei.

(e) It is essential to further improve **high-resolution laser-spectroscopy and mass-spectrometry** techniques, in particular their efficiency for the study of very exotic isotopes and their precision for the application to tests of fundamental symmetries. Improve existing ion-trap setups and manipulation techniques for better suppression of contaminants. Improve synergies between mass spectrometers, lasers and decay stations, either through hybrid setups, mass-tagged or decay-tagged laser spectroscopy, trap-assisted experiments or experiments with laser polarised or selectively re-ionised beams.

(f) **Support for precision measurements**: Experiments designed for studying symmetries and fundamental interactions critically rely on nuclear physics inputs, encompassing **nuclear matrix elements**, cross-section measurements and nuclear form factors. Efforts to measure these quantities should therefore continue to be supported. In addition, the potential for technological synergies with nuclear physics tools should be fully exploited, for example to push current limits for sensing photons, electrons and phonons. Furthermore, some detectors require isotopically enriched or depleted materials, so European facilities for the production of such materials should be strongly supported. Finally, the underground laboratories at Europe's leading institutions such as LNGS, as well as Modane and Canfranc, which host the most sensitive experiments in low-background environments, should receive constant support.

(g) **Polarised frozen spin targets** are an essential tool in measuring nuclear form factors and interactions using spin degrees of freedom. **High-performance internal targets, such as gas-jet, cluster-jet and micro-droplet targets** can be operated with all gaseous materials and are key components for many experiments in hadron physics. It is **of utmost importance to ensure the continued operation** of polarised frozen spin targets and supersonic gas jet targets.

(h) Ensure the continued production of high-quality **lead tungstate crystals** for the EIC, PANDA and upcoming experiments. The **only producer worldwide** of high quality, fast, compact and radiation hard lead tungstate scintillators **is based in Europe** in the Czech Republic. The developments continue in strong cooperation with academia.

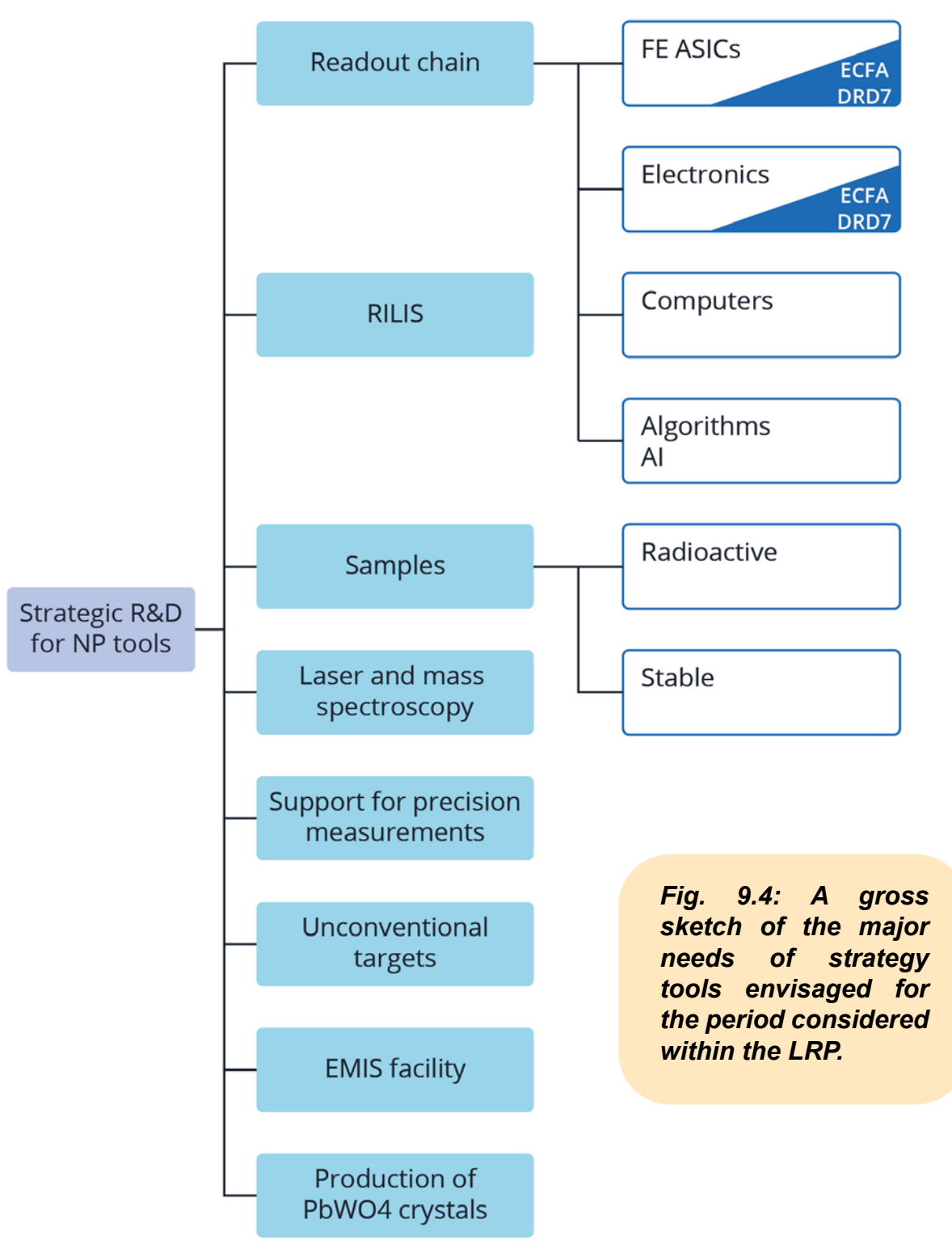

*Fig. 9.4: A gross sketch of the major needs of strategy tools envisaged for the period considered within the LRP.*

**References**

1. https://indico.cern.ch/e/ECFADetectorRDRoadmap.

2. https://snowmass21.org/.

3. https://wiki.bnl.gov/conferences/index.php/EIC R%25D; https://www.jlab.org/research/eic_rd_prgm; https://wiki.bnl.gov/conferences/index.php/General Info.

4. Advancement in Nuclear Instrumentation Measurement Methods and their Applications (ANIMMA2023), Lucca (Italy), 12-16 June 2023, https://animma.com/.

5. EURASIS https://www.nupecc.org/eurasis/.

6. 16th Vienna Conference on Instrumentation (VCI2022, Vienna (Austria), 21-25 February 2022, https://indico.cern.ch/event/1044975/ ;
15th Pisa Meeting on Advanced Detectors (PM2021), La Biodola, Isola d'Elba (Italy), 22-28 May 2022, https://www.pi.infn.it/pm/2021/; Technology and

# Nuclear Physics Tools – Machine Learning, Artificial Intelligence, and Quantum Computing

**Convener:**
**Valerio Bertone** (IRFU CEA Paris-Saclay, France)
**Jana N. Günther** (Univerity of Wuppertal, Germany))

**NuPECC Liaison:**
**Eugenio Nappi** (INFN, Sezione di Bari, Italy)

**WG Members:**
- **Joerg Aichelin** (University of Nantes, France)
- **Mohammad Al-Turany** (GSI, Germany)
- **Sara Collins** (University of Regensburg, Germany)
- **Tommaso Dorigo** (INFN Padova, Italy)
- **Christian S. Fischer** (University of Gießen, Germany)
- **Michele Grossi** (CERN, Switzerland)
- **Philipp Hauke** (University of Trento, Italy)
- **Andreas Ipp** (University of Vienna, Austria)
- **Marc Labiche** (Daresbury Lab., UK)
- **Denis Lacroix** (IJCLab, Orsay, France)
- **Johan Messchendorp** (GSI, Germany)
- **Stefano Piano** (INFN Trieste, Italy)
- **Arnau Rios** (University of Barcelona, Spain)
- **Alessandro Roggero** (University of Trento, Italy)
- **Olga Soloveva** (University of Frankfurt, Germany)
- **Vittorio Somà** (CEA Paris-Saclay, France)
- **Liangxiao Wang** (University of Frankfurt, Germany)

# Introduction

The tremendous progress made in the field of Nuclear Physics has led to the pressing need for appropriate numerical tools aimed at addressing the most relevant experimental, theoretical and technological challenges, such as those encompassed by the Joint ECFA-NuPECC-APPEC (JENA) initiative. To this end, the advent of algorithms based on Machine-Learning (ML) and Artificial-Intelligence (AI) techniques, and the fast progress in the field of Quantum Computing (QC) have opened an entire new world of possibilities. Nuclear and particle physicists from all around the world have turned their attention to these technologies in the quest for more efficient tools to interpret the abundance of experimental data that is currently being delivered at nuclear and high-energy facilities.

However, so far the exploration of possible applications of ML, AI, and QC to nuclear and high-energy physics in Europe has mostly proceeded incoherently with local or at most national initiatives devoted to it. The purpose of this chapter is to provide a broad and as comprehensive as possible overview of the current status of how these techniques are being employed in nuclear physics, to coordinate this effort at a European level.

The width and transversality of this subject make this task complicated to accomplish. To identify the main guidelines for the final recommendations, this chapter is split into four main sections respectively devoted to:
- machine learning and artificial intelligence,
- quantum computing,
- tools and techniques,
- resources and infrastructure.

This organisation will allow us to focus on the specific aspects of these main branches to eventually issue precise recommendations for optimising European resources.

# Machine Learning and Artificial Intelligence

Nuclear physics stands at the forefront of our quest to understand the intricacies of subatomic interactions, playing a pivotal role in deciphering the properties of strongly interacting matter. Parallel to this, the meteoric rise of AI has revolutionised our approach to complex problems. ML, as the most prominent subset of AI, is specifically designed to discern patterns in intricate data and encapsulate them with an optimal set of parameters. ML is suitable for rich data and sophisticated explorations, which promises unprecedented opportunities and insights for understanding nuclear matter in today's AI-driven world.

In the realm of nuclear physics, ML applications have been explored in areas such as nuclear experiments, nuclear astrophysics and various computing-intensive tasks, as shown in Fig. 10.1. Within nuclear physics experiments, ML algorithms have been used to process large datasets, aiding in particle identification, improving event reconstructions and allowing for experiment design and control. In the field of nuclear astrophysics, ML has been applied to analysing signals and is particularly useful in processing data from noisy space environments. It has also assisted in determining the properties of dense matter, which is crucial for understanding certain celestial events. ML has also been beneficial for computing-intensive challenges. It has been applied in hadronic structure and nuclear collisions [see TWGs 1 and 3], astrophysical simulations [see TWG 4], and notably in Lattice QCD [see TWG 1], a first-principles method, to enhance our understanding of nuclear matter.

Thanks to their flexibility as universal approximators, neural networks (NNs) have been proposed as an effective framework to design accurate ansatzes for the wave function of strongly correlated many-body systems. Currently proposed architectures have already shown impressive results for nuclear structure applications on closed shell nuclei where they can match or even surpass the accuracy achieved by state-of-the-art Quantum Monte Carlo simulations. Similarly, the application of NNs to parameterise the hadronic structure of nucleons and nuclei relevant to the physics of high-energy colliders, such as the Large Hadron Collider (LHC), has already been extensively used and proven superior in many respects to more standard approaches. ML techniques on data simulated by various transport models empower us to facilitate conventional numerical simulations. ML methods also enable us to shed light on events and processes during collisions. This inverse problem can help us achieve for the first time real-time event reconstruction or online physics analysis as has been implemented by, for example, the CBM experiment, the STAR experiment at RHIC and the LHCb experiment at the LHC.

The application of ML techniques for real-time event reconstruction will be opportune to efficiently process raw, low-level data from the large variety of detectors. The success of the in-situ data reconstruction and selection will depend crucially on the stability of individual sensor elements and accelerator operations. The usage of ML algorithms and AI in operating nuclear physics experiments and accelerators, enabling fast and vast data processing and analysis, is taking a central role, reducing the traditional dichotomy between offline and online computation.

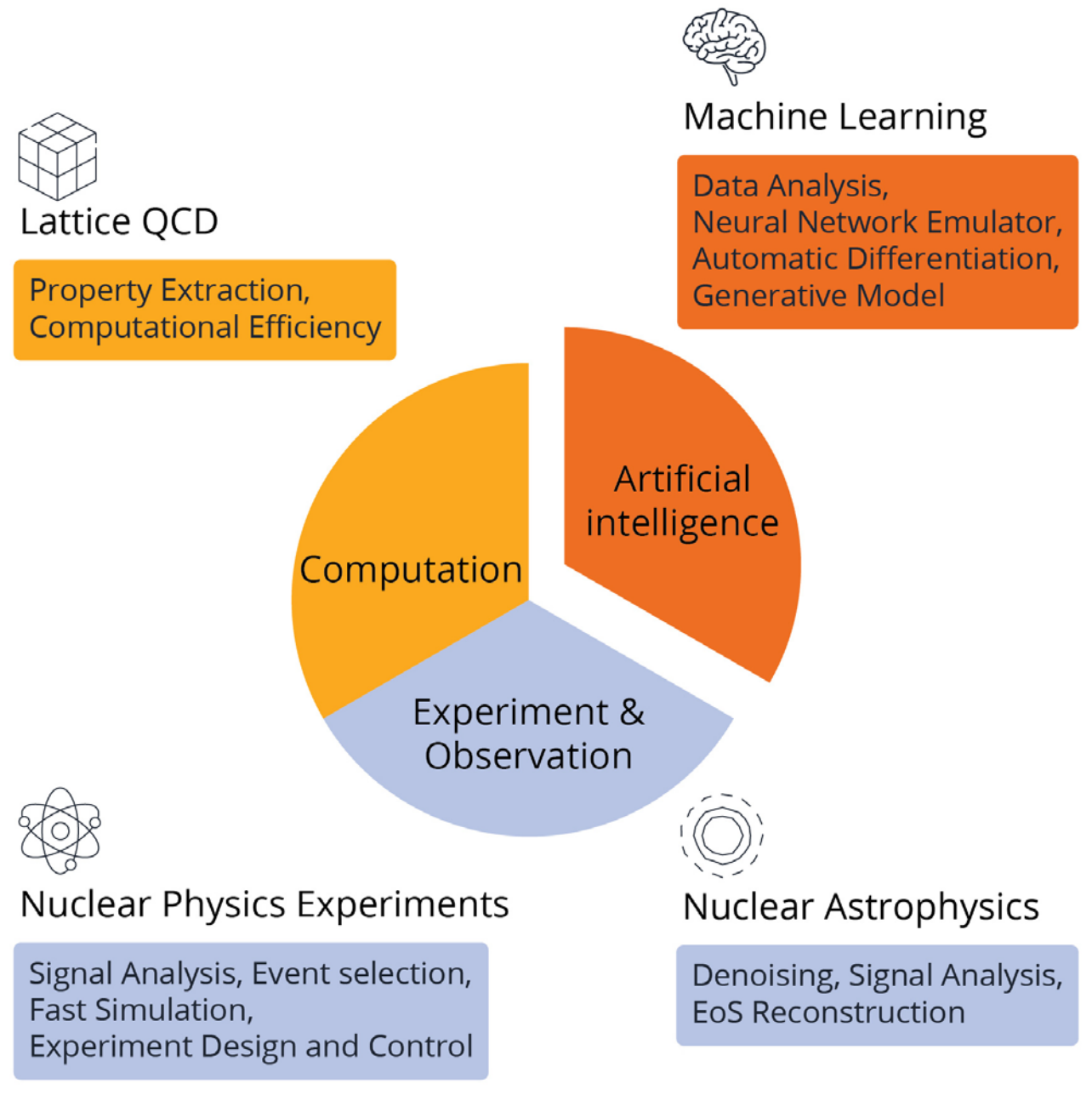

*Fig. 10.1: Chart of ML application in nuclear physics.*

Deep learning (DL), a subset of ML, has proved successful in representing complex processes. Its data-driven nature aligns well with nuclear physics, especially given the massive amount of data from experiments, observations and simulations. Generative models, a prominent subset of DL, are designed to generate data samples resembling the input data. These models have found significant applications across various domains of nuclear physics. In Lattice Quantum Chromodynamics (QCD), flow-based models have now been introduced to improve the efficiency of Monte Carlo algorithms. The incorporation of gauge symmetry, a fundamental principle governing the strong force, has been successfully achieved in lattice-gauge-equivariant convolutional NNs, which can also be applied to real-time glasma simulations, describing the earliest stages of heavy-ion collisions. Generative models have also been instrumental in nuclear experiments, aiding in data augmentation, noise reduction and event simulation, relevant to optimising particle accelerator operations. Currently, as the intersection of ML and nuclear physics deepens, tools like Generative Pretrained Transformer (GPT) models are set to play a pivotal role in both experimental and theoretical advancements.

ML offers numerous benefits but also faces challenges like interpretability which is crucial in nuclear physics. However, physics-informed and physics-driven ML methods are providing promising solutions. Techniques like automatic differentiation are being employed to refine experimental designs, ensuring that the functional designs align with physical principles. It has also been used in tackling inverse problems when seeking to extract spectral functions from correlators calculated in Lattice QCD, build dense nuclear matter equation of states from neutron star observations, and obtain QCD matter properties from heavy-ion collision experiments. Additionally, constraint-based learning ensures that deep models follow physical laws, improving their reliability in, for example, simulating nuclear reactions or predicting material attributes.

## Recommendations Machine Learning and Artificial Intelligence

### Transform ML prototypes into applications for production

Formulate a strategic approach to advance from current short-term, proof-of-concept ML projects in nuclear physics towards practical applications usable in production. This process begins with the identification of specific pilot projects that warrant conversion into long-term projects. Collaborative support from computer science research centres can aid the transformation process. In transitioning from pilot ML projects to production in nuclear physics, leveraging established ML frameworks and codebase management platforms is pivotal.

### Foster data-sharing in nuclear physics

Promote data-sharing initiatives in nuclear physics by establishing a user-friendly hub for open databases, akin to existing platforms in related scientific fields such as the International Lattice Data Grid (ILDG) [https://hpc.desy.de/ildg/], the Gravitational Wave Open Science Center (GWOSC) [https://gwosc.org/], the CERN Open Data Portal [https://opendata.cern.ch/], or the European Open Science Cloud (EOSC) [https://eosc-portal.eu/] and ESCAPE [https://projectescape.eu/].

### Strengthen computational resources

Allocate funding for enhanced GPU clusters within established HPC centres across Europe. With the increasing demand for computational power in emerging fields like large language models, which require extensive computational resources, we anticipate a similar need for substantial computational capabilities in scientific applications. At the same time, it is crucial to support GPU access at various scales, ranging from centralised GPU centres designed for large-scale projects to localised GPU resources for the rapid development needs of individual research groups.

### Train scientific foundation models

Recognising the potential of foundation models in scientific research, we recommend dedicated investments in training and fine-tuning models tailored for scientific purposes, such as GPT models specialised for nuclear science.

## Quantum Computing

Since the recent completion of the first quantum computer demonstrators, the field of quantum QC has seen remarkable growth worldwide. In the physics community, much of the interest is motivated by the exciting potential of these technologies to simulate strongly correlated many-body systems. There are two main platform approaches for this purpose, digital universal quantum computers and analogue special-purpose devices, called quantum simulators.

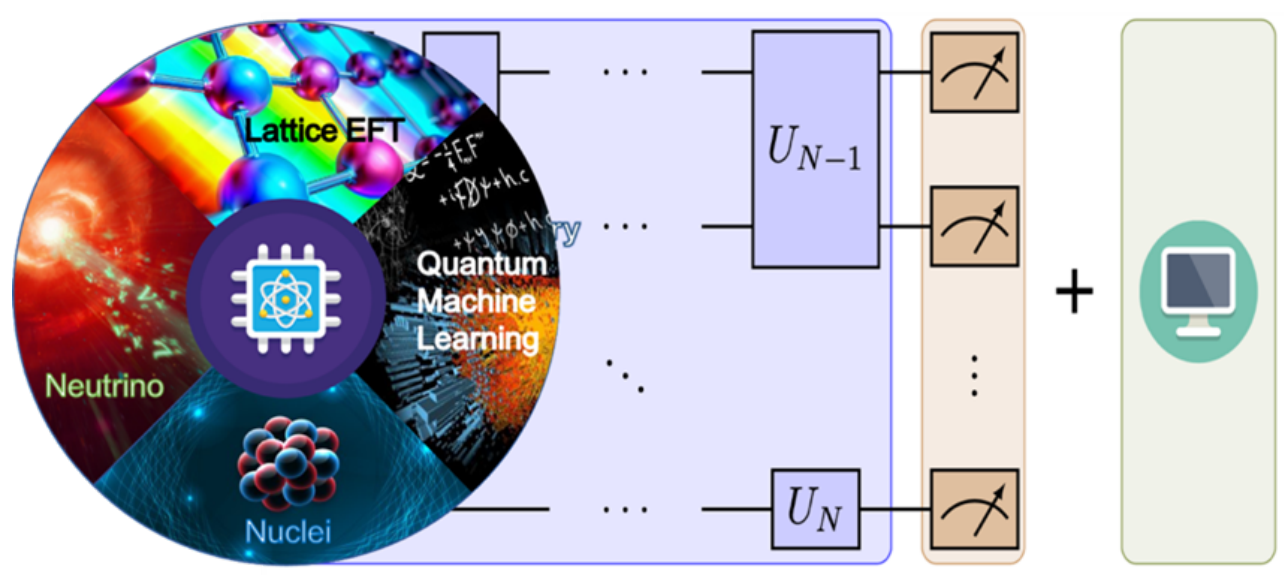

*Fig. 10.2: Schematic illustration of some of the fields where quantum computing is now being explored (left). Examples include physics of quarks, neutrinos and nuclei, as well as a pictorial view of quantum machine learning. All these problems are now being considered as pilot applications that could be treated on digital quantum computers using quantum circuits (right).*

Despite the formidable challenges in designing algorithms suitable to harness the capabilities of these early prototypes, an increasing number of applications has been proposed and tested. This includes efforts from the nuclear theory community to study the structure and dynamics of many-body systems of fermions, such as high- and low-energy nuclear theory, gauge and field theories, and neutrino physics (see Fig. 10.2).

Another important effort is now being made on data mining through the development of advanced Quantum ML (QML) techniques, which blend two unique disciplines: quantum theory and ML. This novel approach provides the possibility of defining quantum and classical algorithms on classic or quantum data. Supervised and unsupervised learning together with reinforcement learning are still valid definitions within the QML framework, supporting the interdisciplinary collaboration of ML and QC experts. QML is thus becoming an excellent candidate to deal with the incoming computational tools outbreak caused by increasingly complex data.

Quantum information processing has conjointly seen rapid advances on the level of devices as well as our theoretical understanding of quantum many-body systems. These hand us paradigmatically novel computational methods, for example in the form of tensor networks, and move fundamental questions such as the role of entanglement in quantum many-body systems into focus. The development of quantum computers and related aspects offer new perspectives and might rapidly revolutionise the way complex problems are treated together with the potential of recent advancements in quantum sensors which open the possibility of looking directly into quantum data.

Small-scale quantum computers are now widely available, often as computational devices accessible as cloud services. Complex quantum computers with hundreds of qubits are being commissioned or built by many academic institutions and industrial partners worldwide. We are currently in a transition period where the size and fidelity of quantum resources are not sufficient for fault tolerance - and decoherence is still an issue. The technological challenges associated with these issues are being addressed by the quantum computing community on the hardware and algorithm level, with substantial progress every year. Despite these difficulties, scientists in different fields have started to seriously consider (noisy) quantum computers as a disruptive technology that opens new horizons.

Internationally, the potential for game-changing applications of quantum technologies and quantum information theory has already been recognised. For example, in the USA it has given rise to a strategic effort towards quantum computing for nuclear theory through an influx of funding and a series of strategic white papers collecting community input and commissioned by government bodies. In Europe, several countries have decided to boost their national strategy on quantum technologies from which one can, in the coming years, kickstart real quantum device applications. Indeed, several relevant initiatives have been put in place, for example by CEA, CERN (CERN Quantum Technology Initiative), CNRS, and INFN. Some of these initiatives aim at establishing joint research, setting up the supporting computing infrastructure and providing dedicated mechanisms for the exchange of knowledge and innovation.

Moreover, several bottom-up initiatives have been made to promote quantum computing in the NuPECC community. These include European collaborative research grants (e.g. Horizon 2020/Europe or Quantera), white papers and the organisation of workshops or doctoral training programmes. Thanks also to such initiatives, quantum computing is currently gaining interest in the NuPECC community, with new groups emerging constantly all across Europe, working on various subjects (see Fig. 10.2). These are encouraging first steps, but a wider consensus with dedicated infrastructure tailored to nuclear physics requirements has not yet been achieved. While on a methodological and technical level Europe is on a par with the best international players, a concerted strategy that bundles and streamlines individual efforts is lacking.

## Recommendations for Quantum Computing

A sustained effort is necessary to advance QC towards realistic applications. Among the requirements for future applications of QC, we highlight the need to:

i. Test and validate the applicability of state-of-the-art algorithms and methods developed in the general theory of quantum computing to fields of direct interest to NuPECC.
ii. Define good benchmarks to leverage cutting-edge QML solutions to problems of interest for NuPECC.
iii. Design new algorithms and new quantum ansatzes tailored, for instance, to complex many-body systems of relevance for nuclear physics, neutrino physics, Lattice field theory, and/or Quantum Machine Learning.
iv. Perform applications on existing quantum devices with an appropriate effort dedicated to contrasting the presence of noise using a variety of error mitigation/correction techniques.

In this respect, a major objective to be addressed soon is the development of new quantum algorithms and error-mitigation techniques that can efficiently exploit the capabilities of current (noisy) and future quantum computers. In parallel to these efforts, a set of recommendations designed to foster collaborations, strengthen expertise, and address emerging challenges within the NuPECC quantum computing community to accelerate advancements in this rapidly evolving field are given below:

### Establish a transnational European network on quantum activities

Create a collaborative network at the European level, focused on QC and quantum information (QI), in alignment with the interests of NuPECC. Foster cooperation and knowledge exchange among researchers from different institutions and countries. Bridge QC theory, QI processing,and machine learning, enabling cross-community collaboration, research and development in future technologies and knowledge-sharing.

### Organise workshops, schools and training programmes

Regularly host workshops, schools and hands-on events to facilitate the transition towards quantum computing [see TWG 10]. Consider partnering with academic institutions to strengthen their role in QC and provide comprehensive training opportunities. Address the emerging topic of QC applications by providing specific training for young and experienced scientists/engineers in quantum mechanics, quantum information, and quantum algorithms. Prioritise the training of researchers in quantum computing. Beyond workshops, explore measures such as student exchanges and joint fellowships to build a strong interdisciplinary network.

### Facilitate access to quantum platforms

Ensure access to state-of-the-art quantum platforms by bridging the gap between academic institutions and private companies. Consider forming agreements with national High-Performance Computing (HPC) centres like CINECA, Jülich, GENCI, etc. to enhance accessibility. These include agreements for the use of classical simulators (e.g. based on tensor network codes) to benchmark quantum hardware. Another key aspect of accessing certain quantum platforms is to promote collaboration between theorists and experimentalists developing current and future quantum devices to collaboratively design application-specific quantum-simulation algorithms, capitalising on their combined expertise.

### Develop strategies for quantum-classical interfaces

Formulate a clear strategy for interfacing quantum and classical machines. As most algorithms will be hybrid quantum/classical, effective integration is crucial for future developments. This includes addressing data mining challenges, given the large flow of data generated by quantum computers with increasing qubits, by developing specific techniques to extract information from the data through classical post-processing. Also, develop common open-source libraries for frictionless integration of classical and quantum software frameworks.

## Numerical Tools and Techniques

A comprehensive and exhaustive treatment of the numerical tools and techniques that have been and are being employed in nuclear and high-energy physics is an enormous task that is impossible to fit exhaustively into this report. However, it remains relevant to discuss some prominent cases that will eventually help us formulate a set of general recommendations concerning the development of new numerical technologies aimed at advancing the field of nuclear and high-energy physics. To be as representative as possible, we will distinguish between computing techniques applied to nuclear and high-energy physics from a theory viewpoint on the one hand, and from an experimental viewpoint on the other. On the theory side, we will consider transport approaches in heavy-ion collisions, the study of the hadronic structure, and computations on the lattice. On the experimental side, we will discuss the development of Monte Carlo generators aimed at developing detector systems.

## Heavy-ion collisions and hadronic structure

Currently, we are observing a distinctive change of paradigm:

● Experiments, especially those at RHIC and the LHC, have moved on from an exploratory phase to a high-precision phase.

● Theoretical approaches evolve from simple models to describe single observables to more complex frameworks aimed at the simultaneous description of all observables.

In several contexts ranging from transport approaches, hydrodynamics and thermalisation in heavy-ion collisions [see TWG 2] to the study of the nucleon structure [see TWG 1], these developments have led to advancements of unprecedented complexity. Developing them further requires software specialists, a new structure of the community (by regrouping and joining efforts) and new computational approaches. This aspect is often not addressed in funding schemes. In addition, the underlying theories, like lattice QCD or perturbative QCD, do not usually provide predictions in the whole kinematic range which is studied in heavy-ion collisions. Forceful extrapolations and approximations are therefore often necessary. To make progress in the understanding of the physics questions at hand, an assessment of the different approaches is necessary.

This is a worldwide problem which has already been successfully addressed in the USA with the creation of structures like JETSCAPE [https://jetscape.org/], MUSES [https://muses.physics.illinois.edu/] or HEFTY [https://hefty.tamu.edu/]. These topical structures typically have a budget of 5 million US dollars for 4 years. In these well-funded structures, several research groups (typically between 10 and 15) collaborate on a given topic. The different approaches to a given topic have been centralised and manpower made available to make the programmes comparable, to benchmark the results, and to subdivide them into different modules. Structures of this type should be created in Europe as well.

The comparison of high-precision data to theoretical predictions of these advanced models needed to answer the essential physics questions at stake and aiming at an accuracy of 10%, must rely on identical treatment of experimental data and numerical results from transport approaches. This needs a forum where the numerical data is made available and where the software tools for such comparisons are developed and maintained.

The study of the hadronic structure in the context of high-energy collisions has also witnessed an impressive development catalysed by experimental, theoretical and technological advances. The quantity and quality of experimental results delivered by past and current facilities have pushed the need for a detailed quantitative understanding of the internal structure of hadrons to an unprecedented level. The community has reacted by producing more accurate theoretical calculations and by developing numerical tools aimed at streamlining the interpretation of experimental results. In many cases, ML techniques have also come to the rescue by enhancing our ability to manage complex data structures.

These successful efforts have been widely recognised and supported. At the European level, the STRONG-2020 project [http://www.strong-2020.eu/scientific-frontiers/high-energy-frontier.html] has invested significant resources in the virtual-access packages NLOAccess and 3DPartons devoted to the development and maintenance of open-source numerical frameworks for high-precision phenomenology at high-energy colliders. Similarly, the Centre for Nuclear Femtography [https://www.femtocenter.org] in the US has been instituted recently with similar purposes. Another stepping stone towards a detailed understanding of the hadronic structure is the recent approval of the Electron-Ion Collider (EIC) to be built in the US within the next decade. This initiative has boosted even further the demand for precision phenomenology at high-energy colliders that now more than ever needs a more incisive organisation and support.

## Recommendations Numerical Tools and Techniques

● The creation of a new, European-wide structure which hosts, maintains and develops open-source numerical tools for the physical interpretation of present and future heavy-ion and collider data. These structures should bring together the community, allow for a detailed comparison of different approaches and provide support in software engineering for future development. US efforts like JETSCAPE could serve as a model.

● Current efforts to provide a solid basis for systematic comparisons of experimental results to theoretical predictions should be intensified. This requires a European-wide platform where theoretical predictions are stored and where open-source analysis tools (like RIVET) are developed to facilitate detailed comparisons with data.

● International collaborations should be maintained and extended.

● Access to substantial computational resources and funding for the development of optimised algorithms and codes should be guaranteed.

● Further support and centralisation of the initiatives aimed at providing open-source numerical frameworks devoted to the study of the hadron structure should be supported.

### Box 10.1: Functional methods

**The functional approach to QCD via Dyson-Schwinger equations (DSEs), functional renormalisation group equations (FRGs) and Bethe-Salpeter equations (BSEs) works at the level of non-perturbative correlation functions. In principle, these objects encode all information on the physics content of QCD. This includes the conventional and exotic hadron spectrum of ground and excited states, form factors, decays and electromagnetic as well as hadronic processes like Compton scattering and pion-nucleon interaction. Approximation schemes at very different levels of sophistication have been developed, approaching a versatile tool: on the one hand, simple approximations allow us to make contact with quark-model calculations; on the other hand, highly sophisticated and numerically demanding schemes allow for direct comparison with lattice QCD. A very recent example is the self-consistent and parameter-free calculation of the glueball spectrum shown in Fig. 10.3 which demonstrates the excellent agreement of both approaches. Functional methods are furthermore fully covariant and do not rely on heavy quark expansions. The approach can be used in the light- and heavy-quark region allowing for fruitful interactions with chiral perturbation theory as well as with heavy-quark effective theory. Systematic comparisons with other approaches have a high potential for the identification of the physical mechanisms behind the observable phenomena.**

**After two decades of advances in terms of both high-quality truncations and applications to three (baryon, hybrid), four (tetraquarks) and multi-body problems including scattering processes, many current projects in the functional community have reached a level where the bottleneck is computational power.**

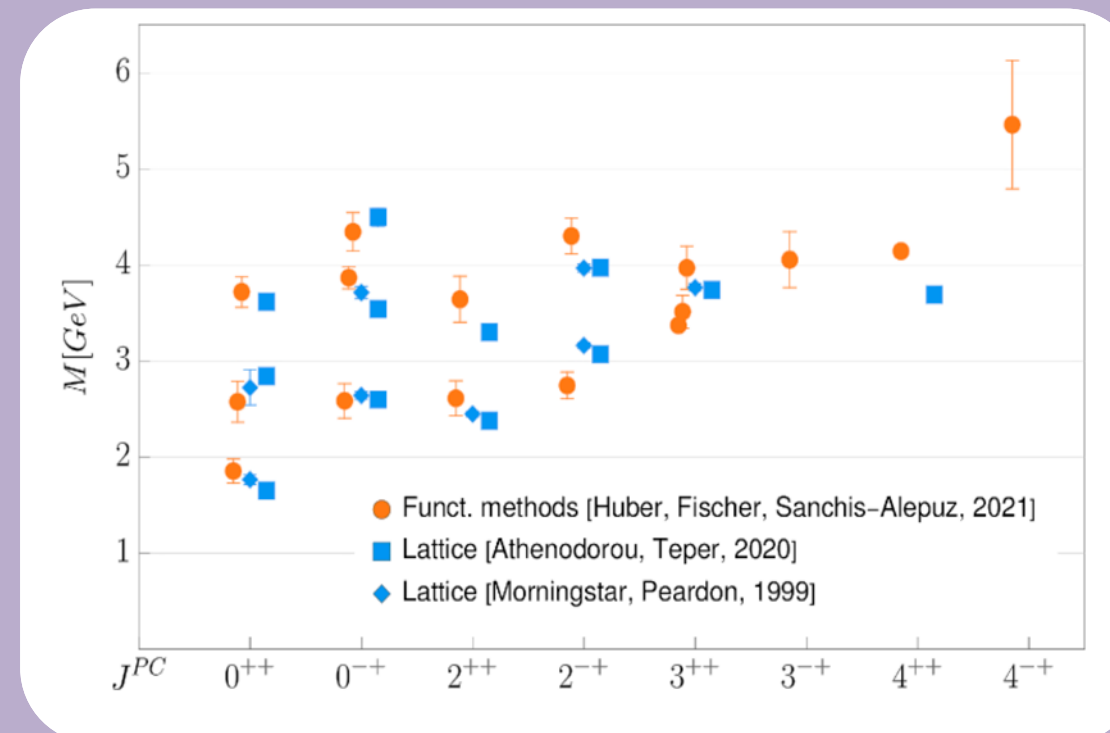

*Fig. 10.3: Glueball spectrum of pure Yang-Mills theory determined on the lattice and with functional methods.*

## Lattice QCD

Numerical simulations of Quantum Chromodynamics (QCD) formulated on a discrete space-time lattice have become an essential tool in strong-interaction physics. Lattice QCD provides a rigorous framework for defining and solving QCD non-perturbatively that is systematically improvable, thus providing stringent tests of QCD [see TWG 1]. Lattice QCD aims to provide quantitative information on nuclear and hadronic properties in terms of the fundamental constituents of matter and their interactions. Physical properties are extracted from correlation functions which are evaluated through high-dimensional integrals over quark and gluon fields (with the number of degrees of freedom depending on the size of the space-time grid). The integrals over the gluon fields are estimated via Markov-chain Monte Carlo such that predictions for physical quantities have an associated statistical error. Representative sets (ensembles) of the gluon fields are generated for lattices with a given lattice spacing and finite volume and, due to the computational expense, sometimes for unphysically large light-quark masses. The simulations are repeated for different lattice spacings, volumes and quark masses, and physically relevant results are obtained by extrapolation to the continuum and infinite volume, and physical quark-mass limits. Controlling the systematics associated with these limits requires, depending on the observable, pushing the simulations towards ever smaller lattice spacings and larger volumes. Current applications include: exploring the phase diagram of strongly interacting matter and studies of hadron spectroscopy and structure as well as of the properties of small nuclei. Since the LRP in 2017, there has been tremendous progress in lattice QCD. This has resulted from the development of new theoretical approaches, simulation algorithms and solvers as well as the possibility to access large pan-European supercomputing resources enabled by PRACE and EuroHPC Joint Undertaking (JU) and supplemented by national computing facilities.

Due to the complexity of the calculations and the huge computational resources required, the lattice community is largely organised into European and worldwide collaborations, each with its code bases, some of which are freely available. Examples are Chroma, the Columbia physics system (CPS) and MIMD Lattice Computation (MILC), which are application packages utilising the USQCD software suite and the GRID, OpenQCD and twisted-mass LQCD (tmLQCD) software packages [see TWG 1 for a list of collaborations]. Such community efforts need to continue to be supported in the future.

Frontier research in lattice QCD necessitates high-performance computing capabilities and continuous access to large computational resources. Thus, for European lattice QCD groups to maintain their competitiveness and achieve their scientific objectives, access to substantial computational resources on GPUs and CPUs (of the order of billions of core hours on conventional CPUs) is an absolute necessity. In particular, in terms of the hardware, the (mostly) double precision floating point performance must be matched with adequate memory and internode communication bandwidth. Furthermore, to effectively harness the available computing power, it is crucial to develop highly optimised algorithms and codes. This means that funding experts to develop appropriate algorithms that can efficiently run on new computer architectures is particularly important, as is the education and training of early-career researchers. Lattice QCD groups have a reputation for being early adopters of cutting-edge computer technologies. We anticipate that these groups will continue to embrace new technologies and become users of exascale infrastructure.

With the advent of exascale facilities, more long-term storage for the gauge ensembles needs to be made available in the future. The need for coordinated data handling has led to the formation of the International Lattice Data Grid (ILDG), with several regional sub-grids, including the European LDG. Support for coordinated data management and storage needs to be intensified in the coming years.

Emerging techniques based on lattice QCD have recently broken through in the field of hadronic structure. Indeed, they offer the long-sought possibility to extract information on partonic distributions defined on the light cone through lattice computations. This challenge has now been taken on by many groups in Europe and all around the world leading to significant advances. This effort needs to be supported and organised at a European level to optimise the large numerical resources required to push it forward and make it effective.

## Recommendations for Lattice QCD

Considering the above, we recommend the following:

- International collaborations and freely available software packages for lattice QCD should be maintained and extended.

- access to substantial computational resources fitted to the need of lattice computations (double precision floating point performance, adequate memory and internode communication bandwidth) is essential.

- it is important to fund the development of optimised algorithms and codes.

- Call for more long-term storage solutions for gauge ensembles, particularly with the advent of exascale facilities.

- Coordinated data management provided by the ILDG and the European LDG should be supported.

- Support for and coordination of the many European groups devoted to the study of hadronic structure on the lattice.

## Low-energy nuclear structure

Nuclear structure theory aims to describe the properties of nuclei and nuclear matter by solving the many-nucleon Schrödinger equations either in an effective or in an ab initio fashion [see TWGs 3 and 4]. The latter strategy has witnessed considerable progress in the last decade, with calculations extending beyond light nuclei to more and more complex systems. Current frontiers include pushing ab initio calculations to heavy and/or deformed isotopes, developing accurate and systematically improvable nuclear interactions and many-body techniques, as well as bridging nuclear structure and reactions.

These open issues involve specific computational challenges. The modelling of nuclear forces requires sophisticated fitting procedures in multi-dimensional parameter spaces. On top of this, meticulous Bayesian analyses are currently deployed to gain insight into the associated statistical and systematic uncertainties. Many-body techniques typically involve large-scale diagonalisations or iterative solutions of tensor networks, where the dimensions of the matrices or tensors increase with the required accuracy. Applications to nuclear reactions further augment the size and complexity of the objects to be manipulated.

This workflow entails CPU- and memory-greedy algorithms required to fully exploit the potentialities of state-of-the-art high-performance clusters. We envisage that in the coming years, many of the existing codes will be ported to exascale machines. Furthermore, in the longer term, the possibilities offered by machine learning and quantum computing will be increasingly explored by the community.

In a context where algorithmic, computational, but also conceptual boundaries are rapidly progressing, collaborative initiatives will be key for systematic and sound advances in the field. In this respect, crucial aspects include the public availability of numerical codes and the involvement of computer scientists and applied mathematicians, possibly leading to additional, dedicated manpower to support nuclear physicists.

## Recommendations Low-energy nuclear structure

- Promote cross-fertilisation from/to other fields using many-body methods and applied mathematics.

- Enhance and systematise collaborations with computer scientists to exploit new computer architectures.

● Support virtual-access facilities such as Theo4Epx of EURO-LABS.

● Encourage the sharing and publication of numerical codes.

## Computing Techniques in Experiments

An indispensable numerical tool in experimental nuclear and high-energy physics is Monte Carlo (MC) event generators. For example, MC simulations of radiation transport are used for the development and optimisation of detector systems and to aid in the interpretation of experimental results. MC generators rely on the generation of pseudo-random numbers which is typically a CPU-bound task. In this respect, a centralised computing server with very high CPU power would be a valuable resource for such applications. Since MC simulations are intrinsically parallelisable, multi-core systems can be exploited. However, this sometimes requires significant work from the user to implement, so such a system would ideally be chosen for per-core CPU performance.

Adding a powerful multi-core CPU and larger memory to the data storage and backup systems mentioned above would allow users to log in remotely and execute data analysis codes directly on the server. Such a system would reduce the need for time-consuming transfers of large data samples between multiple locations and allow sorting to be completed and/or iterated more quickly. Such systems have already been used with great benefits in high-energy physics and could be equally beneficial in nuclear physics.

A centralised data storage and backup facility would help the nuclear physics community secure large data sets collected at great expense from laboratories all around the world. In addition, it would help scientists meet their responsibilities under open-access data-sharing agreements. Such a system would need a very large array of redundant hard discs for storage but would only have basic requirements for CPU and computing power.

## Recommendations

Given the above, we recommend the following:

● Deployment of simulation software such as GEANT4, MNCP and FLUKA on fast computer systems to enable the generation and analysis of large numbers of events and the simulation of complex experimental setups.

● Adaptation of physics models or development of dedicated interfaces to extend the range of physics applications that event generators can currently deal with. Complex theoretical models can be CPU or memory-bound depending on the specific case. A centralised computing resource with high total and per-core CPU performance together with very large RAM would be of great value to the nuclear

# Resources and Infrastructure

To address the challenges posed by the end of Moore's law, the industry has embraced multi-core processors (Microprocessor Trend Data [https://github.com/karlrupp/microprocessor-trend-data]). Modern CPUs commonly have homogeneous multiple cores, enabling parallel processing of tasks. While this approach provides better performance, it has also highlighted the need for software optimisation to fully exploit the potential of multiple cores. More specialised hardware components, the coprocessor units, are needed to perform specific tasks more efficiently than the general-purpose CPU cores.

Historically, CPUs, GPUs and FPGAs have been separate entities within a computing system. As computational demands have grown, the need for efficient and flexible architectures that can handle diverse workloads has become increasingly important. This has led to the development of chiplets, also called tiles, which represent a novel approach to CPU design (Universal Chiplet Interconnect Express (UCIe) [https://www.uciexpress.org/]). Instead of monolithic, single-die processors, chiplets break down the CPU into smaller, specialised components that can be interconnected on a single package. This approach allows for improved modularity, scalability and cost-effective manufacturing. Different types of chiplets, such as CPU, I/O, and GPU ones, are already available or will soon be on the market.

Exploiting the parallel processing capability with such a heterogeneous variety becomes crucial for simulating more and more complex nuclear models, handling the vast amounts of data generated by experiments and training the deep neural networks that are increasingly used in nuclear physics. On the other hand, developing programmes to fully exploit the parallelism and heterogeneity poses several challenges: programming models that abstract the underlying hardware complexity and enable efficient use of resources and portability across different architectures are required. Fortunately, there are several models for writing portable applications on the market such as Alpaka, Kokkos, oneAPI and SYCL.

The advent of HPC exa-scale infrastructure and QC offers European researchers an unprecedented processing capacity. This complements and integrates the capabilities provided by research High-Throughput Computing (HTC) and Cloud facilities at both national and European levels. However, exploiting HPC centres comes with several challenges, such as the diversity in access and usage policies and the heterogeneous computing architectures. To address these challenges and create a cohesive data processing system, the SPECTRUM project, funded by the EU, aims to integrate different European computing resources. This includes on-premises data centres, HPC clusters and Quantum nodes. The ultimate long-term objective is to establish a European exabyte-scale research data federation and a seamless compute continuum.

As the computational demands within various nuclear-physics communities continue to grow, the performance, accessibility and interoperability of IT infrastructures have become indispensable elements for ensuring the success of research endeavours. One of the central challenges to address is the heterogeneity in the usage of computational tools, driven by the evolving technological demands and the diverse needs of research communities.

Future experiments are designed to handle data in situ without the need for hardware triggers, at high rates, while leveraging sophisticated high-level event selections. The effective operation of these experiments hinges on the successful deployment of often intricate data processing algorithms that exhibit a high level of scalability. These algorithms must run seamlessly on a variety of architectures, including FPGA, GPU, ARM and conventional x86 processors. This imposes rigorous demands on the IT infrastructure supporting these experiments, as well as on the quality, reproducibility and portability of the software, frequently developed and maintained by young researchers collaborating in the scientific field.

Similarly, forthcoming theoretical activities will become even more demanding, necessitating precision calculations for complex many-body systems to achieve significant breakthroughs in research. These activities also require the use of a diverse array of IT infrastructures, from high-performance computing centres using conventional x86 processors in tandem with GPU clusters to potential integration with future quantum computers.

The high data rates received (expected) for the next generation of particle-physics experiments (e.g. new experiments at FAIR/GSI and the upgrade of CERN experiments) call for dedicated attention concerning the design of the computing infrastructure needed online and offline [see TWG 6]. The traditional separation between DAQ, online and offline vanishes and one has to consider all these stages in data processing together; i.e. the traditional DAQ/Trigger designs are not able to handle the amount and kind of data which will come from the detection systems, and therefore more data processing is needed online. This change in the role of DAQ/Trigger systems will introduce more software-based components into the DAQ/Trigger systems than ever before. Most of the traditional hardware-based triggers are going to be (being) replaced by systems based on commodity hardware processors and co-processors running standard software algorithms. Handling such systems includes not only the design of efficient and scalable algorithms, but also the development of key software building blocks for ultrafast data processing on large-scale heterogeneous computing infrastructures, synchronisation of multiple data streams, transport services, container orchestration, and efficient binding to storage and network.

## Recommendations for Resources and Infrastructure

● Technology evolution demands significant investment in software frameworks supporting parallelism on multi- or many-core CPU and the execution of algorithms on heterogeneous platforms.

● To harness the synergies among research projects undertaken by diverse collaborations and across various IT infrastructures, it would be extremely beneficial to promote the implementation of a federated computing model. Emphasis should be placed on the accessibility of resources (data, software, and computing) and facilitating interoperability among IT activities within the nuclear-physics community and closely aligned fields, including particle and astroparticle physics.

● Developing interfaces located between the system software layer (i.e. storage, RDMA networks, container runtime, etc.) and the core software developed by the domain scientist implementing the core of the scientific analysis requires collaboration between IT experts and domain scientists; the community should encourage and actively implement such collaborations.

# Open Science and Data

**Convener:**
**Antoine Lemasson** (GANIL, Caen, France)

**NuPECC Liaison:**
**Marek Lewitowicz** (GANIL, Caen, France)

**WG Members:**
- **Hector Alvarez-Pol** (USC, Santiago de Compostela, Spain)
- **Stefano Bianco** (INFN Frascati, Italy)
- **Vivian Dimitriou** (IAEA, Wien, Austria)
- **Xavier Espinal** (CERN, Geneva, Switzerland)
- **Michel Jouvin** (IJCLab, Orsay, France)
- **Antoine Lemasson** (GANIL, Caen, France)
- **Ana Maria Marin Garcia** (GSI, Darmstadt, Germany)
- **Adrien Matta** (LPC, Caen, France)
- **Caterina Michelagnoli** (ILL, Grenoble, France)
- **Andrew Mistry** (GSI/FAIR, Darmstadt, Germany)
- **Panu Rahkila** (JYFL-ACCLAB, Jyväskylä, Finland)
- **Manuela Rodriguez-Gallardo** (Universidad de Sevilla, Spain)
- **Olivier Stezowski** (IP2I, Lyon, France)
- **Enrico Vigezzi** (INFN Milano, Italy)

# Introduction

Within the nuclear physics community and beyond, it is important to recognise the transformative potential of Open Science and reap the benefits it can bring to future advancements and growth within the field. Open Science at its core promotes the open exchange of knowledge, enables transparency, breaks down barriers, enhances collaboration and actively demonstrates sustainability by making research outputs openly accessible, reproducible and reusable.

The fundamental principles of Open Science are commonly stated as the right of access to the outputs of research with as few barriers as possible. These outputs can include scientific articles, research data, software and infrastructure. Key to this endeavour is the setting up of reformed research evaluation criteria with less dependence on journal-related bibliometrics, education of researchers aimed at providing the necessary skills to practise Open Science and the importance of significant contributions from the general public (citizen science). These principles have been adopted in the Horizon Europe funding programme [1] as well as in national Open Science plans so far published. Additional critical issues that must be factored in are the different legislation on copyrights among EU member states and the use of proprietary-owned applications and data repositories.

Embracing Open Science in general can result in several benefits:

● Scientific knowledge advancement through the dissemination of research outputs. Weakening the barriers to access these outputs accelerates the pace of knowledge transfer and means researchers can build on each other's work (and their own, in the future).

● Encourages collaboration both within the community and cross-disciplinary with other fields. The sharing of outputs and the ability to reuse these (e.g. software) can lead to improvements in quality and offer alternative solutions to problems.

● Enhances transparency of research practices, promoting trust between community members on a global research level and to the general public.

● Strengthens and promotes innovation and technology transfer with industrial partners.

● Attracting future researchers: advertising the research outputs as 'open' demonstrates the field as forward-thinking and inclusive. The collaborative nature of Open Science can appeal to the next generation and attract a diverse talent pool to support the future of the community.

● Sustainability is improved by Open Science: rather than repeating the same work tasks, resource-sharing can eliminate waste while preserving scientific competitiveness.

● Research assessment reform: offers an alternative to traditional, outdated metrics.

The purpose of integrating Open Science within the NuPECC Long-range plan is to outline the importance of embracing Open Science principles within the nuclear physics community and the methodology for implementation.

This chapter discusses the benefits and applications of Open Science within the community and explores current and future perspectives for the community. This is divided into the several 'pillars' of Open Science, namely: Open Science developments, Open Access publications, Open Data and lifecycle, Open Software and workflows, Infrastructures for Open Science, and Nuclear data evaluation.

# Open Science Development

## Open science policies

Open science policies form the foundation stone of open science practices, playing a vital role in highlighting the benefits of open science, raising and increasing awareness of ongoing initiatives and laying out a series of best practices. These are present on an international level (e.g. UNESCO recommendation on Open Science - https://doi.org/10.54677/MNMH8546), national policies (examples given below), or on an institutional level, and may be focused on one particular aspect of Open Science. Within the nuclear physics community, these policies should be highlighted and promoted at different levels.

A comprehensive overview of Open Science policies in seven European countries can be found in [4]. These policies typically vary in scope and implementation but have the same underlying message of strong support for open science practices. In particular, emphasising open access to research results, data, software and workflows, while highlighting the benefits to researchers and beyond. Additional insights can be gained from recent statements of intent and plans across Europe: Second French plan for Open Science (2021-2024)" [5]; The Italian "Piano nazionale scienza aperta (2022)" [6]; The Spanish "National Strategy for Open Science (Estrategia Nacional de Ciencia Abierta, 2023-2027)" [7]; The German research foundation (DFG) [8]; and the Finnish Declaration for Open Science and Research [9]. These examples highlight the increasing recognition and significance that Open Science has gained within the community.

While inter/national policies form the basis for promoting and highlighting open science, policies on the institutional level play a crucial role in defining the exact open science procedures and techniques that researchers should use in their projects. These can include policies that state mandatory aspects as well as broader representative guidelines which include specific examples. In addition, institutions may choose to publish several complementary policies focusing on a particular aspect of Open Science. For example, a set of policies and guidelines on research data management which couple data management aspects with open science initiatives.

## Evaluation in the open science and publication era

The rise of open science has led to emerging ways of evaluating research and publications, which are more transparent and inclusive. Traditionally, the impact factor of a published article was used as an indicator of the 'quality' of the research. However, this is deemed to be an insufficient metric, as it does not assess the impact on society or the quality of the underlying research. Open Science enables alternative methods of evaluation, such as open peer review and external collaborative evaluation. As an example, assess the level of 'FAIRness' of a published dataset using metrics such as F-UJI score and then use this to optimise the strategies and infrastructure for publishing open data and software code. These new methods will aim to provide an enhanced evaluation of research while promoting open science values. The community should promote and commit to reforming research assessment through collaborative efforts and shared ethical principles like the COARA European initiative.

## Promotion and recognition of Open Science and data activities

To develop a widespread culture of openness in research, it is important to establish a framework that promotes the benefits of Open Science, rewards researchers who follow its principles and widely communicates these benefits.

Researchers and communities should be encouraged to adopt Open Science practices. As examples, the following strategies could be employed within the NuPECC community:

- Generate award schemes and provide visibility to researchers and institutions that make contributions towards Open Science endeavours.

- Offer training and resources to researchers to develop practices within the community, on an institutional, national and international level (in line with the European Open Science Cloud (EOSC))

- Develope collaborative platforms and tools to enable researchers to work together and provide contact points and networks.

- Set up Open Science policies on an institutional and national level as a guide for best practices and to provide a marker for support.

- Offer public outreach and incorporate Open Science into education to raise awareness and promote understanding of Open Science practices.

- Support Open Science clusters such as ESCAPE and PANOSC along with future activities to ensure that the goals of Open Science are met and receive the necessary resources for implementation within NuPECC.

## Coordinating efforts within the community and across the domains

A wide variety of national and international Open Science projects and initiatives are ongoing, aimed at supporting Open Science practices by offering infrastructure, resources and support to the research community. Currently, the nuclear physics community has limited implication in these initiatives. Another important aspect resides in utilising existing infrastructure such as Open Science platforms and repositories, ensuring sustainability and identifying common standards and best practices among the projects. The EOSC plays a key role towards this through coordination and alignment, and will continue to develop these actions. The current scenario in Europe in the quest for sustainable and efficient Scientific Computing, Open Science and Open Data are boosting collaborations to investigate and test approaches towards common models and tools. The strong implication of stakeholders in the nuclear physics community in present and future projects represents a major asset to accelerate the adoption and implementation of open science practices. An example of this is the ESCAPE project, described in Box 11.1.

## Encouraging Open Access Publication

Open Access (OA) publications are crucial to the active, efficient, fair and sustainable dissemination of academic research. Several models of OA publication are shown in Fig. 11.1: diamond, green, gold or hybrid gold. In Diamond OA, the journal publishes the article and makes it available with no fees involved for the author or reader. The publication costs are covered previously by institutional support. The 'Action Plan for Diamond OA ', launched by Science Europe, is one of the drivers of this, proposing alignment and development of common resources to enable diamond OA publishing.

### Box 11.1: The ESCAPE Open Collaboration

**The European Science Clusters approach is a good example of an institutional effort to connect scientific communities by providing services for interdisciplinary research, in particular the ESCAPE project (https://projectescape.eu/), with the recently announced long-term ESCAPE Open Collaboration Agreement. This collaboration has been signed by the Directors of Landmarks Research Infrastructures, such as the European Organisation for Nuclear Research (CERN), the Cherenkov Telescope Array Observatory (CTAO), the KM3NeT Neutrino Telescope Research Infrastructure (KM3NeT), the European Gravitational-Wave Observatory (EGO-Virgo), the European Southern Observatory (ESO), the European Solar Telescope (EST), the Facility for Antiproton and Ion Research (FAIR), the Joint Institute for VLBI-ERIC (JIV-ERIC) and the Square Kilometer Array Observatory (SKAO).**

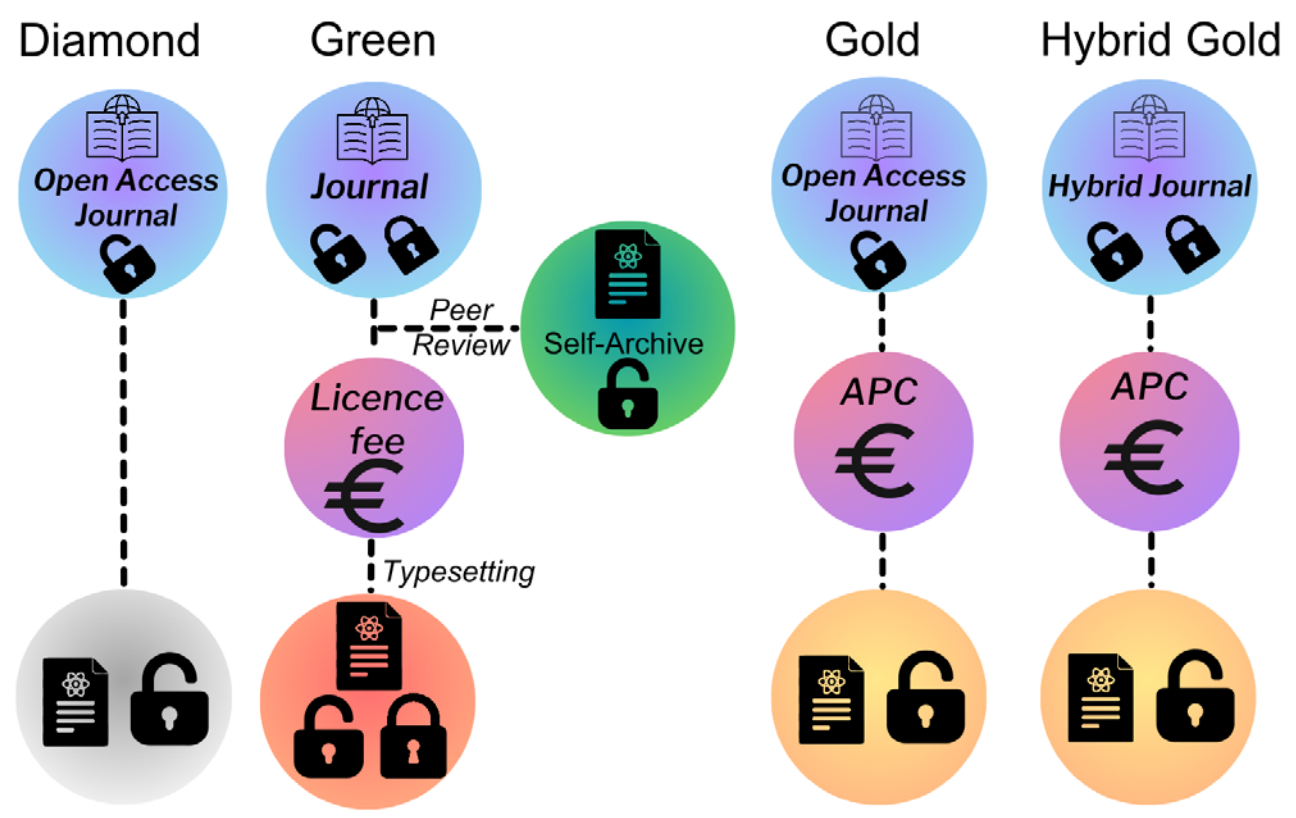

*Fig. 11.1 Depiction of the different categories of open access publications. See text for further details. Diamond OA: content is freely available with no charges to authors or readers. Green OA: allows a version of the work to be deposited in an open-access repository (self-archiving), while the publisher's version may be behind a paywall. Gold OA: content is made available freely, an article processing charge (APC) to be paid by authors or institutions. Hybrid Gold: a traditional subscription-based journal offers the option for open access subject to an APC.*

- Diamond OA should be recognised as the pinnacle of good practice in publishing, and be set as a long-term target across the nuclear physics community. Diamond OA should be strongly supported; resources for community-based diamond access journals are essential. At this time, only a few diamond journals exist and the incentives for researchers to publish in such journals remain limited due to, for example, low low-impact factors.

- As a short-term plan, researchers should be encouraged to publish in green OA when possible.

The scientific community should be aware of the importance of reforming intellectual property and copyright law, maintaining communication with institutions and organisations active in the reform projects at both national and European levels, and possibly participating.

In summary, open access journals should continue to be supported and resources dedicated to support diamond open access journals. Researchers who publish open access should be incentivised to do so.

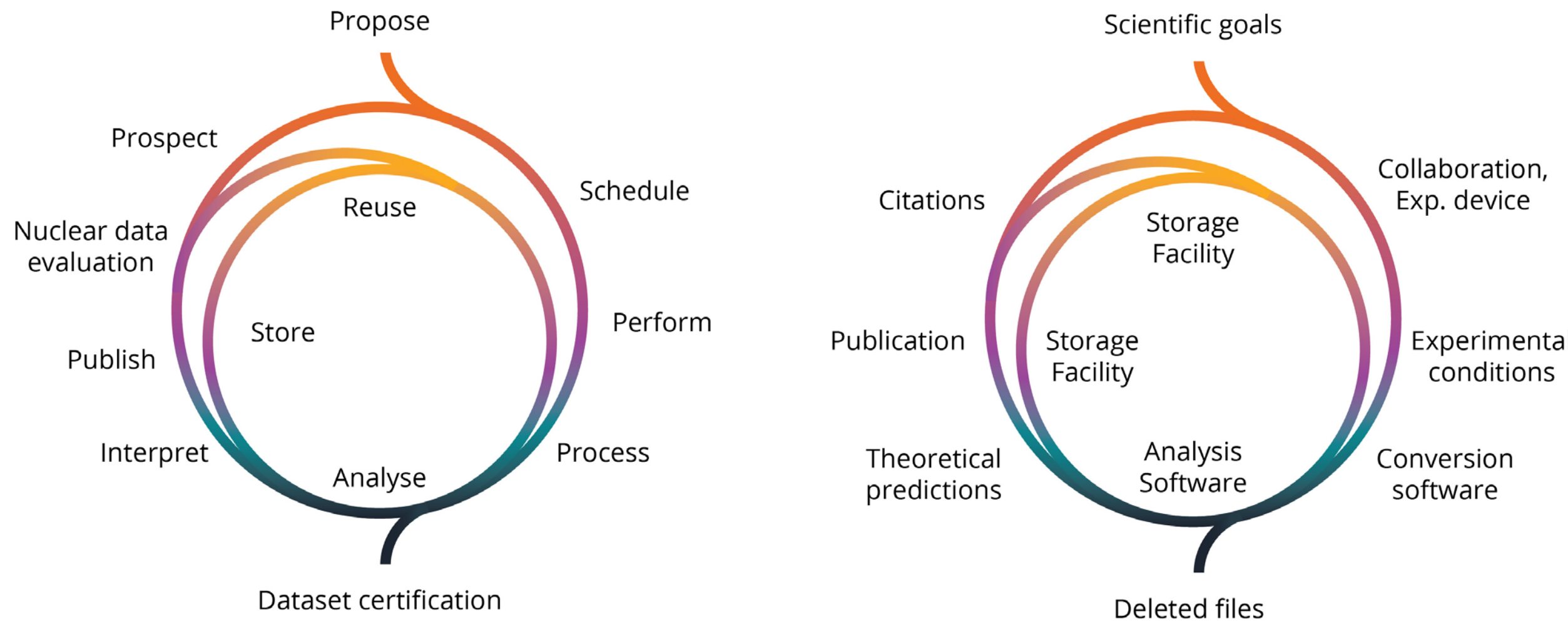

*Fig. 11.2: (left) Schematic view of the Data Life Cycle in nuclear physics and (right) example of associated metadata that could be accumulated during the lifespan of data.*

## Recommendations:

- Advocate for the creation and adoption of open science policies and guidelines within individual institutes.
- Allocate sufficient resources to ensure the successful implementation of these policies.
- Encourage joint initiative across scientific domains.
- The community should promote and commit to reforming research assessment through collaborative efforts and shared ethical principles like the COARA European initiative.
- Encourage the development and support of Diamond OA journals in the field.

## Box 11.2: Open Science in practice at ALICE

**CERN committed to an Open Science policy as a key to maximising the global impact of research conducted at the facility (https://openscience.cern). Among the different actions, CERN established an Open Data Policy, to be followed in a consistent approach by all LHC experiments. This was endorsed by the ALICE Collaboration in November 2020. The policy commits to publicly releasing level three scientific data: namely, the input to most physics studies together with the corresponding Monte Carlo, as well as the software and documentation needed to use the data. The aim is to start data releases within 5 years of the conclusion of the run period, while the full datasets would be made available at the end of the collaboration. CERN has set up a common Open Data Portal (https://opendata.cern.ch) where collaborators can upload their data samples. For example, ALICE uploaded 5% (7%) of the 2010 Pb-Pb (pp) event summary data totalling 6.5 TB. An analysis demonstrator is also available to perform analysis processes directly through the CERN Open Data Portal.**

**ALICE aims to constitute data from the first two runs of publicly Open Data from 2024+ (105 TB/year) using a new data format for resource optimisation. Data corresponding to the third LHC Run will be made Open from 2030+ (2 Pb/year). Dedicated staff and resources have been allocated to achieve this.**

**The international masterclass programme is a successful example of the Open Data policy.**

## Towards Open Data Life Cycle in Nuclear Physics

The nuclear data evaluation community has developed a very efficient and robust evaluation pipeline over several decades. This pipeline allows for the production and maintenance of a variety of high-quality curated nuclear data repositories used in many societal applications. Inspired by this approach and its success, the nuclear physics community aspires to reaching a similar situation when it comes to research dataset repositories.
The concrete implementation of FAIR (Findability, Accessibility, Interoperability and Reusability principles [3] to the large variety of research datasets produced by the diverse nuclear physics community is in itself a challenge. To achieve this goal, it is important to perceive the research data life cycle (DLC) as a production process that demands a commitment to ensuring its quality (See Fig. 11.2).

The DLC should commence upon submission of the experimental proposal and conclude upon the publication of the final physical observables, with proper data storage for reuse and revision and a proper policy for unprofitable data removal. At every step along the cycle, collecting comprehensive metadata is crucial to ensure that FAIR principles are followed. Efforts in standardising metadata are required to ensure reproducibility and interoperability of datasets.

The nuclear physics community faces a few specific challenges: solutions developed by connected fields are not always adapted, as in the heterogeneous sizes of the collaboration, ranging from individuals to hundreds of collaborators (e.g. AGATA); the size of the detection setup from single electronic channel up to complex multidetector system (e.g. R3B-GLAD-NEULAND); and the various lifetimes of an experimental setup from single experiment up to several decades of operation (e.g. INDRA). As a result, it is both unrealistic and impractical to expect uniform technical implementations, such as a common Data Management Plan, common file formats or a common repository across the diverse nuclear physics community. However, the definition of standard requirements, such as the existence of a Data Management Plan, documented file formats or the usage of a repository represents an achievable goal for the community. Such standards should be easily scalable depending on the collaboration size, meaning little to no effort for simple and small experiments. The standard is bound to evolve with practice and the incorporation of new technologies, and a revised standard should be produced according to a planned schedule.

The formalisation of such a production process will require coordination between all actors in the chain. Physicists, IT departments, facilities, data evaluators and funding agencies should be involved in the identification of DLC requirements to demonstrate the quality of the conducted research. The development of common open standardised ontologies frequently used by researchers in the field for the data and metadata description is essential to allow access and sharing of information across the fields.

To solve the complex problem of standardisation, other communities have developed a consortium approach. A good example is the ISO C++ committee overseeing the improvement of the language through a three-stage pipeline involving different working groups. A similar approach could be developed for the research data community wherein domain-specific groups (e.g. beam production, detectors, analysis, etc.) could propose new standardised metadata fields to be included in the collection pipeline in the first stage. In a second stage, the metadata collection operation would be reviewed, to guarantee that proper process and tools exist in the real world. Finally, a wording and consistency stage would take place to enrich the common open vocabulary consistently. The creation of such a metadata collection pipeline would set, and update, standard practices.

Moving towards a functional DLC, a new community of research data specialists should emerge and form the backbone of the thematic working groups of the consortium.

Research data stewardship will be overseen by data officers. The data officer tasks will be handled by members of the collaborations or research groups (e.g. a physicist). His/Her role will consist of defining the collaboration data policy (e.g. DMP) and ensuring that the DMP is properly executed at every step of the DLC, making sure the appropriate tool and infrastructure exists, and that research datasets are readily available.

A second role to emerge is that of data curators, tasked with ensuring that the metadata schemes are interoperable between data sets and that the technical solutions follow the community standards. Typically working at data-producing or processing institutions, these curators will also play an essential role in proposing new data fields to the consortium, as well as educating the community on renewed standards.

During the ‘propose and schedule’ phase of an experiment, an initial metadata pool is built, including scientific motivation, partial authorship and experimental details. This information will help facilities and collaborations to further optimise experimental conditions. In addition, clear policies, using DMP, are integrated into the metadata scheme. Ideally, more advanced information related to the experimental designs (mechanical, acquisition and designs, etc.) should be deposited. This metadata should be published on identified catalogues supported by the community to ensure the ‘findability’ of the datasets.

While the experiment is running, auxiliary data describing the experimental conditions will be gathered. These auxiliary data take various forms today, from physical notebooks to their electronic counterparts. An essential programme needs to be run within the community to create new tools adapted to the reality of this difficult data collection. Interoperability between the machine and the detector is, for instance, a friction point in schemes deployed today. Auxiliary data taken in an archivable form, with the ability to mix automated and human input, is therefore extremely important to be able to analyse the data in new contexts.

The processing step consists of converting the raw data into a more efficient format. Most data acquisition systems produce binary files in native format, with a premium on software efficiency and simplicity. However, this type of format usually performs poorly in size and reading speed-wise. A simple conversion from a native DAQ format to a readout-optimised, higher-level file format (e.g. root tree, hdf5) should be provided. Such software should not alter the data in any form (calibration, selection, etc.) but limit re-organisation and optimisation. The software developed for this process phase will be integrated into the metadata pool from the experiment.

Data analysis requires both analysis and simulation software. An essential change in paradigm is the acknowledgement that software is an integral part of the dataset and should be treated as such. An overlooked issue today is the preservation of analysed and simulated files during the analysis process. Most current DMPs in the field focus only on the raw data with the idea that any other file is in principle reproducible. However, a significant part of the manpower effort of an experiment is assigned to the data analysis process between the experiment and the publication of physical results. It is therefore essential that these datasets be included in the DMPs. All software used for analysis should be open source, documented, preserved and referenced in the metadata. Ideal cases would include the production of running images of the software and its environment (e.g. container, virtual machine) for long-term use. The full panel of technical possibilities will be discussed in the subsequent section on software development and section on shared computing infrastructures.

When the different files produced during this process are stored and curated to form a consistent dataset, the metadata scheme should be enriched with information on storing facilities, file format and embargo information. This will allow proper access, preservation and curation of the data. At this stage, a first selection of the file relevant for future use could be made. This process of data file decimation will contribute to the sustainability of the storage infrastructure with careful and time-sensitive suppression of files no longer relevant to the data set while keeping trace of their past experience. A concrete example could be the choice to store untreated data only in a space-efficient file format and unstage the original DAQ native data file at the end of the initial analysis. Similarly, untreated data files could be removed after a defined period post-publication to retain only refined data. Each of these options should be described in the DMPs.

Long-term data preservation represents a challenge for the scientific communities with fast- evolving environments and large-size datasets. Accumulating all data produced over a long period is neither a sustainable goal nor a pertinent use of the community and societal resources. DMP and catalogues should be the tools of choice for the community to forecast and monitor storage use for research datasets and help design a sustainable and ecology-conscious storage infrastructure in line with our open science objectives.

### Box 11.3: At the forefront of FAIR and Open data at the Institute Laue Langevin

**The Institute Laue Langevin was at the forefront of data management by publishing a ‘Scientific Data Policy’ in November 2011. The text came into force in October 2012. Following the publication of the policy, the ILL created an interdisciplinary working group, the DPP (Data Protection and Processing) to drive the development of the software tools needed to put the policy into practice, with a focus on usability (especially during experiments) and security. The data portal (https://data.ill.eu) was launched in 2014. DOI persistent identifier enables readers to obtain more information about the referenced experiments, access the ILL Data Portal and even request access to the experimental team if the data are not yet publicly available. The DPP continuously upgrades the data policy so as to reflect the evolution of the Data Management tools available at the ILL. Further, under the PANOSC (Photon and Neutron Open Science Cloud) scientific cluster, the VISA platform was developed to provide an integrated environment for data analysis (remote desktops, jupyter notebook) to the scientific community. It has a strong impact on capacity and productivity of scientists by facilitating the access to data and analysis programmes as well as exchanges.**

Publication or citation of the data should be included in the metadata scheme, providing clear metrics for the usefulness of the data. Through dataset citation in publication using DOI or another machine-readable identifier, the metadata scheme will be enriched with this information.

All the produced metadata should be machine-readable, allowing interaction between different catalogues and services, whether at the community-wide or facility-specific level. The richness of the accumulated metadata will provide an unprecedented understanding of the dataset, its scope and limitations, and will help with the 'open review' process to increase the scientific quality, nuclear data curation for societal use, and re-use of datasets in new and unforeseen contexts.

At the final step of scientific production, published results are reviewed and curated by the nuclear data community, thus disseminating incremental and reliable knowledge for societal benefit. The long tradition of a worldwide open collaboration offering the net output of our scientific community to society has a profound impact on the world we live in. From nuclear energy to medical use and going through space exploration, nuclear data play a major role in modern society. Our rich and fruitful community produces a vast number of results, and the curation process of the nuclear data community faces new challenges, as discussed in chapter 7. By producing a catalogue of research datasets to the highest standard, the nuclear physics community will grant access to all necessary information for this review process to take place in the best and most efficient way, increasing our field's impact on society.

The availability of open research datasets and associated analysis and simulation tools will also play a major role in training the next generation of professionals on modern data analysis techniques and simulations. With lectures and schools based on FAIR experimental data, a wide range of students will benefit from the community effort, helping boost the field and its reach within society.
To reach this ambitious goal the community needs to establish new standard practices. This includes a new collaborative framework, letting scientists take stewardship of their data's future. A set of new tools, particularly a comprehensive community-wide data catalogue and a methodical approach to aggregating auxiliary data must be developed.
To this end, we recommend coordinating this effort through the creation of a consortium of the different actors committed to designing, maintaining and updating a 'real-world proven' metadata scheme for the community at large.

## Recommendations:

● Strongly support the application of the FAIR Principles: encourage training and investment in human resources for data management (data officers, data curators etc.) at the various levels (institutions, labs, collaborations) to effectively advance open data practices.

● Support the creation of coordination bodies to pursue standardisation of the DLC to ensure interoperability. Work on tool and guideline development for researchers and collaborations to help real-world applications.

● Engage in active collaboration with other communities (e.g. ESCAPE for HEP/COSMO/ASTRO and PANOSC for photons and neutrons).

## Open software and analysis workflows

In the last two decades, the community has undergone a drastic change in research software development practices. The progressive percolation of industry standards within the research community (in particular, the use of tools improving collaborative development and code quality) has led to the emergence of software specialists along with dedicated infrastructure, such as code repositories. The overall quality of developed software has increased while manpower dedicated to software development is now used much more efficiently.

The community has transformed, progressing from the development of a constellation of small, usually short-lived software packages to the establishment of larger collaborative software packages, maintained over several years by an active group. This has led to a redistribution of development manpower towards the enhancement of new features and the improvement of the scientific quality of the software.

This shift has enabled the community to maintain, and sometimes reduce, the time gap between data taking and publication, while the complexity and data volume of experiments has increased with the widespread use of triggerless and/or digital electronics and multidetector systems.

This rich background and quickly evolving environment pushes the community to the forefront in modular software packages and innovation in the domain. It has become possible to run heterogeneous acquisition systems giving a smooth hardware development in the community. On the analysis and simulation side, the np tool package permits fast and efficient mixing of simulation and analysis of multidetector systems. It has brought modern tools to a large variety of smaller experimental collaborations such as MUST2, SHARC, FATIMA and ISS at a low development cost. FairRoot, on the opposite side of the spectrum, has provided a solid framework for large-scale, fixed-setup experiments such as CBM and R3B and offered a variety of ready-to-use tools for high-performance computing that has led to the O2 collaboration for the ALICE experiment at LHC.

As software packages become more widely adopted, associated collaborations need to be formally built. Those collaborations are today very informal in most cases, with little recognition from their host institution. Two situations typically arise: either the code is limited to a single collaboration and managed as a sub-task of the formal collaboration, or the code is not tied to any collaboration and no formal collaboration is recognised, and therefore no resources are allocated. Institutions should take ownership of the software production on a par with detector manufacturing and manage them as formal and recognised collaborations, effectively promoting the effort made by developers to adopt good practices in software development.

With such formal collaboration, the challenging question of the end of life of software can be managed properly. This involves the careful consideration of options such as maintenance, evolution or replacement by a new project based on more efficient technologies, or even complete retirement should the application not be needed anymore. For the latter, the collaboration should archive and document the code appropriately to guarantee that the community does not lose knowledge in the process.

Software understood as an interface between developers and hardware produced in a given language and programming paradigm may depend on running environments and third-party libraries, but renders some overall performance. All these elements constitute an essential context which should be documented to ensure usability, maintenance and preservation.

In the case of hardware-specific code whenever using FPGA, GPU and co-processor or quantum computers, tension between performance and long-term support is also a challenge. All these promising tools require the maintenance of legacy hardware or virtual emulator environments to keep the data reproducible. FPGAs are critical components of the data acquisition chain, dealing with data readout, treatment and decision. In most cases today, firmware used in embedded electronics is fully closed and unavailable to the community. Finally, quantum computing presents great potential in our field, particularly for theoretical prediction. Early works on the subject have already started; however, the field is still new and no standard practices exist yet.

The Nuclear Physics landscape is intrinsically extremely rich, encompassing a wide variety of physics cases, numerous theoretical models spanning large energy scales, many laboratories and collaborations of various sizes, and many detectors assembled in dedicated and often short-lived configurations. As a consequence, over several decades, countless software projects have been completed for different purposes. While the outcome of developed software has been largely published in papers and many results are available in well-organised, evaluated databases for experimental results, the software itself has not been systematically published and referenced. As a result, much software and its related data is not usable any more, if not outright lost. The consequences of such a state may be different depending on the goal of the application and the availability of similar operational codes. With the trend towards open data, software and science, particular attention should be dedicated to that part of the Nuclear Physics heritage. Currently, operational applications should operate their migration - if not already started.

While still quite fragmented, the Nuclear Physics landscape has evolved over the past decades, notably thanks to projects requiring important resources (large research infrastructures developing high-intensity or radioactive beams, complex travelling arrays and high-computing-demanding applications). Following the example of large projects such as Geant4 or ROOT, some practices have been developed towards collaborative, distributed open software: version control systems (such as git for the most recent one) are now pillars of many collaborations (for instance ALICE, FAIRROOT, …).

Release versions and publications (including in dedicated repositories such as Zenodo, and licences) of well-documented codes tend to be more and more general. For example, the ESCAPE Open-source Software and Service Repository (OSSR) is intended to provide a central location for the dissemination and use of trusted open-source software in the fields of astronomy, astroparticle and particle physics. The repository is regularly updated and maintained by the ESCAPE project team, ensuring that the software and services are reliable and up to date. Moreover, the Matter and Technologies (MT) research programme in the Helmholtz Association has also adopted the OSSR as the standard open-source repository to be used for diverse software projects for simulations and data analysis in the Helmholtz institutes. The OSSR serves as a central location for the dissemination and use of trusted open-source software in various fields. Based on the all-purpose repository Zenodo, the OSSR is designed to make it easy for users to find, access and download the software and services in their community, and to contribute their tools and services if they wish. The curation process taking place during the onboarding of software ensures the quality of the contributions and that they comply with the FAIR principles. This makes the OSSR a source of trusted software that can be used for scientific analysis.

As in the case of data, software can be subject to embargo periods. This might be particularly true for complex theoretical codes since the software itself is the source of scientific progress, contrary to, for instance, data processing code for which the potential of discovery is expected to be inside the data stream. While embargo periods are understandable, there are clear benefits to fully opening any codes since this can also be associated with clear evaluation processes and measurable merits. Thanks to modern tools such as GitHub or GitLab, some collaborations have included in their software production pipelines systematic continuous integration and deployment approaches. This is, however, probably not as widespread as the use of version control systems. It should be noted that testing the reproducibility of codes is common practice even if it is not based on modern systematic ways.

With open software, any researcher has the potential to check any result: however, this is true only if there is access to the necessary data and meta-data. The processing pipeline may in many cases not be as simple as running a single application. This requires not only that the source code should be open but also necessitates a solid description of the workflow that has been set to obtain a result. Only very few collaborations have started to investigate this issue in Nuclear Physics. The systematic adoption of machine-readable workflow descriptions should be made a priority to ensure the reproducibility of research results, for example by the use of containerised applications. Such usage seems marginal in the NP community and more effort should be dedicated to this kind of technology. Ultimately, this path should lead to full environments running reproducible data analysis on shared computing platforms.

Reproducibility of research results is a pillar in open science and this requires, firstly, having open software to fully understand how a particular outcome (experimental or theoretical) has been obtained. Versioning tools provide detailed tracking of changes made in a given software package and automatically generate unique revision identifiers that can be tied to a software output as metadata. As more and more scientific results require complex processing pipelines, descriptions of the processing workflow should also be documented.

Containers or virtual machines offering portability and relatively long-term preservations are likely to be heavily deployed. This allows efficient systematic software quality checks, continuous integration (CI) and deployment (CD). With the wide adoption of institutional software repositories, CI and CD are already becoming mainstream practices within the community. To accelerate the adoption of these practices in all collaborations, the definition of standard practices at the institutional level is the best course of action.

These practices are also mandatory with the rise of AI and Machine Learning (AI/ML) approaches. For instance, Neural Networks do require training phases which are based on input datasets and rely on hyperparameters to be tuned. With this new, additional, link to data, the processing workflow becomes even more complex and should be managed properly with dedicated tools. AI/ML is still in its early stages and the field is rapidly evolving. The coming decade will be decisive in shaping common practices and tools, and the community needs to invest in training to keep up with this evolution. (see also chapter 10)

## Recommendations:

- Encourage formal software collaborations to improve structuration and oversight, and enable institutional recognition through awards and career advancement.

- Invest in software development methodology training: versioning, collaborative development, CI/CD, AI/ML, workflow management

- Formalise the primordial role of software in the reproducibility of scientific results through systematic software publication and software sessions in workshops and conferences.

# Infrastructures for an effective open science

In the nuclear physics community, research activities often lead to the generation and treatment of large volumes of data. It is typically impractical to simply download and process these on single machines. it is necessary, therefore, to develop an ecosystem of infrastructure that allows the end user access to data, software and computing resources, to follow the analysis workflow and visualise the outputs in a meaningful way. Strong synergies with other domains of physical science like particle physics and astrophysics, where significant steps have already been achieved, will strongly benefit the nuclear physics community.

## Authorisation, Authentication and Identity Management (AAI)

The goal is to provide a recognisable infrastructure with a framework of common tools to enable the Nuclear Physics community, experimentalists and theoreticians, to exploit distributed computing resources and enable data management capabilities. The first fundamental step to achieve this is to agree on a common layer of trust, allowing for seamless user integration identities and those of experimental experts with computing resources and experiment workflows. The AAI frameworks should provide the users with a unique entry point and view to perform their work: data analysis, software development and experimental data management. This can be achieved by adopting technology standards promoted in other disciplines and by resource providers. In addition, the chosen federated identity infrastructure should ensure connectivity to other identity providers (IdP) in the community to enable access to computing resources (CPU and storage), data/metadata catalogues or infrastructure providers. The chosen federated identity infrastructure should permit the experiments, laboratories, projects and collaborations to manage user memberships.

## Data Storage Infrastructure and Interconnectivity

The community would greatly benefit from agreeing and adopting a common data model with an agreed set of tools to establish a common and coherent data storage and data management infrastructure for the Research Infrastructures (RI) in the NP community. This would facilitate the administration, operation and manipulation of the experimental data by the RIs, potentially having a positive impact by improving efficiency and costs for the experiments' IT managers and those sites providing and running resources for one (or many) of the RIs.
This data storage infrastructure is commonly referred to as the Data Lake and includes a common set of tools for Data Management, Data Transfer, Data Access and a common AAI framework. The Data Lake displays the dataset of a scientific community as a single repository and eases the implementation of data life cycles, policies and access rules to cater with data FAIR-ness. It provides a common interface and standard protocols to expand data processing capabilities and facilitates access to the data from user laptops to the lab's batch systems or HPC/clouds. A conceptual scheme of the infrastructure is shown in Fig. 11.3.

There are several fundamental aspects to consider, inherent to the data infrastructure:
● The data infrastructure should ensure the accommodation of full life cycles of the data: from raw data recording to data distribution for user access. Also covering long-term needs in terms of Data Management Plans, Data Preservation and Analysis Preservation (Rucio is the data management system widely used in LHC and adopted by ESCAPE).

● A Data Distribution and File Transfer service should be selected to guarantee a high-level transport layer with the required protocols and interfaces with the storage systems in the data infrastructure. This service acts as the 'middleware' to provide third-party copy transfers without proxying the data (i.e. as FTS has been used on ESCAPE).

● The data infrastructure should be flexible and able to be used in a very simple manner for end users, hiding all the complexity. It should also cater for high-level data management experts' needs: access control, data replication and pre-placement life-cycle.

● The data infrastructure should be able to deliver data and provide access from varied types of processing facilities: large batch systems at the experiment's facilities and labs, computing clouds (private and commercial), and potentially to HPC centres. The mechanism to optimise remote data processing is to leverage content delivery networks and latency-hiding mechanisms (data caching services) interconnected with the Data Lake.

## Analysis Platforms

Several nomenclatures are given to the new trends in performing (visual or interactive) data analysis: Analysis Facilities, Analysis Platforms or Virtual Research Environments. The scope of these frameworks is to facilitate the development of end-to-end physics workflows, providing researchers with access to an infrastructure and to the digital content necessary to perform a scientific analysis and preserve the result in compliance with FAIR principles. This is a rapidly evolving field as scientists' methods to perform their analysis are changing. These (usually) containerised environments can hide the infrastructure's complexity from the user and provide: 1) Seamless access to the experiment data by having browsable catalogues and easy (clickable) download and upload options to/from the Data Lake, 2) Access to the latest experiment software (mounted in the framework), 3) access to external repositories for user codes. Recent initiatives within the EURO-LABS and STRONG2020 European projects aim at the deployment of such Virtual Access Theory platforms for the community.

The community should follow up on trends of new analysis tools and methods used by the new generations of scientists and actively participate in the activities and projects, developing prototypes for the new generation frameworks for user data analysis.

## Analysis Preservation and reproducibility

The awareness of 'Analysis Preservation'' and the need to move towards consistent reproducibility should be enforced. The required tools should ideally be embedded and facilitated by the Analysis framework itself, providing researchers with an easy way to package and store the analysis for future and sharing purposes. It is important to highlight that this does not apply only to experimental data, but also to the experiment software, user code and the framework to run the analysis in a stand-alone way (OS, software packages, etc.). Such tools exist and some Funding Agencies, experiments and communities have started to enforce compliance by publications with the Analysis Preservation, given scientific computing FAIR-ness and open data and open science.

The reproducibility must be extended to any theoretical calculation that is published. So the suggestion must be that all publications should include detailed information on the inputs (potentials, integration parameters, etc) necessary to reproduce the calculations.

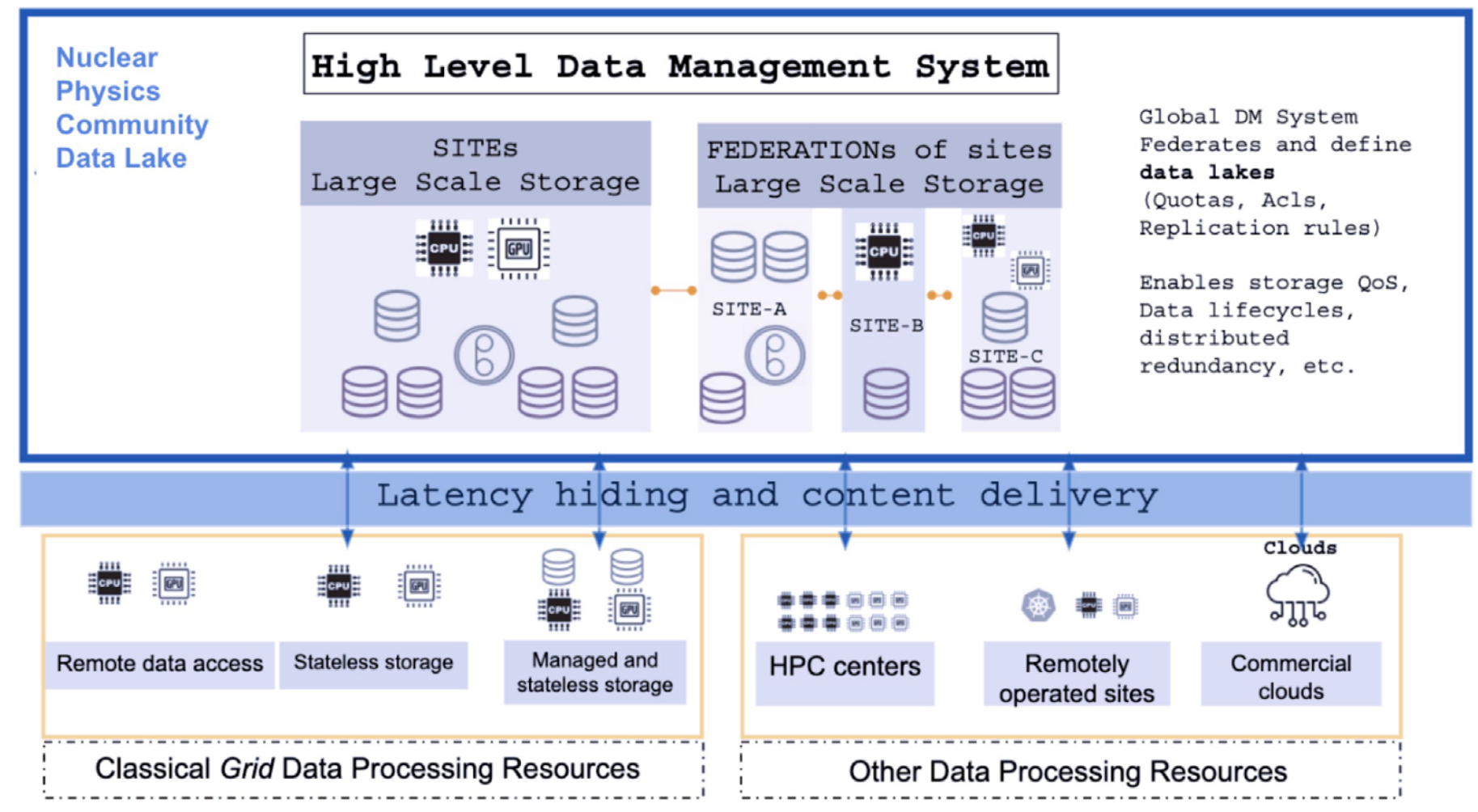

***Fig 11.3: Data Infrastructure (Data Lake) conceptual view, adapted from ESCAPE Data Infrastructure for Open Science report https://arxiv.org/abs/2202.01828v1. A distributed data repository is perceived as a single entity and can deliver the data products to a heterogeneous set of facilities for data processing and data analysis purposes. The ecosystem is formed by labs and sites providing resources to the Nuclear Physics experiments but is also open to embracing punctual contributions from external providers (e.g. clouds and HPCs).***

## Recommendations:

● Investigate the feasibility and interest in the community in having a common distributed data infrastructure (Data Lake) for the Research Infrastructures (RI) in the NP community. To facilitate Data Management, Data Processing and the enforcing of data FAIR-ness policies.

● Strong suggestion to adopt technology standards being promoted in other disciplines and resource providers for Identity Management, Data Management, Data Analysis and Reproducibility/Preservation. To search for commonalities with the potential to bring economies of scale.

● The various stakeholders (researchers, collaborations, research infrastructures and institutes) should actively contribute to joint initiatives and technical developments in coordination with scientific communities for the deployment of scalable infrastructures adapted to their specific needs.

# Nuclear Data (Databases) and Evaluation

Nuclear data refers to curated data describing the properties of atomic nuclei, nuclear decay, cross sections for nuclear reactions and other quantities relevant to nuclear science and engineering. These data are typically obtained through experimental measurements performed in small- and large-scale nuclear physics facilities. Apart from their direct use in a host of applications including nuclear energy, nuclear waste management, nuclear safety, non-proliferation, nuclear medicine, environmental control and cultural heritage, as discussed in TWG 7, nuclear data are also indispensable for basic nuclear physics research. As scientists plan future experimental activities that may lead to discoveries, they seek to improve their knowledge by interrogating the databases. Theoretical developments also rely heavily on the availability of reliable and up-to-date nuclear data.
The research data life cycle can only be complete if the measured data are curated and incorporated into the nuclear data databases and libraries. Only then are the data readily available for use by the broader user community, be it in basic sciences or applications. The main advantages of nuclear data are that they are evaluated, verified and validated before being disseminated as data files, data libraries or databases. As shown schematically in Fig. 11.4, the research data cycle is an iterative cycle with each phase continuously informing the others in a seamless progression.

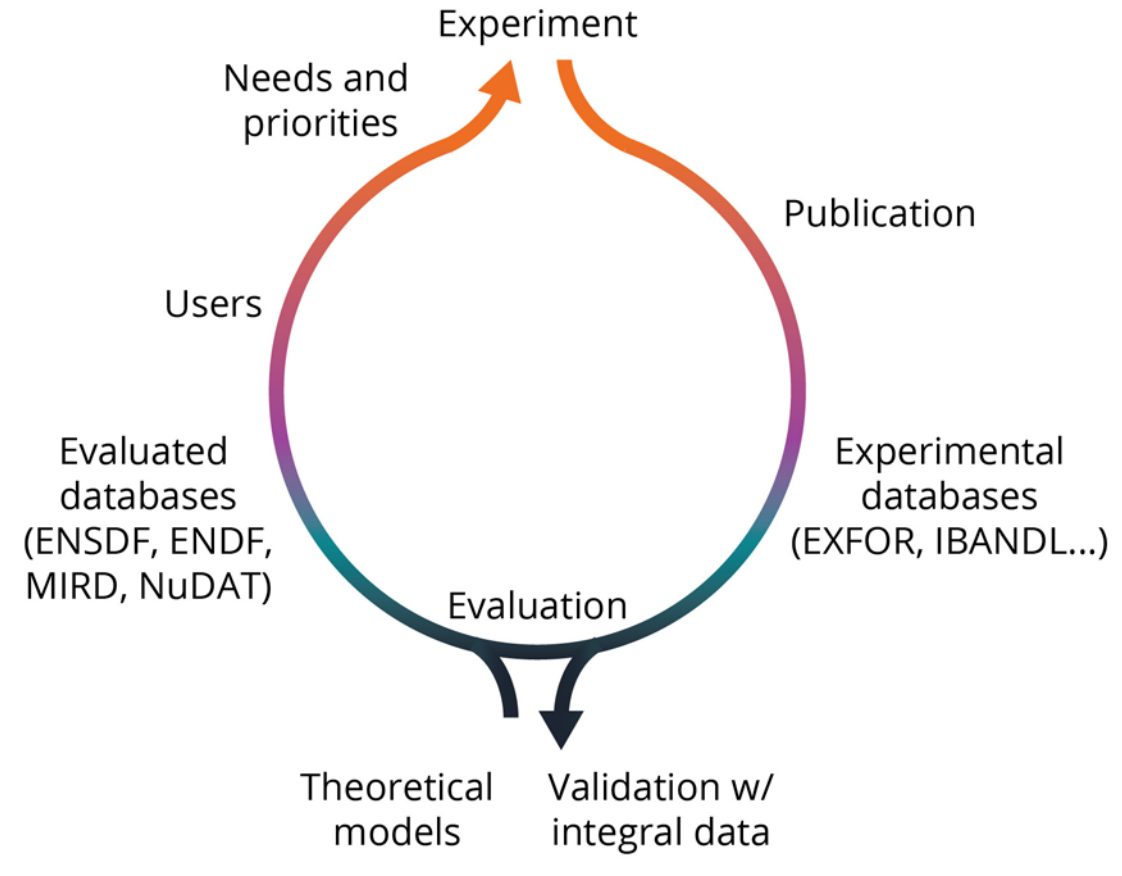

*Fig. 11.4: Schematic representation of the research data cycle including the evaluation process and associated databases.*

The evaluation procedure involves assessment of experimental methods and uncertainty budgets provided by experimentalists, treating discrepant data or outliers and recommending best values for a nuclear parameter using a statistical method or a fitted model with associated uncertainties. The evaluated data are prepared in formatted data files with well-defined formats forming part of a well-organised and documented database, or data library. These libraries are put to various tests to verify the accuracy of the formatted data files, but also to validate the evaluated data against the benchmark of integral experiments before finally being released online. Nuclear data or curated data are essentially the harbingers of FAIR research data as they meet all these criteria through their very definition and construction.

The collection/compilation, evaluation, verification and validation of nuclear data are arduous tasks that rely heavily on contributions from experts in both basic and applied research communities. International organisations, such as the International Atomic Energy Agency (IAEA) and the Nuclear Energy Agency (OECD/NEA), facilitate the curation of nuclear data by bringing together knowledge, expertise and infrastructures from all over the world, thus enhancing progress while reducing the financial burden on national authorities and funding agencies. The major general-purpose nuclear databases that form the foundation for all other derivative databases, interfaces and simulation codes are maintained by international networks and collaborations as follows:

● Exchange Format (EXFOR): compilation of all published measured reaction cross sections, angular distributions, excitation functions, thick-target yields and fission-related data (cross sections, fission yields, etc.) hosted at the IAEA (https://www-nds.iaea.org/exfor). EXFOR is maintained by the international network of Nuclear Reaction Data Centres (NRDC) which comprises 13 Data Centres from 8 countries and 2 international organisations (IAEA, NEA Data Bank), under the auspices of the IAEA.

● Evaluated Nuclear Structure Data File (ENSDF): a unique compilation and evaluation of measured nuclear structure and decay properties across the nuclear chart hosted by the National Nuclear Data Centre, BNL (https://www.nndc.bnl.gov/ensdf). ENSDF is maintained by the International Network of Nuclear Structure and Decay Data Evaluators (NSDD) under the auspices of the IAEA. The NSDD network comprises 17 Data Centres from the USA, Europe, Russia, India, China, Japan and Australia, including one international organisation (IAEA). Derivative products of ENSDF include NuDAT, MIRD, and LiveChart.
● Evaluated Nuclear Data File (ENDF): evaluated nuclear reaction data originally developed for neutrons and later extended to charged-particle/photon transport codes. Different evaluated files are produced and maintained at a national or broader collaborative level (CENDL, China; ENDF/B, USA; JEFF, Europe; JENDL, Japan; INDEN, IAEA; ROSFOND, Russia). The Joint Evaluated Fission and Fusion File (JEFF) (https://www.oecd-nea.org/jcms/pl_20182/jeff) is produced by the collaborative effort of about 100 scientists from 21 countries including 3 international organisations (EU, IAEA, NEA Data Bank) and is coordinated by OECD/NEA Data Bank.

Other horizontal evaluation efforts with predominantly European involvement include the Decay Data Evaluation Project -coordinated by LNHB-CEA- which is responsible for providing recommended decay data for about 230 radionuclides for metrology applications (https://www.lnhb.fr/ddep_wg/).

While there has been significant investment in the development of small- and large-scale experimental facilities and measurements by European member states and EU projects, nuclear data curation activities in Europe have been largely underfunded in the past decades. The main source of funding for nuclear data has been the EURATOM projects such as CHANDA, SANDA, ANDES, ARIEL and ERINDA. These initiatives not only offer financial support but also enable European nuclear data groups to access facilities, share infrastructures and expertise, and play a pivotal role in educating and training the next generation of nuclear data experts. However, this form of funding falls short of what is required for a sustainable European nuclear data curation effort commensurate with the region's rate of production of experimental data.

The absence of a coordinated nuclear data effort in Europe, in conjunction with limited funding, has resulted in a shortage of expert data evaluators with serious repercussions for the international databases. Over the past two decades, Europe's contribution to ENSDF has dwin-dled to less than 20% of the overall. This figure is notably low considering the substantial volume of experimental data generated by the European nuclear physics facilities. Combined with a global decline in the ENSDF effort worldwide, the cumulative ENSDF evaluation effort is no longer adequate to keep the database up to date within a 10-year cycle. To maintain a regular 10-year update cycle, a minimum of 12 full-time evaluators is required, whereas the current global effort amounts to just 5. What this situation implies is that the new, accurate, and precise experimental data generated by state-of-the-art European nuclear physics facilities will not be promptly incorporated into the databases, thus delaying their utilisation in various applications.

Another consequence of the underfunding of nuclear data evaluation in Europe is the loss of expertise in the nuclear data sector. The absence of career paths and promising prospects in the field of nuclear data is discouraging young nuclear physicists and engineers from entering the field, while simultaneously forcing early-career nuclear data scientists to seek opportunities elsewhere. When coupled with the retirement of senior nuclear data experts who may not be immediately replaced, it becomes clear that Europe is at risk of losing its expertise in nuclear data evaluation and hence its capacity to maintain its databases and data libraries up-to-date.
Following a dedicated Consultants' Meeting organised by IAEA and NuPECC in 2023, an IAEA report on the needs for a 'Comprehensive European Plan to acquire and curate nuclear data' (IAEA Technical Report INDC(NDS)-0875) concluded that: **The main challenge facing the European nuclear research and applications community is establishing a sustainable source of funding of measurements and nuclear data evaluation, and ensuring well-defined career paths in nuclear data to maintain and enhance the available expertise.**

Concomitantly, the principles of **OPEN** data, as well as **FAIR** data, are paving the way for new opportunities in nuclear data curation. Both **OPEN** and **FAIR** data promote openness and accessibility in experimental data, while **FAIR** data also focuses on making experimental data more discoverable, interoperable and reusable. Apart from enhancing the development of data-driven technologies that have already been discussed in previous sections of chapter 11, these principles also encourage experimental groups to make available, in an accessible, interoperable and reusable way, a much larger amount of measured data, including raw data, analysed but unpublished data, as well as analysis software and metadata describing the measurements, detector calibration, target preparation and details of the data analysis. This tremendous wealth of experimental information will lead to a new approach to data curation based on new modern tools for retrieving experimental data not limited to the traditional extraction from tables and figures in published articles. The availability of these experimental details in a direct, 'findable', and easily retrievable form would facilitate the evaluation process, and feedback from the evaluation and validation process could be used to reanalyse the raw data without necessitating a new measurement. Creating adequate and open repositories to store this excessive volume of data, along with the necessary metadata for searching and finding, is a necessary condition for this data to be findable and usable in data curation. Another benefit to nuclear data curation from findable, interoperable, and reusable experimental data is that the entire analysis procedure could be verified by independent experimentalists, thus potentially uncovering bugs or errors before publication or evaluation. The push to develop new data-driven technologies will also affect the data curation technologies, introducing automation where possible and Artificial Intelligence/Machine Learning (AI/ML) techniques to compensate for scarce or missing experimental data.

To be able to absorb all these developments into nuclear data curation effectively and seamlessly, and to meet the demands of both basic and applied research, current nuclear data databases need to be modernised and nuclear data experts need to collaborate with data scientists, database programmers and experts in AI/ML. Training the next generation of nuclear data experts to develop and implement these technologies and approaches is of paramount importance for the future of nuclear data curation.

## Recommendations:

To rise to the challenges in nuclear data curation and seize the emerging opportunities, it is recommended that:

- European nuclear physics research and application communities combine forces to establish a comprehensive European nuclear data programme with well-defined priorities defined by stakeholders along with sustainable funding to secure nuclear data career paths.

- Dedicated efforts to train the next generation of nuclear data experts in data evaluation and the use of AI/ML methods and modern data-driven technologies are supported.

- Cooperation between nuclear data curators, data scientists, database programmers and AI/ML experts and international organisations (IAEA, OECD/NEA) is strengthened.

**References**

1. *Horizon Europe funding programme Open science policy*

2. *Towards a reform of the research assessment system*

3. *Wilkinson et al. The FAIR Guiding Principles for scientific data management and stewardship. Sci Data 3, 160018 (2016). https://doi.org/10.1038/sdata.2016.18*

4. *Open science-related policies in Europe https://doi.org/10.1093/scipol/scac082*

5. *Second French plan for Open Science (2021-2024)*

6. *The Italian "Piano nazionale scienza aperta (2022)"*

7. *National Strategy for Open Science (Estrategia Nacional de Ciencia Abierta, 2023-2027)*

8. *Open Science as Part of Research Culture. Positioning of the German Research Foundation*

9. *Policies of open science and research in Finland*

# Nuclear Science – People and Society

## Outreach, Education, Training, Diversity, and Careers

**Conveners:**

**María J.G. Borge** (IEM-CSIC, Madrid, Spain)
**Christian Diget** (University of York, UK)

**NuPECC Liaisons:**

**Rolf-Dietmar Herzberg** (University of Liverpool, UK)
**Yvonne Leifels** (GSI, Darmstadt, Germany)

**WG Members:**

- **Axel Boeltzig** (Helmholtz-Zentrum Dresden-Rossendorf, Germany)
- **Michele Coeck** (SKCEN, Mol, Belgium)
- **Alessandra Fantoni** (INFN Laboratori Nazionali di Frascati, Italy)
- **Fanny Farget** (GANIL, Caen, France)
- **Kathrin Göbel** (GSI, Darmstadt, Germany)
- **Christian Klein-Boesing** (University of Münster, Germany)
- **Agnieszka Korgul** (University of Warsaw, Poland)
- **Arnau Rios** (University of Barcelona, Spain)
- **Magnus Schlösser** (KIT, Karlsruhe, Germany)
- **Annika Thiel** (University of Bonn, Germany)
- **Livius Trache** (IFIN-HH, Bucharest-Magurele, Romania)

# Introduction

Fundamental nuclear science and curiosity-driven research is a rich area of knowledge and development with a broad range of applications and impact on our society. Nuclear experiments and theory have substantial societal impacts in areas of the environment and energy, health, heritage and way beyond, to the exploration of the origins of our universe. NuPECC has explored this throughout the present Long Range Plan, as well as in the recent report on Nuclear Physics in Everyday Life (2022). Our community also has a key economic impact through the training of a highly specialised workforce in nuclear science and technology.

To further develop this pool of knowledge for future generations, however, we must not only explore these areas of knowledge, understanding and development, but communicate them to – and develop them jointly with – the next generations, through outreach, education and training. We must invite both the present and the next generation into the wealth of career options and career development opportunities in the nuclear sciences and across our society, supporting their integration into a diverse and inclusive work environment.

Following the NuPECC charter agreed together with APPEC and ECFA on diversity we must dedicate space, time and effort to consider the diversity, including gender balance, in the nuclear physics community. Diversity should be understood as the acknowledgement, respect and appreciation of the reality that people differ in many ways, visible or invisible. The diversity and gender balance in the nuclear physics community should be a mirror of the European society we live in. Its added value has been demonstrated in industry as well as in research, where most effects of gender and ethnic inclusions have been studied so far.

The present section of the Long-Range Plan therefore outlines a range of current work across outreach, education, training, diversity and careers in European nuclear science. It furthermore details some of the synergies between existing approaches across Europe and explores how the five areas can support each other. Starting with outreach to the general public and the broader society, we move on to considerations of how nuclear science is, can be, and should be integrated into education throughout the school system up to and including undergraduate degrees. Following this, we discuss how specialised training during and following postgraduate studies can utilise synergies across Europe, and we consider the implications for diversity and inclusion, both to enhance them in our community and to see how these are monitored across the international community. We conclude with considerations on practical initiatives to enhance and support career progression in our field. The chapter will, for each of the five sections, include recommendations on how the European nuclear physics community can partner with educational and research institutions as well as with funders of fundamental and applied nuclear science; to reach, inspire and offer opportunities to – particularly – the next generations of scientists, enhancing the impact of Nuclear Physics and the broader nuclear sciences on People and Society.

# Outreach and Education

## Introduction

Scientists acquire knowledge by gathering information and evidence through experiments and direct and indirect observations. These observations are tested through repeatability, using different research methods like descriptive, correlational and experimental setups. This is known as the nature of science and forms the basis of scientific literacy and knowledge.

In the face of numerous challenges such as climate change, pandemics, food security, global competition and inequality, it is critical that we as scientists communicate the nature of science and involve society in the process of acquiring knowledge. This helps people understand scientific concepts, critically evaluate news and information, and make informed decisions regarding topics based on scientific evidence.

We must share the experience of how scientists derive models, laws, mechanisms and theories to describe and explain phenomena. In this process, we trust scientific knowledge while being open to modifying or discarding it when new evidence emerges. The scientific community works together in observation, rational argument, inference, scepticism, peer review and reproducibility of the work. Only when this is clearly communicated and shared with the public and across society can we expect science to be trusted across the societies we are part of.

## State-of-the-art of outreach across the nuclear sciences

Outreach is the interaction of research with society as scientists share their methods, results and open questions in a dialogue with society. Outreach must become an integral building block of both research and culture to fully serve its societal mission. Activities show the benefits of research to the public, spark interest and enthusiasm among young and less young people, connect science and society and allow participation in the process of knowledge gain. The activities require experts from research, education and communication to work together, as well as requiring training in communication and educational theory, embedded in the nuclear and particle physics community. Together, these experts can develop outreach further, evaluate the activities to advance the projects, and build a sustainable network for the future. This requires funding agencies to explicitly sponsor activities and support permanent structures that will allow national and international coordination of outreach projects.

The European nuclear physics community is actively engaged in a wide variety of local outreach and science engagement activities. Many of these activities are highlighted in NuPECC's *Public Awareness of Nuclear Science* (PANS, https://nupecc.org/pans/) programme. Many more resources and tools exist and are continuously being developed for the purpose of engaging the public. Research centres and universities offer a wide range of activities, ranging from open days or lab tours and public lectures to hands-on experiments and science festivals. Beyond that, scientists engage in producing high-quality content on their own or join forces with 'professional content producers' which offer a huge reach to a diverse audience.

Next to these generally non-concerted local and online efforts, several projects with national or international reach have been developed and successfully established, such as the examples we describe below, the collaborations IPPOG, Netzwerk Teilchenwelt or the Binding-Blocks programme shown in the highlight Box 12.1.

IPPOG ("International Particle Physics Outreach Group") is a network of scientists, science educators and communication specialists working across the globe in informal science education and public engagement for fundamental research in particle physics. It is a platform of resources, ideas, inspiration, training and skill-building, providing access to programmes for students, teachers and the public. The international masterclasses, which target high-school students, are IPPOG's flagship programme for tens of thousands of students from around the world, also covering nuclear physics and its applications.

Netzwerk Teilchenwelt is a network that brings together 34 universities and research laboratories in Germany in the field of nuclear, particle and astroparticle physics. The activities are funded to a large extent by the German Ministry of Education and Research (BMBF). Roughly 134 groups are involved in this Germany-wide programme. The activities bring cutting-edge science into the classroom. Each year, about 3,500 high school students take the opportunity to work in Masterclasses with original data from experiments in particle and nuclear physics and nuclear astrophysics or to study cosmic radiation with detectors. On advanced levels, motivated students attend workshops at CERN or the MAinz MIcrotron MAMI (University of Mainz) and conduct their own research projects under the supervision of scientists. Through a Fellowship programme, highly motivated students are offered the opportunity to get in contact with research groups and obtain personal support at an early stage. In 2019, two new hubs at Mainz and Münster Universities were established, focusing especially on integrating nuclear and hadron physics topics, for instance via close collaboration with the experimental programmes at GSI/FAIR.

### Box 12.1: *Binding Blocks*

**Binding Blocks is an initiative started at the University of York and now spread across the UK.**

**Binding Blocks centres around an interactive construction of a seven-metre-long nuclear chart of isotopes built of LEGO(R) bricks, as well as more recent curriculum-linked Online Masterclasses in Nuclear Science. Through engaging with the nuclear chart and online materials, participants get a hands-on experience of key areas of nuclear science, astrophysics and energy. Since 2015, over 30,000 young people and members of the public have taken part in Binding Blocks events and over 400 undergraduates, postgraduates and researchers have taken part in Binding Blocks training. Some of these activities have now been developed for Catalan and Spanish audiences too, and can be easily explored elsewhere in Europe.**

***http://bindingblocks.org.uk/***

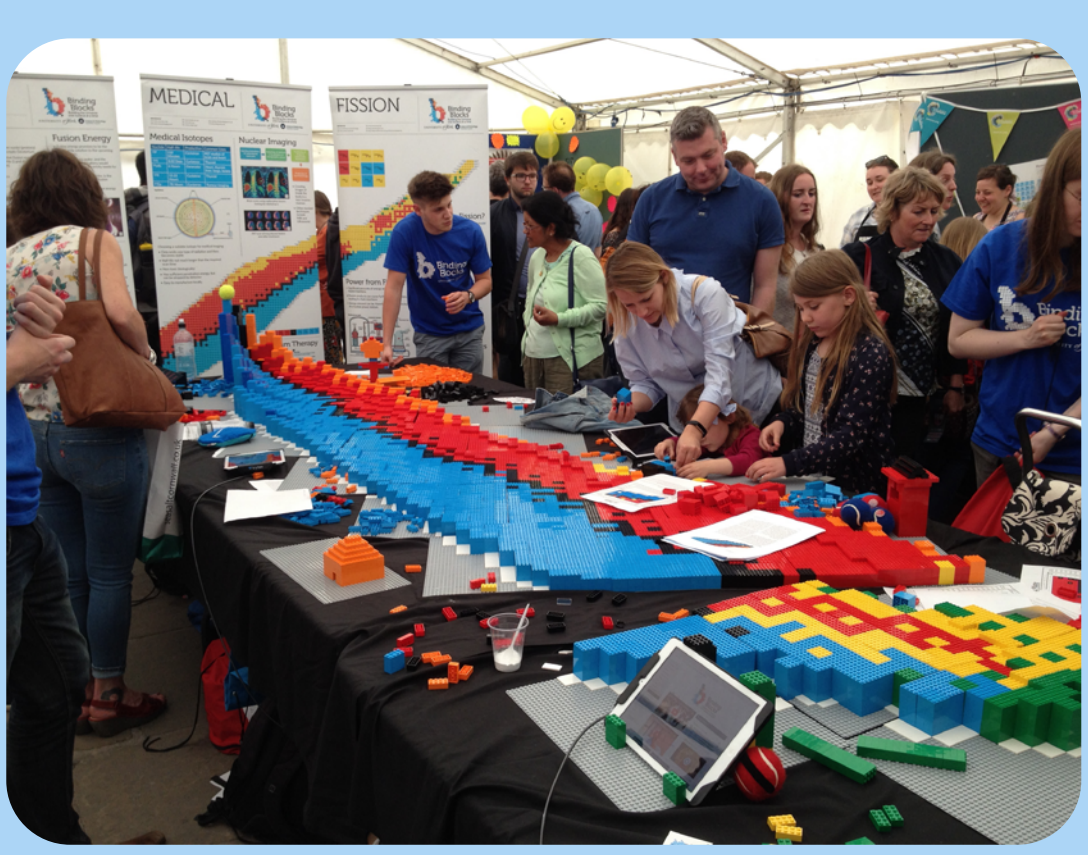

There are many benefits from these large-scale collaborative efforts, such as working out a holistic concept tailored to the target audience by making use of the wide experience of the scientific community involved, with several paid positions fully dedicated to supporting and shaping these efforts. Many more of these networks and activities exist in different countries around the world, connecting outreach across fields including experiments in nuclear, particle and astrophysics.

## Increasing importance and visibility

Outreach provides an opportunity for nuclear scientists to engage with society. Scientists can communicate directly on the scientific method, its inner workings and the relevance of research in the developments and applications in day-to-day use. This becomes increasingly important, if not critical, in the current context where misinformation on key issues like climate change can so easily be spread. Scientists involved in public engagement can help shape a positive image of science to society, but they also often gain a more profound understanding of their work through their interaction with the public.

In addition, outreach activities encourage the development of scientists' soft skills, particularly communication skills with non-scientific audiences. This often results in achieving an advanced insight into societal issues. Scientists improve their presentation and teaching skills through outreach too, which is also helpful in an academic setting (for example, when successfully defending new projects or requesting funding). More importantly, outreach provides an opportunity to promote research and increases its visibility.

The increasing importance of outreach activities and programmes is also visible at conferences. Several international and national conferences have added outreach sessions to their programmes to serve the need for scientists to exchange successful methods and tools. As an example, the National Institute for Science Communication (NaWiK) in Germany has recently launched a network and platform for scientists active in communication, "WissKon", with an annual conference.

## Target groups for outreach activities

The worldwide research field of nuclear and particle physics, through its large international collaborations, connects people from all over the world independently of their age, educational or personal background. However, outreach activities address certain target groups depending on the scope of the activity. To reach underrepresented groups, the activities must focus on their needs and wishes.

### General public

What is the universe made of and how did it evolve to create galaxies, stars, planets and life? Nuclear and particle physics can give the answers to these fundamental questions of humankind. Beyond that, many everyday technologies in energetics, medicine, electronics or information have their origin in fundamental research with large infrastructures from which society profits immensely. Through dialogue and with the participation of society, nuclear physicists communicate the tremendous progress of our understanding of nature and the deep implications nuclear physics has on our view of our world as well as on our culture and philosophy. By obtaining scientific literacy, for example via citizen science initiatives (see Ch 9), people can participate in public and political discussions and contribute to making sound decisions. Reaching the general public is also an effective way to multiply our interests, since people transport their experiences into their peer group, to family and friends.

### Students (secondary schools)

It is important to inspire the next generation of students to pursue a career in science, technology, engineering and mathematics (STEM) research, with nuclear and particle physics as a big attractor towards choosing physics as a major in universities. We recognise a deficit in the number of young women choosing STEM subjects at university. The study: "Why Europe's girls aren't studying STEM" [https://www.stemcoalition.eu/publications/why-europes-girls-arent-studying-stem] shows that young women become attracted to STEM fields between ages 11 and 12. Their interest drops off between ages 15 and 16 and is hardly ever recovered. The scientific community, together with teachers and parents, must encourage girls to consider five factors that drive their interest at this age: female role models, practical experience and hands-on exercises, teacher mentors, real-life applications and confidence in equality. High-school students who are already dedicated to STEM topics may find additional motivation by participating in outreach activities and decide to target their studies to work in nuclear physics later.

### Teachers

Teachers convey the knowledge and methods of science to their students; they are interfaces and multipliers. It is necessary to allow teachers to access state-of-the-art research methods and results of nuclear and particle physics. Special training enables teachers to deal with these topics in their courses by taking opportunities to link to the curriculum and didactic challenges. Dedicated and curious teachers should be encouraged to take part in workshops at research centres and universities to deepen their knowledge, obtain hands-on experience and be inspired for new concepts for classroom activities. The example of a teacher's programme at CERN [https://teacher-programmes.web.cern.ch/] should be repeated at other facilities. Closer integration of research hubs with local teaching communities is highly recommended, as has been done, for instance, at INFN-LNF, Italy (https://edu.lnf.infn.it/programma-docenti/docenti-scuole-secondarie-di-2-grado/).

### University Students

During their Bachelor's and Master's studies, STEM students have to decide in which field to deepen their knowledge and in which research group to carry out their final research and dissertation projects. It is necessary to spark interest in nuclear physics and provide connections and insight into the research groups in our field.

### Policymakers

Nuclear physics is generally an expensive endeavour, which requires a substantial quantity of resources. It is therefore of paramount importance to communicate the importance of science and technology to policymakers. Nuclear physics is a key driver of innovation for modern technologies and for the education of the next generations of scientists and engineers, who need the necessary skills and mindsets to deal with the challenges of the 21st century.

### Industry

Links with industry are an important asset for promoting an image of dynamic careers and diversity in jobs and applications in the field. Communication on the diverse technologies emanating from nuclear science and used in different fields of industries should be enhanced to promote careers beyond academic ones.

## The long-range plan for outreach

### Local – national – international

There is a need to strengthen activities at research centres and universities and increase their visibility and reach. The experiences in Germany, especially the success of the Netzwerk Teilchenwelt initiative for a coordinated effort of outreach in the field of nuclear and particle physics, have shown that this is of utmost importance. Similar success has been obtained in the UK's Binding Block programme mentioned above. These initiatives connect various local activities in a synergetic way and permit the establishment of flagship projects on a national level. It is important to establish similar coordinated structures in other European countries and to combine them on the European level in the future. This requires the awareness of both national funding agencies as well as the European Research Council (ERC) to provide funding opportunities tailored to dissemination.

### Digital outreach

The digital world enables us to move beyond traditional outreach concepts and use the possibilities and methods arising from digitisation. Activities can be transferred into the digital space, such as online tours and courses, digital learning media or virtual realities. Digital activities reach students and people interested in science all over the world, which is indispensable for international research efforts. Access barriers are often lower in digital activities, and this enables participation independently of the socio-economic background and enhances accessibility to a diverse audience (including visually impaired, hearing-impaired and neurodiverse participants).

Digital events such as hackathons connect people worldwide to work on challenges and develop new ideas. Open access resources (see Ch.9) such as Online Masterclasses provide a long-term and sustainable outreach. They are even more impactful when they are designed to support face-to-face immersion in the material in schools and provide direct engagement with researchers. Open data and software mean that the international community can use our tools and even participate in the scientific process.

### Interdisciplinary projects

Nuclear physics research includes strong interdisciplinary links and is, by its nature, cooperative. Modern approaches foster collaborations and exchange across a wide range of disciplines, not limited to the fields of STEM research. The interdisciplinary nature enables one to approach the research fields from different points of view and thus sparks interest in groups who are not primarily involved in physics.

The collaboration between science and art may, for example, open new fields of communication and enable us to reach out to hard-to-reach target groups. Artistic research has established itself as a relevant method. Complex topics and issues are reflected upon, not only in terms of content but also aesthetically, through artistic research practice. Particularly in interdisciplinary collaborations between science and art, the artistic view of scientific processes enriches the discussion between scientists and the public, generates new points of view and gives new impulse on how to present research results.

Outreach activities require collaboration among research, education and communication experts. Together they can develop outreach further, based on experience from education and communication research. They can evaluate activities to advance projects, perform quantitative and qualitative research on outreach and build a sustainable network with best practices for the future.

### Evaluation of the activities

It is necessary to collaborate with experts in education, communication and social studies to establish structures to evaluate outreach activities. Quantitative and qualitative research on outreach must be conducted to analyse a large and complex number of factors: the reach of the activities, the transfer of knowledge and scientific findings, the maintenance of trust and credibility, the acceptance of scientific innovations, changes in attitudes and behaviour, as well as reputation and image.

## Recommendations for outreach

We recommend that funding agencies, national and international bodies and the community of European nuclear physicists emphasise the critical societal investment that is outreach as a tool to inspire the public in nuclear science and its impacts, facilitated through:

- Establishing and equipping a European network for outreach, resourced by national and transnational funding schemes through research-linked and earmarked funding for outreach; and

- Strengthening the provision and support of digital outreach projects and their link to inspiring face-to-face and extracurricular activities.

# Education

## State-of-the-art education in subatomic physics

Nuclear physics offers a very good starting point for a general scientific education and should be embedded from early-age education through to undergraduate university degrees. It is a key scientific endeavour, with a substantial number of relevant day-to-day applications, underpinned by a vibrant research community. From the more fundamental aspects of the building blocks of matter and the understanding of astronomical observations to applications in medicine or carbon-free energy generation, nuclear physics is unique in having a broad interdisciplinary scope and a deep physics basis.

There is currently very little, if any, data on the current impact of the European nuclear physics community on primary, secondary school and University education. It is also difficult to estimate the current level of understanding and awareness of nuclear physics for the average European student. The development of educational resources in nuclear physics is currently not coordinated at the European level. There are no field-specific calls for science education in Horizon Europe. Yet, the expertise of the European nuclear physics community is key in the context of the current climate crisis and in times where Horizon Europe directives aim at a systematic integration of climate change and energy issues into education curricula. The recent NuPECC report "*Nuclear Physics in Everyday Life*" [https://www.nupecc.org/pub/np_life_web.pdf] should be widely distributed to teachers to help them trigger the interest of their students in nuclear issues and to spread state-of-the-art knowledge and applications of nuclear science to society.

## Fundamental education - "laying the fundamentals"

Understanding the principles and concepts of nuclear physics can be an enriching learning experience for students and teachers at all levels of education. At the primary school level, nuclear physics may serve as a foundation for scientific literacy and to encourage critical thinking skills. In particular, applications of nuclear physics like dating, imaging or medical physics can be used in primary education as a means to guide young students towards scientific endeavours. Here, the interplay with outreach is particularly important, and the European nuclear physics community has actively provided useful and inspirational resources for both learners and educators.

## Secondary education - "getting students excited"

Some secondary education curricula in Europe already incorporate topics related to nuclear physics, including elementary particles, radioactivity and the strong force. Nuclear physics can also provide an understanding of the inner workings of atoms, the structure of atomic nuclei, and the forces that govern their interactions. It forms the basis for and links to fields like particle physics, astrophysics, chemistry or medicine. The study of nuclear physics thus prepares learners for specialised scientific pursuits and can be used as an inspirational science engagement tool, beyond the more standard classical physics ideas widespread across European secondary school curricula.

Some examples of secondary education and teaching-orientated activities have been developed in recent years. TU Dresden is, for example, rolling out a masterclass in nuclear astrophysics for secondary school students, as part of the EU-funded ChETEC-INFRA project (http://mc.chetec-infra.eu). The Binding Blocks Nuclear Masterclasses are delivered in the UK, and funded by the UKRI Science and Technology Facilities Council (http://bindingblocks.org.uk/). The CERN-supported BL4S initiative (Beam Line for School, https://beamlineforschools.cern/) engages students through competitions. The INSPYRE school (Int. School on modern PhYsics and Research) is organised by INFN for high school students from all countries around the world (https://edu.lnf.infn.it/inspyre). Such initiatives have, however, not yet fully impacted the European community, and a much more widespread availability of teaching resources would undoubtedly help extending the impact of nuclear physics on the European public.

The following list contains successful examples completed in different NuPECC member states and recommendations for fostering education impact in secondary education:

- STEM lectures held by researchers from universities in the classroom at school. The aim of these lectures is usually to demonstrate how mathematics, physics, chemistry or biology are applied in reality, from technology to medicine and space exploration. It has been noticed that young lecturers (e.g. PhD students or PostDocs) can often bridge the gap to secondary school students more easily.

- Online and digital platforms provide easy and immersive access to contemporary fields in science and technology. Animations, real-time simulations, videos and online experiments make use of the modern means of communication employed by the young generation on an everyday basis. Platforms like Instagram, YouTube or TikTok offer low-threshold access for exposing students to fascinating questions and phenomena in STEM subjects. Care should be taken that the content has sufficient depth and is not limited to superficial explanations and just a "wow effect"!

- Scientific competitions, contests and Olympiads challenge students to excel in specific disciplines, think out of the box and engage in teamwork problem-solving. These skills are highly desirable for a successful career. Besides solving classical physics problems using mathematical methods, it is encouraged to envisage scientific projects with everyday relevance. Valuable rewards, such as prioritised admission to universities or additional points in university admissions can be given to the top performers.

- Extracurricular activities organised in schools can be very stimulating and gratifying for the involved students. Researchers should promote these efforts in school-university partnerships and can help in organising research projects that allow students to independently explore topics related to STEM subjects. Here, university attendance can facilitate access to equipment, knowledge and practical advice. Students will gain hands-on experience in the real research process to develop critical thinking skills and problem-solving abilities.

- Establishing an early mentor-mentee programme for students with STEM aspirations can provide meaningful guidance in higher education and help close the leaky pipeline. Mentors can serve as role models (especially for underrepresented groups) and provide advice in the development of scientific interests.

The role of teachers is capital in passing on the importance of scientific method and nuclear science at the primary and secondary education levels. The European nuclear physics community engages with teachers in a variety of ways, but mostly through Teach the Teachers events that can showcase state-of-the-art nuclear science. More importantly, teachers have access to activities, props and teaching materials generated by the nuclear physics community with current insights on the status of the field. All of this fact-checked, research-informed content is then passed on to a much wider audience through the teachers over a significant period of time, having a lasting impact on many more students than those reached directly in outreach programmes. While many of these teacher events take place across Europe, there is currently no coordinated European strategy to provide high-quality material or share good practice in such teaching events.

## University education - getting students immersed and engaged

Nuclear Physics is traditionally taught in Europe at University level. It is here that students typically access specialised knowledge and can be motivated to engage with state-of-the-art ideas in the field. However, because of the breadth and depth of the discipline, nuclear science subjects are typically relegated to advanced undergraduate or master's level courses. In some cases, this may lead to missing exposure opportunities for younger undergraduates, who may otherwise not know about the field and the associated opportunities it provides. If the structure of the undergraduate degree programme and national accreditation bodies allow for it, there may be opportunities to showcase nuclear physics ideas at the early stages of the higher education curriculum.

Offering nuclear physics lectures for undergraduate students early in their university education enables them to develop interest in related topics. Besides pure nuclear physics, lectures in supporting disciplines such as detector development, electronics, programming or data analysis, including modern machine learning concepts, should be encouraged. Opportunities for conducting research in the field of nuclear science and eased interactions with the broader community, including internationally, should be provided. Importantly, these transformative opportunities for university students are also extremely beneficial for those who eventually join the European industrial workforce, providing them with a first-hand opportunity to experience scientific research.

Several tools exist to facilitate the engagement of university-level students with nuclear science research. These include participation in experimental research conducted at international facilities or in level-specific summer schools, like the well-established International Summer Student Programme at FAIR/GSI (https://theory.gsi.de/stud-pro/). Nuclear physics groups at universities should create opportunities to maximise student engagement in the field and openly discuss and inform on prospects for future careers in the field. Grants are necessary to permit higher education institutions to secure financial resources in order to showcase their research at undergraduate level (for instance, through the organisation of summer schools).

## Postgraduate education - attracting students for PhD projects

A significant number of talented undergraduate or master's students do not consider an academic career through a PhD. A vibrant European research atmosphere in nuclear physics requires the engagement of key undergraduate talent within and outside the EU. The Erasmus Mundus Joint Master Degree on Nuclear Physics (https://www.emm-nucphys.eu/) is a successful example of a coordinated European effort in postgraduate education, attracting around 20 students every year to participate in a highly selective, specialised, cross-European programme. Many other national and European programmes provide an excellent postgraduate offer for students in nuclear science and technology.

A wide offer of research projects, internships or practical training courses that allow students to gain hands-on experience in nuclear science is crucial to the development of a specialised workforce. It is also important to inspire and maintain their passion for science by demonstrating how research has a real-world impact and contributes to progress in the field. Research-orientated postgraduate projects provide a solid foundation for future research careers within and outside of academia.

To encourage students to pursue a scientific career in doctoral studies, the following actions can be taken: inform students about available scholarships and grants for doctoral studies; encourage them to apply for financial support to help cover the costs of their studies and research; raise awareness among students on diverse career possibilities in the scientific field, both in academia and in the industrial sector or research institutions; and finally, help students to know of the professional prospects that await them on completing their doctoral studies, and the possibilities of training along their professional path. This is discussed in the following section. It is also important to monitor the development of Diplomas and PhD theses defended. The figure 12.1 shows their progression in Italy in the last decade.

## Recommendations for next generation nuclear science education

We recommend that national educational accreditation bodies, funding agencies, universities and educational institutions, in collaboration with the community of European nuclear physicists, work to embed nuclear science across all levels of education, highlighting its interdisciplinary nature and impact. This will require:

- The development and resourcing of a European network of science educators across nuclear sciences to showcase the possibilities of the field based on the latest teaching methodologies and guided by research; and

- The development of research-informed and curriculum-linked teaching and training resources for science teachers, funded by dedicated national and transnational funding resources.

# Training

Training new generations of scientists and technologists has always been and remains an important task for the European nuclear physics community across fundamental research and applications. It is approached through several methods and events, building upon teaching in our universities. The increased complexity, size and sophistication of our research infrastructures, experiments, theories and transnational collaborations, has called - and still calls - for increased attention on training. As part of the general task of education and training of the younger generation, the aspect of training has its own specificities. It involves the familiarisation of trainees with the basic methods, devices and instruments used in research, as well as with particular large facilities, setups, data acquisition systems and data analysis codes and protocols.

Although the focus here is on the training of the next generation of scientists, it should be understood that the majority of students will use their skills in a variety of related professions ranging from healthcare to security, cf. the NuPECC report on Nuclear Science in Everyday Life available here: https://www.nupecc.org/pub/np_life_web.pdf.

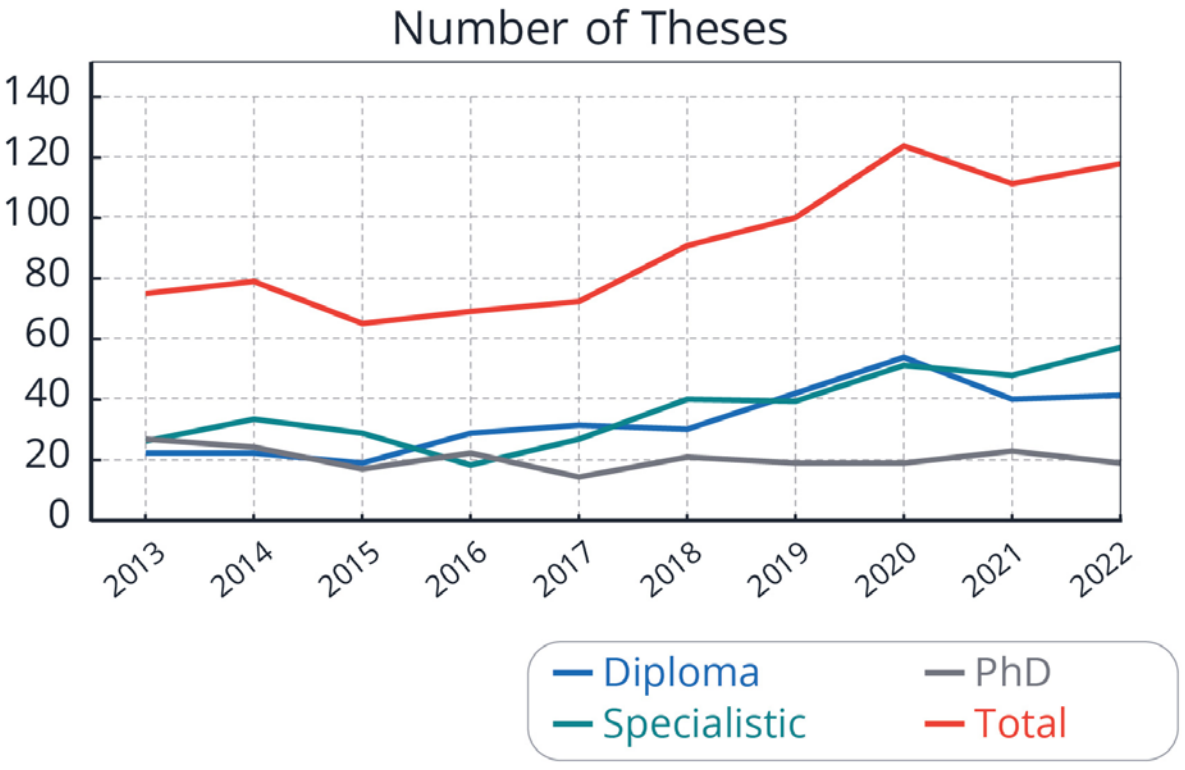

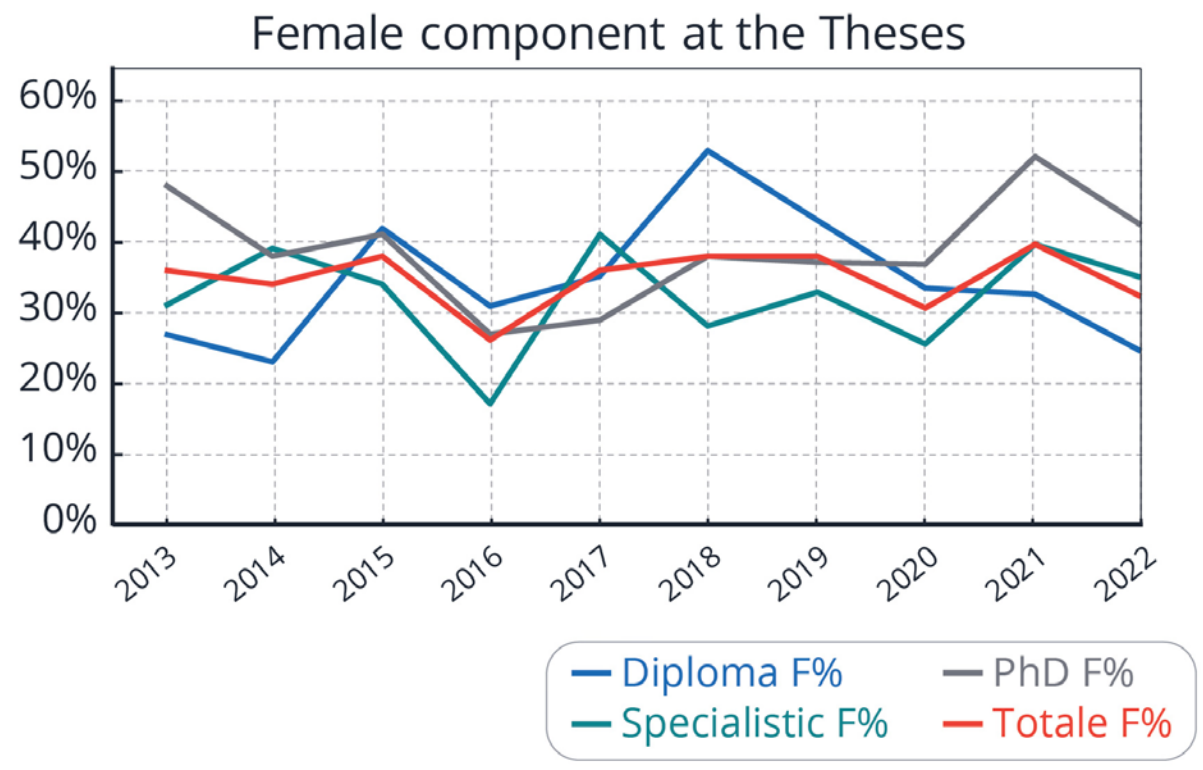

*Fig. 12.1: On the left INFN analysis during the last 10 years of Nuclear Physics Theses, from the Diploma to the PhD. On the right, only the female ratio is illustrated.*

## Box 12.2.: Outreach by the Nuclear Astrophysics community in time of pandemic

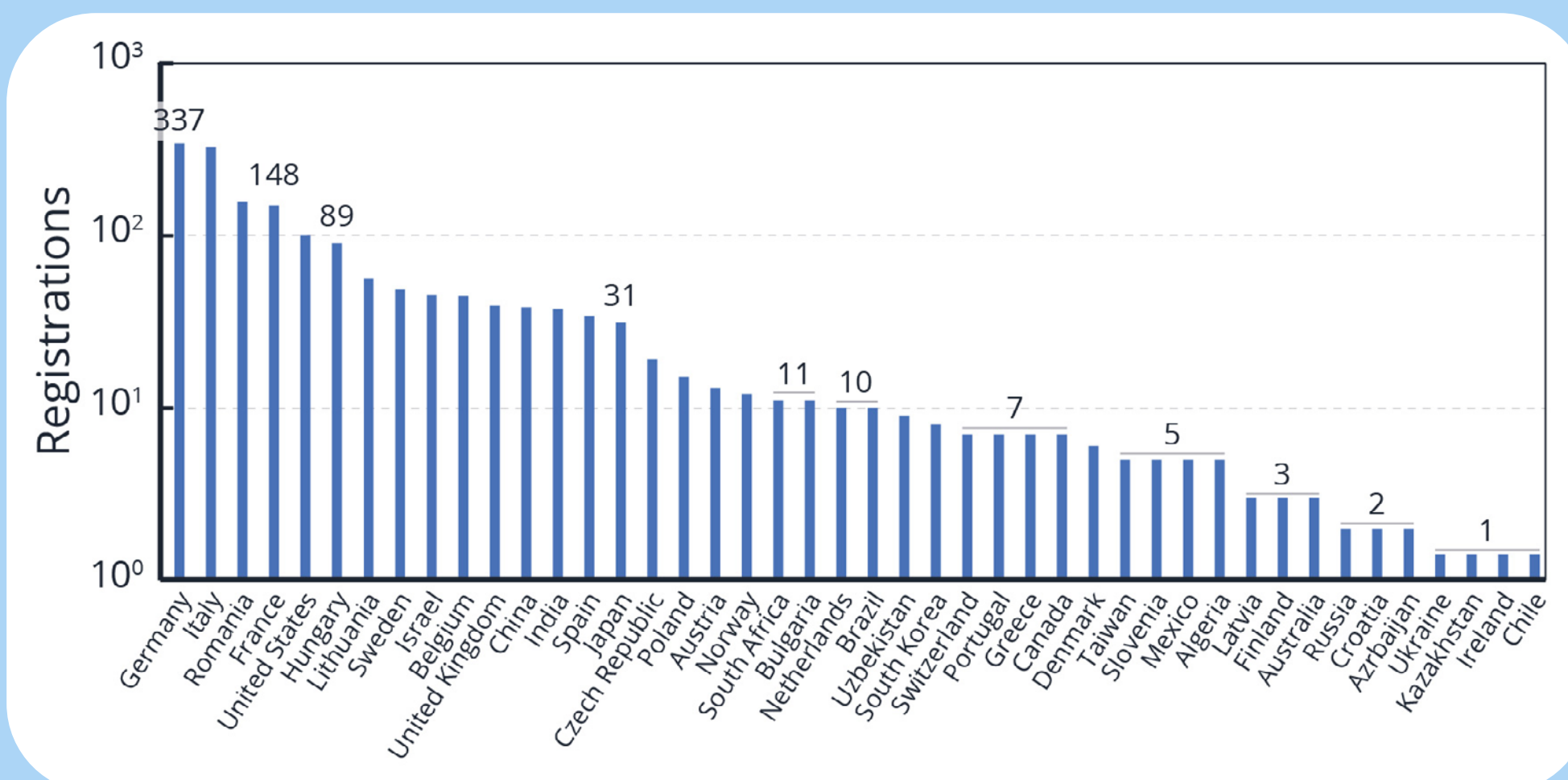

**In response to the Covid-19 pandemic, a monthly online format of Schools on Nuclear Astrophysics Questions (SNAQs) was created as part of the Networking Activities in the European Horizon project ChETEC-INFRA. Each edition focused on a given question in nuclear astrophysics, which was addressed over the course of one afternoon in a combination of lectures from senior scientists and contributed talks by early career researchers. The format received an overwhelming response from the community, with more than 1200 participants from over 40 countries all over the world (see the distribution of registrations per country) in the first 12 editions of SNAQs. The success of SNAQs illustrated the important role of online formats for schools as a flexible tool with low entry barriers for participation which can be established in parallel, or supplementally, to traditional in-person school formats.**

NuPECC recognises the increasing complexity of nuclear physics research and, therefore, the importance of training. The community should forge a **coherent, well-founded and reliable system of teaching and training via schools and events**, using the strengths and capabilities of the European nuclear physics community and its parts. With its declared purpose of maintaining the European nuclear physics community at the forefront of global science, NuPECC acknowledges the importance of successful strategies in developing its human capital and Research Infrastructures (RI).

Starting from the generalisation of the transnational use of facilities, from their inception to their actual use and data analysis, there is a need for a system that contributes to increasing the efficient use of our RIs and the attractiveness of the field to new generations, promoting the societal applications of nuclear physics and technologies.

This training system should contribute to:

● Strengthening the preparation of new generations of scientists through hands-on activities for (i) beginners and for (ii) more advanced students.

● Including the training of technical and engineering staff.

● Working effectively to improve the diversity of new generations, addressing gender, educational background and geographic distribution across Europe.

NuPECC proposes to use and strengthen these aspects, and recommends their continuation, with financial support from European and national programmes. This is particularly the case of the ECT* training infrastructure, complemented by European schools such as the Euroschool on Exotic Beams, the European Network of Nuclear Astrophysics Schools and the TALENT initiative for nuclear theory. Within the EURO-LABS project, training for future scientists and technologists is done (see Box 12.3).

NuPECC proposes to work together with the European Community (EC) and universities to include these training schools in their PhD training curricula. Having a well-founded and reliable system of training schools will contribute to this goal. Our academic research community should strengthen connections and joint training with people working in the fields of nuclear energy and nuclear medicine in order to streamline and increase our contribution to other fields, at the same

## Box 12.3: EURO-LABS Basic Training School BTS23

**The first basic training school BTS23 of the European HORIZON 2020 RI project EURO-LABS used six days of beamtime at the IFIN-HH's tandem accelerators in Bucharest-Magurele for hands-on experiments using state-of-the-art research equipment. The school was attended by 27 students from five continents and seen as a great success by all participants https://indico.nipne.ro/event/246/.**
**The team of early career researchers from IFIN-HH was led by Dr. Razvan Lica, Dr. Alexandra Spiridon, Dr. Dana State and Dr. Constantin Mihai.**

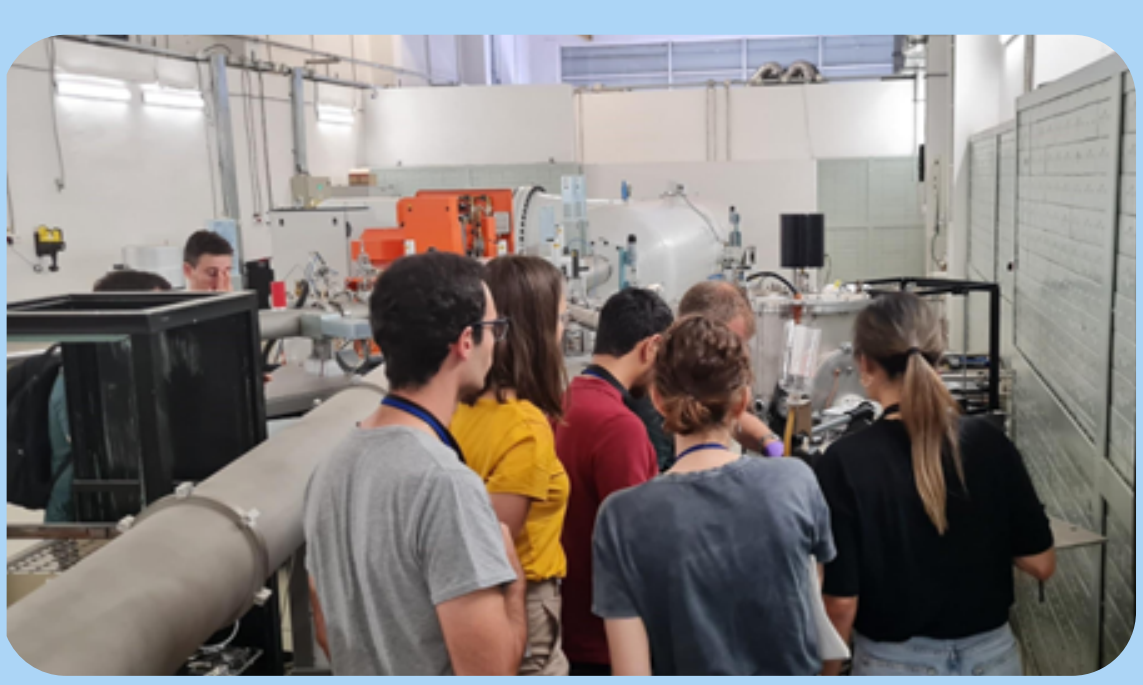

time benefitting from the cross-fertilisation of ideas. For example, we should especially emphasise the interdisciplinary and societal impact of training in biomedical nuclear science or nuclear chemistry, as well as training in machine learning and artificial intelligence.
A joint effort of EC and national funding agencies coordinated by NuPECC is specifically proposed to create a system of fellowships to allow PhD students and early postdocs to work for one year in 2, 3 or 4 of the large European facilities (several months in each institute).
Research and Innovation Action (RIA) -type programmes supported by the European Commission (such as EURO-LABS and ChETEC-INFRA) can be a good mechanism for supporting such training schools. These should be selected based on simple principles.

They should:

● Target the new generations of researchers and nuclear physicists, but also address the need to train the technical and engineering staff – specialisation is the keyword.
● Remain open to the need for cross-disciplinary fertilisation and a wider sharing of information, knowledge and technologies across scientific fields, i.e. address the need for cooperation between fundamental research and applications.

● Utilise the complementarity between larger and smaller nuclear science facilities across Europe to foster hands-on training.

● Use, strengthen and invest in existing traditions: topical and regional schools which are open internationally, avoiding duplication.

● Advertise inclusion in PhD programmes across Europe in cooperation with existing bodies, including the EPS and NuPECC.

## Multi-institutional enhancements of PhD programmes

A crucial institution for successful PhD education is the so-called graduate school (GS) or research training group (RTG). These generally form an umbrella for PhD candidates along a wide topic. Some members are directly paid by the GS/RTG and others are associated with it. All members benefit from generally multidisciplinary research seminars, hard- and soft-skill courses and ample networking opportunities. The inspiration, excitement and mutual support PhD candidates gain from their peers cannot be stressed enough.

1. Mentoring: Provides students with the opportunity to collaborate with mentors who are experienced researchers in their field of interest. Mentors can provide guidance, assist in the development of research projects and share their expertise.

2. Conferences and publications: Establish a culture in which it is the norm that students present their research findings at scientific conferences and publish articles in scientific journals. In this way, they naturally become part of the scientific community.

Today, third-party funding projects require an outreach concept that makes the research accessible to the public. This should be further integrated with high-level training programmes across Europe. It requires a high level of technical quality and clarity in the sense of a target- group-orientated approach and a high degree of motivation and/or involvement of the person concerned.

Coordinated structures in European countries are necessary to implement a long-range plan for training, supported by a committed long-term promotion. These joint efforts permit the development and installation of modern training projects and networks, increase visibility and impact. The interdisciplinary work with arts and social sciences, education and communication encourages new and different approaches to science.

Setting up these permanent and coordinated structures is a sustainable investment in the future of modern science and technology, in dialogue with society and with its participation.

## Capacity-building within and beyond academia

A truly multidisciplinary approach to nuclear science requires access to knowledge that lies beyond our field. Computing and data science, including artificial intelligence or machine learning techniques (see chapter on Nuclear Physics Tools), are becoming increasingly ubiquitous in science. The same is true for quantum computing, nuclear energy (cf. Nuclear Physics Tools) and many other disciplines. The European nuclear science community must have access to training in adjacent fields with the potential to boost our own research capacity.
While academic university settings provide extremely high-quality training opportunities, the whole of the nuclear physics community is not continuously exposed to such training experiences. In addition, when it comes to very specific research knowledge, it may not always be possible to acquire it locally or in an academic setting. In these cases, internships or apprenticeships in specialised industry settings may provide alternative skills useful for nuclear science. In general, a more fluid relationship between nuclear research actors and the nuclear industry would be extremely helpful for training purposes.

## Recommendations for Training

We recommend that the community of European nuclear physicists and technologists, in collaboration with funding bodies and other stakeholders, **resource and support the training of new generations** of nuclear scientists, providing the broad skills required across **experimental and theoretical nuclear physics** research, including **all disciplines and industries in our society** commonly relying on expertise, techniques and skills from the nuclear sciences. This includes the provision of **training for technical and engineering staff** as well as **interdisciplinary researchers**.

# Diversity and Inclusivity

The diversity of people doing research is recognised as an important factor in boosting productivity and innovation. The notion of diversity entails the acknowledgement, respect and appreciation of the fact that people differ in many ways, visible or invisible; mainly in age, gender and sexual orientation, national and ethnic origin, civil status and familial situation, religious convictions, political and philosophical opinions and physical abilities. A diversity charter has been drawn up by APPEC, ECFA and NuPECC (https://nupecc.org/jenaa/docs/Diversity_Charter_of_APPEC__ECFA__NuPECC-9.pdf) specific to our research field, and it should be endorsed at least in the main points by research organisations, scientific collaborations and conference committees. Certainly, some are already working towards achieving these goals.

Including diversity in the roadmap of nuclear physics is a way to position the nuclear physics community against prejudice and discrimination and contribute to the improvement of social and economic standards. Embracing diversity should be considered a necessary but not sufficient step to support and enlarge the community of nuclear scientists actively.

Achieving a diverse community reflective of the society underlying it can only be achieved by identifying and removing barriers to careers. In addition to diversity, progress requires a commitment to equity, justice and fairness. Thus, improving diversity means enhancing inclusion.

Diversity is also to be considered in terms of scientific background, culture and perspective. In this respect, inclusion is also a matter of scientific outreach; applications of basic science make the subject accessible to the broadest public. We should proactively enhance the curiosity of both public and researchers, which is central to the attractivity of the field.

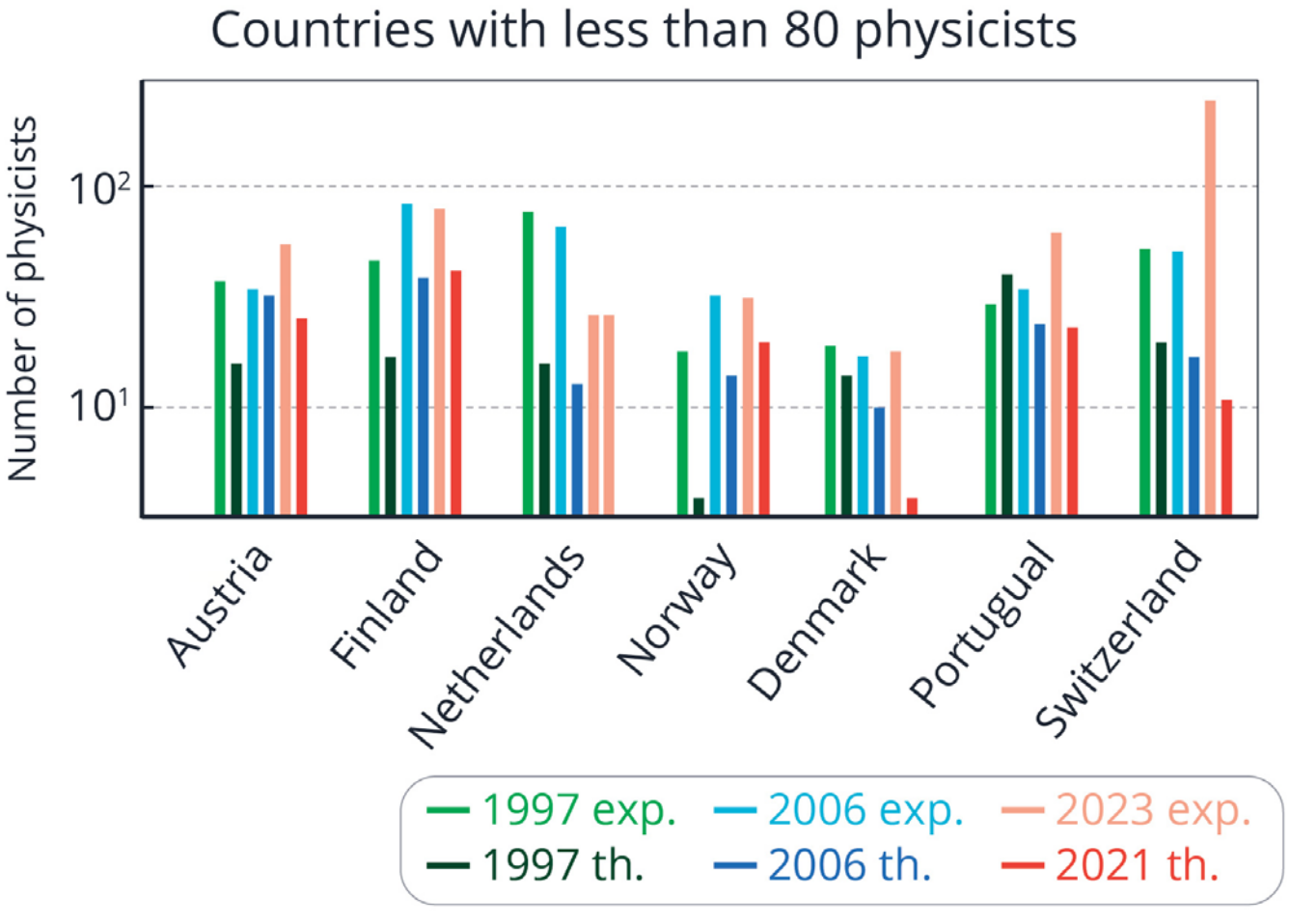

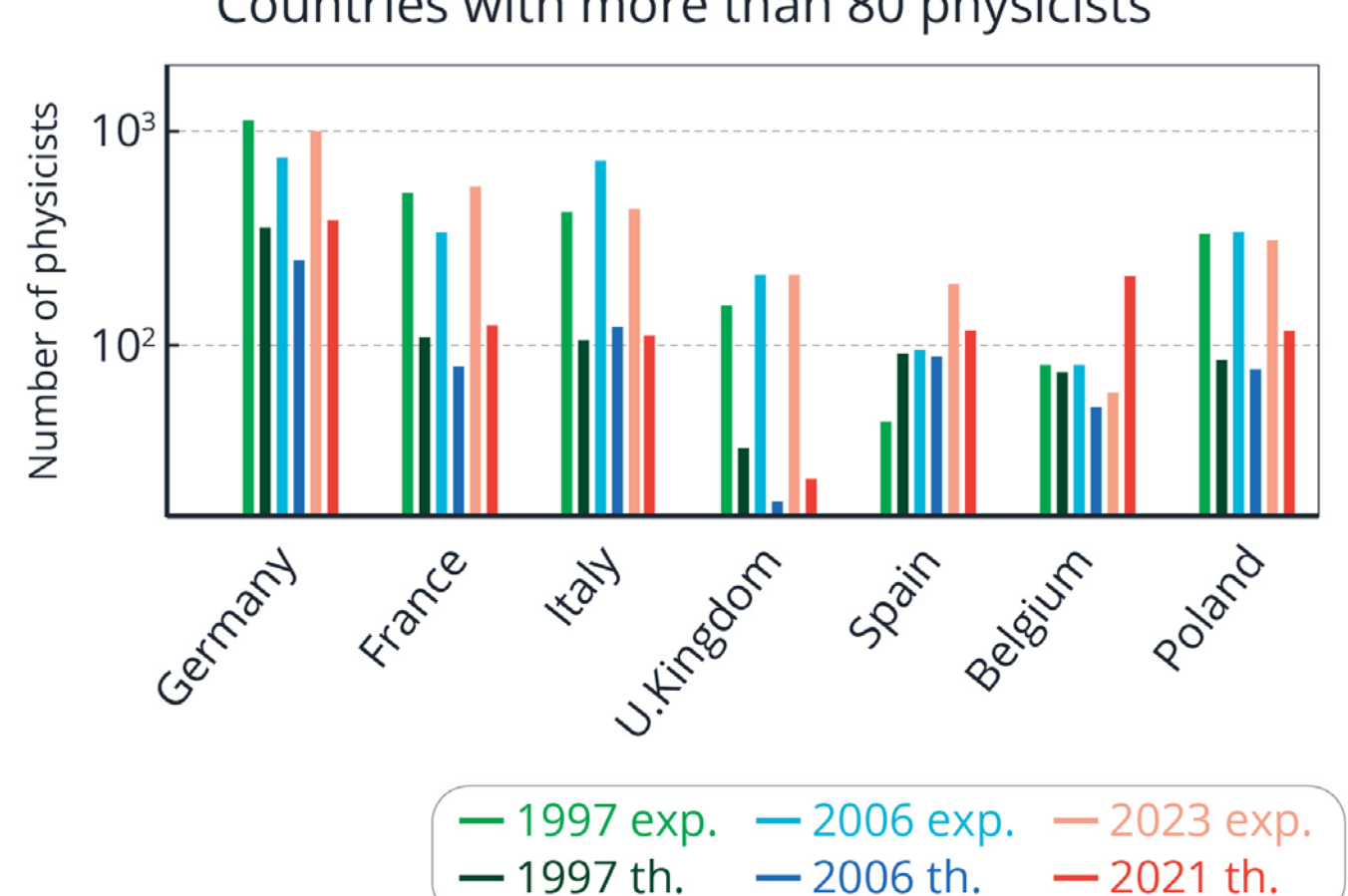

*Fig. 12.2: Evolution of experimental and theory physicists in Europe for countries with less (left) and more (right) than 80 physicists.*

Scientific, outreach and conference materials should stress the importance of our work in society and everyday life (following, for instance, the report Nuclear Physics in Everyday Life https://www.nupecc.org/pub/np_life_print.pdf. More importantly, all the material generated by the community should reflect its diversity. In particular, female and diverse role models should be highlighted whenever possible.

In addition, diversity also includes showing consideration for the propositions of younger physicists, as well as the exchange between the diversity of work expertise encountered by engineers, technicians and researchers. A vast range of competencies is required for nuclear physics besides research, such as those in vacuum, ion beam production, acceleration, electronics, computing and instrumentation, etc.

## Monitoring

Many initiatives to monitor diversity in science and in nuclear science do exist. Monitoring is the basis for understanding where we stand, and a means to increasing diversity and improving its inclusion. It is important to maintain the exercise of monitoring in order to define objectives on diversity and associated actions.

As already mentioned, diversity includes many parameters. Among them, the promotion of gender equality is the one most studied in Europe. In the following link [https://www.provide-education.co.uk/10-ways-to-promote-gender-equality-in-the-classroom/] ten tips to facilitate awareness and learning regarding equality with different gender identities can be found. This is important for pupils throughout various stages of their student lives. Gender equality is a priority of UNESCO for the period 2019-2025. Its strategy [https://www.unesco.org/en/gender-equality/education] focuses on a system-wide transformation to benefit all learners equally in three key areas: better data to inform action, better legal and policy frameworks to advance rights, and better teaching and learning practices to empower.

Women's Day was created 114 years ago, but it is still needed. Universities and research centres hold a unique position in society that makes them critical actors in making changes. A survey that examines the performance of 776 universities across 18 indicators was carried out in May 2022: https://www.iesalc.unesco.org/wp-content/uploads/2022/03/SDG5_Gender_Report-2.pdf. The conclusions are that women outnumber men in university degrees, but they are biased towards Arts and Humanities. On top of this, universities do not usually follow up on their careers and integration in the professional or academic world. In this study, which included 231 European universities, the result indicated that 42% of all students in STEM subjects are female. In nuclear physics, there are over 4100 experimental nuclear physicists in Europe. According to a recent NuPECC study, the ratio of women is very different from country to country. For countries with more than a hundred experimental researchers, the ratio of women varies from 18% in the UK to 41% in Romania

## Communication

There exist many guides and reports on the inclusion of diversity in science, mostly concerned with gender balance issues as indicated above.

Most of these reports conclude on the relevance of communicating the following different issues of diversity on a regular basis:

1. Numbers resulting from monitoring
2. What diversity is, and which people constitute the diversity
3. Use of social networks for wide broadcast and simple messages

It is important to understand the mechanisms by which a favourable diversity distribution at the undergraduate level is lost during further career stages. For example, the gender gap in the workplace widens with maternity. So it is important to know where in Europe it is best to be a working mother: https://www.euronews.com/2023/05/08/where-in-europe-is-it-best-to-be-a-working-mother.

The figure 12.2 shows the monitoring done by NuPECC of experimental (light colours) and theoretical (dark colours) nuclear physicists in three different years: 1997, 2006 and 2023. The grouping of countries is done by the total number of nuclear physicists. The figure 12.3 shows the percentage of women in experimental nuclear physics in the European NuPECC member countries from the 2023 survey.

These reports also point out the importance of raising awareness about diversity and counteracting the unconscious biases that may arise at different stages of scientists' careers (recruitment, promotion, responsibility etc.). This kind of communication needs to be done on a regular basis to familiarise people with the issue.
It is also of the utmost importance that communication and proposed actions not only target minority groups but also the majority group. Responsibility must be shared: individuals in the majority group with more comfortable conditions of work may more easily dedicate time to the improvement of diversity and inclusion in their environment, rather than their minority colleagues who struggle for their own inclusion.

## Recommendations for Diversity and Inclusivity

**The nuclear physics community** supports **respectful, inclusive and safe work and training environments** in academic, industrial and vocational nuclear-science careers.

● We recommend that the network of research organisations, funding agencies, scientific collaborations and conference committees **sign up to and promote a diversity charter,** such as the one prepared by NuPECC together with APPEC and ECFA.
● The nuclear physics community and its stakeholders should further identify a body in Europe that takes charge of collating and providing an overview of the **monitoring of diversity** across nuclear science in Europe. This information should then underpin **recommendations and policies** adapted to the nuclear physics community.

## Careers - A competitive environment for nuclear talent

There are a wealth of career options and career development opportunities in the nuclear sciences including, but not limited to, academic careers. Early career researchers (ECR) provide the backbone of the different research areas of the NuPECC community. The proposed measures are targeted mainly towards early career researchers, including all career stages before a permanent position. All strategies are connected to the different activities listed in the topical sections: “Training” and “Diversity and Inclusivity”.

The nuclear physics workforce, as in other sectors of fundamental research, has been economically affected by the sustained increase in the cost of living for more than a decade. In some EU countries, material and financial stability in the research sector is close to impossible. The situation is particularly critical for the first steps of researchers at the PhD and postdoctoral levels, which often receive too low salaries.

On the one hand, the current job market, with a digital economy sector that thrives on the quantitative skills that physicists bring in, provides good competitive salaries, to the point that nuclear-aware talent may move on to industry positions. This may affect the retention of nuclear researchers, with losses for instance between PhD and postdoctoral positions. More importantly, the competition from the private sector may also have an impact on the entry point of the research workforce, with students not wanting to move into an exciting, but poorly-paid, PhD research career.

On the other hand, research developments and the quality of life of nuclear scientists can be severely impacted by financial precariousness. Financial precariousness most certainly also has an impact on both pre-and post-PhD, limiting the retention capacity and attractivity of EU research institutions. NuPECC recognises the importance of keeping competitive salaries in the European nuclear physics research sector, to attract and keep the best professional talents in nuclear science. We are well aware of the importance of a full career path for researchers. See the example of the progression of nuclear physics researchers at INFN and IN2P3, shown in the Box 12.4 on the situation for female physicists. The situation is very similar in France and Italy.

## Providing resources

One important project to support early career researchers tackles the resources that need to be provided. The NuPECC webpage [https://www.nupecc.org/] collects all useful information, from job opportunities to funding sources. The full community should be engaged in keeping this list as complete and updated as possible. Special emphasis should be put on relevant information for early career researchers. Dedicated sections for job advertisements, depending on country, career stage and job topic should be provided.

Presenting different career options in academia, industry, education or related fields is one important aspect in the support of early career researchers. This can be covered at dedicated sessions at NuPECC-sponsored conferences like, for example, the INPC, which is usually attended by several early career researchers from different fields. These sessions should include presentations on different topics, such as an overview of possibilities for a career in science as well as presentations from professionals from the industry.

## Support for early career researchers

In recent years, NuPECC has participated in the preparation of a document within the JENAA (Joint ECFA-NuPECC-APPEC) activities

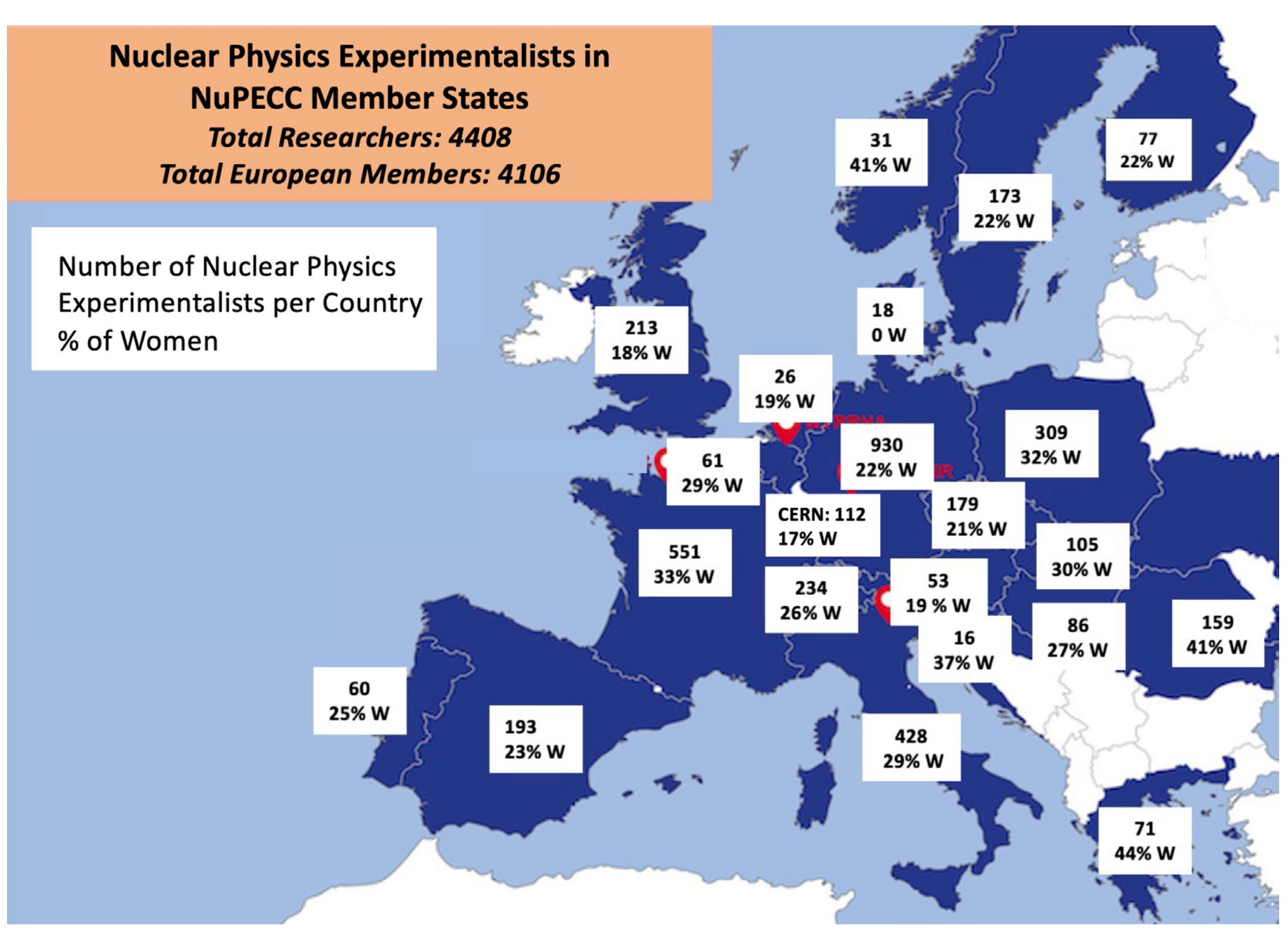

*Fig. 12.3: The map shows the community of experimental researchers within NuPECC. For each Country the total number and the ratio of women are given. The number of women within the experimental physics community is 26%, lower than the average number of female students in physics (over 30%).*

## Box 12.4: Evolution of female researchers in different European Institutions

**Percentage of women in nuclear physics at IN2P3 and INFN for two categories: first level of permanent researcher (in red colours) and the higher level "directors" (blue colours) category. if three categories the numbers corresponding to the two higher ones have been added. Notice the similarity of the numbers and its evolution with time for the two Institutions. It also shows a stabilisation of the ratio in INFN for "directors" while IN2P3 keeps a positive slope.**

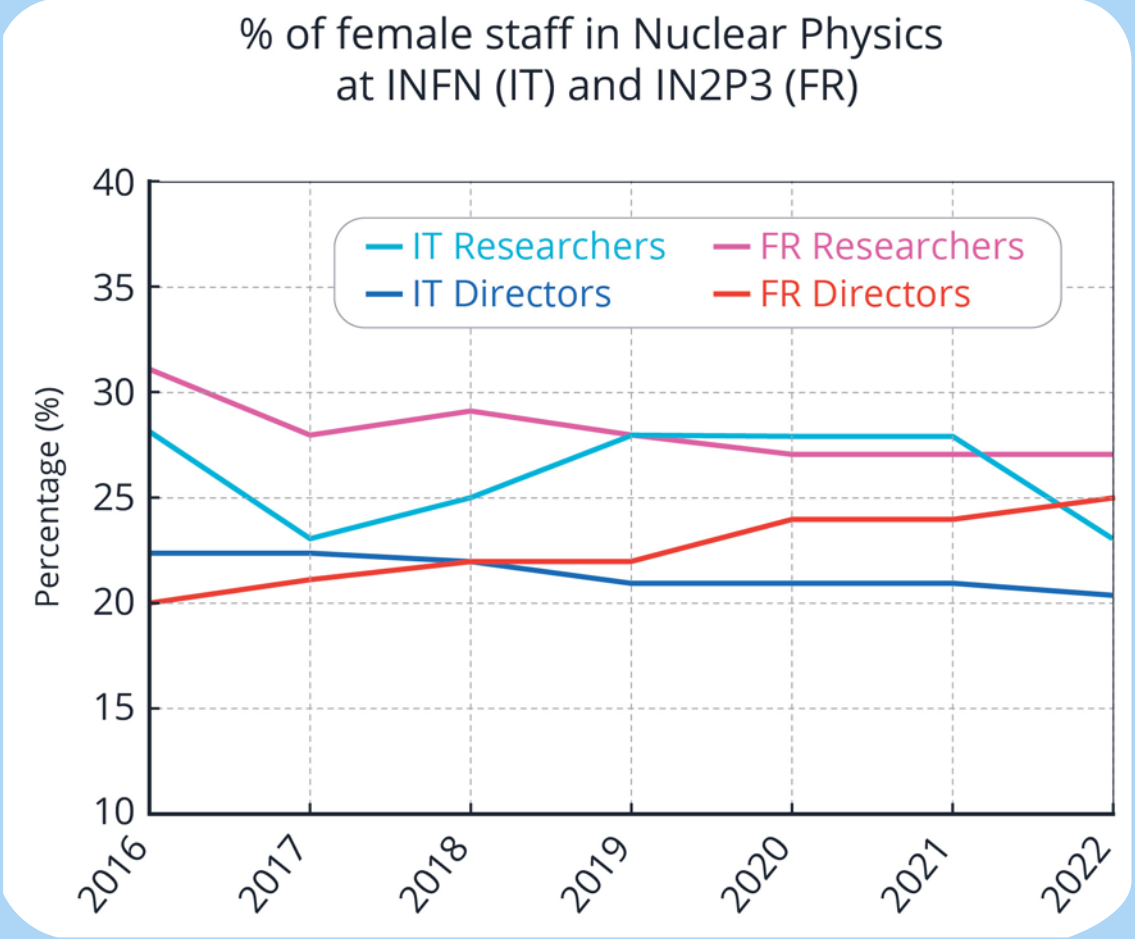

**Another way to monitor the increase of female researchers at the higher level is by the so-called glass-ceiling index that monitor the gender-bias difficulties to reach the highest positions.**
**The Spanish National Research Council, CSIC, has a commission formed to supervise gender and the variation with the career path. They have monitored the effects of the so-called glass-ceiling index, meaning the gender-bias difficulties to reach the highest positions.**
**The figure on the left shows the variation of the glass-ceiling index for female physicists at CSIC from 2001-2021. The glass-ceiling index is given in this case by the percentage of staff researcher female versus the one of female full professors.**
**Notice the slow progress in the last decade. Here a value of 1.0 indicates full parity.**

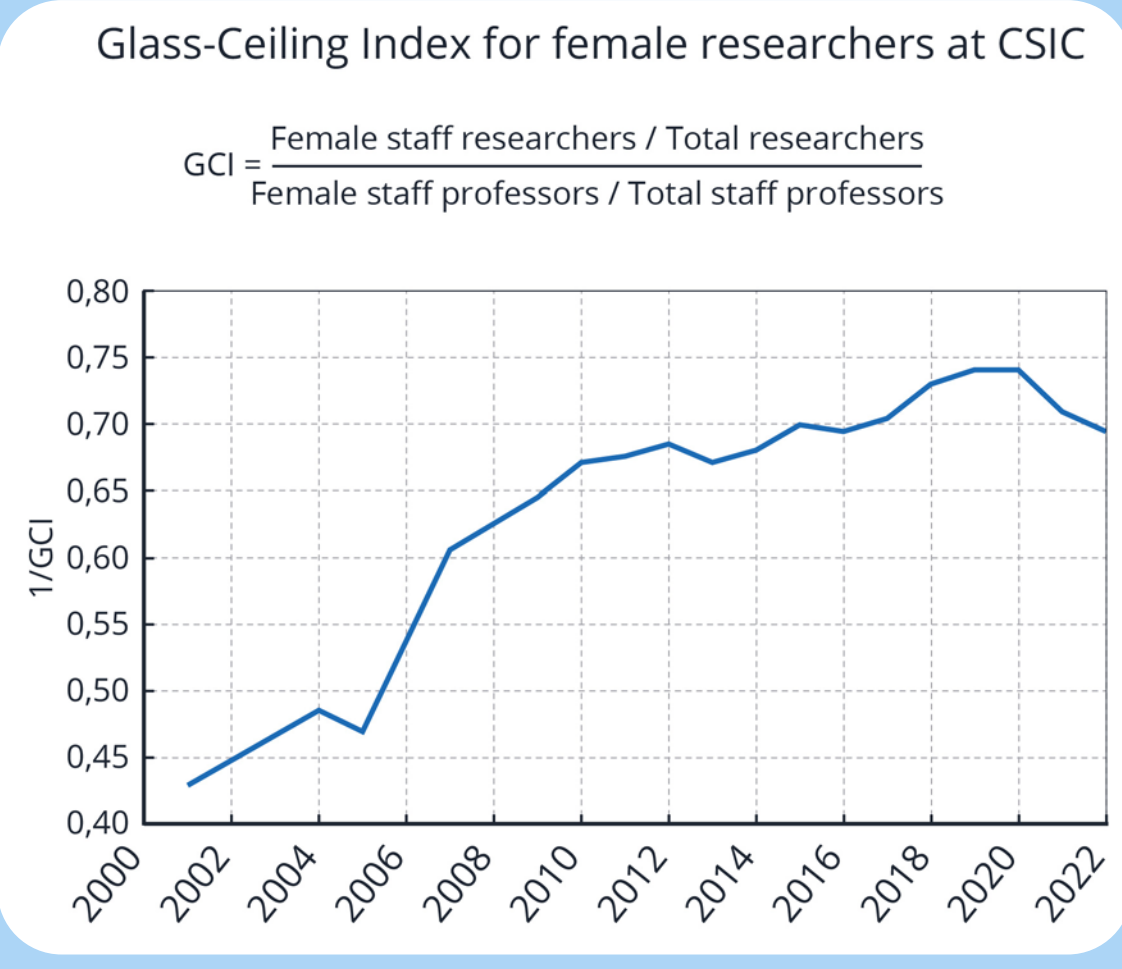

which addresses the difficulties of recognition of individuals in large collaborations. Several collaborations from different fields contributed to this effort.

The strong contribution of Early Career Researchers (ECRs) to all aspects of Nuclear Science is of crucial importance for the field. Yet ECRs still struggle to receive appropriate acknowledgement of their work, support for their careers and support for their scientific work. It is an important goal to ensure that these barriers are further reduced across the board. This will require both local measures and a firm commitment by funders, employers and collaborations to ensure that colleagues in early career stages are nurtured and fully supported for the benefit of all.

These measures include, for example, increased visibility for ECRs at national and international conferences, dedicated prizes and networking opportunities like the APS early career forum (https://engage.aps.org/fecs/home).

Existing First Grant schemes should be fully supported. Networking on a global scale is crucial, and initiatives like the APS-EPS-ICTP Travel Award Fellowship Programme (ATAP), recently established between the American Physical Society (APS), the European Physical Society (EPS) and the International Centre for Theoretical Physics to provide career-enhancing opportunities for active early career scientists from developing countries are very welcome.

In addition, mentoring for young researchers is of the utmost importance. Mentoring should be a widespread and widely valued activity, of great benefit to all.

## Recommendations for Careers

We recommend that **equitable and inclusive career development** be further prioritised by stakeholders across the European nuclear physics community, giving **recognition and visibility to critical contributions from early career researchers** as the future of nuclear physics and its impact on society. This includes efforts to:

● **Support tenure track programmes** giving highly qualified early career researchers the opportunity to lead their own group and establish scientific independence (e.g. ERC Starting Grants);

● **Provide opportunities** for participation in national and transnational **training workshops** for early career researchers, opportunities to **contribute to prestigious committees** and support attendance at **highly visible conferences, including through further early career awards**, as well as to

● Establish career Offices for early career researchers at universities, institutes and research infrastructures to **offer support and training in career development.**

Aside from these suggestions, it would be useful for future planning to distribute a survey on the number of early career researchers in the different sections and their situation.

A solid database covering all participating countries would be helpful for future planning. Topics of interest could, for example, include information about employment status, planned career path and the diversity status of early career researchers. The Fig. 12.4 shows the percentage of PhD students, post-docs and staff within NuPECC members as obtained in the recent surveys of the Nuclear Physics Community in Europe.
In general, and without precise numbers, only one-third of our PhD students who become doctors continue in Academia.

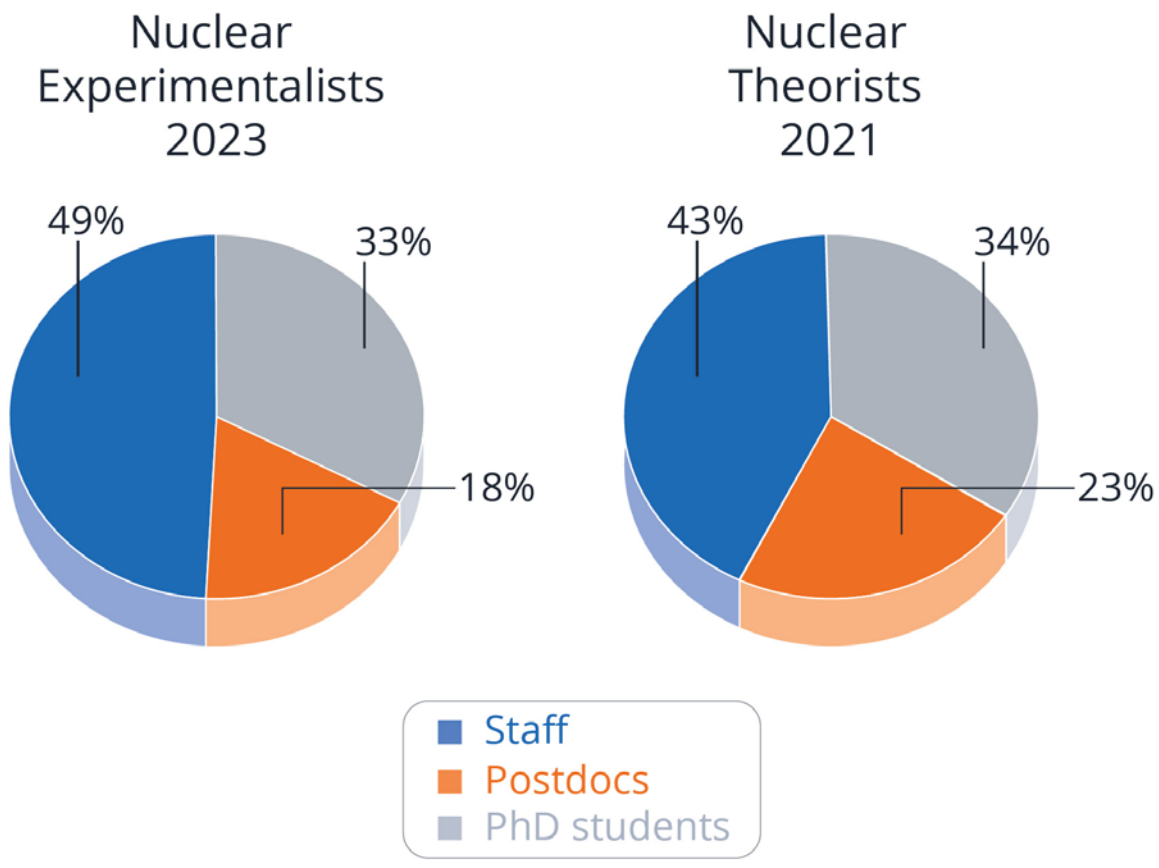

***Fig. 12.4: The pie charts show the distribution of Staff (in blue), Post-docs (in orange) and PhD students (in grey) of the European nuclear physics experimental (2023) and theoretical (2021) communities. On average, we have found that one-third of PhD students continue into Academia.***

## Conclusion

**The prospect is great for nuclear science outreach to both reach and inspire the next generation in our societies across Europe. We, members of the nuclear physics community, together with funders and professional organisations, should invest in this and share experience and resources in an equitable and inclusive way. If, at the same time, the nuclear sciences and their wide-ranging applications and societal impacts are integrated into our educational system, the next generation of nuclear physicists can be nurtured and receive the specialist training they need to excel in their field and advance their careers. With a dedicated effort to ensure that outreach, education, training, and career opportunities are all developed as inclusive and with a focus on the power of enhanced diversity in our field, people and society will benefit hugely from the opportunities offered by nuclear science.**

# List of Acronyms and Abbreviations

| | |
|---|---|
| ACTAR | active target |
| AD | antiproton decelerator |
| ADS | accelerator driven system |
| AGATA | Advanced GAmma Tracking Array |
| AGB | asymptotic giant branch |
| AI | artificial intelligence |
| ALICE | A Large Ion Collider Experiment |
| ALP | axion-like particle |
| ALTO | Accélérateur Linéaire – Tandem Orsay |
| AMADEUS | Experiment at LNF |
| AMBER | Apparatus for Meson and Baryon Experimental Research |
| AMS | accelerator mass spectroscopy |
| ANC | asymptotic normalization coefficient |
| APPA | Atomic, Plasma Physics and Applications |
| APPEC | Astroparticle Physics European Consortium |
| ASIC | application-specific integrated circuit |
| ATLAS | particle physics experiment at LHC |
| BaBar | Experiment at SLAC |
| BAU | baryon asymmetry of the universe |
| BBN | big-bang nucleosynthesis |
| BEC | Bose-Einstein condensate |
| BELLE | experiment at the KEK B-factory |
| BES | Beijing Spectrometer |
| BH | black hole |
| BMBF | German Ministry of Education and Research |
| BMW | Budapest-Marseille-Wuppertal Collaboration |
| BNCT | boron neutron capture therapy |
| BNL | Brookhaven National Laboratory |
| BSM | beyond standard model |
| CBM | Condensed Baryonic Matter |
| CBRNE | chemical, biological, radiological, nuclear, explosives |
| CCB | Cyclotron Center Bronowice |
| CCSN | core collapse supernova |
| CD | central detector |
| CEA | Commissariat à l'Energie Atomique et aux Energies Alternatives |
| CEBAF | Continuous Electron Beam Accelerator Facility |
| CEMP | carbon-enhanced metal-poor |
| CERIC | Central European Research Infrastructure Consortium |
| CERN | Conseil Européen de Recherche Nucléaire |
| CGC | colour glass condensate |
| ChEFT | chiral effective field theory |
| ChETEC | Chemical Elements as Tracers of the Evolution of the Cosmos |
| ChPT | chiral perturbation theory |
| CKM | Cabbibo-Kobayashi-Maskawa |
| CLAS | CEBAF Large Acceptance Spectrometer |
| cLFV | charged lepton flavour violation |
| c.m. | center-of-mass |
| CMAM | Centro de Microanálisis de Materiales |
| CMB | cosmic microwave background |
| CME | coronal mass ejection |
| CMOS | Complementary metal-oxide-semiconductor |
| CMS | particle physics experiment at LHC |
| CN | classical nova |
| CNA | Centro Nacional de Aceleradores |
| CNRS | Centre National de la Recherche Scientifique |
| CODATA | Committee on Data for Science and Technology |
| COMPASS | Common Muon Proton Apparatus for Structure and Spectroscopy |
| COSI | Compton Spectrometer and Imager |
| COSY | COoler Synchrotron |
| CP | charge conjugation and parity |
| CPU | Central processing unit |
| CRYRING@FAIR | Low-energy storage ring for heavy ions |
| CT | computered tomography |

| CUNA | Canfranc Underground Nuclear Astrophyics |
|---|---|
| CVC | conserved vector current |
| CW | continuous wave |
| DAQ | data acquisition |
| DAΦNE | Double Annular Φ Factory for Nice Experiments |
| DESIR | Decay, Excitation and Storage of Radioactive Ions |
| DESY | Deutsches Elektronensynchrotron |
| DFT | density functional theory |
| DIS | deep-inelastic scattering |
| DLC | data life cycle |
| DM | dark matter |
| DMP | data management plan |
| DUNE | Deep Underground Neutrino Experiment |
| DVCS | deeply virtual Compton scattering |
| DVMP | deep virtual meson production |
| DY | Drell-Yan |
| EAF | European Activation File |
| EBIT | electron beam ion trap |
| EBRT | external beam radiation therapy |
| EC | electron capture |
| ECFA | European Committee for Future Accelerators |
| ECR | early career researchers |
| ECT* | European Centre for Theoretical studies in nuclear physics and related areas |
| EDF | energy density functional |
| EDM | electric dipole moment |
| EFT | effective field theory |
| EIC | electron ion collider |
| ELENA | extra low energy antiproton ring |
| ELI-NP | Extreme Light Infrastructure – Nuclear Physics |
| ELSA | Electron Stretcher and Accelerator |
| EMC | European Muon Collaboration |
| EMIS | electromagnetic isotope separator |
| EoS | equation of state |
| ePIC | electron / Proton-Ion Collider |
| EPR | electron paramagnetic resonance |
| EPS | European Physical Society |
| ERA | European Research Area |
| ERC | European Research Council |
| ERL | energy-recovery linac |
| ESA | European Space Agency |
| ESFRI | European Strategic Forum for Research Infrastructures |
| ESR | experimental storage ring |
| ESRF | European Synchrotron Radiation Facility |
| ESS | European Spallation Source |
| EURADOS | EUropean RAdiation DOSimetry group |
| EURISOL | EURopean ISOL facility |
| EURO-LABS | EUROpean Laboratories for Accelerator Based Science |
| EXL | Exotic nuclei studied with Electromagnetic and Light hadronic probes |
| EXOGAM | gamma detector at GANIL |
| FAIR | Facility for Antiproton and Ion Research |
| FAZIA | 4π A and Z Identification Array |
| FCC | Future Circular Collider |
| FENDL | Fusion Evaluated Nuclear Data Library |
| FF | form factor |
| FIB | feebly interacting particle |
| FLAIR | Facility for Low-energy Antiproton and Ion Research |
| FNAL | Fermi National Accelerator Laboratory |
| FOPI | 4π detector at GSI |
| FP7 | EU Framework Programme 7 |
| FPGA | Field programmable gate array |
| FRM-II | Forschungsreaktor München II |
| FRS | FRagment Separator |
| GANIL | Grand Accélérateur National d’Ions Lourds |
| GCR | galactic cosmic rays |
| GDR | giant dipole resonance |
| GEM | gas electron multiplier |
| GENCI | Grand Equipment National de Calcul Intensif |
| GlueX | Experiment at JLab |
| GOTO | Gravitational-wave Optical Transient Observer |
| GPD | generalised parton distributions |
| GPU | Graphics processing unit |
| GRB | gamma ray burst |
| GRS | gamma ray spectroscopy |
| GSI | Gesellschaft für Schwerionenforschung |
| GSR | generalised special relativity |
| GW | gravitational wave |
| HADES | High Acceptance Di-Electron Spectrometer |
| HBB | hot bottom burning |
| HBT | Hanbury Brown-Twiss (interferometry analysis) |
| HCI | highly charged ions |
| HD | hyperdeformed |
| HELIAC | HElmholtz LInear Accelerator |
| HERA | Hadron-Elektron-Ring-Anlage |
| HERMES | HERA experiment for spin physics |
| HESR | high energy storage ring |
| HFS | hyperfine splitting |
| HIE-ISOLDE | high intensity and energy upgrade of ISOLDE |
| HIGS | High Intensity Gamma-Ray Source |
| HIL | Heavy Ion Laboratory |
| HIMB | High-Intensity Muon Beams |
| HITRAP | heavy ion trap |
| HLbL | hadronic light-by-light |
| HPC | high performance computing |
| HPGe | high purity germanium |
| HQET | heavy quark effective theory |
| IA | integrated activity |
| IAEA | International Atomic Energy Agency |
| IAS | isobaric analogue state |
| IBA | ion beam analysis |
| IBER | Investigation on Biological Effects of Radiation |
| ICP | inductively coupled plasma |
| IFIN-HH | Horia Hulubei National Institute of Physics and Nuclear Engineering |
| IFJ-PAN | Institute of Nuclear Physics Polish Academy of Sciences |
| IFMIF | International Fusion Materials Irradiation Facility |
| IFMIF-DONES | IFMIF - DEMO-Oriented Neutron Source |
| IGISOL | ion guide isotope separation on-line |
| ILDG | International Lattice Data Grid |
| ILL | Institut von Laue – Langevin |
| IMR | intermediate mass region |
| IN2P3 | Institut National de Physique Nucléaire et de Physique des Particules |
| INDRA | 4π charged product detection array at GANIL |
| INFN | Istituto Nazionale di Fisica Nucleare |
| INSPYRE | Int. School on modern PhYsics and Research |
| INT | Institute for Nuclear Theory |
| IPPOG | International Particle Physics Outreach Group |
| ISGDR | isoscalar giant dipole resonance |
| ISGMR | isoscalar giant monopole resonance |
| ISGQR | Isoscalar Giant Quadrupole Resonance |
| ISOL | isotope separation on-line |
| ISOL@MYRRHA | isotope separation on-line at MYRRHA |
| ISOLDE | Isotope Mass Separator Online at CERN |
| ITER | International Thermonuclear Experimental Reactor |
| IUPAP | International Union of Pure and Applied Physics |
| IVGDR | Isovector Giant Dipole Resonance |
| JEDI | Jülich Electric Dipole moment Investigations |
| JENAA | Joint ECFA-NuPECC-APPEC Activities |
| JINA | Joint Institute for Nuclear Astrophysics |
| JINR | Joint Institute for Nuclear Research |
| JLab | Thomas Jefferson National Accelerator Facility |
| J-PARC | Japan Proton Accelerator Research Complex |
| JRA | joint research activity |
| JSC | Jülich Supercomputing Centre |
| JWST | James Webb Space Telescope |
| JYFL-ACCLAB | University of Jyväskylä Accelerator Laboratory |
| KAGRA | gravitational wave observatory at Kamioka |
| KATRIN | Karlsruhe Tritium Neutrino Experiment |
| KLOE | Experiment at LNF |
| KOTO | Experiment at JPARC |
| KVI | Kernfysisch Versneller Instituut |
| LABEC | Laboratorio di Tecniche Nucleari per I Beni Culturali |
| LAND | Large Area Neutron Detector |
| LBE | lead-bismuth eutectic |
| LBNL | Lawrence Berkeley National Laboratory |
| LCB | laser Compton backscattering |
| LDG | lattice data grid |
| LEAR | low energy antiproton ring |
| LECs | low-energy constants |
| LET | linear energy transfer |
| LFV | lepton-flavour violation |
| LGAD | low-gain avalanche detector |
| LHC | Large Hadron Collider |
| LHCb | detector at LHC |
| LIGO | Laser Interferometer Gravitational-Wave Observatory |
| LINAC | linear accelerator |
| LINAG | liner accelerator at GANIL |
| LMR | low mass region |
| LNF | Laboratori Nazionali di Frascati |
| LNGS | Laboratory Nazionali di Gran Sasso |
| LNL | Laboratori Nazionali di Legnaro |
| LNS | Laboratori Nazionali del Sud |
| LS3 | Long Shutdown 3 (at CERN) |
| LUNA | Laboratory Underground for Nuclear Astrophysics |
| MA | Minor Actinides |
| MAGIX | Collaboration at MESA |
| MAMI | Mainz Mikrotron |
| MARA | Mass Analysing Recoil Apparatus |
| MAYA | gas-filled active target detection system at GANIL |
| MC | Monte Carlo |
| MEDICIS | MEDical Isotopes Collected from ISOLDE |
| MEGAPIE | Mega Ampere Generator for Plasma Implosion Experiments |
| MEIS | medium-energy ion scattering |
| MELODI | Multidisciplinary European Low Dose Initiative |
| MESA | Mainz Energy-Recovery Superconducting Accelerator |
| ML | Machine learning |
| MORA | Matter's Origin from the RadioActivity of trapped and oriented ions |
| MPGD | micro-pattern gas detectors |
| MRI | magnetic resonance imaging |
| MRT | magnetic resonance tomography |
| MR-ToF-MS | Multi-Reflection Time-of-Flight mass spectrometers |
| MSU | Michigan State University |
| MUSE | MUon proton Scattering Experiment at PSI |
| MWPC | multi-wire proportional counter |
| MYRRHA | multi-purpose research reactor for high-tech applications |
| n_TOF | neutron time-of-flight facility |
| NA | nuclear astrophysics |
| NA(60, 61, 64) | North Area, CERN, Experiment 60, 61, 64 |
| NASA | National Aeronautics and Space Administration |
| NAT | nuclear analytical techniques |
| NEDA | Neutron Detector Array |
| NFS | neutrons for science |
| NICA | Nuclotron-based Ion Collider fAcility |
| NICER | NASA's Neutron star Interior ExploreR |
| NLC | National Laboratory of Cyclotrons (SLCJ and IFJ-PAN consortium) |
| NLC | next linear collider |
| NLD | nuclear level density |
| NLDBD | neutrino-less double beta decay |
| NLO | next-to-leading order |
| NME | nuclear matrix element |
| NN | nucleus-nucleus |
| NRF | nuclear resonance fluorescence |
| NS-BH | neutron star – black hole |
| NSCL | National Superconducting Cyclotron Laboratory |
| NSM | nuclear shell model |
| NS-NS | double neutron star |
| NuSTAR | Nuclear Structure, Astrophysics and Reactions |
| NY | nucleus-hyperon |
| OLYMPUS | Experiment at DESY |
| PAC | perturbed angular correlation |
| PADME | Positron Annihilation into Dark Matter Experiment at LNF |
| PANDA | antiProton ANnihilation at DArmstadt |
| PANS | Public Awareness of Nuclear Science |
| PARIS | Photon Array for studies with Radioactive Ion and Stable beams |
| PDF | parton distribution function |
| PDR | pygmy dipole resonance |
| PEP | Pauli exclusion principle |
| PERC | proton and electron radiation channel |
| PES | potential energy surface |
| PET | positron emission tomography |
| PETRA | Positron-Elektron-Tandem-Ring-Anlage |
| PGCM | projected generator coordinate method |
| PIGE | proton induced gamma emission |
| PIXE | proton induced x-ray emission |
| PM | particulate matter |
| PMNS | Pontecorvo-Maki-Nakagawa-Sakata |
| PRACE | Partnership of Advanced Computing in Europe |
| PRad | Experiment at JLab |
| PRES | Experiment at MAMI |
| PRISMAP | European medical radionuclides programme |
| PSI | Paul Scherrer Institut |
| PSP | permutation-symmetry postulate |
| PUMA | antiProton Unstable Matter Annihilation |
| QC | Quantum computing |
| QCD | quantum chromodynamics |
| QED | quantum electrodynamics |
| QFT | quantum field theory |
| QGP | quark gluon plasma |
| QGR | Giant Quadrupole Resonance |
| QPM | quasiparticle-phonon model |
| QPVC | quasiparticle-vibration coupling |
| R3B | Reactions with Relativistic Radioactive Beams |
| RBE | relative biological effectiveness |
| RBS | Rutherford back scattering |
| REC | radiative electron capture |
| RGM | resonating group method |
| RHIC | Relativistic Heavy Ion Collider |
| RI | research infrastructure |
| RIB | radioactive ion beam |
| RIBF | Radioactive Ion Beam Facilitiy |
| RICH | ring imaging Cherenkov counter |
| RILIS | resonance ionization laser ion source |
| RPA | random phase approximation |
| RSP | relative stopping power |
| RTG | Radioisotope Thermoelectric Generator |
| SAMIRA | strategic agenda for medical ionizing radiation applications |
| SBBN | standard big bang nucleosynthesis |
| SD | superdeformed |
| S-DALINAC | Supraleitender Darmstädter Elektronenlinearbeschleuniger |
| SDG | Sustainable Development Goal |
| SEE | single-event effect |
| SF | spectroscopic factor |
| SFRT | spatially fractionated radiotherapy |
| SHE | superheavy elements |
| SHI | swift heavy ions |
| SHM | statistical hadronisation model |
| SHN | superheavy nuclei |
| SIB | stable ion beam |
| SIDDHARTA | Silicon Drift Detector for Hadronic Atoms Research by Timing Application |
| SIDIS | semi-inclusive deep-inelastic scattering |
| SIMS | secondary-ion mass spectrometry |
| SIS | SchwerIonen Synchrotron |
| SLCJ | Heavy Ion laboratory, University of Warsaw |
| SM | standard model / shell model |
| SME | standard model extension |
| SMEFT | Standard Model Effective Field Theory |
| SMF | stochastic mean field |
| SMMC | shell-model Monte Carlo |
| SMOG | System for Measuring Overlap with Gas |
| SMR | small modular reactor |
| SN | supernova |
| SNETP | European Sustainable Nuclear Energy Technology Platform |
| SPECT | single-photon emission computerized tomography |
| SPES | Selective Production of Exotic Species |
| sPHENIX | Experiment at RHIC |
| SPIRAL | Système de Production d'Ions Radioactifs Accélérés en Ligne |
| SPS | Super Proton Synchrotron |
| SRA | Strategic Research Agency |
| SRC | short range correlation |
| SRG | similarity renormalization group |
| SSF | small scale facilities |
| SST | spin-statistics theorem |
| STAR | Experiment at RHIC |
| STEM | science, technology, engineering, mathematics |
| STRONG2020 | The strong interaction: fundamental research and applications |
| SuperFRS | Super FragmentSeparator |
| SUSY | supersymmetry |
| T2K | Tokai to Kamioka |
| TALENT | Training in Advanced Low-Energy Nuclear Theory |
| TAS | total absorption spectroscopy |
| TASCA | TransActinide Separator and Chemistry Apparatus |
| TAT | targeted alpha therapy |
| TATTOOS@PSI | Targeted Alpha Tumour Therapy and Other Oncological Solutions |
| TD | time-dependent |
| TGCC | Très Grand Centre de calcul du CEA |
| THM | Trojan horse method |
| TJNAF | Thomas Jefferson National Accelerator Facility |
| TMD | transverse momentum dependent distribution function |
| TMEP | Transport Model Evaluation Project |
| TNA | transnational access |
| TOF | time of flight |
| TPC | time projection chamber |
| TRT | targeted radionuclide therapy |
| TUNL | Triangle Universities Nuclear Laboratory |
| UNESCO | United Nations Educational, Scientific and Cultural Organization |
| UNILAC | Universal Linear Accelerator (at GSI) |
| UT | upstream tracker |
| VAMOS | VAriable MOde Spectrometer |
| VCS | virtual Compton scattering |
| WASA | Wide Angle Shower Apparatus |
| WD | white dwarf |
| WIMP | weakly interacting massive particle |
| XRB | X-ray bursts |
| YY | hyperon-hyperon |
| γ-SF | gamma strength function |
| μSR | muon spin resonance / rotation |

# Contributors LRP2024

Gert Aarts (Swansea), Hamid Aït Abderrahim (Mol), Dieter Ackermann (Caen), Jörg Aichelin (Nantes), Navin Alahari (Caen), Constantia Alexandrou (Nicosia), Alejandro Algora (Valencia), Mohammad Al-Turany (Darmstadt), Hector Alvarez-Pol (Santiago de Compostela), Luis Alvarez-Ruso (Valencia), Pietro Antonioli (Bologna), Roberta Arnaldi (Torino), Marlène Assié (Orsay), Michail Athanasakis-Kaklamanakis (Leuven), Sonia Bacca (Mainz), Claude Bailat (Lausanne), Umberto Battino (Keele), Andreas Bauswein (Darmstadt), Saul Becerio Novo (La Coruña), Daniel Bemmerer (Dresden), Valerio Bertone (Saclay), Diego Bettoni (Ferrara), Sayani Biswas (Villigen), Axel Boeltzig (Dresden), Iva Bogdanović Radović (Zagreb), Maria J. G. Borge (Madrid), Sebastian Baunack (Mainz), Sonja Bernitt (Darmstadt), Stefano Bianco (Frascati), Matteo Biassoni (Milano), Bertram Blank (Bordeaux), Michael Block (Mainz), Andrew Boston (Liverpool), Angela Bracco (Milano), Carlo Bruno (Edinburgh), Franco Camera (Milano), Daniel Cano Ott (Madrid), Giovanni Casini (Firenze), Francesca Cavanna (Torino), Michal Ciemala (Kraków), Cristina Chiappini (Potsdam), Thomas Elias Cocolios (Leuven), Michele Coeck (Mol), Gilberto Colangelo (Bern), Sara Collins (Regensburg), Seán Collins (Teddington), Gianluca Colò (Milano), Luigi Coraggio (Napoli), Sandrine Courtin (Strasbourg), Paolo Crivelli (Zürich), Letitia Cunqueiro-Mendez (Roma), Silvia Dalla Torre (Trieste), Andrea Dainese (Padova), Annalisa D'Angelo (Rome), Luigi del Debbio (Edinburgh), Ruben de Groote (Leuven), Pierre Delahaye (Caen), Achim Denig (Mainz), François de Oliveira (Caen), Rosanna Depalo (Milano), Cesar Domingo (Valencia), Nicolas de Séréville (Orsay), Vivian Dimitriou (Vienna), Alessia Di Pietro (Catania), Timo Dickel (Darmstadt), Christian Diget (York), Jacek Dobaczewski (York), Gail Dodge (Norfolk), Tommaso Dorigo (Padova), Charlotte Duchemin (Geneva), Marco Durante (Darmstadt), Giles Edwards (Manchester), Gernot Eichmann (Graz), Evgeny Epelbaum (Bochum), Xavier Espinal (Geneva), Laura Fabbietti (München), Muriel Fallot (Nantes), Alessandra Fantoni (Frascati), Fanny Farget (Caen), Jenny Feige (Berlin), Paolo Finocchiaro (Catania), Christian Fischer (Gießen), Stefan Flörchinger (Jena), Bogdan Fornal (Kraków), Sean Freeman (Manchester), Anne-Marie Frelin (Caen), Zsolt Fülöp (Debrecen), Hans Fynbø (Aarhus), Liam Gaffney (Liverpool), Tetyana Galatyuk (Darmstadt), Piotr Gasik (Darmstadt), Paola Gianotti (Frascati), Kathrin Göbel (Darmstadt), Robin Golser (Wien), Joaquin Gomez Camacho (Sevilla), Paolo Giubellino (Darmstadt), Martín González Alonso (Valencia), Stéphane Goriely (Brussels), Andrea Gottardo (Leggnaro), Jeremy R. Green (Zeuthen), Paul Greenlees (Jyväskylä), Michele Grossi (Geneva), Jana Günther (Wuppertal), Francesca Gulminelli (Caen), Frank Gunsing (Saclay), Philipp Hauke (Trento), Andreas Haungs (Karlsruhe), Marcel Heine (Strasbourg), Fritz-Herbert Heinsius (Bochum), Andreas Heinz (Göteborg), Rolf-Dietmar Herzberg (Liverpool), Morten Hjorth-Jensen (Oslo), Byungsik Hong (Daejeon), Guillaume Hupin (Orsay), Angel Ibarra (Madrid), Andreas Ipp (Wien), Dave Ireland (Glasgow), Karl Jakobs (Freiburg), Tobias Jenke (Grenoble), Ari Jokinen (Jyväskylä), Jordi José (Barcelona), Michel Jouvin (Orsay), Arnd Junghans (Dresden), Beatriz Jurado (Bordeaux), Ihor Kadenko (Kiev), Alexander Philipp Kalweit (Geneva), Anu Kankainen (Jyväskylä), Zsolt Kasztovszky (Budapest), Bernhard Ketzer (Bonn), Alfons Khoukaz (Münster), Klaus Kirch (Villigen), Gabor Kiss (Debrecen), Christian Klein-Boesing (Münster), Andreas Knecht (Villigen), Ulli Köster (Grenoble), Agnieska Korgul (Warsaw), Wolfram Korten (Saclay), Michal Kowal (Warsaw), Magdalena Kowalska (Geneva), Marc Labiche (Daresbury), Denis Lacroix (Orsay), Tuomas Lappi (Jyväskylä), Ann-Cecilie Larsen (Oslo), Barbara Maria Latacz (Geneva), Yen-Jie Lee (MIT), Yvonne Leifels (Darmstadt), Antoine Lemasson (Caen), Silvia Leoni (Milano), Marek Lewitowicz (Caen), Razvan Lica (Bucharest), Armandina Lima Lopes, (Porto), Yuri A. Litvinov (Darmstadt), Maria Lugaro (Budapest), Frank Maas (Mainz), Bastian Märkisch (München), Adam Maj (Kraków), Giulia Manca (Cagliari), Vladimir Manea (Orsay), Ana Maria Marin Garcia (Darmstadt), Jerôme Margueron (Lyon), Silvia Masciocchi (Heidelberg), Jan Matousek (Prague), Adrien Matta (Caen), Ulf Meißner (Bonn), Javier Menendez (Barcelona), Daniele Mengoni (Padova), Johan Messchendorp (Darmstadt), Caterina Michelagnoli (Grenoble), Renata Mikolajczak (Świerk), Alexander Milov (Weizmann), Andrew Mistry (Darmstadt), Hervé Moutarde (Saclay), Carlos Munoz Camacho (Orsay), Luciano Musa (Geneva), Enrique Nacher (Madrid), Eugenio Nappi (Bari), Silvia Niccolai (Orsay), Tamara Niksic (Zagreb), Uwe Oberlack (Mainz), Natalia Oreshkina (Heidelberg), Janne Pakarinen (Jyväskylä), Katia Parodi (München), Sorin Pascu (Bucharest), Vincenzo Patera (Roma), Nikolas Patronis (Ioannina), Nancy Paul (Paris), Ekkehard Peik (Braunschweig), Thomas Peitzmann (Utrecht), Lino Miguel Pereira Da Costa, (Leuven), Elena Perez del Rio (Kraków), Stefano Piano (Geneva), Silvia Piantelli (Firenze), Guillaume Pignol (Grenoble), Rosario Gianluca Pizzone (Catania), Arjan Plompen (Geel), Zsolt Podolyak (Surrey), Tina Pollmann (Amsterdam), Lucia Popescu (Mol), Veronique Puill (Orsay), Catarina Quintans (Lisbon), Riccardo Raabe (Leuven), Marco Radici (Pavia), Panu Rahkila (Jyväskylä), Gerhard Reicherz (Bochum), Arnau Rios (Barcelona), Marco Ripani (Genova), Tomas Rodriguez (Madrid), Manuela Rodriguez-Gallardo (Sevilla), Thomas Roger (Caen), Alessandro Roggero (Trento), Patricia Roussell-Chomaz (Caen), Berta Rubio (Valencia), Hiroyoshi Sakurai (Wako), Piotr Salabura (Kraków), Carlos Salgado (Santiago de Compostela), Magnus Schlösser (Karlsruhe), Konrad Schmidt (Dresden), Karin Schönning (Uppsala), Maria Dorothea Schumann (Villigen), Paul Schuurmans (Hasselt), Jochen Schwiening (Darmstadt), Nathal Severijns (Leuven), Haik Simon (Darmstadt), Monica Sisti (Milano), Raimond Snellings (Utrecht), Olga Soloveva (Frankfurt), Vittorio Soma (Saclay), Olivier Sorlin (Caen), Paul D. Stevenson (Surrey), Olivier Stezowski (Lyon), Thomas Stöhlker (Jena), Joachim Stroth (Frankfurt), Sven Sturm (Heidelberg), Zeynep Talip (Villigen), Annika Thiel (Bonn), Eugenia Toimil-Molares (Darmstadt), Livius Trache (Bucharest), Martino Trassinelli (Paris), Stefan Ulmer (Düsseldorf), José Javier Valiente Dobón (Legnaro), Marine Vandebrouck (Saclay), Charlot Vandevoorde (Darmstadt), Marc Vanderhaeghen (Mainz), Ana Vaniqui (Mol), Ubirajara van Kolck (Trento), Martin Venhart (Bratislava), Jelena Vesić (Ljubljana), Enrico Vigezzi (Milano), Vladimir Wagner (Řež), Clemens Walther (Hannover), Liangxiao Wang (Frankfurt), Carl Wheldon (Birmingham), Urs Wiedemann (Geneva), Eberhard Widmann (Wien), Jonathan Wilson (Orsay), Kathrin Wimmer (Darmstadt), Andreas Zilges (Köln)

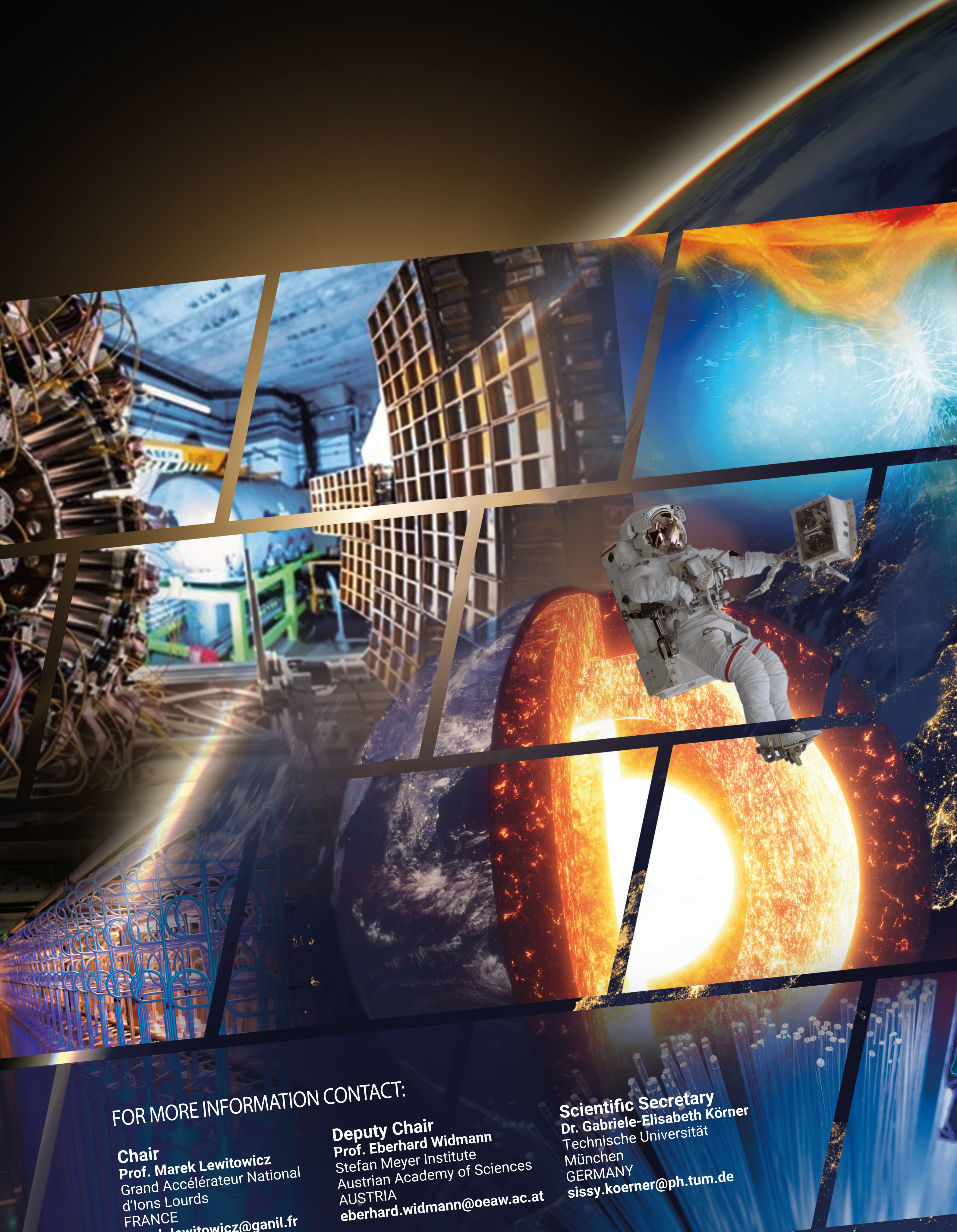

FOR MORE INFORMATION CONTACT:

**Chair**
**Prof. Marek Lewitowicz**
Grand Accélérateur National
d'Ions Lourds
FRANCE
**marek.lewitowicz@ganil.fr**

**Deputy Chair**
**Prof. Eberhard Widmann**
Stefan Meyer Institute
Austrian Academy of Sciences
AUSTRIA
**eberhard.widmann@oeaw.ac.at**

**Scientific Secretary**
**Dr. Gabriele-Elisabeth Körner**
Technische Universität
München
GERMANY
**sissy.koerner@ph.tum.de**

ESF
Your Partner
in Science